\documentclass{kapmono}
\usepackage{curves}
\usepackage{epic}
\usepackage{eepic}
\usepackage{epsfig}

\let\footnote\savefootnote

\let\footnoterule\savefootnoterule 

\normallatexbib

\newcommand{\dd}{\mbox{\rm d}}
\newcommand{\DD}{\mbox{\rm D}}
\newcommand{\cx}{\stackrel{\circ}{X}}
\newcommand{\nnnn}{\noindent}
\newcommand{\oo}{\over}
\newcommand{\p}{\partial}

\newcommand{\be}{\begin{equation}}
\newcommand{\ee}{\end{equation}}
\newcommand{\lbl}{\label}
\newcommand{\bi}{\bibitem}
\newcommand{\ci}{\cite}

\newcommand{\vs}{\vspace}
\newcommand{\hs}{\hspace}
\newcommand{\oklepaj}{
  \qbezier(8,14)(4,34)(10,40)
  \qbezier(6,0)(10,6)(8,14) 
  \qbezier(8,-14)(4,-34)(10,-40)
  \qbezier(6,0)(10,-6)(8,-14) }
\newcommand{\qb}{\qbezier}  

\def\bear{\begin{eqnarray}}
\def\bearr{\begin{eqnarray} &&\nqq}
\def\ear{\end{eqnarray}}
\def\nqq{\hspace{-2em}}
\def\lal{&&\nqq}     
\def\nnn{\nonumber\\ &&\nqq {} }
\def\nq{\hspace{-1em}}

\def\eql{\al =\al}

\def\noi{\noindent}
\def\e{{\rm e}}
\def\nhq{\hspace{-0.5em}}
\def\al{&\nhq}

\begin{document}
\upperandlowerfalse


\

\thispagestyle{empty}

\vs{10mm}

\begin{center}

{\baselineskip .6cm

{\LARGE \bf The Landscape of Theoretical Physics: 

\vs{1mm}

A Global View}

\vs{2mm}

{\Large  \bf From Point Particles to the
Brane World and Beyond,
in Search of a Unifying Principle}

\vs{8mm}

Matej Pav\v si\v c

Jo\v zef Stefan Institute, Jamova 39,
1000 Ljubljana, Slovenia

e-mail: matej.pavsic@ijs.si\,;  http://www-f1.ijs.si/~pavsic/
}

\vs{10mm}

{\bf Abstract}

\end{center}


{\baselineskip .5cm

This is a book for those who would like to learn something
about special and general relativity  beyond the usual textbooks,
about quantum field theory, the elegant Fock-Schwinger-Stueckelberg proper
time formalism, the elegant description of geometry by means of Clifford
algebra, about the fascinating possibilities the latter algebra offers in
reformulating the existing physical theories, and quantizing them in a natural
way. It is shown how Clifford algebra provides much more: it provides
room for new physics, with the prospects of resolving certain long
standing puzzles.
The theory of branes and the idea of how a 3-brane might represent our
world is discussed in detail. Much attention is paid to the elegant geometric
theory of branes which employs the infinite dimensional space of functions
describing branes. Clifford algebra is generalized to the infinite dimensional
spaces. In short, this is a book  for anybody who would like to explore how the
``theory of everything'' might possibly be formulated. The theory that would
describe all the known phenomena, could not be formulated without taking into
account ``all'' the theoretical tools which are available. Foundations of
those tools and their functional interrelations are described in the book.

}

\vs{4mm}

 {\bf Note:} {\it This book was published by Kluwer Academic
Publishers in 2001. The body of the posted version is identical to
the published one, except for corrections of misprints, and some
minor revisions that I have found necessary.}

\newpage

\booktitle[The Landscape of Theoretical Physics: A Global View]
{The Landscape of Theoretical Physics: \\A Global View\\ \ } 

\subtitle{ From Point Particles to the\\
Brane World and Beyond,\\
in Search of a Unifying Principle}

\halftitlepage{The Landscape of Theoretical Physics: \\A Global View\\
\   \\
From Point Particles to the\\
Brane World and Beyond,\\
in Search of a Unifying Principle}



\titlepage{
\author{Matej Pav\v si\v c}
\authoraffil{\ \ \\
Department of Theoretical Physics\\
Jo\v zef Stefan Institute\\
Ljubljana, Slovenia} }


\tableofcontents

\begin{preface}

Today many important directions of research are being pursued
more or less independently of each other.
These are, for instance, strings and
membranes, induced gravity, embedding of spacetime into a 
higher-dimensional space, the brane world scenario, the quantum theory in 
curved spaces, Fock--Schwinger
proper time formalism, parametrized relativistic quantum theory,
quantum gravity, wormholes and the problem of ``time machines",
spin and supersymmetry, geometric calculus based on Clifford algebra,
various interpretations of quantum mechanics
including the Everett interpretation, and the recent important approach
known as ``decoherence".

A big problem, as I see it, is that various people
thoroughly investigate their narrow field without being aware of certain
very close relations to other fields of research. What we need now
is not only to see the trees but also the forest. In the
present book I intend to do just that: to carry out a first approximation
to a synthesis of the related fundamental theories of physics. I sincerely
hope that such a book will be useful to physicists. 

From a certain viewpoint the book could be considered as
a course in theoretical physics in
which the foundations of all those relevant fundamental theories and
concepts are attempted to be thoroughly reviewed. Unsolved problems
and paradoxes
are pointed out. I show that most of those approaches
have a common basis in the theory of unconstrained membranes. The very
interesting
and important concept of membrane space, ${\cal M}$, the tensor calculus
in ${\cal M}$ and functional transformations in ${\cal M}$ are
discussed.
Next I present a theory in which spacetime is
considered as a 4-dimensional unconstrained membrane and discuss how
the usual classical gravity, together with sources, emerges as an
effective theory. Finally, I point out that the Everett
interpretation of quantum mechanics is the natural one in that theory.
Various interpretational issues will be discussed and the relation to
the modern ``decoherence" will be pointed out.

If we look at the detailed structure of a landscape we are
unable to see the connections at a larger scale. We see mountains,
but we do not see the mountain range. A view from afar is as
important as a view from nearby. Every position illuminates
reality from its own perspective. It is analogously so, in my opinion,
in theoretical physics also. Detailed investigations of a certain
fundamental theory are made at the expense of seeing at the same
time the connections with other theories. What we need today is some
kind of atlas of the many theoretical approaches currently under
investigation. During many years of effort I can claim that
I do see a picture which has escaped from attention of other
researchers. They  certainly might profit if they could become
aware of such a more global, though not as detailed, view of
fundamental theoretical physics.

\vs{3mm}

\rightline{MATEJ PAV\v SI\v C}

\end{preface}

\dedication{To my family}

\begin{acknowledgments}

This book is a result of many years of thorough study and research.
I am very grateful to my parents, who supported me and encouraged me
to persist on my chosen path. Also I thank my beloved wife Mojca
for all she has done for me and for her interest in my work.
I have profited a lot from discussions
and collaboration with E. Recami, P. Caldirola, A.O. Barut, W.A.
Rodrigues, Jr., V. Tapia, M. Maia, M. Blagojevi\' c, R. W. Tucker,
A. Zheltukhin, I. Kanatchikov,
L. Horwitz, J. Fanchi, C. Castro, and many others whom I have met on 
various occasions.
The work was supported by the Slovenian Ministry of Science and 
Technology.

\end{acknowledgments}


\introduction{}
The unification of various branches of theoretical physics is a joint
project of many researchers, and everyone contributes as much as
he can. So far we have accumulated
a great deal of knowledge and insight encoded in such marvelous theories as
general relativity, quantum mechanics, quantum field theory,
the standard model of electroweak interaction, and chromodynamics.
In order to obtain a more unified view, various promising theories have
emerged, such as those of strings and ``branes", induced gravity, the
embedding models of gravity, and the ``brane world" models, to mention
just a few. The very powerful Clifford algebra as a useful
tool for geometry and physics is becoming more and more popular. Fascinating
are the ever increasing successes in understanding the foundations of
quantum mechanics and their experimental verification, together with
actual and potential practical applications in cryptography,
teleportation, and quantum computing.

In this book I intend to discuss the conceptual and technical foundations
of those approaches which, in my opinion, are most relevant for
unification of general relativity and quantum mechanics on the one hand,
and fundamental interactions on the other hand. After many years of
active research I have arrived at a certain level of insight into the possible
interrelationship between those theories. Emphases will be on the
exposition and understanding of concepts and basic techniques, at the
expense of detailed and rigorous mathematical development. Theoretical
physics is considered here as a beautiful landscape. A global view
of the landscape will be taken. This will enable us to see forests and
mountain ranges as a whole, at the cost of seeing trees and rocks.

Physicists interested in the foundations of physics, conceptual issues,
and the unification program, as well as those working in a special field
and desiring to broaden their knowledge and see their speciality from
a wider perspective, are expected to profit the most from the present
book. They are assumed to possess a solid knowledge at least of quantum
mechanics, and special and general relativity.

As indicated in the subtitle, I will start from point particles. They
move along geodesics which are the lines of minimal, or, more generally, 
extremal
length, in spacetime. The corresponding action from which the equations of
motion are derived is invariant with respect to reparametrizations of an
arbitrary parameter denoting position on the worldline swept by the
particle. There are several different, but equivalent, reparametrization
invariant point particle actions. A common feature of such an approach is
that actually there is no dynamics in spacetime, but only in space.
A particle's worldline is frozen in spacetime, but from the 3-dimensional
point of view we have a point particle moving in 3-space. This fact is
at the roots of all the difficulties we face when trying to quantize the
theory: either we have a covariant quantum theory but no evolution in
spacetime, or we have evolution in 3-space at the expense of losing manifest
covariance in spacetime. In the case of a point particle this problem is
not considered to be fatal, since it is quite satisfactorily resolved in
relativistic quantum field theory. But when we attempt to quantize
extended objects such as branes of arbitrary dimension, or spacetime
itself, the above problem emerges in its full power: after so many
decades of intensive research we have still not yet arrived at a generally
accepted consistent theory of quantum gravity.

There is an alternative to the usual relativistic point particle action
proposed by Fock \ci{1} and subsequently investigated by Stueckelberg
\ci{2}, Feynman \ci{3},
Schwinger \ci{4}, Davidon \ci{4a}, Horwitz \ci{5,5a} and many 
others \ci{6}--\ci{12b}.  In such a theory a particle or
``event" in spacetime obeys a law of motion analogous to that 
of a nonrelativistic
particle in 3-space. The difference is in the dimensionality and signature
of the space in which the particle moves. None of the coordinates
$x^0, x^1, x^2, x^3$ which parametrize spacetime has the role of
evolution parameter. The latter is separately postulated and is Lorentz
invariant. Usually it is denoted as $\tau$ and evolution goes along
$\tau$. There are no constraints in the theory, which can therefore be
called {\it the unconstrained theory}. First and second quantizations of
the unconstrained theory are straightforward, very elegant, and manifestly
Lorentz covariant. Since $\tau$ can be made to be related to proper time such
a theory is often called a {\it Fock--Schwinger proper time formalism}.
The value and elegance of the latter formalism is widely recognized,
and it is often used, especially when considering quantum fields in
curved spaces \ci{12}. There are two main interpretations of the formalism:

\ (i) According to the first interpretation, it is considered merely as a useful
calculational tool, without any physical significance. Evolution in $\tau$
and the absence of any constraint is assumed to be fictitious and
unphysical. In order to make contact with physics one has to get rid
of $\tau$ in all  the expressions considered by integrating them over $\tau$.
By doing so one projects unphysical expressions onto the physical ones,
and in particular one projects unphysical states onto physical states.

(ii) According to the second interpretation, evolution in $\tau$ is genuine
and physical. There is, indeed, dynamics in spacetime. Mass is a constant
of motion and not a fixed constant in the Lagrangian.

Personally, I am inclined to the interpretation (ii). In the history
of physics it has often happened that a good new formalism also contained 
good new physics waiting to be discovered and identified in suitable
experiments. It is one of the purposes of this book to show a
series of arguments in favor of the interpretation (ii). The first
has roots in geometric calculus based on Clifford algebra \ci{13}

Clifford numbers can be used to represent vectors, multivectors, and,
in general, polyvectors (which are Clifford aggregates). They form a very
useful tool for geometry. The well known equations of physics can be cast
into elegant compact forms by using the geometric calculus based on
Clifford algebra.

These compact forms suggest the generalization that every physical quantity
is a polyvector \ci{14,14a}. For instance, the momentum polyvector in 4-
dimensional spacetime has not only a vector part, but also a scalar,
bivector, pseudovector and pseudoscalar part. Similarly for the velocity
polyvector. Now we can straightforwardly generalize the conventional
constrained action by rewriting it in terms of polyvectors. By doing so
we obtain in the action also a term which corresponds to the pseudoscalar part
of the velocity polyvector. A consequence of this extra term is that,
when confining ourselves, for simplicity, to polyvectors with pseudoscalar and vector
part only, the variables corresponding to 4-vector components can all be
taken as independent. After a straightforward procedure in which we
omit the extra term in the action (since it turns out to be just the
total derivative), we obtain Stueckelberg's unconstrained action! This
is certainly a remarkable result. The original, constrained
action is equivalent to the unconstrained action. 
Later in the book (Sec. 4.2) I show that the analogous procedure
can also be applied to extended objects such as strings,
membranes, or branes in general.

When studying the problem of how to identify points in a generic curved
spacetime, several authors \ci{15}, and, especially recently Rovelli \ci{16},
have
recognized that one must fill spacetime with a reference fluid. Rovelli
considers such a fluid as being composed of a bunch of particles, each particle
carrying a clock on it. Besides the variables denoting positions of particles
there is also a variable denoting the clock. This extra, clock, variable
must enter the action, and the expression Rovelli obtains is formally the
same as the expression we obtain from the polyvector action (in which we
neglect the bivector, pseudovector, and scalar parts). We may therefore
identify the pseudoscalar part of the velocity polyvector with the speed of the
clock variable. Thus have a relation between the polyvector generalization
of the usual constrained relativistic point particle, the Stueckleberg
particle, and the DeWitt--Rovelli particle with clock.

A relativistic particle is known to posses spin, in general. We show
how spin arises from the polyvector generalization of the point particle and
how the quantized theory contains the Dirac spinors together with  the Dirac
equation as a particular case. Namely, in the quantized theory a state is
naturally assumed to be represented as a polyvector wave function $\Phi$,
which, in particular, can be a spinor. That spinors are just a special kind
of polyvectors (Clifford aggregates), namely the elements of the minimal
left or right ideals of the Clifford algebra, is an old observation \ci{17}.
Now, scalars, vectors, spinors, etc., can be reshuffled by the elements of
the Clifford algebra. This means that scalars, vectors, etc., can be
transformed into spinors, and {\it vice versa}. 
Within Clifford algebra thus we have
transformations which change bosons into fermions. In Secs. 2.5 and 2.7  
I discuss the possible relation between the Clifford algebra formulation of the
spinning particle and a more widely used formulation in terms
of Grassmann variables.

A very interesting feature of Clifford algebra concerns the signature of
the space defined by basis vectors which are generators of the
Clifford algebra. In principle we are not confined to choosing just
a particular set of elements as basis vectors; we may choose some
other set. For instance, if $e^0$, $e^1$, $e^2$, $e^3$ are the basis
vectors of a space $M_e$ with signature (+ + + +), then we may declare
the set ($e^0$, $e^0 e^1$, $e^0 e^2$, $e^0 e^3$) as basis vectors
$\gamma^0$, $\gamma^1$, $\gamma^2$, $\gamma^3$ of some other space $M_{\gamma}$
with signature $(+ - - -)$. That is, by suitable choice of basis vectors
we can obtain within the same Clifford algebra a space of arbitrary
signature. This has far reaching implications. For instance, in the case
of even-dimensional space we can always take a signature with an equal number
of pluses and minuses. A harmonic oscillator in such a space has vanishing
zero point energy, provided that we define the vacuum state in a very
natural way as proposed in refs. \ci{18}.
An immediate consequence is that
there are no central terms and anomalies in string theory living
in spacetime with signature $(+ + + ... - - -)$, even if the dimension of such
a space is not critical. In other words, spacetime with such a
`symmetric' signature need not have 26 dimensions \ci{18}. 

The principle of such a harmonic oscillator in a pseudo-Euclidean space
is applied in Chapter 3 to a system of scalar fields. 
The metric in the space of
fields is assumed to have signature $(+ + + ... - - -)$ and it is shown
that the vacuum energy, and consequently the cosmological constant, are then
exactly zero. However, the theory contains some negative energy fields
(``exotic matter") which
couple to the gravitational field in a different way than the usual,
positive energy, fields: the sign of coupling is reversed, which
implies a repulsive gravitational field around such a source. This is
the price to be paid if one wants to obtain a small cosmological
constant in a straightforward way. One can consider this as a prediction
of the theory to be tested by suitably designed experiments.

The problem of the cosmological constant is one of the toughest problems
in theoretical physics. Its resolution would open the door to further
understanding of the relation between quantum theory and general relativity.
Since all more conventional approaches seem to have been more or less
exploited without unambiguous success, the time is right for a more drastic
novel approach. Such is the one which relies on the properties of the harmonic
oscillator in a pseudo-Euclidean space.

In Part II \hs{1mm} I discuss the theory of extended objects,
now known as ``branes" which are membranes of any dimension and are
generalizations of point particles and strings. As in the case of point
particles I pay much attention to the unconstrained theory of membranes.
The latter theory is a generalization of the
Stueckelberg point particle theory. It turns out to be very convenient to
introduce the concept of the infinite-dimensional {\it membrane space}
${\cal M}$. Every point in ${\cal M}$ represents an unconstrained
membrane. In ${\cal M}$ we can define distance, metric, covariant
derivative, etc., in an analogous way as in a finite-dimensional curved space.
A membrane action, the corresponding equations of motion, and other relevant
expressions acquire very simple forms, quite similar to those
in the point particle theory. We may say that a membrane is a point particle
in an infinite dimensional space!

Again we may proceed in two different interpretations of the theory:

\ (i) We may consider the formalism of the membrane space as a useful
calculational tool (a generalization of the Fock--Schwinger proper time
formalism) without any genuine physical significance. Physical quantities
are obtained after performing a suitable projection.

(ii) The points in ${\cal M}$-space are physically distinguishable, that is,
a membrane can be physically deformed in various ways and such a
deformation may change with evolution in $\tau$.

If we take the interpretation (ii) then we have a marvelous connection
(discussed in Sec. 2.8) with the Clifford algebra generalization of the
conventional constrained membrane on the one hand, and the concept of
DeWitt--Rovelli reference fluid with clocks on the other hand.

Clifford algebra in the infinite-dimensional membrane space ${\cal M}$
is described in Sec. 6.1. When quantizing the theory of the unconstrained membrane
one may represent states by wave functionals which are polyvectors
in ${\cal M}$-space. A remarkable connection with quantum field
theory is shown in Sec. 7.2

When studying the ${\cal M}$-space formulation of the membrane theory
we find that in such an approach one cannot postulate the existence of a
background embedding space independent from a membrane configuration.
By ``membrane configuration" I understand a system of (many) membranes,
and the membrane configuration is identified 
with the embedding space. There is no
embedding space without the membranes. This suggests that our spacetime 
is nothing but a membrane configuration. In particular, our spacetime could be
just one 4-dimensional membrane (4-brane) amongst many other membranes
within the configuration. Such a model is discussed in Part III. The
4-dimensional gravity is due to the induced metric on our 4-brane $V_4$, whilst
matter comes from the self-intersections of $V_4$, or the intersections
of $V_4$ with other branes. As the intersections there can occur manifolds of
various dimensionalities. In particular, the intersection or the
self-intersection can be a 1-dimensional worldline. It is shown that such
a worldline is a geodesic on $V_4$. So we obtain in a natural way
four-dimensional gravity with sources. The quantized 
version of such a model is
also discussed, and it is argued that the kinetic term for the
4-dimensional metric $g_{\mu \nu}$ is induced by quantum fluctuations of the
4-brane embedding functions.

In the last part I discuss mainly the problems related to the foundations
and interpretation of quantum mechanics. I show how the brane world view
sheds new light on our understanding of quantum mechanics and the
role of the observer.



\newcommand{\boxA}{
\setlength{\unitlength}{1cm}
\nnnn \begin{picture}(12.5,19)

\put(0,18){\makebox(12.5,1) {\bf Box 1.1}}
\put(0,0){\framebox(12.5,19)}
\put(3.5,15){\vector(-1,0){1} }
\put(2.5,16.5){\vector(0,-1){3} }
\put(4,11){\vector(-1,0){1.5} }
\put(2.5,12){\vector(0,-1){2} }
\put(4,7.7){\vector(-1,0){1.5} }
\put(2.5,8.5){\vector(0,-1){1.5} }
\put(4,4.5){\vector(-1,0){1.5} }
\put(2.5,5.5){\vector(0,-1){2} }

\renewcommand{\xscale}{0.02}
\renewcommand{\yscale}{0.02}
\put(3.7,15){\oklepaj}

\thicklines
\put(1,16.5){\framebox(6,1.5)
    {$G(\Sigma) = \int {\dd} \Sigma_{\mu} \, {\dd} \tau\,{T^{\mu}}_{\nu}
    \, \delta x^{\nu}$}}
\put(5,15){\makebox(5.7,1)
     {$\delta x^{\mu} = \xi^{\mu} \delta s \; , \; \;  \xi^{\mu}$ 
     a vector along $\delta x^{\mu}$-direction}}
\put(5,14){\makebox(6,1)
      {${\dd} \Sigma_{\mu} = n_{\mu} {\dd} \Sigma \; , \; \; n_{\mu}$ 
      a vector along ${\dd} \Sigma_{\mu}$-direction}}

\put(1,12){\framebox(6,1.5)
      {$P(n,\xi) = \int {\dd} \Sigma \, {\dd} \tau \, {T^{\mu}}_{\nu}\,
      n_{\mu} \xi^{\nu}$}}
\put(5,10.5){\makebox(6,1)
     {integrate over ${\dd} s$ and divide by $s_2 - s_1$}}
\put(.5,8.5){\framebox(9,1.5)
      {\begin{minipage} [t] {12.5cm}
      $$P(n, \xi) = \lim_{{s_{1,2}}\to {- \infty , \infty}} \,
          {1\oo {s_2 - s_1}} \int_{s_1}^{s_2} {\dd} \Sigma \,
          {\dd} s \, {\dd} \tau \, {T^{\mu}}_{\nu} \, n_{\mu} \xi^{\nu}$$
       \end{minipage}   
          }}
\put(4,7.3){\makebox(4,1)
      {${\dd} \Sigma \, {\dd} s = {\dd}^D x$}}
\put(0.1,5.5){\framebox(12.3,1.5)
        {\begin{minipage} [t] {12.5cm}         
        $$P(n, \xi) = 
        \lim_{{s_{1,2}}\to {- \infty , \infty}} \,
        \mbox{${{\tau_1 - \tau_2} \oo {s_1 - s_2}}$}
         \int_{s_1}^{s_2} {\dd}^D p \, {\Lambda \oo 2}
        \, (n_{\mu} p^{\mu}) \, p_{\nu} \xi^{\nu} \left ( 
        c^{\dagger} (p) c(p) + c(p) c^{\dagger} (p) \right )$$
        \end{minipage} }}
\put(4,4){\makebox(7,1)
         {$\Lambda = {{\Delta s}\oo {\Delta \tau}} \, {1\oo {\sqrt{p^2}}}\; \; $
        when mass is definite}}
\put(.5,2){\framebox(9,1.5)
           {$P(n,\xi) = \mbox{$1\oo 2$} \int {\dd}^D p \, \epsilon (n p) \, p_{\nu}
           \xi^{\nu} \left ( 
        c^{\dagger} (p) c(p) + c(p) c^{\dagger} (p) \right )$}}
\put(1,0){\makebox(10,1.5)
         {\begin{minipage} [t] {10cm}
         When $\xi^{\nu}$ is time-like $P(n,\xi)$ is projection of {\it energy},
         when $\xi^{\nu}$ is space-like $P(n,\xi)$ is projection of
         {\it momentum} on $\xi^{\nu}$
         \end{minipage}  }}

\end{picture} }

\newcommand{\boxB}{
\setlength{\unitlength}{1cm}
\nnnn \begin{picture}(12.5,19)

\put(0,18){\makebox(12.5,1) {\bf Box 1.2: Momentum on mass shell}}

\put(0,0){\line(0,1){19}}
\put(12.5,0){\line(0,1){19}}
\put(0,19){\line(1,0){12.5}}

\put(1.3,14.8){\vector(-1,0){.5} }
\put(.8,16.5) {\vector(0,-1){3.6} }
\put(2.1,9.4) {\vector(-1,0){1.3} }
\put(.8,11)   {\vector(0,-1){2.6} }
\put(2.4,6.6) {\vector(-1,0){1.3} }
\put(1.1,7.4)   {\vector(0,-1){1.8} }
\put(2.4,1.6) {\vector(-1,0){1.3} }
\put(1.1,3.3)   {\vector(0,-1){3.3} }

\renewcommand{\xscale}{0.060}
\renewcommand{\yscale}{0.030}
\put(1.2,14.8){\oklepaj}

\renewcommand{\xscale}{0.035}
\renewcommand{\yscale}{0.017}
\put(2.2,9.4){\oklepaj}

\renewcommand{\xscale}{0.02}
\renewcommand{\yscale}{0.012}
\put(2.6,6.6){\oklepaj}

\renewcommand{\xscale}{0.05}
\renewcommand{\yscale}{0.03}
\put(2.6,1.6){\oklepaj}

\thicklines
\put(.2,16.5){\framebox(4,1.2)
    {$P^{\mu} = \int T^{\mu \nu} {\dd} \Sigma_{\nu}$}}
\put(2,15.5){\makebox(10.5,1)
       {\begin{minipage} [t] {12.5cm}
        $$T^{\mu \nu} = {\Lambda\oo 2} (\p^{\mu} \Phi^{\dagger} \p^{\nu}
        \Phi + \p^{\mu} \Phi \p^{\nu} \Phi^{\dagger})
        - {\Lambda\oo 2} (\p_{\alpha} \Phi^{\dagger} \p^{\alpha} \Phi -
        M^2 \Phi^{\dagger} \Phi) {\delta^{\mu}}_{\nu}$$
        \end{minipage} }}
\put(2,14.5){\makebox(7.5,1)
       {$\Phi = \int {\dd} p \, \delta 
       (p^2 - M^2) e^{i p x} c(p)$}} 
\put(2,13.8){\makebox(7.5,1)        
        {$ \quad \;  \p_{\mu} \Phi = \int {\dd} p \, \delta 
         (p^2 - M^2) e^{i p x} c(p) i p_{\mu}$}}
\put(2,13){\makebox(7.5,1)
         {$\qquad \qquad {\dd} \Sigma_{\nu} = n_{\nu} {\dd} \Sigma$}}    
         
\put(.2,11){\framebox(10,1.9)
       {\begin{minipage} [t] {10cm}
       $$P^{\mu} = {\Lambda\oo 2} \int {\dd} p \, {\dd} p' \, \delta (p^2 - M^2)
       \delta (p'^2 - M^2) p'^{\mu} p^{\nu} n_{\nu}$$
       $$\; \; \qquad \hs{5mm} \times \left ( c^{\dagger} 
       (p') c(p) + c(p') c^{\dagger} (p) \right ) e^{i(p-p')x} {\dd} \Sigma$$
       \end{minipage} }}

\put(2,9.7){\makebox(9,.7)
        {$\delta(p^2 - M^2) = \int {\dd} s\, e^{i (p^2 - M^2)s} \; , \quad
        {\dd} s \, {\dd} \Sigma = {\dd}^D x$}}

\put(2.8,8.7){\makebox(4,1)
        {\begin{minipage} [t] {4cm}
        $$\Lambda = {1 \oo {\sqrt{p^2}}} \; \qquad {{{\dd} s}\oo {{\dd} \tau}}
        = 1$$
        \end{minipage} }}
        
\put(.2,7.4){\framebox(9.5,1)
        {$P^{\mu} = {1\oo 2} \int {\dd} p \, \delta (p^2 - M^2) \epsilon 
        (pn) p^{\mu}
        \left ( c^{\dagger} (p) c(p) + c(p) c^{\dagger} (p) \right )$ }} 
        
\put(2.8,6.6){\makebox(5,.7)
         {${\dd} p = {\dd} \omega \, {\dd} {\bf p} \; \quad p = 
         (\omega, {\bf p})$ }}
\put(3.2,5.9){\makebox(3,.7)
         {$\omega_{\bf p} = \sqrt{{\bf p}^2 + M^2}$ }} 
\put(.2,3.3){\framebox(11.4,2.3)[b]
         {\begin{minipage} [b] {11cm}
         $$P^{\mu} = {1\oo 2} \int {\dd} {\bf p} \pmatrix{\omega_{\bf p} \cr
         {\bf p} \cr} \, {1 \oo {2 \omega_{\bf p}}} 
         \Bigl( c^{\dagger} (\omega_{\bf p},
         {\bf p}) c(\omega_{\bf p},{\bf p}) +c(\omega_{\bf p},{\bf p})
         c^{\dagger} (\omega_{\bf p},{\bf p})$$
         $$ \; \; \; \quad \quad \hs{7mm}
         + \, c^{\dagger} (-\omega_{\bf p},
         -{\bf p}) c(-\omega_{\bf p},-{\bf p}) +c(-\omega_{\bf p},-{\bf p})
         c^{\dagger} (-\omega_{\bf p},-{\bf p}) \Bigr)$$
         \end{minipage} }}
\put(3.3,2.3){\makebox(9,.7) [l]
          {I. $\qquad \lbrack c(p), c^{\dagger} (p') \rbrack = \delta^D (p - p') \;
          \quad c(p) | 0 \rangle = 0$ }}
\put(3.3,1.6){\makebox(9,.7) [l]
          {$\; \; \; {1\oo {2 \omega_{\bf p}}} \, c^{\dagger}(\omega_{\bf p}, 
          {\bf p} ) =
          a^{\dagger}({\bf p}) \; , \quad  {1\oo {2 \omega_{\bf p}}} 
          c^{\dagger}(-\omega_{\bf p}, -{\bf p} ) = b^{\dagger}(-{\bf p})$ }}        
\put(3.3,.9){\makebox(9,.7) [l]
          {$\; \; \; {1\oo {2 \omega_{\bf p}}} \, c(\omega_{\bf p}, {\bf p} ) =
          a({\bf p}) \; , \quad  {1\oo {2 \omega_{\bf p}}} 
          c(-\omega_{\bf p}, -{\bf p} ) = b(-{\bf p})$ }}
\put(3.3,.1){\makebox(6,.7) [l]
           {$\; \; \; \qquad a({\bf p}) |0 \rangle = 0 \; , \quad b(-{\bf p}) 
           |0 \rangle = 0$}}

\end{picture}

\nnnn \begin{picture}(12.5,9)

\put(0,8){\line(0,-1){7.5}}
\put(12.5,8){\line(0,-1){7.5}}
\put(0,.5){\line(1,0){12.5}}

\put(2.4,6.3) {\vector(-1,0){1.3} }
\put(1.1,8)   {\vector(0,-1){3.7} }

\renewcommand{\xscale}{0.05}
\renewcommand{\yscale}{0.03}
\put(2.6,6.3){\oklepaj}

\put(1.8,6.5){\makebox(.5,.6) [l]
              {or} }

\put(3.3,7){\makebox(9,.7) [l]
          {II. $\; \qquad \lbrack c(p), c^{\dagger} (p') \rbrack = \epsilon (p^0)
          \delta^D (p - p')$ }}
\put(3.3,6.3){\makebox(9,.7) [l]
          {$\; \; \; \; \; {1\oo {2 \omega_{\bf p}}} \, c^{\dagger}(\omega_{\bf p}, 
          {\bf p} ) =
          a^{\dagger}({\bf p}) \; , \quad  {1\oo {2 \omega_{\bf p}}} 
          c^{\dagger}(-\omega_{\bf p}, -{\bf p} ) = b({\bf p})$ }}        
\put(3.3,5.4){\makebox(9,.7) [l]
          {$\; \; \; \; \; {1\oo {2 \omega_{\bf p}}} \, c(\omega_{\bf p}, {\bf p} )
          = a({\bf p}) \; , \quad  {1\oo {2 \omega_{\bf p}}} 
          c(-\omega_{\bf p}, -{\bf p} ) = b^{\dagger}({\bf p})$ }}
\put(3.3,4.7){\makebox(6,.7) [l]
           {$\; \; \; \; \;  \qquad a({\bf p}) |0 \rangle = 0 \; , 
           \quad b({\bf p}) 
           |0 \rangle = 0$}}
\put(.2,2.7){\framebox(12,1.6)[t]
         {\begin{minipage}[t]{12cm}
         $$P^{\mu} = {1\oo 2} \int {\dd} {\bf p} \pmatrix{\omega_{\bf p} \cr
         {\bf p} \cr} \, \left ( a^{\dagger} ({\bf p}) a({\bf p}) +
         a({\bf p}) a^{\dagger} ({\bf p}) + b^{\dagger} ({\bf p}) b({\bf p}) +
         b({\bf p}) b^{\dagger} ({\bf p}) \right ) $$
         \end{minipage} }} 
\put(3,1.9)                   
         {\begin{minipage}[t]{9cm}
         For both definitions I and II the expression for $P_{\mu}$ is the
         same, but the relations of $b({\bf p})$, $b^{\dagger} ({\bf p})$
         to $c(p)$ are different.
        \end{minipage} }
        
\end{picture} }


\newcommand{\boxC}{
\setlength{\unitlength}{1cm}
\nnnn \begin{picture}(12.5,19)

\put(0,18){\makebox(12.5,1) {\bf Box 1.3: Commutation relations}}
\put(0,0){\framebox(12.5,19)}


\put(3.3,15.6){\vector(-1,0){.8} }
\put(2.5,16.5) {\vector(0,-1){1.7} }
\put(2.5,13.8) {\vector(0,-1){.8} }
\put(2.5,12)   {\vector(0,-1){.7} }
\put(4.5,9) {\vector(-1,0){1} }
\put(3.5,10)   {\vector(0,-1){1.8} }
\put(4.5,6.2) {\vector(-1,0){2} }
\put(2.5,10)   {\vector(0,-1){4.7} }

\renewcommand{\xscale}{0.02}
\renewcommand{\yscale}{0.013}
\put(7.7,17.1){\oklepaj}

\thicklines
\put(1,16.5){\framebox(5,1.2)
    {$\lbrack c(p), c^{\dagger} (p') \rbrack = \alpha \delta^D (p - p')  $}}

\put(7,16.7){\makebox(.7,.7)[l]
{$\alpha = $}}
\put(8,17){\makebox(3.5,.7)[l]
{\ \ 1 \hs{6mm} Case I of Box 1.2}}

\put(8,16.4){\makebox(3.5,.7)[l]
{$\epsilon (p^0)$ \hs{2mm} Case II of Box 1.2}}

\put(3.5,15.3){\makebox(8,.7)
{on mass shell we must insert the $\delta$-function}}

\put(1,13.8){\framebox(8,1)
{$\lbrack c(p), c^{\dagger} (p') \rbrack \delta (p^2 - M^2) =
\alpha  \delta^D (p - p')  $}}

\put(1,12){\framebox(8,1)
{$\int \lbrack c(p), c^{\dagger} (p') \rbrack \delta (p^2 - M^2) {\dd} p'^0 =
\alpha  \delta ({\bf p} - {\bf p'})  $}}

\put(.7,10){\framebox(11.5,1.3)
{${1\oo {2 \omega_{\bf p}}} \lbrack c(p^0, {\bf p}), c^{\dagger} (\omega_{\bf p},
{\bf p'}) \rbrack + {1\oo {2 \omega_{\bf p}}} 
\lbrack c(p^0, {\bf p}), c^{\dagger} (-\omega_{\bf p}, {\bf p'}) \rbrack =
\alpha  \delta^D ({\bf p} - {\bf p'})$}}

\put(4.8,8.7){\makebox(8,.7)[l]
{$\lbrack c(p^0, {\bf p}), c^{\dagger} (-\omega_{\bf p}, {\bf p'}) \rbrack = 0
\; , \quad$ when $p^0 > 0$ }}

\put(2.8,7){\framebox(6.5,1.2)
{${1\oo {2 \omega_{\bf p}}} \lbrack c(\omega_{\bf p}, {\bf p}), c^{\dagger} 
(\omega_{\bf p},
{\bf p'}) \rbrack = \delta ({\bf p} - {\bf p'})$}}

\put(4.8,5.9){\makebox(7,.7)[l]
{$\lbrack c(p^0, {\bf p}), c^{\dagger} (\omega_{\bf p},
{\bf p'}) \rbrack = 0 \; , \quad$ when $p^0 < 0$}}

\put(1.2,4){\framebox(8,1.3)
{${1\oo {2 \omega_{\bf p}}} 
\lbrack c(-\omega_{\bf p}, {\bf p}), c^{\dagger} (-\omega_{\bf p}, {\bf p'})
 \rbrack = \alpha  \delta ({\bf p} - {\bf p'})  $ }}

\put(1.4,1.5){\makebox
{$\Rightarrow$}}
\put(2,.5){\framebox(5,2)}
\put(2,2.6)
{\begin{minipage} [t] {5cm}
$$\lbrack a({\bf p}), a^{\dagger} ({\bf p}' \rbrack = 
\delta ({\bf p} - {\bf p'})  $$
$$\lbrack b({\bf p}), b^{\dagger} ({\bf p}' \rbrack = 
\delta ({\bf p} - {\bf p'})  $$
\end{minipage} }

\put(8,1.9)
{\begin{minipage}[t] {4cm}
True either for Case I or Case II with corresponding definitions
of $b({\bf p}), \, b^{\dagger}({\bf p})$
\end{minipage} }

\end{picture} }



\newcommand{\figWiggly}{
\setlength{\unitlength}{.8mm}

\begin{figure}[h]
\hs{3mm} \begin{picture}(120,180)(25,0)
\put(5,118){\epsfig{file=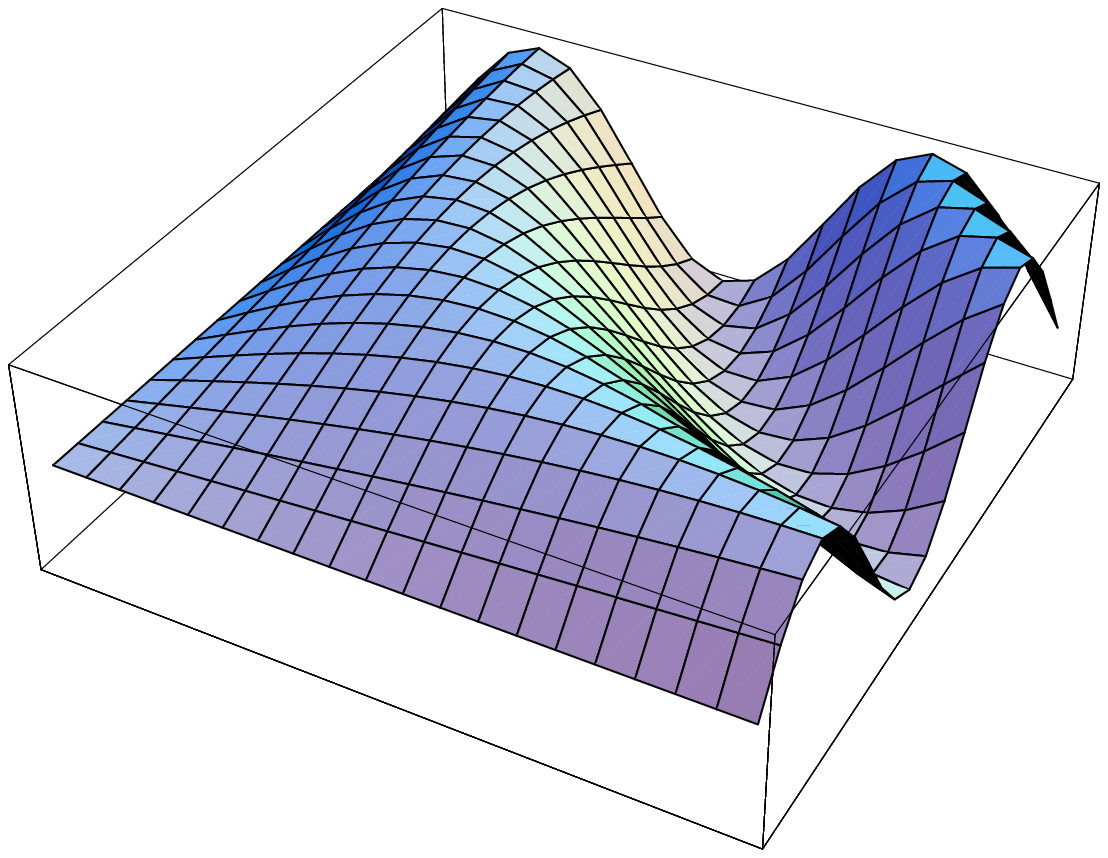,width=78mm}}
\put(78,118){\epsfig{file=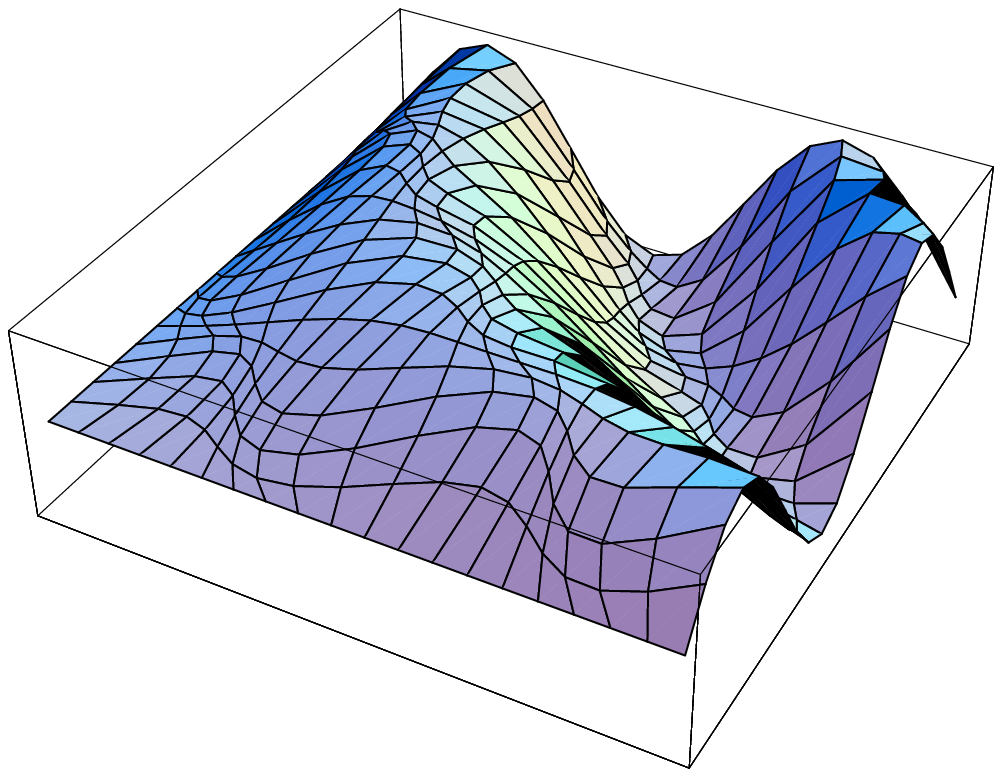,width=78mm}}
\put(0,50){\epsfig{file=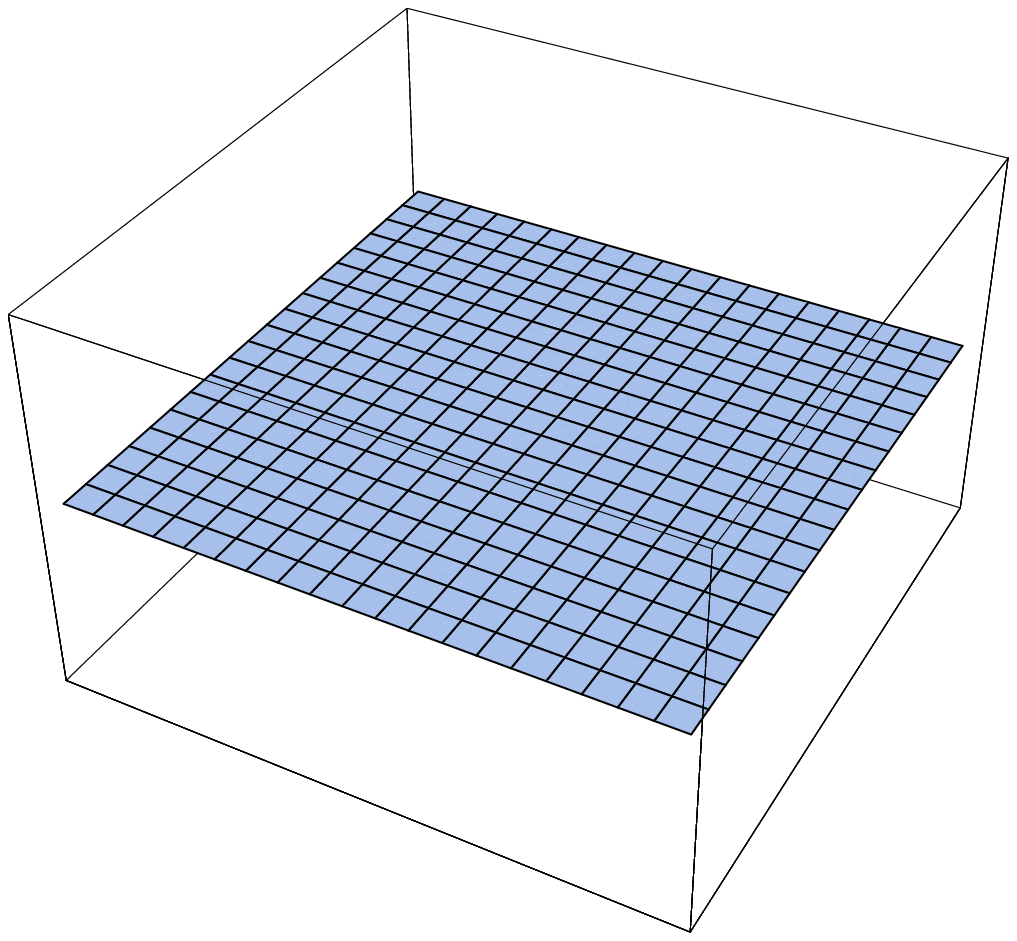,width=78mm}}
\put(73,50){\epsfig{file=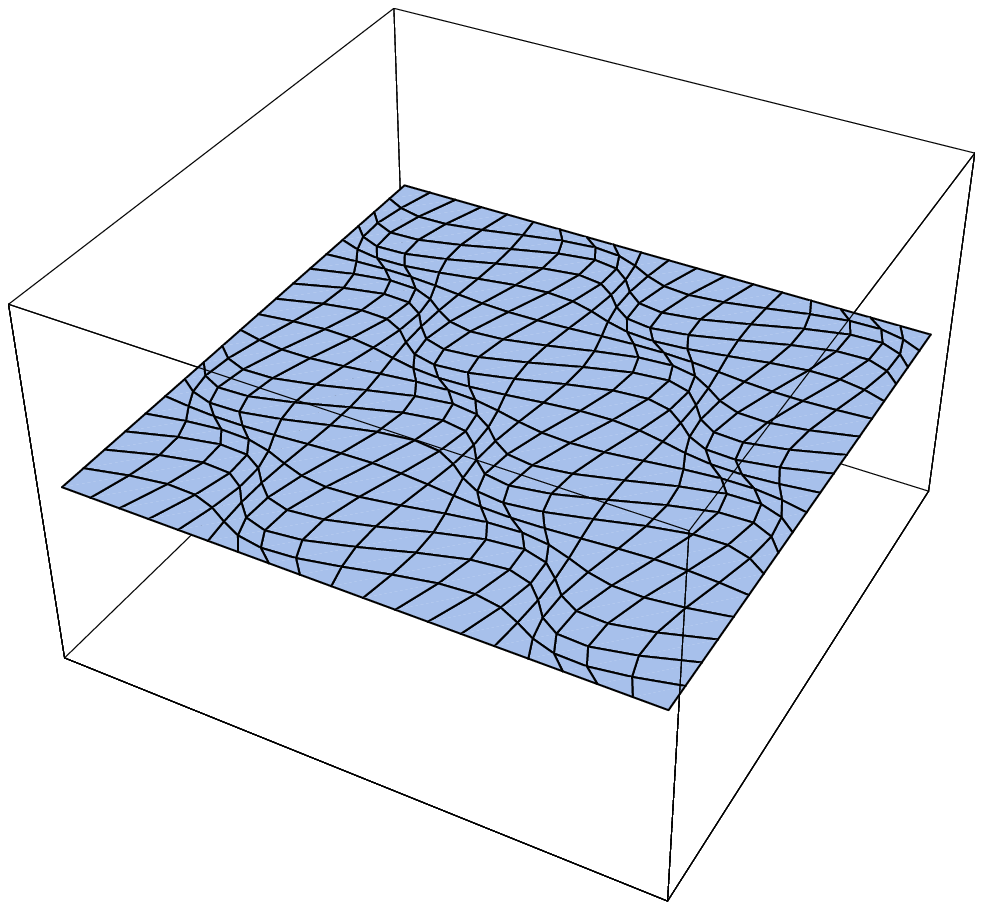,width=78mm}}
\put(0,0){\epsfig{file=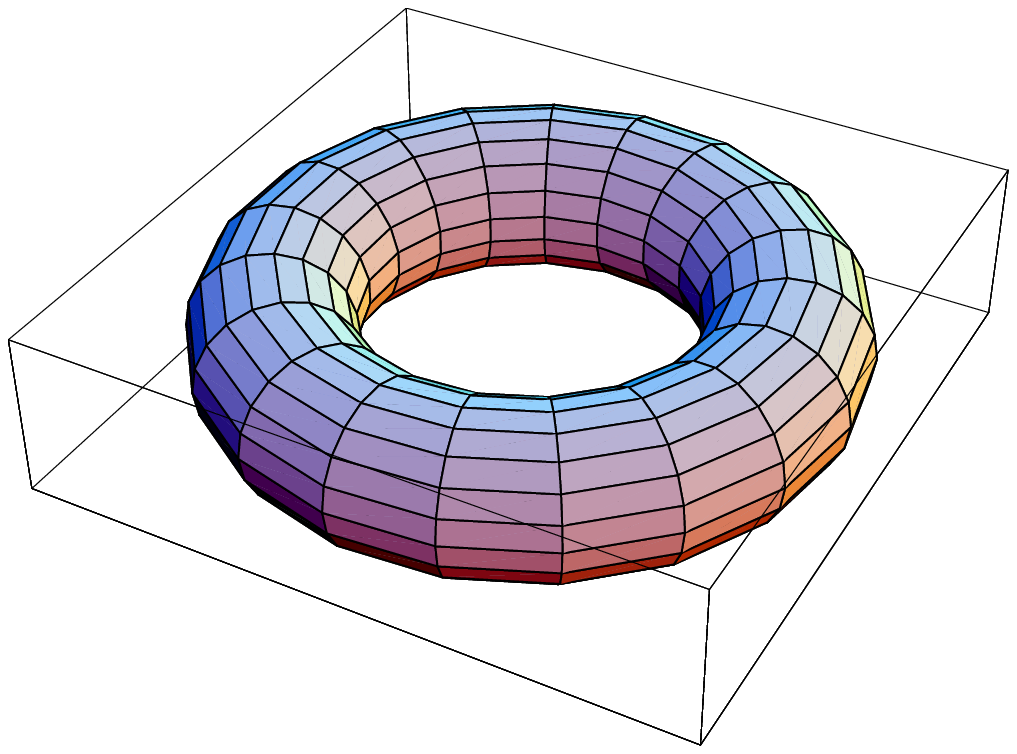,width=78mm}}
\put(73,0){\epsfig{file=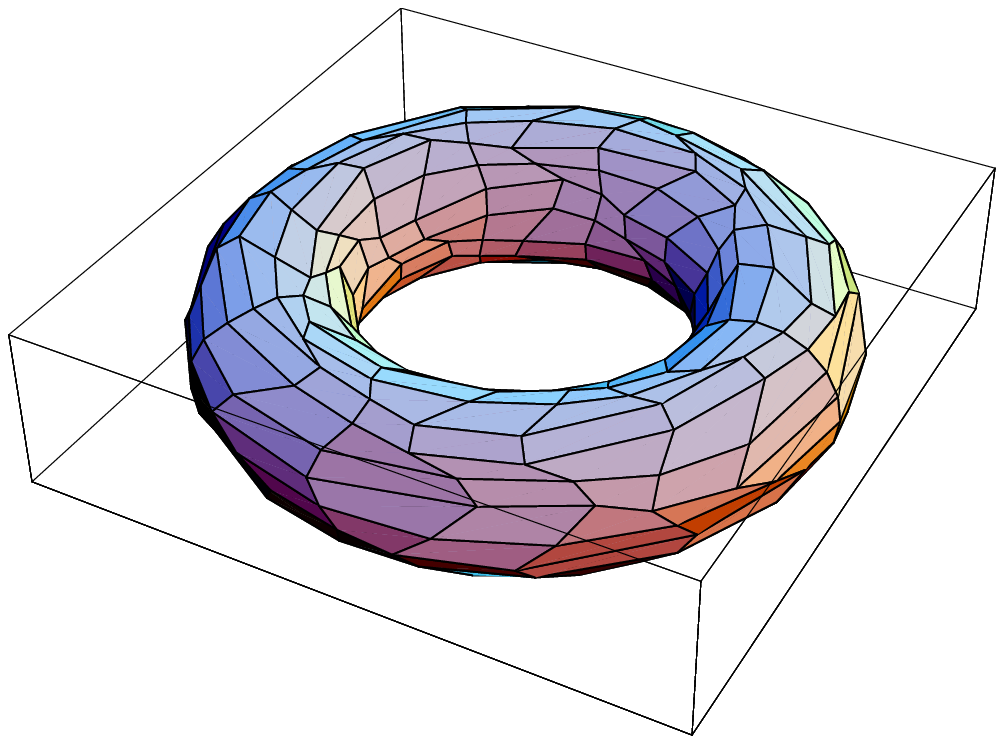,width=78mm}}
\end{picture}

\caption{Examples of tangentially deformed membranes. Mathematically
the surfaces on the right are the same as those on the left, but physically
they are different.}
\lbl{Wiggly}
\end{figure} }


\newcommand{\figIIIA}{
\setlength{\unitlength}{.5mm}
\begin{picture}(120,130)(-60,0)

\thicklines
\qbezier(10,60)(38,67)(60,81)
\qbezier(60,81)(64,55)(61,30)
\qbezier(19,12)(15,50)(10,60)
\qbezier(19,12)(29,20)(61,30)
\qbezier(32,12)(47,26)(81,26)
\qbezier(81,26)(84,33)(85,50)
\qbezier(85,50)(85,72)(91,86)
\qbezier(91,86)(63,81)(32,93)
\qbezier(32,93)(29,79)(30,66.4)
\qbezier(32,12)(33,17)(33,19.5)
\qbezier(84.4,40)(95,42.5)(103,48)
\qbezier(103,48)(96,83)(101,115)
\qbezier(101,115)(86,98)(66,85)

\put(64,54){$V_{p+1}$}
\put(40,95){$V_n$}
\put(103,111){${\hat V}_m$}
\put(125,76){$V_N$}

\end{picture}

\begin{figure}[h]
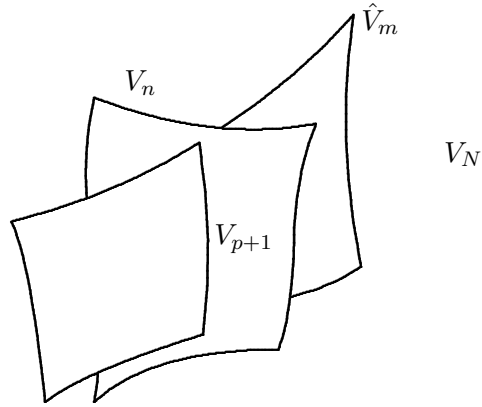

\caption{The intersection between two different branes $V_n$ and ${\hat V}_m$
can be a $p$-brane $V_{p+1}$.}
\end{figure} }


\newcommand{\figIIIB}{
\setlength{\unitlength}{.5mm}
\begin{picture}(120,130)(-30,0)

\thicklines
\qb(20,92)(50,81)(105,116)
\qb(105,116)(96,94)(100,51)
\qb(100,51)(84,39)(63,34)
\qb(63,34)(44,29)(26,15)
\qb(26,15)(14,60)(21,92)

\qb(27,113)(79,79)(131,93)
\qb(131,93)(123,87)(116,34)
\qb(116,34)(65,48)(39,42)
\qb(39,42)(47,65)(27,113)

\qb(80,72)(120,101)(125,120)
\qb(125,120)(121,95)(123,47)
\qb(123,47)(89,32)(76,19)
\qb(76,19)(80,31)(80,72)

\qb(78,123)(100,99)(139,96)
\qb(139,96)(134,72)(135,42)
\qb(135,42)(91,58)(67,55)
\qb(67,55)(76,74)(78,123)


\end{picture}

\begin{figure}[h]
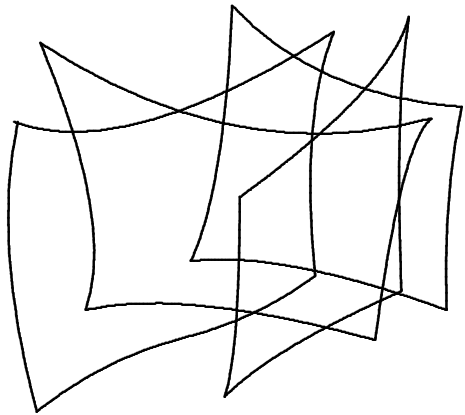

\caption{A system of many intersecting branes creates the bulk metric. In the
absence of the branes there is no bulk (no embedding space).}
\lbl{IIIB}
\end{figure} }


\newcommand{\figIIIC}{
\setlength{\unitlength}{1mm}
\begin{picture}(120,75)(-15,0)
\thicklines

\qb(14,69)(33,40)(51,46)
\qb(51,46)(58,48)(57,55)
\qb(57,55)(56.5,62)(44,60)
\qb(44,60)(24,53)(47,40)
\qb(47,40)(55,36)(66,34)
\qb(66,34)(65,17)(60,0)
\qb(14,69)(14,43)(8,18)
\qb(8,18)(33,16)(60,0)
\qb(38.2,46.2)(40,30)(33,13)
\qb(57,52)(55,45)(54,37)
\curve(34.7,52, 34.6,48)
\put(27,31){$V_{p+1}$}
\put(52,25){$V_n$}
\put(70,50){$V_N$}

\end{picture}
\begin{figure}[h]
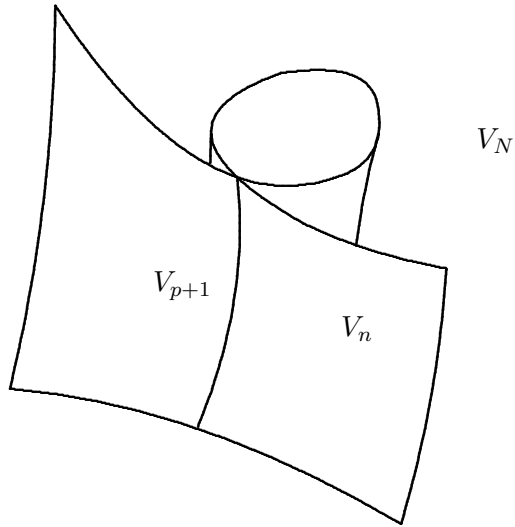

\caption{
Illustration of a self-intersecting brane. At the intersection $V_{p+1}$,
because of the contact interaction the stress--energy tensor on the brane $V_n$
is singular and it manifests itself as matter on $V_n$. The manifold
$V_{p+1}$ is a worldsheet swept by a $p$-brane and it is is minimal surface
(e.g., a geodesic, when $=0$) in $V_n$.}
\lbl{IIIC}
\end{figure}}


\newcommand{\figIIID}{
\setlength{\unitlength}{.7mm}
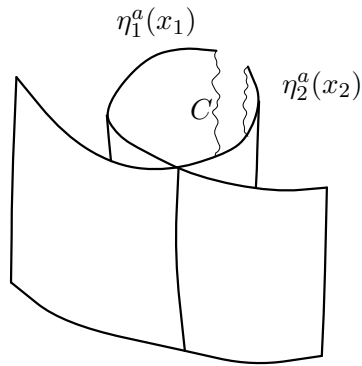
\begin{figure}[h]
\begin{picture}(120,80)(-50,0)

\put(39,53){$C$}
\put(25,70){$\eta_1^a (x_1)$}
\put(56.5,58){$\eta_2^a (x_2)$}

\spline(50,63)(49,62)(50,61)(49.5,60)(50.2,59)(49,58)(50,57)(49.2,56)
(49,55)(50,54)(49.5,53)(49,52)(49.5,51)(49.2,50)(50,49.7)
\spline(44,66)(43.5,65)(44.2,64)(44,63)(45,62)(44,61)(43,60)(43.5,59)
(44,58)(45,57)(44,56)(44.5,55)(43,54)(44,53)(44.6,52)(44,51)(43.6,50)
(44,49)(44.4,48)(45,47)(44.3,46)(43.7,45.8)

\thicklines
\spline(6,61)(15,50)(25,44)(35,43)(45,46)(50,49)(53,57)(50,63)
\spline(65,40)(55,39)(40,42)(30,47)(25,50)(22,55)(30,65)(38,67)(44,66)
\spline(5,22)(15,14)(38,9)
\spline(38,9)(51,6)(64,8)
\spline(36.8,43.5)(36,27)(38,9)
\spline(6,61)(5,40)(5,22)
\spline(65,40)(64.5,24)(64,8)
\spline(52.1,56.4)(51.7,47)(51.5,40)
\spline(23.2,53.7)(24,45)

\end{picture}

\caption{A self-intersecting worldsheet is cut into two pieces, described by
$\eta_1^a (x_1)$ and $\eta_2^a (x_2)$, which are glued together at a
submanifold $C$ where the boundary condition $\eta_1^a (x_1)|_{C} = 
\eta_2^a (x_2)|_{C}$ is imposed.}
\lbl{IIID}
\end{figure} }


\newcommand{\figIIIE}{
\setlength{\unitlength}{.6mm}
\begin{figure}[h]
\begin{picture}(120,130)(-30,-7)

\put(10,115){\large a)}

\thicklines

\qb(23,111)(43,73)(72,72)
\qb(72,72)(92,70)(91,86)
\qb(91,86)(91,96)(78,93)
\qb(78,93)(67,90)(69,80)
\qb(69,80)(71,68)(85,67)
\qb(85,67)(106,66)(99,86)
\qb(99,86)(93,100)(80,100)
\qb(80,100)(62,100)(60,88)
\qb(60,88)(56,71)(82,60)
\qb(82,60)(94,54)(106,69)
\qb(106,69)(111,76)(104,92)
\qb(104,92)(98,105)(80,106)
\qb(80,106)(57,108)(52,91)
\qb(52,91)(46,71)(85,47)
\qb(85,47)(95,40)(111,36)
\qb(111,36)(108,16)(110,-10)
\qb(110,-10)(70,-4)(51,16)
\qb(23,111)(27,80)(17,55)
\qb(17,55)(28,28)(51,16)

\qb(107.5,72)(102,53)(103,39)
\qb(90.8,81)(89.5,76)(90,67.5)
\qb(100,73)(98,69)(97.8,61.7)
\qb(68.7,82)(69,77)(68.6,72.6)
\qb(59.4,83)(60,80)(59.5,74.6)
\qb(51.3,87)(51,82)(50.4,78.8)

\linethickness{.5mm}
\qb(72.6,72)(72.9,68)(72.5,65.2)
\qb(62.8,73)(63.2,69)(62,64.4)
\qb(52.8,77.7)(56,45)(51,16)
\put(72.6,72){\circle*{2}}
\put(62.8,73){\circle*{2}}
\put(52.8,77){\circle*{2}}

\end{picture}

\setlength{\unitlength}{.6mm}
\begin{picture}(120,58)(-35,6)

\put(10,50){\large b)}

\spline(18,42)(16,37)(22,36)(26,38)(23,41)(21,38)(20.5,34)(25,33)(29,35)
(26,36)(24,30)(28,28)(32,29)(33,32)(30,33)(28,30)(30,25)(34,24)(38,26)(36,29)
(34,24)(36,21)(42,21)(43,25)(40,24)(40,19)(45,18.5)(50,22)(47,25)(47,20)
(50,18)(55,19)(58,22)(55,24)(56,18)(60,17)

\put(80,50){\large c)}

\spline(82,38)(82,32)(86,31)(89,34)(86,36)(84,30)(86,27)(91,28)(93,30)(90,31)
(88,27)(90,22)(94,21)(97,23)(95,26)(93,24)(95,20)
(99,19)(105,23)(102,26)(101,23)(104,19)(109,20)(113,25)
(110,28)(112,23)(118,24)(120,29)(118,33)(115,30)(118,27)(123,30)(125,35)(122,
40)(118,38)(121,35)(127,37)(125,45)(120,46)(116,43)(120,42)(122,44)(116,48)
(110,46)(108,42)(110,40)(114,42)(110,45)(105,42)(103,39)(106,35)(110,37)
(106,40)(103,37)(101,31)(105,30)(108,30)(110,33)(106,33)(104,27)(108,26)(111,29)
(108,30)(105,25)(107,20)(113,17)(116,20)(113,21)(111,18)(115,13)(120,14)(122,17)
(119,18)(118,14)(120,11)(126,11)(127,15)(125,16)(123,11)(125,8)(130,7)

\end{picture}

\caption{Some possible self-intersecting branes.}
\lbl{IIIE}
\end{figure}
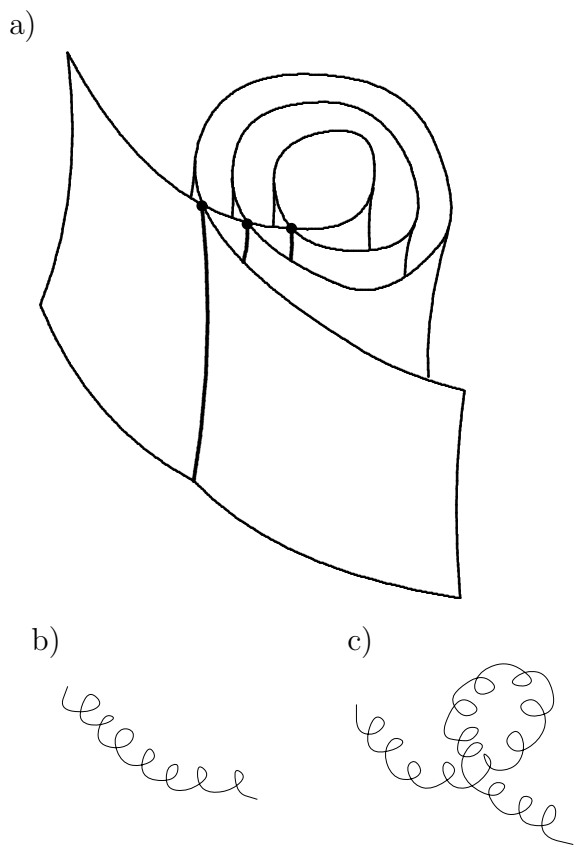 }


\newcommand{\figIIIF}{
\setlength{\unitlength}{1mm}
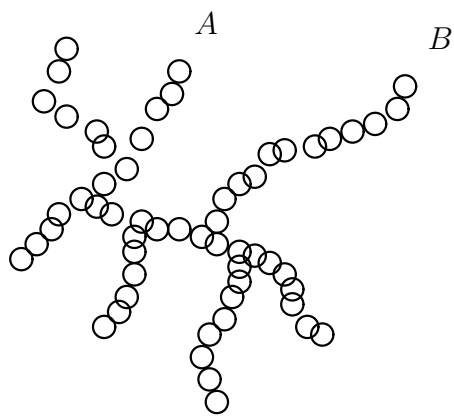
\begin{figure}[h]
\begin{picture}(120,70)(0,0)

\put(48,55){\large $A$}
\put(79,53){\large $B$}

\thicklines

\put(25,25){\circle{3}}\put(27,27){\circle{3}}\put(29,29){\circle{3}}
\put(30,31){\circle{3}}\put(33,33){\circle{3}}\put(36,35){\circle{3}}
\put(39,37){\circle{3}}\put(41,41){\circle{3}}\put(43,45){\circle{3}}
\put(45,47){\circle{3}}\put(46,50){\circle{3}}\put(31,53){\circle{3}}
\put(30,50){\circle{3}}\put(28,46){\circle{3}}\put(31,44){\circle{3}}
\put(35,42){\circle{3}}\put(36,40){\circle{3}}\put(35,32){\circle{3}}
\put(37,31){\circle{3}}\put(41,30){\circle{3}}\put(43,29){\circle{3}}
\put(46,29){\circle{3}}\put(49,28){\circle{3}}\put(51,27){\circle{3}}
\put(54,26){\circle{3}}\put(56,25.5){\circle{3}}\put(58,24.5){\circle{3}}
\put(60,23){\circle{3}}\put(61,21){\circle{3}}\put(61,19){\circle{3}}
\put(63,16){\circle{3}}\put(65,15){\circle{3}}
\put(36,16){\circle{3}}\put(38,18){\circle{3}}\put(39,20){\circle{3}}
\put(40,23){\circle{3}}\put(40,26){\circle{3}}\put(40,28){\circle{3}}
\put(51,6){\circle{3}}\put(50,9){\circle{3}}\put(49,12){\circle{3}}
\put(50,15){\circle{3}}\put(52,17){\circle{3}}\put(53,20){\circle{3}}
\put(54,22){\circle{3}}\put(54,24){\circle{3}}
\put(51,30){\circle{3}}\put(52,33){\circle{3}}\put(54,35){\circle{3}}
\put(56,36){\circle{3}}\put(58,39){\circle{3}}\put(60,39.5){\circle{3}}
\put(64,40){\circle{3}}\put(66,41){\circle{3}}\put(69,42){\circle{3}}
\put(72,43){\circle{3}}\put(75,45){\circle{3}}\put(76,48){\circle{3}}

\end{picture}

\caption{In a space of very large dimension separated point particles do not
feel gravitational interaction, since it is negligible. 
When two particles meet they form a bound
system which grows when it encounters other particles. There is (practically)
no force
between the `tails' (e.g., between the points $A$ and $B$). However, there is
tension within the tail. (The tail, of course, need not be 1-dimensional;
it could be a 2, 3 or higher-dimensional brane.)}
\lbl{IIIF}
\end{figure} }



\newcommand{\figIIIG}{
\setlength{\unitlength}{.5mm}
\begin{figure}[h]
\begin{picture}(210,175)(0,10)

\put(150,25){$\eta_{\rm C}^a (\tau, x)$}
\put(174,140){\vector(1,0){24}}
\put(174,140){\vector(0,1){24}}
\put(174,140){\vector(-1,-1){13}}
\put(156,131){$\eta^1$}\put(201,145){$\eta^2$}\put(177,165){$\eta^0$}
\put(73,22){$\eta_{\rm C}^a (\tau, x)$}

\put(49,68){\vector(1,0){41}}
\put(49,68){\vector(0,1){41}}
\put(51,113){$\eta^0$}
\put(92,72){$\eta^1$}

\spline(64,111)(65,95)(63,84)(65,71)(66,55)(65,38)(65.5,30)
\spline(65.5,30)(68,31)(73,30)(77,32)(80,31)(83,32)
\spline(83,32)(84,52)(82,70)(82.5,91)(82,113)
\spline(82,113)(75,112)(74,110.5)(64,111)

\curvedashes[1mm]{0,2,1}
\curve(123,47, 130,49, 140,54)
\curve(140,54, 143,81, 157,125)
\curve(140,54, 157,50, 187,49)
\curvedashes{}
\curve(123,47, 125,76, 132,110)
\curve(132,110, 159,107, 181,106)
\curve(181,106, 176,86, 169,38)
\curve(123,47, 145,41, 169,38)
\curve(169,38, 180,43, 187,49)
\curve(187,49, 190,75, 201,122)
\curve(201,122, 192,114, 181,106)
\curve(201,122, 177,122, 157,125)
\curve(157,125, 146,116, 132,110)

\thicklines

\curve(143,123, 135,78, 133,35)
\curve(133,35, 153,34, 179,31)
\curve(179,31, 183,71, 191,116)
\curve(191,116, 175,115, 143,123)

\spline(72,115)(74,105)(72,90)(74,75)(73,60)(75,44)(76,28)

\thinlines

\put(78,31){\line(30,17){5}}
\put(68,31){\line(30,17){15.5}}\put(65,35){\line(30,17){18.5}}
\put(65,41){\line(30,17){18.5}}
\put(66,46){\line(30,17){17}}
\put(66,52){\line(30,17){17}}
\put(66,59){\line(30,17){16.5}}\put(65,65){\line(30,17){17}}
\put(65,70){\line(30,17){17}}\put(65,75){\line(30,17){17}}
\put(65,81){\line(30,17){17}}\put(65,86){\line(30,17){17}}
\put(65,92){\line(30,17){17}}\put(65,98){\line(30,17){17}}
\put(64,103){\line(30,17){17}}\put(64,109){\line(30,17){4}}

\put(123,51){\line(30,-32){4.5}}\put(123,54){\line(30,-32){8}}
\put(123,58){\line(30,-32){12}}\put(123.5,62){\line(30,-32){16.5}}
\put(123.6,65){\line(30,-32){21.5}}\put(124,68){\line(30,-32){24}}
\put(125,71){\line(30,-32){27}}\put(125,74){\line(30,-32){30.5}}
\put(125.5,78){\line(30,-32){35}}\put(126,81){\line(30,-32){37}}
\put(126.5,83.5){\line(30,-32){42}}\put(127,86.5){\line(30,-32){43}}
\put(128,90){\line(30,-32){46}}\put(129,94){\line(30,-32){48}}
\put(129.5,99){\line(30,-32){52}}\put(130,102.5){\line(30,-32){52}}
\put(131,105.5){\line(30,-32){53.5}}\put(132,108){\line(30,-32){54}}
\put(133,110.5){\line(30,-32){54.5}}\put(136,112){\line(30,-32){52}}
\put(138.5,113){\line(30,-32){49}}\put(141,114){\line(30,-32){47}}
\put(143,114.5){\line(30,-32){46}}\put(146,116){\line(30,-32){43}}
\put(148,118){\line(30,-32){42}}\put(151,119.5){\line(30,-32){40}}
\put(154,121){\line(30,-32){37}}\put(155.5,122.5){\line(30,-32){36.5}}
\put(157,125){\line(30,-32){35}}\put(163,123){\line(30,-32){30}}
\put(167,123){\line(30,-32){26}}\put(171,122.6){\line(30,-32){24}}
\put(175,122){\line(30,-32){21}}\put(179,122){\line(30,-32){18}}
\put(185,122){\line(30,-32){13}}\put(190,122){\line(30,-32){9}}
\put(198,121.7){\line(30,-32){2.5}}

\end{picture}

\caption{Quantum mechanically a state of our space--time membrane
${\cal V}_4$ is given by a wave packet which is a functional of ${\eta}^a (x)$.
Its ``centre" ${\eta}_{\mbox{\rm c}}^a (\tau,x)$ is the expectation value
$\langle {\eta}^a (x) \rangle$ and moves according to the classical equations of motion
(as derived from the action (\ref{III3.19}).}
\lbl{IIIG}
\end{figure}
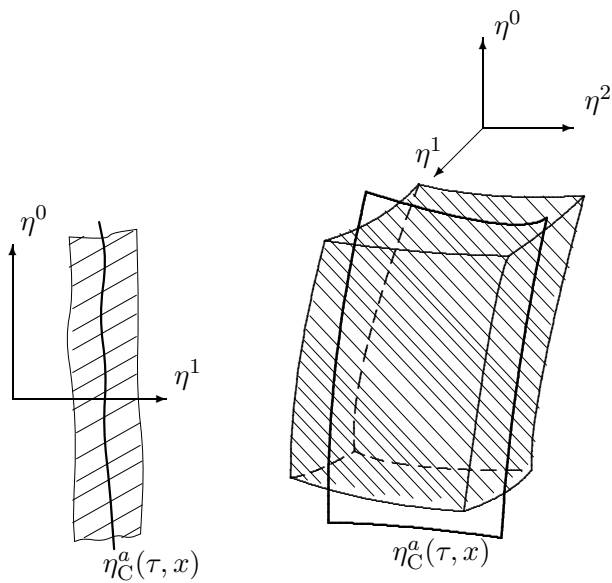 }


\newcommand{\figIIIH}{
\setlength{\unitlength}{.5mm}
\begin{figure}[h]
\begin{picture}(210,175)(0,0)

\put(107,27){\vector(1,0){22}}
\put(107,27){\vector(0,1){22}}
\put(107,27){\vector(-1,-1){15}}
\put(100,50){$\eta^0$}
\put(131,30.5){$\eta^2$}
\put(82.5,15){$\eta^1$}

\put(30,118){\vector(1,0){22}}
\put(30,118){\vector(0,1){22}}
\put(33,142){$\eta^0$}
\put(53,122){$\eta^1$}

\put(57.5,86.5){$B$}
\put(73,80){$\langle V_4 \rangle $}
\put(93,125){$P$}
\put(94,97){$\eta^a (x)$}
\drawline(68.4,94)(92.4,96.4)

\spline(69,158)(69,152)(72,145)(71,134)(77,122)(79,114)(86,107)(87,98)(87,89)
\spline(82,160)(81,150)(78,140)(77,120)(75,107)(72,98)(73,90)
\spline(87,159)(86,149)(83,143)(80,135)(77,126)(73,120)(71,110)(71,102)
(68,96)(68,90)

\curvedashes[1mm]{0,1,1}
\closecurve(77,133, 68,125, 77,118, 85,125)
\closecurve(179,80, 188,90, 197,85, 190,73)

\thicklines

\spline(90,160)(90,132)(88,129)(85,128)(80,125)(85,123)(89,121)(90,119)(90,90)
\spline(65,160)(65,131)(67,129)(70,128)(74.5,125)(70,123)(66,121)(65,119)(65,90)
\spline(77,160)(77,90)

\spline(123,117)(123,100)(123,90)(127,88)(134,85.5)(127,82)(123,79)(123,70)
(123,55)
\spline(123,55)(150,50)(179,48)
\spline(179,48)(189,53)(195,57)
\spline(195,57)(195,70)(194,80)(192,81)(190,83)(193,86)(195,89)(195,100)
(195,124)
\spline(195,124)(188,118)(179,113)
\spline(179,113)(179,90)(180,86)(182,85)(187,81)(184,78)(180,75)(179,72)
(179,62)(179,48)
\spline(123,117)(155,112)(179,113)
\spline(123,117)(131,121)(134,124)
\spline(134,124)(135,134)(135,141)
\spline(135,141)(169,136)(189,139)
\spline(188,139)(188,119)(188.4,105)(188.2,82)(189.5,65)(188,40)
\spline(188,40)(160,41)(136,42)
\spline(136,42)(136,52)
\spline(134,124)(160,121)(188,119)
\spline(195,124)(188.2,124.5)

\put(200,83){$P$}
\put(148,81){present}
\put(150,35){past}
\put(150,142){future}
\put(115,66){$B$}
\put(188,142){$\langle V_4 \rangle$}

\end{picture}

\caption{The wave packet representing a quantum state of a space--time membrane
is localized within an effective boundary $B$. The form of the latter may be
such that the localization is significantly sharper around a space-like surface
$\Sigma$.}
\lbl{IIIH}
\end{figure}
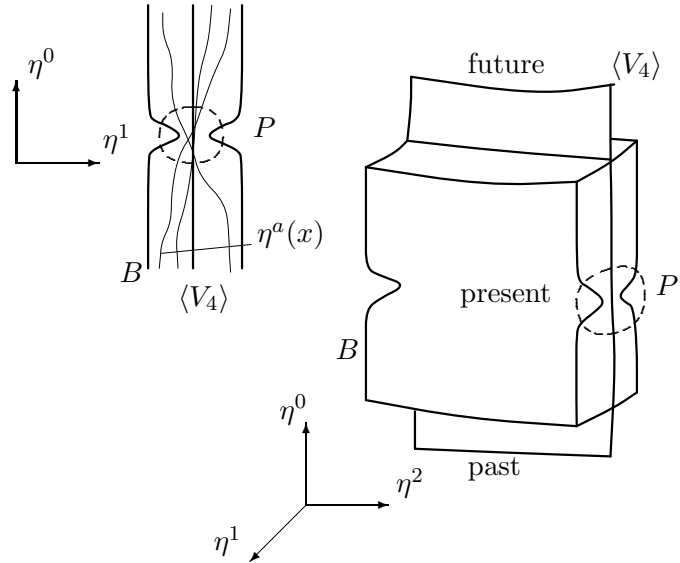 }


\newcommand{\figIIII}{
\setlength{\unitlength}{.8mm}
\begin{figure}[h]
\begin{picture}(120,90)(-5,0)
\put(6,40){$V_N$}
\put(67,62){$P$}
\put(90,7){$B$}
\put(0,0){\epsfig{file=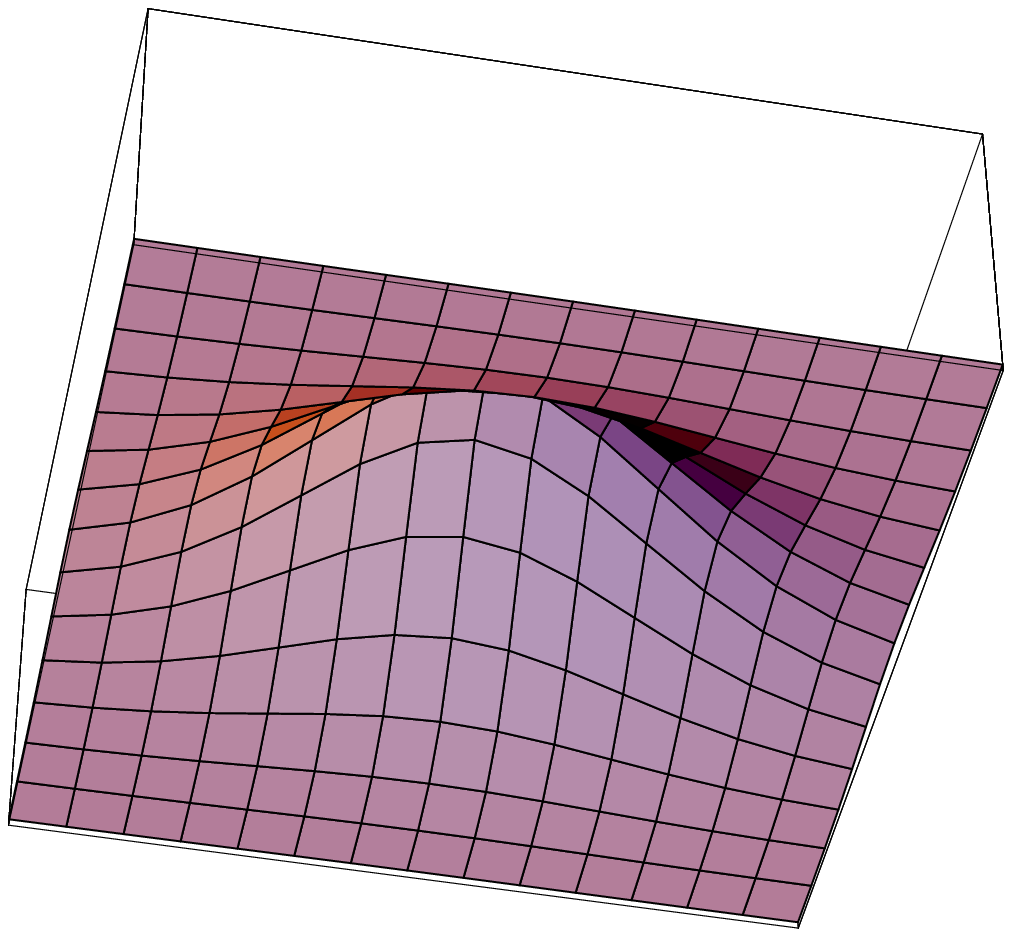,width=87mm,angle=8.5}}
\end{picture}

\caption{Illustration of a wave packet with a region of sharp localization $P$.}
\end{figure} }



\newcommand{\boksIVA}{
\setlength{\unitlength}{1cm}

\nnnn \begin{picture}(12.5,13.5)

\put(0,0){\framebox(12.5,13.5){
\begin{minipage}[t]{11.5cm}

\centerline{\bf Box 12.1: Human language and multiverse\footnote{
This idea was earlier discussed in ref. \cite{71}. Later it was also mentioned
by Deutsch \cite{86}.}}

\vs{5mm}

\hs{4mm}In the proposed brane world model spacetime, together with matter, is
represented by a 4-dimensional self-intersecting surface $V_4$. An
observer associated with a $V_4$ distinguishes present, past, and
future events. Because of the quantum principle an observer is, in fact,
associated not with a definite $V_4$, but with a corresponding wave function.
The latter takes into account all {\it possible} $V_4$s entering the
superposition.

\hs{4mm}We see that within the conceptual scheme of the proposed brane world
model all the principal tenses of human language ---present, past, future
tenses, and {\it conditional}--- are taken into account. In our human
conversations we naturally talk not only about the actual events
(present, past, future), but also about possible events, i.e., those
which could have occurred (conditional). According to Piaget \cite{105}
a child acquires the ability of formal logical thinking, which includes
use of alternatives and conditional, only at an advanced stage in his
mental development. Reasoning in terms of possible events is a sign
that an individual has achieved the highest stage on the Piaget
ladder of conceptual development.

\hs{4mm}Now, since the emergence of quantum mechanics, even in physics, we are
used to talking about possible events which are incorporated in the
wave function. According to the Everett interpretation of quantum
mechanics as elaborated by Deutsch, those possible events (or better
states) constitute the {\it multiverse}.

\end{minipage} }}
\end{picture} }


\newcommand{\figIVA}{

\setlength{\unitlength}{.8mm}
\begin{figure}[h]
\begin{picture}(120,80)(-12,10)

\curve(30,55, 40,54, 50,50, 60,45, 70,40, 80,37, 90,36, 100,37)
\spline(30,65)(40,64)(42,58)(45,58)(50,59)(55,57)(56,52)(53,55)(60,54)
(65,50)(70,46)(75,51)(80,44)(85,46)(90,42)(95,43)(100,47)
\spline(30,46)(35,47)(40,45)(45,48)(50,40)(55,42)(60,35)(65,36)(70,37)
(75,39)(80,32)(85,31)(90,34)(95,35)(100,36)
\thicklines
\closecurve(30,40, 36,51, 36,61, 30,70, 22,61, 21,50)
\curve(30,70, 40,69, 50,65, 60,60, 70,55, 80,52, 90,51, 100,52)
\closecurve(100,22, 106,33, 106,43, 100,52, 92,43, 91,32)
\curve(30,40, 40,39, 50,35, 60,30, 70,25, 80,22, 90,21, 100,22)
\put(80,70){$V_N$}

\end{picture}

\caption{An illustration of a wave packet describing a 3-brane.
Within the effective region of localization any brane configuration
is possible. The wavy lines indicate such possible configurations.}
\label{12.1}
\end{figure} }


\newcommand{\figIVB}{
\setlength{\unitlength}{.8mm}
\begin{figure}[h]
\begin{picture}(120,90)(-10,10)

\spline(40,80)(40,72)(44,69)(47,70)(44,71)(43,66)(46,62)(43,58)(45,53)
(46,56)(42,54)(43,48)(45,45)(41,40)(44,35)(38,32)(39,27)
\spline(61,76)(65,65)(64,58)(61,52)(65,45)(58,38)(55,28)
\spline(72,77)(77,68)(75,65)(80,60)(75,56)(79,53)(74,49)(78,45)
(75,41)(76,36)(74,33)
\thicklines
\spline(26,84)(30,84.5)(35,85)(40,86)(45,88)
\spline(45,88)(50,86)(55,84)(60,82)(65,81)(70,80.5)(75,80)(80,80)(90,79)
\spline(90,79)(85,76)(80,73)(74,69)
\spline(74,69)(65,70)(55,71)(45,73)(35,76)(26,84)
\spline(26,84)(27,70)(27,59)(25,50)(24,40)(22,30)
\spline(22,30)(30,26)(40,22)(50,20)(63,19)
\spline(63,19)(67,30)(68,40)(70,50)(72,60)(74,69)
\spline(63,19)(70,23)(78,25)
\spline(78,25)(79,35)(82,45)(85,60)(90,79)
\spline(45,88)(45,81)(45,73)

\put(20,66){$B$}
\put (100,50){$V_N$}

\end{picture}

\caption{A 4-brane wave packet localized within an effective
boundary $B$. A wavy line represents a possible 4-brane.}
\lbl{12.2}
\end{figure} }

\part[Point Particles]
{Point Particles}

\chapter[The spinless point particle]%
{The spinless point particle}%




\section{Point particles versus worldlines}%

The simplest objects treated by physics are point particles. Any object, if
viewed from a sufficiently large scale, is approximately a point particle.
The concept of an exact point particle is an idealization which holds in
the limit of infinitely large scale from which it is observed. Equivalently,
if observed from a finite scale its size is infinitely small. According
to the special and general theory of relativity the arena in which
physics takes place is not a 3-dimensional space but the 4-dimensional
space with the pseudo--Euclidean signature, say (+ - - -), called
{\it spacetime}. In the latter space the object of question is actually
a worldline, a 1-dimensional object, and it appears as a point particle
only from the point of view of a space-like 3-surface which intersects
the worldline.

Kinematically, any worldline (i.e., a curve in spacetime) is possible, but not
all of them can be realized in a dynamical situation. An obvious question
arises of which amongst all kinematically possible worldlines in spacetime
are actually possible. The latter worldlines are solutions of certain
differential equations.

An action from which one can derive the equations of a worldline is
proportional to the length of the worldline:
\be
    I[X^{\mu}] = m \int {\dd} \tau ({\dot X}^{\mu} {\dot X}^{\nu} g_{\mu \nu}
    )^{1/2} .
\lbl{2.1.1}
\ee
Here $\tau$ is an arbitrary parameter associated with a point on the
worldline, the variables $X^{\mu}$ denote the position of that point in spacetime,
the fixed constant $m$ is its mass, and $g_{\mu \nu} (x)$ is the metric tensor
which depends on the position $x \equiv x^{\mu}$ in spacetime
\inxx{mass,fixed}.

Variation of the action (\ref{2.1.1}) with respect to $X^{\mu}$ gives
{\it the geodesic equation}\inxx{geodesic equation}
\be
   E^{\mu} \equiv {1\over {\sqrt{{\dot X}^2}}} {{\dd}\over {{\dd} \tau}}
   \left ( {{{\dot X}^{\mu}}\over {\sqrt{{\dot X}^2}}} \right ) +
   \Gamma_{\alpha \beta}^{\mu} {{{\dot X}^{\alpha} {\dot X}^{\beta}}\over
   {{\dot X}^2}} = 0
\lbl{2.1.2}
\ee
where
\be
 \Gamma_{\alpha \beta}^{\mu} \equiv \mbox{$1\over 2$} g^{\mu \rho}
(g_{\rho \alpha , \beta} + g_{\rho \beta , \alpha} - g_{\alpha \beta,
\rho})  
\lbl{2.1.3}
\ee
is the affinity \inxx{affinity} of spacetime. We often use the shorthand notation
${\dot X}^2 \equiv {\dot X}^{\mu} {\dot X}_{\mu}$, where indices are
lowered by the metric tensor $g_{\mu \nu}$.

The action (\ref{2.1.1}) is invariant with respect to transformations
of spacetime coordinates $x^{\mu} \rightarrow {x'}^{\mu} = f^{\mu} (x)$
(diffeomorphisms) and with respect to reparametrizations $\tau \rightarrow
f(\tau)$. A consequence of the latter invariance is that the equations of
motion (\ref{2.1.2}) are not all independent, since they satisfy the identity
\be
    E^{\mu} {\dot X}_{\mu} = 0
\lbl{2.1.3a}
\ee
Therefore the system is under-determined: there are more variables
$X^{\mu}$ than available equations of motion.

From (\ref{2.1.1}) we have
\be
    p_{\mu} {\dot X}^{\mu} - m ({\dot X}^{\mu} {\dot X}_{\mu})^{1/2} = 0
\lbl{2.1.4}
\ee
or
\be
    H \equiv p_{\mu} p^{\mu} - m^2 = 0 ,
\lbl{2.1.5}
\ee
where
\be
   p_{\mu} = {{\p L} \over {\p {\dot X}^{\mu}}} = {{m {\dot X}_{\mu}} \over
   {({\dot X}^{\nu} {\dot X}_{\nu})^{1/2}}} .
\lbl{2.1.6}
\ee
This demonstrates that the canonical momenta
\inxx{canonical momentum} $p_{\mu}$ are not
all independent, but are subjected to the constraint 
\inxx{constraint,for point particle} (\ref{2.1.5}). A general
theory of constrained systems is developed by Dirac \ci{19}. An alternative
formulation is owed to Rund \ci{20}.

We see that the Hamiltonian
\inxx{Hamiltonian,for point particle}  for our system  ---which is a worldline in
spacetime--- is zero. This is often interpreted as there being no
evolution  in spacetime: worldlines are frozen and they exist `unchanged'
in spacetime.

Since our system is under-determined we are free to choose a relation
between the $X^{\mu}$. In particular, we may choose $X^0 = \tau$, then the
action (\ref{2.1.1}) becomes a functional of the reduced
number of variables\footnote{
In flat spacetime we may have $g_{00} = 1$, $g_{rs} = -\delta_{rs}$, 
and the action
(\ref{2.1.7}) takes the usual special relativistic form $I = m \int {\dd} \tau
\, \sqrt{1 - {\dot {\bf X}}^2}$, where ${\dot {\bf X}} \equiv  {\dot X}^i$.}
\be
   I[X^i] = m \int {\dd} \tau (g_{00} + 2 {\dot X}^i g_{0i} + {\dot X}^i
   {\dot X}^i g_{ij})^{1/2} .
\lbl{2.1.7}
\ee
All the variables $X^i$, $i = 1,2,3$, are now independent and they
represent {\it motion}
\inxx{motion,in 3-space} of a point particle in 3-space. From the 3-dimensional
point of view the action (\ref{2.1.1}) describes a dynamical system, whereas
from the 4-dimensional point of view there is no dynamics.

Although the reduced action (\ref{2.1.7}) is good as far as dynamics is concerned,
it is not good from the point of view of the theory of relativity: it is
not manifestly covariant with respect to the coordinate transformations of
$x^{\mu}$, and in particular, with respect to Lorentz transformations.

There are several well known classically equivalent forms of the point particle
action. One of them is the second order or Howe--Tucker action \ci{21}
\inxx{Howe--Tucker action,for point particle}
\be
     I[X^{\mu},\lambda] = {1\over 2} \int {\dd} \tau \left (
     {{{\dot X}^{\mu} {\dot X}_{\mu}} \over {\lambda}} + \lambda m^2
     \right )
\lbl{2.1.8}
\ee
which is a functional not only of the variables $X^{\mu}(\tau)$ but
also of $\lambda$ which is the Lagrange multiplier giving the relation
\be
   {\dot X}^{\mu} {\dot X}_{\mu} = \lambda^2 m^2 .
\lbl{2.1.9}
\ee
Inserting (\ref{2.1.9}) back into the Howe--Tucker action (\ref{2.1.8}) we obtain
the minimal length action (\ref{2.1.1}). The canonical momentum
\inxx{canonical momentum,for point particle} is $p_{\mu} =
{\dot X}_{\mu}/\lambda$ so that (\ref{2.1.9}) gives the constraint (\ref{2.1.5}).

Another action is the first order, or phase space action,
\be
   I[X^{\mu}, p_{\mu}, \lambda] = \int
   {\dd} \tau  \left ( p_{\mu} {\dot X}^{\mu}
   - {{\lambda}\over 2} (p_{\mu} p^{\mu} - m^2) \right ) 
\lbl{2.1.10}
\ee       
which is also a functional of the canonical momenta $p_{\mu}$. Varying
(\ref{2.1.10}) with respect to $p_{\mu}$ we have the relation between
$p^{\mu}$, ${\dot X}^{\mu}$, and $\lambda$:
\be
    p^{\mu} = {{{\dot X}^{\mu}}\over {\lambda}}
\lbl{2.1.10a}
\ee
which, because of (\ref{2.1.9}), is equivalent to (\ref{2.1.6}).

The actions (\ref{2.1.8}) and (\ref{2.1.10}) are invariant under reparametrizations
$\tau' = f(\tau)$, provided $\lambda$ is assumed to transform according
to $\lambda' = ({\dd} \tau/{\dd} \tau') \, \lambda$.

The Hamiltonian \inxx{Hamiltonian,for point particle}
corresponding to the action (\ref{2.1.8}),
\be
     H = p_{\mu} {\dot X}^{\mu} - L ,
\lbl{2.1.10b}
\ee
is identically zero, and therefore it does not generate any genuine
evolution in $\tau$. 

Quantization of the theory goes along several possible lines \ci{22}.
Here let me mention\inxx{Gupta--Bleuler quantization}
the Gupta--Bleuler quantization.
Coordinates and momenta become operators satisfying the commutation relations
\be
    [X^{\mu}, p_{\nu}] = i {\delta^{\mu}}_{\nu} \; , \quad 
    [X^{\mu}, X^{\nu}] = 0 \; , \quad [p_{\mu}, p_{\nu}] = 0 .
\lbl{2.1.11}
\ee
The constraint (\ref{2.1.5}) \inxx{constraint,for point particle}
is imposed on state vectors
\be
    (p^{\mu} p_{\mu} - m^2) |\psi \rangle = 0 .
\lbl{2.1.12}
\ee

Representation of the operators is quite straightforward if
spacetime is {\it flat}, so we can use a coordinate system
in which the metric tensor is everywhere of the Minkowski type, $g_{\mu \nu}
= \eta_{\mu \nu}$ with signature (+ - - -). A useful representation is that
in which the coordinates $x^{\mu}$ are diagonal and $p_{\mu} = - i \p_{\mu}$,
where $\p_{\mu} \equiv \p/\p x^{\mu}$. Then the constraint relation
(\ref{2.1.12}) becomes the Klein--Gordon equation \inxx{Klein--Gordon equation}
\be
   (\p_{\mu} \p^{\mu} + m^2) \psi = 0 .
\lbl{2.1.13}
\ee

Interpretation of the wave function $\psi(x) \equiv \langle x| \psi \rangle$,
$x \equiv x^{\mu}$, is not straightforward. It cannot be interpreted as the
one particle probability amplitude. Namely, if $\psi$ is complex valued then to
(\ref{2.1.13}) there corresponds the non-vanishing current
\be
   j_{\mu} = {i\over {2 m}} (\psi^* \p_{\mu} \psi - \psi \p_{\mu} \psi^*) .
\lbl{2.1.14}
\ee
The components $j_{\mu}$ can all be positive or negative, and there is
the problem of which quantity then serves as the probability density.
Conventionally the problem is resolved by switching directly into the
second quantized theory and considering $j_{\mu}$ as the charge--current
operator. I shall not review in this book the conventional relativistic
quantum field theory, since I am searching for an alternative approach,
much better in my opinion, which  has actually already been proposed and
considered in the literature \ci{1}--\ci{12b} under various names
such as the Stueckelberg
theory, the unconstrained theory, the parametrized theory, etc.\,. In the rest
of this chapter I shall discuss various aspects of the {\it unconstrained
point particle theory}.

\section{Classical theory}

It had been early realized that instead of the action (\ref{2.1.1}) one can use
the action
\be
   I[X^{\mu}] = {1\over 2} \int {\dd} \tau {{{\dot X}^{\mu} {\dot X}_{\mu}} 
   \over {\Lambda}} ,
\lbl{2.2.1}
\ee
where $\Lambda(\tau)$ is a fixed function  of $\tau$, and can be a
constant. The latter action, since being quadratic in velocities, is much more
suitable to manage, especially in the quantized version of the theory.

There are two principal ways of interpreting (\ref{2.2.1}).

{\it Interpretation (a)} We may consider (\ref{2.2.1}) as a gauge fixed action
(i.e., an action in which reparametrization of $\tau$ is fixed), equivalent
to the constrained action (\ref{2.1.1}).

{\it Interpretation (b)} Alternatively, we may consider (\ref{2.2.1}) as an
unconstrained action.

The equations of motion obtained by varying (\ref{2.2.1}) with respect to
$X^{\mu}$ are
\be   
      {1\over {\Lambda}} {{\dd}\over {\dd \tau}} \left ( 
      {{{\dot X}^{\mu}}\over {\Lambda}}\right )
      + \Gamma_{\alpha \beta}^{\mu} \, {{{\dot X}^{\alpha}
     {\dot X}^{\beta}} \over {\Lambda^2}}  = 0 .
\lbl{2.2.2}
\ee
This can be rewritten as
\be
    {{\sqrt{{\dot X}^2}}\oo {\Lambda^2}} {{\dd} \over {{\dd} \tau}} \left (
     { {{\dot X}^{\mu}} \over {\sqrt{{\dot X}^2}} }  \right ) +
     \Gamma_{\alpha \beta}^{\mu} { { {\dot X}^{\alpha} {\dot X}^{\beta}}
     \over {\Lambda^2}} + {{{\dot X}^{\mu}}\oo {\sqrt{{\dot X}^2}}}
     \, {{\dd} \oo {{\dd} \tau}} \, \left ( {{\sqrt{{\dot X}^2}} \oo 
     {\Lambda}} \right ) = 0 .
\lbl{2.2.3}
\ee
Multiplying the latter equation by ${\dot X}^{\mu}$, summing over $\mu$
and assuming ${\dot X}^2 \neq 0$  we have
\be
   {{\sqrt{{\dot X}^2}}\oo {\Lambda}} \, {{\dd} \oo {{\dd} \tau}} \, \left ( 
   {{\sqrt{{\dot X}^2}} \oo {\Lambda}} \right )  
   = {1\oo 2} {{\dd} \oo {{\dd} \tau}} \left ( {{{\dot X}^2}\oo {\Lambda^2}}
   \right ) =  0 .
\lbl{2.2.4}   
\ee
Inserting (\ref{2.2.4}) into (\ref{2.2.3}) we obtain that (\ref{2.2.2})
is equivalent to the
geodetic equation (\ref{2.1.1}).

According to {\it Interpretation (a)} the parameter $\tau$ in (\ref{2.2.2}) is not
arbitrary, but it satisfies eq.\,(\ref{2.2.4}). In other words,
eq.\,(\ref{2.2.4}) is 
{\it a gauge fixing equation} telling us the ${\dot X}^2 = constant \times
\Lambda^2 $,
which means that $({\dd} X^{\mu} {\dd} X_{\mu})^{1/2} \equiv {\dd} s =constant 
\times \Lambda \,  {\dd} \tau$, or, when ${\dot \Lambda} = 0$ that
$s = constant \times \tau$; the parameter $\tau$ is thus proportional to the
proper time $s$.

According to {\it Interpretation (b)} eq.\,(\ref{2.2.4}) tells us that mass is a
constant of motion. Namely, the canonical momentum
\inxx{canonical momentum,point particle} belonging to the
action (\ref{2.2.1}) is
\be
    p_{\mu} = {{\p L} \oo {\p {\dot X}^{\mu}}} = {{{\dot X}_{\mu}}\oo {\Lambda}}
\lbl{2.2.5}
\ee
and its square is
\be 
      p^{\mu} p_{\mu} \equiv M^2 = {{{\dot X}^{\mu} {\dot X}_{\mu}}\oo 
      {\Lambda^2}} .
\lbl{2.2.6}
\ee
From now on I shall assume {\it Interpretation (b)} and consider (\ref{2.2.1}) as
{\it the unconstrained action}. It is very convenient to take $\Lambda$
as a constant. Then $\tau$ is just proportional to the proper time $s$;
if $\Lambda = 1$, then it is the proper time.   However, instead of $\tau$, we
may use another parameter $\tau' = f(\tau)$. In terms of the new parameter
the action (\ref{2.2.1}) reads
\be  
  {1\oo 2} \int {\dd} \tau' \, {{{\dd} X^{\mu}}\oo {{\dd} \tau'}}\, 
   {{{\dd} X_{\mu}}\oo {{\dd} \tau'}} {1\oo {\Lambda'}} = I'[X^{\mu}] ,
\lbl{2.2.7}
\ee
where
\be
      \Lambda' = {{{\dd}\tau}\oo {{\dd} \tau'}} \, \Lambda   .
\lbl{2.2.8}
\ee
The form of the transformed action is the same as that of
the `original' action (\ref{2.2.1}): our unconstrained action is {\it covariant}
under reparametrizations. But it is not {\it invariant}
under repara\-metrizations,
since the transformed action is a different functional of the variables
$X^{\mu}$ than the original action. The difference comes from $\Lambda'$,
which, according to the assumed eq.\,(\ref{2.2.8}), does not transform as a scalar,
but as a 1-dimensional vector field. The latter vector field $\Lambda(\tau)$
is taken to be a fixed, background, field; it is not a dynamical
field in the action (\ref{2.2.1}).

At this point it is important to stress that in a given parametrization
the ``background"  field $\Lambda(\tau)$ can be arbitrary in principle. If the
parametrizations changes then $\Lambda(\tau)$  also changes according to
(\ref{2.2.8}). This is in contrast with the constrained action (\ref{2.1.8}) where
$\lambda$ is a `dynamical' field (actually the Lagrange multiplier) giving
the constraint (\ref{2.1.9}). In eq.\,(\ref{2.1.9}) $\lambda$ is arbitrary, it can
be freely chosen. This is intimately connected with the choice of parametrization
(gauge): choice of $\lambda$ means choice of gauge. Any change of $\lambda$
automatically means change of gauge. On the contrary, in the action
(\ref{2.2.1}) and in eq.\,(\ref{2.2.6}), since $M^2$ is not prescribed but it is a
constant of motion, $\Lambda(\tau)$ is not automatically connected to
choice of parametrization. It does change under a reparametrization, but
in a given, fixed parametrization (gauge), $\Lambda(\tau)$ can still be
different in principle. This reflects that $\Lambda$ is assumed
here to be a physical field associated to the particle. Later we shall find
out (as announced already in Introduction) that physically $\Lambda$ is
a result either of (i) {\it the ``clock variable"} sitting on the particle,
or (ii) of the {\it scalar part of the velocity polyvector} occurring in
the Clifford algebra generalization of the theory.

I shall call (\ref{2.2.1}) the {\it Stueckelberg action},
\inxx{Stueckelberg action,for point particle} since Stueckelberg was
one of the first protagonists of its usefulness. The Stueckelberg action
is not invariant under reparametrizations, therefore it does not imply any
constraint. All $X^{\mu}$ are independent dynamical variables. The situation
is quite analogous to the one of a non-relativistic particle in 3-space. The 
difference is only in the number of dimensions and in the signature, otherwise
the mathematical expressions are the same. Analogously to the evolution
in 3-space, we have, in the unconstrained theory, evolution in 4-space, and
$\tau$ is the evolution parameter.\inxx{evolution parameter}
Similarly to 3-space, where a particle's
trajectory is ``built up" point by point while time proceeds, so in 4-space
a worldline is built up point by point (or event by event) while $\tau$
proceeds. The Stueckelberg theory thus implies genuine {\it dynamics }
in spacetime: a particle is a point-like object not only from the
3-dimensional but also from the 4-dimensional point of view, and it {\it
moves} in spacetime.\footnote{Later, we shall see that realistic particles
obeying the usual electromagnetic interactions should actually be modeled by
time-like strings evolving in spacetime. But at the moment, in order to set
the concepts and develop the theory we work with the idealized,
point-like objects or\inxx{motion,in spacetime}
{\it events} moving in spacetime.}

A trajectory in $V_4$ has a  status analogous that of the usual trajectory in
$E_3$. The latter trajectory is not an existing object in $E_3$; it is
a mathematical line obtained after having collected all the points in
$E_3$ through which the particle has passed during its motion. The same
is true for a moving particle in $V_4$. The unconstrained relativity is thus
just a theory of {\it relativistic dynamics}. In the conventional,
constrained, relativity there is no dynamics in $V_4$;
events in $V_4$ are considered as frozen.

\paragraph{Non-relativistic limit}  We have seen that the unconstrained
equation of motion (\ref{2.2.2}) is equivalent to the equation of geodesic. The
trajectory of a point particle or ``event" \ci{5} moving in spacetime is
a geodesic. In this respect the unconstrained theory gives the same
predictions as the constrained theory. Now let us assume that spacetime
is flat and that the particle ``spatial" speed ${\dot X}^r$, $r = 1,2,3$ is
small in comparison to its ``time" speed ${\dot X}^0$, i.e.,
\be
   {\dot X}^r {\dot X}_r \ll ({\dot X}^0)^2 .
\lbl{2.2.5a}
\ee
Then the constant of motion $M = \sqrt{{\dot X}^{\mu} {\dot X}_{\mu}}/
\Lambda$ (eq.\,(\ref{2.2.6})) is approximately equal to the time component
of 4-momentum:
\be
     M = {{\sqrt{({\dot X}^0)^2 + {\dot X}^r {\dot X}_r} }\oo
     {\Lambda}} \approx {{{\dot X}^0} \oo {\Lambda}} \equiv p^0 .
\lbl{2.2.8a}
\ee
Using (\ref{2.2.8}) the action (\ref{2.2.1}) becomes
\be
   I = {1\oo 2} \int {\dd} \tau \int \left [ \left ( {{{\dot X}^0}\oo 
   {\Lambda}} \right )^2 + {{{\dot X}^r {\dot X}_r}\oo {\Lambda}} \right ]
   \approx {1\oo 2} \int {\dd} \tau \, \left [ M^2 + {M\oo {{\dot X}^0}} \,
    {\dot X}^r {\dot X}_r  \right ]
\lbl{2.2.9}
\ee
Since $M$ is a constant the first term in (\ref{2.2.9}) has no influence on the
equations of motion and can be omitted. So we have
\be
     I \approx {M\oo 2} \int {\dd} \tau \, {{{\dot X}^r {\dot X}_r} \oo
     {{\dot X}^0}} .
\lbl{2.2.10}
\ee
which can be written as
\be
    I \approx {M \oo 2} \int {\dd} t \, {{{\dd} X^r}\oo {{\dd} t}} \,
     {{{\dd} X_r}\oo {{\dd} t}} \; , \qquad t \equiv X^0
\lbl{2.2.11}
\ee
This is the usual non-relativistic free particle action.

We see that the non-relativistic theory examines changes of the particle
coordinates $X^r$, $r = 1,2,3$, with respect to $t \equiv X^0$. The fact that
$t$ is yet another coordinate of the particle (and not an 
``evolution parameter"), that the space is actually not three- but
four-dimensional, and that $t$ itself can change during evolution is
obscured in the non relativistic theory.

\paragraph{The constrained theory within the unconstrained theory}
\hs{4mm} Let us assume that the basic theory is the unconstrained
theory\footnote{Later we shall see that there are more fundamental 
theories which contain the unconstrained theory formulated in this section.}.
A particle's mass is a constant of motion given in (\ref{2.2.6}). Instead of
(\ref{2.2.1}), it is often more convenient to introduce an arbitrary fixed
constant $\kappa$ and use the action
\be
     I[X^{\mu}] = {1\over 2} \int {\dd} \tau \left (
     {{{\dot X}^{\mu} {\dot X}_{\mu}} \over {\Lambda}} + \Lambda \kappa^2
     \right ) ,
\lbl{2.2.12}
\ee
which is equivalent to (\ref{2.2.1})  since $\Lambda(\tau)$ can be written
as a total derivative and thus the second term in (\ref{2.2.12}) has no influence
on the equations of motion for the variables $X^{\mu}$. We see that 
(\ref{2.2.12})
is analogous to the unconstrained action (\ref{2.1.8}) except for the fact that
$\Lambda$ in (\ref{2.2.12}) is not the Lagrange multiplier.

Let us now choose a constant $M = \kappa$ and write
\be
    \Lambda M = \sqrt{{\dot X}^{\mu} {\dot X}_{\mu}} .
\lbl{2.2.13}
\ee
Inserting the latter expression into (\ref{2.2.12}) we have
\be
   I[X^{\mu}] = M \int {\dd} \tau \, ({\dot X}^{\mu} {\dot X}_{\mu})^{1/2} .
\lbl{2.2.14}
\ee
This is just the constrained action (\ref{2.1.1}), proportional to the
length of particle's trajectory (worldline) in $V_4$. In other words, fixing
or choosing a value for the \inxx{mass,constant of motion}
constant of motion $M$ yields the theory which looks like
the constrained theory. The fact that mass $M$ is actually not a prescribed
constant, but results dynamically as a constant of motion in an unconstrained
theory is obscured within the formalism of the constrained theory. This can
be shown also by considering the expression for the 4-momentum
$p^{\mu} = {\dot X}^{\mu}/\Lambda$ (eq.\,(\ref{2.2.5})). Using (\ref{2.2.13}) 
we find
\be
     p^{\mu} = {{M {\dot X}^{\mu}}\oo {({\dot X}^{\nu} {\dot X}_{\nu})^{1/2}}} ,
\lbl{2.2.15}
\ee
which is the expression for 4-momentum of the usual, constrained
relativity. For a chosen $M$ it appears as if $p^{\mu}$ are constrained
to a mass shell:
\be
     p^{\mu} p_{\mu} = M^2
\lbl{2.2.16}
\ee
and are thus not independent. This is shown here to be just an illusion:
within the unconstrained relativity all 4 components $p^{\mu}$ are
independent, and $M^2$ can be arbitrary; which particular value of $M^2$
a given particle possesses depends on its initial 4-velocity ${\dot X}^{\mu}
(0)$.

Now a question arises. If all $X^{\mu}$ are considered as independent
dynamical variables which evolve in $\tau$, what is then the meaning of
the equation $x^0 = X^0 (\tau)$? If it is not a gauge fixing equation as it
is in the constrained relativity, what is it then? In order to understand this,
one has to abandon certain deep rooted concepts learned when studying the
constrained relativity and to consider spacetime $V_4$ as a higher-dimensional
analog of the usual 3-space $E_3$. A particle moves in $V_4$ in the
analogous way as it moves in $E_3$. In $E_3$ all its three coordinates are
independent variables; in $V_4$ all its four coordinates are independent
variables. Motion along $x^0$ is assumed here as a physical fact. Just think
about the well known observation that today we experience a different value
of $x^o$ from yesterday and from what we will tomorrow. The quantity $x^0$
is just a coordinate (a suitable number) given by our calendar and our clock.
The value of $x^0$ was different yesterday, is different today, and will be
different tomorrow. But wait a minute! What is `yesterday', `today' and
`tomorrow'? Does it mean that we need some additional parameter which is
related to those concepts. Yes, indeed! The additional parameter is just
$\tau$ which we use in the unconstrained theory. After thinking hard enough
for a while about the fact that we experience yesterday, today and
tomorrow, we become gradually accustomed to the idea of motion along
$x^0$. Once we are prepared to accept this idea intuitively we are ready
to put it into a more precise mathematical formulation of the theory,
and in return we shall then get an even better intuitive understanding 
of `motion along  $x^0$'. In this book we will return several times to
this idea and formulate it at subsequently higher levels of sophistication.

\paragraph{\bf Point particle in an electromagnetic field} 
We have seen that
a free particle dynamics in $V_4$ is given by the action (\ref{2.2.1}) or by
(\ref{2.2.12}). There is yet another, often very suitable, form, namely
{\it the phase space}
\inxx{phase space action,for point particle} or {\it the first order}
action which is the unconstrained version of (\ref{2.1.10}):
\be
   I[X^{\mu}, p_{\mu}] = \int
   {\dd} \tau  \left ( p_{\mu} {\dot X}^{\mu}
   - {{\Lambda}\over 2} (p_{\mu} p^{\mu} - \kappa^2) \right )  .
\lbl{2.2.16a}
\ee       
Variation of the latter action with respect to $X^{\mu}$ and $p_{\mu}$
gives:
\be
     \delta X^{\mu} \; : \qquad {\dot p}_{\mu} = 0 ,
\lbl{2.2.17}
\ee
\be
     \delta p_{\mu} \; :  \qquad p_{\mu} = {{{\dot X}_{\mu}}\oo {\Lambda}} .
\lbl{2.2.18}
\ee
We may add a total derivative term to (\ref{2.2.16a}):    
\be
   I[X^{\mu}, p_{\mu}] = \int
   {\dd} \tau  \left ( p_{\mu} {\dot X}^{\mu}
   - {{\Lambda}\over 2} (p_{\mu} p^{\mu} -\kappa^2)  + 
   {{{\dd}\phi}\oo {{\dd}\tau}}
   \right ) 
\lbl{2.2.18a}
\ee
where $\phi = \phi(\tau,x)$ so that ${\dd}\phi/{\dd}\tau = 
\p \phi/\p \tau + \p_{\mu} \phi \, {\dot X}^{\mu}$. 
The equations of motion derived from (\ref{2.2.18a}) remain the same, since
\bear
   \delta X^{\mu} \; : \quad {{\dd}\oo {{\dd} \tau }} 
   {{\p L'}\oo {\p {\dot X}^{\mu}}} - {{\p L'} \oo {\p X^{\mu}}}
   &=& {{\dd}\oo {{\dd} \tau }} (p_{\mu} + \p_{\mu} \phi ) - \p_{\mu} \p_{\nu}
   \phi \, {\dot X}^{\nu} - {\p \oo {\p \tau}} \p_{\mu} \phi \nonumber \\
   \nonumber \\
     &=& {\dot p}_{\mu} = 0  ,
\lbl{2.2.19}
\ear
\be
  \delta p_{\mu} \; :  \; \; \quad p_{\mu} = {{{\dot X}_{\mu}}\oo {\Lambda}} .
   \hs{6.85cm}
\lbl{2.2.20}
\ee
But the canonical momentum \inxx{canonical momentum,point particle}
is now different:
\be
     p'_{\mu} = {{\p L} \oo {\p {\dot X}^{\mu}}} =
      p_{\mu} + \p_{\mu} \phi .
\lbl{2.2.21}
\ee
The action (\ref{2.2.16a}) is not covariant under such a transformation
(i.e., under addition of the term ${\dd \phi}/{\dd} \tau$).

In order to obtain a covariant action we introduce  compensating  vector
and scalar fields $A_{\mu}$, $V$ which transform according to
\bear
      &&e A'_{\mu} = e A_{\mu} + \p_{\mu} \phi ,   
\lbl{2.2.21a}\\
    &&eV' = e V + {{\p \phi} \oo {\p \tau}} ,
\lbl{2.2.21b}
\ear    
and define
\be
   I[X^{\mu}, p_{\mu}] = \int
   {\dd} \tau  \left ( p_{\mu} {\dot X}^{\mu}
   - {{\Lambda}\over 2} (\pi_{\mu} \pi^{\mu} -\kappa^2)  + e V \right ) ,
\lbl{2.2.22}
\ee
where $\pi_{\mu} \equiv p_{\mu} - e A_{\mu}$ is
\inxx{kinetic momentum,for point particle} {\it the kinetic momentum}.
If we add the term  ${\dd} \phi/{\dd} \tau = \p \phi/\p \tau +
\p_{\mu} \phi \, {\dot X}^{\mu}$ 
to the latter action we obtain
\be
   I[X^{\mu}, p_{\mu}] = \int
   {\dd} \tau  \left ( p'_{\mu} {\dot X}^{\mu}
   - {{\Lambda}\over 2} (\pi_{\mu} \pi^{\mu} -\kappa^2)  + eV' \right ) ,
\lbl{2.2.23}
\ee
where $p'_{\mu} = p_{\mu} + \p_{\mu} \phi$. The transformation 
(\ref{2.2.21a})
is called
\inxx{gauge transformation} gauge transformation, $A_{\mu}$ is a
\inxx{gauge field} gauge field, identified
with the electromagnetic field potential, and $e$ the electric charge.
In addition we have also a scalar gauge field $V$.

In eq.\,(\ref{2.2.22}) $\Lambda$ is a fixed function and there is no constraint. 
Only variations of $p_{\mu}$ and $X^{\mu}$ are allowed to be performed.
The resulting equations of motion in flat spacetime are
\be
    {\dot \pi}^{\mu} = e {F^{\mu}}_{\nu} \, {\dot X}^{\nu} +
    e \left ( \p_{\mu} V - {{\p A_{\mu}}\oo {\p \tau}} \right ) ,
\lbl{2.2.24}
\ee
\be
    \pi_{\mu} = {{{\dot X}_{\mu}} \oo {\Lambda}} \; \; , \qquad \qquad
    \pi^{\mu} \pi_{\mu} = {{{\dot X}^{\mu} {\dot X}_{\mu}}\oo {\Lambda^2}} ,
\lbl{2.2.25}
\ee
where $F_{\mu \nu} = \p_{\mu} A_{\nu} - \p_{\nu} A_{\mu}$ is the
electromagnetic field tensor.

All components $p_{\mu}$ and $\pi_{\mu}$ are independent. Multiplying 
(\ref{2.2.24}) by $\pi_{\mu}$ (and summing over $\mu$) we find
\be
   {\dot \pi}^{\mu} \pi_{\mu} = \mbox{$1\oo 2$} {{\dd}\oo {{\dd} \tau}} (\pi^{\mu}
   \pi_{\mu}) = e \left ( \p_{\mu} V - {{\p A_{\mu}}\oo {\p \tau}} \right )
   {{{\dot X}^{\mu}}\oo {\Lambda}} ,
\lbl{2.2.26}
\ee
where we have used $F_{\mu \nu} {\dot X}^{\mu} {\dot X}^{\nu} = 0$.
In general, $V$ and $A_{\mu}$ depend on $\tau$. In a special case when they
do not depend on $\tau$, the right hand side of eq.\,(\ref{2.2.26}) becomes
$e \p_{\mu} V {\dot X}^{\mu} = e {\dd} V/{\dd} \tau$. Then eq.\,(\ref{2.2.26}) 
implies
\be
      {{\pi^{\mu} \pi_{\mu}}\oo 2} - e V = {\rm constant} .
\lbl{2.2.26a}
\ee
In the last equation $V$ depends on the spacetime point $x^{\mu}$.
In particular, it can be independent of $x^{\mu}$. Then eqs.(\ref{2.2.24}),
(\ref{2.2.26}) become
\be
  {\dot \pi}^{\mu} = e {F^{\mu}}_{\nu} \, {\dot X}^{\nu} ,
\lbl{2.2.26b} 
\ee
 \be
   {\dot \pi}^{\mu} \pi_{\mu} = \mbox{$1\oo 2$} {{\dd}\oo {{\dd} \tau}} (\pi^{\mu}
   \pi_{\mu})  = 0 .
\lbl{2.2.26c}
\ee
      The last result implies
\be
    \pi^{\mu} \pi_{\mu} = {\rm constant} = M^2 .
\lbl{2.2.27} 
\ee
We see that in such a particular case when $\p_{\mu} V = 0$ and
$\p A_{\mu}/ \p \tau = 0$, the square of the kinetic momentum
\inxx{kinetic momentum,for point particle} is a
constant of motion.
As before, let this this constant of motion
\inxx{mass,constant of motion} be denoted
$M^2$. It can be positive, zero or negative. We restrict ourselves
in this section to the case of bradyons ($M^2 > 0$) and leave a discussion
of tachyons
\inxx{tachyon} ($M^2 < 0$) to Sec. 13.1. By using (\ref{2.2.27}) 
and (\ref{2.2.25})
we can rewrite the equation of motion (\ref{2.2.26b}) in the form
\be
    {1\oo {\sqrt{{\dot X}^2}}} \, {{\dd}\oo {{\dd} \tau}} \left (
    {{{\dot X}^{\mu}}\oo {\sqrt{{\dot X}^2}}} \right ) = {e\oo M} \,
    {F^{\mu}}_{\nu} \, {{{\dot X}^{\nu}}\oo {\sqrt{{\dot X}^2}}} .
\lbl{2.2.28}
\ee
In eq.\,(\ref{2.2.28}) we recognize the familiar Lorentz force law
\inxx{Lorentz force} of motion
for a point particle in an electromagnetic field. A worldline which is a
solution to the usual equations of motion for a charged particle is also
a solution to the unconstrained equations of motion (\ref{2.2.26b}).
        
From (\ref{2.2.28}) and the expression for the kinetic momentum
\be
    \pi^{\mu} = {{M\, {\dot X}^{\mu}}\oo {({\dot X}^{\alpha} {\dot X}_{\alpha}
    )^{1/2}}}
\lbl{2.2.29}
\ee
it is clear that, since mass $M$ is a constant of motion,
\inxx{mass,constant of motion} a particle
cannot be accelerated by the electromagnetic field beyond the speed of
light. The latter speed is a limiting speed, at which the particle's 
energy and momentum become infinite.

On the contrary, when $\p_{\mu} V \neq 0$ and $\p A_{\mu}/ \p \tau \neq 0$
the right hand side of eq.\,(\ref{2.2.26}) is not zero and consequently
$\pi^{\mu} \pi_{\mu}$ is not a constant of motion.
Then each of the components of the kinetic momentum $\pi^{\mu} = {\dot X}^{\mu}
/ \Lambda$ could be independently accelerated to arbitrary value; there
would be no speed limit. A particle's trajectory in $V_4$ could even turn
backwards in $x^0$, as shown in Fig. \ref{sl1}. In other words, 
a particle would first
overcome the speed of light, acquire infinite speed and finally start
traveling ``backwards" in the coordinate time $x^0$. However, such a scenario
is classically not possible by the familiar electromagnetic force alone.

\setlength{\unitlength}{1mm}
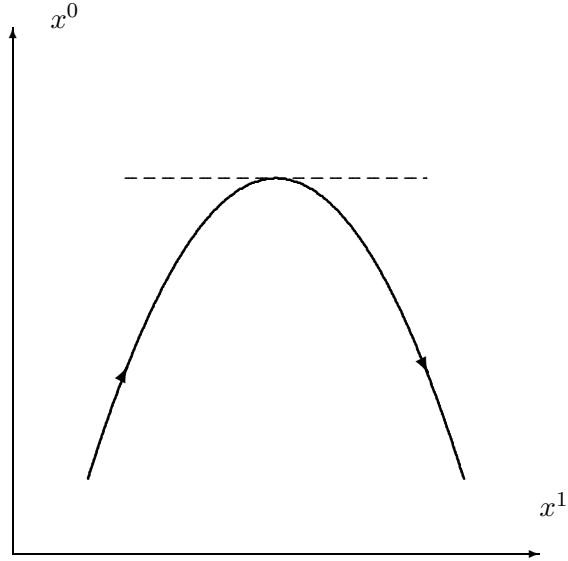
\begin{figure}[h]
\begin{picture}(95,95)(-10,0)

\put(10,10){\vector(1,0){70}}
\put(10,10){\vector(0,1){70}}

\put(80,15){$x^1$}
\put(15,80){$x^0$}

\thicklines

\curve(20,20, 45,60, 70,20)

\Thicklines
\put(25,34.5){\vector(1,2){0.2}}
\put(65,34.5){\vector(1,-2){.2}}

\thinlines
\curvedashes[1mm]{0,1.5,1}
\curve(25,60, 65,60)

\end{picture}

\caption{A possible trajectory of a particle accelerated by an exotic
4-force $f^{\mu}$ which does not satisfy $f^{\mu} {\dot X}_{\mu} = 0$.}
\label{sl1}
\end{figure}

To sum up, we have formulated a classical unconstrained theory of a point
particle in the presence of a gravitational and electromagnetic field. This
theory encompasses the main requirements of the usual constrained relativity.
It is covariant under general coordinate transformations, and locally
under the Lorentz transformations. Mass normally remains 
constant during the  motion
and particles cannot be accelerated faster than the speed of light.
However, the unconstrained theory goes beyond the usual theory of relativity.
In principle, {\it the particle is not constrained to a mass shell}. It
only appears to be constrained, since normally its mass does not change;
this is
true even if the particle moves in a gravitational and/or electromagnetic
field. Release of the mass shell constraint has far reaching consequences
for the quantization.

Before going to the quantized theory let us briefly mention that
the Hamiltonian
\inxx{Hamiltonian,for point particle} belonging to the unconstrained 
action (\ref{2.2.12})
\be
    H = p_{\mu} {\dot X}^{\mu} - L = {\Lambda \oo 2} (p^{\mu} p_{\mu} - 
    \kappa^2)
\lbl{2.2.30}
\ee
is different from zero and it is the generator of the genuine
$\tau$-evolution. As in the non-relativistic
mechanics one can straightforwardly derive the Hamilton equations of
motion, Poisson brackets, etc.. We shall not proceed here with such a
development. 

\vs{1cm}   

\section{First quantization}

Quantization of the unconstrained relativistic particle is straightforward.
Coordinates $X^{\mu}$ and momenta $p_{\mu}$ become
\inxx{momentum operator,for point particle} operators satisfying
\be
    [x^{\mu}, \, p_{\nu}] = i {\delta^{\mu}}_{\nu} \; , \quad 
    [x^{\mu}, \, x^{\nu}] = 0 \; , \quad [p_{\mu}, \, p_{\nu}] = 0 ,
\lbl{2.2.31}
\ee
which replace the corresponding Poisson brackets of the classical
unconstrained theory.

\nnnn \subsection[\ \ Flat spacetime]{Flat spacetime}
    
Let us first consider the situation
\inxx{first quantization,in flat spacetime}
in flat spacetime. The operators $x^{\mu}$,
$p_{\mu}$ act on the state vectors which are vectors in Hilbert space.
For the basis vectors we can choose $|x' \rangle$ which are eigenvectors
of the operators $x^{\mu}$:
\be
         x^{\mu} |x' \rangle = x'^{\mu} |x' \rangle  .
\lbl{2.2.32}
\ee
They are normalized according to
\be
   \langle x' | x'' \rangle = \delta (x' - x'') ,
\lbl{2.2.33}
\ee
where 
$$\delta (x' - x'') \equiv \delta^4 (x' - x'') \equiv
\prod_{\mu} \delta (x'^{\mu} - x''^{\mu}).$$
The eigenvalues of $x^{\mu}$ are
spacetime coordinates. A generic vector $|\psi \rangle$ can be expanded
in terms of $|x \rangle $:
\be
    |\psi \rangle = \int |x \rangle {\dd} \langle x|\psi \rangle ,
\lbl{2.2.33a}
\ee
where $\langle x|\psi \rangle \equiv \psi(x)$ is {\it the wave
function}.

The matrix elements of the operators $x^{\mu}$ in the coordinate representation
\inxx{coordinate representation} are
\be
      \langle x'|x^{\mu}|x'' \rangle = x'^{\mu} \, \delta (x' - x'') =
      x''^{\mu} \,  \delta (x' - x'') .
\lbl{2.2.33b}
\ee         
The matrix elements $\langle x' |p_{\mu}|x'' \rangle$ of the momentum operator
$p_{\mu}$ can be calculated from the commutation relations (\ref{2.2.31}):
\be
    \langle x'|[x^{\mu}, \, p_{\nu}]|x'' \rangle = (x'^{\mu} - x''^{\mu})
    \langle x'|p_{\nu} |x'' \rangle = i \, {\delta^{\mu}}_{\nu} \, 
    \delta (x' - x'').
\lbl{2.2.34}
\ee
Using 
  $$(x'^{\mu} - x''^{\mu})\p_{\alpha}\, \delta (x' - x'') =
- {\delta^{\mu}}_{\alpha} \, \delta (x' - x'')$$
(a 4-dimensional analog
of $f(x) {\dd} \delta(x)/ {\dd} x = - ({\dd} f (x)/{\dd} x) \,
\delta (x)$) we have
\be
    \langle x' |p_{\mu}|x'' \rangle = - i \, {\p'}_{\mu} \delta (x' - x'').
\lbl{2.2.34a}
\ee
Alternatively,
\begin{eqnarray}
    \langle x'|p_{\mu}|x'' \rangle  & = & \int \langle x'|p' \rangle
    {\dd} p' \langle p'|p|p''\rangle {\dd} p'' \langle p''| x'' \rangle
    \nonumber \\
    & = & \int \left ({1\oo {2 \pi}} \right )^{D/2} e^{i p'_{\mu} x'^{\mu}} \,
    {\dd} p' \delta (p' - p'') p'' \, {\dd} p'' \, 
    \left ({1\oo {2 \pi}} \right )^{D/2} e^{- i p''_{\mu} x''^{\mu}} \nonumber \\
    & = & \left ({1\oo {2 \pi}} \right )^D {\dd} p' \, {\dd} p' \, 
    e^{i p'_{\mu}
    (x'^{\mu} - x''^{\mu} ) } \nonumber \\
     &=& - i \, {{\p} \oo {\p x'}} \, \delta (x' - x'')
\lbl{2.2.34b}
\end{eqnarray}   
This enables us to calculate
\bear   
   \langle x'|p_{\mu}|\psi \rangle &=& \int \langle x'|p_{\mu}|x'' \rangle
    {\dd} x'' \langle x''|\psi \rangle = \int -i \, \p'_{\mu} \delta (x' - x'')
    {\dd} x'' \, \psi (x'') \nonumber \\   
     &=& \int i \p''_{\mu} \delta (x' - x'') {\dd} x''
    \psi(x'') \nonumber \\
     &=& - \int i \, \delta (x' - x'') {\dd} x'' {\p''}_{\mu} \psi(x'')
     \nonumber \\
    &=& - i \, \p'_{\mu} \psi (x')
\lbl{2.2.35}
\ear
where ${\dd} x'' \equiv {\dd}^4 x'' \equiv \prod_{\mu} {\dd} x''^{\mu}$.
The last equation can be rewritten as
\be
   p_{\mu} \psi (x) = - i \, \p_{\mu} \psi (x)
\lbl{2.2.36}
\ee
which implies that $p_{\mu}$ acts as the differential operator $-i \p_{\mu}$
on the wave function.

We work in the Schr\" odinger picture in which operators are independent of the
evolution parameter $\tau$, while the actual evolution is described by a
$\tau$-dependent state vector $|\psi \rangle $. A state can be represented
by the wave function $\psi (\tau, x^{\mu})$ which depends on four spacetime
coordinates and the evolution parameter $\tau$.

The quantity $\psi^* \psi$ is {\it the probability density}
\inxx{probability density} in spacetime. The
probability of finding a particle within a spacetime region $\Omega$ is
$\int {\dd}^4 x \, \psi^* (\tau, x) \psi (\tau, x)$ where the integration
is performed in spacetime (and not in space).

A wave function is assumed to satisfy the evolution equation
\be
       i \, {{\p \psi} \oo {\p \tau}} = H \psi \; , \quad 
       H = {\Lambda \oo 2} (-i)^2( \p_{\mu} \p^{\mu} - \kappa^2) ,
\lbl{2.2.37}
\ee
which is just the {\it Schr\" odinger equation}
\inxx{Schr\" odinger equation} in spacetime.

For a constant $\Lambda$ a general solution is a superposition
\be
    \psi (\tau,x) = \int {\dd}^4 p \, c(p) \, {\rm exp} \left [ 
    i p_{\mu} x^{\mu}
    - i {\Lambda \oo 2} (p^2 - \kappa^2) \tau \right ] ,
\lbl{2.2.38}
\ee
where $p_{\mu}$ are now the eigenvalues of the momentum operator. In
general (\ref{2.2.38}) represents a particle with {\it indefinite}
$p^2$, that is with
\inxx{indefinite mass} {\it indefinite mass}. A wave packet (\ref{2.2.38})
\inxx{wave packet} is
localized in spacetime (Fig. \ref{sl2}a) and it moves in 
spacetime \inxx{motion,in spacetime}
as the parameter $\tau$ increases (see Fig. \ref{sl2}b).

\vs{4mm}

\newcommand{\obmocje}{
    \thinlines
    \curve(4,-21, 22,-12) \curve(-4,-19, 23,-5) \curve(-9,-17, 22,-1) 
    \curve(-14,-13, 21,5) \curve(-16,-8.6, 19,9) \curve(-19,-4, 16,14)
    \curve(-19,2, 12.7,17.7) \curve(-19.7,8.3, 8,21) \curve(-18.6,14.7, 1,23)
    \thicklines
    \closecurve(0,23, 23,-5, 5,-21, -20,0)}

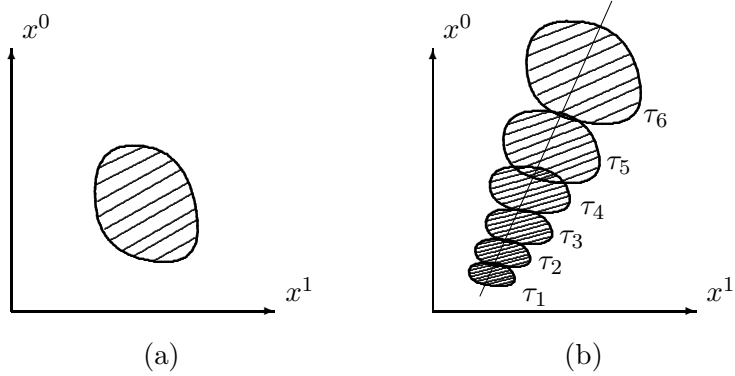
\begin{figure}[h]
\setlength{\unitlength}{.7mm}

\nnnn \begin{picture}(150,80)(-5,0)

\put(10,10){\vector(1,0){50}}
\put(10,10){\vector(0,1){50}}
\put(90,10){\vector(1,0){50}}
\put(90,10){\vector(0,1){50}}

\put(12,62){$x^0$}
\put(62,12){$x^1$}
\put(92,62){$x^0$}
\put(142,12){$x^1$}
\put(35,0){(a)}
\put(115,0){(b)}

\put(107,12){$\tau_1$}
\put(110,18){$\tau_2$}
\put(114,23){$\tau_3$}
\put(118,28.5){$\tau_4$}
\put(123,37){$\tau_5$}
\put(130,46){$\tau_6$}

\renewcommand{\xscale}{0.45}
\renewcommand{\yscale}{0.5}
\put(35,30){\obmocje}

\renewcommand{\xscale}{0.2}
\renewcommand{\yscale}{0.1}
\put(101,17){\obmocje}

\renewcommand{\xscale}{0.24}
\renewcommand{\yscale}{0.12}
\put(103,21){\obmocje}

\renewcommand{\xscale}{0.29}
\renewcommand{\yscale}{0.15}
\put(106,26){\obmocje}

\renewcommand{\xscale}{0.35}
\renewcommand{\yscale}{0.20}
\put(108,33){\obmocje}

\renewcommand{\xscale}{0.42}
\renewcommand{\yscale}{0.31}
\put(112,41){\obmocje}

\renewcommand{\xscale}{0.50}
\renewcommand{\yscale}{0.44}
\put(118,55){\obmocje}

\thinlines
\spline(99,13)(124,69)

\end{picture}

\caption{ (a) Illustration of a wave packet which is localized in spacetime.
(b) As the para\-meter $\tau$ increases, position of the wave packet in spacetime
is moving so that its centre describes a world line.}
\label{sl2}
\end{figure}

\vs{4mm}

The continuity equation \inxx{continuity equation} 
which is admitted by the Schr\" odinger equation
(\ref{2.2.37}) is
\be
    {{\p \rho}\oo {\p \tau}} + \p_{\mu} J^{\mu} = 0,
\lbl{2.2.39}
\ee
\be
   \rho \equiv \psi^* \psi \; ; \qquad J^{\mu} \equiv - {i\oo 2} (\psi^* \p_{\mu}
   \psi - \psi \p_{\mu} \psi ).
\lbl{2.2.40}
\ee
The probability density $\rho$ \inxx{probability density,in spacetime} 
is positive by definition, whilst 
components of the probability current $J^{\mu}$ \inxx{probability current} 
(including the time-like
component $J^0$) can be positive or negative. On the contrary, during
the development of the constrained theory (without the parameter $\tau$)
$J^0$ had been expected to have the meaning of the probability density
\inxx{probability density,in 3-space} in
3-space. The occurrence of negative $J^0$ had been a nuisance; it was
concluded that $J^0$ could not be the probability density but a charge
density, and $J^{\mu}$ a charge current. In {\it the unconstrained, or
parametrized theory}, the probability density is $\rho \equiv \psi^* \psi$
and there is no problem with $J^{\mu}$ being the probability current. A
wave function which has positive $J^0$ evolves into the future (as
$\tau$ increases, the time coordinate $t\equiv x^0$ of the center of the wave
packet also increases), whilst a wave function with negative $J^0$ evolves
into the past (as $\tau$ increases, $t$ decreases). The sign of $J^0$ thus
distinguishes a particle from its antiparticle, as in the usual theory. Since
particles and antiparticles have different charges ---the electric or any
other charge--- it is obvious that $J^0$ is proportional to a charge, and
$J^{\mu}$ to a charge current.

We normalized the wave function in {\it spacetime} so that for arbitrary
$\tau$ we have
\be
    \int {\dd}^4 x \, \psi^* \psi = 1.
\lbl{2.2.41}
\ee
Such a normalization is consistent with the continuity equation 
\inxx{continuity equation}
(\ref{2.2.40}) and consequently the evolution operator
\inxx{evolution operator} $U \equiv
{\rm exp} [i H \tau]$ which brings $\psi (\tau) \rightarrow \psi (\tau')
= U\, \psi (\tau)$ is {\it unitary}. The generator $H$ of the unitary
transformation $\tau \rightarrow \tau + \delta \tau$ is the Hamiltonian
(\ref{2.2.30}).

Instead of a function $\psi(\tau, x)$, we can consider its Fourier
transform\footnote{We assume here that the coefficient $c(p)$ is suitably
redefined so that it absorbs the constant resulting from the integration
over $\tau$.}
\be
    \phi (\mu, x) = \int_{-\infty}^{\infty} {\dd} \tau \, e^{i \mu^2
    \Lambda \tau} \, \psi (\tau, x) = \int {\dd}^4 \, p \, c(p) \, e^{i
    p_{\mu} x^{\mu}} \, \delta (p^2 - \kappa^2- \mu^2)
\lbl{2.2.42}
\ee
which, for a chosen value of $\mu$,
has {\it definite mass} $p^2 = M^2 =  \kappa^2 + \mu^2$. A general
function $\psi (\mu, x)$ solves the Klein-Gordon equation
\inxx{Klein--Gordon equation} with mass
square $\kappa^2 + \mu^2$:
\be
   \p_{\mu} \p^{\mu} \phi + (\kappa^2 + \mu^2) \phi = 0.
\lbl{2.2.43}
\ee
Here $\kappa^2$ is arbitrary fixed constant, while $\mu$ is a variable in the
Fourier transformed function $\phi (\mu, x)$. Keeping $\kappa^2$ 
in our equations is only a matter  of notation. We could have set 
$\kappa^2 = 0$ in the action (\ref{2.2.12}) and the Hamiltonian 
(\ref{2.2.30}). Then the Fourier transform
(\ref{2.2.42}) would simply tell us that mass is given by
$p^2 = \mu^2$.

If the limits of integration over $\tau$ in (\ref{2.2.42}) are from
$-\infty$ to $\infty$, we obtain a {\it wave function with definite
mass}. But if we take the limits from zero to infinity, then instead of
(\ref{2.2.42}) we obtain 
\be
    \Delta (\mu, x) = \int_0^\infty {\dd} \tau \, e^{i \mu^2 \Lambda \tau}
    \, \psi (\tau, x) = i \int {\dd}^4 p \, c(p) \,  {{{\rm exp} [i p_{\mu}
    x^{\mu}]} \oo {p^2 - \kappa^2 - \mu^2}} .
\lbl{2.2.44}
\ee
For the choice $c(p) = 1$ the above equation becomes the well known
\inxx{Feynman propagator} {\it Feynman propagator}.    

Quantization can be performed also in terms of the \inxx{Feynman path integral}
Feynman path integral. By
using (\ref{2.2.12}) with fixed $\Lambda$ we can calculate the transition
amplitude or propagator
   $$K(\tau,x; \tau',x') = \langle x,\tau|x',\tau' \rangle 
   = \int {\cal D} X(\tau) \, e^{i I[X(\tau)]}$$  
\be
   = \left ( {1 \oo {2 \pi i \Lambda \, 
   (\tau' - \tau)}} \right )^{D/2} \, {\rm exp} \left [ {i \oo {2 \Lambda}} \,
   {{(x' - x)^2} \oo {\tau' - \tau}} \right ] ,
\lbl{2.2.45}
\ee
where $D$ is the dimension of spacetime ($D = 4$ in our case).  In
performing the above functional integration there is no need for ghosts. The
latter are necessary in a constrained theory in order to cancel out the
unphysical states related to reparametrizations. There are no unphysical
states of such a kind in the unconstrained theory.

In eq.\,(\ref{2.2.45}) $K(\tau,x; \tau',x')$ is the probability amplitude
to find the particle in a spacetime point $x^{\mu}$ at a value of the evolution
parameter $\tau$ when it was previously found in the point $x'^{\mu}$ at
$\tau'$. The propagator has also the role of the Green's function
\inxx{Green's function} satisfying the
equation
\be
    \left ( i \, {\p \oo {\p \tau}} - H \right ) K(\tau,x; \tau',x') =
    \delta (\tau - \tau') \delta^D (x - x') .
\lbl{2.2.46}
\ee
Instead of $K(\tau,x; \tau',x')$ we can use its Fourier transform
\be
     \Delta (\mu, x-x') = \int {\dd} \tau \, e^{i \mu^2 \Lambda \tau}
     K(\tau,x; \tau',x') ,
\lbl{2.2.47}
\ee
which is just (\ref{2.2.44}) with $c(p) = 1$, $x^{\mu}$ and $\tau$
being replaced by $x^{\mu} - x'^{\mu}$ and $\tau - \tau'$, respectively.     

\subsection[\ \ Curved spacetime]
{Curved spacetime}

In order to take into account the gravitational field one has to consider
curved spacetime. Dynamical theory in curved spaces has been developed
by DeWitt \ci{23}. Dimension and signature of the space has not been specified;
it was only assumed that the metric is symmetric and nonsingular. Therefore
DeWitt's theory holds also for a 4-dimensional spacetime, and its flat space
limit is identical to the unconstrained theory discussed in Secs. 1.2 and 1.4.
\inxx{first quantization,in curved spacetime}
\inxx{DeWitt,Dynamical theory in curved spaces}

Coordinates $x^{\mu}$ and momenta $p_{\mu}$ are Hermitian operators satisfying
the commutation relations which are the same as those in flat spacetime
(see eq.\,(\ref{2.2.31})). Eigenvectors $|x' \rangle$ satisfy
\be
     x^{\mu} |x' \rangle = x'^{\mu} |x' \rangle
\lbl{2.2.49}
\ee
and they form a complete eigenbasis. A state vector $|\psi \rangle$ can
be expanded in terms of $|x' \rangle $ as follows:
\be
    |\psi \rangle = \int |x' \rangle \, \sqrt{|g(x')|} \, {\dd} x'\
    \langle x' |\psi \rangle .
\lbl{2.2.50}
\ee
The quantity $\langle x'|\psi \rangle \equiv \psi (\tau, x')$ is the
wave function and $|\psi|^2 \sqrt{|g|} \, {\dd} x'$, ${\dd} x' \equiv
{\dd}^n x$, is the probability that the coordinates of the system will be
found in the volume $\sqrt{|g|} {\dd} x'$ in the neighborhood of the point
$x'$ at the evolution time $\tau$. The probability to find the particle anywhere
in spacetime is 1, hence\footnote{We shall often omit the prime when it is
clear that a symbol without a prime denotes the eigenvalues of the corresponding
operator.}
\be
   \int \sqrt{|g|} {\dd} x \, \psi^* (\tau,x) \psi(\tau,x) = 1 .
\lbl{2.2.51}
\ee
Multiplying (\ref{2.2.50}) by $\langle x''|$ (that is projecting
the state $|\psi \rangle$ into an eigenstate $\langle x'' |$) we have
\be
   \langle x''|\psi \rangle = \psi (\tau, x'') = \int \langle x''|x' \rangle
   \sqrt{|g(x')|} {\dd} x \, \langle x' |\psi \rangle .
\lbl{2.2.52}
\ee
This requires the following normalization condition for the eigenvectors
\be
    \langle x'|x'' \rangle = {{\delta (x' - x'')}\oo {\sqrt{|g(x')|}}} =
       {{\delta (x' - x'')}\oo {\sqrt{|g(x'')|}}} \equiv \delta(x',x'') .
\lbl{2.2.53}
\ee
From (\ref{2.2.53}) we derive
\be
     \p'_{\mu} \delta(x',x'') = - \p''_{\mu} \delta(x',x'') - {1\oo 
     {\sqrt{|g(x'')|}}} \, \p''_{\mu} \sqrt{|g(x'')|} \, \delta (x',x'') ,
\lbl{2.2.54}
\ee
\be
    (x'^{\mu} - x''^{\mu}) \p'_{\alpha} \delta (x',x'') = - {\delta^{\mu}}_
    {\alpha} \delta (x',x'') .
\lbl{2.2.55}
\ee
We can now calculate the matrix elements $\langle x' |p_{\mu}|x'' \rangle$
according to
\be
    \langle x'|[x^{\mu}, p_{\nu}]|x'' \rangle = (x'^{\mu} - x''^{\mu})
     \langle x' |p_{\nu}|x'' \rangle = i {\delta^{\mu}}_{\nu} \delta(x',x'').
\lbl{2.2.56}
\ee
Using the identities (\ref{2.2.54}), (\ref{2.2.55}) we find
\be
   \langle x' |p_{\mu}|x'' \rangle  = (- i \p'_{\mu} + F_{\mu} (x'))\,
   \delta (x',x'')
\lbl{2.2.56a}
\ee
where $F_{\mu} (x')$ is an arbitrary function. If we take into account
the commutation relations $[p_{\mu}, p_{\nu}] = 0$ and the Hermitian
condition
\be
     \langle x' |p_{\mu}|x'' \rangle = \langle x'' |p_{\mu}|x' \rangle ^*
\lbl{2.2.57}
\ee
we find that $F_{\mu}$ is not entirely arbitrary, but can be of the form
\be
      F_{\mu} (x) = - i {|g|}^{-1/4} \, \p_{\mu} |g|^{1/4} .
\lbl{2.2.58}
\ee
Therefore
\bear
    p_{\mu} \psi (x') \equiv \langle x' |p_{\mu}|\psi \rangle &=& \int
    \langle x'|p_{\mu}|x'' \rangle \sqrt{|g(x'')|} \, {\dd} x'' \, \langle x''|
    \psi \rangle  \nonumber \\ 
     &=& - i (\p_{\mu} + |g|^{-1/4} \p_{\mu} |g|^{1/4}) \psi (x')
\lbl{2.2.59}
\ear
or
\be
   p_{\mu} = - i (\p_{\mu} + |g|^{-1/4} \p_{\mu} |g|^{1/4})
\lbl{2.2.59a}
\ee
\be
       p_{\mu} \psi = - i |g|^{-1/4} \p_{\mu} (|g|^{1/4} \psi) . \hs{7mm}
\lbl{2.2.59b}
\ee       
   
The expectation value
\inxx{expectation value} of the momentum operator is
   $$\langle p_{\mu} \rangle = \int \psi^* (x')\sqrt{|g(x')|} {\dd} x' \, 
    \langle x' |p_{\mu}|x'' \rangle \sqrt{|g(x'')|} \, {\dd} x'' \, \psi (x'')$$
\be    
    = - i \int \sqrt{|g|} {\dd} x \, \psi^* (\p_{\mu} + 
    {|g|}^{-1/4} \, \p_{\mu} |g|^{1/4}) \psi . \hs{.8cm}
\lbl{2.2.60}
\ee
Because of the Hermitian condition (\ref{2.2.57}) the expectation value
of $p_{\mu}$ is real. This can be also directly verified from (\ref{2.2.60}):
\bear
   \langle p_{\mu} \rangle^* &=& i \int \sqrt{|g|} \, {\dd} x \, \psi
   (\p_{\mu} + {|g|}^{-1/4} \, \p_{\mu} |g|^{1/4}) \psi^* \nonumber \\
    &=& - i \int \sqrt{|g|} \, {\dd} x \, \psi^* 
    (\p_{\mu} + {|g|}^{-1/4} \, \p_{\mu} |g|^{1/4}) \psi \nonumber \\
    & &+ \, i \int {\dd} x \, \p_{\mu} (\sqrt{|g|} \psi^* \psi) .
\lbl{2.2.61}
\ear
The surface term in (\ref{2.2.61}) can be omitted and we find
$\langle p_{\mu} \rangle^* = \langle p_{\mu} \rangle$.     

{\it Point transformations}\inxx{point transformations} 
\hs{4mm} In classical mechanics we can
transform the generalized coordinates according to
\be
      x'^{\mu} = x'^{\mu} (x) .
\lbl{2.2.61a}
\ee
The conjugate momenta then transform as covariant components of a vector:
\be
     p'_{\mu} = {{\p x^{\nu}}\oo {\p x'^{\mu}}} \, p_{\nu} .
\lbl{2.2.61b}
\ee
The transformations (\ref{2.2.61a}), (\ref{2.2.61b}) preserve the
canonical nature of $x^{\mu}$ and $p_{\mu}$, and define what is called
{\it a point transformation}.

According to DeWitt, point transformations may also be defined in quantum
mechanics in an unambiguous manner. The quantum analog of eq.\,(\ref{2.2.61a})
retains the same form. But in eq.\,(\ref{2.2.61b}) the right hand side has to
be symmetrized so as to make it Hermitian:
\be
     p'_{\mu} = \mbox{$1\oo 2$} \left ( {{\p x^{\nu}}\oo {\p x'^{\mu}}} \, p_{\nu}
     + p_{\nu} \, {{\p x^{\nu}}\oo {\p x'^{\mu}}} \right ) .
\lbl{2.2.61c}
\ee
Using the definition (\ref{2.2.59a}) we obtain by explicit calculation         
\be
    p'_{\mu} = -i \, \left ( \p'_{\mu} + |g'|^{-1/4} \, \p'_{\mu} |g'|^{1/4}
    \right ) .
\lbl{2.2.61d}
\ee
Expression (\ref{2.2.59a}) is therefore covariant under point
transformations.    

Quantum dynamical theory is based on the following postulate:

{\it The temporal behavior of the operators representing the observables
of a physical system is determined by the unfolding-in-time of a unitary
transformation.}

`Time' in the above postulate stands for the evolution time (the evolution
parameter $\tau$). As in flat spacetime, the unitary transformation is
\be
      U = e^{i H \tau}
\lbl{2.2.48}
\ee
where the generator of evolution is the Hamiltonian \inxx{Hamiltonian,in curved
space}
\be
     H = {\Lambda \oo 2} (p^{\mu} p_{\mu} - \kappa^2) .
\lbl{2.2.49a}
\ee
We shall consider the case when $\Lambda$ and the metric $g_{\mu \nu}$
are independent of $\tau$.

Instead of using the Heisenberg picture in which operators evolve,
we can use the Schr\" odinger picture in which the states evolve:
\be
     |\psi (\tau) \rangle = e^{- i H \tau} |\psi (0) \rangle .
\lbl{2.2.48a}
\ee
From (\ref{2.2.48}) and (\ref{2.2.48a}) we have
\be 
     i \, {{\p |\psi \rangle }\oo {\p \tau}} = H |\psi \rangle \; ,
     \qquad i \, {{\p \psi}\oo {\p \tau}} = H \psi ,
\lbl{2.2.48b}
\ee
which is {\it the Schr\" odinger equation}.
\inxx{Schr\" odinger equation} It governs the state or
the wave function defined over the entire spacetime.          
       
In the expression (\ref{2.2.49a}) for the Hamiltonian there is an ordering
ambiguity in the definition of $p^{\mu} p_{\mu} = g_{\mu \nu} p^{\mu} p^{\nu}$.
Since $p_{\mu}$ is a differential operator, it matters at which place one puts
$g_{\mu \nu} (x)$; the expression $g_{\mu \nu} p^{\mu} p^{\nu}$ is not
the same as $p^{\mu} g_{\mu \nu} p^{\nu}$ or $p^{\mu} p^{\nu} g_{\mu \nu}$.

Let us use the identity
\be
     |g|^{1/4} p_{\mu} |g|^{-1/4} = - i \, \p_{\mu}
\lbl{2.2.62}
\ee
which follows immediately from the definition (\ref{2.2.59a}), and define
\bear
    p^{\mu} p_{\nu} \psi &=& |g|^{-1/2} \left ( |g|^{1/4} p_{\mu} |g|^{-1/4}
     \right ) |g|^{1/2} \, g^{\mu \nu} \left ( |g|^{1/4} p_{\nu} |g|^{-1/4}
     \right ) \psi \nonumber \\
     \nonumber \\
      &=& - |g|^{-1/2} \p_{\mu} (|g|^{1/2} g^{\mu \nu} \, \p_{\nu} \psi ) 
      \nonumber \\
      \nonumber \\
     &=& - {\rm D}_{\mu} {\rm D}^{\mu} \psi ,
\lbl{2.2.63}
\ear
where ${\rm D}_{\mu}$ is covariant derivative with respect to the
metric $g_{\mu \nu}$.
Expression (\ref{2.2.63}) is nothing but an ordering prescription:
if one could neglect  that $p_{\mu}$ and $g_{\mu \nu}$ (or $|g|$)
do not commute, then the factors $|g|^{-1/2}$, $|g|^{1/4}$, etc., 
in eq.\,(\ref{2.2.63}) altogether would give 1.

One possible definition of the square of the momentum operator
\inxx{momentum operator,in curved spacetime} is thus
\be
    p^2 \psi = |g|^{-1/4} p_{\mu} g^{\mu \nu} |g|^{1/2} p_{\nu} |g|^{-1/4}
    \psi = - {\rm D}_{\mu} {\rm D}^{\mu} \psi .
\lbl{2.2.63a}
\ee
Another well known definition is
  $$p^2 \psi = \mbox{$1\oo 4$} (p_{\mu} p_{\nu} g^{\mu \nu} + 2 p_{\mu} g^{\mu \nu}
    p_{\nu} + g^{\mu \nu} p_{\mu} p_{\nu}) \psi $$ 
\be
   = (- {\rm D}_{\mu} {\rm D}^{\mu} + \mbox{$1\oo 4$} R + g^{\mu \nu}
   \Gamma_{\beta \mu}^{\alpha} \Gamma_{\alpha \nu}^{\beta} ) \psi .
\lbl{2.2.63b}
\ee
 Definitions (\ref{2.2.63a}), (\ref{2.2.63b}) can be combined according to
  $$ p^2 \psi = \mbox{$1\oo 6$} (2 |g|^{-1/4} p_{\mu} g^{\mu \nu}
  |g|^{1/2} p_{\nu} |g|^{-1/4} + p_{\mu} p_{\nu} g^{\mu \nu} + 
  2 p_{\mu} g^{\mu \nu}  p_{\nu} + g^{\mu \nu} p_{\mu} p_{\nu}) \psi$$
\be
    = (- {\rm D}_{\mu} {\rm D}^{\mu} + \mbox{$1\oo 6$} R + \mbox{$2\oo 3$}   
    g^{\mu \nu}
   \Gamma_{\beta \mu}^{\alpha} \Gamma_{\alpha \nu}^{\beta} ) \psi .
\lbl{2.2.63c}
\ee
One can verify that the above definitions all give the Hermitian
operators $p^2$. Other, presumably infinitely many, Hermitian combinations
are possible. Because of such an ordering ambiguity the quantum
Hamiltonian
\be
     H = {\Lambda\oo 2} p^2
\lbl{2.2.63d}
\ee
is undetermined up to the terms like $(\hbar^2 \kappa/2)(R + 4 g^{\mu \nu}
\Gamma_{\beta \mu}^{\alpha} \Gamma_{\alpha \nu}^{\beta} ) $, (with $\kappa
= 0$, $\mbox{$1\oo 4$}$, ${1\oo 6}$, etc., which disappear 
in the classical approximation
where $\hbar \rightarrow 0$).
 
The Schr\" odinger equation (\ref{2.2.48b})
\inxx{Schr\" odinger equation,in curved spacetime} 
admits the following continuity equation \inxx{continuity equation}
\be
   {{\p \rho} \oo {\p \tau}} + {\rm D}_{\mu} j^{\mu} = 0
\lbl{2.2.64}
\ee
where $\rho = \psi^* \psi$ and
\be
     j^{\mu} = {\Lambda\oo 2} \left [ \psi^* p^{\mu} \psi + (\psi^* p^{\mu} 
     \psi )^* \right ]
      = - i {{\Lambda}\oo 2} (\psi^* \p^{\mu} \psi - \psi \, \p^{\mu}
     \psi^*) .
\lbl{2.2.65}
\ee
\paragraph{The stationary Schr\" odinger equation} \hs{4mm} 
If we take the ansatz
\be
     \psi = e^{-i E \tau} \phi (x) ,
\lbl{2.2.65a}
\ee
where $E$ is a constant, then we obtain the stationary Schr\" odinger
equation
\be
     {\Lambda \oo 2} (- {\rm D}_{\mu} {\rm D}^{\mu} - \kappa^2) \phi = E \phi ,
\lbl{2.2.65b}
\ee
which has the form of the Klein--Gordon equation in curved spacetime
\inxx{Klein--Gordon equation,in curved spacetime}
with squared mass  $M^2 = \kappa^2 + 2 E/\Lambda$.

Let us derive {\it the classical limit}
\inxx{classical limit} of eqs.\,(\ref{2.2.59b}) and
(\ref{2.2.48b}). For this purpose we rewrite those equations by including the
Planck constant $\hbar$\footnote{Occasionally we use the hat sign over
the symbol in order to distinguish operators from their eigenvalues.}:
\be
   {\hat p}_{\mu} \psi = - i \hbar \, |g|^{-1/4} \, \p_{\mu} (|g|^{1/4} \psi ) ,
\lbl{2.2.66}
\ee
\be
     {\hat H} \psi = i\, \hbar \, {{\p \psi}\oo {\p \tau}} .
\lbl{2.2.67}
\ee

For the wave function we take the expression
\be
     \psi = A(\tau, x) {\rm exp} \left [ {i\oo \hbar} \, S(\tau, x) \right ]
\lbl{2.2.68}
\ee
with real $A$ and $S$.

Assuming (\ref{2.2.68}) and taking the limit $\hbar \rightarrow 0$, 
eq.\,(\ref{2.2.66}) becomes
\be
     {\hat p}_{\mu} \psi = \p_{\mu} S \, \psi .
\lbl{2.2.69}
\ee

Inserting (\ref{2.2.68}) into eq.\,(\ref{2.2.67}), taking the limit
$\hbar \rightarrow 0$ and writing separately the real and imaginary part
of the equation, we obtain
\be
    - \, {{\p S}\oo {\p \tau}} = {\Lambda\oo 2} \, (\p_{\mu} S \p^{\mu} S 
    - \kappa^2) ,
\lbl{2.2.70}
\ee
\be
    {{\p A^2}\oo {\p \tau}} + {\rm D}_{\mu} (A^2 \p^{\mu} S) = 0 .
\lbl{2.2.71}
\ee

Equation (\ref{2.2.70}) is just the 
Hamilton--Jacobi equation\inxx{Hamilton--Jacobi equation} of the classical
point particle theory in curved spacetime, where $E = - \p S/\p \tau$ is
"energy" (the spacetime analog of the non-relativistic energy) and
$p_{\mu} = \p_{\mu} S$ the classical momentum.

Equation (\ref{2.2.71}) is the continuity equation,
\inxx{continuity equation} where $\psi^* \psi =
A^2$ is {\it the probability density} and
\be
        A^2 \p^{\mu} S = j^{\mu}
\lbl{2.2.72}
\ee
is {\it the probability current}.

\vs{5mm}

\paragraph{The expectation value}
\inxx{expectation value} If we calculate the expectation value
of the momentum operator $p_{\mu}$ we face a problem, since the quantity
$\langle p_{\mu} \rangle $ obtained from (\ref{2.2.60}) is not, in general,
a vector in spacetime. Namely, if we insert the expression (\ref{2.2.68})
into eq.\,(\ref{2.2.60}) we obtain
\bear
    \int \sqrt{|g|} {\dd} x  \, \psi^* p_{\mu} \psi &=& \int 
    \sqrt{|g|} {\dd} x \, A^2 \p_{\mu} S - i\, \hbar \int \sqrt{|g|} {\dd} x \,
    (A \p_{\mu} A \nonumber \\
    \nonumber \\
    & & + \,  A^2 |g|^{-1/4} \p_{\mu} |g|^{1/4} \nonumber \\
    \nonumber \\
   &=& \int  \sqrt{|g|} {\dd} x \, A^2 \p_{\mu} S  - {{i \hbar}\oo 2} \int
   {\dd} x \, \p_{\mu} (\sqrt{|g|} A^2) \nonumber \\
   \nonumber \\
    &=& \int \sqrt{|g|} {\dd} x \, A^2 \p_{\mu} S ,
\lbl{2.2.73}
\ear
where in the last step we have omitted the surface term.

If, in particular, we choose a wave packet localized around a classical
trajectory $x^{\mu} = X_c^{\mu} (\tau)$, then for a certain period of $\tau$
(until the wave packet spreads too much), the amplitude is approximately
\be
      A^2 = {{\delta (x - X_c)} \oo {\sqrt{|g|}}} .
\lbl{2.2.73a}
\ee
Inserting (\ref{2.2.73a}) into (\ref{2.2.73}) we have
\be
    \langle p_{\mu} \rangle = {\p_{\mu} S \vert}_{X_c} = p_{\mu} (\tau)
\lbl{2.2.73b}
\ee
That is, the expectation value is equal to the classical momentum 4-vector
of the center of the wave packet. \inxx{wave packet}

But in general there is a problem. In the expression
\be
      \langle p_{\mu} \rangle = \int \sqrt{|g|} {\dd} x \, A^2 \p_{\mu} S  
\lbl{2.2.73c}
\ee
we integrate a vector field over spacetime. Since one cannot covariantly
sum vectors at different points of spacetime, $\langle p_{\mu} \rangle$ is
not a geo\-metric object: it does not transform as a 4-vector. The result
of integration depends on which coordinate system (parametrization) one
chooses. It is true that the expression (\ref{2.2.73})
is covariant under the point transformations (\ref{2.2.61a}), (\ref{2.2.61c}),
but the expectation value is not a classical geometric object: it is
neither a vector nor a scalar.

One way to resolve the problem is in considering $\langle p_{\mu} \rangle$
as a generalized geometric object which is a functional of 
parametrization. If the parametrization changes, then also
$\langle p_{\mu} \rangle$ changes in a well defined manner. For the ansatz
(\ref{2.2.68}) we have
\be
    \langle p'_{\mu} \rangle = \int \sqrt{|g'|} \, {\dd} x' \, A'^2 (x')\,
    \p'_{\mu} S = \int \sqrt{|g|} \, {\dd} x \, A^2 (x) {{\p x^{\nu}}\oo
    {\p x'^{\mu}}} \, \p_{\nu} S .
\lbl{2.2.74}
\ee
The expectation value is thus a parametrization dependent generalized
geometric object. The usual classical geometric object $p_{\mu}$ is also
parametrization dependent, since it transforms according to (\ref{2.2.61b}).

\paragraph{Product of operators.}  Let us calculate the product of two
operators.\inxx{product,of operators} We have
\bear   
   \langle x | p_{\mu} p_{\nu} |\psi \rangle &=& \int \langle x | p_{\mu}| x'
  \rangle \sqrt{|g(x')|} {\dd} x' \langle x'| p_{\nu}|\psi \rangle \nonumber \\ 
   &=& \int (-i)(\p_{\mu} + \mbox{$1\oo 2$} \Gamma_{\mu \alpha}^{\alpha})
   {{\delta(x - x')}\oo {\sqrt{|g(x')|}}} \, (-i) \p'_{\nu} \psi \,
   \sqrt{|g(x')|} \, {\dd} x' \nonumber \\
     &=& - \p_{\mu} \p_{\nu} \psi - \mbox{$1\oo 2$}  \Gamma_{\mu \alpha}^{\alpha} 
      \p_{\nu} \psi ,
\lbl{2.2.75}
\ear
where we have used the relation
\be
      {1\oo {|g|^{1/4}}} \, \p_{\mu} |g|^{1/4} = {1\oo {2 \sqrt{|g|}}}\,
      \p_{\mu} \sqrt{|g|} = \mbox{$1\oo 2$} \Gamma_{\mu \alpha}^{\alpha} .
\lbl{2.2.75a}
\ee                
Expression (\ref{2.2.75}) is covariant with respect to point
transformations, but it is not a geometric object in the usual sense.
If we contract (\ref{2.2.75}) by $g^{\mu \nu}$ we obtain $g^{\mu \nu}
(- \p_{\mu} \p_{\nu} \psi - \mbox{$1\oo 2$}  \Gamma_{\mu \alpha}^{\alpha} 
\p_{\nu} \psi)$
which is not a scalar under reparametrizations (\ref{2.2.61a}). Moreover,
as already mentioned, there is an ordering ambiguity where to insert
$g^{\mu \nu}$.

\subsubsection{Formulation in terms of a local Lorentz frame}
\inxx{local Lorentz frame}

Components $p_{\mu}$ are projections of the vector $p$ into basis vectors
$\gamma_{\mu}$ of a coordinate frame.
\inxx{coordinate frame} A vector can be expanded as
$p = p_{\nu} \gamma^{\nu}$ and the components are given by
$p_{\mu} = p \cdot \gamma_{\mu}$ (where the dot 
means the scalar product\footnote{
For a more complete treatment see the section on Clifford algebra.}).
Instead of a {\it coordinate frame} in which
\be
        \gamma_\mu \cdot \gamma_{\nu} = g_{\mu \nu}
\lbl{2.2.76}
\ee
we can use a local Lorentz frame $\gamma_a (x)$ in which
\be
     \gamma_a \cdot \gamma_b = \eta_{a b}
\lbl{2.2.77}
\ee
where $\eta_{a b}$ is the Minkowski tensor. The scalar product
\be
      \gamma^{\mu} \cdot \gamma_a = {e_a}^{\mu}
\lbl{2.2.78}
\ee
is {\it the tetrad}, or {\it vierbein, field}.

With the aid of the vierbein we may define the operators
\be
    p_a = \mbox{$1\oo 2$} ({e_a}^{\mu} p_{\mu} + p_{\mu} {e_a}^{\mu})
\lbl{2.2.79}
\ee
with matrix elements
\be
    \langle x|p_a| x' \rangle = - i \left ( {e_a}^{\mu} \p_{\mu} + 
    {1\oo {2 \sqrt{|g|}}}   \, \p_{\mu} (\sqrt{|g|} \, {e_a}^{\mu}) \right )
     \delta(x,x')
\lbl{2.2.80}
\ee
and
\be
     \langle x|p_a|\psi \rangle =  - i \left ( {e_a}^{\mu} \p_{\mu} + 
    {1\oo {2 \sqrt{|g|}}}   \, \p_{\mu} (\sqrt{|g|} \, {e_a}^{\mu}) \right ) . 
     \psi
\lbl{2.2.81}
\ee
In the coordinate representation we thus have 
\be
     p_a = - i ({e_a}^{\mu} \p_{\mu} + \Gamma_a) ,
\lbl{2.2.82}
\ee
where
\be
  \Gamma _a \equiv {1\oo {2 \sqrt{|g|}}}\, \p_{\mu} (\sqrt{|g|} \, {e_a}^{\mu}).
\lbl{2.2.83}
\ee

One can easily verify that the operator $p_a$ is invariant under the point
transformations (\ref{2.2.61b}).

The expectation value \inxx{expectation value}
\be
    \langle p_a \rangle = \int {\dd} x \sqrt{|g|} \, \psi^* p_a \psi
\lbl{2.2.84}
\ee
has some desirable properties. First of all, it is {\it real},
$\langle p_a \rangle^* =  \langle p_a \rangle$, which reflects 
that $p_a$ is a Hermitian operator. Secondly, it is invariant under
coordinate transformations (i.e., it transforms as a scalar).

On the other hand, since $p_a$ depends on a chosen local Lorentz frame,
the integral $\langle p_a \rangle$ is a functional of the frame field
${e_a}^{\mu} (x)$. In other words, the expectation value is an object which
is defined with respect to a chosen local Lorentz frame field.

In {\it flat spacetime} we can choose a constant frame field $\gamma_a $,
and the components ${e_a}^{\mu} (x)$ are then the transformation coefficients
into a curved coordinate system. Eq. (\ref{2.2.84}) then means that
after having started from the momentum $p_{\mu}$ in arbitrary coordinates,
we have calculated its expectation value in a Lorentz frame.
Instead of a constant frame field $\gamma_a$, we can choose an
$x$-dependent frame field $\gamma_a (x)$; the expectation value of momentum
will then be different from the one calculated with respect to the
constant frame field. In curved spacetime there is no constant frame field.
First one has to choose a frame field $\gamma_a (x)$, then define
components $p_a (x)$ in this frame field, and finally calculate the expectation
value $\langle p_a \rangle$.

Let us now again consider the ansatz (\ref{2.2.68}) for the wave function.
For the latter ansatz we have
\be
     \langle p_a \rangle = \int {\dd} x \, \sqrt{|g|} \, A^2 \, {e_a}^{\mu}
     \p_{\mu} S ,
\lbl{2.2.85}
\ee
where the term with total divergence has been omitted. If the wave
packet is localized around a classical world line $X_c (\tau)$, so that
for a certain $\tau$-period $A^2 = \delta (x - X_c)/\sqrt{|g|}$ is a
sufficiently good approximation, eq.\,(\ref{2.2.85}) gives
\be
       \langle {\hat p}_a  \rangle = 
       { {e_a}^{\mu} \, \p_{\mu} S \vert}_{X_c(\tau)} = p_a (\tau) ,
\lbl{2.2.86}
\ee
where $p_a (\tau)$ is the classical momentum along the world line
$X_c^{\mu}$ in a local Lorentz frame.

Now we may contract (\ref{2.2.86}) by ${e^a}_{\mu} $ and we obtain
$p_{\mu} (\tau) \equiv {e^a}_{\mu} p_a (\tau)$, which are components
of momentum along $X_c^{\mu}$ in a coordinate frame.

\paragraph{Product of operators} \inxx{product,of operators}
The product of operators in a local Lorentz frame is given by    
\bear   
    \langle x|p_a p_b|\psi \rangle &=& \int \langle x|p_a|x' \rangle
    \sqrt{|g(x')|} {\dd} x' \langle x' |p_b|\psi \rangle \nonumber \\
    \nonumber \\
     &=& \int (-i) \left ( {e_a}^{\mu} (x) \p_{\mu} + \Gamma_a (x) \right )
    \, {{\delta (x - x')}\oo {\sqrt{|g(x')|}}} \, \sqrt{|g(x')|} {\dd} x'
    \nonumber \\
    \nonumber \\
     & & \times (-i) \left ( {e_b}^{\nu} (x') \p'_{\nu} + 
    \Gamma_a (x') \right ) \psi (x') \nonumber \\
    \nonumber \\
    &=& - ({e_a}^{\mu} \p_{\mu} + \Gamma_a)({e_b}^{\nu} \p_{\nu} + \Gamma_b)
    \psi = p_a p_b \psi .
\lbl{2.2.87}
\ear
This can then be mapped into the coordinate frame:
\be
    {e^a}_{\alpha} {e^b}_{\beta} p_a p_b \psi = - {e^q}_{\alpha} {e^b}_{\beta}
    ({e^{\mu}}_a \p_{\mu} + \Gamma_a)({e^{\nu}}_b \p_{\nu} + \Gamma_b) \psi .
\lbl{2.2.88}
\ee

At this point let us use
\be
    \p_{\mu} {e^a}_{\nu} - \Gamma_{\mu \nu}^{\lambda} {e^a}_{\lambda} 
    + {\omega^a}_{b \mu} {e^b}_{\nu} = 0 ,
\lbl{2.2.89}
\ee
from which we have
\be
     {\omega^{ab}}_{\mu} = e^{a \nu} {e^b}_{\nu ; \mu}
\lbl{2.2.91}
\ee
and
\be
   \Gamma_a \equiv \mbox{$1\oo 2$} {1\oo {\sqrt{|g|}}} \, \p_{\mu} ( \sqrt{|g|} \,
   {e_a}^{\mu}) = \mbox{$1\oo 2$} {{e_a}^{\mu}}_{; \mu} = \mbox{$1\oo 2$} {\omega_{ca}}^{\mu}
   {e^c}_{\mu} .
\lbl{2.2.92}
\ee
Here
\be
    {e^a}_{\nu ; \mu} \equiv  \p_{\mu} {e^a}_{\nu} - \Gamma_{\mu \nu}^{\lambda}
    {e^a}_{\lambda}
\lbl{2.2.90}
\ee
is the covariant derivative
\inxx{covariant derivative} of the vierbein
\inxx{vierbein} with respect to the metric.

The relation (\ref{2.2.89}) is a consequence of the relation which tells us how
the frame field $\gamma^a (x)$ changes with position:
\be
    \p_{\mu} \gamma_a = {{\omega^a}_b}_{\mu} \gamma^b .
\lbl{2.2.93}
\ee

It is illustrative to calculate the product (\ref{2.2.88}) first in
{\it flat spacetime}. Then one can always choose a constant frame field,
so that ${\omega_{ab}}_{\mu} = 0$. Then eq.\,(\ref{2.2.88}) becomes
\be
   {e^a}_{\alpha} {e^b}_{\beta} p_a p_b \psi = - {e^b}_{\beta} \p_{\alpha}
   ({e_b}^{\nu} \p_{\nu} \psi ) = - \p_{\alpha} \p_{\beta} \psi +
   \Gamma_{\alpha \beta}^{\nu} \p_{\nu} \psi = - {\rm D}_{\alpha} {\rm D}_
   {\beta} \psi ,
\lbl{2.2.94}
\ee
where we have used
\be
   {e^b}_{\beta} {e_b}^{\nu} = {\delta_b}^{\nu} \; \; , \qquad
   {e^b}_{\beta , \alpha} {e_b}^{\nu} = - {e^b}_{\beta} {e_b}^{\nu}_{,\alpha} ,
\lbl{2.2.95}
\ee
and the expression for the affinity
\be
     \Gamma_{\alpha \beta}^{\mu} = e^{a \mu} e_{a \alpha , \beta}
\lbl{2.2.96}
\ee
which holds in flat spacetime where
\be
    {e^a}_{\mu , \nu} - {e^a}_{\nu , \mu} = 0 .
\lbl{2.2.96a}
\ee
We see that the product of operators
\inxx{product,of covariant derivatives} is just the product of the
covariant derivatives.

Let us now return to curved spacetime and consider the expression
(\ref{2.2.88}). After expansion we obtain
 $$ - {e^a}_{\alpha} {e^b}_{\beta} p_a p_b \psi = {\rm D}_{\alpha} {\rm D}_
   {\beta} \psi + {\omega^{ab}}_{\alpha} e_{a \beta} {e_b}^{\nu} \p_{\nu}
   \psi + \mbox{$1\oo 2$} {e^c}_{\mu} {\omega_{ca}}^{\mu} 
   ({e^a}_{\alpha} \p_{\beta}
   \psi + {e^a}_{\beta} \p_{\alpha} \psi)$$
   $$ + \mbox{$1\oo 2$} {e^a}_{\beta} {e^c}_{\mu} 
   {{\omega_{ca}}^{\mu}}_{; \alpha} \psi
    - \mbox{$1\oo 2$} {e^a}_{\beta} {e^b}_{\mu} 
    {\omega_{ca}}^{\mu} {{\omega^c}_b}_
    {\alpha} \psi \hs{.1cm}$$
\be    
     + \mbox{$1\oo 4$} {e^a}_{\alpha} {e^b}_{\beta} {e^c}_{\mu}
    {e^d}_{\nu} {\omega_{ca}}^{\mu} {\omega_{db}}^{\nu} \, \psi . \hs{1.94cm}
\lbl{2.2.97}
\ee

This is not a Hermitian operator. In order to obtain a Hermitian operator one
has to take a suitable symmetrized combination.

The expression (\ref{2.2.97}) simplifies significantly if we
contract it by $g^{\alpha \beta}$ and use the relation (see Sec. 6.1) for
more details)
\be
    R = {e_a}^{\mu} {e_b}^{\nu} ({\omega^{ab}}_{\nu; \mu} - 
    {\omega^{ab}}_{\mu; \nu} + {\omega^{ac}}_{\mu} {{\omega_c}^b}_{\nu}
   - {\omega^{ac}}_{\nu} {{\omega_c}^b}_{\mu}) .
\lbl{2.2.98}
\ee
We obtain
    $$ - g^{\alpha \beta} {e^a}_{\alpha} {e^b}_{\beta} p_a p_b \psi =
     - \eta^{ab} p_a p_b \psi = - p^a p_a \psi$$
\be
    = \left ( {\rm D}_{\alpha} {\rm D}^{\alpha} - \mbox{$1\oo 4$} R - 
    \mbox{$1\oo 4$} {e^a}^{\mu}
     e_{b \nu} {\omega_{ca}}^{\nu} {\omega^{cb}}_{\mu} \right )
\lbl{2..2.99}
\ee
which is a Hermitian operator. If one tries another possible
Hermitian combination, 
\be
   \mbox{$1\oo 4$} \left ( {e^a}_{\mu} p_a {e^b}^{\mu} p_b + e^{b \mu} p_a
   {e^a}_{\mu} p_b + p_a {e^a}_{\mu} p_b e^{b \mu} + p_a e^{b \mu}
   p_b {e^a}_{\mu} \right ) ,
\lbl{2.2.100}
\ee
one finds, using $[p_a , e^{b \mu}] = - i {e_a}^{\nu} \p_{\nu}
e^{b \mu}$, that it is equal to $\eta^{ab} p_a p_b$ since the extra
terms cancel out.
               
\section{Second quantization} \inxx{second quantization}
In the first quantized theory we arrived at the Lorentz invariant
Schr\" odin\-ger equation. Now, following refs. \ci{6}--\ci{11b},\,\ci{12a,12b}
we shall treat the wave function $\psi( \tau, x^{\mu})$
as a field and the Schr\" odinger equation as a field
equation which can be derived from an action. First, we shall consider
the classical field theory, and then the quantized field theory of
$\psi(\tau, x^{\mu})$. We shall construct a Hamiltonian and find out that
the equation of motion is just the Heisenberg equation. Commutation
relations for our field $\psi$ and its conjugate canonical momentum
$i \psi^{\dagger}$ are defined in a straightforward way and are not
quite the same as in the conventional relativistic field theory. The
norm of our states is thus preserved, although the states are localized
in spacetime. Finally, we point out that for the states with definite
masses the expectation value of the
energy--momentum and the charge operator coincide with the
corresponding expectation values of the conventional field theory.
We then compare the new theory with the conventional one.

\subsection[\ \ Classical field theory with invariant evolution parameter]
{Classical field theory with invariant evolution parameter}
\inxx{field theory} \inxx{evolution parameter}

The wave function $\psi$ in the Schr\" odinger equation (\ref{2.2.37})
\inxx{Schr\" odinger equation}
can be considered as a field derived from the following action
\be
   I[\psi] = \int {\dd} \tau \, {\dd}^D x \, \left ( i \psi^* {{\p \psi}
   \over {\p \tau}} - {{\Lambda}\over 2} (\p_{\mu} \psi^* \p^{\mu}
   \psi - \kappa^2 \psi^* \psi) \right ) ,
\lbl{1}
\ee
where the Lagrangian density ${\cal L}$ depends also on the
$\tau$-derivatives. By using the standard techniques of field theory
(see, e.g., \cite{24}) we may vary the action (\ref{1}) and the
boundaries of the region of integration $R$:
\bear
    \delta I &=& \int_{R} {\dd} \tau \, {\dd}^D x \, \delta {\cal L} +
    \int_{R-R'} {\dd} \tau \, {\dd}^D x \, {\cal L} \nonumber \\   
     &=& \int_{R}
    {\dd} \tau \, {\dd}^D x \, \delta {\cal L} + \int_{B} {\dd} \tau
    {\dd} \Sigma_{\mu} \, {\cal L} \delta x^{\mu} + \int {\dd}^D x\,
    {\cal L} \delta \tau {\Biggl\vert}_{\tau_1}^{\tau_2} .
\lbl{5}
\ear
Here ${\dd} \Sigma_{\mu}$ is an element of a $(D-1)$-dimensional
closed hypersurface $B$ in spacetime. If we assume that the equations
of motion are satisfied then the change of the action is 
    $$\delta I = \int {\dd} \tau \, {\dd}^D x \, \Biggl[\p_{\mu} \left (
    {{\p {\cal L}} \over {\p \p_{\mu} \psi } } \delta \psi +
    {{\p {\cal L}} \over {\p \p_{\mu} \psi^*} } \delta \psi^* +
    {\cal L} \delta x^{\mu} \right ) \hs{3cm}$$
\be
     \hs{3cm} \,  +  \, {{\p} \over {\p \tau}} \left ( {{\p {\cal L}} \over {\p 
    \p \psi/\p \tau }} \delta \psi +
    {{\p {\cal L}} \over {\p 
    \p \psi^* /\p \tau }} \delta \psi^* + {\cal L} \delta \tau
    \right ) \Biggr] ,
\lbl{6}
\ee
where $\delta \psi \equiv \psi' (\tau,x) - \psi (\tau,x)$ is
the variation of the field at fixed $\tau$ and $x^{\mu}$. By introducing
the total variation
\be
   {\bar \delta} \psi \equiv \psi'(\tau',x') - \psi (\tau,x) =
   \delta \psi + \p_{\mu} \psi \delta x^{\mu} + {{\p \psi} \over {\p \tau}} 
   \delta \tau
\lbl{6a}
\ee
eq.\,(\ref{6}) becomes\footnote{
J. R. Fanchi \ci{8} considered only the terms ${T^{\mu}}_{\nu}$ and
${T^{\mu}}_{\tau}$, but he omitted the terms $\Theta$ and $\Theta_{\mu}$.
R. Kubo \ci{9} considered also $\Theta$, but he ommitted $\Theta_{\mu}$.}
 \cite{11a}
\bear
   &&\delta I =\int {\dd} \tau \, {\dd}^D x \, \Biggl[ \p_{\mu} \left (
    {{\p {\cal L}} \over {\p \p_{\mu} \psi } } {\bar \delta} \psi +
    {{\p {\cal L}} \over {\p \p_{\mu} \psi^*} } {\bar \delta} \psi^* +           
    {T^{\mu}}_{\nu} \delta x^{\nu} + {T^{\mu}}_{\tau} \delta \tau 
    \right ) \hs{9mm} \nonumber \\
     & & \hs{2.7cm}+ {{\p} \over {\p \tau}} \left ( {{\p {\cal L}} \over {\p 
    \p \psi/\p \tau }} {\bar \delta} \psi + {{\p {\cal L}} \over {\p 
    \p \psi^* /\p \tau }} {\bar \delta} \psi^* + \Theta_{\mu}
    \delta x^{\mu} + \Theta \delta \tau \right ) \Biggr] , \nonumber \\
\lbl{7}     
\ear
where    
\bear   
   \Theta &\equiv&  {\cal L} - {{\p {\cal L}} \over {\p \p \psi/\tau } } 
   {{\p \psi} \over {\p \tau}} - 
     {{\p {\cal L}} \over {\p \p \psi^*/\tau} } {{\p \psi^*} \over {\p \tau}}
    = {{\Lambda}\over 2} (\p_{\mu} \psi^* \p^{\mu} \psi - \kappa^2 \psi^* \psi) ,
\hs{12mm} \lbl{8} \\
\nonumber \\
   \Theta_{\mu} &\equiv& - {{\p {\cal L}} \over {\p \p \psi/\p \tau }} 
   \p_{\mu} \psi -
   {{\p {\cal L}} \over {\p \p \psi^*/\p \tau }} \p_{\mu} \psi^* =
   -i \psi^* \p_{\mu} \psi ,
\lbl{9} \\
\nonumber \\
    {T^{\mu}}_{\tau} &\equiv& - {{\p {\cal L}} \over {\p \p_{\mu} \psi } }
    {{\p \psi} \over {\p \tau}} -
     {{\p {\cal L}} \over {\p \p_{\mu} \psi^* } }
    {{\p \psi^*} \over {\p \tau}} =
    {{\Lambda} \over 2} \left ( \p^{\mu} \psi^* \, {{\p \psi} \over {\p \tau}} +
    \p^{\mu} \psi \, {{\p \psi^*} \over {\p \tau}} \right ) , \; \; 
\lbl{10} \\
\nonumber \\
  {T^{\mu}}_{\nu} &\equiv& {\cal L} {\delta^{\mu}}_{\nu} -
    {{\p {\cal L}} \over {\p \p_{\mu} \psi } } \p_{\nu} \psi -
    {{\p {\cal L}} \over {\p \p_{\mu} \psi^* } } \p_{\nu} \psi^* \nonumber \\
\nonumber \\    
  &=& \left [ i \psi^* {{\p \psi} \over {\p \tau}} - {{\Lambda}\over 2}
   (\p_{\alpha} \psi^* \p^{\alpha} \psi - \kappa^2 \psi^* \psi) \right ] 
   {\delta^{\mu}}_{\nu} \nonumber \\
   \nonumber \\
  & &+ {\Lambda\over 2} (\p^{\mu} \psi^* \, \p_{\nu} \psi + \p^{\mu} \psi
   \, \p_{\nu} \psi^* ) .
\lbl{11}
\ear
Besides the usual stress--energy tensor ${T^{\mu}}_{\nu}$
\inxx{stress--energy tensor} we obtain
the additional terms due to the presence of the $\tau$-term in the
action (\ref{1}).
The  quantities $\Theta$, $\Theta_{\mu}$, ${T^{\mu}}_{\tau}$ are
generalizations of ${T^{\mu}}_{\nu}$. 

A very important aspect of this parametrized field theory is that
{\it the generator of $\tau$ evolution, or the Hamiltonian}, \inxx{generator}
\inxx{generator,of evolution} \inxx{Hamiltonian} is
\be
      H = \int {\dd}^D x \, \Theta = - {\Lambda \over 2} \int {\dd}^D x
      (\p_{\mu} \psi^* \p^{\mu} \psi - \kappa^2 \psi^* \psi)
\lbl{24}
\ee
and {\it the generator of spacetime translations}
\inxx{generator,of spacetime translations} is
\be
    H_{\mu} = \int {\dd}^D x \, \Theta_{\mu} = -i \int {\dd}^D x \,
    \psi^* \p_{\mu} \psi .
\lbl{25}
\ee
This can be straightforwardly verified. Let us take into account
that the canonically conjugate variables are
\be
       \psi \; , \quad \Pi = {{\p {\cal L}} \over {\p \p \psi/\p \tau }}
       = i \psi^* ,
\lbl{12}
\ee
and they satisfy the following equal $\tau$ Poisson bracket relations
\inxx{Poisson brackets}
\be
         \lbrace \psi (\tau,x), \Pi (\tau',x')
          {\rbrace \vert}_{\tau'=\tau} = \delta (x - x') ,
\lbl{13}
\ee
or
\be  
          \lbrace \psi (\tau,x), \psi^* {\rbrace
          \vert}_{\tau'=\tau} = -i\, \delta (x - x') ,
\lbl{14}
\ee
and
\be        
     \lbrace \psi (\tau,x), \psi \rbrace {\vert}_{\tau'=\tau} = 0 \; ,
     \quad \quad
     {\lbrace \psi^* (\tau,x), \psi^* \rbrace \vert}_{\tau'=\tau} = 0 . 
\lbl{15}
\ee

Let us consider the case ${\bar \delta} \psi = {\bar \delta} \psi^* = 0$
and assume that the action does not change under variations $\delta x^{\nu}$
and $\delta \tau$. Then from (\ref{7}) we obtain the following conservation
law: \inxx{conservation law}
\be
   \oint {\dd} \Sigma_{\mu} ( {T^{\mu}}_{\nu} \delta x^{\nu}+
   {T^{\mu}}_{\tau} \delta \tau) + {\p \over \p \tau} \int {\dd}^D x\, 
   (\Theta_{\mu} \delta x^{\mu} + \Theta \delta \tau) = 0 .
\lbl{16}
\ee

In general $\psi(\tau,x)$ has indefinite mass
\inxx{indefinite mass} and is localized in spacetime.
Therefore the boundary term in eq.\,(\ref{16}) vanishes and we find
that the generator \inxx{generator}
\be
     G(\tau) = \int {\dd}^D x 
     (\Theta_{\mu} \delta x^{\mu} + \Theta \delta \tau)
\lbl{17}
\ee
is conserved at all values of $\tau$ (see Fig. 1.3a).

In particular $\psi(\tau,x)$ may have definite mass $M$, and then from
(\ref{8}--\ref{10}) we have
$$\p\Theta/\p \tau = 0, \quad
i \psi^* \p \psi / \p \tau + \kappa^2 \psi^* \psi = M^2 \psi^* \psi , \quad
{T^{\mu}}_{\tau} = 0 , \quad\p \Theta_{\mu} / \p \tau = 0$$
and the
conservation law (\ref{16}) becomes in this particular case
\be
    \oint {\dd} \Sigma_{\mu} \, {T^{\mu}}_{\nu} \, \delta x^{\nu} = 0 ,
\lbl{18}
\ee
which means that the generator    
\be
      G(\Sigma) = \int {\dd} \Sigma_{\mu} \, {T^{\mu}}_{\nu} 
      \delta x^{\nu} 
\lbl{19}
\ee
is conserved at all space-like hypersurfaces $\Sigma$ (Fig. 1.3b).
The latter case just corresponds to the conventional field theory.

\vs{4mm}

\setlength{\unitlength}{.7mm}
\begin{picture}(180,90)(7,0)

\put(10,10){\vector(1,0){65}}
\put(10,10){\vector(0,1){65}}
\put(100,10){\vector(1,0){65}}
\put(100,10){\vector(0,1){65}}

\put(14,75){$x^0$}
\put(78,12){$x^1$}
\put(72.5,22){$\Sigma_1$}
\put(73,63){$\Sigma_2$}
\put(55,40){$\Omega$}

\put(104,75){$x^0$}
\put(168,12){$x^1$}
\put(155,21){$\Sigma_1$}
\put(160,63){$\Sigma_2$}

\put(40,0){(a)} \put(130,0){(b)}

\spline(14,61)(20,60)(30,59)(50,61)(71,66)
\spline(17,18)(23,16)(35,19)(53,21)(70,20)
\spline(111,69)(118,69.5)(130,66)(140,64)(150,63)(157,65)
\spline(108,16)(118,18)(130,17)(145,16)(156,19)

\put(44,34){\line(1,1){6}}\put(41,34){\line(1,1){10}}\put(38,35){\line(1,1){13}}
\put(36,36){\line(1,1){14}}\put(34,39){\line(1,1){13}}\put(33,40){\line(1,1){12}}
\put(32,43){\line(1,1){9}}\put(33,46){\line(1,1){6}}

\put(137,7){\line(2,1){3}}\put(131,8){\line(2,1){10}}
\put(124,8){\line(2,1){18}}\put(118,10){\line(2,1){24}}
\put(119,15){\line(2,1){23}}\put(120,20){\line(2,1){23}}
\put(122,26.5){\line(2,1){22}}\put(122,32){\line(2,1){22}}
\put(123,38){\line(2,1){21}}\put(123,44){\line(2,1){22}}
\put(122,55){\line(2,1){23}}\put(122,59){\line(2,1){19}}
\put(121.5,63.5){\line(2,1){13}}\put(121.5,67){\line(2,1){6}}
\put(122,50){\line(2,1){22}}
\thicklines

\closecurve(40,52, 50,50, 51,42, 46,33, 36,36, 32,45)
\spline(121,72)(121.5,70)(122,50)(122.5,40)(122,30)(120,19)(116,6)
\spline(146,70)(145,60)(144,30)(141,6)

\end{picture}

\begin{figure}[h]
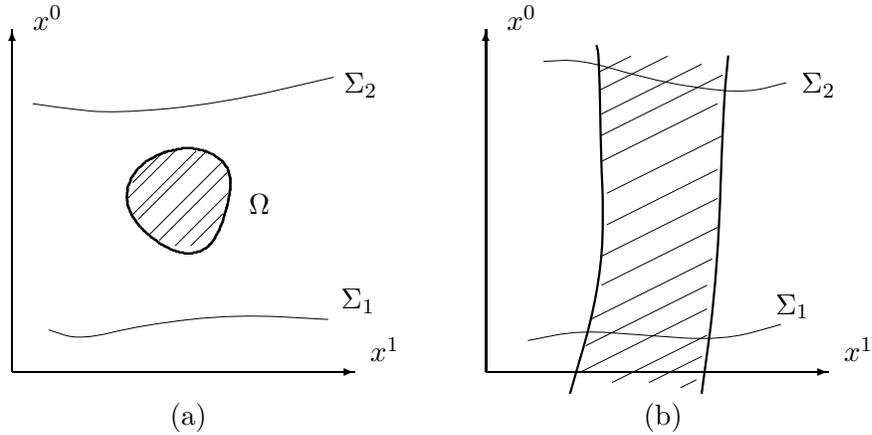

\caption{ A field $\psi(\tau,x)$ is localized within a spacetime region
$\Omega$. The field at the boundaries $\Sigma_1$ and $\Sigma_2$ (which
may be at infinity) is zero (a). A field $\psi$ which has definite mass
can be localized only in space but not in time (b).
}
\label{sl3}
\end{figure}

\vs{4mm} 

Variation of an arbitrary functional $A[\psi(x), \Pi(x)]$ is given by 
$G(\tau)$ and can be expressed in terms of the Poisson bracket
\be
      \delta A = \lbrace A, G(\tau) \rbrace .
\lbl{20}
\ee
The Poisson bracket
\inxx{Poisson brackets} is defined according to
    $$\lbrace A[\psi(x'),\Pi(x')], B[\psi(x),\Pi(x)]\rbrace \mbox{\hs{4.5cm}}$$
\be
      \qquad \quad \hs{1.5cm} = \int {\dd}^D x'' \, \left (
    {{\delta A}\over {\delta \psi(x'')}} {{\delta B} \over {\delta
    \Pi(x'')}} - 
    {{\delta B}\over {\delta \psi(x'')}} {{\delta A} \over {\delta
    \Pi(x'')}} \right ) ,
\lbl{21}
\ee
where $\delta/\delta \psi(x'')$, $\delta/\delta \Pi(x'')$ are
functional derivatives. \inxx{functional derivative}

If in eq.\,(\ref{20}) we put $A = \psi$ and take into account eq.\,(\ref{6a})
with ${\bar \delta} \psi = 0$, we obtain
\be
       {{\p \psi}\over {\p \tau}}   = - \, \lbrace \psi,H \rbrace ,
\lbl{22}
\ee
\be
     \p_{\mu} \psi = - \, \lbrace \psi,H_{\mu} \rbrace .
\lbl{23}
\ee
Eq.(\ref{22}) is equivalent to the Schr\" odinger equation (\ref{2.2.37}).
\inxx{Schr\" odinger equation}

A generic field $\psi$ which satisfies the Schr\" odinger equation (\ref{2.2.37})
can be written as a superposition of the states with definite $p_{\mu}$:
\be
    \psi(\tau,x) = \int {\dd}^D p \, c(p) \, {\rm exp} \left [
    i p_{\mu} x^{\mu} \right ] {\rm exp} \left [
    -i {\Lambda\over 2} (p^2 - \kappa^2) \tau \right ] .
\lbl{25a}
\ee
Using (\ref{25a}) we have
\be
    H = - \, {\Lambda\over 2} \int {\dd}^D p \, (p^2 - \kappa^2) c^*(p) c(p) ,
\lbl{26}
\ee
\be
      H_{\mu} = \int {\dd}^D p \, p_{\mu} \, c^*(p) c(p) .
\lbl{27}
\ee

On the other hand, from the spacetime stress--energy tensor
\inxx{stress--energy tensor} (\ref{11}) we
find the total energy--momentum \inxx{energy--momentum}
\be
        P^{\mu} = \int {\dd} \Sigma_{\nu} T^{\mu \nu} .
\lbl{28}
\ee
It is not the generator of spacetime translations. Use of $P^{\mu}$
as generator of spacetime translations results in all conceptual and
technical complications (including the lack of manifest Lorentz
covariance) of the conventional relativistic field theory.

\subsection[\ \ The canonical quantization]
{The canonical quantization}
Quantization of the parametrized field theory is straightforward. The field
$\psi (\tau,x)$ and its \inxx{momentum,of the Stueckelberg field}
canonically conjugate momentum $\Pi = i \psi^
{\dagger}(\tau,x)$ become operators satisfying the following equal $\tau$
commutation relations\footnote{
Enatsu  \cite{6} has found that the commutation relation (\ref{30}) is
equivalent to that of the conventional, on shell, quantum field
theory, provided that $\epsilon(x-x') \epsilon(\tau-\tau') = 1$. However,
since $x^{\mu}$ and $\tau$ are independent, the latter relation is not
satisfied in general, and the commutation relation (\ref{30}) is not
equivalent to the usual on shell commutation relation. And yet, as far
as the expectation values of certain operators are concerned,
such as the energy--momentum
are concerned, the predictions of both theories are the same; and this
is all that matters.}  :
\be
    {[\psi(\tau,x),\Pi(\tau',x')] \vert}_{\tau'=\tau} = i \delta (x - x') ,
\lbl{29}
\ee
or
\be  
  [\psi(\tau,x),\psi^{\dagger}(\tau',x')] {\vert}_{\tau'=\tau} = 
  \delta (x - x') ,  
\lbl{30}
\ee
and
\be
  [\psi(\tau,x),\psi(\tau',x'] {\vert}_{\tau'=\tau} =
  [\psi^{\dagger}(\tau,x),\psi^{\dagger}(\tau',x')] {\vert}_{\tau'=\tau} = 0 .
\lbl{31}
\ee

In momentum space the above commutation relations read
\be
      [c(p),c^{\dagger}(p')] = \delta (p - p') ,
\lbl{32}
\ee
\be
  [c(p),c(p')] = [c^{\dagger}(p),c^{\dagger}(p')] = 0 .
\lbl{33}
\ee
The general commutator (for $\tau' \neq \tau$) is:
\be
   [\psi(\tau,x),\psi^{\dagger}(\tau',x')] = \int {\dd}^D p \, e^{i p_{\mu}
   (x^{\mu}-x'^{\mu})} \, e^{-{\Lambda\over 2} (p^2-\kappa^2)(\tau - \tau')} .
\lbl{33a}
\ee

The operators $\psi^{\dagger}(\tau,x)$ and $\psi (\tau, x)$
are creation and annihilation operators, \inxx{creation operators}
\inxx{annihilation operators} respectively. The vacuum is defined
as a state which satisfies
\be
    \psi (\tau, x) |0 \rangle = 0 .
\lbl{38a}
\ee
If we act on the vacuum
\inxx{vacuum} by $\psi^{\dagger} (\tau, x)$ we obtain
a single particle state with definite position:
\be
    \psi^{\dagger} (\tau, x) |0 \rangle = |x, \tau \rangle .
\lbl{38b}
\ee
An arbitrary 1-particle state is a superposition
\be
     |\Psi^{(1)} \rangle = \int {\dd} x \, f(\tau, x) \, \psi^{\dagger}
     (\tau, x) |0 \rangle ,
\lbl{38c}
\ee
where $f(\tau, x)$ is the wave function.

A 2-particle state with one particle at $x_1$ and another at $x_2$ is
obtained by applying the creation operator twice:
\be
    \psi^{\dagger} (\tau, x_1) \psi^{\dagger} (\tau, x_2) |0 \rangle
    = |x_1, x_2, \tau \rangle .
\lbl{38d}
\ee
In general
\be
   \psi^{\dagger} (\tau, x_1) ... \psi^{\dagger} (\tau, x_n) |0 \rangle =
   |x_1, ..., x_n, \tau \rangle ,
\lbl{38e}
\ee
and an arbitrary $n$-particle state is the superposition
\be
    |\Psi^{(n)} \rangle = \int {\dd} x_1 ... {\dd} x_n \, f(\tau, x_1, ...
    , x_n) \psi^{\dagger} (\tau, x_1) ... 
    \psi^{\dagger} (\tau, x_n) |0 \rangle ,
\lbl{38f}
\ee
where $f(\tau, x_1,...,x_n)$ is
\inxx{wave function,of many particles} the wave function the $n$ particles
are spread with.

In momentum space
\inxx{momentum space} the corresponding equations are
\be
   c(p)|0 \rangle = 0 ,
\lbl{38g}
\ee
\be
      c^{\dagger} (p) |0 \rangle = |p \rangle ,
\lbl{38h}
\ee
\be
      c^{\dagger} (p_1)... c^{\dagger} (p_n) |0 \rangle = |p_1,...,p_n \rangle ,
\lbl{38j}
\ee

\be
    |\Psi^{(n)} \rangle = \int {\dd} p_1 ... {\dd} p_n \, g(\tau, p_1,...,
    p_n) \, c^{\dagger} (p_1) ... c^{\dagger} (p_n) |0 \rangle
\lbl{38k}
\ee

The most general state is a superposition of the states $|\Psi^{(n)} \rangle$
with definite numbers of particles.

According to the commutation relations (\ref{30})--(\ref{33}) the states
\newline
$|x_1,...,x_n \rangle $ and $|p_1,...,p_n \rangle $ satisfy
\be
      \langle x_1,...,x_n|x'_1,...,x'_n \rangle = \delta (x_1 - x'_1)...
      \delta (x_n - x'_n) ,
\lbl{38l}
\ee
\be
      \langle p_1,...,p_n|p'_1,...,p'_n \rangle = \delta (p_1 - p'_1)...
      \delta (p_n - p' _n) ,
\lbl{38m}
\ee
which assures that the norm of an arbitrary state $|\Psi \rangle$ is
always positive.

We see that the formulation of the unconstrained relativistic quantum field
theory goes along the same lines as the well known second quantization of 
a non-relativistic particle. Instead of the 3-dimensional space we have
now a $D$-dimensional space whose signature is arbitrary. In particular,
we may take a 4-dimensional space with Minkowski signature and identify it
with {\it spacetime}.            
             
\subsubsection{The Hamiltonian and the generator of spacetime translations}
The generator $G(\tau)$ defined
\inxx{generator} in (\ref{17}) is now an operator. An
infinitesimal change of an arbitrary operator $A$ is given by the
commutator
\be
    \delta A = - \, i [A, G(\tau)] ,
\lbl{41}
\ee
which is the quantum analog of eq.\,(\ref{20}). If we take $A=\psi$,
then eq.\,(\ref{41}) becomes
\be
     {{\p \psi} \over {\p \tau}} = i [\psi,H] ,
\lbl{42}
\ee
\be
         {{\p \psi} \over {\p x^{\mu}}} = i [\psi, H_{\mu}] ,
\lbl{43}
\ee
where
\bear
   H &=& \int {\dd}^D x \, \Theta = - {\Lambda \over 2} \int {\dd}^D x
      (\p_{\mu} \psi^{\dagger} \p^{\mu} \psi - \kappa^2 \psi^{\dagger} \psi) ,
\lbl{44} \\                  
   H_{\mu} &=& \int {\dd}^D x \, \Theta_{\mu} = -i \int {\dd}^D x \,
    \psi^{\dagger} \p_{\mu} \psi .
\lbl{45}
\ear       
Using the commutation relations (\ref{29})--(\ref{31}) we find that
eq.\,(\ref{42}) is equivalent to the field equation (\ref{2.2.37}) (the
Schr\" odinger equation). Eq. (\ref{42}) is thus {\it the Heisenberg
equation} \inxx{Heisenberg equation}
for the field operator $\psi$. We also find that eq.\,(\ref{43})
gives just the identity $\p_{\mu} \psi = \p_{\mu} \psi$.

In momentum representation \inxx{momentum representation}
the field operators are expressed in terms
of the operators $c(p)$, $c^{\dagger}(p)$ according to eq.\,(\ref{25a}) and
we have
\bear
     H &=& -\, {\Lambda\over 2} \int {\dd}^D p \, (p^2 - \kappa^2)
     c^{\dagger}(p) c(p_) ,
\lbl{46}\\
      H_{\mu} &=& \int {\dd}^D p \, p_{\mu} c^{\dagger}(p) c(p) .
\lbl{47}
\ear

The operator $H$ is the Hamiltonian \inxx{Hamiltonian}
and it generates the $\tau$-evolution, 
whereas $H_{\mu}$ is the generator of spacetime translations.
In particular, $H_0$ generates translations along the axis $x^0$
and can be either positive or negative definite.

\subsubsection{Energy--momentum operator} \inxx{energy--momentum operator}
Let us now consider the generator $G(\Sigma)$ defined in eq.\,(\ref{19})
with ${T^{\mu}}_{\nu}$ given in eq.\,(\ref{11}) in which the classical
fields $\psi$, $\psi^*$ are now replaced by the operators $\psi$,
$\psi^{\dagger}$. The total energy--momentum $P_{\nu}$ of the field is
given by the integration of ${T^{\mu}}_{\nu}$ over a space-like hypersurface:
\be
        P_{\nu} = \int {\dd} \Sigma_{\mu} \, {T^{\mu}}_{\nu} .
\lbl{48}
\ee
Instead of $P_{\nu}$ defined in (\ref{48}) it is convenient to
introduce
\be
      {\widetilde P}_{\nu} = \int {\dd} s \, P_{\nu} ,
\lbl{49}
\ee
where ${\dd} s$ is a distance element along the direction $n^{\mu}$
which is orthogonal to the hypersurface element ${\dd} \Sigma_{\mu}$.
The latter can be written as ${\dd} \Sigma_{\mu} = n_{\mu} {\dd} \Sigma$.
Using ${\dd} s {\dd} \Sigma = {\dd}^D x$ and integrating out $x^{\mu}$ in
(\ref{49}) we find that $\tau$-dependence disappears and we obtain (see Box 1.1)
\be
     {\widetilde P}_{\nu} = \int {\dd}^D p {\Lambda\over 2} (n_{\mu} p^{\mu})\,
     p_{\nu} (c^{\dagger}(p) c(p) + c(p) c^{\dagger}(p)) .
\lbl{50}
\ee

\boxA

From (\ref{50}) we have that the projection of ${\widetilde P}_{\nu}$ on a
time-like vector $n^{\nu}$ is always positive, while the projection on
a space-like vector $N^{\nu}$ can be
positive or negative, depending on
signs of $p_{\mu} n^{\mu}$ and $p_{\mu} N^{\mu}$. In a special reference
frame in which $n^{\nu} = (1,0,0,...,0)$ this means that $P_0$ is
always positive, whereas $P_r$, $r=1,2,...,D-1$, can be positive or
negative definite.

What is the effect of the generator ${\widetilde P}_{\nu} \delta x^{\nu}$ on
a field $\psi(\tau,x)$. From (\ref{41}),\,(\ref{50}) we have
\bear    
    \delta \psi &=& - i [\psi,{\widetilde P}_{\nu}] \delta x^{\nu} 
     \lbl{51}\\
     \nonumber \\    
    &=& -i \int {\dd}^D p \, \Lambda (n_{\mu} p^{\mu}) c(p) {\rm exp}
    \left [ i p_{\mu} x^{\mu} \right ] {\rm exp} \left [ - {{i \Lambda}
    \over 2} (p^2 - \kappa^2) \tau \right ] p_{\nu} \delta x^{\nu} \nonumber
\ear

Acting on a Fourier component of  $\psi$ with definite $p_{\mu} p^{\mu}
= M^2$ the gene\-rator ${\widetilde P}_{\nu} \delta x^{\nu}$ gives
 $$\delta \phi = - i [\phi, {\widetilde P}_{\nu}] \delta x^{\nu} = - i \int
  {\dd}^D p \, \Lambda\,  M \epsilon (np) c(p) e^{i p_{\mu} x^{\mu}}
  \delta (p^2 - M^2) p_{\nu} \delta x^{\nu}$$
\be
        = -\, \p_{\nu} \phi^{(+)} \delta x^{\nu} + \p_{\nu} \phi^{(-)}
        \delta x^{\nu} , \hs{4mm}
\lbl{53}
\ee
\be
    \epsilon (np) \equiv {{n^{\mu} p_{\mu}}\over {\sqrt{|p^2|}}} =
    \left \{ \begin{array}{c}
                    + 1 \; , \quad n^{\mu} p_{\mu} > 0 \\
                    - 1 \; , \quad n^{\mu} p_{\mu} < 0
               \end{array}
     \right. \              
\lbl{54}
\ee
where
\bear
    \phi(x) &=& \int_{-\infty}^{\infty} {\dd} \tau\, e^{i \mu \tau} \,
    \psi(\tau, x) \nonumber \\
    \nonumber \\
     &=& \int_{p_0=-\infty}^{\infty} {\dd}^D p \, \delta
    (p^2 - M^2)c(p) e^{i p_{\mu} x^{\mu}} \nonumber \\
     \nonumber \\   
     &=& \int_{p_0=0}^{\infty} {\dd}^D p \, \delta
    (p^2 - M^2)c(p) e^{i p_{\mu} x^{\mu}} \nonumber \\
     \nonumber \\ 
    & & + \int_{p_0=-\infty}^{0} {\dd}^D p \, \delta
    (p^2 - M^2)c(p) e^{i p_{\mu} x^{\mu}} \nonumber \\
    \nonumber \\
     &=& \phi^{(+)} + \phi^{(-)}
\lbl{55}
\ear
As in Sec.\,1.3.  we identify $\mu^2 + \kappa^2 \equiv M^2$.

We see that the action of ${\widetilde P}_{\nu} \delta x^{\nu}$ on
$\phi$ differs from
the action of $H_{\nu} \delta x^{\nu}$. The difference is in the step
function $\epsilon (np)$. When acting on $\phi^{(+)}$ the generator
${\widetilde P}_{\nu} \delta x^{\nu}$ gives the same result as
$H_{\nu} \delta x^{\nu}$. On the contrary, when acting on $\phi^{(-)}$ it
gives the opposite sign. This demonstrates that
${\widetilde P}_{\nu} \delta x^{\nu}$ does not generate translations and
Lorentz transformations in Minkowski space $M_D$. This is a consequence
of the fact that $P_0$ is always positive definite.

\subsubsection{Phase transformations and the charge operator}
Let us now consider the case when the field is varied at the boundary,
while $\delta x^{\mu}$ and $\delta \tau$ are kept zero. Let the field
transform according to
\be
    \psi' = e^{i \alpha} \psi \> , \quad \delta \psi = i \alpha \psi
    \> , \quad \delta \psi^{\dagger} = - i \alpha \psi^{\dagger} .
\lbl{56}
\ee
Eq. (\ref{7}) then reads
\be
    \delta I = \oint {\dd} \tau \, {\dd} \Sigma_{\mu} \left ( - {{i \Lambda
    \alpha}\over 2} \right ) (\p^{\mu} \psi^{\dagger} \psi - \p^{\mu} \psi
    \, \psi^{\dagger}) + \int {\dd}^D x \, (-\alpha) \psi^{\dagger} \psi
    {\Biggl \vert}_{\tau_1}^{\tau_2}
\lbl{57}
\ee
Introducing {\it the charge density} \inxx{charge density}
\be
    \rho_c = - \alpha \psi^{\dagger} \psi
\lbl{58}
\ee
and {\it the charge current density} \inxx{charge current density}
\be
     j_c^{\mu} = - {{i \Lambda \alpha}\over 2} (\p^{\mu} \psi^{\dagger} 
     \psi - \p^{\mu} \psi \, \psi^{\dagger}) ,
\lbl{58a}
\ee
and assuming that the action is invariant under
the phase transformations \inxx{phase transformations}
(\ref{56}) we obtain from (\ref{57})
\inxx{conservation law} the following conservation law:
\be
    \oint {\dd} \Sigma_{\mu} \, j_c^{\mu} + {{\dd}\over {{\dd} \tau}}
    \int {\dd}^D x \, \rho_c = 0 ,
\lbl{59}
\ee
or
\be
      \p_{\mu} j_c^{\mu} + {{\p \rho_c}\over {\p \tau}} = 0 .
\lbl{60}
\ee

In general, when the field has indefinite mass it is localized in spacetime
and the surface term in (\ref{59}) vanishes. So we have that the
generator
\be
      C = - \int {\dd}^D x \, \rho_c
\lbl{61}
\ee
is conserved at all $\tau$-values.

In particular, when $\psi$ has definite mass, then $\psi^{\dagger} \psi$
is independent of $\tau$ and the conservation of $C$ is trivial. Since
$\psi$ is not localized along a time-like direction
\inxx{current density} the current density
$j_c^{\mu}$ does not vanish at the space-like hypersurface even if taken
at infinity. It does vanish, however, at a sufficiently far away
time-like hypersurface. Therefore the generator \inxx{generator}
\be
     Q = \int {\dd} \Sigma_{\mu} j_c^{\mu}
\lbl{62}
\ee
is conserved at all space-like hypersufaces. The quantity $Q$
corresponds to the charge operator
\inxx{charge operator} of the usual field theory. However,
it is not generator of
\inxx{phase transformations} the phase transformations (\ref{56}).

By using the field expansion (\ref{25a}) we find
\be
     C = - \alpha \int {\dd}^D p \, c^{\dagger} (p) c(p) .
\lbl{63}
\ee
From
\be
        \delta \psi = - i [\psi,C]
\lbl{65}
\ee
and the commutation relations (\ref{29})--(\ref{33}) we have
\be
   \delta \psi = i \alpha \psi \> , \quad  \delta \psi^{\dagger} = -i
   \alpha \psi^{\dagger} ,
\lbl{66}
\ee
\be
    \delta c(p) = i \alpha c(p) \> , \quad  \delta c^{\dagger} (p) =
    -i \alpha c^{\dagger}(p) ,
\lbl{67}
\ee
which are indeed the phase transformations (\ref{56}). Therefore
$C$ is {\it the generator of phase transformations}.

On the other hand, from $\delta \psi = -i [\psi,Q]$ we do not obtain
the phase transformations. But we obtain from $Q$ something which is
very close to the phase transformations, if we proceed as follows. Instead of
Q we introduce
\be
        {\widetilde Q} = \int {\dd} s \, Q ,
\lbl{68}
\ee
where ${\dd} s$ is an infinitesimal interval along a time-like
direction $n^{\mu}$ orthogonal to ${\dd} \Sigma_{\mu} = n_{\mu}
{\dd} \Sigma$. Using ${\dd} s {\dd} \Sigma = {\dd}^D x$ and integrating out
$x^{\mu}$ eq.\,(\ref{68}) becomes
\be
   {\widetilde Q} = \int \left ( -{{\alpha \Lambda}\over 2} \right ) \, {\dd}
   p \, (p^{\mu} n_{\mu}) (c^{\dagger} (p) c(p) + c(p) c^{\dagger}(p)) .
\lbl{69}
\ee
The generator ${\widetilde Q}$ then gives \inxx{generator}
\be
    \delta \psi = -i [\psi,{\widetilde Q}] = \int i \alpha \Lambda {\dd}^D p \,
    (p^{\nu} n_{\nu}) c(p) {\rm exp} \left [ i p_{\mu} x^{\mu} \right ]
    {\rm exp} \left [ - {{i \Lambda}\over 2} (p^2 - \kappa^2) \tau \right ] .
\lbl{70}
\ee
For a Fourier component with definite $p_{\mu} p^{\mu}$ we have
\bear
  \delta \phi = -i [\phi, {\widetilde Q}] &=& \int {\dd}^D p \, i \alpha
     \Lambda m \, \epsilon (np)c(p) \, e^{i p_{\mu} x^{\mu}} \delta (p^2 - M^2)
     \nonumber \\
        &=& i \alpha \phi^{(+)} - i \alpha \phi^{(-)}
\lbl{71}
\ear
where $\epsilon (np)$ and $\phi^{(+)}$, $\phi^{(-)}$ are defined in
(\ref{54}) and (\ref{55}), respectively.

\vs{1cm}

\paragraph{Action of operators on states} From the commutation relations
(\ref{30})--(\ref{30}) one finds that when the operators $c^{\dagger} (p)$
and $c(p)$ act on the eigenstates of the operators $H$, $H_{\mu}$,
${\widetilde P}_{\mu}$, $C$, ${\widetilde Q}$ they increase and decrease,
respectively, the corresponding eigenvalues. For instance, if $H_{\mu}
|\Psi \rangle = p_{\mu} |\Psi \rangle$ then
\be
     H_{\mu} c^{\dagger} (p') |\Psi \rangle = (p_{\mu} + p'_{\mu}) |\Psi
     \rangle ,
\lbl{71a}
\ee
\be
      H_{\mu} c(p') |\Psi \rangle = (p_{\mu} - p'_{\mu}) |\Psi \rangle ,
\lbl{71b}
\ee
and similarly for other operators.                
     
\subsubsection{The expectation values of the operators}
\inxx{expectation value}
Our parametrized field theory is just a straightforward generalization of
the non-relativistic field theory in $E_3$. Instead of the 3-dimensional
Euclidean space $E_3$ we have now the 4-dimensional Minkowski space $M_4$
(or its $D$-dimensional generalization). Mathematically the theories are
equivalent (apart from the dimensions and signatures of the corresponding
spaces). A good exposition of the unconstrained field theory is given in
ref.\,\cite{25}

A state $|\Psi_n \rangle $ of $n$ identical free particles
can be given in terms of the generalized 
\inxx{wave packet profile} wave packet profiles
$g^{(n)}(p_1, ...,p_n)$ and the generalized state vectors
$|p_1, ...,p_n \rangle $:
\be
      |\Psi_n \rangle  = \int {\dd} p_1 ... {\dd} p_n \, g^{(n)}(p_1, ...,p_n)   
                         |p_1,...,p_n \rangle .
\lbl{72}
\ee
The action of the operators $c(p)$, $c^{\dagger}(p)$
on the state vector is the following:
\bear
    c(p) |p_1, ...,p_n \rangle &=& {1\over {\sqrt{n}}} 
    \sum_{i=1}^n \delta (p-p_i)
|p_1, ..,{\mbox{\it \v p}}_i ,..,p_n \rangle , 
\lbl{73}\\
  c^{\dagger}(p) |p_1, ...,p_n \rangle &=& \sqrt{n+1} |p,p_1,...,p_n \rangle .
\lbl{74}
\ear
The symbol ${\mbox{\it \v p}}_i$ means that the quantity
$p_i$ is not present in the expression. Obviously
\be
    |p_1, ...,p_n \rangle = {1\over {\sqrt{n!}}} 
    c^{\dagger}(p_n)...c^{\dagger} (p_1) |0 \rangle .
\lbl{75}
\ee
For the product of operators we have
\be
    c^{\dagger} (p) c(p)|p_1, ...,p_n \rangle = 
\sum_{i=1}^n \delta  (p-p_i) |p_1, ...,p_n \rangle
\lbl{76}
\ee
The latter operator obviously counts how many there are particles
with momentum $p$. Integrating over $p$ we obtain
\be
    \int {\dd}^D p \, c^{\dagger} (p) c(p)  |p_1, ...,p_n \rangle = 
    n |p_1, ...,p_n \rangle .
\lbl{77}
\ee

The expectation value of the operator ${\widetilde P}^{\mu}$
(see eq.\,(\ref{50})) in the state $|\Psi_n \rangle $ is
\bear
   \langle |\Psi_n | {\widetilde P}^{\mu} | \Psi \rangle &=& \langle \Psi_n | 
\int {\dd}^D p \, {\Lambda\over 2} (n_{\nu} p^{\nu}) p^{\mu} 
(2 c^{\dagger} (p) c(p) + \delta (0) ) |\Psi_n \rangle \nonumber \\
     &=& \Lambda \sum_{i=1}^n \left < (n_{\nu} p_i^{\nu}) p_i^{\mu} \right > + 
{\Lambda\over 2} \delta (0) \int  {\dd}^D p \, (n_{\nu} p^{\nu}) p^{\mu} ,
\lbl{78}
\ear
where 
$$ \left < (n_{\nu} p_i^{\nu}) p_i^{\mu} \right >= 
\int {\dd} p_1 ... {\dd} p_n g^{* (n)} (p_1,..., p_n)
 g^{(n)}(p_1,..., p_n) (p_i^{\nu} n_{\nu})  p_i^{\mu}.$$

In particular, the wave packet profile $g^{(n)}(p_1,..., p_n)$ 
can be such that $|\Psi_n \rangle =
|p_1,...,p_n \rangle$. Then the expectation value (\ref{78}) becomes
\be
    \langle p_1,...,p_n | {\widetilde P}^{\mu} |p_1,...,p_n \rangle = 
\Lambda \sum_{i=1}^n  (n_{\nu} p_i^{\nu}) p_i^{\mu}  + 
{\Lambda\over 2} \delta (0) \int  {\dd}^D p \, (n_{\nu} p^{\nu}) p^{\mu} .
\lbl{79}
\ee
The extra term ${\Lambda\over 2} \delta (0) \int  
{\dd}^D p \, (n_{\nu} p^{\nu}) p^{\mu}$
vanishes for a space-like component of $p^{\mu}$, 
but it does not vanish for a time-like component.

So far the momenta $p_1, ..., p_n$ of $n$ identical particles 
have not been constrained to a mass shell. Let us now consider 
the case where all the momenta $p_1, ..., p_n$  are constrained to the 
mass shell, so that $p_i^{\mu} p_{i \mu} = p_i^2 = p^2 = M^2$, $i=1,2,...,n$. 
From the classical equation of motion $p_i^{\mu} = 
{\dot x}_i^{\mu}/\Lambda$ we have
\be
      \Lambda = {{{\dd} s_i}\over {{\dd} \tau}} {1 \over {\sqrt{p_i^2}}} ,
\lbl{80}
\ee
where ${\dd} s_i = \sqrt{{\dd} x_i^{\mu} {\dd} x_{i \mu}}$. Now, 
$\Lambda$ and $\tau$ being the same for all particles we see that 
if $p_i^2$ is the same for all particles also the 4-dimensional speed 
${\dd} s_i/{\dd} \tau$ is the same. Inserting (\ref{80})
into (\ref{79}) and
chosing parametrization such that ${\dd} s_i/{\dd} \tau = 1$ 
we obtain the following expectation value of ${\widetilde P}^{\mu}$ in 
a state on the mass shell:
\be
     {\langle {\widetilde P}^{\mu} \rangle}_m  = \langle p_1,...,p_n | 
{\widetilde P}^{\mu} |p_1,...,p_n \rangle_m =
    \sum_{i=1}^n \epsilon (n p_i) p_i^{\mu} + 
{{\delta (0)}\over 2} \int {\dd}^D p \, \epsilon
   (np) p^{\mu} .
\lbl{81}
\ee
Taking $n_{\nu} = (1,0,0,....,0)$ we have
\be
     \langle {\widetilde P}^0 \rangle_m = \sum_{i=1}^n \epsilon (p_i^0) p_i^0 + 
{{\delta (0)}\over 2} \int {\dd}^D p \, \epsilon (p^0) p^0 ,
\lbl{82}
\ee
\be
     \langle {\widetilde P}^r \rangle_m = \sum_{i=1}^n \epsilon 
(p_i^0) p_i^r  \; \; , \quad r = 1,2,...,D-1 .
\lbl{83}
\ee

So far we have considered momenta as continuous. However, 
if we imagine a large box and fix the
boundary conditions, then the momenta are discrete. Then in 
eqs.\,(\ref{82}),\,(\ref{83}) particles
$i = 1,2,...,n$ do not necessarily all have different momenta 
$p_i^{\mu}$. There can be
$n_p$ particles with a given discrete value of momentum $p$. 
Eqs.\,(\ref{82}),\,(\ref{83}) can then be
written in the form
\bear
    \langle {\widetilde P}^0 \rangle_m &=& \sum_{\bf p} 
    (n_{\bf p}^+ + n_{\bf p}^- +  {1\over 2}) \omega_{\bf p} ,
\lbl{84}\\
 \langle {\tilde {\bf P}} \rangle_m &=& \sum_{\bf p} (n_{\bf p}^+ + 
  n_{\bf p}^- ) {\bf p}
\lbl{85}
\ear
Here 
$$\omega_{\bf p} \equiv |\sqrt{{\bf p}^2 + M^2}| = \epsilon (p^0) p^0 ,$$
whereas
$n_{\bf p}^+ = n_{\omega_{\bf p},{\bf p}}$ is the number of 
particles with positive $p^0$ at a given value of momentum and 
$n_{\bf p}^- = n_{- {\omega}_{\bf p}, -{\bf p}}$ .
In eq.\,(\ref{85}) we have
used 
$$- \sum_{{\bf p} = - \infty}^{\infty} n_{- {\omega}_{\bf p},
{\bf p}} {\bf p} =
\sum_{{\bf p} = - \infty}^{\infty} n_{- {\omega}_{\bf p}, -{\bf p}} {\bf p} .$$
Instead of $\delta (0)$ in eq.\,(82) we have written 1, since in 
the case of discrete momenta
$\delta (p - p')$ is replaced by $\delta_{p p'}$, and $\delta_{pp} = 1$.

Equations (\ref{84}),\,(\ref{85}) are just the same expressions 
as obtained in the conventional,
on shell quantized field theory of a non-Hermitian scalar field. 
In an analogous way we also find that
the expectation value of the electric charge operator 
${\widetilde Q}$, when taken on the mass
shell, is identical to that of the conventional field theory.

\subsection[\ \ Comparison with the conventional relativistic quantum field
theory]
{Comparison with the conventional relativistic quantum field
theory}

In the unconstrained relativistic quantum field theory mass is
indefinite, in general. But in particular it can be definite. The
field is then given by the expression (\ref{2.2.42}) for a fixed
value of $\mu$, say $\mu = 0$, in which momentum is constrained to
a mass shell. If we calculate the commutator $[\phi(x), \phi^{\dagger}
(x')]$ using the commutation relations (\ref{32}),\,(\ref{33}) we obtain
\be
     [\phi(x),\phi(x')] = \int {\dd}^D p \, e^{i p_{\mu} (x^{\mu} - x'^{\mu})}
     \delta (p^2 - M^2) ,
\lbl{86}
\ee
which differs from zero both for time-like and for space-like
separations between $x^{\mu}$ and $x'^{\mu}$. In this respect the new
theory differs significantly from the conventional theory in which
the commutator is zero for space-like separations and which assures
that the process
   $$ \langle 0|\phi(x) \phi^{\dagger} (x')|0 \rangle $$
has vanishing amplitude. No faster--than--light propagation is possible
in the conventional relativistic field theory.

The commutation relations leading to the conventional field theory are
\be
   [c(p), c^{\dagger} (p')] = \epsilon(p^0) \delta^D(p - p') ,
\lbl{89}
\ee
where $\epsilon(p^0) = +1$ for $p^0 > 0$ and $\epsilon (p^0) = -1$.

By a direct calculation we then find
\be
   [\phi(x), \phi^{\dagger}(x')] = \int {\dd}^D p \, e^{i p (x - x')} 
   \epsilon (p^0) \delta (p^2 - M^2) \equiv D(x - x') ,
\lbl{90}
\ee
which is indeed different from zero when $(x - x')^2 > 0$ and zero
when $(x - x')^2 < 0$.

The commutation relations (\ref{89}),\,(\ref{90}) also assure that the
Heisenberg equation \inxx{Heisenberg equation}
\be
     {{\p \phi}\oo {\p x^0}} = i [\psi, P^0]
\lbl{91}
\ee
is equivalent to the field equation
\be
      (\p_{\mu} \p^{\mu} + M^2) \phi = 0 .
\lbl{92}
\ee
Here $P^0$, which now serves as Hamiltonian, is the 0-th component of
the momentum operator defined in terms of the definite
mass fields $\phi(x)$, $\phi^{\dagger} (x)$ (see Box 1.2).

If vacuum is defined according to \inxx{vacuum}
\be
       c(p) |0 \rangle = 0 ,
\lbl{93}
\ee
then because of the commutation relation (\ref{89})
 the scalar product is
\be
   \langle p|p' \rangle = \langle 0|c(p) c^{\dagger}|0 \rangle
   \epsilon(p^0)\delta^D(p - p') , 
\lbl{94}
\ee
from which it follows that states with negative $p^0$ have
negative norms.

The usual remedy is in the redefinition of vacuum. Writing
\be
    p^{\mu} \equiv p = ({\bf p}, \omega_{\bf p}) \; \; , \qquad
    \omega_{\bf p} = |\sqrt{M^2 + {\bf p}^2}| ,
\lbl{95}
\ee

\boxB

\nnnn and denoting
\be
     {1\oo {\sqrt{2 \omega_{\bf p}}}} c({\bf p}, \omega_{\bf p})\equiv 
     a({\bf p}) \; \; ,  \qquad
     {1\oo {\sqrt{2 \omega_{\bf p}}}} c^{\dagger} ({\bf p}, \omega_{\bf p}) 
     \equiv a^{\dagger} ({\bf p}) ,
\lbl{96}
\ee
\be
     {1\oo {\sqrt{2 \omega_{\bf p}}}} c^{\dagger} (-{\bf p}, - 
     \omega_{\bf p}) \; \; , \qquad 
     {1\oo {\sqrt{2 \omega_{\bf p}}}} c(-{\bf p}, - \omega_{\bf p}) = 
     b^{\dagger} ({\bf p}) ,
\lbl{97}
\ee
let us define the vacuum according to
\be
       a({\bf p}) |0 \rangle = 0 \; \; , \qquad b({\bf p}) |0 \rangle = 0 .
\lbl{98}
\ee
If we rewrite the commutation relations (\ref{89}) in terms of the
operators $a({\bf p})$, $b({\bf p})$, we find (see Box 1.3)
\be
      [a({\bf p}), a^{\dagger}({\bf p}')] = \delta ({\bf p} - {\bf p}') ,
\lbl{99}
\ee
\be
      [b({\bf p}), b^{\dagger}({\bf p}')] = \delta ({\bf p} - {\bf p}') ,
\lbl{100}
\ee
          \be
[a({\bf p}), b({\bf p}')] = [a^{\dagger}({\bf p}), b^{\dagger}({\bf p}')] =
 [a({\bf p}), b^{\dagger}({\bf p}')]  = [a^{\dagger}({\bf p}), b({\bf p}')] = 0. 
\lbl{101}
\ee

\boxC

\nnnn Using the operators (\ref{96}), (\ref{97}) and the vacuum
\inxx{vacuum} definition (\ref{98})
we obtain that the scalar products between the states ${|{\bf p} \rangle}_+
= a^{\dagger} ({\bf p})|0 \rangle$, $|{\bf p} \rangle_-
= b^{\dagger} ({\bf p})|0 \rangle$ are non-negative:
\bear
  \langle {\bf p}|_+ |{\bf p}' \rangle_+ &=&  
  \langle {\bf p}|_- |{\bf p}' \rangle_- = \delta({\bf p} - {\bf p}') ,
  \nonumber \\
    {\langle {\bf p}|}_+ |{\bf p}' \rangle_- &=& 
    \langle {\bf p}|_- |{\bf p}' \rangle_+ = 0    
\lbl{102}
\ear

If we now calculate,  in the presence
of the commutation relations (\ref{89}) and the vacuum definition (\ref{98}),
the eigenvalues of the operator $P^0$ we find that they are all positive.
Had we used the vacuum definition (\ref{93}) we would have found that $P^0$
can have negative eigenvalues.

On the other hand, in {\it the unconstrained
theory} the commutation relations (\ref{32}), (\ref{33}) and the vacuum
definition (\ref{93}) are valid. Within such a framework
the eigenvalues of $P^0$ are also all positive, if the space-like hypersurface
$\Sigma$ is oriented along a positive time-like direction, otherwise they are
all negative. The creation operators are $c^{\dagger}(p) = c^{\dagger}(p^0,
{\bf p})$, and they create states $|p \rangle = |p^0, {\bf p} \rangle$
where $p^0$ can be positive or negative. Not only the states with
positive, but also the states with negative 0-th component of {\it momentum}
(equal to frequency if units are such that $\hbar = 1)$ have positive
energy $P^0$, regardless of the sign of $p^0$. One has to be careful not
to confuse $p^0$ with the energy.

That energy \inxx{energy}
is always positive is clear from the following classical example.
If we have a matter continuum (fluid or dust) then the energy--momentum
is defined as the integral over a hypersurface $\Sigma$ of the
stress--energy tensor: \inxx{stress--energy tensor}
\be
       P^{\mu} = \int T^{\mu \nu} {\dd} \Sigma_{\nu}.
\lbl{103}
\ee
For dust we have $T^{\mu \nu} = \rho \, u^{\mu} u^{\nu}$, where $u^{\mu}=
{\dd} x^{\mu}/{\dd} s$ and $s$ is the proper time. The dust energy is
then $P^0 = \int \rho \, u^0 u^{\nu} n_{\nu} \, {\dd} \Sigma$,
where the hypersurface
element has been written as ${\dd} \Sigma_{\nu} = n_{\nu} {\dd} \Sigma$.
Here $n^{\nu}$ is a time-like vector field orthogonal to the
hypersurface, pointing along a positive time-like direction
(into the ``future"), such that there exists a coordinate system in which
$n_{\nu} = (1,0,0,0,...)$. Then $P^0 = \int \dd \Sigma \, \rho \, u^0 u^0$
is obviously positive, even if $u^0$ is negative. If the dust consists of
only one massive particle, then the density is singular on the
particle worldline:
\be
     \rho (x)  = m \int  {\dd} s \, \delta (x - X(s)) ,
\lbl{104}
\ee
and 
$$P^0 = \int \dd \Sigma \, \dd s \, m u^0 u^{\mu} n_{\mu}
\, \delta (x - X(s)) = m u^0 \, (u^{\mu} n_{\nu}) .$$ 
We see that the point particle energy,
defined by means of $T^{\mu \nu}$, differs from $p^0 = m u^0$. The difference
is in the factor $u^{\mu} n_{\mu}$, which in the chosen coordinate system
may be plus or minus one.

\paragraph{Conclusion}
Relativistic quantum field theory is one of the most successful physical
theories. And yet it is not free from serious conceptual and technical
difficulties. This was especially clear to Dirac, who expressed his
opinion that quantum field theory, because it gives infinite results,
should be taken only as a provisional theory waiting to be replaced
by a better theory.

In this section I have challenged some of the cherished basic assumptions
which, in my opinion, were among the main stumbling blocks preventing
a further real progress in quantum field theory and its relation to
gravity. These assumptions are:

\vs{2mm}

1) {\it Identification of negative frequency with negative
     energy.} When a field is expanded both positive and negative
     $p^0 = \hbar \omega
     = \omega$, $\hbar = 1$, occur. It is then taken for granted that,
     by the correspondence principle, negative frequencies mean negative
     energies. {\it A tacit assumption is that the quantity $p^0$
     is energy, while actually it is not energy}, as it is shown
     here. \inxx{negative frequencies} \inxx{negative energies}

\vs{2mm}
     
2)     {\it Identification of causality violation with
     propagation along space-like separations.}
     \inxx{causality violation} It is widely believed
     that faster--than--light propagation violates causality. A series
     of thought experiments with faster--than--light particles has been
     described \ci{Tolman}
     and concluded that they all lead to causal loops which
     are paradoxical. All those experiments are classical and tell us
     nothing about how the situation would change if quantum mechanics
     were taken into account\footnote{Usually it is argued that tachyons are
     even worse in quantum field theory, because their negative energies
     would have caused vacuum instability. Such an argument is valid in
     the conventional quantum field theory, but not in its generalization
     based on the invariant evolution parameter $\tau$.}.
     In Sec. 13.1 I show that quantum mechanics,
     properly interpreted, eliminates the causality paradoxes
     \inxx{causality paradoxes} of tachyons
     (and also of worm holes, as already stated by Deutsch). In this
     section, therefore, I have assumed that amplitudes do not need to
     vanish at space-like separations. What I gain is a very elegant,
     manifestly covariant quantum field theory based on the straightforward
     commutation relations (\ref{31})--(\ref{33}) in which fields depend
     not only on spacetime coordinates $x^{\mu}$, but also on the
     Poincar\' e invariant parameter $\tau$ (see Sec. 1.2). Evolution and
     causality are related to $\tau$. The coordinate $x^0$ has nothing
     to do with evolution and causality considerations.

To sum up, we have here a consistent classical and quantum unconstrained
(or ``parametrized") field theory which is manifestly Lorentz
covariant and in which the
fields depend on an invariant parameter $\tau$. Although the quantum
states are localized in spacetime there is no problem with
negative norm states \inxx{negative norm states} \inxx{unitarity}
and unitarity. This is a result of the fact that the
commutation relations between the field operators are not quite the
same as in the conventional relativistic quantum field theory. While in
the conventional theory evolution of a state goes along the coordinate $x^0$,
and is governed by the components $P^0$ of the momentum operator, in the
unconstrained theory evolutions goes along $\tau$ and is governed by
the covariant Hamiltonian \inxx{Hamiltonian}
$H$ (eq.\,(\ref{24}).  The commutation
relations between the field operators are such that the Heisenberg equation
of motion determined by $H$ is equivalent to the field equation which is
just the Lorentz covariant Schr\" odinger equation (\ref{2.2.37}).
Comparison of the parametrized quantum field theory
\inxx{parametrized quantum field theory} with the conventional
relativistic quantum field theory reveals that the expectation values of
energy--momentum and charge operator in the states with definite masses are the
same for both theories. Only the free field case has been considered here.
I expect that inclusion of interactions will be straightforward (as
in the non-relativistic quantum field theory), very instructive and will
lead to new, experimentally testable predictions. Such a development
waits to be fully worked out, but many partial results have been reported
in the literature \ci{12a,12b}. Although very interesting, it is beyond
the scope of this book.

We have seen that the conventional field theory is obtained from
the unconstrained theory
if in the latter we take the definite mass fields $\phi (x)$ and treat
negative frequencies differently from positive one. Therefore, strictly
speaking, the conventional theory is not a special case of the unconstrained
theory. When the latter theory is taken on mass shell then the commutator
$[\phi(x),\phi(x')]$ assumes the form (\ref{86}) and thus differs from
the conventional commutator (\ref{90}) which has the function
$\epsilon(p^0) = (1/2) (\theta(p^0) - \theta(-p^0))$ under the integral.

Later we shall argue that the conventional relativistic quantum field
theory is a special case, not of the unconstrained point particle,
but of the unconstrained string theory, where the considered objects
are the time-like strings (i.e. worldlines) moving in spacetime.
     
\newpage

\chapter[Point particles and Clifford algebra]
{Point particles \\and Clifford algebra}

Until 1992 I considered the tensor calculus of general relativity to be the
most useful language by which to express physical theories. Although I was aware
that differential forms were widely considered to be superior to tensor
calculus I had the impression that they did not provide a sufficiently
general tool for all the cases. Moreover, whenever an actual calculation
had to be performed, one was often somehow forced to turn back to
tensor calculus. A turning point in my endeavors to understand better 
physics and its mathematical formalization
was in May 1992 when I met professor Waldyr Rodrigues, Jr.\footnote{We were both
guests of Erasmo Recami at The Institute of Theoretical Physics,
Catania, Italia.}, who introduced
me into the subject of {\it Clifford algebra}. After the one or two week
discussion I became a real enthusiast of the geometric calculus\
based on Clifford algebra. This was the tool I always missed and that
enabled me to grasp geometry from a wider perspective. 

In this chapter I would like to forward my enthusiasm to those readers
who are not yet enthusiasts themselves. I will describe the subject
from a physicist's point of view and concentrate on the usefulness of
Clifford algebra as a language for doing physics and geometry. I am
more and more inclined towards the view that Clifford algebra is not
only {\it a} language but but {\it the} language of a ``unified theory"
which will encompass all our current knowledge in fundamental theoretical
physics.

I will attempt to introduce the ideas and the concepts in such a way as
to satisfy one's intuition and facilitate an easy understanding. For 
more rigorous mathematical treatments the reader is advised to look at
the existing literature \ci{13,14,26}.

\section{Introduction to geometric calculus based on Clifford algebra}

We have seen that point particles move in some kind of space. In non relativistic
physics the space is 3-dimensional and Euclidean, while in the theory of 
relativity space has 4-dimensions and pseudo-Euclidean signature,
and is called spacetime. Moreover, in general relativity spacetime is curved,
which provides gravitation. If spacetime has even more dimensions ---as in
Kaluza--Klein theories--- then such a higher-dimensional gravitation
contains 4-dimensional gravity and Yang--Mills fields (including the fields
associated with electromagnetic, weak, and strong forces). Since physics
happens to take place in a certain space which has the role of a stage or
arena, it is desirable to understand its geometric properties as deeply
as possible.

Let $V_n$ be a continuous space of arbitrary dimension $n$. To every point\
of $V_n$ we can ascribe $n$ {\it parameters} $x^{\mu}$, $\mu = 1,2,...,n$,
which are also called {\it coordinates}. Like house numbers they can be
freely chosen, and once being fixed they specify points of the space\footnote{
See Sec. 6.2, in which the subtleties related to specification 
of spacetime points are discussed.}.

When considering points of a space we ask ourselves what are the
distances \inxx{distance}
between the points. {\it The distance} between two infinitesimally
separated points is given by
\be
     {\dd} s^2 = g_{\mu \nu} \, {\dd} x^{\mu} \, {\dd} x^{\nu} .
\lbl{3.1}
\ee
Actually, this is the square of the distance, and $g_{\mu \nu}(x)$ is
the metric tensor. The quantity ${\dd} s^2$ is {\it invariant} with respect
to general coordinate transformations $x^{\mu} \rightarrow x'^{\mu} = f^{\mu}
(x)$.

Let us now consider {\it the square root} of the distance. Obviously it is
$\sqrt{g_{\mu \nu} {\dd} x^{\mu} \, {\dd} x^{\nu} }$. But the latter expression
is not linear in ${\dd} x^{\mu}$. We would like to define an object which
is {\it linear} in ${\dd} x^{\mu}$ and whose square is eq.\,(\ref{3.1}).
Let such object be given by the expression
\be
     {\dd} x = {\dd} x^{\mu} \, e_{\mu}
\lbl{3.2}
\ee
It must satisfy
\be
    {\dd} x^2 = e_{\mu} e_{\nu} \, {\dd} x^{\mu} {\dd} x^{\nu} =
    \mbox{$1\oo 2$} (e_{\mu} e_{\nu} + e_{\nu} e_{\mu})\, {\dd} x^{\mu} {\dd} x^{\nu} =
    g_{\mu \nu} \, {\dd} x^{\mu} \, {\dd} x^{\nu} = {\dd} s^2 ,
\lbl{3.3}
\ee
from which it follows that
\be
     \mbox{$1\oo 2$} (e_{\mu} e_{\nu} + e_{\nu} e_{\mu}) =  g_{\mu \nu}
\lbl{3.4}
\ee
The quantities $e^{\mu}$ so introduced are a new kind of number,
called {\it Clifford numbers}. \inxx{Clifford numbers}
They do not commute, but satisfy eq.\,(\ref{3.4})
which is a characteristic of {\it Clifford algebra}. \inxx{Clifford algebra}

In order to understand what is the meaning of the object ${\dd} x$ introduced
in (\ref{3.2}) let us study some of its properties. For the sake of avoiding
use of differentials let us write (\ref{3.2}) in the form
\be
    {{{\dd} x}\oo {{\dd} \tau}} = {{{\dd} x^{\mu} }\oo {{\dd} \tau}} \,
    e_{\mu} \, ,
\lbl{3.5}
\ee
where $\tau$ is an arbitrary parameter invariant under general
coordinate transformations. Denoting ${\dd} x/{\dd} \tau \equiv a$,
${\dd} x^{\mu}/{\dd} \tau = a^{\mu}$, eq.\,(\ref{3.5}) becomes
\be
        a = a^{\mu} e_{\mu} \, .
\lbl{3.6}
\ee
Suppose we have two such objects $a$ and $b$. Then
\be
      (a + b)^2 = a^2 + ab + ba + b^2
\lbl{3.7}
\ee
and
\be 
         \mbox{$1\oo 2$}(ab + ba) =
         \mbox{$1\oo 2$} (e_{\mu} e_{\nu} + e_{\nu} e_{\mu}) a^{\mu} b^{\nu} =           
         g_{\mu \nu} a^{\mu} b^{\nu} .
\lbl{3.8}
\ee
The last equation algebraically corresponds to \inxx{inner product}
{\it the inner product}
of two {\it vectors} with components $a^{\mu}$ and $b^{\nu}$.
Therefore we denote
\be
      a \cdot b \equiv \mbox{$1\oo 2$}(ab + ba) .
\lbl{3.8a}
\ee      

From (\ref{3.7})--(\ref{3.8}) we have that the sum 
$a + b$ is an object whose square is also a scalar.

What about the antisymmetric combinations? We have
\be
     {1\oo2}(ab - ba) =
         \mbox{$1\oo 2$} (a^{\mu} b^{\nu} - a^{\nu} b^{\mu}) e_{\mu} e_{\nu}
\lbl{3.9}
\ee  
This is nothing but {\it the outer product} \inxx{outer product}
of the vectors.
Therefore we denote it as 
\be
      a \wedge b \equiv \mbox{$1\oo 2$}(ab - ba)
\lbl{3.9a}
\ee     
In 3-space
this is related to the familiar vector product $a \times b$ which is the
dual of $a \wedge b$.

The object $a = a^{\mu} e_{\mu}$ is thus nothing but a \inxx{vectors}
{\it vector}\ : $a^{\mu}$
are its components and $e^{\mu}$ are $n$ linearly independent basis vectors
of $V_n$. Obviously, if one changes parametrization, $a$ or ${\dd} x$ 
remains the same. Since under a general coordinate transformation the
components $a^{\mu}$ and ${\dd} x^{\mu}$ do change, $e_{\mu}$ should also 
change in such a way that the vectors $a$ and ${\dd} x$ remain invariant.

An important lesson we have learnt so far is that
\begin{itemize}
  
  \item the ``square root" of the distance is a vector;
  \item vectors are Clifford numbers;
  \item vectors are objects which, like distance, are {\it invariant}
   under general coordinate transformations.
\end{itemize}
  
\setlength{\unitlength}{1cm}

\vs{8mm}
\nnnn \begin{picture}(12.5,8)

\put(0,0){\framebox(12.5,8){
\begin{minipage}[t]{11.5cm}

\centerline{\bf Box 2.1: Can we add apples and oranges?}

\vs{5mm}

When I asked my daughter Katja, then ten years old, how much is 3 apples and
2 oranges plus 1 apple and 1 orange, she immediately replied 
``4 apples and 3 oranges". If a child has no problems with adding apples and
oranges, it might indicate that contrary to the common wisdom, often
taught at school, such an addition has mathematical sense after all.
The best example that this is indeed the case is complex numbers.
Here instead of `apples' we have real and, instead of `oranges',
imaginary numbers. The sum of a real and imaginary number is a
complex number, and summation of complex numbers is a mathematically well
defined operation. Analogously, in Clifford algebra we can sum Clifford
numbers of different degrees. In other words, summation of scalar, vectors,
bivectors, etc., is a well defined operation.

\end{minipage} }}
\end{picture}

\vs{8mm}

The basic operation in Clifford algebra is \inxx{Clifford product}
{\it the Clifford product}
$a b$. It can be decomposed into the symmetric part $a \cdot b$ (defined
in (\ref{3.8a}) and the antisymmetric part $a \wedge b$ (defined in
(\ref{3.9a})):
\be
     a b = a \cdot b + a \wedge b
\lbl{3.10}
\ee
We have seen that $a \cdot b$ is a scalar. On the contrary, eq.\,(\ref{3.9})
shows that $a \wedge b$ is not a scalar. Decomposing the product $e_{\mu}
e_{\nu}$ according to (\ref{3.10}),
 $$e_{\mu} e_{\nu} = e_{\mu} \cdot
e_{\nu} + e_{\mu} \wedge e_{\nu} = g_{\mu \nu} + e_{\mu} \wedge e_{\nu} \, ,$$
we can rewrite (\ref{3.9}) as
\be
        a \wedge b = {1 \oo 2} (a^{\mu} b^{\nu} - a^{\nu} b^{\mu})\,
        e_{\mu} \wedge e_{\nu} ,
\lbl{3.13}
\ee
which shows that $a \wedge b$ is a new type of geometric object,
called {\it bivector}, \inxx{bivector}
which is neither a scalar nor a vector.

The geometric product (\ref{3.10}) is thus the sum of a scalar and a bivector.
The reader who has problems with such a sum is advised to read Box 2.1.

A {\it vector} is an algebraic representation of {\it direction} in a
space $V_n$; it is associated with an oriented line.

A {\it bivector} is an algebraic representation of an oriented plane.

This suggests a generalization to trivectors, quadrivectors, etc. It
is convenient to introduce the name $r$-{\it vector} \inxx{$r$-vector}
and call $r$ its
{\it degree} or {\it grade}: 

\setlength{\unitlength}{1cm}
\begin{picture}(12.5,4.5)

\put(1,0){\makebox(1.8,4.5)
{\begin{minipage} [t] {1.8cm}
0-vector\\
1-vector\\
2-vector\\
3-vector\\
\centerline{.\ \ }
\centerline{.\ \ }
\centerline{.\ \ }
$r$-vector
\end{minipage} }}

\put(4,0){\makebox(5,4.5)
{\begin{minipage} [t] {5cm}
\centerline{$s$} 
\centerline{$a$}
\centerline{$a \wedge b$}
\centerline{$a \wedge b \wedge c$}
\centerline{.}
\centerline{.}
\centerline{.}
\centerline{$A_r = a_1 \wedge a_2 \wedge ... \wedge a_r$}

\end{minipage} }}

\put(9.5,0){\makebox(2,4.5)
{\begin{minipage} [t] {2cm}
scalar\\
vector\\
bivector\\
trivector\\
\centerline{.\ \ \ \ }
\centerline{.\ \ \ \ }
\centerline{.\ \ \ \ }
multivector

\end{minipage} }}

\end{picture}

In a space of finite dimension this cannot continue indefinitely: an
$n$-vector is the highest $r$-vector in $V_n$ and an $(n+1)$-{\it vector} is
identically zero. An $r$-vector $A_r$ represents an oriented $r$-volume
(or $r$-direction) in $V_n$.

Multivectors $A_r$ are elements \inxx{multivector}
of the {\it Clifford algebra} ${\cal C}_n$ of
$V_n$. An element of ${\cal C}_n$ will be called a {\it Clifford number}.
Clifford numbers \inxx{Clifford number}
can be multiplied amongst themselves and the results are
Clifford numbers of mixed degrees, as indicated in the basic equation
(\ref{3.10}). The theory of multivectors, based on Clifford algebra, was
developed by Hestenes \ci{13}. \inxx{Hestenes}
In Box 2.2 some useful formulas are displayed
without proofs.

Let $e_1, \, e_2, \, ..., \, e_n$ be linearly independent vectors,
and $\alpha, \, \alpha^i, \, \alpha^{i_1 i_2}, ... \, $ scalar coefficients.
A generic Clifford number can then be written as
\be
       A = \alpha + \alpha^i e_i + {1\oo {2!}} \, \alpha^{i_1 i_2}
       \,e_{i_1} \wedge e_{i_2} + ... {1\oo {n!}} \, \alpha^{i_1 ... i_n}
       e_{i_1} \wedge ... \wedge e_{i_n} \, .
\lbl{3.14}
\ee

Since it is a superposition of multivectors of all possible grades
it will be called \inxx{polyvectors}
{\it polyvector}.\footnote{
Following a suggestion by Pezzaglia \ci{14} I call a generic Clifford number
{\it polyvector} and reserve the name {\it multivector} for an $r$-vector,
since the latter name is already widely used for the corresponding object
in the calculus of differential forms.} Another name, also often used
in the literature,
is {\it Clifford aggregate}.\inxx{Clifford aggregates}These mathematical objects
have far reaching geometrical and physical implications which will be
discussed and explored to some extent in the rest of the book.

\setlength{\unitlength}{1cm}
\begin{picture}(12.5,19)

\put(0,0){\framebox(12.5,19){
\begin{minipage}[t]{11.5cm}

\centerline{\bf Box 2.2: Some useful basic equations}

\vs{5mm}

For a vector $a$ and an $r$-vector $A_r$ the inner and the
outer product are defined according to \inxx{inner product} \inxx{outer product}
\be
   a \cdot A_r \equiv \mbox{$1\oo 2$} \left ( a A_r - (-1)^r A_r a \right ) =
   - (-1)^r A_r \cdot a ,
\lbl{3.13a}
\ee
\be
      a \wedge A_r = \mbox{$1\oo 2$} \left ( a A_r + (-1)^r A_r a \right )  =
      (-1)^r A_r \wedge a .
\lbl{3.13b}
\ee
The inner product has symmetry opposite to that of the outer pro\-duct,
therefore the signs in front of the second terms in the above equations
are different.      

Combining (\ref{3.13a}) and (\ref{3.13b}) we find
\be
    a A_r = a \cdot A_r + a \wedge A_r .
\lbl{3.13g}
\ee
For $A_r = a_1 \wedge a_2 \wedge ... \wedge a_r$ eq.\,(\ref{3.13a}) can be
evaluated to give the useful expansion
\be
   a \cdot (a_1 \wedge ... \wedge a_r) = \sum_{k=1}^r (-1)^{k+1}(a \cdot a_k)
   a_1 \wedge ... a_{k-1} \wedge a_{k+1} \wedge ... a_r .
\lbl{3.13h}
\ee
In particular,
\be
      a \cdot (b \wedge c) = (a \cdot b)c - (a \cdot c) b .
\lbl{3.13i}
\ee        
It is very convenient to introduce, besides the basis vectors $e_{\mu}$,
another set of basis vectors $e^{\nu}$ by the condition
\be
    e_{\mu} \cdot e^{\nu} = {\delta_{\mu}}^{\nu} .
\lbl{ 3.13c}
\ee
Each $e^{\mu}$ is a linear combination of $e_{\nu}$:
\be
     e^{\mu} = g^{\mu \nu} e_{\nu} ,
\lbl{3.13d}
\ee
from which we have
\be
    g^{\mu \alpha} g_{\alpha \nu} = {\delta_{\mu}}^{\nu}
\lbl{3.13e}
\ee
and 
\be
     g^{\mu \nu} = e^{\mu} \cdot e^{\nu} = \mbox{$1\oo 2$} (e^\mu e^\nu +
     e^{\nu} e^{\mu} ) .
\lbl{3.13f}
\ee         

\end{minipage} }}
\end{picture}

\section{Algebra of spacetime}

In spacetime we have 4 linearly independent vectors $e_{\mu}$, $\mu =
0,1,2,3$. Let us consider {\it flat} spacetime. It is then convenient to
take orthonormal basis vectors $\gamma_{\mu}$ \inxx{basis vectors}
\be
       \gamma_{\mu} \cdot \gamma_{\nu} = \eta_{\mu \nu} ,
\lbl{3.14a}
\ee
where $\eta_{\mu \nu}$ is the diagonal metric tensor with signature
(+ - - -). 

\vs{1mm}

The Clifford algebra \inxx{Clifford algebra} \inxx{Dirac algebra}
in $V_4$ is called the {\it Dirac algebra}. Writing
$\gamma_{\mu \nu} \equiv \gamma_{\mu} \wedge \gamma_{\nu}$ for a basis
bivector, $\gamma_{\mu \nu \rho} \equiv  \gamma_{\mu} \wedge \gamma_{\nu} 
\wedge \gamma_{\rho}$ for a basis trivector, and $\gamma_{\mu \nu \rho
\sigma} \equiv  \gamma_{\mu} \wedge \gamma_{\nu} \wedge \gamma_{\rho} 
\wedge \gamma_{\sigma}$ for a basis quadrivector we can express
an arbitrary number of the Dirac algebra as
\be
     D = \sum_r D_r = d + d^{\mu} \gamma_{\mu} + {1 \oo {2!}} \, d^{\mu \nu}
     \gamma_{\mu \nu} + {1\oo {3!}} \, d^{\mu \nu \rho} \gamma_{\mu \nu \rho} +
     {1\oo {4!}} \, d^{\mu \nu \rho \sigma} \gamma_{\mu \nu \rho \sigma} \, ,
\lbl{3.15}
\ee
where $d, \, d^{\mu}, \, d^{\mu \nu}, ...$ are scalar coefficients.

\vs{1mm}

Let us introduce
\be
   \gamma_5 \equiv \gamma_0 \wedge \gamma_1 \wedge \gamma_2 \wedge \gamma_3
   = \gamma_0 \gamma_1 \gamma_2 \gamma_3 \; \; , \qquad \gamma_5^2 = - 1 ,
\lbl{3.16}
\ee
which is the unit element of 4-dimensional volume and is called a
{\it pseudoscalar}. Using the relations \inxx{pseudoscalar}
\be
    \gamma_{\mu \nu \rho \sigma} = \gamma_5 \epsilon_{\mu \nu \rho \sigma} \, ,
\lbl{3.17}
\ee
\be
    \gamma_{\mu \nu \rho} = \gamma_{\mu \nu \rho \sigma} \gamma^{\rho} \, ,
\lbl{3.18}
\ee

\nnnn where  $\epsilon_{\mu \nu \rho \sigma}$ is the totally
antisymmetric tensor  and introducing the new coefficients

  $$ S \equiv d \; , \quad V^{\mu} \equiv d^{\mu} \; , \quad T^{\mu \nu}
  \equiv\mbox{$1\oo 2$} d^{\mu \nu}\, , $$
\be
    C_{\sigma} \equiv {1\oo {3!}} d^{\mu \nu \rho} 
    \epsilon_{\mu \nu \rho \sigma} \; , \qquad P \equiv {1\oo {4!}}
    d^{\mu \nu \rho \sigma} \epsilon_{\mu \nu \rho \sigma}  \; ,
\lbl{3.19}
\ee

\nnnn we can rewrite $D$ of eq.\,(\ref{3.15}) as the sum of scalar, vector,
bivector, pseudovector and pseudoscalar parts:\inxx{pseudovector}

\be
    D = S + V^{\mu} \gamma_{\mu} + T^{\mu \nu} \gamma_{\mu \nu} + 
    C^{\mu} \gamma_5 \gamma_{\mu} + P \gamma_5 .
\lbl{3.20}
\ee

\newpage

\subsubsection{Polyvector fields}

A polyvector may depend on spacetime points. Let $A = A(x)$ be an $r$-vector
field. Then one can define the {\it gradient operator} 
\inxx{gradient operator} according to
\be
     \p = \gamma^{\mu} \p_{\mu} \; ,
\lbl{3.21}
\ee
where $\p_{\mu}$ is the usual partial derivative. The gradient
operator $\p$ can act on any $r$-vector field. Using (\ref{3.13g}) we
have \inxx{gradient operator}
\be
      \p A = \p \cdot A + \p \wedge A .
\lbl{3.22}
\ee

{\it Example}. Let $A = a = a_{\nu} \gamma^{\nu}$ be a 1-vector field.
Then
\begin{eqnarray}
    \p a &=& \gamma^{\mu} \p_{\mu} (a_{\nu} \gamma^{\nu}) = \gamma^{\mu}
    \cdot \gamma^{\nu} \, \p_{\mu} a^{\nu} + \gamma^{\mu} \wedge \gamma^{\nu}
    \p_{\mu} a_{\nu} \nonumber \\
    &=& \p_{\mu} a^{\mu} + \mbox{$1\oo 2$} (\p_{\mu} a_{\nu} - \p_{\nu} a_{\mu})
    \gamma^{\mu} \wedge \gamma^{\nu} \; .
\lbl{3.23}
\end{eqnarray}
The simple expression $\p a$ thus contains a scalar and a bivector part,
the former being the usual divergence and the latter the usual curl of
a vector field.

\paragraph{Maxwell equations} \inxx{Maxwell equations}
 We shall now demonstrate by a concrete physical
example the usefulness of Clifford algebra. Let us consider the electromagnetic
field which, in the language of Clifford algebra, is a bivector field $F$.
The source of the field is the electromagnetic current
\inxx{electromagnetic current} $j$ which is a 1-vector
field. Maxwell's equations read
\be 
    \p F = - 4 \pi j .
\lbl{3.24}
\ee
The grade of the gradient operator $\p$ is 1. Therefore we can use
the relation (\ref{3.22}) and we find that eq.\,(\ref{3.24}) becomes
\be
    \p \cdot F + \p \wedge F = - 4 \pi j ,
\lbl{3.25}
\ee
which is equivalent to
\be
    \p \cdot F = - 4 \pi j ,
\lbl{3.25a}
\ee
\be
     \p \wedge F = 0 ,
\lbl{3.25b}
\ee
since the first term on the left of eq.\,(\ref{3.25}) is a vector and the second
term is a bivector. This results from the general relation (\ref{3.25} ).
It can also be explicitly demonstrated. Expanding
\be
    F = \mbox{$1\oo 2$} F^{\mu \nu} \, \gamma_{\mu} \wedge \gamma_{\nu} \; ,
\lbl{3.26}
\ee
\be
   j = j^{\mu} \gamma_{\mu} \; ,
\lbl{3.27a}
\ee
we have
\begin{eqnarray}
      \p \cdot F & = & \gamma^{\alpha} \p_{\alpha} \cdot (\mbox{$1\oo 2$} F^{\mu \nu}
      \gamma_{\mu} \wedge \gamma_{\nu}) = \mbox{$1\oo 2$} \gamma^{\alpha} \cdot
      (\gamma_{\mu} \wedge \gamma_{\nu}) \p_{\alpha} F^{\mu \nu} \nonumber \\
      \nonumber \\
       & = & \mbox{$1\oo 2$} \left ( (\gamma^{\alpha} \cdot \gamma_{\mu}) \gamma_{\nu} -
       (\gamma^{\alpha} \cdot \gamma_{\nu})\gamma_{\mu} \right ) \p_{\alpha}
       F^{\mu \nu} = \p_{\mu} F^{\mu \nu} \, \gamma_{\nu} \; ,
\lbl{3.27} \\
     \nonumber \\ 
    \p \wedge F &=& \mbox{$1\oo 2$} \gamma^{\alpha} \wedge \gamma_{\mu} \wedge
     \gamma_{\nu} \, \p_{\alpha} F^{\mu \nu} = \mbox{$1\oo 2$} {\epsilon^{\alpha}}_{
     \mu \nu \rho} \, \p_{\alpha} F^{\mu \nu} \gamma_5 \gamma^{\rho} \; ,
\lbl{3.28}
\ear
where we have used (\ref{3.13i}) and eqs.(\ref{3.17}), (\ref{3.18}).
From the above considerations it then follows that the compact equation 
(\ref{3.24}) is equivalent to the usual tensor form of
\inxx{Maxwell equations} Maxwell equations
\be
    \p_{\nu} F^{\mu \nu} = - 4 \pi j^{\mu} \; ,
\lbl{3.29}
\ee
\be
    {\epsilon^{\alpha}}_{\mu \nu \rho} \, \p_{\alpha} F^{\mu \nu} = 0 .
\lbl{3.30}
\ee
Applying the gradient operator \inxx{gradient operator}
$\p$ to the left and to the right side
of eq.\,(\ref{3.24}) we have
\be
     \p^2 F = - 4 \pi \, \p j .
\lbl{3.31}                
\ee
Since $\p^2 = \p \cdot \p + \p  \wedge \p = \p \cdot \p$ is a scalar
operator, $\p^2 F$ is a bivector. The right hand side of eq.\,(\ref{3.31}) gives
\be
    \p j = \p \cdot j + \p \wedge j .
\lbl{3.32}
\ee

Equating the terms of the same grade on the left and the right hand side of
eq.\,(\ref{3.31}) we obtain
\be
     \p^2 F = - 4 \pi \, \p \wedge j ,
\lbl{3.33}
\ee
\be
     \p \cdot j = 0 .
\lbl{3.34}
\ee
The last equation expresses the conservation of
\inxx{electromagnetic current} the electromagnetic current.

\paragraph{Motion of a charged particle} \inxx{charged particle}
In this example we wish to go a step
forward. Our aim is not only to describe how a charged particle moves in an
electromagnetic field, but also include a particle' s(classical) spin. Therefore,
following Pezzaglia \ci{14}, we define the
\inxx{momentum polyvector} {\it momentum polyvector} $P$ as the
{\it vector momentum} $p$ plus the bivector
\inxx{spin angular momentum} {\it spin angular momentum}
$S$,
\be
    P = p + S ,
\lbl{3.35}
\ee
or in components
\be
    P = p^{\mu} \gamma_{\mu} + \mbox{$1\oo 2$} S^{\mu \nu} \, \gamma_{\mu} \wedge
    \gamma_{\nu} .
\lbl{3.36}
\ee
We also assume that the condition $p_{\mu} S^{\mu \nu} = 0$ is
satisfied. The latter condition ensures the spin to be a simple bivector,
which is purely space-like in the rest frame of the particle. The polyvector
equation of motion is
\be {\dot P} \equiv {{{\dd} P} \oo {{\dd} \tau}} = {e \oo {2m}} \, [P, F] ,
\lbl{3.37}
\ee
where $[P,F] \equiv PF - FP$. The vector and bivector parts of 
eq.\,(\ref{3.37}) are
\be
     {\dot p}^{\mu} = {e\oo m} \, {F^{\mu}}_{\nu} p^{\nu} \; ,
\lbl{3.38}
\ee
\be
     {\dot S}^{\mu \nu} = {e\oo {2m}} ({F^{\mu}}_{\alpha} S^{\alpha \nu}
     - {F^{\nu}}_{\alpha} S^{\alpha \mu}) .
\lbl{3.39}
\ee
These are just the equations of motion for linear momentum and spin,
respectively.
          
\section{Physical quantities as polyvectors}

The compact equations at the end of the last section suggest
a generalization that every physical quantity is
\inxx{polyvector} a polyvector. We shall
explore such an assumption and see how far we can come.

In 4-dimensional spacetime
\inxx{momentum polyvector} {\it the momentum polyvector} is
\be
    P = \mu + p^{\mu} e_{\mu} + S^{\mu \nu} e_{\mu} e_{\nu} +
    \pi^{\mu} e_5 e_{\mu} + m e_5 \; ,
\lbl{3.40}
\ee
and {\it the velocity polyvector} \inxx{velocity polyvector} is
\be
   {\dot X} = {\dot \sigma} + {\dot x}^{\mu} e_{\mu} + {\dot \alpha}^{\mu \nu}
   e_{\mu} e_{\nu} + {\dot \xi}^{\mu} e_5 e_{\mu} + {\dot s} e_5 \; ,
\lbl{3.41}
\ee
where $e_{\mu}$ are four basis vectors satisfying
\be
       e_{\mu} \cdot e_{\nu} = \eta_{\mu \nu} \; ,
\lbl{3.41a}
\ee
and $e_5 \ \equiv e_0 e_1 e_2 e_3$ is the pseudoscalar. For the purposes
which will become clear later we now use the symbols $e_{\mu}$, $e_5$
instead of $\gamma_{\mu}$ and $\gamma_5$.

We associate with each particle the velocity polyvector ${\dot X}$ and
its conjugate momentum polyvector $P$. These quantities are  generalizations
of the point particle 4-velocity ${\dot x}$ and its conjugate momentum $p$.
Besides a vector part we now include the scalar part ${\dot \sigma}$, the
bivector part ${\dot \alpha}^{\mu \nu} e_{\mu} e_{\nu}$, the pseudovector
part ${\dot \xi}^{\mu} e_5 e_{\mu}$ and the pseudoscalar part ${\dot s} e_5$
into the definition of the particle's velocity, and analogously 
for the particle's
momentum. We would now like to derive the equations of motion which will
tell us how those quantities depend on the evolution parameter $\tau$. For
simplicity we consider a free particle.

Let the action be a straightforward generalization of the first order or
phase space action (\ref{2.1.10}) of the usual constrained point particle
relativistic theory:
\be
    I[X,P,\lambda] = {1 \oo 2} \int {\dd} \tau \, 
    \left ( P {\dot X} + {\dot X} P
    - \lambda (P^2 - K^2) \right ) ,
\lbl{3.42}
\ee
where $\lambda$ is a scalar Lagrange multiplier and $K$ a polyvector
constant\footnote{
   The scalar part is not restricted to positive values, but for later
   convenience we write it as $\kappa^2$, on the understanding that $\kappa^2$
   can be positive, negative or zero.}:
\be
     K^2 = \kappa^2 + k_{\mu} e_{\mu} + K^{\mu \nu} e_{\mu} e_{\nu} +
     K^{\mu} e_5 e_{\mu} + k^2 e_5 .
\lbl{3.42a}
\ee
It is a generalization of particle's mass squared. In the usual,
unconstrained, theory, mass squared was a scalar constant, but here we
admit that, in principle, mass squared is a polyvector. Let us now insert
the explicit expressions (\ref{3.40}),(\ref{3.41}) and (\ref{3.42a})
into the Lagrangian
\be
     L = \mbox{$1\oo 2$} \left ( P {\dot X} + {\dot X} P - \lambda (P^2 - K^2) \right )
     = \sum_{r=0}^4 {\langle L \rangle}_r \; ,
\lbl{3.43}
\ee
and evaluate the corresponding multivector parts ${\langle L \rangle}_r$.
Using
\be
    e_{\mu} \wedge e_{\nu} \wedge e_{\rho} \wedge e_{\sigma} = e_5 \, 
    \epsilon_{\mu \nu \rho \sigma} \; ,
\lbl{3.44}
\ee
\be
    e_{\mu} \wedge e_{\nu} \wedge e_{\rho} = 
    (e_{\mu} \wedge e_{\nu} \wedge e_{\rho} \wedge e_{\sigma})   e^{\sigma} =
    e_5 \, \epsilon_{\mu \nu \rho \sigma} e^{\sigma} \; ,
\lbl{3.45}
\ee
\be     e_{\mu} \wedge e_{\nu} = - \mbox{$1\oo 2$} (
     e_{\mu} \wedge e_{\nu} \wedge e_{\rho} \wedge e_{\sigma})(e^{\rho} \wedge 
     e^{\sigma}) = -\mbox{$1\oo 2$} e_5 \, \epsilon_{\mu \nu \rho \sigma}
     e^{\rho}\wedge e_{\sigma} \; ,
\lbl{3.46}
\ee
we obtain
   $$\langle L \rangle_0 = \mu {\dot \sigma} - m {\dot s} 
     + p_{\mu} {\dot x}^{\mu} + {\pi}_{\mu} {\dot \xi}^{\mu} + S^{\mu \nu}
     {\dot \alpha}^{\rho \sigma} \eta_{\mu \sigma} \eta_{\nu \rho}
     \hs{5cm}$$
\be
     \hs{3cm} - \, {\lambda \oo 2} (\mu^2 + p^{\mu} 
     p_{\mu} + {\pi}^{\mu} {\pi}_{\mu}
     - m^2 - 2 S^{\mu \nu} S_{\mu \nu} - \kappa^2) \; ,
\lbl{3.47}
\ee
$$\langle L \rangle_1 = \left [ {\dot \sigma} p_{\sigma} + \mu {\dot x}_{\sigma}
-({\dot \xi}^{\rho} S^{\mu \nu} + \pi^{\rho} {\dot \alpha}^{\mu \nu})
  \epsilon_{\mu \nu \rho \sigma} \right ] e^{\sigma} \hs{5cm}$$
\be
  \hs{3cm} - \, \lambda (\mu p_{\sigma} - S^{\mu \nu} \pi^{\rho} 
   \epsilon_{\mu \nu \rho \sigma} - \mbox{$1\oo 2$} k_{\sigma} ) e^{\sigma} \; ,
\lbl{3.48}
\ee
$$ \langle L \rangle_2 = \mbox{\hspace{11.5cm}}$$
$$\left [ \mbox{$1\oo 2$} (\pi^{\mu} {\dot x}^{\nu} -
p^{\mu} {\dot \xi}^{\nu} + {\dot s} S^{\mu \nu} + m {\dot \alpha}^{\mu \nu})
\epsilon_{\mu \nu \rho \sigma} + {\dot \sigma} S_{\rho \sigma} + \mu
{\dot \alpha}_{\rho \sigma} + 2 S_{\rho \nu} {{\dot \alpha}^{\nu}}_{\sigma}
\right ] e^{\rho} \wedge e^{\sigma}$$
\be
   - {\lambda \oo 2} \left [ (\pi^{\mu} p^{\nu} + m S^{\mu \nu}) 
   \epsilon_{\mu \nu \rho \sigma} + 2 \mu S_{\rho \sigma} 
    - K_{\rho \sigma} \right ] e^{\rho} \wedge e^{\sigma} \; ,
\lbl{3.49}
\ee
\be \langle L \rangle_3 = \biggl[ {\dot \sigma} \pi^{\sigma} + \mu {\dot \xi}^
{\sigma} + (S^{\mu \nu} {\dot x}^{\rho} + {\dot \alpha}^{\mu \nu} p^{\rho})
{\epsilon_{\mu \nu \rho}}^{\sigma} - \lambda (\mu \pi^{\sigma} + S^{\mu \nu}
p^{\rho} {\epsilon_{\mu \nu \rho}}^{\sigma} - \mbox{$1\oo 2$} \kappa_{\sigma})
\biggr] e_5 e^{\sigma} \; ,
\lbl{3.50}
\ee
\be
   \langle L \rangle_4 =
   \left [ m {\dot \sigma} + \mu {\dot s} - \mbox{$1\oo 2$}\, S^{\mu \nu} \,
   {\dot \alpha}^{\rho \sigma} \epsilon_{\mu \nu \rho \sigma} - 
   {\lambda \oo 2} ( 2 \mu m + S^{\mu \nu} S^{\rho \sigma} 
   \epsilon_{\mu \nu \rho \sigma} - k^2) \right ] e_5 . \hs{7mm}
\lbl{3.51}
\ee

The equations of motion are obtained for each pure grade multivector
$\langle L \rangle_r$ separately. That is, when varying the polyvector
action $I$, we vary each of its $r$-vector parts separately. From the
scalar part $\langle L \rangle_0$ we obtain
\begin{eqnarray}
\delta \mu \; &:& \qquad {\dot \sigma} - \lambda \mu = 0 , 
\lbl{3.52} \\
\delta m \; &:& \qquad - {\dot s} + \lambda m = 0 , 
\lbl{3.53} \\
\delta s \; &:& \qquad {\dot m} = 0 \lbl{3.54} \\
\delta \sigma \; &:& \qquad {\dot \mu} = 0 , 
\lbl{3.55} \\
\delta p_{\mu} \; &:& \qquad {\dot x}^{\mu} - \lambda p^{\mu} = 0 , 
\lbl{3.56} \\
\delta \pi_{\mu} \; &:& \qquad {\dot \xi}^{\mu} - \lambda \pi^{\mu} = 0 , 
\lbl{3.57} \\
\delta x^{\mu} \; &:& \qquad {\dot p}^{\mu} = 0 \lbl{3.58} \\
\delta \xi^{\mu} \; &:& \qquad {\dot \pi}_{\mu} = 0 , 
\lbl{3.59} \\
\delta \alpha^{\mu \nu} \; &:& \qquad {\dot S}_{\mu \nu} = 0 , 
\lbl{3.60}  \\
\delta S_{\mu \nu} \; &:& \qquad {\dot \alpha}^{\mu \nu} - \lambda S^{\mu \nu}
= 0 . \lbl{3.61}
\end{eqnarray}

From the $r$-vector parts $\langle L \rangle_r$ for $r = 1$, 2, 3, 4 we
obtain the same set of equations (\ref{3.52})--(\ref{3.61}). Each individual
equation results from varying a different variable in $\langle L \rangle_0$,
$\langle L \rangle_1$, etc.. Thus, for instance, the $\mu$-equation of motion
(\ref{3.52}) from $\langle L \rangle_0$ is the same as the $p_{\mu}$
equation from $\langle L \rangle_1$ and the same as the $m$-equation from
$\langle L \rangle_4$, and similarly for all the other equations 
(\ref{3.52})--(\ref{3.61}). Thus, as far as the variables $\mu$, $m$, $s$,
$\sigma$, $p_{\mu}$, $\pi_{\mu}$, $S_{\mu \nu}$, $\xi^{\mu \nu}$ and $\alpha^
{\mu \nu}$ are considered, the higher grade parts $\langle L \rangle_r$ of
the Lagrangian $L$ contains the same information about the equations of
motion. The difference occurs if we consider the {\it Lagrange multiplier}
$\lambda$. Then every $r$-vector part of $L$ gives a different equation of
motion:
\begin{eqnarray}
     {{\p \langle L \rangle_0}\oo {\p \lambda}} &=& 0 \; : \quad
     \mu^2 + p^{\mu} p_{\mu} + \pi^{\mu} \pi_{\mu} - m^2 - 2 S^{\mu \nu}
     S_{\mu \nu} - \kappa^2 = 0 , \lbl{3.62} \\
     {{\p \langle L \rangle_1}\oo {\p \lambda}} &=& 0 \; : \quad
     \mu \, \pi_{\sigma} - S^{\mu \nu} \pi^{\rho} \epsilon_{\mu \nu \rho \sigma}
     - \mbox{$1\oo 2$} k_{\sigma} = 0 , \lbl{3.63} \\
     {{\p \langle L \rangle_2}\oo {\p \lambda}} &=& 0 \; : \quad   
      (\pi^{\mu} \pi^{\nu} + m S^{\mu \nu})\epsilon_{\mu \nu \rho \sigma} +
      2 \mu S_{\rho \sigma} - K_{\rho \sigma} = 0  , \lbl{3.64} \\
      {{\p \langle L \rangle_3}\oo {\p \lambda}} &=& 0 \; : \quad
      \mu \pi_{\sigma} + S^{\mu \nu} p^{\rho} \epsilon_{\mu \nu \rho \sigma}
     + \mbox{$1\oo 2$} \kappa_{\sigma} = 0 , \lbl{3.65} \\
     {{\p \langle L \rangle_4}\oo {\p \lambda}} &=& 0 \; : \quad
     2 \mu m + S^{\mu \nu} S^{\rho \sigma} \epsilon_{\mu \nu \rho \sigma} 
     - k^2 = 0  . \lbl{3.66}
\end{eqnarray}
The above equations represent constraints among the dynamical variables.
Since our Lagrangian is not a scalar but a polyvector, we obtain more than
one constraint.

Let us rewrite eqs.(\ref{3.65}),(\ref{3.63}) in the forms
\bear
      \pi_{\sigma} - {{\kappa_{\sigma}}\oo {2 \mu}} &=& - {1\oo \mu} \,
      S^{\mu \nu} p^{\rho} \epsilon_{\mu \nu \rho \sigma} \; ,
\lbl{3.69}\\
    p_{\sigma} - {{k_{\sigma}}\oo {2 \mu}} &=& {1 \oo \mu} S^{\mu \nu}
    \pi^{\rho} \epsilon_{\mu \nu \rho \sigma}  .
\lbl{3.70}
\ear
We see from (\ref{3.69}) that the vector momentum $p_{\mu}$ and its
pseudovector partner $\pi_{\mu}$ are related in such a way that $\pi_{\mu} -
\kappa_{\mu}/2\mu$ behaves as the well known {\it Pauli-Lubanski spin
pseudo vector}. \inxx{Pauli--Lubanski pseudo-vector}
A similar relation (\ref{3.70}) holds if we interchange
$p_{\mu}$ and $\pi_{\mu}$.

Squaring relations (\ref{3.69}), (\ref{3.70}) we find
\bear
    (\pi_{\sigma} - {{\kappa_{\sigma}}\oo {2 \mu}})
     (\pi^{\sigma} - {{\kappa^{\sigma}}\oo {2 \mu}}) &=&
     - \, {2 \oo {\mu^2}} \, p_{\sigma} p^{\sigma} \, S_{\mu \nu} S^{\mu \nu}
     + {4\oo \mu^2} \, p_{\mu} p^{\nu} S^{\mu \sigma} S_{\nu \sigma} \; ,
\hs{.5cm}\lbl{3.71} \\
    (p_{\sigma} - {{k_{\sigma}}\oo {2 \mu}})
     (p^{\sigma} - {{k^{\sigma}}\oo {2 \mu}}) &=&
  - \, {2 \oo {\mu^2}} \, \pi_{\sigma} \pi^{\sigma} \, S_{\mu \nu} S^{\mu \nu}
  + {4\oo \mu^2} \, \pi_{\mu} \pi^{\nu} S^{\mu \sigma} S_{\nu \sigma} .
\hs{.5cm}\lbl{3.71a}
\ear

From (\ref{3.69}), (\ref{3.70}) we also have
\be
     \left ( \pi_{\sigma} - {{\kappa_{\sigma}}\oo {2 \mu}} \right ) p^{\sigma} 
     = 0 \; , \quad
     \left ( p_{\sigma} - {{k_{\sigma}}\oo {2 \mu}} \right ) \pi^{\sigma} = 0 .
\lbl{3.72}
\ee

Additional interesting equations which follow from (\ref{3.69}),
(\ref{3.70}) are
\be
   p^{\rho} p_{\rho} \left ( 1 + {2 \oo {\mu^2}} \, S_{\mu \nu} S^{\mu \nu}
   \right )
   - {4\oo \mu^2} \, p_{\mu} p^{\nu} S^{\mu \sigma} S_{\nu \sigma} =
   - \, {{\kappa_{\rho} \kappa^{\rho}}\oo {4 \mu^2}} + {1 \oo {2 \mu}} 
   (p_{\rho} k^{\rho} + \pi_{\rho} \kappa^{\rho}) ,
\lbl{3.73}
\ee
\be
   \pi^{\rho} \pi_{\rho} \left ( 1 + {2 \oo {\mu^2}} \, S_{\mu \nu} S^{\mu \nu}
   \right ) 
   - {4\oo \mu^2} \, \pi_{\mu} \pi^{\nu} S^{\mu \sigma} S_{\nu \sigma} =
   - \, {{k_{\rho} k^{\rho}}\oo {4 \mu^2}} + {1 \oo {2 \mu}} 
   (p_{\rho} k^{\rho} + \pi_{\rho} \kappa^{\rho}) .
\lbl{3.74}
\ee

Contracting (\ref{3.69}), (\ref{3.70}) by $\epsilon^{\alpha_1 \alpha_2
\alpha_3 \sigma}$ we can express $S^{\mu \nu}$ in terms of $p^{\rho}$ and
$\pi^{\sigma}$  
\be
    S^{\mu \nu} = {\mu \oo {2 p^{\alpha} p_{\alpha}}} \, \epsilon^{\mu \nu \rho
    \sigma} p_{\rho} \left ( \pi_{\sigma} - {{\kappa_{\sigma}}\oo {2 \mu}}
    \right ) =
    - \, {\mu \oo {2 \pi^{\alpha} \pi_{\alpha}}} \, \epsilon^{\mu \nu \rho
    \sigma} \pi_{\rho} \left ( p_{\sigma} - {{k_{\sigma}}\oo {2 \mu}}
    \right ) ,
\lbl{3.75}
\ee
provided that we assume the following extra condition:
\be
       S^{\mu \nu} p_{\nu} = 0 \;  \; , \qquad S^{\mu \nu} \pi_{\nu} = 0 .
\lbl{3.75a}
\ee       
Then for positive $p^{\sigma} p_{\sigma}$ it follows from (\ref{3.71})
that $(\pi_{\sigma} - \kappa_{\sigma}/2 \mu)^2$ is negative, i.e.,
$\pi_{\sigma} - \kappa_{\sigma}/2 \mu$ are components of a space-like
(pseudo-) vector. Similarly, it follows from (\ref{3.71a}) that when
$\pi^{\sigma} \pi_{\sigma}$ is negative, $(p_{\sigma} - k_{\sigma}/2 \mu)^2$
is positive, so that $p_{\sigma} - k_{\sigma}/2 \mu$ is a time-like vector.
Altogether we thus have that $p_{\sigma}$, $k_{\sigma}$ are time-like and
$\pi_{\sigma}$, $\kappa_{\sigma}$ are space-like.
Inserting (\ref{3.75}) into the remaining constraint (\ref{3.66})
and taking into account the condition (\ref{3.75a}) we obtain
\be
    2 m \mu - k^2 = 0 .
\lbl{3.76}
\ee

The polyvector action (\ref{3.42}) is thus shown 
to represent a very interesting classical 
dynamical system with spin. \inxx{spin}
The interactions could be included by generalizing
the minimal coupling prescription. {\it Gravitational interaction} is
included by generalizing (\ref{3.41a}) to
\be
     e_{\mu} \cdot e_{\nu} = g_{\mu \nu} \; ,
\lbl{3.77}
\ee
where $g_{\mu \nu} (x)$ is the spacetime metric tensor. {\it A gauge
interaction} is included by introducing a polyvector gauge field $A$,
a polyvector coupling constant $G$, and assume an action of the kind
\be
   I[X,P,\lambda] = \mbox{$1\oo 2$} \, \int {\dd} \tau \, \left [ P {\dot X} +
   {\dot X} P - \lambda \left ( (P - G \star A)^2 - K^2 \right ) \right ] ,
\lbl{3.78}
\ee
where `$\star$' means the scalar product \inxx{scalar product}
between Clifford numbers, so that
$G \star A \equiv \langle G A \rangle_0$.
The polyvector equations of motion can be elegantly obtained by using
the Hestenes formalism for multivector derivatives. We shall not go into
details here, but merely sketch a plausible result,
\be
    {\dot \Pi} = \lambda \lbrack G \star \p_X A , P \rbrack \; , \qquad
    \Pi \equiv P - G \star A \; ,
\lbl{3.79}
\ee
which is a generalized Lorentz force equation of motion, a more
particular case of which is given in (\ref{3.37}).

After this short digression let us return to our free particle case. One
question immediately arises, namely, what is the physical meaning of the
polyvector mass squared $K^2$. Literally this means that a particle is
characterized not only by a scalar and/or a pseudoscalar mass squared, but also
by a vector, bivector and pseudovector mass squared. To a particle are thus
associated a constant vector, 2-vector, and 3-vector which point into fixed
directions in spacetime, regardless of the direction of particle's motion.
For a given particle the Lorentz symmetry is thus broken, since there
exists a preferred direction in spacetime. This cannot be true at the
fundamental level. Therefore the occurrence of the polyvector $K^2$ in the
action must be a result of a more fundamental dynamical principle, presumably
an action in a higher-dimensional spacetime without such a fixed term $K^2$.
It is well known that the scalar mass term in 4-dimensions can be considered as
coming from a massless action in 5 or more dimensions. Similarly, also the
1-vector, 2-vector, and 3-vector terms of $K^2$ can come from a 
higher-dimensional action without a $K^2$-term. Thus {\it in 5-dimensions}:

\vs{2mm}

\hs{5mm} \ (i) {\it the scalar constraint} will contain the term
$p^A p_A = p^{\mu} p_{\mu} + p^5 p_5$, and the constant $- p^5 p_5$ takes the
role of the scalar mass term in 4-dimensions;

\hs{5mm} (ii) {\it the vector constraint} will contain a term like
$P_{ABC} S^{AB} e^C$, $A, B = 0,1,2,3,5$, containing the term $P_{\mu \nu
\alpha} S^{\mu \nu} e^{\alpha}$ (which, since $P_{\mu \nu \alpha} =
\epsilon_{\mu \nu \alpha \beta} \pi^{\beta}$, corresponds to the term
$S^{\mu \nu} \pi^{\rho} \epsilon_{\mu \nu \rho \sigma} e^{\sigma})$ plus
an extra term $P_{5 \nu \alpha} S^{5 \alpha} e^{\alpha}$ which corresponds
to the term $k^{\alpha} e_{\alpha}$.

\vs{2mm}

In a similar manner we can generate the 2-vector term $K_{\mu \nu}$ and the
3-vector term $\kappa_{\sigma}$ from 5-dimensions.

The polyvector mass term $K^2$ in our 4-dimensional action (\ref{3.78}) is
arbitrary in principle. Let us find out what happens if we set $K^2 = 0$.
Then,  in the presence of the condition
(\ref{3.75a}), eqs.\,(\ref{3.73}) or (\ref{3.74}) imply
\be
    S_{\mu \nu} S^{\mu \nu} = - {\mu^2 \oo 2} ,
\lbl{3.80}
\ee
that is $S_{\mu \nu} S^{\mu \nu} < 0$. On the other hand
$S_{\mu \nu} S^{\mu \nu}$ in the presence of the condition (\ref{3.75a})
can only be positive (or zero), as can be straightforwardly verified.
In 4-dimensional spacetime $S_{\mu \nu} S^{\mu \nu}$ were to be negative only
if in the particle's rest frame the spin components
$S^{0r}$  were different from zero which would be the
case if (\ref{3.75a}) would not hold.

Let us assume that $K^2 = 0$ and that condition (\ref{3.75a}) does hold.
Then the constraints (\ref{3.63})--(\ref{3.66}) 
have a solution\footnote{This holds
even if we keep $\kappa^2$ different from zero, but take
vanishing values for $k^2$, $\kappa_{\mu}$, $k_{\mu}$ and $K_{\mu \nu}$.}
\be
     S^{\mu \nu} = 0 \; , \quad \pi^{\mu} = 0 \; , \quad \mu = 0 .
\lbl{3.81}
\ee

The only remaining constraint is thus
\be
       p^{\mu} p_{\mu} - m^2  = 0 ,
\lbl{3.82}
\ee
and the polyvector action (\ref{3.42}) is simply\inxx{polyvector action}
\bear
     I[X,P,\lambda] &=& I[s,m,x^{\mu},p_{\mu},\lambda] \nonumber \\
      &=& \int {\dd} \tau \left [ - m {\dot s} + p_{\mu} {\dot x}^{\mu} -
     {\lambda \oo 2} (p^{\mu} p_{\mu} - m^2) \right ] ,
\lbl{3.83}
\ear
in which the mass $m$ is a dynamical variable conjugate to $s$.
In the action (\ref{3.83}) mass is thus just a 
pseudoscalar component of the polymomentum \inxx{polymomentum}
\be
        P = p^{\mu} e_{\mu} + m e_5 \; ,
\lbl{3.83a}
\ee
and ${\dot s}$ is a  pseudoscalar component of the
velocity polyvector \inxx{velocity polyvector}
\be
    {\dot X} = {\dot x}^{\mu} e_{\mu} + {\dot s} e_5 .
\lbl{3.83b}
\ee
Other components of the polyvectors ${\dot X}$ and $P$ (such as
$S^{\mu \nu}$, $\pi^{\mu}$, $\mu$), when $K^2 =0$ (or more weakly, when
$K^2 = \kappa^2$), are automatically eliminated by the
constraints (\ref{3.62})--(\ref{3.66}).

From a certain point of view this is very good, since our
analysis of the polyvector action
(\ref{3.42}) has driven us close to the conventional point particle theory,
with the exception that mass is now a dynamical variable. This reminds us
of the Stueckelberg point particle theory \ci{2}--\ci{11} 
in which mass is a constant of
motion. This will be discussed in the next section. We have here demonstrated
in a very elegant and natural way that the Clifford algebra generalization
of the classical point particle in four dimensions tells us that a fixed mass term
in the action cannot be considered as fundamental. This is not so obvious for
the scalar (or pseudoscalar) part of the polyvector mass squared term $K^2$,
but becomes dramatically obvious for the 1-vector, 2-vector and 4-vector parts,
because they imply a preferred direction in spacetime, and such a preferred
direction cannot be fundamental if the theory is to be Lorentz covariant.

This is a very important point and I would like to rephrase it. We start
with the well known relativistic constrained  action
\be
     I[x^{\mu}, p_{\mu}, \lambda] = \int {\dd} \tau \, \left ( p_{\mu}
     {\dot x}^{\mu} - {\lambda \oo 2} (p^2 - \kappa^2) \right ) .
\lbl{3.83c}
\ee
Faced with the existence of the geometric calculus based on Clifford
algebra, it is natural to generalize this action to polyvectors. Concerning
the fixed mass constant $\kappa^2$ it is natural to replace it by a fixed
polyvector or to discard it. If we discard it we find that mass is
nevertheless present, because now momentum is a polyvector and as such it
contains a pseudoscalar part $m e_5$. If we keep the fixed mass term
then we must also keep, in principle, its higher grade parts, but this is in
conflict with Lorentz covariance. Therefore the fixed mass term in the
action is not fundamental but comes, for instance, from higher
dimensions. Since, without the $K^2$ term, in the presence of the condition
$S^{\mu \nu} p_{\nu} = 0$ we cannot have classical spin in \inxx{classical spin}
four dimensions (eq.\,(\ref{3.80}) is inconsistent), this points to the
existence of higher dimensions. Spacetime must have more than four dimensions,
where we expect that the constraint $P^2 = 0$
(without a fixed polyvector mass squared term $K$) allows for nonvanishing
classical spin.

The ``fundamental" classical action  is thus a polyvector action in higher
dimensions without a fixed mass term. Interactions are associated with the metric
of $V_N$. Reduction to four dimensions gives us gravity plus gauge interactions,
such as the electromagnetic and Yang--Mills interactions, and also the 
classical spin which is associated with the bivector dynamical degrees of
freedom sitting on the particle, for instance the particle's finite
extension, magnetic moment, and similar.

There is a very well known problem with \inxx{Kaluza--Klein theory}
Kaluza--Klein theory, since
in four dimensions a charged particle's mass cannot be smaller that the
Planck mass. \inxx{Planck mass}
Namely, when reducing from five to four dimensions mass is given by
$p^{\mu} p_{\mu} = {\hat m}^2 + {\hat p}_5^2$, where ${\hat m}$ is the
5-dimensional mass.
Since ${\hat p}_5$ has the role of electric charge $e$, the latter relation is 
problematic for the electron: in the units in which $\hbar = c = G =1$ the charge
$e$ is of the order of the Planck mass, so  $p^{\mu} p_{\mu}$ is also of
the same order of magnitude. There is no generally accepted mechanism
for solving such a problem. In the polyvector generalization of the
theory, the scalar
constraint is (\ref{3.62}) and in five or more dimensions it assumes an even
more complicated form. The terms in the constraint have different signs, and
the 4-dimensional mass $p^{\mu} p_{\mu}$ is not necessarily of the order of the
Planck mass: there is a lot of room to ``make" it small.

All those considerations clearly illustrate why the polyvector generalization
of the point particle theory is of great physical interest.

\section{The unconstrained action from the polyvector action}

\subsection[\ \ Free particle]
{Free particle}

In the previous section we have found that when the polyvector fixed
mass squared $K^2$ is zero then a possible solution of the equations of motion 
satisfies (\ref{3.81}) and the generic action (\ref{3.42}) simplifies to
\be
     I[s,m,x^{\mu}, p_{\mu}, \lambda] =
     \int {\dd} \tau \left [ - m {\dot s} + p_{\mu} {\dot x}^{\mu} -
     {\lambda \oo 2} (p^{\mu} p_{\mu} - m^2) \right ] .
\lbl{3.84}
\ee
       
At this point let us observe that a similar action, with a scalar variable
$s$, has been considered by DeWitt \ci{15} and Rovelli \ci{16}. 
They associate the 
variable $s$ with the clock carried by the particle. We shall say more about that
in Sec. 6.2.

We are now going to show that the latter action is equivalent to the
Stueckelberg action discussed in Chapter 1. \inxx{Stueckelberg action}

The equations of motion resulting from (\ref{3.84}) are
\begin{eqnarray}
    \delta s \, &:& \qquad {\dot m} = 0  ,   \lbl{3.85} \\
    \delta m \, &:& \qquad {\dot s} - \lambda m = 0  , \lbl{3.86} \\
    \delta x^{\mu} \, &:& \qquad {\dot p}_{\mu} = 0 \lbl{3.87} , \\
    \delta p_{\mu}  \, &:& \qquad {\dot x}^{\mu} - \lambda p^{\mu} = , 0
       \lbl{3.88} \\
    \delta \lambda  \, &:& \qquad  p^{\mu} p_{\mu} - m^2 = 0  .\lbl{3.89} \\
\end{eqnarray}
We see that in this dynamical system mass $m$ is one of the dynamical
variables; it is canonically conjugate to the variable $s$. From the
equations of motion we easily read out that $s$ is the proper time. Namely,
from (\ref{3.86}), (\ref{3.88}) and (\ref{3.89}) we have
\be
     p^{\mu} = {{{\dot x}^{\mu}} \oo \lambda} = m \, {{{\dd} x^{\mu}}\oo {{\dd}
     s}} \; ,
\lbl{3.90}
\ee
\be
    {\dot s}^2 = \lambda^2 m^2 = {\dot x}^2 \; ,  \quad {\rm i.e} \quad \quad
    {\dd} s^2 = {\dd} x^{\mu} {\dd} x_{\mu} .
\lbl{3.91}
\ee
Using eq.\,(\ref{3.86}) we find that
\be
    - m {\dot s} + {\lambda \oo 2} \kappa^2 = - {{m {\dot s}}\oo 2} = 
    - \, {1\oo 2} {{{\dd}(m s)}\oo {{\dd} \tau}} . 
\lbl{3.92}
\ee
The action (\ref{3.84}) then becomes
\be
    I = \int {\dd} \tau \left ( {1\oo 2} {{{\dd}(m s)}\oo {{\dd} \tau}} +
    p_{\mu} {\dot x}^{\mu} - {\lambda \oo 2} \, p^{\mu} p_{\mu} \right ) \; ,
\lbl{3.93}
\ee
where $\lambda$ should be no more considered as a quantity to be varied, 
but it is
now fixed: $\lambda = \Lambda(\tau)$. The total derivative in (\ref{3.93})
can be omitted, and the action is simply
\be
    I[x^{\mu}, p_{\mu}] = \int \dd \tau (p_{\mu} {\dot x}^{\mu} - 
    {\Lambda \oo 2}\, p^{\mu} p_{\mu}) .
\lbl{3.93a}
\ee
This is just {\it the Stueckelberg action} \inxx{Stueckelberg action}
(\ref{2.2.16a}) with $\kappa^2 = 0$.
The equations of motion derived from (\ref{3.93a}) are
\be
     {\dot x}^{\mu} - \Lambda p^{\mu} = 0 ,
\lbl{3.94}
\ee
\be
      {\dot p}_{\mu} = 0 .
\lbl{3.95}
\ee
From (\ref{3.95}) it follows that $p_{\mu} p^{\mu}$ is a constant of
motion. Denoting the latter constant of motion as $m$ and using (\ref{3.94})
we obtain that momentum can be written as
\be
    p^{\mu} = m \, {{{\dot x}^{\mu}}\oo {\sqrt{{\dot x}^{\nu}{\dot x}_{\nu}}}}
    = m \, {{{\dd} x^{\mu}}\oo {{\dd} s}} \; , \quad {\dd} s = {\dd} x^{\mu}
    {\dd} x_{\mu} \; ,
\lbl{3.96}
\ee
which is the same as in eq.\,(\ref{3.90}). The equations of motion for
$x^{\mu}$ and $p_{\mu}$ derived from the Stueckelberg action (\ref{3.83a})
are the same as the equations of motion derived from the action (\ref{3.84}).
{\it A generic Clifford algebra action} (\ref{3.42}) {\it thus leads directly
to the Stueckleberg action}.

The above analysis can be easily repeated for a more general case where the
scalar constant $\kappa^2$ is different from zero, so that
instead of (\ref{3.83}) or (\ref{3.84}) we have
\be
     I[s,m,x^{\mu}, p_{\mu}, \lambda] =
     \int {\dd} \tau \left [ - m {\dot s} + p_{\mu} {\dot x}^{\mu} -
     {\lambda \oo 2} (p^{\mu} p_{\mu} - m^2 - \kappa^2) \right ] .
\lbl{3.97}
\ee
Then instead of (\ref{3.93a}) we obtain
\be
     I[x^{\mu}, p_{\mu}] = \int {\dd} \tau \left ( p_{\mu} {\dot x}^{\mu}
     - {\Lambda \oo 2} \, (p^{\mu} p_{\mu} - \kappa^2) \right ) .
\lbl{3.98}
\ee
The corresponding Hamiltonian is \inxx{Hamiltonian}
\be
    H = {\Lambda \oo 2} \, (p^{\mu} p_{\mu} - \kappa^2)  ,
\lbl{3.98a}
\ee
and in the quantized theory the \inxx{Schr\" odinger equation}
Schr\" odinger equation reads
\be
     i {{\p \psi} \oo {\p \tau}} =  
     {\Lambda \oo 2} \, (p^{\mu} p_{\mu} - \kappa^2) \psi .
\lbl{3.98b}
\ee

Alternatively, in the action (\ref{3.84}) or (\ref{3.97}) we can first eliminate
$\lambda$ by the equation of motion (\ref{3.86}). So we obtain
\be
   I[s,m,x^{\mu},p_{\mu}] = \int {\dd} \tau \left [ - {{m {\dot s}}\oo 2} +
  p_{\mu} {\dot x}^{\mu} - {{\dot s} \oo {2 m}} (p^{\mu} p_{\mu} - \kappa^2)
   \right ] .
\lbl{3.99}
\ee
The equations of motion are
\begin{eqnarray}
     \delta s \, &:& \quad - {{\dot m} \oo 2} - {{\dd} \oo {{\dd} \tau}}
     \left ( {{p^{\mu} p_{\mu} - \kappa^2}\oo {2 m}} \right ) = 0 \;
     \Rightarrow \; {\dot m} = 0 , \lbl{3.100}  \\
     \nonumber \\
     \delta m \, &:& \; \; - {1\oo 2} + {1\oo {2 m^2}} \, (p^{\mu} p_{\mu} -
     \kappa^2) = 0 \; \Rightarrow \; p^{\mu} p_{\mu} - m^2 - \kappa^2 = 0 ,
     \hs{11mm}  \lbl{3.101} \\
     \nonumber \\
     \delta x^{\mu} \, &:& \quad  {\dot p}_{\mu} = 0 , \lbl{101a} \\
     \nonumber \\
     \delta p_{\mu} \, &:& \quad {\dot x}^{\mu} - {{\dot s}\oo m} \, p^{\mu} = 0
     \; \Rightarrow \; p^{\mu} = {{m {\dot x}^{\mu}}\oo {\dot s}} =
     m\, {{{\dd} x^{\mu}} \oo {{\dd} s}} . \lbl{3.102}
\end{eqnarray}
Then we can choose a ``solution" for $s(\tau)$, 
write ${\dot s}/m = \Lambda$, and
omit the first term, since in view of (\ref{3.100}) it is a total derivative.
So again we obtain the Stueckelberg action (\ref{3.98}).

The action that we started from, e.g., (\ref{3.99}) or (\ref{3.84}) has a
constraint on the variables $x^{\mu}$, $s$ or on the $p_{\mu}$, $m$, but
the action (\ref{3.98}) which we arrived at contains only the variables
$x^{\mu}$, $p_{\mu}$ and has no constraint.

In the action (\ref{3.99}) we can use the relation ${\dot s} = {\dd} s/{\dd}
\tau$ and write it as
\be
    I[m,p_{\mu}, x^{\mu}] = \int {\dd} s \, \left [ - {m\oo 2} + p_{\mu}
    {{{\dd} x^{\mu}}\oo {{\dd} s}} - {1\oo {2 m}} (p^{\mu} p_{\mu} -
    \kappa^2 ) \right ] .
\lbl{3.103}
\ee
The evolution parameter is now $s$, and again variation with respect to
$m$ gives the constraint $p^{\mu} p_{\mu} - m^2 - \kappa^2 = 0$. Eliminating
$m$ from the action (\ref{3.103}) by the the latter constraint, written in
the form
\be
        m = \sqrt{p^{\mu} p_{\mu} - \kappa^2}
\lbl{3.104}
\ee
we obtain {\it the unconstrained action}
\be
      I[x^{\mu},p_{\mu}] = \int {\dd} s \, \left ( p_{\mu}
    {{{\dd} x^{\mu}}\oo {{\dd} s}} - \sqrt{p^{\mu} p_{\mu} - \kappa^2} 
    \right ) ,
\lbl{3.105}
\ee
which is also equivalent to the original action (\ref{3.97}). The
Hamiltonian corresponding to (\ref{3.105}) is
\be
    H =   p_{\mu} {{{\dd} x^{\mu}}\oo {{\dd} s}} - L = 
    \sqrt{p^{\mu} p_{\mu} - \kappa^2}  \, .      
\lbl{3.106}
\ee
Such a Hamiltonian is not very practical for quantization, since the
Schr\" odin-ger equation contains the square root of operators
\be
      i \, {{\p \psi} \oo {\p s}} = \sqrt{p^{\mu} p_{\mu} - \kappa^2} \, \psi .    
\lbl{3.107}
\ee
In order to perform the quantization properly one has to start directly
from the original polyvector action (\ref{3.42}). This will be discussed
in Sec. 2.5.

However, in the approximation\footnote{
Remember that $\kappa^2$ comes from the scalar part of the polyvector
mass squared  term (\ref{3.42a}) and that it was a matter of our convention
of writing it in the form $\kappa^2$. We could have used another
symbol without square, e.g., $\alpha$, and then it would be manifestly
clear that $\alpha$ can be negative.}
$p_{\mu} p^{\mu} \ll - \kappa^2$ eq.\,(\ref{3.105}) becomes
\be
    I[x^{\mu}, p_{\mu}] \approx \int {\dd} s \, \left ( p_{\mu} {{{\dd} x^{\mu}}
    \oo {{\dd} s}} - {1 \oo {2 \sqrt{- \kappa^2}}} p^{\mu} p_{\mu} -
    \sqrt{- \kappa^2} \right )
\ee    
which is again the Stueckelberg action, \inxx{Stueckelberg action}
but with $1/\sqrt{- \kappa^2} =
\Lambda$. It is very interesting that on the one hand the Stueckelberg
action arises exactly from the polyvector action, and on the other
hand it arises as an approximation.

\subsection[\ \ Particle in a fixed background field]
{Particle in a fixed background field}

Let us now consider the action (\ref{3.97}) and modify it so that it will
remain covariant under the transformation
\be
      L \rightarrow L' = L + {{{\dd} \phi} \oo {{\dd} \tau}} \; ,
\lbl{3.107a1}
\ee      
where
\be
        \phi = \phi(s,x^{\mu})
\lbl{3.107a2}        
\ee
For this purpose we have to introduce the gauge fields $A_{\mu}$ and
$V$ which transform according to
\be
       eA'_{\mu} = e A_{\mu} + \p_{\mu} \phi ,
\lbl{3.107a3}
\ee
\be
       e V' = e V + {{\p \phi}\oo {\p s}}.
\lbl{3.107a4}
\ee
The covariant action is then
\be
   I = \int {\dd} \tau \left [ - m {\dot s} + p_{\mu} {\dot x}^{\mu} -
   {\lambda \oo 2} (\pi_{\mu} \pi^{\mu} - \mu^2 - \kappa^2) \right ] ,
\lbl{3.107a5}
\ee
where we have introduced the kinetic momentum \inxx{kinetic momentum}
\be
     \pi_{\mu} = p_{\mu} - e A_{\mu}
\lbl{3.107a6}
\ee
and its pseudoscalar counterpart
\be
     \mu = m + e V .
\lbl{3.107a7}
\ee
The symbol `$\mu$' here should not be confused with the same symbol used
in Sec. 2.3 for a completely different quantity.

From (\ref{3.107a5}) we derive the following equations of motion:
\begin{eqnarray}
   \delta x^{\mu} &:& \quad {\dot \pi}_{\mu} = e F_{\mu \nu}
   {\dot x}^{\nu} - {\dot s} e \left ( {{\p A_{\mu}}\oo {\p s}} - \p_{\mu} V
   \right ) , \lbl{3.107a8} \\
   \delta s &:& \quad {\dot \mu} = - {\dot x}^{\nu} e 
   \left ( {{\p A_{\nu}}\oo {\p s}} - \p_{\nu} V \right ) ,  \lbl{3.107a9} \\
   \delta p_{\mu} &:& \quad \lambda \pi_{\mu} = {\dot x}_{\mu}  \; ,
   \lbl{3.107a10} \\ \\
   \delta m &:& \quad {\dot s} = \lambda \mu  .\lbl{3.107a11}
\end{eqnarray}
These equations of motion are the same as those from the Stueckelberg
action (\ref{3.107a15}).

From (\ref{3.107a9}) and (\ref{3.107a11}) we have
\be
    - m {\dot s} + {\lambda \oo 2} \mu^2 = - \mbox{$1\oo 2$} {{\dd}\oo {{\dd} \tau}}
    (\mu s) + e {\dot s} V - {1 \oo 2} e {\dot x}^{\nu} 
    \left ( {{\p A_{\nu}}\oo {\p s}} - \p_{\nu} V \right ) s .
\lbl{3.107a12}
\ee
Inserting the latter expression into the action (\ref{3.107a5}) we obtain
\bear
     &&I = \int {\dd} \tau \Biggl[ - \mbox{$1\oo 2$} {{{\dd} (\mu s)}\oo {{\dd} \tau}}
     + p_{\mu} {\dot x}^{\mu} 
     -  {\lambda \oo 2} (\pi^{\mu} \pi_{\mu} -
     \kappa^2) + e {\dot s} V \hs{2.5cm} \nonumber \\
   && \hs{4.7cm}  - {1 \oo 2} e {\dot x}^{\nu} 
    \left ( {{\p A_{\nu}}\oo {\p s}} - \p_{\nu} V \right ) s \Biggr] ,
\lbl{3.107a13}
\ear
which is analogous to eq.\,(\ref{3.93}). However, in general ${\dot \mu}$
is now not zero, and as a result we cannot separate the variables $m, \, s$ into
a total derivative term as we did in (\ref{3.97}).

Let us consider a particular case when the background fields $A_{\mu}, \, V$
satisfy
\be
     {{\p A_{\mu}}\oo {\p s}} = 0 \; , \qquad \p_{\mu} V = 0 \; .
\lbl{3.107a14}
\ee
Then the last term in (\ref{3.107a13}) vanishes; in addition we may
set $V = 0$. Omitting the total derivative term, eq.\,(\ref{3.107a13}) becomes
\be
     I[x^{\mu}, p_{\mu}] = \int {\dd} s \left [ p_{\mu} {{{\dd} x^{\mu}} \oo
     {{\dd} s}} - {\Lambda \oo 2} ( \pi^{\mu} \pi_{\mu} - \kappa^2) \right ] 
     \; ,
\lbl{3.107a15}
\ee
where $\Lambda = \lambda/{\dot s}$ is now fixed. This is precisely the
Stueckelberg action \inxx{Stueckelberg action}
in the presence of a fixed electromagnetic field, and
$s$ corresponds to the Stueckelberg Lorentz invariant parameter $\tau$.

However, when we gauged the free particle Stueckelberg action we obtained
in general a $\tau$-dependent gauge field $A_{\mu}$ and also a scalar
field $V$. We shall now see that such a general gauged Stueckelberg
action is an approximation to the action (\ref{3.107a5}). For this purpose
we shall repeat the procedure of eqs.\,(\ref{3.99})--(\ref{3.105}). Eliminating
$\lambda$ from the action (\ref{3.107a5}) by using the equation of motion
(\ref{3.107a11}) we obtain an equivalent action
\be
     I[x^{\mu},p_{\mu}, s, m]
      = \int {\dd} \tau \left [ - m {\dot s} + p_{\mu} {\dot x}^{\mu} -
     {{\dot s}\oo {2 \mu}} (\pi^{\mu} \pi_{\mu} - \mu^2 - \kappa^2) \right ]
\lbl{3.107a16}
\ee
whose variation with respect to $m$ again gives the constraint
$\pi^{\mu} \pi_{\mu} - \mu^2 - \kappa^2 = 0$. From (\ref{3.107a16}),
using (\ref{3.107a7}) we have
\begin{eqnarray}
    I &=& \int {\dd} \tau \left [ p_{\mu} {\dot x}^{\mu}  - 
    {{\dot s}\oo {2 \mu}} (\pi^{\mu} \pi_{\mu} - \kappa^2) + e {\dot s} V
    - {{\mu {\dot s}} \oo 2} \right ] \nonumber \\
    &=& \int {\dd} \tau \left [ p_{\mu} {\dot x}^{\mu}  - {\dot s}
    (\pi^{\mu} \pi_{\mu} - \kappa^2)^{1/2} + e {\dot s} V \right ] \nonumber \\
    & \approx & \int {\dd} \tau \left [ p_{\mu} {\dot x}^{\mu} -
    {{\dot s}\oo {2 \sqrt{- \kappa^2}}} \, \pi^{\mu} \pi_{\mu} -
    {\dot s} \sqrt{- \kappa^2} + e{\dot s} V \right ] .
\lbl{3.107a17}
\end{eqnarray}
Thus
\be
   I[x^{\mu}, p_{\mu}] = \int {\dd} s \left [ p_{\mu} \, {{{\dd} x^{\mu}} \oo
   {{\dd} s}} - {1 \oo {2 \sqrt{- \kappa^2}}} \, \pi^{\mu} \pi_{\mu} -
   \sqrt{- \kappa^2} + e V \right ] .
\lbl{3.107a18}
\ee
The last step in eq.\,(\ref{3.107a17}) is valid under the approximation
$\pi^{\mu} \pi_{\mu} \ll - \kappa^2$, where we assume $-\kappa^2 > 0$.
In (\ref{3.107a18}) we indeed obtain an action which is equivalent to the
gauged Stueckelberg action (\ref{3.107a15}) if we make the correspondence 
$1\sqrt{- \kappa^2} \rightarrow \Lambda$. The constant terms $- \sqrt{- \kappa^2}$
in (\ref{3.107a18})  and $\Lambda \kappa^2/2$ in eq.\,(\ref{3.107a15})
have no influence on the equations of motion.

We have thus found a very interesting relation between the Clifford algebra
polyvector action and the Stueckelberg action in the presence of an
electromagnetic and pseudoscalar field. If the electromagnetic field $A_{\mu}$
does not depend on the pseudoscalar parameter $s$ and if there is no force owed
to the pseudoscalar field $V$, then the kinetic momentum squared $\pi^{\mu}
\pi_{\mu}$ is a {\it constant of motion}, and the gauged Clifford algebra action
(\ref{3.107a5}) is exactly equivalent to the Stueckelberg action. In the
presence of a pseudoscalar force, i.e., when $\p_{\mu} V \neq 0$ and/or when
$\p A_{\mu}/\p s \neq 0$, the action (\ref{3.107a5}) is approximately
equivalent to the gauged Stueckelberg action (\ref{3.107a15}) 
if the kinetic momentum
squared $\pi^{\mu} \pi_{\mu}$ is much smaller than the scalar mass constant
squared $- \kappa^2$.

\section{Quantization of the polyvector action}
We have assumed that a point particle's classical motion is governed by
the polyvector action (\ref{3.42}). Variation of this action with respect
to $\lambda$ gives the polyvector constraint \inxx{polyvector constraint}
\be
      P^2 - K^2 = 0 .
\lbl{3.108}
\ee

In the quantized theory the position and momentum polyvectors
$X = X^J e_J$ and $P = P^J e_J$, where $e_J = (1, e_{\mu}, e_{\mu} e_{\nu},
e_5 e_{\mu}, e_5)\; $, $\mu < \nu$, become the operators
\be
    {\widehat X} = {\widehat X}^J e_J \; , \qquad {\widehat P} = {\widehat P}^J 
    e_J ; ,
\lbl{3.109}
\ee
satisfying
\be
           [{\widehat X}^J , {\widehat P}_K ] = i {\delta^J}_K \; .
\lbl{3.110}
\ee
Using the explicit expressions like (\ref{3.40}),(\ref{3.41}) the above 
equations imply
\be
   [{\hat \sigma},{\hat \mu} ] = i \; , \quad [{\hat x}^{\mu}, {\hat p_{\nu}}]
   = i {\delta^{\mu}}_{\nu} \; , \quad [{\hat \alpha}^{\mu \nu}, {\hat
   S}_{\alpha, \beta}] = i {\delta^{\mu \nu}}_{\alpha \beta} \; ,
\lbl{3.111}
\ee
\be
    [{\hat \xi}^{\mu}, {\hat \pi}_{\nu}] = i {\delta^{\mu}}_{\nu} \; , \quad
    [{\hat s}, {\hat m}] = i .
\lbl{3.113}
\ee
In a particular representation in which ${\widehat X}^J$ are diagonal,
the momentum polyvector operator is represented by the multivector
derivative (see Sec. 6.1).
\be
      {\widehat P}_J = - i {\p \oo {\p X^J}}
\lbl{3.114}
\ee
Explicitly, the later relation means
\be
     {\hat \mu} = - i {\p \oo {\p \sigma}} \; , \; \; {\hat p}_{\mu} = - i
     {\p \oo {\p x^{\mu}}} \; , \; \; {\hat S}_{\mu \nu} = - i {\p \oo
     {\p \alpha^{\mu \nu}}} \; , \; \; {\hat \pi}_{\mu} = - i {\p \oo
     {\p \xi^{\mu}}} \; , \; \; {\hat m} = - i {\p \oo {\p s}} .
\lbl{3.115}
\ee

Let us assume that a quantum state can be represented by a polyvector-valued
wave function $\Phi (X)$ \inxx{wave function,polyvector valued}
of the position polyvector $X$. A possible physical
state is a solution to the equation
\be
    ({\widehat P}^2 - K^2) \Phi = 0 ,
\lbl{3.116}
\ee
which replaces the classical constraint (\ref{3.108}).

When $K^2 = \kappa^2 = 0$ eq.\,(\ref{3.116}) becomes
\be
         {\widehat P}^2 \Phi = 0 .
\lbl{3.117}
\ee

Amongst the set of functions $\Phi(X)$ there are some such that satisfy
\be
         {\widehat P} \Phi = 0 .
\lbl{3.118}
\ee

Let us now consider a special case where $\Phi$ has definite values of
the operators ${\hat \mu}$, ${\hat S}_{\mu \nu}$, ${\hat \pi}_{\mu}$:
\be
    {\hat \mu} \Phi = 0 \; , \quad {\hat S}_{\mu \nu} \Phi = 0 \; ,  \quad
    {\hat \pi}_{\mu} \Phi = 0
\lbl{3.119}
\ee
Then 
\be
    {\widehat P} \Phi = ({\hat p}^{\mu} e_{\mu} + {\hat m} e_5 )
    \Phi = 0 .
\lbl{3.120}
\ee
or
\be
    ({\hat p}^{\mu} \gamma_{\mu} - {\hat m}) \Phi = 0 ,
\lbl{3.121}
\ee
where
\be
       \gamma_{\mu} \equiv e_5 e_{\mu} \; , \quad \gamma_5 = \gamma_0 \gamma_1
       \gamma_2 \gamma_3 = e_0 e_1 e_2 e_3 = e_5 .
\lbl{3.122}
\ee
When $\Phi$ is an eigenstate of ${\hat m}$
with definite value $m$, i.e., when ${\hat m} \phi = m \Phi$, then
eq.\,(\ref{3.121}) becomes the familiar {\it Dirac equation} \inxx{Dirac equation}
\be
     ({\hat p}_{\mu} \gamma^{\mu} - m) \Phi = 0 .
\lbl{3.123}
\ee

A polyvector wave function which satisfies eq.\,(\ref{3.123}) is a {\it
spinor}. We have arrived at the very interesting result that {\it spinors
can be represented by particular polyvector wave functions}. \inxx{spinor}

\vs{6mm}

{\it 3-dimensional case}
\vs{2mm}
 
To illustrate this let us consider the 3-dimensional space $V_3$. Basis
vectors are $\sigma_1$, $\sigma_2$, $\sigma_3$ and they satisfy the
Pauli algebra \inxx{basis vectors} \inxx{Pauli algebra}
\be
     \sigma_i \cdot \sigma_j \equiv \mbox{$1\oo 2$} (\sigma_i \sigma_j +
     \sigma_j \sigma_i) = \delta_{i j} \; , \qquad i , j = 1, 2, 3 .
\lbl{3.124}
\ee

The unit pseudoscalar
\be
    \sigma_1 \sigma_2 \sigma_3 \equiv I
\lbl{3.125}
\ee
commutes with all elements of the Pauli algebra \inxx{Pauli algebra}
and its square is
$I^2 = -1$. It behaves as the ordinary imaginary unit $i$. Therefore,
in 3-space, we may identify the imaginary unit $i$ with the unit pseudoscalar
$I$.

An arbitrary polyvector in $V_3$ can be written in the form
\be
     \Phi = \alpha^0 + \alpha^i \sigma_i + i \beta^i \sigma_i + i \beta
     = \Phi^0 + \Phi^i \sigma_i \; ,
\lbl{3.126}
\ee
where $\Phi^0$, $\Phi^i$ are formally complex numbers.

We can decompose \cite{13}:
\be
    \Phi = \Phi \mbox{$1\oo 2$} (1 + \sigma_3) + \Phi \mbox{$1\oo 2$} (1 - \sigma_3) =
    \Phi_+ + \Phi_- \; ,
\lbl{3.127}
\ee
where $\Phi_+ \in {\cal I}_+$ and $\Phi_- \in {\cal I}_-$ are
independent minimal {\it left ideals} (see Box 3.2). \inxx{ideal}

\setlength{\unitlength}{1cm}
\nnnn \begin{picture}(12.5,5.5)

\put(0,0){\framebox(12.5,4.8){
\begin{minipage}[t]{11.5cm}

\centerline{\bf Box 3.2: Definition of ideal}

\vs{5mm}

A left ideal ${\cal I}_L$ in an algebra $C$ is a set of elements such
that if $a \in {\cal I}_L$ and $c \in C$, then $c a \in {\cal I}_L$.
If $a \in {\cal I}_L$, $b \in {\cal I}_L$, then $(a + b) \in {\cal I}_L$.
A right ideal ${\cal I}_R$ is defined similarly except that $a c \in {\cal I}_R$.
A left (right) minimal ideal is a left (right) ideal which contains no
other ideals but itself and the null ideal.

\end{minipage} }}
\end{picture}

\vs{3mm}

A basis in ${\cal I}_+$ is given by two polyvectors
\be
       u_1 = \mbox{$1\oo 2$} (1 + \sigma_3) \; \; , \quad u_2 = 
       (1 - \sigma_3) \sigma_1 \; ,
\lbl{3.128}
\ee
which satisfy
\begin{eqnarray}
    &&\sigma_3 u_1 = \; \; u_1 \; ,\qquad \, \sigma_1 u_1 = u_2 \; , 
    \qquad \sigma_2 u_1 = 
    \; i u_2  \; ,\nonumber \\
    &&\sigma_3 u_2 = - u_2 \; , \qquad \sigma_1 u_2 = u_1 \; , \qquad
    \sigma_2 u_2 = - i u_1 .
\lbl{3.129}
\end{eqnarray}
These are precisely the well known relations \inxx{spinor}
for basis spinors. Thus
we have arrived at the very profound result that the polyvectors $u_1$, $u_2$
behave as basis spinors. \inxx{basis spinors}

Similarly, a basis in ${\cal I}_-$ is given by
\be
    v_1 = \mbox{$1\oo 2$} (1 + \sigma_3) \sigma_1 \; , \quad 
    v_2 = \mbox{$1\oo 2$} (1 - \sigma_3)
\lbl{3.130}
\ee
and satisfies    
\begin{eqnarray}
    &&\sigma_3 v_1 = \; \; v_1 \; , \qquad \, \sigma_1 v_1 = v_2 \; , 
    \qquad \sigma_2 v_1 = \; i v_2 \; ,\nonumber \\
   &&\sigma_3 v_2 = - v_2 \; , \qquad \sigma_1 v_2 = v_1 \; , \qquad
    \sigma_2 v_2 = - i v_1 .
\lbl{3.131}
\end{eqnarray}

A polyvector $\Phi$ can be written in {\it spinor basis} as
\be
    \Phi = \Phi_+^1 u_1 + \Phi_+^2 u_2 + \Phi_-^1 v_1 + \Phi_-^2 v_2 \; ,
\lbl{3.132}
\ee
where
\begin{eqnarray}
   \Phi_+^1 &=& \Phi^0 + \, \Phi^3  \; , \qquad \, \Phi_-^1 = \Phi^1 - i \Phi^2
   \nonumber \\
   \Phi_+^2 &=& \Phi^1 + i \Phi^2 \; , \qquad \Phi_-^2 = \Phi^0 - \, \Phi^3
\lbl{3.133}
\end{eqnarray}
Eq. (\ref{3.132}) is an alternative expansion of a polyvector. We can
expand the same polyvector $\Phi$ either according to (\ref{3.126}) or
according to (\ref{3.132}).

Introducing the matrices
\be
     \xi_{a b} = \pmatrix{
     u_1 & v_1 \cr
     u_2 & v_2 \cr } \; , \qquad \Phi^{a b} = \pmatrix{
     \Phi_+^1 & \Phi_-^1 \cr
     \Phi_+^2 & \Phi_-^2 \cr }
\lbl{3.134}
\ee
we can write (\ref{3.132}) as
\be
      \Phi = \Phi^{a b} \xi_{a b} .
\lbl{3.135}
\ee

Thus a polyvector can be represented as a {\it matrix} $\Phi^{a b}$. The
decomposition (\ref{3.127}) then reads
\be
    \Phi = \Phi_+ + \Phi_- = (\Phi_+^{a b} + \Phi_-^{a b}) \xi_{a b} \; ,
\lbl{3.136}
\ee
where
\be
  \Phi_+^{a b} = \pmatrix{
         \Phi_+^1 & 0 \cr
         \Phi_+^2 & 0 \cr } \; ,
\lbl{3.137}
\ee
\be
     \Phi_-^{a b} = \pmatrix{
     0 & \Phi_-^1 \cr
     0 & \Phi_-^2 \cr } .
\lbl{3.138}
\ee

From (\ref{3.135}) we can directly calculate the matrix elements $\Phi^{ab}$.
We only need to introduce the new elements $\xi^{\dagger ab}$ which
satisfy
\be
     ({\xi^{\dagger}}^{a b} \xi_{c d} )_S = {\delta^a}_c {\delta^b}_d .
\lbl{3.139}
\ee
The superscript $\dagger$ means Hermitian conjugation \cite{13}. If
\be
      A = A_S + A_V +A_B + A_P
\lbl{3.139a}
\ee
is a Pauli number, then
\be
     A^{\dagger} = A_S + A_V - A_B - A_P .
\lbl{3.139b}
\ee
This means that the order of basis vectors $\sigma_i$ in the expansion
of $A^{\dagger}$ is reversed. Thus $u_1^{\dagger} = u_1$, but
$u_2^{\dagger} = \mbox{$1\oo 2$} (1 + \sigma_3) \sigma_1$. Since $(u_1^{\dagger}
u_1)_S = \mbox{$1\oo 2$}$, $(u_2^{\dagger} u_2)_S = \mbox{$1\oo 2$}$, it is convenient
to introduce ${u^{\dagger}}^{1} = 2 u_1$ and ${u^{\dagger}}^2 = 2 u_2$
so that $({u^{\dagger}}^1 u_1)_S = 1$, $({u^{\dagger}}^2 u_2)_S = 1$. If we
define similar relations for $v_1$, $v_2$ then we obtain (\ref{3.139}).

From (\ref{3.135}) and (\ref{3.139}) we have
\be
     \Phi^{a b} = ({\xi^{\dagger}}^{a b} \Phi )_I \; .
\lbl{3.140}
\ee
Here the subscript $I$ means {\it invariant part}, i.e., scalar plus
pseudoscalar part (remember that pseudoscalar unit has here the role of
imaginary unit and that $\Phi^{a b}$ are thus complex numbers).

The relation (\ref{3.140}) tells us how from an arbitrary polyvector $\Phi$
(i.e., a Clifford number) can we obtain its {\it matrix representation}
$\Phi^{a b}$.

$\Phi$ in (\ref{3.140}) is an arbitrary Clifford number. In particular, $\Phi$
may be any of the basis vectors $\sigma_i$.

{\it Example}  $\Phi = \sigma_1$:
\begin{eqnarray}
     \Phi^{11} &=& ({\xi^{\dagger}}^{11} \sigma_1)_I = ({u^{\dagger}}^1
     \sigma_1)_I = \left ( (1 + \sigma_3)\sigma_1 \right )_I = 0, \nonumber \\
     \Phi^{12}  &=& ({\xi^{\dagger}}^{12} \sigma_1)_I = ({v^{\dagger}}^1
     \sigma_1)_I = \left ( (1 - \sigma_3)\sigma_1 \sigma_1 \right )_I = 
     1, \nonumber \\
    \Phi^{21}  &=& ({\xi^{\dagger}}^{21} \sigma_1)_I = ({u^{\dagger}}^2
     \sigma_1)_I = \left ( (1 + \sigma_3)\sigma_1 \sigma_1 \right )_I = 
     1, \nonumber \\
    \Phi^{22}  &=& ({\xi^{\dagger}}^{22} \sigma_1)_I = ({v^{\dagger}}^2
     \sigma_1)_I = \left ( (1 - \sigma_3)\sigma_1 \right )_I = 0 .
\lbl{3.141}
\end{eqnarray}
Therefore
\be
     (\sigma_1)^{a b} = \pmatrix{
     0 & 1 \cr
     1 & 0 \cr } .
\lbl{3.142}
\ee
Similarly we obtain from (\ref{3.140}) when $\Phi = \sigma_2$ and
$\Phi = \sigma_3$, respectively, that
\be
     (\sigma_2)^{a b} = \pmatrix{
     0 & -i \cr
     i & 0 \cr }  \; \; , \qquad
     (\sigma_3)^{a b} = \pmatrix{
     1 & 0 \cr
     0 & -1 \cr } .
\lbl{3.143}
\ee

So we have obtained the matrix representation of the basis vectors $\sigma_i$.
Actually (\ref{3.142}), (\ref{3.143}) \inxx{Pauli matrices}
are the well known {\it Pauli matrices}.

When $\Phi = u_1$ and $\Phi = u_2$, respectively, we obtain
\be
     (u_1)^{a b} = \pmatrix{
     1 & 0 \cr
     0 & 0 \cr }  \; \; , \qquad
     (u_2)^{a b} = \pmatrix{
     0 & 0 \cr
     1 & 0 \cr }
\lbl{3.144}
\ee
which are a matrix representation \inxx{basis spinors,matrix representation of}
of the {\it basis spinors} $u_1$ and $u_2$.

Similarly we find
\be
     (v_1)^{a b} = \pmatrix{
     0 & 1 \cr
     0 & 0 \cr }  \; , \qquad
     (v_2)^{a b} = \pmatrix{
     0 & 0 \cr
     0 & 1 \cr }
\lbl{3.145}
\ee

In general a {\it spinor} is a superposition
\be
     \psi = \psi^1 u_1 + \psi^2 u_2 \; ,
\lbl{3.146}
\ee
and its matrix representation is
\be
       \psi \rightarrow \pmatrix{
       \psi^1 & 0 \cr
       \psi^2 & 0 \cr} .
\lbl{3.147}
\ee

Another independent spinor is
\be
     \chi = \chi^1 v_1 + \chi^2 v_2 \; ,
\lbl{3.148}
\ee
with matrix representation
\be
     \chi \rightarrow \pmatrix{
     0 & \chi^1 \cr
     0 & \chi^2 \cr} .
\lbl{3.149}
\ee

If we multiply a spinor $\psi$ from the left by any element $R$ of the
Pauli algebra we obtain another spinor
\be
     \psi' = R \psi \rightarrow \pmatrix{
     \psi'^1 & 0 \cr
     \psi'^2 & 0\cr}
\lbl{3.150}
\ee
which is an element of the same minimal left ideal. Therefore,
if only multiplication from the left is considered, a spinor can be
considered as a column matrix
\be
     \psi \rightarrow \pmatrix{
     \psi^1 \cr
     \psi^2 \cr} .
\lbl{3.151}
\ee
This is just the common representation of spinors. But it is not
general enough to be valid for all the interesting situations which
occur in the Clifford algebra.

We have thus arrived at a very important finding. {\it Spinors} are just
particular {\it Clifford numbers}: they belong to a left or right minimal
ideal. For instance, a generic {\it spinor} is
\be
     \psi = \psi^1 u_1 + \psi^2 u_2  \quad {\rm with} \quad
     \Phi^{ab} = \pmatrix{
     \psi^1 & 0 \cr
     \psi^2 & 0 \cr} .
\lbl{3.152}
\ee
A {\it conjugate spinor} is \inxx{spinor,conjugate}
\be
     \psi^{\dagger} = \psi^{1*} u_1^{\dagger} + \psi^{2*} u_2^{\dagger}
     \quad {\rm with} \quad
     {(\Phi^{ab})}^* = \pmatrix{
     {\psi^1}^* & {\psi^2}^* \cr
     0 & 0 \cr}
\lbl{3.153}
\ee
and it is an element of a minimal {\it right ideal}.

\vs{6mm}

{\it 4-dimensional case}

\vs{2mm}

The above considerations can be generalized to 4 or more dimensions.
Thus
\be
       \psi = \psi^0 u_0 + \psi^1 u_1 + \psi^2 u_2 + \psi^3 u_3 \rightarrow
      \pmatrix{
      \psi^0 & 0 & 0 & 0 \cr
      \psi^1 & 0 & 0 & 0 \cr
      \psi^2 & 0 & 0 & 0 \cr
      \psi^3 & 0 & 0 & 0 \cr}
\lbl{3.153a}
\ee
and
\be
       \psi^{\dagger} = {\psi^*}^0 u_0^{\dagger} + {\psi^*}^1 u_1^{\dagger} + 
       {\psi^*}^2 u_2^{\dagger} + {\psi^*}^3 u_3^{\dagger} \rightarrow
      \pmatrix{
      {\psi^*}^0 & {\psi^*}^1 & {\psi^*}^2 & {\psi^*}^3  \cr
      0 & 0 & 0 & 0 \cr
      0 & 0 & 0 & 0 \cr 
      0 & 0 & 0 & 0 \cr} ,
\lbl{3.153b}
\ee
where $u_0$, $u_1$, $u_2$, $u_3$ are four basis spinors in spacetime,
and $\psi^0$, $\psi^1$, $\psi^2$, $\psi^3$ are complex scalar
coefficients.

In 3-space the pseudoscalar unit can play the role of the imaginary unit $i$.
This is not the case of the 4-space $V_4$, since $e_5 = e_0 e_1 e_2 e_3$ does not
commute with all elements of the Clifford algebra in $V_4$. Here the
approaches taken by different authors differ. A straightforward possibility
\ci{27} is just to use the complex Clifford algebra with complex
coefficients of expansion in terms of multivectors. Other authors prefer to
consider real Clifford algebra ${\cal C}$ and ascribe  the role of the imaginary
unit $i$ to an element of ${\cal C}$ which commutes with all other elements of 
${\cal C}$ and  whose square is $-1$. Others \ci{13,26} explore 
the possibility of using a
non-commuting element as a substitute for the imaginary unit. I am not going
to review all those various approaches, but I shall simply assume that
the expansion coefficients are in general complex numbers. In Sec. 7.2 I
explore the possibility that such complex numbers which occur in the
quantized theory originate from the Clifford algebra description of the
$(2 \times n)$-dimensional phase space $(x^{\mu}, \, p_{\mu})$. In such a way
we still conform to the idea that complex numbers are nothing but special
Clifford numbers.

A Clifford number $\psi$ expanded according to (\ref{3.153a}) is an
element of a left minimal ideal if the four elements $u_0, \, u_1, \, u_2, \,
u_3$ satisfy
\be
   C u_{\lambda} = C_{0 \lambda} u_0 + C_{1 \lambda} u_1 + C_{2 \lambda} u_2
   + C_{3 \lambda} u_3
\lbl{3.153c}
\ee
for an arbitrary Clifford number $C$. General properties of $u_{\lambda}$
were investigated by Teitler \ci{27}. In particular, he found the following
representation for $u_{\lambda}$:
\begin{eqnarray}
    u_0 &=& \mbox{$1\oo 4$} (1 - e^0 + i e^{12} - i e^{012}) , \nonumber \\
    u_1 &=& - e^{13} u_0 = \mbox{$1\oo 4$} (- e^{13} + e^{013} + i e^{23} - 
    i e^{023}), \nonumber \\
    u_2 &=& - i e^3 u_0 = \mbox{$1\oo 4$} (-i e^3 - i e^{03} + e^{123} +
     e^{0123}) , \nonumber \\
    u_3 &=& - i e^1 u_0 = \mbox{$1\oo 4$} (- i e^1 - i e^{01} - e^2 - e^{02}) ,
\lbl{3.153d}
\end{eqnarray}
from which we have
\begin{eqnarray}
    e^0 u_0 &=& - u_0  \; ,\nonumber \\
    e^1 u_0 &=& i u_3 \; , \nonumber \\
    e^2 u_0 &=& - u_3 \; , \nonumber \\
    e^3 u_0 &=& i u_2 .
\lbl{3.153e}
\end{eqnarray}
Using the representation (\ref{3.153d}) we can calculate from 
(\ref{3.153c})  the matrix elements $C_{\rho \lambda}$ of any Clifford
number. For the spacetime basis vectors $e^{\mu} \equiv (e^0, e^i), 
\, i = 1,2,3$, we obtain
\be
      e^0 = \pmatrix{
      -{\bf 1} & 0 \cr
      0 & {\bf 1} \cr} \; , \qquad e^i = \pmatrix{
      0 & i {\bf \sigma^i} \cr
      i {\bf \sigma^i} & 0 \cr} \; ,
\lbl{3.153f}
\ee
which is one of the standard matrix representations of $e^{\mu}$
(the Dirac matrices). \inxx{Dirac matrices}

If a spinor is multiplied from the left by an arbitrary Clifford number,
it remains a spinor. But if is multiplied from the right, it in general
transforms into another Clifford number which is no more a spinor.
{\it Scalars, vectors, bivectors, etc., and spinors can be reshuffled by the
elements of Clifford algebra: scalars, vectors, etc., can be transformed
into spinors, and vice versa.}

Quantum states are assumed to be represented by polyvector wave functions
(i.e., Clifford numbers). If the latter are pure scalars, vectors, bivectors,
pseudovectors, and pseudoscalars they describe {\it bosons}.
If, on the contrary, wave functions are spinors, then they describe
{\it fermions}. {\it Within Clifford algebra we have thus transformations
which change bosons into fermions!} It remains to be investigated whether
this kind of ``supersymmetry" is related to the well known supersymmetry.

\section{On the second quantization of the polyvector action}

\inxx{second quantization,of the polyvector action}
If we first quantize the polyvector action (\ref{3.97}) we obtain the
wave equation
\be
    ({\hat p}^{\mu} {\hat p}_{\mu} - {\hat m}^2 - \kappa^2 ) \phi = 0 ,
\lbl{5.1}
\ee
where 
$${\hat p}_{\mu} = - \p/\p x^{\mu} \equiv \p_{\mu}\; , {\hat m} =
- i \p/ \p s \, ,$$ 
and $\kappa$ is a fixed constant. The latter wave equation
can be derived from the action
\bear
     I[\phi] &=& \mbox{$1\oo 2$} \int {\dd} s \, {\dd}^d x \, \phi ( - \p^{\mu}
     \p_{\mu} + {{\p^2}\oo {\p s^2}} - \kappa^2) \phi \nonumber \\
        &=& {1 \oo 2} \int {\dd} s \, {\dd}^d x \left ( \p_{\mu} \phi 
        \p^{\mu} \phi - \left ( {{\p \phi}\oo {\p s}} \right )^2 - 
        \kappa^2 \phi^2 \right ) ,
\lbl{5.2}
\ear
where in the last step we have omitted the surface and the total derivative
terms.

The canonical momentum is
\be
     \pi(s,x) = {{\p {\cal L}}\oo {\p \p \phi/s}} = - {{\p \phi}\oo {\p s}}
\lbl{5.3}
\ee
and the Hamiltonian is \inxx{Hamiltonian}
\be
       H[\phi,\pi] = \int {\dd}^d x \, \left ( \pi \, 
       {{\p \phi}\oo {\p s}} - {\cal L} \right ) = \mbox{$1\oo 2$} \int 
       {\dd}^d x \, (- \pi^2 - \p_{\mu} \phi
       \p^{\mu} \phi + \kappa^2 \phi^2 ) .
\lbl{5.4}
\ee
If instead of one field $\phi$ there are two fields $\phi_1$, $\phi_2$ we have
     $$I[\phi_1, \phi_2] =      
     \int {\dd} s \, {\dd}^d x \, \Biggl[ \p_{\mu} \phi_1 \p
     ^{\mu} \phi_1 - \left ( {{\p \phi_1}\oo {\p s}} \right )^2 - 
     \kappa^2 \phi_1^2 \hs{3cm}$$
\be     
      \hs{2cm} + \; \p_{\mu} \phi_2 \p
     ^{\mu} \phi_2 - \left ( {{\p \phi_2}\oo {\p s}} \right )^2 - 
     \kappa^2 \phi_2^2 \Biggr] .   
\lbl{5.5}
\ee
The canonical momenta are
\be
   \pi_1 = {{\p {\cal L}}\oo  {\p \p \phi_1/s}} = - {{\p \phi_1}\oo {\p s}}  
   \;, \qquad
   \pi_2 = {{\p {\cal L}}\oo  {\p \p \phi_2/s}} = - {{\p \phi_2}\oo {\p s}} 
\lbl{5.6}
\ee
and the Hamiltonian is
   $$ H[\phi_1,\phi_2,\pi_1,\pi_2] = \int {\dd}^d x \, \left ( \pi_1 \, 
       {{\p \phi_1}\oo {\p s}} + \pi_2 {{\p \phi_2}\oo {\p s}}
       - {\cal L} \right ) \hs{3cm}$$       
       $$\hs{3mm} = \mbox{$1\oo 2$} \int 
       {\dd}^d x \, (- \pi_1^2 - \p_{\mu} \phi_1 \p^{\mu} \phi_1  
       + \kappa^2 \phi_1^2$$
\be       
       \hs{26mm} - \; \pi_2^2 - \p_{\mu} \phi_2 \p^{\mu} \phi_2  
       + \kappa^2 \phi_2^2)  .
\lbl{5.7}
\ee        

Introducing the complex fields
\begin{eqnarray}
     \phi &=& \phi_1 + i \phi_2 \; , \qquad \; \pi \, = \pi_1 + i \pi_2 \nonumber \\
     \phi^* &=& \phi_1 - i \phi_2 \; , \qquad \pi^* = \pi_1 - i \pi_2            
\lbl{5.8}
\end{eqnarray}
we have
\be
    I[\phi,\phi^*] = \mbox{$1\oo 2$} \int {\dd} s \, {\dd}^d x \, 
    \left ( {{\p \phi^*}\oo
    {\p s}} {{\p \phi}\oo {\p s}} - \p_{\mu} \phi^* \p^{\mu} \phi -
    \kappa^2 \phi^* \phi \right )
\lbl{5.9}
\ee
and
\be
    H[\phi, \phi^*, \pi, \pi^*] = \mbox{$1\oo 2$} \int {\dd}^d x (- \pi^* \pi -
    \p_{\mu} \phi^* \p^{\mu} \phi + \kappa^2 \phi^* \phi ) .
\lbl{5.10}
\ee
Comparing the latter Hamiltonian with the one of the Stueckelberg field theory
(\ref{24}), we see that it is the same, except for the additional term
$-\pi^* \pi$ which is absent in the Stueckelberg field theory. We see also
that the theory described by (\ref{5.9}) and (\ref{5.10}) has the same
structure as the conventional field theory, except for the number of dimensions.
In the conventional theory we have time $t$ and three space-like coordinates
$x^i$, $i = 1,2,3$, while here we have $s$ and four or more coordinates $x^{\mu}$,
one of them being time-like.

As the non-relativistic field theory is an approximation to the relativistic
field theory, so the field theory derived from the Stueckelberg action is
an approximation to the field theory derived from the polyvector action.

On the other hand, at the classical level (as we have seen in Sec. 2.4)
Stueckelberg action, in the absence of interaction, arises exactly from
the polyvector action. Even in the presence of the electromagnetic
interaction both actions are equivalent, since they give the same
equations of motion.
However, the field theory based on the latter action
differs from the field theory based on the former action by the term
$\pi^* \pi$ in the Hamiltonian (\ref{5.10}). While at the classical level
Stueckelberg and the polyvector action are equivalent, at the first and
the second quantized level differences arise which need further investigation.

Second quantization then goes along the usual lines: $\phi_i$ and $\pi_i$
become operators satisfying the equal $s$ commutation relations:
   $$\lbrack \phi_i (s,x), \phi_j (s,x') \rbrack = 0 \; , \qquad
   \lbrack \pi_i (s,x), \pi_j (s,x') \rbrack = 0 \; ,$$
\be
    \lbrack \phi_i (s,x), \pi_j(s,x') \rbrack = i \delta_{ij} \delta (x - x') .
\lbl{5.11}
\ee
The field equations are then just the Heisenberg equations
\inxx{Heisenberg equations}
\be
      {\dot \pi}_i = i \lbrack \pi_i, H \rbrack .
\lbl{5.12}
\ee

We shall not proceed here with formal development, since it is in many
respects just a repetition of the procedure expounded in Sec. 1.4. But we shall
make some remarks. First of all it is important to bear in mind
that the usual arguments about causality, unitarity, negative energy, etc.,
do not apply anymore, and must all be worked out again. Second, whilst in
the conventional quantum field theory the evolution parameter
\inxx{evolution parameter} is a time-like
coordinate $x^0 \equiv t$, in the field theory based on (\ref{5.9}),
(\ref{5.10}) the evolution parameter is the pseudoscalar variable $s$. 
In even-dimensions it is invariant
with respect to the Poincar\' e and the general coordinate transformations
of $x^{\mu}$, including the inversions. And what is very nice here is that 
the (pseudo)scalar parameter $s$
naturally arises from the straightforward polyvector extension of the
conventional reparametrization invariant theory.

Instead of the {\it Heisenberg picture}
\inxx{Heisenberg picture} we can use the {\it Schr\" odinger
picture} \inxx{Schr\" odinger picture}
and the coordinate representation in which the operators 
$\phi_i (0,x) \equiv \phi_i (x)$ are diagonal, i.e., they are just
ordinary functions. The momentum operator is represented by functional
derivative\footnote{
A detailed discussion of the Schr\" odinger representation in field
theory is to be found in ref. \cite{28}.}
\be
     \pi_j = - i {\delta \oo {\delta \phi_j (x)}} .
\lbl{5.13}
\ee
A state $| \Psi \rangle$ is represented by a {\it wave functional}
\inxx{wave functional}
$\Psi [\phi (x)] = \langle \phi (x) | \Psi \rangle$ and satisfies
the Schr\" odinger equation
\be
      i {{\p \Psi}\oo {\p s}} = H \Psi
\lbl{5.14}
\ee
in which the evolution parameter is $s$. Of course $s$ is invariant under
the Lorentz transformations, and $H$ contains all four components of
the 4-momentum. Equation (\ref{5.14}) is just like the Stueckelberg equation,
the difference being that $\Psi$ is now not a wave function, but a
wave functional.

We started from the {\it constrained} polyvector action (\ref{3.97}).
Performing the
{\it first quantization} we obtained the wave equation (\ref{5.1}) which follows
from the action (\ref{5.2}) for the field $\phi (s, x^{\mu})$.
The latter action is
{\it unconstrained}. Therefore we can straightforwardly quantize it, and thus 
perform the {\it second quantization}. The state vector $|\Psi \rangle$
in the Schr\" odinger picture evolves in $s$ which is a {\it Lorentz
invariant evolution parameter}. $|\Psi \rangle$ can be represented by
a wave functional $\Psi [s, \phi(x)]$ which satisfied the functional
Schr\" odinger equation. Whilst upon the first quantization the equation of
motion for the field $\phi (s, x^{\mu})$ contains the second order derivative 
with respect 
to $s$, upon the second quantization only the first order $s$-derivative remains
in the equation of motion for the state functional $\Psi [s, \phi (x)]$.

An analogous procedure is undertaken in the usual approach to quantum field
theory (see, e.g., \cite{28}), with the difference that the evolution 
parameter becomes one of the space time coordinates, namely $x^0 \equiv t$.
When trying to quantize the gravitational field it turns out that the 
evolution parameter $t$ does not occur at all in the Wheeler--DeWitt equation!
\inxx{Wheeler--DeWitt equation}
This is the well known problem of time
\inxx{problem of time} in quantum gravity. We anticipate
that a sort of polyvector generalization of the Einstein--Hilbert action
should be taken, which would contain the scalar or pseudoscalar parameter $s$,
and retain it in the generalized Wheeler--DeWitt equation. Some important
research in that direction has been pioneered by Greensite and Carlini
\cite{29}

\section{Some further important consequences of Clifford algebra}

\subsection[\ \ Relativity of signature]
{Relativity of signature}

\inxx{relativity of signature}
In previous sections we have seen how Clifford algebra can be used 
in the formulation of the point particle classical and quantum theory.
The metric of spacetime was assumed, as usually, to have the Minkowski signature,
and we have used the choice $(+ - - -)$. We are now going to 
find out that within
Clifford algebra the signature is a matter of choice of basis vectors amongst
available Clifford numbers.

Suppose we have a 4-dimensional space $V_4$ with signature (+ + + +). Let
$e_{\mu},\, \mu = 0,1,2,3$, be basis vectors satisfying
\be
     e_{\mu} \cdot e_{\nu} \equiv \mbox{$1\oo 2$} (e_{\mu} e_{\nu} + e_{\nu} e_{\mu})
     = \delta_{\mu \nu} \; ,
\lbl{4.1}
\ee
where $\delta_{\mu \nu}$ is the {\it Euclidean signature}
\inxx{Euclidean signature} of $V_4$.
The vectors $e_{\mu}$ can be used as generators of Clifford algebra
${\cal C}$ over $V_4$ with a generic Clifford number (also called
polyvector or Clifford aggregate) expanded in term of $e_J = (1,e_{\mu},e_{\mu
\nu}, e_{\mu \nu \alpha}, e_{\mu \nu \alpha \beta})$,
$\; \mu < \nu < \alpha < \beta$,
\be
        A = a^J e_J = a + a^{\mu} e_{\mu} + a^{\mu \nu} e_{\mu} e_{\nu} +
        a^{\mu \nu \alpha} e_{\mu} e_{\nu} e_{\alpha} + a^{\mu \nu \alpha \beta}
        e_{\mu} e_{\nu} e_{\alpha} e_{\beta} .
\lbl{4.2}
\ee
Let us consider the set of four Clifford numbers $(e_0,\, e_i e_0)$,
$i = 1,2,3$, and denote them as
\begin{eqnarray}
         e_0 &\equiv& \gamma_0  ,        \nonumber \\
         e_i e_0 &\equiv& \gamma_i .
\lbl{4.3}
\end{eqnarray}
The Clifford numbers $\gamma_{\mu}$, $\mu = 0,1,2,3$, satisfy
\be
         \mbox{$1\oo 2$} (\gamma_{\mu} \gamma_{\nu} + \gamma_{\nu} \gamma_{\mu})
         = \eta_{\mu \nu} \; ,
\lbl{4.4}
\ee
where $\eta_{\mu \nu} = {\rm diag} (1, -1, -1 ,-1)$ is the {\it Minkowski
tensor}. We see that the $\gamma_{\mu}$ behave as basis vectors in a
4-dimensional space $V_{1,3}$ with signature $(+ - - -)$. We can form a
Clifford aggregate
\be
     \alpha = \alpha^{\mu} \gamma_{\mu}
\lbl{4.5}
\ee
which has the properties of a {\it vector} in $V_{1,3}$. From the point of
view of the space $V_4$ the same object $\alpha$ is a linear combination of
a vector and bivector:
\be
       \alpha = \alpha^0 e_0 + \alpha^i e_i e_0 .
\lbl{4.6}
\ee
We may use $\gamma_{\mu}$ as
\inxx{generators,of the Clifford algebra} generators of the Clifford algebra
${\cal C}_{1,3}$ defined over the pseudo-Euclidean space $V_{1,3}$. The
basis elements of ${\cal C}_{1,3}$ are $\gamma_J = (1,\gamma_{\mu},\gamma_{\mu
\nu}, \gamma_{\mu \nu \alpha}, \gamma_{\mu \nu \alpha \beta})$, with
$\mu < \nu < \alpha < \beta$. A generic  Clifford aggregate in ${\cal C}_{1,3}$
is given by
\be
        B = b^J \gamma_J = b + b^{\mu} \gamma_{\mu} + 
        b^{\mu \nu} \gamma_{\mu} \gamma_{\nu} +
        b^{\mu \nu \alpha} \gamma_{\mu} \gamma{\nu} \gamma_{\alpha} + 
        b^{\mu \nu \alpha \beta}
        \gamma_{\mu} \gamma_{\nu} \gamma_{\alpha} \gamma_{\beta} .
\lbl{4.7}
\ee
With suitable choice of the coefficients $b^J = (b, b^{\mu},
b^{\mu \nu}, b^{\mu \nu \alpha}, b^{\mu \nu \alpha \beta})$
we have that $B$ of eq.\,(\ref{4.7}) is equal to $A$ of eq.(\ref{4.2}). Thus the
same number $A$ can be described either within ${\cal C}_4$ or within
${\cal C}_{1,3}$. The expansions (\ref{4.7}) and (\ref{4.2}) exhaust all possible
numbers of the Clifford algebras ${\cal C}_{1,3}$ and ${\cal C}_4$.
The algebra ${\cal C}_{1,3}$ is isomorphic to the
algebra ${\cal C}_4$, and actually they are just two different representations
of the same set of Clifford numbers (also being called polyvectors or Clifford
aggregates).

As an alternative to (\ref{4.3}) we can choose
\begin{eqnarray}
          e_0 e_3 &\equiv& {\tilde \gamma}_0 , \nonumber \\
          e_i &\equiv& {\tilde \gamma}_i ,
\lbl{4.8}
\end{eqnarray}
from which we have
\be
      \mbox{$1\oo 2$} ({\tilde \gamma}_{\mu} {\tilde \gamma}_{\nu} + {\tilde 
      \gamma}_{\nu} {\tilde \gamma}_{\mu} ) = {\tilde \eta}_{\mu \nu}
\lbl{4.9}
\ee
with ${\tilde \eta}_{\mu \nu} = {\rm diag} (- 1, 1, 1, 1)$. Obviously
${\tilde \gamma}_{\mu}$ are basis vectors of a pseudo-Euclidean space
${\widetilde V}_{1,3}$ and they generate the Clifford algebra over 
${\widetilde V}_{1,3}$ which is
yet another representation of the same set of objects (i.e., polyvectors). But
the spaces $V_4$, $V_{1,3}$ and ${\widetilde V}_{1,3}$ are not the same and
they span different subsets of polyvectors. In a similar way we can obtain
spaces with signatures $(+ - + +)$, $(+ + - +)$, $(+ + + -)$, $(- + - -)$, 
$(- - + -)$,
$(- - - +)$ and corresponding higher dimensional analogs. But we cannot obtain 
signatures of the type $(+ + - -)$, $(+ - + -)$, etc. In order to obtain such
signatures we proceed as follows.

{\it 4-space.}\hs{2mm}
First we observe that the bivector ${\bar I} = e_3 e_4$ satisfies ${\bar I}^2
= -1$, commutes with $e_1,\, e_2$ and anticommutes with $e_3,\, e_4$. So we
obtain that the set of Clifford numbers $\gamma_{\mu} = (e_1 {\bar I}, e_2
{\bar I}, e_3, e_3)$ satisfies
\be
      \gamma_{\mu} \cdot \gamma_{\nu}  = {\bar \eta}_{\mu \nu}   \; ,
\lbl{4.9a}
\ee
where ${\bar \eta} = {\rm diag} ( -1,-1,1,1)$.

{\it 8-space.} \hs{2mm} Let $e_A$ be basis vectors of 8-dimensional 
vector space with signature (+ + + + + + + +). Let us decompose
\begin{eqnarray}
      e_A = (e_{\mu}, e_{\bar \mu})\; , \; \; 
      \qquad \mu &=& 0,1,2,3, \nonumber \\
      {\bar \mu} &=& {\bar 0}, {\bar 1}, {\bar 2}, {\bar 3} .
\lbl{4.10}
\end{eqnarray}
The inner product of two basis vectors
\be
      e_A \cdot e_B = \delta_{AB}  ,    \nonumber
\ee
then splits into the following set of equations:
\begin{eqnarray}
     e_{\mu} \cdot e_{\nu} &=& \delta_{\mu \nu} \; ,  \nonumber \\
     e_{\bar \mu} \cdot e_{\bar \nu} &=& \delta_{{\bar \mu} {\bar \nu}} \; , 
     \nonumber \\
     e_{\mu} \cdot e_{\bar \nu} &=& 0 . 
\lbl{4.11}
\end{eqnarray}
The number ${\bar I} = e_{\bar 0} e_{\bar 1} e_{\bar 2} e_{\bar 3}$ has
the properties
\begin{eqnarray}
      {\bar I}^2 &=& 1 , \nonumber \\
      {\bar I} e_{\mu} &=& e_{\mu} {\bar I} , \nonumber \\
      {\bar I} e_{\bar \mu} &=& - e_{\bar \mu} {\bar I} .
\lbl{4.12}
\end{eqnarray}
The set of numbers
\begin{eqnarray}
       \gamma_{\mu} &=& e_{\mu}    \; ,\nonumber \\
       \gamma_{\bar \mu} &=& e_{\bar \mu} {\bar I}
\lbl{4.13}
\end{eqnarray}
satisfies
\begin{eqnarray}
      \gamma_{\mu} \cdot \gamma_{\nu} &=& \delta_{\mu \nu} \; ,\nonumber \\
      \gamma_{\bar \mu} \cdot \gamma_{\bar \nu} &=& - 
      \delta_{{\bar \mu} {\bar \nu}} \; ,
      \nonumber \\
      \gamma_{\mu} \cdot \gamma_{\bar \mu} &=& 0 .
\lbl{4.14}
\end{eqnarray}
The numbers $(\gamma_{\mu},\, \gamma_{\bar \mu})$ thus form a set of
basis vectors of a vector space $V_{4,4}$ with signature $(+ + + + - - - -)$.

{\it 10-space.} \hs{2mm} Let $e_A = (e_{\mu}, e_{\bar \mu})$,
$\mu = 1,2,3,4,5$; ${\bar
\mu} = {\bar 1}, {\bar 2}, {\bar 3}, {\bar 4}, {\bar 5}$ be basis vectors of a 
10-dimensional Euclidean space $V_{10}$ with signature (+ + + ....).
We introduce ${\bar I} = e_{\bar 1} e_{\bar 2} e_{\bar 3} e_{\bar 4} 
e_{\bar 5}$ which satisfies
\begin{eqnarray}
         {\bar I}^2 &=& 1 \; , \nonumber \\
         e_{\mu} {\bar I} &=& - {\bar I} e_{\mu} \; ,\nonumber \\
         e_{\bar \mu} {\bar I} &=& {\bar I} e_{\bar \mu} .
\lbl{4.15}
\end{eqnarray}
Then the Clifford numbers
\begin{eqnarray}
       \gamma_{\mu} &=& e_{\mu} {\bar I}  \; ,  \nonumber \\
       \gamma_{\bar \mu} &=& e_{\mu}
\lbl{4.16}
\end{eqnarray}
satisfy
\bear
     \gamma_{\mu} \cdot \gamma_{\nu} &=& - \delta_{\mu \nu} \; , \nonumber \\
      \gamma_{\bar \mu} \cdot \gamma_{\bar \nu} &=&
       \delta_{{\bar \mu} {\bar \nu}} \; ,\nonumber \\      
       \gamma_{\mu} \cdot \gamma_{\bar \mu} &=& 0 .
\lbl{4.17}
\ear
The set $\gamma_A = (\gamma_{\mu}, \gamma_{\bar \mu})$ therefore spans
the vector space of signature $(- - - - - + + + + +)$.

The examples above demonstrate how vector spaces of various signatures are
obtained within a given set of polyvectors. Namely, vector spaces of
different signature are different subsets of polyvectors within the same
Clifford algebra.

This has important physical implications. We have argued that physical
quantities are polyvectors (Clifford numbers or Clifford aggregates). Physical
space is then not simply a vector space (e.g., Minkowski space), but a 
space of polyvectors. The latter is a
\inxx{pandimensional continuum} pandimensional continuum
${\cal P}$ \cite{14} of points, 
lines, planes, volumes, etc., altogether. Minkowski
space is then just a subspace with
\inxx{pseudo-Euclidean signature} pseudo-Euclidean signature. Other subspaces
with other signatures also exist within the pandimensional continuum
${\cal P}$ and they all have physical significance. If we describe a particle
as moving in Minkowski spacetime $V_{1,3}$ we consider only certain
physical aspects of the object considered. We have omitted its other
physical properties like spin, charge, magnetic moment, etc.. We can as well
describe the same object as moving in an Euclidean space $V_4$. Again such
a description would reflect only a part of the underlying physical
situation described by Clifford algebra.

\subsection[\ \ Grassmann numbers from Clifford numbers]
{Grassman numbers from Clifford numbers} \inxx{Grassmann numbers}

In  Sec. 2.5 we have seen that certain
\inxx{Clifford aggregates} Clifford aggregates are \inxx{spinors} spinors. Now
we shall find out that also Grassmann (anticommuting) numbers are Clifford
aggregates. As an example let us consider 8-dimensional space $V_{4,4}$
with signature $(+ - - - - + + +)$ spanned by basis vectors $\gamma_A =
(\gamma_{\mu}, \gamma_{\bar \mu})$. The numbers
\begin{eqnarray}
        \theta_{\mu} &=& \mbox{$1\oo 2$} (\gamma_{\mu} + \gamma_{\bar \mu}) ,
        \nonumber \\
        \theta_{\mu}^{\dagger} &=& \mbox{$1\oo 2$} (\gamma_{\mu} - \gamma_{\bar \mu})
\lbl{4.18}
\end{eqnarray}
satisfy
\be
         \lbrace \theta_{\mu}, \theta_{\nu} \rbrace = \lbrace 
         \theta_{\mu}^{\dagger}, \theta_{\nu}^{\dagger} \rbrace = 0 \; ,
\lbl{4.19}
\ee
\be
        \lbrace \theta_{\mu} , \theta_{\nu}^{\dagger}
        \rbrace = \eta_{\mu \nu} \; ,
\lbl{4.20}
\ee
where $\lbrace A, B \rbrace \equiv A B + B A$. From (\ref{4.19}) we 
read out that $\theta_{\mu}$ anticommute among themselves and are thus
Grassmann numbers. \inxx{Grassmann numbers}
 Similarly $\theta_{\mu}^{\dagger}$ form a set of
Grassmann numbers. But, because of (\ref{4.20}), $\theta_{\mu}$ and
$\theta_{\mu}^{\dagger}$ altogether do not form a set of Grassmann
numbers. They form yet another set of basis elements which generate
Clifford algebra.

A Clifford number in $V_{4,4}$ can be expanded as
\be
      C = c + c^{A_1} \gamma_{A_1} + c^{A_1 A_2} \gamma_{A_1} \gamma_{A_2} +
      ... + c^{A_1 A_2 ...A_8} \gamma_{A_1} \gamma_{A_2} ... \gamma_{A_8} .
\lbl{4.21}
\ee
Using (\ref{4.18}), the same Clifford number $C$ can be expanded in
terms of $\theta_{\mu}$, $\theta_{\mu}^{\dagger}$:
\begin{eqnarray}
    C = c &+& a^{\mu} \theta_{\mu} + a^{\mu \nu} \theta_{\mu} \theta_{\nu} +
        a^{\mu \nu \alpha} \theta_{\mu} \theta_{\nu} \theta_{\alpha} +
        a^{\mu \nu \alpha \beta} \theta_{\mu} \theta_{\nu} \theta_{\alpha}
        \theta_{\beta} \nonumber \\
         &+& {\bar a}^{\mu} \theta_{\mu}^{\dagger} + 
         {\bar a}^{\mu \nu} \theta_{\mu}^{\dagger} \theta_{\nu}^{\dagger} +
        {\bar a}^{\mu \nu \alpha} \theta_{\mu}^{\dagger} \theta_{\nu}^{\dagger}
         \theta_{\alpha}^{\dagger} + {\bar a}^{\mu \nu \alpha \beta} 
         \theta_{\mu}^{\dagger} \theta_{\nu}^{\dagger} \theta_{\alpha}^{\dagger}
        \theta_{\beta}^{\dagger} \nonumber \\
        &+& ({\rm mixed \; terms\;  like}\; \theta_{\mu}^{\dagger} \theta_{\nu}, 
        \; {\rm etc.})
\lbl{4.22}
\end{eqnarray}
where the coefficients $a^{\mu}, \, a^{\mu \nu}, \, ..., {\bar a}^{\mu},
{\bar a}^{\mu \nu} ...$  are linear  combinations of coefficients 
$c^{A_i}, \, c^{A_i A_j}$,...

In a particular case, coefficients $c$, ${\bar a}^{\mu}, \, {\bar a}^{\mu \nu}$,
etc., can be zero and our Clifford number is then a {\it Grassmann number}
in 4-space:
\be
      \xi = a^{\mu} \theta_{\mu} + a^{\mu \nu} \theta_{\mu} \theta_{\nu} +
        a^{\mu \nu \alpha} \theta_{\mu} \theta_{\nu} \theta_{\alpha} +
        a^{\mu \nu \alpha \beta} \theta_{\mu} \theta_{\nu} \theta_{\alpha}
        \theta_{\beta} .
\lbl{4.23}
\ee
Grassmann numbers expanded according to (\ref{4.23}), or analogous
expressions in dimensions other than 4, are much used in contemporary
theoretical physics. Recognition that Grassmann numbers can be considered
as particular numbers within a more general set of numbers, namely
Clifford numbers (or polyvectors), leads in my opinion to further progress in
understanding and development of the currently fashionable supersymmetric
theories, including superstrings, $D$-branes and $M$-theory.

We have seen that a Clifford number $C$ in 8-dimensional space can be
expanded in terms of the basis vectors $(\gamma_{\mu}, \gamma_{\bar \mu})$
or $(\theta_{\mu}, \theta_{\mu}^{\dagger})$. Besides that, one can expand
$C$ also in terms of $(\gamma_{\mu}, \theta_{\mu})$:
\begin{eqnarray}
C = c &+& b^{\mu} \gamma_{\mu} + b^{\mu \nu} \gamma_{\mu} \gamma_{\nu} +
        b^{\mu \nu \alpha} \gamma_{\mu} \gamma_{\nu} \gamma_{\alpha} +
        b^{\mu \nu \alpha \beta} \gamma_{\mu} \gamma_{\nu} \gamma_{\alpha}
        \gamma_{\beta} \nonumber \\
        &+& \beta^{\mu} \theta_{\mu} + \beta^{\mu \nu} \theta_{\mu} \theta_{\nu} +
        \beta^{\mu \nu \alpha} \theta_{\mu} \theta_{\nu} \theta_{\alpha} +
        \beta^{\mu \nu \alpha \beta} \theta_{\mu} \theta_{\nu} \theta_{\alpha}
        \theta_{\beta} \nonumber \\
        &+& ({\rm mixed\; terms\; such\; as}\; \theta_{\mu} \gamma_{\nu} , 
        {\rm \; etc.}) .
\lbl{4.24}
\end{eqnarray}
The basis vectors $\gamma_{\mu}$ span the familiar 4-dimensional
spacetime, while $\theta_{\mu}$ span an extra space, often called
Grassmann space. Usually it is stated that besides four spacetime coordinates
$x^{\mu}$ there are also four extra Grassmann coordinates $\theta_{\mu}$ and
their conjugates $\theta_{\mu}^{\dagger}$ or ${\bar \theta}_{\mu} = \gamma_0
\theta_{\mu}^{\dagger}$. This should be contrasted with the picture above in
which $\theta_{\mu}$ are basis {\it vectors} of an extra space, and not
{\it coordinates}.            
        
\section[The polyvector action and De Witt--Rovelli material reference system]
{The polyvector action and \\
De Witt--Rovelli material reference system}

Following an argument by Einstein \ci{30}, that points of
spacetime are not {\it a priori}
distinguishable, DeWitt \ci{15} introduced a concept of reference fluid. 
Spacetime
points are then defined with respect to the reference fluid. The idea
that we can localize points by means of some matter has been further
elaborated by Rovelli \ci{16}. As a starting model he considers a reference
system consisting of a single particle and a clock attached to it. Besides
the particle coordinate variables $X^{\mu} (\tau)$ there is also the
clock variable $T(\tau)$, attached to the particle, which grows monotonically
along the particle trajectory. Rovelli then assumes the following action for
the variables $X^{\mu}(\tau)$, $T(\tau)$:
\be
     I[X^{\mu},T] = m \int {\dd} \tau \left ( {{{\dd} X^{\mu}} \oo {{\dd}
     \tau}} \, {{{\dd} X_{\mu}}\oo {{\dd} \tau}} - {1\oo {\omega^2}} \left (
     {{{\dd} T} \oo {{\dd} \tau}} \right )^2 \right )^{1/2} .
\lbl{DW.12} 
\ee
If we make replacement $m \rightarrow \kappa$ and $T/\omega \rightarrow s$
the latter action reads
\be
    I[X^{\mu}, s] = \int \dd \tau \kappa \left ( {\dot X}^{\mu} 
    {\dot X}_{\mu} - {\dot s}^2 \right )^{1/2} .
\lbl{DW.12a}
\ee    
If, on the other hand, we start from the polyvector action (\ref{3.97})
and eliminate $m, \, p_{\mu}, \, \lambda$ by using the equations of motion, we
again obtain the action (\ref{DW.12a}). {\it Thus the pseudoscalar variable
$s(\tau)$ entering the polyvector action may be identified with Rovelli's
clock variable}. Although Rovelli starts with a single particle and clock,
he later fills space with these objects. We shall return to Rovelli's
reference systems when we discuss extended objects.

\chapter[Harmonic oscillator in pseudo-Euclidean space]
{Harmonic oscillator \\in pseudo-Euclidean space}

One of the major obstacles to future progress
in our understanding of the relation between quantum field theory and
gravity is the problem of the\inxx{cosmological constant}
cosmological constant \cite{31}. Since a quantum
field is an infinite set of harmonic oscillators, each having a
non-vanishing zero point energy,\inxx{zero point energy} 
a field as a whole has infinite
vacuum energy density.\inxx{vacuum energy density} 
The latter can be considered,
when neglecting the gravitational field, just as an additive constant
with no influence on dynamics. However, the situation changes
dramatically if one takes into account the gravitational field
which feels the absolute energy density. As a consequence the
infinite (or, more realistically, the Planck scale cutoff)
cosmological constant is predicted which is in drastic contradiction
with its observed small value. No generally accepted solution of
this problem has been found so far.

In this chapter I study the system of uncoupled\inxx{harmonic oscillator} 
harmonic oscillators
in a space with arbitrary metric signature. One interesting consequence
of such a model is vanishing zero point energy of the system when
the number of positive and negative signature coordinates is the
same, and so there is no cosmological constant problem. As an example
I first solve a system of two oscillators in the space with signature
$(+ -)$ by using a straightforward, though surprisingly unexploited,
approach based on the energy being here just a quadratic
form consisting of positive and negative terms \cite{32,33}.
Both positive and negative
energy components are on the same footing, and the difference in sign does
not show up until we consider gravitation. In the quantized theory
the vacuum state can be defined straightforwardly and the zero point
energies cancel out. If the action is written in a covariant notation
one has to be careful of how to define the vacuum state. Usually
it is required that energy be positive, and then, in the absence
of a cutoff, the formalism contains an infinite vacuum energy and
negative norm states \cite{34}. 
In the formalism I am going to adopt energy is not necessarily
positive, negative energy states are also stable, and there are no negative
norm states \cite{3}.

The formalism will be then applied to the case of scalar fields. 
The extension of the spinor--Maxwell field is also discussed. Vacuum
energy density is zero and the cosmological constant vanishes.
However, the theory contains the negative energy fields which
couple to the gravitational field in a different way than the usual,
positive energy, fields: the sign of coupling is reversed. This is
the price to be paid if one wants to obtain a small cosmological
constant in a straightforward way. One can consider this as a prediction
of the theory to be tested by suitably designed experiments.
Usually, however, the argument is just the opposite of the one
proposed here and classical matter is required to satisfy
certain (essentially positive) energy conditions  \cite{35} which can
only be violated by quantum field theory.

\section[The 2-dimensional pseudo-Euclidean harmonic oscillator]
{The 2-dimensional pseudo-Euclidean harmonic oscillator}

\inxx{harmonic oscillator,pseudo-Euclidean}
Instead of the usual harmonic oscillator in 2-dimensional space
let us consider that given by the Lagrangian
\be
     L = \mbox{$1\oo 2$} ({\dot x}^2 - {\dot y}^2) - \mbox{$1\oo 2$} {\omega^2}
     (x^2 - y^2) .
\lbl{H.1}
\ee

The corresponding equations of motion are
\be
       {\ddot x} + \omega^2 x = 0 \> , \qquad {\ddot y} + \omega^2 y = 0 .
\lbl{H.2}
\ee
Note that in spite of the minus sign in front of the $y$-terms
in the Lagrangian (\ref{H.1}), the $x$ and $y$ components satisfy
the same type equations of motion.

The canonical momenta conjugate to $x$,$y$ are
\be
    p_x = {{\p L}\oo {\p \dot x}} = \dot x \> , \qquad
       p_y =  {{\p L}\oo {\p \dot y}} = - \dot y .
\lbl{H.5}   
\ee
The Hamiltonian is \inxx{Hamiltonian}
\be
       H = p_x \dot x + p_y \dot y - L = \mbox{$1\oo 2$} (p_x^2 - p_y^2)
       + \mbox{$1\oo 2$} \omega^2 (x^2 - y^2)
\lbl{H.5a}
\ee
We see immediately that the energy so defined may have positive
or negative values, depending on the initial conditions. Even if the system
happens to have negative energy it is stable, since the particle
moves in a closed curve around the point (0,0). The motion of the harmonic
oscillator based on the Lagrangian (\ref{H.1}) does not differ from that
of the usual harmonic oscillator. The difference occurs when one
considers the gravitational fields around the two systems.

The Hamiltonian equations of motion are
\be
       \dot x = {{\p H}\oo {\p p_x}} =  \lbrace x, H \rbrace = p_x
       \; , \quad
       \dot y = {{\p H}\oo {\p p_y}} = \lbrace y, H \rbrace  = - p_y \; ,
\lbl{H.7}
\ee
\be
      {\dot p}_x = - {{\p H}\oo {\p x}} = \lbrace p_x,H \rbrace =
      - \omega^2 x \; , \quad
      {\dot p}_y = - {{\p H}\oo {\p y}} = \lbrace p_y,H \rbrace =
      \omega^2 y \; ,
\lbl{H.9}
\ee
where the basic Poisson brackets
\inxx{Poisson brackets} are $\lbrace x, p_x \rbrace = 1$
and $\lbrace y, p_y \rbrace = 1$.

Quantizing our system we have
\be
       \lbrack x, p_x \rbrack = i \; , \; \; \quad \quad [y, p_y] = i
\lbl{H.10}
\ee
Introducing the non-Hermitian operators according to
\be
    c_x = {1\oo \sqrt{2}} \left ( \sqrt{\omega} x + {i\oo {\sqrt{\omega}}} p_x
     \right ) \> , \qquad c_x^{\dagger} = 
     {1\oo \sqrt{2}} \left ( \sqrt{\omega} x - 
     {i\oo {\sqrt{\omega}}} p_x \right ) \; ,
\lbl{H.11}
\ee        
         \be
    c_y = {1\oo \sqrt{2}} 
    \left ( \sqrt{\omega} x + {i\oo {\sqrt{\omega}}} p_y \right ) \> , \qquad
    c_y^{\dagger} = {1\oo \sqrt{2}} \left ( \sqrt{\omega} x - 
    {i\oo {\sqrt{\omega}}} p_y \right ) \; ,
\lbl{H.12}
\ee        
we have
\be
      H = {\omega \oo 2} (c_x^{\dagger} c_x + c_x c_x^\dagger -
          c_y^{\dagger} c_y - c_y c_y^{\dagger}) .
\lbl{H.13}
\ee
From the commutation relations (\ref{H.10}) we obtain
\be
     [c_x, c_x^{\dagger}] = 1 \> , \qquad [c_y, c_y^{\dagger}] = 1 \; ,
\lbl{H.14}
\ee
and the normal ordered Hamiltonian then becomes
\be
      H= \omega (c_x^{\dagger} c_x - c_y^{\dagger} c_y) .
\lbl{H.15}
\ee
The vacuum state is defined as \inxx{vacuum state}
\be
      c_x|0 \rangle = 0 \> , \qquad c_y |0 \rangle = 0 .
\lbl{H.16}
\ee
The eigenvalues of $H$ are
\be
      E = \omega (n_x - n_y) ,
\lbl{H.16a}
\ee
where $n_x$ and $n_y$ are eigenvalues of the operators
$c_x^{\dagger} c_x$ and $c_y^{\dagger} c_y$, respectively.

The zero point energies belonging to the $x$ and $y$
components cancel. {\it Our 2-dimensional pseudo harmonic
oscillator has vanishing zero point energy}!. This is a result we
obtain when applying the standard Hamilton procedure to the
Lagrangian (\ref{H.1}).

In the $(x,y)$ representation the vacuum state $\langle x,y|0 \rangle
\equiv \psi_0 (x,y)$ satisfies
\be
      {1\oo \sqrt{2}} \left ( \sqrt{\omega} \, x + {1\oo \sqrt{\omega}} 
      {\p \oo {\p x}} \psi_0 (x,y) \right ) = 0
\; , \; \; \; 
     {1\oo \sqrt{2}} \left ( \sqrt{\omega} \, y + {1\oo \sqrt{\omega}} 
      {\p \oo {\p y}} \psi_0 (x,y) \right ) = 0 \; ,
\lbl{H.18}
\ee
which comes straightforwardly from (\ref{H.16}). A solution which
is in agreement with the probability interpretation,
\be
     \psi_0 = {{2 \pi} \oo \omega} {\rm exp}[- \mbox{$1\oo 2$}{\omega} (x^2 + y^2)]
\lbl{H.19}
\ee
is normalized according to $\int \psi_0^2 \, {\dd} x \,{\dd} y = 1$.

We see that our particle is localized around the origin. The excited
states obtained by applying $c_x^{\dagger}$, $c_y^{\dagger}$ to the
vacuum state are also localized. This is in agreement with the property
that also according to the classical equations of motion (\ref{H.2}),
the particle is localized in the vicinity of the origin.
All states $|\psi \rangle$ have positive norm. For instance,
$$\langle 0| c c^{\dagger}|0 \rangle = \langle 0|[c,c^{\dagger}]|0 \rangle
= \langle 0|0 \rangle = \int \psi_0^2 {\dd} x \, {\dd} y = 1.$$

\vs{2mm}

\upperandlowertrue
\section[Harmonic oscillator in $d$-dimensional pseudo-Euclidean space]
{HARMONIC OSCILLATOR IN $d$-DIMENSIONAL PSEUDO-EUCLIDEAN SPACE}
\upperandlowerfalse

Extending (\ref{H.1}) to arbitrary dimension it is convenient to use the
compact (covariant) index notation
\be
     L = {1 \oo 2} {\dot x}^{\mu} {\dot x}_{\mu} -\mbox{$1\oo 2$} {{\omega^2}}
     x^{\mu} x_{\mu} \; ,
\lbl{H.20}
\ee
where for arbitrary vector $A^{\mu}$ the quadratic form is
$A^{\mu} A_{\mu} \equiv \eta_{\mu \nu} A^{\mu} A^{\nu}$. The metric
tensor $\eta_{\mu \nu}$ has signature $(+ + + ... - - - ...)$. The
Hamiltonian is
\be
      H= \mbox{$1\oo 2$} p^{\mu} p_{\mu} + \mbox{$1\oo 2$} {{\omega^2}} x^{\mu} x_{\mu}
\lbl{H.21}
\ee
Conventionally one introduces
\be
       a^{\mu} = {1 \oo \sqrt{2}} \left ( \sqrt{\omega} \, x^{\mu} + {i \oo 
       \sqrt{\omega}} \, p^{\mu} \right )
\> , \qquad
       {a^{\mu}}^{\dagger} = {1 \oo \sqrt{2}} \left ( \sqrt{\omega} x^{\mu} - 
       {i \oo \sqrt{\omega}} p^{\mu} \right ) .
\lbl{H.23}
\ee
In terms of $a^{\mu}$, $a^{\mu \dagger}$ the Hamiltonian reads
\be
      H = {\omega \oo 2} (a^{\mu \dagger} a_{\mu} + a_{\mu} a^{\mu \dagger}) .
\lbl{H.24}
\ee
Upon quantization we have
\be  [x^{\mu}, p_{\nu}] = i {\delta^{\mu}}_{\nu} \qquad {\rm or}
    \qquad [x^{\mu}, p^{\nu}] = i \eta^{\mu \nu}
\lbl{H.25}
\ee
and
\be
    [a^{\mu}, a_{\nu}^{\dagger}] = {\delta^{\mu}}_{\nu} \qquad {\rm or} 
    \qquad  [a^{\mu}, a^{\nu \dagger}] = \eta^{\mu \nu} .
\lbl{H.25a}
\ee

We shall now discuss two possible definitions of
\inxx{vacuum state} the vacuum state. The first
possibility is the one usually assumed, whilst the second 
possibility \cite{32,33} is the one I am going to adopt.

\vs{2mm}

{\it Possibility I.} The vacuum state can be defined according to
\be
        a^{\mu} |0 \rangle = 0
\lbl{H.26}
\ee
and the Hamiltonian, normal ordered with respect to the vacuum
definition (\ref{H.26}), becomes, after using (\ref{H.25a}),
 \be
       H = \omega \left ( a^{\mu \dagger} a_{\mu} + {d \oo 2}
       \right ) \; ,  \qquad d = \eta^{\mu \nu} \eta_{\mu \nu} .
\lbl{H.27}
\ee
Its eigenvalues are all positive and there is the non-vanishing
zero point energy $\omega d/2$. In the $x$ representation the vacuum
state is
\be
  \psi_0 = \left ({{2 \pi} \oo \omega} \right )^{d/2} 
  {\rm exp}[- \mbox{$1\oo 2$}{\omega} x^{\mu} x_{\mu}]
\lbl{H.28}
\ee
It is a solution of the Schr\" odinger equation
$- \mbox{$1\oo 2$} \p^{\mu} \p_{\mu} \psi_0 + (\omega^2/2) x^{\mu} x_{\mu} \psi_0
= E_0 \psi_0$ with positive $E_0 = \omega (\mbox{$1\oo 2$} + \mbox{$1\oo 2$} + ....)$.
The state $\psi_0$ as well as excited states can not be normalized to 1.
Actually, there exist negative norm states. \inxx{negative norm states}
For instance, if $\eta^{33}
= -1$, then 
$$\langle 0|a^3 a^{3 \dagger}|0 \rangle = \langle 0|
[a^3, a^{3 \dagger}]|0 \rangle = - \langle 0|0 \rangle .$$

{\it Possibility II.} Let us split $a^{\mu} = (a^{\alpha}, 
a^{\bar \alpha})$, where the indices
$\alpha$, ${\bar \alpha}$ refer to the components with positive and
negative signature, respectively, and define the vacuum according to\footnote{
Equivalently, one can define annihilation and creation operators in terms
of $x^{\mu}$ and the canonically conjugate momentum $p_{\mu} = 
\eta_{\mu \nu} p^{\nu}$ according to $c^{\mu} = (1/\sqrt{2}) (\sqrt{
\omega} x^{\mu} + (i/\sqrt{\omega}) p_{\mu})$ and $c^{\mu \dagger} =
(1/\sqrt{2}) (\sqrt{ \omega} x^{\mu} - (i/\sqrt{\omega}) p_{\mu})$,
satisfying $[c^{\mu}, c^{\nu \dagger}] = \delta^{\mu \nu}$. The vacuum is
then defined as $c^{\mu} |0 \rangle = 0$. This is just the 
higher-dimensional generalization of $c_x$, $c_y$ (eq.\,(\ref{H.11}),(\ref{H.12})
and the vacuum definition (\ref{H.16}). }
\be
    a^{\alpha} |0 \rangle = 0  \; , \; \qquad \qquad a^{{\bar \alpha} \dagger}
    |0 \rangle = 0 .
\lbl{H.29}
\ee
Using (\ref{H.25a}) we obtain the normal ordered Hamiltonian with
respect to the vacuum definition (\ref{H.29})
\be
      H = \omega \left ( a^{\alpha \dagger} a_{\alpha} + {r \oo 2} +
           a_{\bar \alpha} a^{{\bar \alpha} \dagger} - {s \oo 2} \right ) ,
\lbl{H.30}
\ee
where ${\delta_{\alpha}}^{\alpha} = r$ and ${{\delta}_{\bar \alpha}}^
{\bar \alpha} = s$. If the number of positive and negative signature
components is the same, i.e., $r = s$, then the Hamiltonian (\ref{H.30})
has vanishing zero point energy:
\be
H = \omega (a^{\alpha \dagger} a_{\alpha} + 
           a_{\bar \alpha} a^{{\bar \alpha} \dagger}) .
\lbl{H.31}
\ee
Its eigenvalues are positive or negative, depending on which components
(positive or negative signature) are excited. In the $x$-representation
the vacuum state (\ref{H.29}) is
\be
     \psi_0 = {\left ( {{2 \pi} \oo \omega} \right ) }^{d/2} 
     {\rm exp}[- \mbox{$1\oo 2$} \omega \delta_{\mu \nu} x^{\mu} x^{\nu}] ,
\lbl{H.32}
\ee
where the Kronecker symbol $\delta_{\mu \nu}$ has the values +1 or 0.
It is a solution of the Schr\" odinger equation
$-\mbox{$1\oo 2$} \p^{\mu} \p_{\mu} \psi_0 + (\omega^2/2) x^{\mu} x_{\mu} \psi_0
= E_0 \psi_0$ with $E_0 = \omega (\mbox{$1\oo 2$} + \mbox{$1\oo 2$} + .... - \mbox{$1\oo 2$} -
\mbox{$1\oo 2$}- ...)$. One can also easily verify that there are no negative
norm states.

Comparing {\it Possibility I} with {\it Possibility II} we observe that the
former has positive energy vacuum invariant under pseudo-Euclidean
rotations, whilst the latter has the vacuum invariant under Euclidean
rotations and having vanishing energy (when $r=s$). In other words,
we have: either (i) non-vanishing energy and pseudo-Euclidean invariance or
(ii) vanishing energy and Euclidean invariance of the vacuum state.
In the case (ii) the vacuum state $\psi_0$ changes under 
the pseudo-Euclidean rotations, but its energy remains zero.

The invariance group of our Hamiltonian (\ref{H.21}) and the
corresponding Schr\" odinger equation consists of
pseudo-rotations. Though a solution of the Schr\" odinger
equation changes under a pseudo-rotation, the theory is 
covariant under the pseudo-rotations, in the sense that the
set of all possible solutions does not change under the pseudo-rotations.
Namely, the solution $\psi_0 (x')$ of the Schr\" odinger equation
\be
- \mbox{$1\oo 2$} {\p'}^{\mu} {\p'}_{\mu} \psi_0(x') + 
(\omega^2/2) x'^{\mu} x'_{\mu} \psi_0(x') = 0
\lbl{H.32a}
\ee
in a pseudo-rotated
frame  $S'$ is
\be
\psi_0 (x') = {{2\pi} \oo \omega}^{d/2} {\rm exp} [- \mbox{$1\oo 2$}\omega
\delta_{\mu \nu} x'^{\mu} x'^{\nu}] .
\lbl{H.32b}
\ee
If observed from the frame
$S$ the latter solution reads 
\be
\psi'_0 (x) =
{{2 \pi} \oo \omega}^{d/2} \times {\rm exp} 
[- \mbox{$1\oo 2$} \omega\delta_{\mu \nu} 
{L^\mu}_\rho x^\rho {L^{\nu}}_\sigma x^\sigma] ,
\lbl{H.32c}
\ee
where $x'^\mu = {L^\mu}_\rho x^\rho$. One finds
that $\psi'_0 (x)$ as well as $\psi_0(x)$ (eq.\,(\ref{H.32})) are solutions
of the Schr\" odinger equation in $S$ and they both have the same
vanishing energy. In general, in a given reference frame we have thus
a degeneracy of solutions with the same energy \cite{32}.
This is so also in the case of excited states.

In principle it seem more natural to adopt {\it Possibility II},
because the classically energy of our harmonic oscillator is nothing but
a quadratic form $E = \mbox{$1\oo 2$} (p^\mu p_\mu + \omega^2 x^\mu x_\mu)$, which
in the case of a metric of pseudo-Euclidean signature can be positive,
negative, or zero.

\section{A system of scalar fields}

Suppose we have a system of two scalar \inxx{system of two scalar fields}
fields described by the action\footnote
{Here, for the sake of demonstration, I am using the formalism
of the conventional field theory, though in my opinion a better
formalism involves an invariant evolution parameter, as discussed in Sec. 1.4}
\be
    I = \mbox{$1\oo 2$} \int {\dd}^4 x \, (\p_{\mu} \phi_1 \, \p^\mu \phi_1 - 
    m^2 \phi_1^2 - \p_\mu \phi_2 \, \p^\mu \phi_2 + m^2 \phi_2^2) .
\lbl{H.36}
\ee
This action differs from the usual action for a charged field
in the sign of the $\phi_2$ term. It is a field generalization of
our action for the point particle harmonic oscillator in 2-dimensional
pseudo-Euclidean space.

The canonical momenta are
\be
        \pi_1 = {\dot \phi}_1 \; , \qquad \pi_2 = - {\dot \phi}_2
\lbl{H.37}
\ee
satisfying
\be
   [\phi_1({\bf x} ), \pi_1({\bf x}')] = i \delta^3 ({\bf x} - {\bf x} ')
   \> , \qquad
   [\phi_2({\bf x}), \pi_2({\bf x}')] = i \delta^3 ({\bf x} - {\bf x}') .
\lbl{H.39}
\ee
The Hamiltonian is
\be
    H = \mbox{$1\oo 2$} \int {\dd}^3 {\bf x} \, ({\bf \pi}_1^2 + m^2 \phi_1^2 -
    \p_i \phi_1 \, \p^i \phi_1 - {\bf \pi}_2^2 - m^2 \phi_2^2 +
    \p_i \phi_2 \p^i \phi_2) .
\lbl{H.40}
\ee
We use the spacetime metric with signature $(+ - - -)$ so that
$-\p_i \phi_1 \, \p^i \phi_1$ $= ({\bf \nabla} \phi)^2$, $i = 1,2,3$. Using the
expansion ($\omega_{\bf k} = (m^2 + {\bf k} ^2)^{1/2}$)
\be
    \phi_1 = \int {{{\dd}^3 {\bf k}}\oo {(2 \pi)^3}} \, 
    {1\oo {2 \omega_{\bf k}}}
    \left (c_1 ({\bf k} ) e^{- i k x} + c_1^{\dagger} ({\bf k} ) e^{i k x}
    \right ) ,
\lbl{H.41}
\ee
\be
    \phi_2 = \int {{{\dd}^3 {\bf k}} \oo {(2 \pi)^3}} \,
    {1\oo {2 \omega_{\bf k}}}
    \left (c_2 ({\bf k} ) e^{- i k x} + c_2^{\dagger} ({\bf k} ) e^{i k x}
    \right )
\lbl{H.42}
\ee
we obtain
\be
    H = {1\oo 2} \int {{\dd}^3 {\bf k} \oo {(2 \pi)^3}} \, 
    {\omega_{\bf k} \oo 2 \omega_{\bf k}} \left ( c_1^{\dagger}({\bf k} ) 
     c_1({\bf k}) + c_1 ({\bf k}) c_1^{\dagger} ({\bf k}) -
     c_2^{\dagger}({\bf k} ) c_2({\bf k}) -
     c_2 ({\bf k}) c_2^{\dagger} ({\bf k}) \right ) .
\lbl{H.43}
\ee

The commutation relations are
\be
     [c_1 ({\bf k}), c_1^{\dagger}({\bf k}'] = (2 \pi)^3 2 \omega_{\bf k}
     \, \delta^3 ({\bf k} - {\bf k}') ,
\lbl{H.44}
\ee
\be
     [c_2 ({\bf k}), c_2^{\dagger}({\bf k}'] = (2 \pi)^3 2 \omega_{\bf k}
     \, \delta^3 ({\bf k} - {\bf k}') .
\lbl{H.45}
\ee
The Hamiltonian can be written in the form \inxx{Hamiltonian}
\be
    H = {1\oo 2} \int {{\dd}^3 {\bf k} \oo {(2 \pi)^3}} \, 
    {\omega_{\bf k} \oo 2 \omega_{\bf k}} \left ( c_1^{\dagger}({\bf k} ) 
     c_1({\bf k})  - c_2^{\dagger}({\bf k} ) c_2({\bf k}) \right )
\lbl{H.47}
\ee
If we define the vacuum according to \inxx{vacuum}
\be
    c_1({\bf k}) |0 \rangle = 0 \; , \qquad c_2({\bf k}) |0 \rangle = 0 \; ,
\lbl{H.48}
\ee
then the Hamiltonian (\ref{H.47}) contains the creation operators on the left
and has no zero point energy. However, it is not positive definite: it
may have positive or negative eigenvalues. But, as is obvious from our
analysis of the harmonic oscillator (\ref{H.1}), negative energy states in
our formalism are not automatically unstable; they can be as stable as
positive energy states.

Extension of the action (\ref{H.36}) to arbitrary number os fields $\phi^a$ is
straightforward. Let us now include also the gravitational field $g_{\mu \nu}$.
The action is then \inxx{gravitational field}
\be
    I = {1\oo 2} \int {\dd}^4 x \, \sqrt{-g} \,
    \left ( g^{\mu \nu} \p_\mu \phi^a \, \p_\nu \phi^b
    \, \gamma_{ab} - m^2 \, \phi^a \phi^b \gamma_{ab} + 
    {1 \oo {16 \pi G}} \, R \right ) ,
\lbl{H.49}
\ee
where $\gamma_{ab}$ is the metric tensor in the space of $\phi^a$.
Variation of (\ref{H.49}) with respect to $g^{\mu \nu}$ gives the Einstein
equations
\be
     R_{\mu \nu} - \mbox{$1\oo 2$} g_{\mu \nu} R = - 8 \pi G \, T_{\mu \nu} \; ,
\lbl{H.50}
\ee
where the stress--energy tensor is \inxx{stress--energy tensor}
\be
     T_{\mu \nu} = {2 \oo {\sqrt{-g}}} \,  {{\p {\cal L}} \oo {\p g^{\mu \nu}}}
     = \left [\p_{\mu} \phi^a \, \p_{\nu} \phi^b - \mbox{$1\oo 2$} \, g_{\mu \nu} \, 
     (g^{\rho \sigma} \p_{\rho } \phi^a \, \p_{\sigma} \phi^b 
     - m^2 \phi^a \phi^b ) \right ]\gamma_{ab} .
\lbl{H.51}
\ee
If $\gamma_{ab}$ has signature $(+ + + ... - - \, -)$, with the same number of 
plus and minus signs, then the vacuum contributions to $T_{\mu \nu}$ cancel,
so that the expectation value $\langle T_{\mu \nu} \rangle$ remains finite.
In particular, we have
\be
     T_{00} = \mbox{$1\oo 2$} ({\dot \phi}^a {\dot \phi}^b - \p_i 
     \phi^a \, \p^i \phi^b + m^2 \phi^a \phi^b ) \gamma_{ab} \; ,
\lbl{H.52}
\ee
which is just the Hamiltonian $H$ of eq.\,(\ref{H.40}) generalized to
an arbitrary number of fields $\phi^a$.

A procedure analogous to that before could be carried out 
for other types of fields such as charged scalar, spinor and gauge fields.
The notorious cosmological constant problem does not arise in our
model, since the vacuum expectation value $\langle 0| T_{\mu \nu}|0 \rangle = 0$.
We could reason the other way around: since experiments clearly show that the
cosmological constant is small, this indicates (especially in the
absence of any other acceptable explanation) that to every field there
corresponds a companion field with the signature opposite of the metric
eigenvalue in the space of fields. The companion field need not be excited
---and thus observed--- at all. Its mere existence is sufficient to be
manifested in the vacuum energy.

However, there is a price to be paid.\inxx{negative signature fields} If 
negative signature fields are
excited then $\langle T_{00} \rangle$ can be negative, which implies a repulsive
gravitational field \inxx{gravitational field,repulsive}
around such a source. Such a prediction of the theory
could be considered as an annoyance, on the one hand, or a virtue, on the
other. If the latter point of view is taken then we have here so
called exotic matter with negative energy density, \inxx{exotic matter}
which is necessary for \inxx{negative energy density}
the construction of the stable wormholes with time machine properties \cite{36}.
Also the Alcubierre warp drive \cite{37} which enables faster than light
motion with respect to a distant observer requires matter of negative energy
density. \inxx{wormholes} \inxx{time machines} \inxx{warp drive}

As an example let me show how the above procedure works for spinor
and gauge fields. Neglecting gravitation the action is
\be
    I = \int {\dd}^4 x\, \left [ i {\bar \psi}^a \gamma^{\mu}
    (\p_{\mu} \psi^b + i e 
    {{A_{\mu}}^b}_c \psi^c) - m {\bar \psi}^a \psi^b +{1 \oo {16 \pi}}
    {F_{\mu \nu}}^{ac} {{F^{\mu \nu}}_c}^b \right ] \gamma_{ab} ,
\label{H.53}
\ee    
where 
\be
{F_{\mu \nu}}^{ab} = \p_{\mu} {A_{\nu}}^{ab} - 
\p_{\nu} {A_{\mu}}^{ab} - ({A_{\mu}}^{ac} {A_{\nu}}^{db} - 
{A_{\nu}}^{ac} {A_{\mu}}^{db}) \gamma_{cd} .
\lbl{H.54}
\ee
This action is invariant
under local rotations 
\be
\psi'^a = {U^a}_b \phi^b \; , \; \; {\bar \psi}'^a = {U^a}_b
\psi^b \; , \; \; {{A'_{\mu}}^a}_b = {{U^*}^c}_b {{A_{\mu}}^d}_c {U^a}_d +
i {{U^*}^c}_b \p_{\mu} {U^a}_c ,
\lbl{H.55}
\ee
which are generalization of the usual
$SU(N)$ transformations to the case of the metric $\gamma_{ab} =
{\rm diag} (1,1,..., -1, -1)$.

In the case in which there are two spinor fields $\psi_1$, $\psi_2$, and
$\gamma_{ab} = {\rm diag} (1,-1)$, the equations of motion derived from
(\ref{H.53}) admit a solution $\psi_2 =0$,
$A_{\mu}^{12}=A_{\mu}^{21}=A_{\mu}^{22}=0$.
In the quantum field theory such a solution can be interpreted as that when
the fermions of type $\psi_2$ (the companion or negative signature fields)
are not excited (not present) the gauge fields $A_{\mu}^{12}$,
$A_{\mu}^{21}$, $A_{\mu}^{22}$ are also not excited (not present).
What remains are just the ordinary
$\psi_1 \equiv \psi$ fermion quanta and $U(1)$ gauge field $A_{\mu}^{11}
\equiv A^{\mu}$ quanta. The usual spinor--Maxwell electrodynamics is just
a special solution of the more general system given by (\ref{H.53}).

Although having vanishing vacuum energy, such a model is consistent with
the well known experimentally observed effects which are manifestations
of vacuum energy. Namely, the companion particles $\psi_2$ are expected
to be present in the material of the Earth  in small amounts at most, because
otherwise the gravitational field around the Earth would be repulsive.
So, when considering vacuum effects, there remain only (or predominantly)
the interactions between the fermions $\psi_1$ and the virtual photons
$A_{\mu}^{11}$. For instance, in the case of the
\inxx{Casimir effect} Casimir effect \cite{38}
the fermions $\psi_1$ in the two conducting plates interact with the
virtual photons $A_{\mu}^{11}$ in the vacuum, and hence impose the boundary
conditions on the vacuum modes of $A_{\mu}^{11}$ in the presence of the plates.
As a result we have a net force between the plates, just as in the
usual theory.

When gravitation is not taken into account, the fields within the doublet
$(\psi_1, \, \psi_2)$ as described by the action (\ref{H.53}) are not easily
distinguishable, since they have the same mass, charge, and spin. They mutually
interact only through the mixed coupling terms in the action (\ref{H.53}),
and unless the effects of this mixed coupling are specifically measured
the two fields can be mis-identified as a single field.
Its double character could manifest itself straightforwardly in the presence
of a gravitational field to which the members of a doublet couple with the
opposite sign. In order to detect such doublets (or perhaps multiplets)
of fields one has to perform suitable experiments. The description of such
experiments is beyond the scope of this book, which only aims to bring
attention to such a possibility. Here I only mention that difficulties
and discrepancies in measuring the precise value of the gravitational constant
might have roots in negative energy matter. The latter would affect
the measured value of the effective gravitational constant, but would
leave the equivalence principle untouched.

\section{Conclusion}

The problem of the cosmological constant is one of the toughest problems
in theoretical physics. Its resolution would open the door to further
understanding of the relation between quantum theory and general relativity.
Since all more conventional approaches seem to have been more or less
exploited without unambiguous success, the time is right for a more drastic
novel approach. Such is the one which relies on the properties of the harmonic
oscillator in a pseudo-Euclidean space. This can be applied to the field
theory in which the fields behave as components of a harmonic oscillator. If the
space of fields has a metric with signature $( + + + ... - - -)$ then the
vacuum energy can be zero in the case in which the number of plus and minus signs
is the same. As a consequence the expectation value of the stress--energy tensor,
the source of the gravitational field, is finite, and there is no cosmological
constant problem. However, the stress--energy tensor can be negative in
certain circumstances and the matter then acquires exotic properties
which are desirable for certain very important theoretical constructions,
such as\inxx{time machines} \inxx{warp drive}time machines \cite{36} or a 
faster than light warp drive \cite{37}.
Negative energy matter, with repulsive gravitational field, is considered here
as a prediction of the theory. On the contrary, in a more conventional approach
just the opposite point of view is taken. It is argued that, since for all
known forms of matter gravitation is attractive, certain energy conditions
(weak, strong and dominant) must be satisfied \cite{36}. But my point of view,
advocated in this book, is that the existence of negative energy matter is
necessary in order to keep the cosmological constant small (or zero). In
addition to
that, such exotic matter, if indeed present in the Universe, should
manifest itself in various phenomena related to gravitation. 
Actually, we cannot
claim to possess a complete knowledge and understanding of all those
phenomena, especially when some of them are still waiting for a generally
accepted explanation.  

The theory of the pseudo-Euclidean signature harmonic oscillator is
important for\inxx{string} \inxx{zero point energy}strings also. 
Since it eliminates the zero point energy
it also eliminates the need for a critical dimension, as indicated in
ref. \ci{18}.  We 
thus obtain a consistent string theory in an arbitrary
even dimensional spacetime with suitable signature. Several exciting
new possibilities of research are thus opened.

The question is only whether \inxx{pandimensional continuum}
an $n$-dimensional spacetime $V_n$ has indeed an equal number of time-like
and space-like dimensions. Here we bring into the game the assumption
that physical quantities are Clifford aggregates, or polyvectors,
and that what we observe as
spacetime is just a segment of a higher geometrical structure, called the
{\it pandimensional continuum} \cite{14}. Among the
available elements of Clifford algebra one can choose $n$ elements such
that they serve as basis vectors of a pseudo-Euclidean space with
arbitrary signature. Harmonic oscillator in pseudo-Euclidean space is only a tip
of an iceberg. The iceberg is geometric calculus based on Clifford algebra
which allows for any signature. Actually a signature is relative to a chosen
basis $e_{\mu}, \, e_{\mu \nu}, \, e_{\mu \nu \alpha},...$ of Clifford algebra.

On the other hand, in conventional theory a signature with two or more time-like
dimensions is problematic from the point of view of initial conditions:
predictability\inxx{predictability}
is not possible. \inxx{invariant evolution parameter} 
But such a problem does not arise in the
theory with an invariant evolution parameter $\tau$, where initial conditions
are given in the entire spacetime, and the evolution then goes along
$\tau$. This is yet another argument in favor of a Stueckelberg-like
theory. However, a question arises of how a theory in which initial conditions
are given in the entire spacetime can be reconciled with the fact that
we do not observe the entire spacetime ``at once". This question will be
answered in the remaining parts of the book.

\part[Extended Objects]
{Extended objects}

\chapter[General principles of membrane \\ \  kinematics and dynamics]
{General principles of membrane \\ kinematics and dynamics}

We are now going to extend our system. Instead of point particles\inxx{string}
we shall consider strings\inxx{higher-dimensional membranes}
and higher-dimensional membranes. These
objects are nowadays amongst the hottest topics in fundamental
theoretical physics. Many people are convinced that strings and
accompanying higher-dimensional membranes provide a clue to unify
physics \ci{39}. In spite of many spectacular successes in unifying gravity
with other interactions, there still remain open problems. Amongst the
most serious is perhaps the 
problem of a\inxx{geometric principle,behind string theory} 
geometrical principle
behind string theory \ci{40}. The approach pursued in this book aims to shed
some more light on just that problem. We shall pay much attention to the
treatment of membranes as points in an infinite-dimensional space,
called membrane space ${\cal M}$.\inxx{infinite-dimensional space}
When pursuing such\inxx{membrane space} an approach the researchers 
usually try to build in
right from the beginning \inxx{reparametrization invariance} 
a complication which arises from reparametrization
invariance (called also 
diffeomorphism invariance).\inxx{diffeomorphism invariance} 
Namely, the same
$n$-dimensional membrane can be represented by different sets of parametric
equations $x^{\mu} = X^{\mu} (\xi^a)$, where functions $X^{\mu}$ map
a membrane's parameters (also called coordinates) $\xi^a , \, a = 1,2,...,n$,
into spacetime coordinates $x^{\mu} , \, \mu = 0,1,2,..., N-1$. The problem
is then what are coordinates of the membrane space ${\cal M}$? If there
were no complication caused by reparametrization invariance, then the $X^{\mu} (
\xi^a)$ would be coordinates of ${\cal M}$-space. But because of
reparametrization invariance such a mapping from a point in ${\cal M}$
(a membrane) to its coordinates $X^{\mu} (\xi)$ is one-to-many. So far there is
no problem: a point in any space can be represented by many different
possible sets of coordinates, and all those different sets are related by
coordinate transformations or diffeomorphisms.
Since the latter transformations\inxx{coordinate transformations,passive} 
refer to the same point they are called \inxx{passive diffeomorphisms}
{\it passive coordinate transformations} or {\it passive diffeomorphisms}.
The problem occurs when one brings into the game {\it active diffeomorphisms}
which refer to different points\inxx{coordinate transformations,active}
of the space in question. In the case of\inxx{diffeomorphisms,active}
membrane space ${\cal M}$ 
active diffeomorphisms\inxx{tangentially deformed membranes} 
would imply the existence of
{\it tangentially deformed membranes}. But such objects are not present
in a relativistic theory of membranes, described by the minimal surface
action which is invariant under reparametrizations of $\xi^a$.

The approach pursued here is the following. We shall assume that {\it
at the kinematic level such tangentially deformed membranes do exist}
\ci{41}.
When considering dynamics it may happen that a certain action and its
equations of motion exclude tangential motions within the membrane. This
is precisely what happens with membranes obeying the relativistic minimal
surface action. But the latter dynamical principle is not the most general
one. We can extend it according to geometric calculus based on Clifford
algebra. We have done so in Chapter 2 for point particles, and now we shall
see how the procedure can be generalized to membranes of arbitrary dimension.
\inxx{membrane action,polyvector generalization of}
And we shall find a remarkable result that {\it the polyvector generalization}
of the membrane action allows for tangential motions of membranes. Because
of the presence of an extra pseudoscalar variable entering the polyvector action,
the membrane variables $X^{\mu}(\xi)$ and the corresponding momenta become
{\it unconstrained}; tangentially deformed membranes are thus present
not only at the kinematic, but also at the dynamical level. In other words,
such a generalized dynamical principle allows for tangentially deformed
membranes.

In the following sections I shall put the above description into a more
precise form. But in the spirit of this book I will not attempt to achieve
complete mathematical rigor, because for most readers this would be
at the expense of seeing the main outline of my proposal of how to
formulate membrane's theory

\upperandlowertrue
\section[Membrane space ${\cal M}$]
{MEMBRANE SPACE ${\cal M}$} 
\upperandlowerfalse
\inxx{membrane space}

The basic kinematically possible objects of the theory
we are going to discuss are
$n$-dimensional, arbitrarily deformable, and hence unconstrained, membranes
${\cal V}_n$ living in an $N$-dimensional space $V_N$.
The dimensions $n$ \inxx{membranes,unconstrained}
and $N$, as well as the corresponding signatures, are left unspecified at this
stage. An unconstrained membrane ${\cal V}_n$ is represented by the
embedding functions $X^{\mu} (\xi^a)$, $\mu = 0,1,2,...,N-1$, where
$\xi^a$, $a = 0,1,2,...,n-1$, are local parameters (coordinates) on
${\cal V}_n$. The set of all possible membranes ${\cal V}_n$, with $n$ fixed,
forms an infinite-dimensional space ${\cal M}$. A membrane ${\cal V}_n$
can be considered as a point in ${\cal M}$ parametrized by coordinates
$X^{\mu} (\xi^a) \equiv X^{\mu (\xi)}$
which bear a discrete index $\mu $ and $n$ continuous
indices $\xi^a$. To the discrete index $\mu$ we can ascribe arbitrary numbers:
instead of $\mu = 0,1,2,...,N-1$ we may set $\mu ' = 1,2,...,N$ or
$\mu ' = 2,5,3,1,...,$ etc.. In general,
\be
    \mu' = f(\mu) ,
\label{A1}
\ee
where $f$ is a transformation. Analogously, a continuous index
$\xi^a$ can be given arbitrary continuous values. Instead of
$\xi^a$ we may take ${\xi'}^a$ which are functions of
$\xi^a$ :
\be
       {\xi'}^a = f^a (\xi) .
\label{A2}
\ee
As far as we consider, respectively, $\mu$ and $\xi^a$ as a
discrete and a continuous index of coordinates $X^{\mu (\xi)}$
in the infinite-dimensional space ${\cal M}$, reparametrization of
$\xi^a$ is analogous to a renumbering of $\mu$. Both kinds of
transformations, (\ref{A1}) and (\ref{A2}), refer to the same point
\inxx{passive transformations}
of the space ${\cal M}$; they are {\it passive transformations}. For
instance, under the action of (\ref{A2}) we have
\be
     {X '}^{\mu} (\xi') = {X '}^{\mu} \left ( f(\xi) \right ) = X^{\mu} (\xi)
\label{A3}
\ee
which says that the same point ${\cal V}_n$ can be described
either by functions $X^{\mu} (\xi)$ or ${X'}^{\mu} (\xi)$ (where we may
write ${X '}^{\mu} (\xi)$ instead of ${X '}^{\mu} (\xi')$
since $\xi '$ is a running parameter and can be
renamed as $\xi$).

Then there also exist the {\it active transformations}, which transform
\inxx{active transformations}
one point of the space ${\cal M}$ into another. Given a parametrization
of $\xi^a$ and a numbering of $\mu$, a point ${\cal V}_n$ of
${\cal M}$ with coordinates $X^{\mu} (\xi)$ can be transformed into
another point ${\cal V '}_n$ with coordinates ${X '}^{\mu} (\xi)$. Parameters
$\xi^a$ are now considered as ``body fixed", so that distinct
functions $X^{\mu} (\xi)$, ${X'}^{\mu} (\xi)$ represent distinct
points ${\cal V}_n$, ${\cal V'}_n$ of ${\cal M}$. Physically these are
distinct membranes which may be deformed one with respect to the
other. Such a membrane is {\it unconstrained}, since all coordinates
$X^{\mu} (\xi)$ are necessary for its description \cite{41}--\cite{43}.
In order to \inxx{membranes,unconstrained}
distinguish an unconstrained membrane ${\cal V}_n$ from the corresponding
mathematical manifold $V_n$ we use different symbols ${\cal V}_n$
and $V_n$.

It may happen, in particular, that two distinct membranes
${\cal V}_n$ and ${\cal V '}_n$ both
lie on the same mathematical surface $V_n$, and yet they are physically
distinct objects, represented by different points in ${\cal M}$.

The concept of an unconstrained membrane can be illustrated by imagining
a rubber sheet spanning a surface $V_2$. The sheet can be deformed
from one configuration (let me call it ${\cal V}_2$) into another
configuration ${\cal V'}_2$ in such a way that both configurations
${\cal V}_2$, ${\cal V'}_2$ are spanning the same surface $V_2$.
The configurations ${\cal V}_2$, ${\cal V'}_2$ are described by functions
$X^i (\xi^1, \xi^2)$, $X'^i (\xi^1, \xi^2)$ ($i = 1,2,3$), respectively.
The latter functions, from the mathematical point of view, both represent
the same surface $V_2$, and can be transformed one into the other by a
reparametrization of $\xi^1, \xi^2$. But from the physical point
of view,
$X^i (\xi^1, \xi^2)$ and $X'^i (\xi^1, \xi^2)$ represent two different
configurations of the rubber sheet (Fig. \ref{Wiggly}).

\figWiggly

The reasoning presented in the last few paragraphs implies that,
since our membranes are assumed to be arbitrarily deformable,
different functions $X^{\mu} (\xi)$
can always represent physically different membranes. This
justifies use of the coordinates $X^\mu (\xi)$ for the
description of points in ${\cal M}$. Later, when we consider a
membrane's kinematics and dynamics we shall admit $\tau$-dependence
of coordinates $X^\mu (\xi)$. In this section all expressions
refer to a fixed value of $\tau$, therefore we omit it from the notation.

In analogy with the finite-dimensional case we can introduce the
\inxx{distance} distance \inxx{infinite-dimensional space}
${\dd} {\ell}$ in our infinite-dimensional space ${\cal M}$ :
\bear
    {\dd} {\ell}^2 &=& \int {\dd} \xi \, {\dd} \zeta {\rho}_{\mu \nu}
    (\xi,\zeta) \,
    {\dd} X^{\mu} (\xi) \, {\dd} X^{\nu} (\zeta) \nonumber \\
     &=& {\rho}_{\mu(\xi )
    \nu(\zeta)} \, 
    {\dd} X^{\mu (\xi)} \, {\dd} X^{\nu (\zeta)} =
    {\dd} X^{\mu (\xi)} {\dd} X_{\mu(\xi)} ,
\label{A4}
\ear
where ${\rho}_{\mu \nu} (\xi ,\zeta) = {\rho}_{\mu(\xi)\nu(\zeta)}$
is the metric in
${\cal M}$. In eq.\ (\ref{A4}) we use a notation, similar to one that is
usually used when dealing with more evolved functional expressions
\cite{44}, \cite{44a}. In order to distinguish
continuous indices from  discrete indices, the former are written
within parentheses. When we write $\mu(\xi)$ as a subscript or superscript
this denotes a pair of indices $\mu$ and $(\xi)$ (and not that $\mu$ is a
function of $\xi$). We also use the convention that summation is
performed over repeated indices (such as $a , b$) and integration over
repeated continuous indices (such as $(\xi), (\zeta)$). 

The tensor calculus \inxx{tensor calculus,in ${\cal M}$-space}
in ${\cal M}$ \cite{42,43} is analogous to that in a finite-dimensional 
space. The differential of coordinates ${\dd} X^{\mu} (\xi)
\equiv {\dd} X^{\mu(\xi)}$ is a vector in ${\cal M}$. The coordinates
$X^{\mu(\xi)}$ can be transformed into new coordinates
${X '}^{\mu(\xi)}$ which are functionals of $X^{\mu(\xi)}$ :
\be
       {X'}^{\mu (\xi)} = F^{\mu (\xi)} [X] .
\label{A5}
\ee
The transformation (\ref{A5}) is very important. It says that if
functions $X^{\mu} (\xi)$ represent a membrane ${\cal V}_n$ then
any other functions $X'^{\mu} (\xi)$ obtained from $X^{\mu} (\xi)$
\inxx{functional transformations} \inxx{reparametrizations}
by a functional transformation also represent the same membrane
${\cal V}_n$. In particular, under a reparametrization of $\xi^a$
the functions $X^{\mu} (\xi)$ change into new functions; a
reparametrization thus manifests itself as a special functional
transformation which belongs to a subclass of the general
functional transformations (\ref{A5}).

Under a general coordinate transformation (\ref{A5}) a generic
vector $A^{\mu(\xi)} \equiv A^{\mu} (\xi)$ transforms as\footnote{
A similar formalism, but for a specific type of the functional
transformations (\ref{A5}), namely the reparametrizations which
functionally depend on string coordinates, was developed by
Bardakci \cite{44} }
\be
      A^{\mu(\xi)} = {{\p {X'}^{\mu(\xi)}} \over {\p X^{\nu(\zeta)}}}
     A^{\nu (\zeta)} \equiv \int {\dd} \zeta 
     {{\delta {X'}^{\mu} (\xi)} \over {\delta X^{\nu} (\zeta)}} A^{\nu} (\zeta)
\label{A6}
\ee
where $\delta/\delta X^{\mu} (\xi)$ denotes \inxx{functional derivative}
the functional derivative (see Box 4.1).
Similar transformations hold for a covariant vector $A_{\mu(\xi)}$,
a tensor $B_{\mu(\xi)\nu(\zeta)}$, etc.. Indices are
lowered and raised, respectively, by ${\rho}_{\mu(\xi)\nu(\zeta)}$
and ${\rho}^{\mu(\xi)\nu(\zeta)}$, the latter being
the inverse metric tensor satisfying
\be
     {\rho}^{\mu(\xi) \alpha (\eta)} {\rho}_{\alpha (\eta) \nu (\zeta)} =
     {{\delta}^{\mu(\xi)}}_{\nu(\zeta)} .
\label{A8}
\ee
\setlength{\unitlength}{1cm}

\vs{8mm}
\nnnn \begin{picture}(12.5,15)

\put(0,0){\framebox(12.5,15){
\begin{minipage}[t]{11.5cm}

\centerline{\bf Box 4.1: Functional derivative}

\vs{5mm}

Let $X^{\mu} (\xi)$ be a function of $\xi \equiv \xi^a$. The functional
derivative of a functional $F[X^{\mu} (\xi)]$ is defined according to
\be
    {{\delta F}\oo {\delta X^{\nu} (\xi')}} = \lim_{\epsilon\to 0}
    {{F[X^{\mu} (\xi) + \epsilon \delta (\xi - \xi') {\delta^{\mu}}_{\nu}]
    - F[X^{\mu} (\xi)]}\oo \epsilon}
\lbl{fun1}
\ee

\noindent {\bf Examples}

\bear
1)& &  F = X^{\mu} (\xi) \nonumber \\
  & & {{\delta F}\oo {\delta X^{\nu} (\xi)}} = \delta (\xi - \xi')
        {\delta^{\mu}}_{\nu}\hs{5cm} \nonumber
\ear
\bear
2)& &  F = \p_a X^{\mu} (\xi) \nonumber \\
  & &  {{\delta F}\oo {\delta X^{\nu} (\xi)}} = \p_a \delta(\xi - \xi')
     {\delta^{\mu}}_{\nu}\hs{4.6cm} \nonumber
\ear     
\bear     
3)& &  F = \lambda (\xi) \, \p_a X^{\mu} (\xi)\nonumber \\
  & &  {{\delta F}\oo {\delta X^{\nu} (\xi)}}  = \lambda (\xi) \,
   {{\delta \p_a X^{\mu} (\xi)}\oo {\delta X^{\nu} (\xi')}} \nonumber \\
  & &  \left ( \mbox{\rm in general} \hs{2mm} 
  {\delta \oo {\delta X^{\nu} (\xi')}} (\lambda (\xi) F[X]) =
  \lambda (\xi) {{\delta F}\oo {\delta X^{\nu} (\xi)}} \right ) 
  \hs{.7cm} \nonumber
\ear  
\bear
4)& &  F = \p_a X^{\mu} (\xi) \p_b X_{\mu} (\xi) \nonumber \\
  & &  {{\delta F}\oo {\delta X^{\nu} (\xi)}} = \p_a \delta (\xi - \xi')
    \p_b X_{\nu} + \p_b \delta (\xi - \xi') \p_a X_{\nu} 
    \hs{1.2cm} \nonumber
\ear    

\end{minipage} }}
\end{picture}

\vs{8mm}

A suitable choice of the metric --- assuring the invariance of the line
\inxx{metric,of ${\cal M}$-space}
element (\ref{A4}) under the transformations (\ref{A2}) and (\ref{A5}) ---
is, for instance,
\be
       {\rho}_{\mu(\xi) \nu (\zeta)} = \sqrt{|f|} \, \alpha \, g_{\mu \nu}
       \delta (\xi - \zeta) ,
\label{A9}
\ee
where $f \equiv {\rm det} \, f_{ab}$ is the determinant of the
induced metric 
\be
    f_{ab} \equiv {\p}_a X^{\alpha} {\p}_b X^{\beta} \, 
g_{\alpha \beta}
\lbl{A10}
\ee
on the sheet $V_n$, $g_{\mu \nu}$ is the metric
tensor of the embedding space $V_N$, and $\alpha$ an arbitrary function
of $\xi^a$. 

With the metric (\ref{A9}) the line element (\ref{A4}) becomes
\be
    {\dd} {\ell}^2 = \int  {\dd} \xi \sqrt{|f|} \, \alpha \, 
    g_{\mu \nu} \, {\dd} X^{\mu(\xi)} {\dd} X^{\nu(\xi)} .
\label{A11}
\ee
Rewriting the abstract formulas back into
the usual notation, with explicit integration, we have
\begin{eqnarray}
    A^{\mu(\xi)} &=& A^{\mu} (\xi) ,  \\
    A_{\mu(\xi)} &=& \rho_{\mu (\xi) \nu (\zeta)} A^{\nu (\zeta)}\nonumber \\
      &=& \int {\dd} \zeta \,  {\rho}_{\mu \nu} (\xi,\zeta)
    A^{\nu} (\xi)  =
    \sqrt{|f|} \, \alpha \, g_{\mu \nu} A^{\nu} (\xi) .
\label{A12}
\end{eqnarray}
The inverse metric is
\be
    {\rho}^{\mu(\xi)\nu(\zeta)} = {1 \over {\alpha \sqrt{|f|}}} \, \,
    g^{\mu \nu} \delta (\xi - \zeta) .
\label{A13}
\ee
Indeed, from (\ref{A8}), (\ref{A9}) and (\ref{A13}) we obtain
\be
     {{\delta}^{\mu(\xi)}}_{\nu(\zeta)} = \int  {\dd} \eta \, g^{\mu 
     \sigma} g_{\nu \sigma}
     \, \delta (\xi - \eta) \delta (\zeta - \eta) =
     {{\delta}^{\mu}}_{\nu} \delta (\xi - \zeta) .
\label{A14}
\ee

The invariant volume element (measure) of our membrane space
${\cal M}$ is \cite{45} \inxx{invariant volume element,in membrane space}
\be
    {\cal D} X = {\left ( \mbox{\rm Det} \, {\rho}_{\mu \nu} (\xi,\zeta) 
    \right )}
    ^{1/2} \prod_{\xi,\mu} {\dd} X^{\mu} (\xi) .
\label{A15}
\ee
Here Det denotes a continuum determinant taken over $\xi ,\zeta$
as well as over $\mu, \nu$. In the case of the diagonal metric (\ref{A9})
we have
\be
      {\cal D} X = \prod_{\xi ,\mu} \left ( \sqrt{|f|} \, \alpha 
       \, |g|\right ) ^{1/2}
       {\dd} X^{\mu} (\xi)
\label{A16}
\ee
As can be done in a finite-dimensional space, we can now also 
define the covariant
derivative in ${\cal M}$. For a scalar functional $A[X (\xi)]$
\inxx{covariant derivative,in membrane space}
the covariant functional derivative coincides with the ordinary
functional derivative:
\be
    A_{;\mu(\xi)} = {{\delta A} \over {\delta X^{\mu} (\xi)}}
    \equiv A_{,\mu(\xi)} .
\label{A16a}
\ee
But in general a geometric object in ${\cal M}$ is a tensor of 
\inxx{tensors,in membrane space}
arbitrary rank, ${A^{{\mu}_1 (\xi_1) \mu_2 (\xi_2)...}}
_{\nu_1 (\zeta_1) \nu_2 (\zeta_2)...}$,
which is a functional of $X^{\mu} (\xi)$, and its covariant derivative
contains the affinity ${\Gamma}_{\nu(\zeta)\sigma(\eta)}^{\mu(\xi)}$
composed of the metric (\ref{A9}) \cite{42,43}. For instance, for a vector
we have \inxx{affinity,in ${\cal M}$-space}
\be 
    {A^{{\mu}(\xi)}}_{; \nu(\zeta)} = {A^{\mu(\xi)}}_{, \nu(\zeta)}
    + {\Gamma}_{\nu(\zeta)\sigma(\eta)}^{\mu(\xi)}
    A^{\sigma(\eta)}.
\label{A17}
\ee

Let the alternative notations for ordinary and covariant functional
\inxx{functional derivative,covariant}
derivative be ana\-logous to those used in a finite-dimensional space:
\be
     {{\delta } \over {\delta X^{\mu} (\xi)}} \equiv
     {{\p} \over {\p X^{\mu (\xi)}}} \equiv {\p}_{\mu(\xi)} 
\quad , \quad \quad {{\DD} \over {{\DD} X^{\mu} (\xi)}} \equiv
     {{\DD} \over {{\DD} X^{\mu (\xi)}}} \equiv {\DD}_{\mu(\xi)} .
\label{A18}
\ee

\section{Membrane dynamics}

In the previous section I have considered arbitrary deformable membranes as
{\it kinematically possible objects} \inxx{kinematically possible objects}
of a membrane theory. A membrane,
in general, is not static, but is assumed to move in an embedding space
$V_N$. The parameter of evolution (``time") will be denoted $\tau$. Kinematically
every continuous trajectory $X^{\mu} (\tau, \xi^a) \equiv X^{\mu(\xi)}
(\tau)$ is possible in principle. A particular dynamical theory then
selects which amongst those kinematically possible membranes and 
trajectories are also dynamically possible. In this section I am going to
describe the theory in which a dynamically possible trajectory
$X^{\mu(\xi)}(\tau)$ is a {\it geodesic} in the membrane space ${\cal M}$.
\inxx{membranes,kinematically possible} \inxx{membranes,dynamically possible}
\inxx{geodesic,in ${\cal M}$-space}

\upperandlowertrue
\subsection[\ \ Membrane theory as a free fall in ${\cal M}$-space]
{MEMBRANE THEORY AS A FREE FALL IN ${\cal M}$-SPACE}

Let $X^{\alpha (\xi)}$ be $\tau$-dependent {\it coordinates} of a point in
${\cal M}$-space and $\rho_{\alpha(\xi') \beta(\xi'')}$ an arbitrary fixed
metric in ${\cal M}$. From the point of view of a finite-dimensional
space $V_N$ the symbol$\; X^{\alpha (\xi)} \equiv X^{\alpha} (\xi)$
represents an $n$-dimensional membrane embedded in $V_N$. We assume that
every dynamically possible trajectory $X^{\alpha (\xi)} (\tau)$ satisfies
the variational principle given by the action
\be
     I[X^{\alpha (\xi)}] = \int {\dd} \tau' \left (\rho_{\alpha (\xi')
     \beta (\xi'')} {\dot X}^{\alpha (\xi')} {\dot X}^{\beta (\xi'')} 
     \right )^{1/2} .
\lbl{II1.1}
\ee
This is just the action for a {\it geodesic} in ${\cal M}$-space.

The equation of motion is obtained if we functionally differentiate
(\ref{II1.1}) with respect to $X^{\alpha (\xi)} (\tau)$:
    $${{\delta I}\oo {\delta X^{\mu (\xi)} (\tau)}} = \int {\dd} \tau' \,
    {1\oo{\mu^{1/2}}} \, \rho_{\alpha (\xi') \beta (\xi'')} 
    {\dot X}^{\alpha (\xi'')}
    \, {{\dd}\oo {{\dd} \tau'}} \delta (\tau - \tau') {\delta_{(\xi)}}^{(\xi')}
    $$
\be \quad     
    \qquad + \, {1\oo 2} \int {\dd} \tau' \, {1\oo {\mu^{1/2}}} \left (
    {{\delta} \oo {\delta
    X^{\mu (\xi)} (\tau)}} \, \rho_{\alpha (\xi') \beta (\xi'')} \right )
    {\dot X}^{\alpha (\xi'')} {\dot X}^{\beta (\xi'')} = 0 ,
\lbl{II1.2}
\ee
where
\be
          \mu \equiv  \rho_{\alpha (\xi') \beta (\xi'')}
          {\dot X}^{\alpha (\xi')} {\dot X}^{\beta (\xi'')}
\lbl{II1.3}
\ee
and
\be
     {\delta_{(\xi)}}^{(\xi')} \equiv \delta (\xi - \xi') .
\lbl{II1.3a}
\ee     
     The integration over $\tau$ in the first term of eq.\ (\ref{II1.2}) can be
easily performed and eq.\ (\ref{II1.2}) becomes
    $${{\delta I}\oo {\delta X^{\mu (\xi)} (\tau)}} =     
    - {{\dd}\oo {{\dd} \tau}}
    \left ( {{\rho_{\alpha (\xi') \mu (\xi)} {\dot X}^{\alpha (\xi')} } \oo
    {\mu^{1/2}} } \right ) {\hs{6.2cm}}$$
\be    
    \qquad \qquad \quad + \; {1\oo 2} \int {\dd} \tau' {1\oo {\mu^{1/2}}} \,
    \left ( {{ \delta } \oo {\delta X^{\mu (\xi)} (\tau) }}
     \rho_{\alpha (\xi') \beta (\xi'')} \right )
    {\dot X}^{\alpha (\xi')} {\dot X}^{\beta (\xi'')} = 0.
\lbl{II1.2a}
\ee
Some exercises with such a variation are performed in Box 4.2, where we use
the notation $\p_{\mu (\tau,\xi)} \equiv \delta/ \delta X^{\mu} (\tau,\xi)$.

If the expression for the metric $\rho_{\alpha (\xi') \beta (\xi'')}$ does not
contain the velocity ${\dot X}^{\mu}$, then eq.\ (\ref{II1.2a}) further
simplifies to
\be
     -{{\dd} \oo {{\dd} \tau}} \left ({\dot X}_{\mu (\xi)} \right ) + {1\oo 2}
     \p_{\mu (\xi)} \rho_{\alpha (\xi') \beta (\xi'')} {\dot X}^{\alpha (\xi')}
     {\dot X}^{\beta (\xi'')} = 0 .
\lbl{II1.2b}
\ee
This can be written also in the form
\be
     {{{\dd} {\dot X}^{\mu (\xi)} }\oo {{\dd} \tau}} + {\Gamma^{\mu (\xi)}}_{
     \alpha (\xi') \beta (\xi'')}
     {\dot X}^{\alpha (\xi')} {\dot X}^{\beta (\xi'')} = 0 ,
\lbl{II1.2c}
\ee
which is a straightforward generalization of the usual geodesic equation from a
finite-dimensional space to an infinite-dimensional ${\cal M}$-space.
\inxx{geodesic equation,in ${\cal M}$-space}

The metric $\rho_{\alpha (\xi') \beta (\xi'')}$ is arbitrary fixed background
metric of ${\cal M}$-space. Choice of the latter metric determines, from
the point of view of the embedding space $V_N$, a particular membrane
theory. But from the viewpoint

\setlength{\unitlength}{1cm}

\vs{8mm}

\nnnn \begin{picture}(12.5,19)

\put(0,0){\framebox(12.6,19){
\begin{minipage}[t]{12.3cm}

\centerline{\bf Box 4.2: Excercises with variations and functional derivatives}

$$1) \hs{.7cm} I[X^{\mu} (\tau)] = {1\oo 2} \int \dd \tau' \, {\dot X}^{\mu}
(\tau') {\dot X}^{\nu} (\tau') \eta_{\mu \nu} \hs{4cm}$$
\bear
{{\delta I}\oo {\delta X^{\alpha} (\tau)}} &=& \int \dd \tau' \, {\dot X}^{\mu}
(\tau') {{\delta {\dot X}^{\nu} (\tau')}\oo {\delta X^{\alpha} (\tau)}}
\eta_{\mu \nu} \nonumber \\
 &=& \int \dd \tau' \, {\dot X}_{\alpha} (\tau') \, 
{\dd \oo {\dd \tau'}} \delta (\tau - \tau') = - {\dd \oo {\dd \tau'}}
{\dot X}_{\alpha} \nonumber
\ear

\vs{2mm}

$$2) \hs{.6cm} I[X^{\mu} (\tau,\xi)] = {1\oo 2} \int \dd \tau' \dd \xi' \,
\sqrt{|f(\xi')|} \, {\dot X}^{\mu} (\tau',\xi') {\dot X}^{\nu} (\tau',\xi')
\eta_{\mu \nu} \hs{1cm}$$
$$\hs{2mm}{{\delta I}\oo {\delta X^{\alpha}(\tau,\xi)}} = 
{1\oo 2} \int \dd \tau'\,
\dd \xi' \, {{\delta \sqrt{|f(\tau',\xi')|}}\oo {\delta X^{\alpha}(\tau',\xi')}}
\, {\dot X}^2 (\tau',\xi')\hs{1.5cm}$$
$$ \hs{21mm}+ \, {1\oo 2} \int \dd \tau'\, \dd \xi' \, 
\sqrt{|f(\tau',\xi')|} 2 {\dot X}^{\mu} {{\delta {\dot X}^{\nu}}\oo {\delta
{\dot X}^{\alpha}(\tau,\xi)}} \eta_{\mu \nu}$$
$$= {1\oo 2} \int \dd \tau' \dd \xi' \, \sqrt{|f(\tau',\xi')|} \p'^a X_{\alpha}
\p'_a \delta (\xi - \xi') \delta (\tau - \tau') {\dot X}^2 (\tau',\xi')$$
$$ +
\int \dd \tau' \dd \xi' \, \sqrt{|f(\tau',\xi')|} {\dot X}_{\alpha} (\tau',\xi')
\, {\dd \oo {\dd \tau'}} \delta (\tau - \tau') \delta (\xi - \xi')\hs{4mm}$$
$$ \hs{1cm}=  - {1\oo 2} \p_a \left ( \sqrt{|f|} \p^a X_{\alpha} \, {\dot X}^2
\right ) - 
{\dd \oo {\dd \tau}} \left ( \sqrt{|f|} {\dot X}_{\alpha} \right )$$

\vs{3mm}

$$3) \hs{2mm} I = \int \dd \tau' \, (\rho_{\alpha (\xi') \beta (\xi'')}
{\dot X}^{\alpha (\xi')} {\dot X}^{\beta (\xi'')} + K)\hs{4cm}$$
$${{\delta I}\oo {\delta X^{\mu (\xi)} (\tau)}} = {1\oo 2} \int \dd \tau'
\Biggl[\p_{\mu (\tau,\xi)} \rho_{\alpha (\xi') \beta (\xi'')} 
{\dot X}^{\alpha (\xi')}
{\dot X}^{\beta (\xi'')}\hs{1.5cm}$$
$$ + 2 \rho_{\alpha (\xi') \beta (\xi'')}
{\dd \oo {\dd \tau'}} \delta (\tau - \tau') \delta (\xi - \xi') 
{\delta_{\mu}}^{\alpha} {\dot X}^{\beta (\xi'')} + \p_{\mu (\tau, \xi)} K
\Biggr]$$
$$ = - {\dd \oo {\dd \tau}} (\rho_{\mu (\xi) \beta (\xi'')} 
{\dot X}^{\beta (\xi'')}) + {1\oo 2} \int \dd \tau'
\Biggl[\p_{\mu (\tau,\xi)} \rho_{\alpha (\xi') \beta (\xi'')} 
{\dot X}^{\alpha (\xi')}
{\dot X}^{\beta (\xi'')} + \p_{\mu (\tau, \xi)} K \Biggr]$$
\rightline{({\it continued})}

\end{minipage} }}
\end{picture}

\nnnn \begin{picture}(12.5,19)

\put(0,0){\framebox(12.5,19){
\begin{minipage}[t]{11.5cm}

{\bf Box 4.2 ({\it continued}) }

$$a) \hs{7mm} \rho_{\alpha (\xi') \beta (\xi'')} = {{\kappa \sqrt{|f(\xi')|}}
\oo {\lambda (\xi')}} \delta (\xi' - \xi'') \eta_{\alpha \beta} \; , \hs{1cm}
K = \int \dd \xi \sqrt{|f|} \, \kappa \lambda$$
$${{\delta I}\oo {\delta X^{\mu (\xi)} (\tau)}} = - {\dd \oo {\dd \tau}} 
\left ( {{\kappa \sqrt{|f|}}\oo {\lambda }} {\dot X}_{\mu} \right ) -
{1\oo 2} \, \p_a \left ( {{\kappa \sqrt{|f|} \p^a X_{\mu} \, {\dot X}^2}\oo
\lambda} \right )$$
$$\hs{4cm} - {1\oo 2 } \, \p_a (\kappa \sqrt{|f|} \p^a X_{\mu} \, \lambda )$$
$${{\delta I}\oo {\lambda (\xi)}} = 0 \qquad \Rightarrow \qquad \lambda^2 =
{\dot X}^{\alpha} {\dot X}_{\alpha} \hs{2cm}$$
$$ \Rightarrow {\dd \oo {\dd \tau}} \left ( {{\kappa \sqrt{|f|}}\oo {\sqrt{
{\dot X}^2}}} \, {\dot X}_{\mu} \right ) + \p_a (\kappa \, \sqrt{|f|} \p^a
X_{\mu} \, \sqrt{{\dot X}^2} ) = 0$$

\vs{3mm}

$$b) \hs{8mm} \rho_{\alpha (\xi') \beta (\xi'')} = {{\kappa \sqrt{|f(\xi')|}}
\oo {\sqrt{{\dot X}^2 (\xi')}}} \, \delta (\xi' - \xi'') \eta_{\alpha \beta} \; , 
\hs{8mm}
K = \int \dd \xi \sqrt{|f|} \, \kappa \, \sqrt{{\dot X}^2}$$
$$\p_{\mu (\tau,\xi)} \rho_{\alpha (\xi') \beta (\xi'')} = \kappa \, 
{{\delta \sqrt{|f(\tau',\xi')|}}\oo {\delta X^{\mu}(\tau,\xi)}}\, 
{1\oo {\sqrt{{\dot X}^2} (\tau',\xi')}}  \eta_{\alpha \beta}
\delta(\xi' - \xi'')$$
$$ \hs{1cm} + \, \kappa \sqrt{|f(\tau',\xi')|}\,
{\delta \oo {\delta X^{\mu} (\tau,\xi)}} 
\left ({1\oo {\sqrt{{\dot X}^2} (\tau',\xi')}} \right ) \eta_{\alpha \beta}
\delta(\xi' - \xi'')$$
$$= \kappa \, \sqrt{|f(\tau',\xi')|} \, \p'^a X_{\mu} \, {{\p'_a \delta 
(\xi - \xi') \delta (\tau-\tau')}\oo {\sqrt{{\dot X}^2}}} \, \eta_{\alpha \beta}
\, \delta (\xi' - \xi'')$$
$$ \hs{10mm} - \, \kappa \, \sqrt{|f(\tau',\xi')|} \, {{{\dot X}_{\mu}
(\xi')}\oo {{\dot X}^2 (\xi'))^{3/2}}} \delta(\xi - \xi') \delta (\xi' - \xi'')
{\dd \oo {\dd \tau'}} \delta (\tau - \tau') \eta_{\alpha \beta}$$

\vs{2mm}

$$\int \dd \tau' \, {{\delta \rho_{\alpha (\xi') \beta (\xi'')}}\oo {\delta 
X^{\mu (\xi)} (\tau)}} \, {\dot X}^{\alpha (\xi')} {\dot X}^{\beta (\xi'')}
\hs{5cm}$$
$$ =
- \kappa \, \p_a (\sqrt{|f|} \p^a X_{\mu} \, \sqrt{{\dot X}^2} ) +
\kappa \, {\dd \oo {\dd \tau}} \left ( \sqrt{|f|} \, {{{\dot X}_{\mu}}\oo {\sqrt
{{\dot X}^2}}} \right )$$
$$\int \dd \tau' \, {{\delta K}\oo {\delta X^{\mu (\xi)} (\tau)}} =
- \kappa \, \p_a (\sqrt{|f|} \p^a X_{\mu} \, \sqrt{{\dot X}^2} ) -
\kappa {\dd \oo {\dd \tau}} \left ( \sqrt{|f|} \, {{{\dot X}_{\mu}}\oo {\sqrt
{{\dot X}^2}}} \right )$$

\end{minipage} }}
\end{picture}

\nnnn of ${\cal M}$-space there is just one membrane
theory in a background metric $\rho_{\alpha (\xi') \beta (\xi'')}$ which is
an arbitrary functional of $X^{\mu (\xi)} (\tau)$. 
\inxx{background metric,in ${\cal M}$-space}

Suppose now that the metric is given by the following expression:
\be
     \rho_{\alpha (\xi') \beta (\xi'')} = \kappa {{\sqrt{|f(\xi')|}} \oo
     {\sqrt{{\dot X}^2 (\xi')}} } \, \delta (\xi' - \xi'') \eta_{\alpha \beta}
      \; ,
\lbl{II1.4}
\ee
where ${\dot X}^2 (\xi') \equiv {\dot X}^{\mu} (\xi') {\dot X}_{\mu} (\xi')$,
and $\kappa$ is a constant. If we insert the latter expression into the
equation of geodesic (\ref{II1.2}) and take into account the

\nnnn prescriptions of
Boxes 4.1 and 4.2, we immediately obtain the following equations of motion:
\be
    {{\dd}\oo {{\dd} \tau}} \left ({1\oo {\mu^{1/2}}} {{\sqrt{|f|}}\oo
    {\sqrt{{\dot X}^2}}} \, {\dot X}_{\mu} \right ) + {1\oo {\mu^{1/2}}}
    \p_a \left ( \sqrt{|f|} \sqrt{{\dot X}^2} \p^a X_{\mu} \right ) = 0 .
\lbl{II1.5}
\ee
The latter equation can be written as
\be
    \mu^{1/2} {{\dd}\oo {{\dd} \tau}} \left ({1\oo {\mu^{1/2}}} \right ) \,
    {{\sqrt{|f|}}\oo {\sqrt{{\dot X}^2}}} \, {\dot X}_{\mu} +
    {{\dd} \oo {{\dd} \tau}} \left ( {{\sqrt{|f|}}\oo {\sqrt{{\dot X}^2}}} \, 
    {\dot X}_{\mu} \right ) +  
    \p_a \left ( \sqrt{|f|} \sqrt{{\dot X}^2} \p^a X_{\mu} \right ) = 0 .
\lbl{II1.6}
\ee
If we multiply this by ${\dot X}^{\mu}$, sum over $\mu$, and integrate over
$\xi$, we obtain
\begin{eqnarray}
      &&{1\oo 2} \int \dd \xi \, \sqrt{|f|} \sqrt{{\dot X}^2} \, \frac{1}{\mu}
       {{{\dd} \mu} \oo {{\dd} \tau}}  \nonumber \\
       && \hs{2cm} = \kappa \int {\dd} \xi \,
      \left [ {{\dd}\oo {{\dd} \tau}} \left ( 
      {{\sqrt{|f|}}\oo {\sqrt{{\dot X}^2}}} \, {\dot X}_{\mu} \right ) 
      {\dot X}^{\mu} +
      \p_a (\sqrt{|f|} \p^a X_{\mu} \sqrt{{\dot X}^2} )  {\dot X}^{\mu}
      \right ] \nonumber \\
      &&\hs{2cm} = \kappa \int {\dd} \xi \,
      \left [ {{\dd}\oo {{\dd} \tau}} \left ( 
      {{\sqrt{|f|}}\oo {\sqrt{{\dot X}^2}}} \, {\dot X}_{\mu} \right ) 
      {\dot X}^{\mu} - \sqrt{|f|} \sqrt{{\dot X}^2} \p^a X_{\mu} \p_a
      {\dot X}^{\mu} \right ] \nonumber \\
       &&\hs{2cm} = \kappa \int {\dd} \xi \, \Biggl[ \sqrt{{\dot X}^2} \,
      {{{\dd} \sqrt{|f|}}\oo {{\dd} \tau}} + \sqrt{|f|} \, 
      {{\dd}\oo {{\dd}\tau}} 
      \left ( {{{\dot X}_{\mu}} \oo {\sqrt{{\dot X}^2}}} \right ) 
      \dot X^{\mu} \nonumber \\
      & & \hs{5.5cm}- \sqrt{{\dot X}^2} \, {{\dd}\oo {{\dd} \tau}} \sqrt{|f|}\Biggr]
      \nonumber \\ 
        &&\hs{2cm}= 0   .
\lbl{II1.7}
\ear
In the above calculation we have used the relations
\be
    {{{\dd} \sqrt{|f|}}\oo {{\dd} \tau}} = {{\p \sqrt{|f|} } \oo {\p f_{ab}}}\,
    {\dot f}_{ab} = \sqrt{|f|} \, f^{ab} \p_a {\dot X}^{\mu} \p_b X_{\mu} =
    \sqrt{|f|} \, \p^a X_{\mu} \p_a {\dot X}^{\mu}
\lbl{II1.8}
\ee
and
\be
     {{{\dot X}_{\mu}} \oo {\sqrt{{\dot X}^2}}}             
    {{{\dot X}^{\mu}} \oo {\sqrt{{\dot X}^2}}} = 1 \quad \Rightarrow \quad
    {{\dd}\oo {{\dd} \tau}} \left ( {{{\dot X}_{\mu}} \oo {\sqrt{{\dot X}^2}}}
    \right ) {\dot X}^{\mu} = 0 .
\lbl{II1.9}
\ee
Since eq.\,(\ref{II1.7}) holds for arbitrary $\sqrt{|f|} \sqrt{{\dot X}^2} >0$,
assuming $\mu>0$, it follows that $\dd \mu/\dd \tau = 0$.

We have thus seen that the equations of motion (\ref{II1.5}) automatically imply
\be
    {{{\dd} \mu} \oo {{\dd} \tau}} = 0  \qquad {\rm or} \qquad
    {{{\dd} \sqrt{\mu}} \oo {{\dd} \tau}} = 0 \; , \; \; \mu \ne 0 .
\lbl{II1.10}
\ee
Therefore, instead of (\ref{II1.5}) we can write
\be
    {{\dd} \oo {{\dd} \tau}} \left ( {{\sqrt{|f|}}\oo {\sqrt{{\dot X}^2}}} \, 
    {\dot X}_{\mu} \right ) +  
    \p_a \left ( \sqrt{|f|} \sqrt{{\dot X}^2} \p^a X_{\mu} \right ) = 0 .
\lbl{II1.11}
\ee   
This is precisely the equation of motion of the Dirac-Nambu-Goto membrane of
\inxx{Dirac--Nambu--Goto membrane}
arbitrary dimension. The latter objects are nowadays known as
{\it $p$-branes}, \inxx{$p$-brane}
and they include point particles (0-branes) and
strings (1-branes). It is very interesting that the conventional theory of
$p$-branes is just a particular case ---with the metric (\ref{II1.4})---
of the membrane dynamics given by the action (\ref{II1.1}).

The action (\ref{II1.1}) is by definition invariant under reparametrizations
of $\xi^a$. In general, it is not invariant under reparametrization of
the evolution parameter $\tau$. If the expression for the metric
\inxx{evolution parameter}
$\rho_{\alpha (\xi') \beta(\xi'')}$ does not contain the velocity
${\dot X}^{\mu}$ then the invariance of (\ref{II1.1}) under reparametrizations
of $\tau$ is obvious. On the contrary, if $\rho_{\alpha (\xi') \beta(\xi'')}$
contains ${\dot X}^{\mu}$ then the action (\ref{II1.1}) is not
invariant under reparametrizations of $\tau$. For instance, if
$\rho_{\alpha (\xi') \beta(\xi'')}$ is given by eq.\ (\ref{II1.4}), then,
as we have seen, the equation of motion automatically contains the
relation
\be
     {{\dd}\oo{{\dd} \tau}} \left ( {\dot X}^{\mu (\xi)} {\dot X}_{\mu (\xi)}
     \right )
     \equiv {{\dd}\oo{{\dd} \tau}} \int {\dd} \xi \, \kappa \sqrt{|f|}
     \sqrt{{\dot X}^2} = 0 .
\lbl{II1.12}
\ee
The latter relation is nothing but {\it a gauge fixing relation},
where by ``gauge" we mean here a choice of parameter $\tau$. The action
(\ref{II1.1}), which in the case of the metric (\ref{II1.4}) is not
reparametrization invariant, contains the gauge fixing term. The latter
term is not added separately to the action, but is implicit by the exponent
${1 \oo 2}$ of the expression ${\dot X}^{\mu (\xi)} {\dot X}_{\mu (\xi)}$.

In general the exponent in the Lagrangian is not necessarily
${1\oo 2}$, but can be arbitrary:
\be
     I[X^{\alpha (\xi)}] = \int {\dd} \tau \, \left (
     \rho_{\alpha (\xi') \beta(\xi'')} {\dot X}^{\alpha (\xi')}
     {\dot X}^{\beta (\xi'')} \right )^a .
\lbl{II1.13}
\ee
For the metric (\ref{II1.4}) the corresponding equation of motion is
\be
    {{\dd}\oo{{\dd} \tau}} \left ( a \mu^{a-1} {{\kappa \sqrt{|f|}}\oo
    {\sqrt{{\dot X}^2}}} \, {\dot X}_{\mu} \right ) + a \mu^{a - 1} \p_a
    \left ( \kappa \sqrt{|f|} \sqrt{{\dot X}^2} \p^a X_{\mu} \right ) = 0 .
\lbl{II1.14}
\ee
For any $a$ which is different from 1 we obtain a gauge fixing relation
\inxx{gauge fixing relation}
which is equivalent to (\ref{II1.10}), and the same equation of motion
(\ref{II1.11}). When $a = 1$ we obtain directly the equation of motion
(\ref{II1.11}), and no gauge fixing relation (\ref{II1.10}). For $a = 1$ and
the metric (\ref{II1.4}) the action (\ref{II1.13}) is invariant under
reparametrizations of $\tau$.

We shall now focus our attention to the action
\be
    I[X^{\alpha(\xi)}] = \int {\dd} \tau \, \rho_{\alpha (\xi') \beta (\xi'')}
    {\dot X}^{\alpha (\xi')} {\dot X}^{\beta (\xi')} = \int {\dd} \tau \,
    {\dd} \xi \, \kappa \sqrt{|f|} \sqrt{{\dot X}^2}
\lbl{II1.14a}
\ee
with the metric (\ref{II1.4}). It is invariant under the transformations
\be
     \tau \rightarrow \tau' = \tau' (\tau) ,
\lbl{II1.15a}
\ee
\be
      \xi^a \rightarrow \xi'^a = \xi'^a (\xi^a)
\lbl{II1.15b}
\ee
in which $\tau$ and $\xi^a$ do not mix.

Invariance of the action (\ref{II1.14a}) under reparametrizations (\ref{II1.15a})
of the evolution parameter $\tau$ implies the existence of a constraint
among the canonical momenta $p_{\mu (\xi)}$ and coordinates $X^{\mu (\xi)}$.
Momenta are given by \inxx{canonical momentum}
\bear
    p_{\mu (\xi)} &=& {{\p L}\oo {\p {\dot X}^{\mu (\xi)}}} = 2 \rho_{\mu (\xi)
    \nu (\xi')} {\dot X}^{\nu (\xi')} + 
    {{\p \rho_{\alpha (\xi') \beta (\xi'')}} \oo
    {\p {\dot X}^{\mu (\xi)}}} {\dot X}^{\alpha (\xi')} {\dot X}^{\beta 
    (\xi'')} \nonumber \\    
     &=& {{\kappa \sqrt{|f|}}\oo {\sqrt{{\dot X}^2}}} \, {\dot X}_{\mu} .
\lbl{II1.15c}
\ear
By distinsguishing covariant and contravariant components one finds
\be
    p_{\mu (\xi)} = {\dot X}_{\mu (\xi)} \; , \quad 
    p^{\mu (\xi)} = {\dot X}^{\mu (\xi)}  .
\lbl{II1.15d}
\ee
We define
\be
   p_{\mu (\xi)} \equiv p_{\mu} (\xi) \equiv p_{\mu} \; ,
   \quad {\dot X}^{\mu (\xi)} \equiv {\dot X}^{\mu} (\xi) \equiv {\dot X}^{\mu}.
\lbl{II1.15e}
\ee
Here $p_{\mu}$ and ${\dot X}^{\mu}$ have the meaning of the usual finite
dimensional vectors whose components are lowered  raised by
the finite-dimensional metric tensor $g_{\mu \nu}$ and its inverse
$g^{\mu \nu}$:
\be
       p^{\mu} = g^{\mu \nu} p_{\nu} \; , \quad {\dot X}_{\mu} = g_{\mu \nu}
       {\dot X}^{\nu}
\lbl{II1.15ee}
\ee
Eq.(\ref{II1.15c}) implies
\be
      p^{\mu} p_{\mu} - \kappa^2 |f| = 0
\lbl{II1.15f}
\ee
which is satisfied at every $\xi^a$.

Multiplying (\ref{II1.15f}) by $\sqrt{{\dot X}^2}/(\kappa \sqrt{|f|}\, )$ and
integrating over $\xi$ we have
\be
   {1\oo 2} \int \dd \xi \, {{\sqrt{{\dot X}^2}}\oo {\kappa \sqrt{|f|} }} \,
    (p^{\mu} p_{\mu} - \kappa^2 |f|) =
    p_{\mu (\xi)} {\dot X}^{\mu (\xi)} - L = H = 0
\lbl{II1.15g}
\ee
where $L = \int {\dd} \xi \, \kappa \sqrt{|f|} \sqrt{{\dot X}^2}$.

We see that the Hamiltonian \inxx{Hamiltonian}
belonging to our action (\ref{II1.14a}) is
identically zero. This is a well known consequence of the reparametrization
invariance (\ref{II1.15a}). The relation (\ref{II1.15f}) is a constraint at
$\xi^a$ and the Hamiltonian (\ref{II1.15g}) is a linear superposition of the
constraints at all possible $\xi^a$.

An action which is equivalent to (\ref{II1.14a}) is
\be
    I[X^{\mu (\xi)}, \lambda] = {1\oo 2} \int {\dd} \tau 
    \dd \xi \, \kappa \sqrt{|f|} \left ( {{{\dot X}^{\mu} {\dot X}_{\mu}} \oo
    {\lambda}} + \lambda \right ) ,
\lbl{II1.16}
\ee
where $\lambda$ is a Lagrange multiplier.

In the compact notation of ${\cal M}$-space eq.\ (\ref{II1.16}) reads
\be
     I[X^{\mu (\xi)}, \lambda] = {1\oo 2} \int \dd \tau \left (\rho_{\alpha
     (\xi') \beta (\xi'')} {\dot X}^{\alpha (\xi')} {\dot X}^{\beta (\xi'')}
     + K \right ) ,
\lbl{II1.18}
\ee         
where
\be
     K = K[X^{\mu (\xi)},\lambda] = \int {\dd} \xi \kappa \sqrt{|f|} \lambda 
\lbl{II1.19}
\ee
and
\be
     \rho_{\alpha (\xi') \beta (\xi'')} =  \rho_{\alpha (\xi') \beta (\xi'')}
     [X^{\mu (\xi)},\lambda] = {{\kappa \sqrt{|f(\xi')|} }\oo {\lambda (\xi')}}
     \delta (\xi' - \xi'') \eta_{\alpha \beta} .
\lbl{II1.20}
\ee
Variation of (\ref{II1.18}) with respect to $X^{\mu (\xi)} (\tau)$ and
$\lambda$ gives
\begin{eqnarray}       
       {{\delta I}\oo {\delta X^{\mu (\xi)} (\tau)}} &=&
       - {\dd \oo {\dd \tau}}
       \left ({{\kappa \sqrt{|f|}}\oo
       {\lambda}} \, {\dot X}_{\mu} \right ) \nonumber \\       
       & & - {1\oo 2} \p_a \left (\kappa
       \sqrt{|f|} \p^a X_{\mu} \left ({{\sqrt{{\dot X}^2}}\oo {\lambda}} 
       + \lambda \,  \right ) \right ) = 0 \lbl{II1.21} , \\
    {{\delta I}\oo {\delta \lambda (\tau, \xi)}} &=& - {{{\dot X}^{\mu} 
    {\dot X}_{\mu}}\oo {\lambda^2}} + 1 = 0 .
\lbl{II1.22}
\end{eqnarray}
The system of equations (\ref{II1.21}), (\ref{II1.22}) is equivalent to
(\ref{II1.11}). This is in agreement with the property that after inserting the
$\lambda$ ``equation of motion" (\ref{II1.22}) into the action (\ref{II1.16})
one obtains the action (\ref{II1.14a}) which directly leads to the equation
of motion (\ref{II1.11}).

The invariance of the action (\ref{II1.16}) under reparametrizations 
(\ref{II1.15a}) of the evolution parameter $\tau$ is assured if $\lambda$
transforms according to
\be
    \lambda \rightarrow \lambda' = {{\dd \tau'}\oo {\dd \tau}} \, \lambda .
\lbl{II1.23}
\ee
This is in agreement with the relations (\ref{II1.22}) which says that 
$$\lambda = ({\dot X}^{\mu} {\dot X}_{\mu})^{1/2} .$$
\setlength{\unitlength}{1cm}

\nnnn \begin{picture}(12.5,19)

\put(0,0){\framebox(12.5,19){
\begin{minipage}[t]{11.5cm}

\centerline{\bf Box 4.3: Conservation of the constraint}
\inxx{conservation,of the constraint}
\vs{2mm}

\hs{4mm} Since the Hamiltonian $H = \int {\dd} \xi \lambda {\cal H}$
in eq.\ (\ref{II1.26}) is zero for any $\lambda$, it follows that
the Hamiltonian density
\be
       {\cal H} [X^{\mu},p_{\mu}]
        = {1\oo {2 \kappa}} \left ( {{p_{\mu} p^{\mu}}\oo
       {\sqrt{|f|}}} - \kappa^2 \sqrt{|f|} \right )
\lbl{BII3.1}
\ee
vanishes for any $\xi^a$. The requirement that the constraint
(\ref{II3.1}) is conserved in $\tau$ can be written as
\be
     {\dot {\cal H}} = \lbrace {\cal H}, H \rbrace = 0 ,
\lbl{BII3.2}
\ee
which is satisfied if
\be
   \lbrace {\cal H} (\xi), {\cal H} (\xi') \rbrace = 0 .
\lbl{BII3.3}
\ee

\hs{4mm} That the Poisson bracket (\ref{BII3.3}) indeed vanishes can be found
\inxx{Poisson brackets} \inxx{Hamilton--Jacobi functional}
as follows. Let us work in the language of the Hamilton--Jacobi functional
$S[X^{\mu} (\xi)]$, in which one considers the momentum vector field
$p_{\mu (\xi)}$ to be a function of position $X^{\mu (\xi)}$ in
${\cal M}$-space, i.e., a functional of $X^{\mu} (\xi)$ given by
\be
    p_{\mu (\xi)} = p_{\mu (\xi)} (X^{\mu (\xi)} ) \equiv p_{\mu}
    [X^{\mu} (\xi)] = {{\delta S} \oo {\delta X^{\mu} (\xi)}} .
\lbl{BII3.4}
\ee
Therefore ${\cal H} [X^{\mu} (\xi), p_{\mu} (\xi)]$ is a functional of
$X^{\mu} (\xi)$. Since ${\cal H} = 0$, it follows that its functional
derivative also vanishes:
     $${{{\dd} {\cal H}}\oo {\dd X^{\mu (\xi)}}} = {{\p {\cal H}}\oo
     {\p X^{\mu (\xi)}}} + {{\p {\cal H}}\oo {\p p_{\nu (\xi')}}} 
     {{\p p_{\nu (\xi')}}\oo {\p X^{\mu (\xi)}}} \hs{2cm}$$
\be     
      \hs{3.2cm} \equiv
     {{\delta {\cal H}}\oo {\delta X^{\mu} (\xi)}} + \int \dd \xi' \,
     {{\delta {\cal H}}\oo {\delta p_{\nu} (\xi')}} {\delta p_{\nu} (\xi')
     \oo {\delta X^{\mu} (\xi)}} = 0 .
\lbl{BII3.5}
\ee
Using (\ref{BII3.5}) and (\ref{BII3.4}) we have
       $$\lbrace {\cal H} (\xi), {\cal H} (\xi') \rbrace = \int {\dd} \xi'' \,
       \left ( {{\delta {\cal H} (\xi)}\oo {\delta X^{\mu} (\xi'')}}
       {{\delta {\cal H} (\xi')}\oo {\delta p_{\mu} (\xi'')}} -     
      {{\delta {\cal H} (\xi')}\oo {\delta X^{\mu} (\xi'')}}
       {{\delta {\cal H} (\xi)}\oo {\delta p_{\mu} (\xi'')}} \right )$$
\be
           \qquad = - \int {\dd} \xi'' \, {\dd} \xi''' \, {{\delta {\cal H}
           (\xi)} \oo {\delta p_{\nu} (\xi''')}} {{\delta {\cal H} (\xi')} \oo
           {\delta p_{\mu} (\xi'')}} \left ( 
           {{\delta p_{\nu} (\xi''')}\oo {\delta X^{\mu} (\xi'')}} - 
           {{\delta p_{\mu} (\xi''')}\oo {\delta X^{\mu} (\xi'')}}
           \right ) = 0 .
\lbl{BII3.6}
\ee
\rightline{({\it continued})}

\end{minipage} }}
\end{picture}

\nnnn \begin{picture}(12.5,14)
\put(0,0){\framebox(12.5,14){
\begin{minipage}[t]{11.5cm}

{\bf Box 4.3 ({\it continued}) }

\vs{3mm}

Conservation of the constraint (\ref{BII3.1}) is thus shown to be
automatically sastisfied. \inxx{conservation,of the constraint}

\hs{4mm} On the other hand, we can calculate the Poisson bracket (\ref{BII3.3}) by
using the explicit expression (\ref{BII3.1}). So we obtain
    $$\lbrace {\cal H} (\xi), {\cal H} (\xi') \rbrace = \hs{5cm} $$
    $$ - {\sqrt{|f(\xi)|}\oo
    {\sqrt{|f(\xi')|}}} \, \p_a \delta (\xi - \xi') p_{\mu} (\xi') \p^a
    X^{\mu} (\xi) + {\sqrt{|f(\xi')|}\oo {\sqrt{|f(\xi)|}}}
    \p'_a \delta (\xi - \xi') p_{\mu} (\xi) \p'^a X^{\mu} (\xi')$$
\be
       \qquad = - \left ( p_{\mu} (\xi) \p^a X^{\mu} (\xi) +     
       p_{\mu} (\xi') \p'^a X^{\mu} (\xi') \right ) \p_a \delta (\xi - \xi')
       = 0 ,
\lbl{BII3.7}
\ee
where we have used the relation
    $$F(\xi') \p_a \delta (\xi - \xi') = \p_a \left [ F(\xi') \delta 
    (\xi - \xi') \right ] = \p_a \left [ F(\xi) \delta 
    (\xi - \xi') \right ]$$
\be
       \qquad \; \; = F(\xi) \p_a \delta (\xi - \xi') + \p_a F(\xi) \delta
       (\xi - \xi') .
\lbl{BII3.8}
\ee
Multiplying (\ref{BII3.7}) by an arbitrary ``test" function $\phi (\xi')$
and integrating over $\xi'$ we obtain
\be
       2 p_{\mu} \p^a X^{\mu} \p_a \phi  + \p_a (p_{\mu} \p^a X^{\mu})
       \phi = 0 .
\lbl{BII3.9}
\ee
Since $\phi$ and $\p_a \phi$ can be taken as independent at any point $\xi^a$,
it follows that
\be
      p_{\mu} \p_a X^{\mu} = 0 .
\lbl{BII3.10}
\ee
The ``momentum" constraints (\ref{BII3.10}) are thus shown to be automatically
satisfied as a consequence of the conservation of the ``Hamiltonian" constraint
(\ref{BII3.1}). This procedure was been discovered in ref.\ \cite{46}. Here I
have only adjusted it to the case of membrane theory. \inxx{`momentum' constraints}
\inxx{`Hamiltonian' constraint}

\end{minipage} }}
\end{picture}

\hs{8mm}

If we calculate the Hamiltonian belonging to (\ref{II1.18}) we find
\be
    H = (p_{\mu (\xi)} {\dot X}^{\mu (\xi)} - L) = \mbox{$1\oo 2$}
    (p_{\mu (\xi)} p^{\mu (\xi)} - K) \equiv 0 ,
\lbl{II1.24}
\ee
where the canonical momentum is
\be
     p_{\mu (\xi)} = {{\p L}\oo {\p {\dot X}^{\mu (\xi)}}} = {{\kappa \sqrt{
     |f|}}\oo {\lambda }} {\dot X}_{\mu} \; .
\lbl{II1.25}
\ee
Explicitly (\ref{II1.24}) reads
\be
     H = {1\oo 2} \int {\dd} \xi \, {\lambda \oo {\kappa \sqrt{|f|}}}
     (p^{\mu} p_{\mu} - \kappa^2 |f|) \equiv 0 .
\lbl{II1.26} 
\ee

The Lagrange multiplier $\lambda$ is arbitrary. The choice of $\lambda$
determines the choice of parameter $\tau$. Therefore (\ref{II1.26})
holds for every $\lambda$, which can only be satisfied if we have
\be
       p^{\mu} p_{\mu} - \kappa^2 |f| = 0
\lbl{II1.27}
\ee
at every point $\xi^a$ on the membrane. Eq.\ (\ref{II1.27}) is a constraint
at $\xi^a$, and altogether there are infinitely many constraints.

In Box 4.3 it is shown that the constraint (\ref{II1.27}) is conserved in
$\tau$ and that as a consequence we have
\be
     p_{\mu} \p_a X^{\mu} = 0 .
\lbl{II1.28}
\ee
The latter equation is yet are another set of constraints\footnote{
Something similar happens in canonical gravity. Moncrief and Teitelboim \ci{46}
have
shown that if one imposes the Hamiltonian constraint on the Hamilton
functional then the momentum constraints are automatically satisfied.}
which are satisfied at any point $\xi^a$ of
the membrane manifold $V_n$

\paragraph{First order form of the action} Having the constraints 
(\ref{II1.27}), (\ref{II1.28}) one can easily write the first 
order, or phase space action, \inxx{action,first order form}
\be
    I[X^{\mu},p_{\mu},\lambda,\lambda^a] = \int {\dd} \tau \, {\dd} \xi \,
    \left ( p_{\mu} {\dot X}^{\mu} - {\lambda \oo {2 \kappa \sqrt{|f|}}}
    (p^{\mu} p_{\mu} - \kappa^2 |f|) - \lambda^a p_{\mu} \p_a X^{\mu}
    \right ) ,
\lbl{II1.29}
\ee
where $\lambda$ and $\lambda^a$ are Lagrange multipliers.

The equations of motion are
\begin{eqnarray}
    \delta X^{\mu}  &:& \quad {\dot p}_{\mu} + \p_a \left ( \kappa \lambda
    \sqrt{|f|} \p^a X_{\mu} - \lambda^a p_{\mu} \right ) = 0 ,\lbl{II1.30} \\
    \nonumber \\
    \delta p_{\mu}  &:& \quad {\dot X}^{\mu} - {\lambda \oo 
    {\kappa \sqrt{|f|}}}
    \,p_{\mu} - \lambda^a \p_a X^{\mu} = 0, \lbl{II1.31} \\
    \nonumber \\
    \delta \lambda  &:& \quad p^{\mu} p_{\mu} - \kappa^2 |f| = 0 ,
    \lbl{II1.32} \\
    \nonumber \\
    \delta \lambda^a  &:& \quad p_{\mu} \p_a X^{\mu} = 0.
    \lbl{II1.33}
\end{eqnarray}
Eqs.\ (\ref{II1.31})--(\ref{II1.33}) can be cast into the following form:
\bear
    p_{\mu} &=& {{\kappa \sqrt{|f|}}\oo {\lambda}} ({\dot X}_{\mu} -
    \lambda^a \p_a X^{\mu}),
\lbl{II1.34} \\
\nonumber \\
    \lambda^2 &=& ({\dot X}^{\mu} - \lambda^a \p_a X^{\mu})
    ({\dot X}_{\mu} - \lambda^b \p_b X_{\mu})
\lbl{II1.35}\\
\nonumber \\
      \lambda_a &=& {\dot X}^{\mu} \p_a X_{\mu} .
\lbl{II1.36}
\ear

Inserting the last three equations into the phase space action (\ref{II1.29})
we have
\be
     I[X^{\mu}] = \kappa \int {\dd} \tau \, {\dd} \xi \sqrt{|f|} \,
     \left [{\dot X}^{\mu} {\dot X}^{\nu} (\eta_{\mu \nu} - \p^a X_{\mu}
     \p_a X_{\nu}) \right ]^{1/2} .
\lbl{II1.37}
\ee
The vector ${\dot X}(\eta_{\mu \nu} - \p^a X_{\mu} \p_a X_{\nu})$ is normal
to the membrane $V_n$; its scalar product with tangent vectors $\p_a
X^{\mu}$ is identically zero. The form ${\dot X}^{\mu} {\dot X}^{\nu} 
(\eta_{\mu \nu} - \p^a X_{\mu} \p_a X_{\nu})$ can be considered as a
1-dimensional metric, equal to its determinant, on a line which is orthogonal
to $V_n$. The product
\be
        f{\dot X}^{\mu} {\dot X}^{\nu} (\eta_{\mu \nu} - \p^a X_{\mu}
     \p_a X_{\nu}) = {\rm det} \, \p_A X^{\mu} \p_B X_{\mu}
\lbl{II1.38}
\ee
is equal to the determinant of the induced metric $\p_A X^{\mu} \p_B
X_{\mu}$ on the $(n+1)$-dimensional surface $X^{\mu} (\phi^A)$, $\phi^A =
(\tau, \xi^a)$, swept by our membrane $V_n$. The action (\ref{II1.37}) is then
{\it the minimal surface action} for the $(n+1)$-dimensional worldsheet
$V_{n+1}$: \inxx{world sheet}
\be
      I[X^{\mu}] = \kappa \int {\dd}^{n+1} \phi \, ({\rm det} \, 
      \p_A X^{\mu} \p_B X_{\mu})^{1/2} .
\lbl{II1.39}
\ee
This is the conventional Dirac--Nambu--Goto action, and (\ref{II1.29}) is
one of its equivalent forms. \inxx{Dirac--Nambu--Goto action}

We have shown that from the point of view of ${\cal M}$-space a
membrane of any dimension is just a point moving along a geodesic
in ${\cal M}$. The metric of ${\cal M}$-space is taken to be an
arbitrary fixed background metric. For a special choice of the
metric we obtain the conventional $p$-brane theory. The latter theory
is thus shown to be a particular case of the more general theory,
based on the concept of ${\cal M}$-space.

Another form of the action is obtained if in (\ref{II1.29}) we use
the replacement
\be
     p_{\mu} = {{\kappa \sqrt{|f|}}\oo {\lambda}} \, ({\dot X}_{\mu} -
     \lambda^a \p_a X_{\mu})
\lbl{II1.39a}
\ee
which follows from ``the equation of motion" (\ref{II1.31}). Then instead
of (\ref{II1.29}) we obtain the action
\be
    I[X^{\mu}, \lambda, \lambda^a] = {\kappa\oo 2} \int {\dd} \tau \, {\dd}^n
    \xi \sqrt{|f|} \, \left ( {{({\dot X}^{\mu} - \lambda^a \p_a X^{\mu})    
    ({\dot X}_{\mu} - \lambda^b \p_b X_{\mu})} \oo {\lambda}} + \lambda
    \right ) .
\lbl{II1.39b}
\ee
If we choose a gauge such that $\lambda^a = 0$, then (\ref{II1.39b})
coincides with the action (\ref{II1.16}) considered before.

\paragraph{The analogy with the point particle} The action (\ref{II1.39b}),
and espec\-ially (\ref{II1.16}), looks like the well known Howe--Tucker 
action \ci{21} for \inxx{Howe--Tucker action}
a point particle, apart from the integration over coordinates $\xi^a$
of a space-like hypersurface $\Sigma$ on the worldsheet $V_{n+1}$. Indeed,
a worldsheet can be considered as a continuum collection or a
bundle of worldlines $X^{\mu} (\tau, \xi^a)$, and (\ref{II1.39b}) is an
action for such a bundle. Individual worldlines are distinguished by the
values of parameters $\xi^a$.

We have found a very interesting inter-relationship between various concepts:
\begin{description}
    \item{1)} membrane as a ``point particle" moving along a geodesic in an
    infinite-dimensional membrane space ${\cal M}$;
    
    \item{2)} worldsheet swept by a membrane as a minimal surface in a
    finite-dimensional embedding space $V_N$;
    
    \item{3)} worldsheet as a bundle of worldlines swept by point particles
    moving in $V_N$.
\end{description}

\upperandlowerfalse
\subsection[\ \ Membrane theory as a minimal surface in an embedding space]
{Membrane theory as a minimal surface in an embedding space}

In the previous section we have considered a membrane as a point in an
infinite-dimensional membrane space ${\cal M}$. Now let us change
our point of view and consider a membrane as a surface in a 
finite-dimensional embedding space $V_D$ When moving, a $p$-dimensional 
membrane sweeps a $(d = p+1)$-dimensional surface which I shall call
a {\it worldsheet}\footnote{
In the literature on $p$-branes such a surface is often called
"world volume" and sometimes world surface.}.
What is an action which determines the membrane dynamics, i.e.,
a possible worldsheet? Again the analogy with the point particle
provides a clue. Since a point particle sweeps a worldline whose action is
{\it the minimal length action}, it is natural to postulate that a
\inxx{action,minimal surface}
membrane's worldsheet satisfies {\it the minimal surface action}:
\be
     I[X^{\mu}] = \kappa \int {\dd}^d \phi \, ({\rm det} \, \p_A
     X^{\mu} \p_B X_{\mu})^{1/2} .
\lbl{II1.40}
\ee
This action, called also the Dirac--Nambu--Goto action, is invariant under
\inxx{Dirac--Nambu--Goto action}
reparametrizations of the worldsheet coordinates $\phi^A$, $A = 0,1,2,...,
d - 1$. Consequently the dynamical variables $X^{\mu}$, $\mu = 0,1,2,...,
D$, and the corresponding momenta are subjected to $d$ primary constraints.     

Another suitable form of the action (equivalent to (\ref{II1.40})) is 
the Howe--Tucker action \ci{21} generalized to a membrane of
arbitrary dimension $p$ ($p$-brane): \inxx{Howe--Tucker action}
\be
  I[X^{\mu},{\gamma}^{AB}] = {{{\kappa}_0} \over 2} \int \sqrt
  {|\gamma|} ({\gamma}^{AB} {\p}_A X^{\mu} {\p}_B X_{\mu} + 2 - d ).
\label{spl3.1}
\ee
Besides the variables $X^{\mu} (\phi)$, $\mu = 0,1,2,...,D-1$,
which denote the position of a $d$-dimensional ($d=p+1$) worldsheet
$V_d$ in the embedding spacetime $V_D$, the above action also contains
the auxiliary variables ${\gamma}^{AB}$ (with the role of
Lagrange multipliers) which have to be varied independently from
$X^{\mu}$.

By varying (\ref{spl3.1}) with respect to ${\gamma}^{AB}$ we arrive at
the equation for the induced metric on a worldsheet:
\be
    {\gamma}_{AB} = {\p}_A X^{\mu} {\p}_B X_{\mu} .
\label{spl3.2} 
\ee
Inserting (\ref{spl3.2}) into (\ref{spl3.1}) we obtain the 
Dirac--Nambu--Goto action (\ref{II1.40}).

In eq.\ (\ref{spl3.1}) the ${\gamma}^{AB}$ are the Lagrange multipliers, but they
are not all independent. The number of worldsheet constraints is $d$, which is
also the number of independent Lagrange multipliers. In order to
separate out of ${\gamma}^{AB}$ the independent multipliers we proceed
as follows. Let $\Sigma$ be a space-like hypersurface on the worldsheet,
and $n^A$ the normal vector field to $\Sigma$. Then the worldsheet metric
tensor can be written as
\be
  {\gamma}^{AB} = {{n^A n^B} \over n^2} +  {\bar \gamma}^{AB} \; ,
  \qquad {\gamma}_{AB} = {{n_A n_B} \over n^2} +  {\bar \gamma}_{AB} \; ,
\label{spl3.4}
\ee
where ${\bar \gamma}^{AB}$ is projection tensor, satisfying
\be
   {\bar \gamma}^{AB} n_B = 0 , \; \; {\bar \gamma}_{AB} n^B = 0 .
\label{spl3.4a}
\ee
It projects any vector into the hypersurface to which $n^a$ is
the normal. For instance, using (\ref{spl3.4}) we can introduce the
tangent derivatives
\be
   {\bar \p}_A X^{\mu} = {{\bar \gamma}_A}^{\, B} \, {\p}_B X^{\mu}
   = {{\gamma}_A}^{\, B} {\p}_B X^{\mu} - 
     {{n_A n^B} \over {n^2}} {\p}_B X^{\mu} .
\label{spl3.4b}
\ee
An arbitrary derivative ${\p}_A X^{\mu}$ is thus decomposed into
a normal and tangential part (relative to $\Sigma$):
\be
    {\p }_A X^{\mu} = n_A {\p} X^{\mu} + {\bar \p}_A X^{\mu} \; ,
\label{spl3.4c}
\ee
where
\be
  \p X^{\mu} \equiv {{n^A {\p}_A X^{\mu}} \over {n^2}} , \; \; \; \; \; \; \;
  n^A {\bar \p}_A X^{\mu} = 0 .
\label{spl3.4d}
\ee
Details about using and keeping the $d$-dimensional covariant notation
as far as possible are given in ref. \cite{49}. Here, following
ref. \ci{50}, I shall present a shorter
and more transparent procedure, but without the covariant notation in
$d$-dimensions.

Let us take such a class of coordinate systems in which covariant components
of normal vectors are
\be
     n_A = (1, 0, 0, . . . , 0) .
\label{spl3.5}
\ee
From eqs.\ (\ref{spl3.4}) and (\ref{spl3.5}) we have
\be
   n^2 = {\gamma}_{AB} n^A n^B = {\gamma}^{AB} n_A n_B = n^0 = {\gamma}^{00} ,
\label{spl3.6}
\ee
\be
  {\bar \gamma}^{00} = 0 \; , \; \; \;  {\bar \gamma}^{0a} = 0 ,
\label{spl3.7}
\ee
and
\begin{eqnarray}
{\gamma}_{00} & = & {1 \over {n^0}} + {\bar \gamma}_{ab} {{n^a n^b}
  \over {{(n^0)}^2}}  , \label{spl3.8a} \\
{\gamma}_{0a} & = & - {{{\bar \gamma}_{ab} n^b} \over {n^0}} ,
 \label{spl3.8b} \\
 \nonumber \\
{\gamma}_{ab} & = & {\bar \gamma}_{ab}  , \label{spl3.8c} \\
\nonumber \\
{\gamma}^{00} & = & n^0   , \label{spl3.8d} \\
\nonumber \\
{\gamma}^{0a} & = & n^a  , \label{spl3.8e} \\
\nonumber \\
{\gamma}^{ab} & = & {\bar \gamma}^{ab} + {{n^a n^b} \over {n^0}} \label{spl3.8f} \; ,
\;  \;  \;  a,b = 1,2,...,p .
\end{eqnarray}
The decomposition (\ref{spl3.4c}) then becomes
\be
   {\p}_0 X^{\mu} = \p X^{\mu} + {\bar \p}_0 X^{\mu} ,
\label{spl3.9}
\ee
\be
   {\p}_a X^{\mu} = {\bar \p}_a X^{\mu} ,
\label{spl3.10}
\ee
where
\be
  \p X^{\mu} = {\dot X}^{\mu} + {{n^a {\p}_a X^{\mu}} \over {n^0}} \; ,
  \; \; \; {\dot X}^{\mu} \equiv {\p}_0 X^{\mu} \equiv {{\p X^{\mu}}
  \over {\p {\xi}^0}} \; , \; \; \; \p_a X^{\mu} \equiv {{\p x^{\mu}}\oo
  {\p \xi^a}} ,
\label{spl3.11}
\ee
\be
   {\bar \p}_0 X^{\mu} = - {{n^a {\p}_a X^{\mu}} \over {n^0}} .
\label{spl3.12}
\ee
The $n^A = (n^0, n^a)$ can have the role of $d$ independent 
Lagrange multipliers.
We can now rewrite our action in terms of $n^0$, $n^a$, and ${\bar \gamma}^{
ab}$ (instead of $\gamma^{AB}$). We insert
(\ref{spl3.8d})--(\ref{spl3.8f}) into (\ref{spl3.1}) and take into account that
\be
     |\gamma| = {{\bar \gamma} \over {n^0}} ,
\label{spl3.13}
\ee
where $\gamma = \mbox{\rm det} \, {\gamma}_{AB}$ is the determinant of
the worldsheet metric and ${\bar \gamma} = \mbox{\rm det} \, {\bar \gamma}_{ab}$
the determinant of the metric ${\bar \gamma}_{ab} = {\gamma}_{ab} \, , \; 
a, b = 1,2,...,p$ on the hypersurface $\Sigma$.

So our action (\ref{spl3.1}) after using (\ref{spl3.8d})--(\ref{spl3.8f}) becomes
$$[X^{\mu},n^A,{\bar \gamma}^{ab}] = 
{{{\kappa}_0} \over 2} \int {\dd}^d \phi {{\sqrt{\bar \gamma}} \over
{\sqrt{n^0}}}\hs{6cm}$$
\be
 \times \left ( n^0 {\dot X}^{\mu} {\dot X}_{\mu} + 2 n^a 
{\dot X}^{\mu} {\p}_a X_{\mu} + ({\bar \gamma}^{ab} + {{n^a n^b} \over
{n^0}}) {\p}_a X^{\mu} {\p}_b X_{\mu} + 2 - d \right ) .
\lbl{spl3.14}
\ee
Variation of the latter action with respect to ${\bar \gamma}^{ab}$
gives the expression for the induced metric on the surface $\Sigma$ :
\inxx{induced metric,on hypersurface}
\be
  {\bar \gamma}_{ab} = {\p}_a X^{\mu} {\p}_b X_{\mu} \; , \qquad
  {\bar \gamma}^{ab} {\bar \gamma}_{ab} = d - 1 .
\label{spl3.15}
\ee
We can eliminate ${\bar \gamma}^{ab}$ from the action (\ref{spl3.14})
by using the relation (\ref{spl3.15}):
\be
  I[X^{\mu},n^a] = {{{\kappa}_0} \over 2} \int {\dd}^d \phi 
  {{\sqrt{|f|}} \over {\sqrt{n^0}}} \left ( {1 \over {n^0}}
  (n^0 {\dot X}^{\mu} + n^a {\p}_a X^{\mu})
  (n^0 {\dot X}_{\mu} + n^b {\p}_b X_{\mu}) + 1 \right ) ,
\label{spl3.16}
\ee
where $\sqrt{|f|} \equiv {\rm det} \, \p_a X^{\mu} \p_b X_{\mu}$.
The latter action is a functional of the worldsheet variables
$X^{\mu}$ and $d$ independent Lagrange multipliers $n^A = (n^0,n^a)$.
Varying (\ref{spl3.16}) with respect to $n^0$ and $n^a$ we obtain the
worldsheet constraints:
\be
     \delta n^0 \, :  \; \; \; ({\dot X}^{\mu} + {{n^b {\p}_b X^{\mu}}
  \over {n^0}} ) {\dot X}_{\mu} = {1 \over {n^0}} ,
\label{spl3.17}
\ee
\be
     \delta n^a \, :  \; \; \; ({\dot X}^{\mu} + {{n^b {\p}_b X^{\mu}}
  \over {n^0}} ) {\p}_a X_{\mu} = 0 .
\label{spl3.18}
\ee
Using (\ref{spl3.11}) the constraints can be written as
\be
  \p X^{\mu} \p X_{\mu} = {1 \over {n^0}} ,
\label{spl3.19}
\ee
\be
   \p X^{\mu} {\p}_i X_{\mu} = 0.
\label{spl3.20}
\ee
The action (\ref{spl3.16}) contains the expression for the normal
derivative $\p X^{\mu}$ and can be written in the form
\be
  I = {{{\kappa}_0} \over 2} \int \mbox{\rm d} \tau 
  \mbox{\rm d}^p \xi
  \sqrt {|f|} \left( {{{\p X}^{\mu }{\p X}_{\mu}}
  \over {\lambda}} + \lambda \right) , \; \; \;  \lambda \equiv
  {1 \over \sqrt{n^0}} ,
\label{spl3.21}
\ee
where we have written ${\dd}^d \phi = \dd \tau \ {\dd}^p \xi$,
since ${\phi}^A = (\tau, {\xi}^a)$ .

So we arrived at an action which looks like the well known Howe--Tucker
action for a point particle, except for the integration over a space-like
\inxx{Howe--Tucker action}
hypersurface $\Sigma$, parametrized by coordinates ${\xi}^a,
a = 1,2,...,p$. Introducing $\lambda^a = - n^a/n^0$ the normal derivative
can be written as $\p X^{\mu} \equiv {\dot X}^{\mu} - \lambda^a \p_a
X_{\mu}$. Instead of $n^0$, $n^a$ we can take $\lambda \equiv 1/\sqrt{n^0}$,
$\ \lambda^a \equiv - n^a/n^0$ as the Lagrange multipliers. In eq.
(\ref{spl3.21}) we thus recognize the action (\ref{II1.39b}).

\upperandlowertrue
\subsection[\ \ Membrane theory based on the geometric calculus in
${\cal M}$-space]
{MEMBRANE THEORY BASED ON THE GEOMETRIC CALCULUS IN
${\cal M}$-SPACE} \inxx{geometric calculus,in ${\cal M}$-space}
\upperandlowerfalse

We have seen that a membrane's velocity ${\dot X}^{\mu (\xi)}$ and
momentum $p_{\mu (\xi)}$ can be considered as components of vectors
in an infinite-dimensional membrane space ${\cal M}$ 
in which every point\inxx{membrane space}\inxx{vector,in membrane space} can be 
parametrized by coordinates $X^{\mu (\xi)}$ which represent a membrane.
In analogy with the finite-dimensional case considered in Chapter 2
we can introduce the concept of a vector in ${\cal M}$ and a set of
basis vectors $e_{\mu (\xi)}$, such that any vector $a$ can be expanded
according to 
\be
      a = a^{\mu (\xi)} e_{\mu (\xi)} .
\lbl{II1.41}
\ee
From the requirement that
\be
     a^2 = a^{\mu (\xi)} e_{\mu (\xi)} a^{\nu (\xi')} e_{\nu (\xi')} =
     \rho_{\mu (\xi) \nu (\xi')} a^{\mu (\xi)} a^{\nu (\xi')}
\lbl{II1.42}
\ee
we have
\be
    {1\oo 2} (e_{\mu (\xi)} e_{\nu (\xi')} + e_{\nu (\xi')} e_{\mu (\xi)}) 
    \equiv  e_{\mu (\xi)} \cdot e_{\nu (\xi')} = \rho_{\mu (\xi) \nu (\xi')} .
\lbl{II1.43}
\ee
This is the definition of {\it the inner product} and $e_{\mu (\xi)}$ are
generators of {\it Clifford algebra} in ${\cal M}$-space.

A more complete elaboration of geometric calculus based on Clifford algebra
in ${\cal M}$ will be provided in Chapter 6. Here we just use (\ref{II1.41}),
(\ref{II1.43}) to extend the point particle polyvector action (\ref{3.42})
to ${\cal M}$-space.

We shall start from the first order action (\ref{II1.29}). First we rewrite
\inxx{action,first order,in ${\cal M}$-space}
the latter action in terms of the compact ${\cal M}$-space notation:
\be
     I[X^{\mu}, p_{\mu}, \lambda, \lambda^a] = \int {\dd} \tau \,
     \left ( p_{\mu (\xi)} {\dot X}^{\mu (\xi)} - {1\oo 2}(p_{\mu (\xi)} p^{\mu
     (\xi)} - K) - \lambda^a p_{\mu (\xi)} \p_a X^{\mu (\xi)} \right ) .
\lbl{II1.44}
\ee
In order to avoid introducing a new symbol, it is understood that
the pro\-duct $\lambda^a p_{\mu (\xi)}$ denotes covariant components of
an ${\cal M}$-space vector. The Lagrange multiplier $\lambda$
is included in the metric $\rho_{\mu (\xi) \nu (\xi')}$.

According to (\ref{II1.41}) we can write the momentum and velocity vectors as
\be
     p = p_{\mu (\xi)} e^{\mu (\xi)} , \lbl{II1.45}
\ee
\be
     {\dot X} = {\dot X}^{\mu (\xi)} e_{\mu (\xi)} ,
\lbl{II1.46}
\ee
where 
\be
    e^{\mu (\xi)} = \rho^{\mu (\xi) \nu (\xi')} e_{\nu (\xi')}
\lbl{Ii1.47}
\ee
and
\be
       e^{\mu (\xi)} \cdot e_{\nu (\xi')} = {\delta^{\mu (\xi)}}_{\nu (\xi')} .
\lbl{II1.48}
\ee
The action (\ref{II1.44}) can be written as 
\be
    I(X,p,\lambda, \lambda^a) = \int {\dd} \tau \, \left [ p \cdot {\dot X}
    - {1\oo 2} (p^2 - K) - \lambda^a \p_a X \cdot p \right ] ,
\lbl{II1.49}
\ee
where
\be
      \p_a X = \p_a X^{\mu (\xi)} e_{\mu (\xi)}
\lbl{II1.50}
\ee
are tangent vectors. We can omit the dot operation in (\ref{II1.49}) and
write the action \inxx{tangent vectors}
\be
       I(X,p,\lambda,\lambda^a) = \int {\dd} \tau \left [ p {\dot X} -
       {1\oo 2} (p^2 - K) - \lambda^a p \, \p_a X \right ]
\lbl{Ii1.49a}
\ee
which contains the scalar part and the bivector part. It is straightforward to
show that the bivector part contains the same information about the equations
of motion as the scalar part.

Besides the objects (\ref{II1.41}) which are 1-vectors in ${\cal M}$ we
can also form 2-vectors, 3-vectors, etc., according to the analogous
procedures as explained in Chapter 2. For instance, a 2-vector is
\be
        a \wedge b = a^{\mu (\xi)} b^{\nu (\xi')} e_{\mu (\xi)} \wedge
        e_{\nu \xi')} ,
\lbl{II1.51)}
\ee
where $e_{\mu (\xi)} \wedge e_{\nu \xi')}$ are basis 2-vectors. Since
the index $\mu (\xi)$ has the discrete part $\mu$ and the continuous part
$(\xi)$, the wedge product
\be
      e_{\mu (\xi_1)} \wedge e_{\mu (\xi_2)} \wedge ... \wedge e_{\mu (\xi_k)}
\lbl{II1.52}
\ee
can have any number of terms with different values of $\xi$ and the same
value of $\mu$. The number of terms in the wedge product 
\be
       e_{\mu_1 (\xi)} \wedge e_{\mu_2 (\xi)} \wedge ... 
       \wedge e_{\mu_k (\xi)} ,
\lbl{Ii1.53}
\ee
with the same value of $\xi$, but with different values of ${\mu}$, 
is limited by the number of discrete dimensions of ${\cal M}$-space.
At fixed $\xi$ the Clifford algebra of ${\cal M}$-space behaves as the
Clifford algebra of a finite-dimensional space.

Let us write the pseudoscalar unit of the finite-dimensional subspace
$V_n$ of ${\cal M}$ as
\be
      I_{(\xi)} = e_{\mu_1 (\xi)} \wedge e_{\mu_2 (\xi)} 
      \wedge ... \wedge e_{\mu_n (\xi)}
\lbl{II1.54}
\ee

A generic polyvector in ${\cal M}$ 
is a superposition \inxx{polyvectors,in ${\cal M}$-space}
     $$A = a_0 + a^{\mu (\xi)} e_{\mu (\xi)} + a^{\mu_1 (\xi_1) \mu_2 (\xi_2)}
     e_{\mu_1 (\xi_1)} \wedge e_{\mu_2 (\xi_2)} + ... \hs{2cm}$$
\be     
    \hs{1.8cm} + \ a^{\mu_1 (\xi_1) \mu_2 (\xi_2)... \mu_k (\xi_k)} 
     e_{\mu_1 (\xi_1)} \wedge e_{\mu_2 (\xi_2)}\wedge ... \wedge 
     e_{\mu_k (\xi_k)} + ... \; .
\lbl{II1.55}
\ee

As in the case of the point particle I shall follow the principle that
the most general physical quantities related to membranes, such as
momentum $P$ and velocity ${\dot X}$, are polyvectors in ${\cal M}$-space.
I invite the interested reader to work out as an exercise (or perhaps as
a research project) what physical interpretation\footnote{
As a hint the reader is advised to look at Secs. 6.3  and 7.2.}
could be ascribed
to all possible multivector terms of $P$ and ${\dot X}$. 
For the finite-dimensional case I have already worked out in Chapter 2, Sec 3, to 
certain extent such a physical interpretation. We have also seen that at
the classical level, momentum and velocity polyvectors which solve
the equations of motion can have all the multivector parts vanishing 
except for the vector and pseudoscalar part. Let us assume a
similar situation for the membrane momentum and velocity:
\be
        P = P^{\mu (\xi)} e_{\mu (\xi)} + m^{(\xi)} I_{(\xi)} ,
\lbl{II1.56}
\ee
\be
      {\dot X} = {\dot X}^{\mu (\xi)} e_{\mu (\xi)} + {\dot s}^{(\xi)} .
      I_{(\xi)}
\lbl{II1.57}
\ee
In addition let us assume
\be
     \p_a X = \p_a X^{\mu (\xi)} e_{\mu (\xi)} + \p_a s^{(\xi)} I_{(\xi)} .
\lbl{II1.58}
\ee

Let us assume the following general membrane action:
\be
    I(X,P,\lambda, \lambda^a) = \int {\dd} \tau \, \left [ P {\dot X} -
    {1\oo 2} (P^2 - K) - \lambda^a \p_a X \, P \right ] .
\lbl{II1.59}
\ee
On the one hand the latter action is a generalization of the action (\ref
{II1.49}) to arbitrary polyvectors ${\dot X}, \, P, \, \lambda^a
\p_a X$. On the other hand, (\ref{II1.59}) is a generalization of the
point particle polyvector action \inxx{polyvector action}
(\ref{3.42}), where the polyvectors in
a finite-dimensional space $V_n$ are replaced by polyvectors in the
infinite-dimensional space ${\cal M}$.

Although the polyvectors in the action (\ref{II1.59}) are arbitrary
in principle (defined according to (\ref{II1.55})), we shall from now 
on restrict our consideration to a particular case in which the polyvectors
are given by eqs.\ (\ref{II1.56})-(\ref{II1.58}). Rewriting the action
action (\ref{II1.59}) in the component notation, that is, by inserting
(\ref{II1.56})-(\ref{II1.58}) into (\ref{II1.59}) and by taking into
account (\ref{II1.43}), (\ref{II1.48}) and
\bear
     e_{\mu (\xi)} \cdot e_{\nu (\xi')} &=& \rho_{\mu (\xi) \nu (\xi')} =
     {{\kappa \sqrt{|f|}}\oo {\lambda}} \, \delta (\xi - \xi') \eta_{\mu \nu} ,
\lbl{II1.61}\\
     I_{(\xi)} \cdot I_{(\xi')} &=& \rho_{(\xi) (\xi')} = 
     - {{\kappa \sqrt{|f|}}\oo {\lambda}} \, \delta (\xi - \xi')  ,
\lbl{II1.62}\\
      I_{(\xi)} \cdot e_{\nu (\xi')} &=& 0 ,
\lbl{II1.62a}
\ear
we obtain
    $$\langle I \rangle_0 = \int \dd \tau \Biggl[ p_{\mu (\xi)} {\dot X}^{\mu
    (\xi)} - {\dot s}^{(\xi)} m_{(\xi)} - {1\oo 2} (p^{\mu (\xi)} p_{\mu (\xi)}
    + m^{(\xi)} m_{(\xi)} - K)$$
\be
      - \lambda^a (\p_a X^{\mu (\xi)} p_{\mu (\xi)} - \p_a s^{(\xi)}
      m_{(\xi)} ) \Biggr] .
\lbl{II1.60}
\ee          
    By the way, let us observe that using (\ref{II1.61}), (\ref{II1.62})
we have
\be
     - K = - \int {\dd} \xi \, \kappa \sqrt{|f|} \lambda = \lambda^{(\xi)}
     \lambda_{(\xi)} \; , \qquad \lambda^{(\xi)} \equiv \lambda (\xi) ,
\lbl{II1.63}
\ee
which demonstrates that the term $K$ can also be written in the elegant
tensor notation.

More explicitly, (\ref{II1.60}) can be written in the form
     $$I[X^{\mu},s, p_{\mu}, m,\lambda, \lambda^a ] = \int
     {\dd} \tau \, {\dd}^n \xi \Biggl[ p_{\mu} {\dot X}^{\mu} - m {\dot s}
     - {{\lambda}\oo {2 \kappa \sqrt{|f|}}} \, \left ( p^{\mu} p_{\mu}
      - m^2 - \kappa^2 |f| \right )$$
\be     
      \hs{7cm} - \; \lambda^a (\p_a X^{\mu} p_{\mu} - \p_a \, s \, m ) \Biggr]
\lbl{II1.64}
\ee
This is a generalization of the membrane action (\ref{II1.29}) considered in
Sec. 4.2. Besides the coordinate variables $X^{\mu} (\xi)$ we have now an
additional variable $s(\xi)$. Besides the momentum variables $p_{\mu} (\xi)$
we also have the variable $m(\xi)$. We retain the same symbols $p_{\mu}$
and $m$ as in the case of the point particle theory, with understanding that
those variables are now $\xi$-dependent densities of weight 1.

The equations of motion derived from (\ref{II1.64}) are:
\begin{eqnarray}
     \delta s & : &  \;  - {\dot m} + \p_a (\lambda^a m) = 0 ,\lbl{II1.65}\\
     \delta X^{\mu} & : & \; \; {\dot p}_{\mu} + 
     \p_a \left ( \kappa \lambda \sqrt{|f|}
     \p^a X_{\mu} - \lambda ^a p_{\mu} \right ) = 0 , \lbl{II1.66}\\
     \delta m & : & \; \;  - {\dot s} + \lambda^a \p_a s + 
     {{\lambda}\oo {\kappa
     \sqrt{|f|}}} m = 0 ,\lbl{II1.67}\\
     \delta p_{\mu} & : & \; \; {\dot X}^{\mu} - \lambda^a \p_a X^{\mu}
     - {\lambda \oo {\kappa \sqrt{|f|}}} \, p_{\mu} = 0 ,\lbl{II1.68} \\
     \delta \lambda & : & \; \; p^{\mu} p_{\mu} - m^2 - \kappa^2 |f| = 0 ,
     \lbl{II1.69} \\
     \nonumber \\
     \delta \lambda^a & : & \; \; \p_a X^{\mu} p_{\mu} - \p_a s \, m = 0.
     \lbl{II1.70}
\end{eqnarray}

Let us collect in the action those term which contain $s$ and $m$ and
re-express them by using the equations of motion (\ref{II1.67}). We obtain
\be
    - m {\dot s} + {{\lambda}\oo {2 \sqrt{|f|}}} \, m^2 + \lambda^a \p_a s \, m
    = - {\lambda\oo {2 \kappa \sqrt{|f|}}} m^2 
\lbl{II1.71}
\ee
Using again (\ref{II1.67}) and also (\ref{II1.65}) we have
\be
    - {\lambda\oo {2 \kappa \sqrt{|f|}}} m^2  = {1\oo 2} ( - m {\dot s} +
    \lambda^a \p_a s \, m ) = {1\oo 2} \left ( - {{{\dd} (m s)}\oo {{\dd} 
    \tau }} + \p_a (\lambda^a s \, m) \right ) .
\lbl{II1.72}
\ee
Inserting (\ref{II1.71}) and (\ref{II1.72}) into the action (\ref{II1.64})
we obtain
     $$I =
   \int {\dd} \tau \, {\dd}^n \xi \, \Biggl[ - {1\oo2} {{\dd}\oo {{\dd} \tau }}
   (m s) + {1\oo 2} \p_a (m s \lambda^a) \hs{4.5cm}$$
\be   
    \hs{3.2cm} + \  p_{\mu} {\dot X}^{\mu} -
   {\lambda\oo {2 \kappa \sqrt{|f|}}} (p^{\mu} p_{\mu} - \kappa^2 |f|) -
   \lambda^a \p_a X^{\mu} p_{\mu} \Biggr] .
\lbl{II1.73}
\ee
We see that the extra variables $s, \, m$ occur only in the terms which
are total derivatives. Those terms have no influence on the equations of
motion, and can be omitted, so that
\be
   I[X^{\mu}, p_{\mu}] = \int {\dd} \tau \, {\dd}^n \xi \, \left [
   p_{\mu} {\dot X}^{\mu} -
   {\lambda\oo {2 \kappa \sqrt{|f|}}} (p^{\mu} p_{\mu} - \kappa^2 |f|) -
   \lambda^a \p_a X^{\mu} p_{\mu} \right ] .
\lbl{II1.74}
\ee
This action action looks like the action (\ref{II1.29}) considered in
Sec 2.2. However, now $\lambda$ and $\lambda^a$ are no 
longer Lagrange multipliers.
They should be considered as fixed since they have already been ``used" when
forming the terms $- ({\dd}/{\dd} \tau) (m s)$ and $\p_a (m s \lambda^a)$ in
(\ref{II1.73}). Fixing of $\lambda, \, \lambda^a$ means fixing the gauge,
that is the choice of parameters $\tau$ and $\xi^a$. In (\ref{II1.74}) we have
\inxx{reduced action}
thus obtained a {\it reduced action} which is a functional of the reduced
number of variables $X^{\mu}$, $p_{\mu}$. All $X^{\mu}$ or all $p_{\mu}$ are
independent; there are no more constraints.

However, a choice of gauge (the fixing of $\lambda,\, \lambda^a$) 
must be such that
the equations of motion derived from the reduced action are consistent with
the equations of motion derived from the original constrained action. In our
case we find that an admissible choice of gauge is given by
\be
     {\lambda\oo {\kappa \sqrt{|f|}}} = \Lambda \; , \quad \lambda^a = 
     \Lambda^a ,
\lbl{II1.75}
\ee
where $\Lambda, \, \Lambda^a$ are arbitrary fixed functions of $\tau, \,
\xi^a$. So we obtained the following \inxx{unconstrained action}
{\it unconstrained action}:
\be
    I[X^{\mu}, p_{\mu}] = \int \dd \tau \, {\dd}^n \xi \, \left [
    p_{\mu} {\dot X}^{\mu} - {\Lambda \oo 2} (p^{\mu} p_{\mu} - \kappa^2
    |f| ) - \Lambda^a \p_a X^{\mu} p_{\mu} \right ] .
\lbl{II1.76}
\ee

The fixed function $\Lambda$ does not transform as a scalar 
under reparametrizations of $\xi^a$, but a scalar density of weight $-1$, 
whereas $\Lambda^a$ transforms as a vector. Under reparametrizations
of $\tau$ they are assumed to transform according to $\Lambda' =
(\dd \tau/ \dd \tau') \Lambda$ and $\Lambda'^a = 
(\dd \tau/ \dd \tau') \Lambda^a$.
The action (\ref{II1.76}) is then {\it covariant} under reparametrizations
of $\tau$ and $\xi^a$, i.e., it retains the same form. However, it is
not {\it invariant} ($\Lambda$ and $\Lambda^a$ in a new parametrization
are different functions of the new parameters), therefore there are 
no constraints.

Variation of the action (\ref{II1.76}) with respect to $X^{\mu}$ and $p_{\mu}$
gives
\begin{eqnarray}
    \delta X^{\mu} &:& \; \; {\dot p}_{\mu} + \p_a \left ( \Lambda \kappa^2 |f|
    \p^a X_{\mu} - \Lambda^a p_{\mu} \right ) = 0 ,\lbl{II1.77} \\
    \nonumber \\
    \delta p_{\mu} &:& \; \; {\dot X}^{\mu} - \Lambda^a \p_a X^{\mu} -
    \Lambda p^{\mu} = 0 .
\lbl{II1.78}
\end{eqnarray}
The latter equations of motion are indeed equal to the equations of motion
(\ref{II1.66}),(\ref{II1.68}) in which gauge is fixed according to
(\ref{II1.75}).

Eliminating $p_{\mu}$ in (\ref{II1.76}) by using eq (\ref{II1.78}), we
obtain
\be
    I[X^{\mu}] = {1\oo 2} \dd \tau \, {\dd}^n \xi\, \left [ {{({\dot X}^{\mu}
    - \Lambda^a \p_a X^{\mu})({\dot X}_{\mu} - \Lambda^b \p_b X^{\mu})}\oo
    \Lambda} + \Lambda \kappa^2 |f| \right ] .
\lbl{II1.78a}
\ee
If $\Lambda^a = 0$ this simplifies to
\be
      I[X^{\mu}] = {1\oo 2} \int \dd \tau \, {\dd}^n \xi \, \left (
      {{{\dot X}^{\mu} {\dot X}_{\mu}}\oo \Lambda} + \Lambda \kappa^2 |f|
      \right )  .   
\lbl{II1.78b}
\ee
In the static case, i.e., when ${\dot X}^{\mu} = 0$, we have
\be
       I[X^{\mu}] = {1 \oo 2} \int \dd \tau \, {\dd}^n \xi \, 
       \Lambda \kappa^2 |f| ,
\lbl{II1.78c}
\ee
which is the well known Schild action \cite{51}. \inxx{Schild action }       

\paragraph{Alternative form of the ${\cal M}$-space metric}

Let us now again consider the action (\ref{II1.60}). Instead of (\ref{II1.61}),
(\ref{II1.62}) let us now take the following form of the metric:
\be
       \rho_{\mu (\xi) \nu (\xi')} = {1\oo {\tilde \lambda}} \delta (\xi - \xi')
       \eta_{\mu \nu} \; ,
\lbl{II1.79}
\ee
\be
        \rho_{(\xi) (\xi')} = - {1\oo {\tilde \lambda}} \delta (\xi - \xi') \; ,
\lbl{II1.80}                  
\ee
and insert it into (\ref{II1.60}). Then we obtain the action
\bear
    \lefteqn{I[X^{\mu}, s, p_{\mu}, m, \lambda', \lambda^a] =} 
    \hs{11cm}\lbl{II1.81}
\ear    
   $$\int \dd \tau \, {\dd}^n \xi \, \left [ - m {\dot s} + 
     p_{\mu} {\dot X}^{\mu}
     - {{\tilde \lambda}\oo 2} (p^{\mu} p_{\mu} - m^2 - \kappa^2 |f|) -
     \lambda^a (\p_a X^{\mu} p_{\mu} - \p_a s \, m) \right ] ,$$
which is equivalent to (\ref{II1.64}). Namely, we can easily verify that the
corresponding equations of motion are equivalent to the equations of motion
(\ref{II1.65})--(\ref{II1.70}). From the action (\ref{II1.81}) we then obtain
the unconstrained action (\ref{II1.76}) by fixing ${\tilde \lambda} = \Lambda$
and $\lambda^a = \Lambda^a$.

We have again found (as in the case of a point particle) that the polyvector
generalization of the action naturally contains ``time" and
evolution of the membrane variables $X^{\mu}$. Namely, in the theory there
occurs an extra variable $s$ whose derivative ${\dot s}$ with
respect to the worldsheet parameter $\tau$ is the pseudoscalar part of
the velocity polyvector. This provides a mechanism of obtaining 
the Stueckelberg action from a more basic principle. \inxx{Stueckelberg action} 
            
\paragraph{Alternative form of the constrained action}

Let us again consider the constrained action (\ref{II1.64}) which is a
functional of the variables $X^{\mu}, \, s$ and the canonical momenta $p_{\mu}$,
$m$. We can use equation of motion (\ref{II1.67}) in order to eliminate
the Lagrange multipliers $\lambda$ from the action. By doing so we obtain
   $$I = \int {\dd} \tau \, {\dd}^n \xi \Biggl[ - m ({\dot s} - \lambda^a \p_a
   s) + p_{\mu} ({\dot X}^{\mu} - \lambda^a \p_a X^{\mu} ) \hs{3cm}$$
\be
    \hs{6cm} - \ {{{\dot s} - \lambda^a \p_a s }\over {2 m}} \,
    (p^2 - m^2 - \kappa^2 |f|) \Biggr] .
\lbl{II1.87}
\ee
We shall now prove that
\be
     {{\dd s}\oo {\dd \tau}} = {\dot s} - \lambda^a \p_a s \quad {\rm and}
     \quad {{\dd X^{\mu}}\oo {\dd \tau}} = {\dot X}^{\mu} - \lambda^a
     \p_a X^{\mu} ,
\lbl{II1.88}
\ee
where ${\dot s} \equiv \p s/\p \tau$ and ${\dot X}^{\mu} = \p X^{\mu}/\p \tau$
are partial derivatives. The latter relations follow from the definitions of 
the total derivatives
\be
       {{\dd s}\oo {\dd \tau}} = {{\p s}\oo {\p \tau}} +
       {{\dd \xi^a}\oo {\dd \tau}}
        \quad {\rm and} \quad {{\dd X^{\mu}}\oo {\dd \tau}} =
       {{\p X^{\mu}}\oo {\p \tau}} + \p_a X^{\mu} \, 
       {{\dd \xi^a}\oo {\dd \tau}} ,
\lbl{II1.89}
\ee
and the relation $\lambda^a = - \dd \xi^a/\dd \tau$, which comes from
the momentum constraint $p_{\mu} \p_a X^{\mu} = 0$. 

Inserting (\ref{II1.88}) into eq.\ (\ref{II1.87}) we obtain yet another
equivalent classical action
\be
    I[X^{\mu},p_{\mu},m] = \int {\dd} s \, {\dd}^n \xi \left [ p_{\mu}
    {{\dd X^{\mu}}\oo {\dd s}} - {m\oo 2} - {1\oo {2 m}} (p^2 - \kappa^2
    |f|) \right ] ,
\lbl{II1.90}
\ee
in which the variable $s$ has disappeared from the Lagrangian, and it has
instead become the evolution parameter. Alternatively, if in the action
(\ref{II1.87}) we choose a gauge such that ${\dot s} = 1$, $\lambda^a = 0$,
then we also obtain the same action (\ref{II1.90}).

The variable $m$ in (\ref{II1.90}) acquired the status of a Lagrange
multiplier leading to the constraint \inxx{constraint}
\be
     \delta m : \quad p^2 - \kappa^2 |f| - m^2 = 0 .
\lbl{II1.90a}
\ee
Using the constraint (\ref{II1.90a}) we can eliminate $m$ from the action
(\ref{II1.90}) and we obtain the following reduced action \inxx{reduced action}
\be
       I[X^{\mu},p_{\mu}] = \int \dd s \, {\dd}^n \xi \left ( p_{\mu}
       {{\dd X^{\mu}}\oo {\dd s}} - \sqrt{p^2 - \kappa^2 |f|} \, \right )
\lbl{II1.91}
\ee
which, of course, is unconstrained. It is straightforward to verify that
the equations of motion derived from the unconstrained action (\ref{II1.91})
are the same as the ones derived from the original constrained action
(\ref{II1.64}).

The extra variable $s$ in the reparametrization invariant constrained action
(\ref{II1.64}), after performing reduction of variables by using the
constraints, has become the evolution parameter.

There is also a more direct derivation of the unconstrained action
which will be provided in the next section.

\paragraph{Conclusion}
Geometric calculus based on Clifford algebra in a finite-dimensional space
\inxx{geometric calculus,in ${\cal M}$-space} \inxx{Clifford numbers}
can be generalized to the infinite-dimensional membrane space ${\cal M}$.
Mathematical objects of such an algebra are Clifford numbers, also called
Clifford aggregates, 
or polyvectors.\inxx{Clifford aggregates}\inxx{polyvectors} It seems natural 
to assume that
physical quantities are in general polyvectors in ${\cal M}$. Then, for
instance, the membrane velocity ${\dot X}^{\mu}$ in general is not a
vector, but a polyvector, and hence it contains all other possible
$r$-vector parts, including a scalar and a pseudoscalar part. As a preliminary
step I have considered here a model in which velocity is the sum of 
a vector and a pseudoscalar. The pseudoscalar component is ${\dot s}$, i.e.,
the derivative of an extra variable $s$. Altogether we thus have the
variables $X^{\mu}$ and $s$, and the corresponding canonically conjugate
momenta $p_{\mu}$ and $m$. The polyvector action is reparametrization
invariant, and as a consequence there are constraints on those variables.
Therefore we are free to choose appropriate number of extra relations
which eliminate the redundancy of variables. We may choose relations such  
that we get rid of the extra variables $s$ and $m$, but then the
remaining variables $X^{\mu}$, $p_{\mu}$ are {\it unconstrained}, and
they evolve in the evolution parameter $\tau$ which, by choice of a gauge,
can be made proportional to $s$.

Our model with the polyvector action thus allows for {\it dynamics}
in spacetime. \inxx{dynamics,in spacetime}
{\it It resolves the old problem of the conflict between
our experience of the passage of time on the one hand, and the fact
that the theory of relati\-vity seems incapable of describing the flow
of time at all:} past, present and future coexist in a four- 
(or higher-dimensional) ``block" spacetime, with objects corresponding to 
worldlines
(or worldsheets) within this block. And what, in my opinion, is very nice,
the resolution is not a result of an {\it ad hoc procedure}, but is a necessary
consequence of the existence of Clifford algebra as a general tool for the
description of the geometry of spacetime! 

Moreover, when we shall also consider dynamics of spacetime itself, we
shall find out that the above model with the polyvector action, when
suitably generalized, will provide a natural resolution of the notorious
``problem of time" in quantum gravity. \inxx{dynamics,of spacetime}
\inxx{problem of time}

\section{More about the interconnections among various membrane actions}

In the previous section we have considered various membrane actions. One action
was just that of a {\it free fall in ${\cal M}$-space} (eq.\ (\ref{II1.1})).
For a special metric (\ref{II1.4}) which contains the membrane velocity we
have obtained the equation of motion (\ref{II1.11}) 
which is identical to that of
the {\it Dirac--Nambu--Goto membrane} described by {\it the minimal surface
action} (\ref{II1.39}).

Instead of the free fall action in ${\cal M}$-space we have considered some 
{\it equivalent forms} such as {\it the quadratic actions} (\ref{II1.14a}),
(\ref{II1.18}) and the corresponding {\it first order} or {\it phase space
action} (\ref{II1.29}).

Then we have brought into the play {\it the geometric
calculus based on Clifford algebra} and applied it to ${\cal M}$-space. 
The membrane velocity\inxx{membrane velocity} 
and momentum are promoted to {\it polyvectors}. The\inxx{membrane momentum}
latter variables were then used to construct {\it the polyvector phase
space action} (\ref{II1.59}), and its more restricted form in which the
polyvectors contain the vector and the pseudoscalar parts only.

Whilst all the actions described in the first two paragraphs were equivalent
to the usual minimal surface action which describes {\it the constrained
membrane}, we have taken with the polyvectors a step beyond the conventional
membrane theory. We have seen that the presence of a pseudoscalar variable
results in {\it unconstraining} the rest of the membrane's variables which are
$X^{\mu} (\tau, \xi)$. This has important consequences.

If momentum and velocity polyvectors are given by expressions 
(\ref{II1.56})--(\ref{II1.58}), then the polyvector action 
(\ref{II1.59}) becomes 
(\ref{II1.60}) whose more explicit form is (\ref{II1.64}). Eliminating from the
latter phase space action the variables $P_{\mu}$ and $m$ by using their
equations of motion (\ref{II1.65}), (\ref{II1.68}), we obtain
  $$I[X^{\mu},s,\lambda, \lambda^a] \hs{10cm}$$
\bear  
    &=& {\kappa \oo 2} \int \dd \tau \, {\dd}^n
   \xi  \, \sqrt{|f|} \lbl{II1.92}\\   
  & & \hs{8mm} \times \left ( {{({\dot X}^{\mu} - \lambda^a \p_a X^{\mu})
   ({\dot X}_{\mu} - \lambda^b \p_b X_{\mu}) - ({\dot s} - \lambda^a \p_a s)^2}
   \oo \lambda} + \lambda \right ) . \nonumber
\ear 
The choice of the Lagrange multipliers $\lambda , \; \lambda^a$ fixes the 
parametrization $\tau$ and $\xi^a$. We may choose $\lambda^a = 0$ and  
action (\ref{II1.92}) simplifies to
\be
   I[X^{\mu}, s ,\lambda] = {\kappa \oo 2} \int \dd \tau \, {\dd}^n \xi 
   \, \sqrt{|f|} \left ( {{{\dot X}^{\mu} {\dot X}_{\mu} - {\dot s}^2 } \oo
   \lambda} + \lambda \right ) ,
\lbl{II1.92a}
\ee
which is an extension of the Howe--Tucker-like action (\ref{II1.16}) or    
(\ref{3.21}) considered in the first two sections.

Varying (\ref{II1.92a}) with respect to $\lambda$ we have
\be
     \lambda^2 = {\dot X}^{\mu} {\dot X}_{\mu} - {\dot s}^2 .
\lbl{II1.93}
\ee
Using relation (\ref{II1.93}) in eq.\ (\ref{II1.92a}) we obtain
\be
    I[X^{\mu}, s] = \kappa \int \dd \tau \, {\dd}^n \xi \, \sqrt{|f|}
    \sqrt{{\dot X}^{\mu} {\dot X}_{\mu} - {\dot s}^2} .
\lbl{II1.94}
\ee
This reminds us of the relativistic point particle action (\ref{II1.1}). The
difference is in the extra variable $s$ and in that the variables
depend not only on the parameter $\tau$ but also on the parameters $\xi^a$,
hence the integration over $\xi^a$ with the measure ${\dd}^n \xi \sqrt{|f|}$
(which is invariant under reparametrizations of $\xi^a$).

Bearing in mind ${\dot X} =\p X^{\mu}/\p \tau$, ${\dot s} = \p s/\p \tau$,
and using the relations (\ref{II1.88}), (\ref{II1.89}),
we can write (\ref{II1.94}) as
\be
    I[X^{\mu}] = \kappa \int \dd s \, {\dd}^n \xi \, \sqrt{|f|}
    \sqrt{ {{\dd X^{\mu}}\oo {\dd s}} {{\dd X_{\mu}}\oo {\dd s}} - 1 } .
\lbl{II1.95}
\ee
The step from (\ref{II1.94}) to (\ref{II1.95}) is equivalent to choosing
the parametrization of $\tau$ such that ${\dot s} =1$ for any $\xi^a$, which
means that $\dd s = \dd \tau$.

We see that in (\ref{II1.95}) the extra variable $s$ takes the role of the
evolution parameter and that the variables $X^{\mu}(\tau, \xi)$ and the conjugate
\inxx{evolution parameter}
momenta $p_{\mu} (\tau, \xi)= \p {\cal L}/\p {\dot X}^{\mu}$ 
are {\it unconstrained}\ \footnote{The invariance of action (\ref{II1.95})
under reparametrizations of $\xi^a$ brings no constraints amongst
the dynamical variables $X^{\mu}(\tau, \xi)$ and $p_{\mu} (\tau, \xi)$
which are related to motion in $\tau$ (see also \cite{41}-\ci{43}).}.

In particular, a membrane ${\cal V}_n$ which solves the variational principle
(\ref{II1.95}) can have vanishing velocity
\be
      {{\dd X^{\mu}}\oo {\dd s}} = 0 .
\lbl{II1.96}
\ee
Inserting this back into (\ref{II1.95}) we obtain the action\footnote{The factor
$i$ comes from our inclusion of a {\it pseudoscalar} in the
velocity polyvector. Had we instead included a scalar, the corresponding
factor would then be 1.}
\be
     I[X^{\mu}] = i \kappa \int \dd s \, {\dd}^n \xi \, \sqrt{|f|} ,
\lbl{II1.97}
\ee
which governs the shape of such a {\it static} membrane ${\cal V}_n$.
\inxx{static membrane}

In the action (\ref{II1.94}) or (\ref{II1.95}) the dimensions and
signatures of the corresponding manifolds $V_n$ and $V_N$ are left unspecified.
So action (\ref{II1.95}) contains many possible particular cases.
Especially interesting are the following cases:
      
\vs{2mm}

{\it Case 1}. The manifold $V_n$ belonging to an unconstrained
membrane ${\cal V}_n$ has the signature $(+ - - - ...)$ and corresponds to
an {\it n-dimensional worldsheet} with one time-like and $n-1$ space-like
dimensions. The index of the worldsheet
coordinates assumes the values $a = 0,1,2,...,n-1$.

{\it Case 2}. The manifold $V_n$ belonging to our membrane
${\cal V}_n$ has the signature $(- - - - ...)$ and corresponds to a
space-like {\it p-brane}; therefore we take $n = p$. The index of
the membrane's coordinates $\xi^a$ assumes the values $a = 1,2,...,p$.

\vs{2mm}

Throughout the book we shall often use the single formalism and apply it,
when convenient, either to the {\it Case 1} or to the {\it Case 2}.

When the dimension of the manifold $V_n$ belonging to
${\cal V}_n$ is $n = p + 1$
and the signature is $(+ - - - ...)$, i.e. when we consider {\it Case 1},
then the action (\ref{II1.97}) is just that of the usual 
{\it Dirac--Nambu--Goto
$p$-dimensional membrane} (well known under the name {\it $p$-brane})
\be
      I = i {\tilde \kappa} \int {\dd}^n \xi \, \sqrt{|f|}
\lbl{II1.98}
\ee
with ${\tilde \kappa} = \kappa \int \dd s$.

The usual $p$-brane is considered here as a particular case of a more general
membrane\footnote{By using the name {\it membrane} we distinguish our moving
extended object from the static object, which is called $p$-{\it brane}.}
which can {\it move} in the embedding spacetime (target space) according to
the action (\ref{II1.94}) or (\ref{II1.95}). Bearing in mind two
particular cases described above, our action (\ref{II1.95}) describes either

\vs{2mm}

\ (i) {\it a moving worldsheet}, in the {\it Case I}; or

(ii) {\it a moving space like membrane}, in the {\it Case II}.

\vs{2mm}

Let us return to the action (\ref{II1.92a}). We can write it in the form
\be
      I[X^{\mu},s,\lambda] = {\kappa \oo 2} \int \dd \tau \, {\dd}^n \xi
      \left [ \sqrt{|f|} \left ( {{{\dot X}^{\mu} {\dot X}_{\mu}}\oo \lambda}
      +\lambda \right ) - {\dd \oo {\dd \tau}} \left ( {{\kappa \sqrt{|f|}
      {\dot s} s }\oo \lambda} \right ) \right ],
\lbl{II1.99}
\ee
where by the equation of motion
\be
    {\dd \oo {\dd \tau}} \left ( {{\kappa \sqrt{|f|}}\oo \lambda} \, {\dot s}
    \right ) = 0
\lbl{II1.99a}
\ee
we have    
\be
    {\dd \oo {\dd \tau}} \left ( {{\kappa \sqrt{|f|} {\dot s} s}\oo \lambda}
    \right ) =  {{\kappa \sqrt{|f|} {\dot s}^2}\oo \lambda}  \, .
\lbl{II1.100}
\ee
The term with the total derivative does not contribute to the equations
of motion and we may omit it, provided that we fix $\lambda$ in such
a way that the $X^{\mu}$-equations of motion derived from the reduced action
are consistent with those derived from the original constrained action
(\ref{II1.92}). This is indeed the case if we choose
\be
   \lambda = \Lambda \kappa \sqrt{|f|} ,
\lbl{II1.101}
\ee
where $\Lambda$ is arbitrary fixed function of $\tau$.

Using (\ref{II1.93}) we have
\be
     \sqrt{{\dot X}^{\mu} {\dot X}_{\mu} - {\dot s}^2} = \Lambda \kappa
     \sqrt{|f|} .
\lbl{II1.102}
\ee
Inserting into (\ref{II1.102}) the relation
\be
     {{\kappa \sqrt{|f|} {\dot s}}\oo {\sqrt{{\dot X}^{\mu} {\dot X}_{\mu} - 
     {\dot s}^2}}} = {1\oo C} = {\rm constant} ,
\lbl{II1.103}
\ee
which follows from the equation of motion (\ref{II1.65}), we obtain 
\be
      {\Lambda \oo C} = {{\dd s}\oo {\dd \tau}} \quad {\rm or} \quad
      \Lambda \, \dd \tau = C \, \dd s
\lbl{II1.104}
\ee
where the differential ${\dd s} = (\p s/\p \tau) \dd \tau + 
\p_a s \dd \xi^a$ is taken along a curve on the membrane along which
$\dd \xi^a = 0$ (see also eqs.\ (\ref{II1.88}), (\ref{II1.89})).
Our choice of parameter $\tau$ (given by a  choice of $\lambda$ in 
eq.\ (\ref{II1.101})) is related to the variable $s$ by the simple 
proportionality relation (\ref{II1.104}).

Omitting the total derivative term in action (\ref{II1.99}) and using
the gauge fixing (\ref{II1.101}) we obtain
\be
      I[X^{\mu}] = {1\oo 2} \int \dd \tau \, {\dd}^n \xi \, \left (
      {{{\dot X}^{\mu} {\dot X}_{\mu}}\oo \Lambda} + \Lambda \kappa^2 |f|
      \right ) .
\lbl{II1.105}
\ee
This is the {\it unconstrained} membrane action that was already derived in
previous section, eq.\ (\ref{II1.78b}). \inxx{membrane action,unconstrained}

Using (\ref{II1.104}) we find that action (\ref{II1.105}) can be written in
terms of $s$ as the evolution parameter:
\be
      I[X^{\mu}] = {1\oo 2} \int \dd s \, {\dd}^n \xi \, \left (
      {{{\cx}^{\mu} {\cx}_{\mu}}\oo C} + C \kappa^2 |f|
      \right )
\lbl{II1.105a}
\ee		      
where $\cx^{\mu} \equiv {\dd} X^{\mu}/\dd s$.

The equations of motion derived from the {\it constrained action} 
(\ref{II1.92a}) are
\begin{eqnarray}
    \delta X^{\mu} :&&  {\dd\oo {\dd \tau}} \left ( {{\kappa \sqrt{|f|}
    {\dot X}_{\mu}}\oo {\sqrt{{\dot X}^2 - {\dot s}^2}}} \right ) +
    \p_a \left ( \kappa \sqrt{|f|} \sqrt{{\dot X}^2 - {\dot s}^2} \, \p^a X_{\mu}
    \right ) = 0, \hs{9mm}  \lbl{II1.106} \\
    \delta s :&&  {\dd\oo {\dd \tau}} \left ( {{\kappa \sqrt{|f|}
    {\dot s}}\oo {\sqrt{{\dot X}^2 - {\dot s}^2}}} \right ) = 0,
    \lbl{II1.107}
\end{eqnarray}
whilst those from the {\it reduced} or {\it unconstrained action} 
(\ref{II1.105}) are
\be
      \delta X^{\mu} : \quad {\dd \oo {\dd \tau}} \left ( {{{\dot X}_{\mu}}
      \oo {\Lambda}} \right ) + \p_a \left ( \kappa^2 |f| 
      \Lambda \p^a X_{\mu} \right ) = 0 .
\lbl{II1.108}
\ee
By using the relation (\ref{II1.102}) we verify the equivalence of (\ref{II1.108})
and (\ref{II1.106}).

The original, constrained action (\ref{II1.94}) implies the constraint
\inxx{constraint}
\be
      p^{\mu} p_{\mu} - m^2 - \kappa^2 |f| = 0 ,
\lbl{II1.109}
\ee
where
$$p_{\mu} - \kappa \sqrt{|f|} {\dot X}_{\mu}/\lambda\; ,\quad 
m = \kappa \sqrt{
|f|} {\dot s}/\lambda \; , \quad \lambda = 
\sqrt{{\dot X}^{\mu} {\dot X}_{\mu} - {\dot s}^2} .$$
According to the equation of motion (\ref{II1.99a}) ${\dot m}
= 0$, therefore
\be
      p^{\mu} p_{\mu} - \kappa^2 |f| = m^2 = {\rm constant} .
\lbl{II1.110}
\ee

The same relation (\ref{II1.110}) also holds in the reduced, {\it unconstrained
theory} based on the action (\ref{II1.105}). If, in particular, $m = 0$, then
the corresponding solution $X^{\mu} (\tau, \xi)$ is identical with that for
the ordinary Dirac--Nambu--Goto membrane described by the minimal surface
action which, in a special parametrization, is \inxx{Dirac--Nambu--Goto membrane}
\be
    I[X^{\mu}] = \kappa \int \dd \tau {\dd}^n \, \xi \sqrt{|f|} 
    \sqrt{{\dot X}^{\mu} {\dot X}_{\mu}}
\lbl{II1.111}
\ee
This is just a special case of (\ref{II1.94}) for ${\dot s} = 0$.

To sum up, {\it the constrained action} (\ref{II1.94}) has the two limits:

\vs{2mm}

\ (i) {\it Limit} ${\dot X}^{\mu} = 0$. Then 
\be
I[X^{\mu} (\xi)] = i {\tilde
\kappa} \int {\dd}^n \xi \sqrt{|f|} .
\lbl{II1.112}
\ee
This is the minimal\inxx{evolution parameter}
surface action. Here the $n$-dimensional membrane (or the worldsheet in the
{\it Case I}\,) is {\it static} with respect to the evolution\footnote{
The evolution parameter $\tau$ should not be confused with one of the
worldsheet parameters $\xi^a$.}
parameter $\tau$.

(ii) {\it Limit} ${\dot s} = 0$. Then 
\be
I[X^{\mu} (\tau,\xi)] = \kappa \int
\dd \tau \, \dd^n \xi \, \sqrt{|f|} \sqrt{{\dot X}^{\mu} {\dot X}_{\mu}}.
\lbl{II1.113}
\ee
This is an action for a {\it moving} $n$-dimensional membrane which sweeps an
$(n+1)$-dimensional surface $X^{\mu} (\tau, \xi)$ subject to the constraint
$p^{\mu} p_{\mu} - \kappa^2 |f| = 0$. Since the latter constraint is conserved
in $\tau$ we have automatically also the constraint $p_{\mu} {\dot X}^{\mu} =
0$ (see Box 4.3). Assuming the {\it Case II} we have thus the motion of a
conventional constrained $p$-brane, with $p = n$.

\vs{2mm}

In general none of the limits (i) or (ii) is satisfied, and our membrane moves
according to the action (\ref{II1.94}) which involves the constraint 
(\ref{II1.109}). From the point of view of the variables $X^{\mu}$ 
and the conjugate
momenta $p_{\mu}$ there is no constraint, and instead of (\ref{II1.94}) we
can use {\it the unconstrained action} (\ref{II1.105}) or (\ref{II1.105a}),
where {\it the extra variable $s$ has become the parameter of evolution }.

\upperandlowertrue
\chapter[More about physics in ${\cal M}$-space]
{MORE ABOUT PHYSICS IN ${\cal M}$-SPACE}

\vs{18mm}

In the previous chapter we have set up the general principle that a membrane
of any dimension can be considered as a point in an infinite-dimensional
membrane space ${\cal M}$. The metric of ${\cal M}$-space is arbitrary
in principle and a membrane traces 
out a geodesic in ${\cal M}$-space.\inxx{metric,of ${\cal M}$-space} 
For a particular choice of 
${\cal M}$-space metric,\inxx{geodesic,in ${\cal M}$-space}
 the geodesic in
${\cal M}$-space, when considered from the spacetime point of view,
is a worldsheet obeying the dynamical equations derived from the
Dirac--Nambu--Goto minimal surface action. We have thus arrived very
close at a geometric principle behind the string (or, in general, the
membrane) theory. In this chapter we shall explore how far such a
geometric principle can be understood by means of the tensor calculus
in ${\cal M}$. \inxx{geometric principle,behind string theory}
\inxx{metric,fixed}

From the ${\cal M}$-space point of view a membrane behaves as a point
particle. It is minimally coupled to a fixed background metric field
$\rho_{\mu (\xi) \nu (\xi')}$ of ${\cal M}$-space. The latter metric field
has the role of a generalized gravitational field. Although at the moment
I treat $\rho_{\mu (\xi) \nu (\xi')}$ as a fixed background field, I
already have in mind a later inclusion of a kinetic term for
$\rho_{\mu (\xi) \nu (\xi')}$. The membrane theory will thus become
analogous to general relativity. We know that in the case of the usual
point particles the gravitational interaction is not the only one; there
are also other interactions which are elegantly described by
gauge fields. One can naturally generalize such a gauge principle to
the case of a point particle in ${\cal M}$ (i.e., to membranes).

\section[Gauge fields in ${\cal M}$-space]
{GAUGE FIELDS IN ${\cal M}$-SPACE}

\upperandlowerfalse
\subsection[\ \ General considerations]
{General Considerations}

One of the possible classically equivalent actions which describe the
\inxx{gauge field,in ${\cal M}$-space}
Dirac--Nambu--Goto membrane is the phase space action (\ref{II1.29}). In
${\cal M}$-space notation the latter action reads
\be
    I = \int \dd \tau \left [ p_{\mu (\sigma)} {\dot X}^{\mu (\sigma)}
    - \mbox{$1\oo 2$} (p_{\mu (\sigma)} p^{\mu (\sigma)} - K) - \lambda^a
    \p_a X^{\mu (\sigma)} p_{\mu (\sigma)} \right ],
\lbl{II2.1}
\ee
where the membrane's parameters are now denoted as $\sigma \equiv \sigma^a
, \; a = 1,2,...,p$.

Since the action (\ref{II2.1}) is just like the usual point particle action,
the inclusion of a gauge field is straightforward. First, let us add to
(\ref{II2.1}) a term which contains a total derivative. The new action
is then
\be
      I' = I + \int \dd \tau \, {{\dd W} \oo {\dd \tau}} ,
\lbl{II2.2}
\ee
where $W$ is a functional of $X^{\mu}$. More explicitly,
\be
    {{\dd W} \oo {\dd \tau}}  = {{\p W} \oo {\p X^{\mu (\sigma)}}} {\dot
    X}^{\mu (\sigma)} .
\lbl{II2.3}
\ee
If we calculate the canonical momentum belonging to $I'$ we find
\be
     p'_{\mu (\sigma)} = p_{\mu (\sigma)} + \p_{\mu (\sigma)} W .
\lbl{II2.4}
\ee
Expressed in terms of $p'_{\mu (\sigma)}$ the transformed action $I'$ is a
completely new functional of $p'_{\mu (\sigma)}$, depending 
on the choice of $W$.
In order to remedy such an undesirable situation and obtain an action
functional which would be invariant under the transformation (\ref{II2.2}) we
introduce a compensating gauge field ${\cal A}_{\mu (\sigma)}$ in
${\cal M}$-space and, following the minimal coupling prescription, instead of
(\ref{II2.1}) we write
\be
    I = \int \dd \tau \left [ p_{\mu (\sigma)} {\dot X}^{\mu (\sigma)}
    - \mbox{$1\oo 2$} (\pi_{\mu (\sigma)} \pi^{\mu (\sigma)} - K) - \lambda^a
    \p_a X^{\mu (\sigma)} \pi_{\mu (\sigma)} \right ] ,
\lbl{II2.5}
\ee
where
\be
        \pi_{\mu (\sigma)} \equiv p_{\mu (\sigma)} - q {\cal A}_{\mu (\sigma)} .
\lbl{II2.6}
\ee
$q$ being the total electric charge of the membrane.
If we assume that under the transformation (\ref{II2.2}) 
\be
      {\cal A}_{\mu (\sigma)} \rightarrow {\cal A}'_{\mu (\sigma)} =
      {\cal A}_{\mu (\sigma)} + q^{-1} \p_{\mu (\sigma)} W ,
\lbl{II2.7}
\ee
the transformed action $I'$ then has the same form as $I$:
\be
    I' = \int \dd \tau \left [ p'_{\mu (\sigma)} {\dot X}^{\mu (\sigma)}
    - \mbox{$1\oo 2$} (\pi'_{\mu (\sigma)} \pi'^{\mu (\sigma)} - K) - \lambda^a
    \p_a X^{\mu (\sigma)} \pi'_{\mu (\sigma)} \right ] ,
\lbl{II2.8}
\ee
where 
$$\pi'_{\mu (\sigma)} \equiv p'_{\mu (\sigma)} - q {\cal A}'_{\mu (\sigma)} =
     p_{\mu (\sigma)} - q {\cal A}_{\mu (\sigma)} \equiv \pi_{\mu (\sigma)}.$$

Varying (\ref{II2.5}) with respect to $p_{\mu (\sigma)}$ we find
\be
       \pi^{\mu (\sigma)} = {\dot X}^{\mu (\sigma)} - \lambda^a \p_a X^{
       \mu (\sigma)} .
\lbl{II2.9}
\ee
Variation with respect to the Lagrange multipliers $\lambda, \, \lambda^a$
gives the constraints\footnote{Remember that $\pi_{\mu (\sigma)} \pi^{\mu
(\sigma)} = \rho_{\mu (\sigma) \nu (\sigma')} \pi^{\mu (\sigma)} \pi^{\nu
(\sigma')}$ and $\rho_{\mu (\sigma) \nu (\sigma')} = (\kappa \sqrt{|f|}/
\lambda) \eta_{\mu \nu} \delta (\sigma - \sigma')$.}, 
and the variation with respect to $X^{\mu (\sigma)}$ gives the equation of
motion
    $${\dot \pi}_{\mu (\sigma)} - \p_a (\lambda^a \pi_{\mu (\sigma)}) -
    \mbox{${1\oo 2}$} \p_{\mu (\sigma)} K - \mbox{${1\oo 2}$} 
    \p_{\mu (\sigma)} \rho_{
    \alpha (\sigma') \beta (\sigma'')} \pi^{\alpha (\sigma')} \pi^{\beta
    (\sigma'')}$$ 
\be    
   \hs{2.6cm} - \int \dd \tau' \, {\cal F}_{\mu (\sigma) \nu (\sigma')}
    (\tau, \tau') {\dot X}^{\nu (\sigma')} (\tau') = 0 ,
\lbl{II2.10}
\ee
where
\be
   {\cal F}_{\mu (\sigma) \nu (\sigma')}(\tau, \tau') \equiv
   {{\delta {\cal A}_{\nu (\sigma')} (\tau')}\oo {\delta X^{\mu (\sigma)}
   (\tau)}}
   - {{\delta {\cal A}_{\mu (\sigma)} (\tau)}\oo {\delta X^{\nu (\sigma')}
   (\tau')}}
\lbl{II2.11}
\ee
is the ${\cal M}$-space gauge field strength. A gauge field ${\cal A}_{\mu
(\sigma)}$ in ${\cal M}$ is, of course, a function(al) of ${\cal M}$-space
coordinate variables $X^{\mu (\sigma)} (\tau)$. In the derivation above
I have taken into account the possibility that the expression for
the functional ${\cal A}_{\mu (\sigma)}$ contains the velocity ${\dot X}^{
\mu (\sigma)} (\tau)$ (see also Box 4.2).

For
\be
  K = \int {\dd}^p \sigma \, \kappa \sqrt{|f|} \lambda
\lbl{II2.11a}
\ee  
and
\be
   \rho_{\alpha
(\sigma') \beta (\sigma'')} = (\kappa \sqrt{|f|}/
\lambda) \eta_{\alpha \beta} \delta (\sigma' - \sigma'')
\lbl{II2.11b}
\ee
 eq.\,(\ref{II2.10})
becomes
\be
    {\dot \pi}_{\mu} + \p_a (\kappa \sqrt{|f|} \p^a X_{\mu} - \lambda^a \pi_{
    \mu}) - q \int {\dd} \tau' \, {\cal F}_{\mu (\sigma) \nu (\sigma')} (\tau,
    \tau') {\dot X}^{\nu (\sigma')} (\tau') = 0
\lbl{II2.12}
\ee
If we make use of (\ref{II2.9}) we find out that eq.\ (\ref{II2.12}) is indeed
a generalization of the equation of motion (\ref{II1.30}) derived in the
absence of a gauge field.

\subsection[\ \ A specific case]
{A Specific Case}

Now we shall investigate the transformation (\ref{II2.2}) in some more
detail. Assuming
\be     W = \int {\dd}^p \sigma \, \left ( w(x) + \p_a w^a (x) \right )
\lbl{II2.13}
\ee
the total derivative (\ref{II2.3}) of
the functional $W$ can be written as
\bear
    {{\dd W}\oo {\dd \tau}} = \p_{\mu (\sigma)} W {\dot X}^{\mu (\sigma)}
    &=& \int {\dd}^p \sigma (\p_{\mu} w - \p_a \p_{\mu} w^a) {\dot X}^{\mu}
    \nonumber \\
      &=& \int {\dd}^p \sigma \, (\p_{\mu} w {\dot X}^{\mu} - \p_a {\dot w}^a)
      \nonumber \\
      &=& \int {\dd}^p \sigma \, (\p_{\mu} w {\dot X}^{\mu} - \p_{\mu}
      {\dot w}^a \p_a X^{\mu}) .
\lbl{II2.14}
\ear
In the above calculation we have
used 
$$\p_a \p_{\mu} w^a {\dot X}^{\mu} = \p_{\mu} \p_a w^a {\dot X}^{\mu}
= (\dd /\dd \tau) \p_a w^a = \p_a {\dot w}^a.$$
In principle $W$ is an
arbitrary functional of $X^{\mu (\sigma)}$. Let us now choose the
ansatz with $\chi = \chi (X^{\mu})$:
\be
      w = e \chi \; , \qquad - {\dot w}^a = e^a \chi \; ,
\lbl{II2.15}
\ee
subjected to the condition
\be    {\dot e} + \p_a e^a = 0
\lbl{II2.16}
\ee
Then (\ref{II2.14}) becomes
\be
     {{\dd W}\oo {\dd \tau}} = \int {\dd}^p \sigma \, (e {\dot X}^{\mu} +
     e^a \p_a X^{\mu}) \p_{\mu} \chi
\lbl{II2.17}
\ee
and, under the transformation (\ref{II2.2}) which implies (\ref{II2.4}),
(\ref{II2.7}), we have
\bear
     p'_{\mu (\sigma)} {\dot X}^{\mu (\sigma)} &=& p_{\mu (\sigma)} {\dot X}^{
     \mu (\sigma)} + \int {\dd}^p \sigma (e {\dot X}^{\mu} + e^a \p_a X^{\mu})
     \p_{\mu} \chi ,
\lbl{II2.18} \\
   q {\cal A}'_{\mu (\sigma)} {\dot X}^{\mu (\sigma)} &=& q {\cal A}_{\mu (
   \sigma)} {\dot X}^{
   \mu (\sigma)} + \int {\dd}^p \sigma (e {\dot X}^{\mu} + e^a \p_a X^{\mu})
     \p_{\mu} \chi  . 
\lbl{II2.19}
\ear
The momentum $p_{\mu (\sigma)}$ thus consists of two terms
\be
   p_{\mu (\sigma)} = {\hat p}_{\mu (\sigma)} + {\bar p}_{\mu (\sigma)}
\lbl{II2.20}
\ee
which transform according to
\bear
   {\hat p}'_{\mu (\sigma)} {\dot X}^{\mu (\sigma)} &=& {\hat p}_{\mu (\sigma)}
   {\dot X}^{\mu (\sigma)} + \int {\dd}^p \sigma \, e {\dot X}^{\mu}
   \p_{\mu} \chi ,
\lbl{II2.21}\\
    {\bar p}'_{\mu (\sigma)} {\dot X}^{\mu (\sigma)} &=& {\bar p}_{\mu (\sigma)}
    {\dot X}^{\mu (\sigma)} + \int {\dd}^p \sigma \, e^a \p_a X^{\mu} \p_{\mu}
    \chi .
\lbl{II2.22}
\ear
Similarly for the gauge field ${\cal A}_{\mu (\sigma)}$:
\bear
     {\cal A}_{\mu (\sigma)} &=& {\widehat {\cal A}}_{\mu (\sigma)} +  
     {\bar {\cal A}}_{\mu (\sigma)} ,
\lbl{II2.23} \\
  q {\widehat {\cal A}}'_{\mu (\sigma)} {\dot X}^{\mu (\sigma)} &=& 
  {\widehat {\cal A}}_{\mu
  (\sigma)} {\dot X}^{\mu (\sigma)} + \int {\dd}^p \sigma
  \, e {\dot X}^{\mu} \p_{\mu} \chi  ,
\lbl{II2.24}\\
   q {\bar {\cal A}}'_{\mu (\sigma)} {\dot X}^{\mu (\sigma)} &=& {\bar {\cal A}}_
   {\mu (\sigma)} {\dot X}^{\mu (\sigma)}
   + \int {\dd}^p \sigma \, e^a \p_a X^{\mu} \p_{\mu} \chi .
\lbl{II2.25}
\ear
The last two relations are satisfied if we introduce a field $A_{\mu}$
which transforms according to
\be
      A'_{\mu} = A_{\mu} + \p_{\mu} \chi
\lbl{II2.28}
\ee
and satisfies the relations
\bear
     q {\widehat {\cal A}}_{\mu (\sigma)} {\dot X}^{\mu (\sigma)} &=&
     \int {\dd}^p \sigma \, e {\dot X}^{\mu} A_{\mu} ,
\lbl{II2.26} \\
    q {\bar {\cal A}}_{\mu (\sigma)} {\dot X}^{\mu (\sigma)} &=&
    \int {\dd}^p \sigma \, e^a \p_a X^{\mu} A_{\mu} ,
\lbl{II2.27}
\ear
or
\be
    q {\cal A}_{\mu (\sigma)} {\dot X}^{\mu (\sigma)} =
    \int {\dd}^p \sigma \, (e {\dot X}^{\mu} + e^a \p_a X^{\mu}) A_{\mu} .
\lbl{II2.27a}
\ee

We now observe that from  (\ref{II2.9}), which can be written as
\be
   \pi_{\mu} = {{\kappa \sqrt{|f|}}\oo {\lambda}} ({\dot X}_{\mu} -
   \lambda^a \p_a X_{\mu}),
\lbl{II2.29}
\ee
and from eq.\ (\ref{II2.6}) we have
\be
    p_{\mu} =  {{\kappa \sqrt{|f|}}\oo {\lambda}} ({\dot X}_{\mu} -
   \lambda^a \p_a X_{\mu}) +q {\cal A}_{\mu} .
\lbl{II2.30}
\ee
Using the splitting (\ref{II2.20}) and (\ref{II2.23}) we find that the
choice
\begin{eqnarray}
   {\hat p}_{\mu} &=&  {{\kappa \sqrt{|f|}}\oo {\lambda}} ({\dot X}_{\mu} -
   \lambda^a \p_a X_{\mu}) + q {\widehat {\cal A}}_{\mu} , \lbl{II2.31} \\
   {\bar p}_{\mu} &=& q {\bar A}_{\mu} \lbl{II2.32} \lbl{II2.32}
\end{eqnarray}
is consistent with the transformations (\ref{II2.21}), (\ref{II2.22}) and
the relations (\ref{II2.26}), (\ref{II2.27}).

\vs{2mm}

Thus the phase space action can be rewritten in terms of ${\hat p}_{\mu}$ by
using (\ref{II2.20}), (\ref{II2.32}) and (\ref{II2.27}):
   $$I = \int \dd \tau \, \Biggl[ {\hat p}_{\mu (\sigma)} {\dot X}^{\mu (\sigma)}
   + e^a \p_a X^{\mu (\sigma)} A_{\mu (\sigma)}\hs{4cm}$$
\be
      \hs{3.5cm}- \, \mbox{${1\oo 2}$} ({\hat \pi}_{\mu (\sigma)} {\hat \pi}^{\mu
      (\sigma)} - K) - \lambda^a \p_a X^{\mu (\sigma)} {\hat \pi}_{\mu (\sigma)}
      \Biggr] .
\lbl{II2.33}      
\ee
In a more explicit notation eq.\ (\ref{II2.33}) reads
\newpage
   $$I = \int \dd \tau \, {\dd}^p {\sigma} \Biggl[ {\hat p}_{\mu} {\dot X}^{\mu}
   + e^a \p_a X^{\mu} A_{\mu} \hs{4cm}$$
\be   
    \hs{4cm}- \, {1\oo 2} {\lambda\oo
   {\kappa \sqrt{|f|}}} ({\hat \pi}_{\mu} {\hat \pi}^{\mu} - \kappa^2 |f|)
   - \lambda^a \p_a X^{\mu} {\hat \pi}_{\mu} \Biggr] .
\lbl{II2.34}
\ee
This is an action \ci{52} for a $p$-dimensional membrane in the presence 
of a fixed
electromagnetic field $A_{\mu}$. The membrane bears {\it the electric charge
density} $e$ and {\it the electric current density}  $e^a$ which satisfy
the conservation law (\ref{II2.16}). \inxx{electric charge density}
\inxx{electric current density} 

The Lorentz force term then reads
  $$q \int {\dd} \tau' \, {\cal F}_{\mu (\sigma) \nu (\sigma')} (\tau,
    \tau') {\dot X}^{\nu (\sigma')} (\tau')\hs{6cm}$$    
    $$\hs{3cm} = \, q \int \dd \tau' \,
    \left ( {{\delta {\cal A}_{\nu (\sigma')} (\tau')}\oo {\delta X^{\mu
    (\sigma)} (\tau)}} - {{\delta {\cal A}_{\mu (\sigma)} (\tau)}\oo
    {\delta X^{\nu (\sigma')} (\tau')}} \right ) {\dot X}^{\nu (\sigma')}$$
 $$= q \int {\dd} \tau' \Biggl[ {{\delta \left ({\cal A}_{\nu (\sigma')} (\tau')
 {\dot X}^{\nu (\sigma')} (\tau') \right ) }\oo {\delta X^{\mu (\sigma)}
 (\tau)} } - {\cal A}_{\nu (\sigma')} {\delta^{\nu (\sigma')}}_{\mu (\sigma)}
 {{\dd}\oo {\dd \tau'}} \delta (\tau - \tau')\hs{14mm}$$
 $$\hs{7cm} - {{\delta {\cal A}_{\mu (\sigma)}
 (\tau)}\oo {\delta X^{\nu (\sigma')} (\tau')}} {\dot X}^{\nu (\sigma')}
 (\tau') \Biggr]$$
  $$  = \int {\dd} \tau' \, {\dd}^p \sigma' \, \left ( {{\delta A_{\nu} (\tau',
    \sigma')}\oo {\delta X^{\mu} (\tau, \sigma)}} - {{\delta A_{\mu} (\tau,
    \sigma)}\oo {\delta X^{\nu} (\tau',\sigma')}} \right ) \left (
    e {\dot X}^{\nu} (\tau', \sigma') + e^a \p_a X^{\nu} (\tau',\sigma') 
    \right ) $$
\be    
     = F_{\mu \nu} (e {\dot X}^{\nu} + e^a \p_a X^{\nu}) ,
\lbl{II2.28a}
\ee
where we have used eq.\ (\ref{II2.27a}).
In the last step of the above derivation we have assumed that
the expression for $A_{\mu} (\tau, \sigma)$ satisfies
\be
     {{\delta A_{\mu} (\tau, \sigma)}\oo {\delta X^{\nu} (\tau', \sigma')}}
     - {{\delta A_{\nu} (\tau', \sigma')}\oo {\delta X^{\mu} (\tau, \sigma)}}
     = \left ( {{\p A_{\mu}}\oo {\p X^{\nu}}} -
        {{\p A_{\nu}}\oo {\p X^{\mu}}} \right )
     \delta (\tau - \tau') \delta (\sigma - \sigma')  ,  
\lbl{II2.28b}
\ee
where $F_{\mu \nu} = \p_{\mu} A_{\nu} - \p_{\nu} A_{\mu}$.

Let us now return to the original phase space action (\ref{II2.5}). It is
a functional of the canonical momenta $p_{\mu (\sigma)}$. The latter
quantities can be eli\-minated from the action by using the equations of
"motion" (\ref{II2.9}), i.e., for $p_{\mu}$ we use the substitution
(\ref{II2.30}). After also taking into account the constraints
$\pi^{\mu} \pi_{\mu} - \kappa^2 |f| = 0$ and $\p_a X^{\mu} \pi_{\mu} = 0$, the
action (\ref{II2.5}) becomes
\be
   I[X^{\mu}] = \int \dd \tau \, {\dd}^p \sigma \, \left [ {{\kappa \sqrt{|f|}}\oo
   {\lambda}} ({\dot X}_{\mu} - \lambda^a \p_a X_{\mu}) {\dot X}^{\mu} +
   q {\cal A}_{\mu} {\dot X}^{\mu} \right ] .
\lbl{II2.35}
\ee
Using (\ref{II2.27a}) we have
\be
   I[X^{\mu}] = \int \dd \tau {\dd}^p \sigma \, \left [ {{\kappa 
   \sqrt{|f|}}\oo
   {\lambda}} ({\dot X}_{\mu} - \lambda^a \p_a X_{\mu}) {\dot X}^{\mu} +
   (e {\dot X}^{\mu} + e^a \p_a X^{\mu}) A_{\mu} \right ] .
\lbl{II2.37}
\ee
From the constraints and from (\ref{II2.29}) we find the expressions
(\ref{II1.35}), (\ref{II1.36}) for $\lambda$ and $\lambda^a$ which we insert
into eq.\ (\ref{II2.37}). Observing also that
$$({\dot X}_{\mu} - 
\lambda^a \p_a X_{\mu}) {\dot X}^{\mu} = ({\dot X}_{\mu} - \lambda^a \p_a
X_{\mu}) ({\dot X}^{\mu} - \lambda^b \p_b X^{\mu})$$
we obtain
   $$I[X^{\mu}] = \int \dd \tau \, {\dd}^p \sigma \biggl( \kappa \sqrt{|f|}
   \left [ {\dot X}^{\mu} {\dot X}^{\nu} (\eta_{\mu \nu} - \p^a X_{\mu}
   \p_a X_{\nu}) \right ]^{1/2}\hs{3cm}$$
\be   
   \hs{7cm} + \, (e {\dot X}^{\mu} + e^a \p_a X^{\mu}) A_{\mu}
   \biggr) ,
\lbl{II2.37a}
\ee
or briefly (see eq.\ (\ref{II1.38}))
\be
I[X^{\mu}] = \int {\dd}^{p+1} \phi \left [ \kappa ({\rm det} \, \p_A X^{\mu}
\p_B X_{\mu})^{1/2} + e^A \p_A X^{\mu} \, A_{\mu} \right ] ,
\lbl{II2.38}
\ee
where
\be
    \p_A \equiv {{\p}\oo {\p \phi^A}} \; , \quad \phi^A \equiv (\tau, \sigma^a)
    \quad {\rm and} \quad e^A \equiv (e, e^a) .
\lbl{II2.38a}
\ee
This form of the action was used in refs. \cite{53}, and is a generalization
of the case for a 2-dimensional membrane considered by Dirac \cite{54}.

Dynamics for the gauge field $A_{\mu}$ can be obtained by adding the 
corresponding kinetic term to the action:
  $$I[X^{\mu} (\phi), A_{\mu}] \hs{10cm}$$
  $$\hs{5mm}= \int {\dd}^d \phi \left [ \kappa ({\rm det} \p_A
  X^{\mu} \p_B X_{\mu} )^{1/2} + e^A \p_A \, A_{\mu} \right ]
  \delta^D \left ( x - X(\phi) \right ) \, {\dd}^D x$$
\be
       + \, {1\oo {16 \pi}} \int F_{\mu \nu} F^{\mu \nu} \, {\dd}^D x , 
       \hs{4.7cm}
\lbl{II2.39}
\ee
where $F_{\mu \nu} = \p_{\mu} A_{\nu} - \p_{\nu} A_{\mu}$. Here $d = p+1$
and $D$ are the worldsheet and the spacetime dimensions, respectively.
The $\delta$-function was inserted in order to write the corresponding
Lagrangian as the {\it density} in the embedding spacetime $V_D$.

By varying (\ref{II2.39}) with respect to $X^{\mu}(\phi)$ we obtain the
Lorentz force law for the membrane:
\be
    \kappa \, \p_A (\sqrt{{\rm det} \p_C X^{\nu} \p_D X_{\nu}} \, \p^A X^{\mu})
    + e^A \p_A X^{\nu} F_{\nu \mu} = 0 .
\lbl{II2.40}
\ee
The latter equation can also be derived directly from the previously
considered equation of motion (\ref{II2.12}).           

If, on the contrary, we vary (\ref{II2.39}) with respect to the
electromagnetic potential $A_{\mu}$, we obtain the Maxwell equations
\inxx{Maxwell equations}
\bear
         {F^{\mu \nu}}_{\nu} &=& - 4 \pi j^{\mu} ,
\lbl{II2.41}\\
     j^{\mu} &=& \int e^A \p_A X^{\mu} \delta^D (x - X(\phi)) {\dd}^d \phi .
\lbl{II2.42}
\ear
The field around our membrane can be expressed by the solution\footnote{
This expression contains both the retarded and the advanced part,
$A^{\mu} = {1\oo 2} (A_{\rm ret}^{\mu} + A_{\rm adv}^{\mu})$. When 
initial and boundary conditions require so, we may use only the retarded part.}
\be
    A^{\mu}(x) = \int e^A (\phi') \, \delta [\left ( x - X(\phi') \right )^2 ]
    \, \p_A X^{\mu} (\phi') \, {\dd}^d \phi' ,
\lbl{II2.43}
\ee
which holds in the presence of the gauge condition $\p_{\mu} A^{\mu} = 0$.
This shows that $A^{\mu} (x)$ is indeed {\it a functional} of $X^{\mu} (\phi)$
as assumed in previous section. It is defined at every spacetime point
$x^{\mu}$. In particular, $x^{\mu}$ may be positioned on the membrane.
Then instead of $A^{\mu} (x)$ we write $A^{\mu} (\phi)$, or $A^{\mu} (\tau,
\sigma^a) \equiv A^{\mu (\sigma)} (\tau)$. If we use the splitting
$\phi^A = (\tau, \sigma^a)$, $e^A = (e, e^a)$, $a = 1,2,...,p$,
(\ref{II2.43}) then reads
\be
     A^{\mu} (\tau, \sigma) = \int {\dd} \tau' \, {\dd}^p \sigma' \,
     \delta^D \left [ \left( X(\tau, \sigma) - X(\tau',\sigma') \right )^2
     \right ] (e {\dot X}^{\mu} + e^a \p_a X^{\mu}) ,
\lbl{II2.44a}
\ee
which demonstrates that the expression for $A^{\mu (\sigma)} (\tau)$ indeed
contains the velocity ${\dot X}^{\mu}$ as admitted in the derivation of the 
general equation of motion (\ref{II2.10}).

Moreover, by using the expression (\ref{II2.44a}) for $A^{\mu (\sigma)} (\tau)$
we find that the assumed expression (\ref{II2.28b}) is satisfied exactly.

\upperandlowertrue
\subsection[\ \ ${\cal M}$-space point of view again]
{${\cal M}$-SPACE POINT OF VIEW AGAIN}

In the previous subsection we have considered the action (\ref{II2.39}) which
described the membrane dynamics from the point of view of spacetime. Let
us now return to considering the \inxx{membrane dynamics}
membrane dynamics from the point of view
of ${\cal M}$-space, as described by the action (\ref{II2.5}), or,
equivalently, by the action
\be
      I = \int \dd \tau \, \left ( {\dot X}^{\mu (\sigma)} {\dot X}_{\mu
      (\sigma)} + q {\cal A}_{\mu (\sigma)} {\dot X}^{\mu (\sigma)} \right ) ,
\lbl{II2.44}
\ee
which is a generalization of (\ref{II1.14a}). In order to include the
dynamics of the gauge field ${\cal A}_{\mu (\sigma)}$ itself we have to add its
kinetic term. A possible choice for the total action is then
    $$I = \int \dd \tau \, \left ( {\dot {\tilde X}}^{\mu (\sigma)} 
    {\dot {\tilde X}}_{\mu (\sigma)} + q {\cal A}_{\mu (\sigma)} [X] 
    {\dot {\tilde X}}^{\mu (\sigma)} \right ) \delta^{({\cal M})} (X -
    {\tilde X} (\tau)) {\cal D} X \hs{12mm}$$
\be
     \hs{12mm} + \, {1\oo {16 \pi}} \int \dd \tau \, 
     \dd \tau' \, {\cal F}_{\mu (\sigma)
     \nu (\sigma')} (\tau, \tau') {\cal F}^{\mu (\sigma)
     \nu (\sigma')} (\tau, \tau') \sqrt{|\rho|} {\cal D} X \; ,
\lbl{II2.45}
\ee
where
\be
    \delta^{({\cal M})} \left ( X - {\tilde X} (\tau) \right )  
    \equiv \prod_{\mu (\sigma)} \delta \left ( X^{\mu (\sigma)} -
    {\tilde X}^{\mu (\sigma)} (\tau) \right )
\lbl{II2.45a}
\ee
and
\be
       \sqrt{|\rho|} {\cal D} X \equiv \sqrt{|\rho|} \prod_{\mu (\sigma)}
       \dd X^{\mu (\sigma)} = \prod_{\mu (\sigma)} {{\kappa \sqrt{|f|}}\oo
       {\sqrt{{\dot X}^2}}} \, \dd X^{\mu (\sigma)}
\lbl{II2.45b}
\ee
are respectively the $\delta$-function and the volume element in 
${\cal M}$-space.\inxx{volume element in ${\cal M}$-space}

Although (\ref{II2.45}) is a straightforward generalization of the point
particle action in spacetime (such that the latter space is replaced by
${\cal M}$-space), I have encountered serious difficulties in attempting
to ascribe a physical meaning to (\ref{II2.45}). In particular, it is not
clear to me what is meant by the distinction between a point $X$ at which the
field ${\cal A}_{\mu (\sigma)}[X]$ is calculated and the point ${\tilde X}$ at
which there is a ``source" of the field. One might think that the  problem
would disappear if instead of a single membrane which is a source of its
own field, a system of membranes is considered. Then, in principle one
membrane could be considered as a ``test membrane" (analogous to a ``test
\inxx{test membrane}
particle") to probe the field ``caused" by all the other membranes within
our system. The field ${\cal A}_{\mu (\sigma)} [X]$ is then a functional of
the test membrane coordinates $X^{\mu (\sigma)}$ (in analogy with $A_{\mu} (x)$
being a function of the test particle coordinates $x^{\mu}$). Here there is
a catch: how can a field ${\cal A}_{\mu (\sigma)}$  be a functional of a single
membrane's coordinates, whilst it would be much more sensible (actually correct)
to consider it as a functional of {\it all} membrane coordinates within
our system. But then there would be no need for a $\delta$-function in the
action (\ref{II2.45}).

Let us therefore write the action without the $\delta$-function.
Also let us simplify our notation by writing
\be
    X^{\mu} (\tau , \sigma) \equiv X^{\mu (\tau ,\sigma)} 
    \equiv X^{\mu (\phi)} , \quad (\phi) \equiv (\tau, \sigma) .
\lbl{II2.46}
\ee
Instead of the space of $p$-dimensional membranes with coordinates $X^{\mu (
\sigma)}$ we thus obtain the space of $(p+1)$-dimensional membranes\footnote{
When a membrane of dimension $p$ moves it sweeps a wordlsheet $X^{\mu (\tau ,
\sigma)}$, which is just another name for a {\it membrane} of dimension
$p+1$. }
with coordinates $X^{\mu (\phi)}$. The latter space will also  be called 
${\cal M}$-space, and its metric is
\be
    \rho_{\mu (\phi) \nu (\phi')} = {{\kappa \sqrt{|f|}} \oo {\sqrt{{\dot X}^2}
    }} \, \eta_{\mu \nu} \delta (\phi - \phi') ,
\lbl{II2.47}
\ee
where $\delta (\phi - \phi') = \delta (\tau - \tau') \delta^p (\sigma -
\sigma')$. A natural choice for the action is
\be
    I = 
    {\dot X}_{\mu (\phi)} {\dot X}^{\mu (\phi)} + q {\cal A}_{\mu (\phi)}
    [X] {\dot X}^{\mu (\phi)} + {\epsilon \oo {16 \pi}} {\cal F}_{\mu (\phi)
    \nu (\phi')} {\cal F}^{\mu (\phi) \nu (\phi')} ,
\lbl{II2.48}
\ee
where
\be
     {\cal F}_{\mu (\phi) \nu (\phi')} = \p_{\mu (\phi)} {\cal A}_{\nu (\phi')}
     - \p_{\nu (\phi')} {\cal A}_{\mu (\phi)} .
\lbl{II2.49} 
\ee
A constant $\epsilon$ is introduced for dimensional reasons (so that the
units for ${\cal A}_{\mu (\phi)}$ in both terms of (\ref{II2.48}) will be
consistent). Note that the integration over $\tau$ is implicit by the
repetition of the index $(\phi) \equiv (\tau, \sigma^a)$.

Variation of the action (\ref{II2.48}) with respect to $X^{\mu (\phi)}$
gives
    $$ - 2 \, {\dd \oo {\dd \tau}} {\dot X}_{\mu (\phi)} + \p_{\mu (\phi)}
     \rho_{\alpha (\phi') \beta (\phi'')} {\dot X}^{\alpha (\phi')} {\dot X}^{
     \beta (\phi '')} + q {\cal F}_{\mu (\phi) \nu (\phi')} {\dot X}^{\nu
     (\phi')}$$
\be
       \hs{2cm} + {\epsilon \oo {8 \pi}} \left ({\rm D}_{\mu (\phi)} {\cal
       F}_{\alpha (\phi') \beta (\phi'') } \right ) {\cal F}^{\alpha (\phi')
        \beta (\phi'') }    = 0
\lbl{II2.50}
\ee
If we take the metric (\ref{II2.47}) we find that, apart from the last term,
the latter equation of motion is equivalent to (\ref{II2.12}). Namely, as
shown in Chapter 4, Sec. 2, the second term in eq.\ (\ref{II2.50}), after
performing the functional derivative of the metric (\ref{II2.47}), brings an
extra term $(\dd {\dot X}_{\mu}/\dd \tau)$ and also the term
$\p_A (\kappa \sqrt{|f|} {\dot X}^2 \p^A X_{\mu})$.

If we raise the index $\mu (\phi)$ in eq.\ (\ref{II2.50}) we obtain
   $$  {{\dd {\dot X}^{\mu (\phi)}}\oo {\dd \tau}} + \Gamma_{\alpha (\phi')
   \beta (\phi'')}^{\mu (\phi)} {\dot X}^{\alpha (\phi')} {\dot X}^{\beta
   (\phi'')} \hs{7cm}$$
\be
    \hs{1.3cm} - \, {q \oo 2} {{\cal F}^{\mu (\phi)}}_{\nu (\phi')} 
    {\dot X}^{\nu (\phi')} 
   - {\epsilon \oo {4 \pi}} \left ( {\rm D}^{\mu (\phi)} {\cal F}_{\alpha
     (\phi') \beta (\phi'')} {\cal F}^{\alpha (\phi') \beta (\phi'')} \right )
     = 0
\lbl{II2.51a}
\ee
which contains the familiar term of the geodetic equation plus the terms
resulting from the presence of the gauge field. 
The factor ${1\oo 2}$ in front of
the Lorentz force term compensates for such a factor also accourring
in front of the acceleration after inserting the metric (\ref{II2.47}) into
the connection  and summing over the first two terms of eq.\ (\ref{II2.51a}).

If, on the contrary, we vary (\ref{II2.48}) with respect to the field
${\cal A}_{\mu (\phi)} [X]$, we obtain the equations which contain the
${\cal M}$-space $\delta$-function. The latter is eliminated after
integration over ${\cal D} X$. So we obtain the following field equations
\be
      - 4 \pi q {\dot X}^{\mu (\phi)} = \epsilon {\rm D}_{\nu (\phi')}
      {\cal F}^{\mu (\phi) \nu (\phi')} .
\lbl{II2.51}
\ee
They imply the ``conservation" law ${\rm D}_{\mu (\phi)} {\dot X}^{\mu (\phi)}
= 0$. 

These equations describe the motion of a membrane interacting with its own
electromagnetic field. First let us observe that the electromagnetic field
${\cal A}_{\mu (\phi)} [X]$ is a functional of the whole membrane. No
explicit spacetime dependence takes place in ${\cal A}_{\mu (\phi)} [X]$.
Actually, the concept of a field defined at spacetime points does not occur
in this theory. However, besides the discrete index $\mu$, the field bears
the set of continuous indices $(\phi) = (\tau, \sigma^a)$, and ${\cal A}_{
\mu (\phi)}$ means ${\cal A}_{\mu} (\phi) \equiv {\cal A}_{\mu} (\tau, 
\sigma^a)$, that is, the field is defined for all values of the worldsheet
parameters $\tau, \; \sigma^a$. At any values of  $\tau, \; \sigma^a$ we
can calculate the corresponding position $x^{\mu}$ in spacetime by making use
of the mapping $x^{\mu} = X^{\mu} (\tau, \sigma)$ which, in principle, is
determined after solving the system of equations (\ref{II2.50}), (\ref{II2.51}).
The concept of a field which is a functional of the system configuration is 
very important and in our discussion we shall return to it whenever
appropriate.

\upperandlowerfalse
\subsection[\ \ A system of many membranes]
{A System of Many Membranes} \inxx{system of many membranes}

We have seen that that within the ${\cal M}$-space description a gauge field
is defined for all values of membrane's parameters $\phi = (\tau, \sigma^a)$,
or, in other words, the field is defined only at those points which are
located on the membrane. If there are more than one membrane then the field is
defined on all those membranes.

One membrane is described by the parametric equation $x^{\mu} = X^{\mu} (\tau,
\sigma^a)$, where $x^{\mu}$ are target space coordinates and $X^{\mu}$
are functions of continuous parameters $\tau, \; \sigma^a$. In order to describe
a system of membranes we add one more parameter $k$, which has {\it discrete}
values. The parametric equation for {\it system of membranes} is thus
\be
     x^{\mu} = X^{\mu} (\tau, \sigma^a, k) .
\lbl{II2.52}
\ee
A single symbol $\phi$ may now denote the set of those parameters:
\be
     \phi \equiv (\tau, \sigma^a, k) ,
\lbl{II2.53}
\ee
and (\ref{II2.52}) can be written more compactly as
\be
   x^{\mu} = X^{\mu} (\phi) \qquad {\rm or} \qquad x^{\mu} = X^{\mu (\phi)} .
\lbl{II2.54}
\ee
With such a meaning of the index $(\phi)$ the equations (\ref{II2.50}),
(\ref{II2.51}) describe the motion of a system of membranes and the
corresponding gauge field. \inxx{gauge field,in ${\cal M}$-space}

The gauge field ${\cal A}_{\mu (\phi)} [X]$ is a {\it functional} of
$X^{\mu (\phi)}$, i.e., it is a functional of the system's configuration,
described by $X^{\mu (\phi)}$.

The gauge field ${\cal A}_{\mu (\phi)} [X]$ is also a function of the
parameters $(\phi) \equiv  (\tau, \sigma^a, k)$ which denote position on
the $k$-th membrane. If the system of membranes ``nearly" fills the target
space, then the set of all the points $x^{\mu}$ which, by the mapping
(\ref{II2.52}), corresponds to the set of all the parameters $\tau, \sigma^a,
k$ approximately (whatever this words means) fills the target space.
\inxx{target space} \inxx{field equation,in ${\cal M}$-space}
\inxx{field equation,in target space}

Eq.\,(\ref{II2.51}) is a field equation in ${\cal M}$-space. We can transform it
into a field equation in target space as follows. Multiplying (\ref{II2.51})
by $\delta(x - X (\phi))$ and integrating over $\phi$ we have
\be
     - 4 \pi q \int \dd \phi \, {\dot X}^{\mu (\phi)} \delta (x - X (\phi)
     ) = \epsilon \int \dd \phi \, \delta (x - X (\phi)) {\rm D}_{\nu
     (\phi')} {\cal F}^{\mu (\phi) \nu (\phi')} .
\lbl{II2.55}
\ee
The left hand side of the latter equation can be written explicitly as
\be
      - 4 \pi q \sum_k \int \dd \tau \, {\dd}^p \sigma \, {\dot X}^{\mu} 
      (\tau, \sigma, k) \delta \left ( x - X (\tau, \sigma, k) \right ) 
      \equiv  - 4 \pi j^{\mu} (x) .
\lbl{II2.55a}
\ee

In order to calculate the right hand side of eq.\ (\ref{II2.55}) we write
\be
      {\rm D}_{\nu(\phi')} {\cal F}^{\mu (\phi) \nu (\phi)} =
      \rho^{\mu (\phi) \alpha (\phi'')} \rho^{\nu (\phi') \beta (\phi''')}
      {\rm D}_{\nu (\phi')} {\cal F}_{\alpha (\phi'') \beta (\phi''')}
\lbl{II2.56}
\ee
and
      $${\rm D}_{\nu (\phi')} {\cal F}_{\alpha (\phi'') \beta (\phi''')} =
      \p_{\nu (\phi')} {\cal F}_{\alpha (\phi'') \beta (\phi''')} \hs{4cm}$$
\be
     \hs{3cm} \,   -
      \Gamma_{\nu (\phi') \alpha (\phi'')}^{\epsilon ({\tilde \phi})}
      {\cal F}_{\epsilon ({\tilde \phi}) \beta (\phi''')} - 
      \Gamma_{\nu (\phi') \beta (\phi''')}^{\epsilon ({\tilde \phi})}
      {\cal F}_{\alpha (\phi'') \epsilon ({\tilde \phi})} .
\lbl{II2.57}
\ee
We also assume\footnote{
If ${\cal A}^{\mu} (\phi)$ is given by an expression of the
form (\ref{II2.43}) then eq.\ (\ref{II2.59}) is indeed satisfied.}
\be
     {\cal F}_{\mu (\phi) \nu (\phi')} = {\cal F}_{\mu \nu} \delta (\phi -
     \phi') \; , \quad {\cal F}_{\mu \nu} = {{\p {\cal A}_{\nu}}\oo {\p X^{\mu}
     }} - {{\p {\cal A}_{\mu}}\oo {\p X^{\nu}}} .
\lbl{II2.59}
\ee
Then, for the metric (\ref{II2.47}), eq.\ (\ref{II2.55}) becomes
\bear
    - 4 \pi j^{\mu} &=& \epsilon \delta (0) \sum_k \int \dd \tau {\dd}^p \sigma
    \, \delta \left ( x - X(\tau, \sigma, k) \right ) {{{\dot X}^2}\oo 
    {\kappa^2 |f|}} \, \p_{\nu} {\cal F}^{\mu \nu} (\tau, \sigma, k)
    \nonumber \\
   & & \hs{2cm} + \, (\mbox{covariant derivative terms})  \nonumber \\   
    &=& \p_{\nu} F^{\mu \nu} (x) + (\mbox{covariant derivative terms}),
\lbl{II2.60}
\ear                    
where we have defined
\be
     \epsilon \delta (0) \sum_k \int \dd \tau {\dd}^p \sigma
    \, \delta \left ( x - X(\tau, \sigma, k) \right ) {{{\dot X}^2}\oo 
    {\kappa^2 |f|}} {\cal A}^{\mu} (\tau, \sigma, k) \equiv A^{\mu} (x)
\lbl{II2.60a}
\ee
and taken the normalization\footnote{
The infinity $\delta (0)$ in an expression such as (\ref{II2.60}) can be
regularized by taking into account the plausible assumption that a generic
physical object is actually a fractal (i.e., an object with detailed structure
at all scales). The coordinates $X^{\mu (\phi)}$
which we are using in our calculations are well behaved functions with
definite derivatives and should be considered as an approximate
description of the actual
physical object. This means that a description with a given $X^{\mu (\phi)}$
below a certain scale has no physical meaning. In order to make a physical
sense of the expressions leading to (\ref{II2.60}), the $\delta$ function
$\delta (\phi - \phi')$ should therefore be replaced by a function $F(a,
\phi - \phi')$ which in the limit $a \rightarrow 0$ becomes $\delta (\phi -
\phi')$. Instead of $\delta (0)$ we thus have  $F(a,0)$ which is finite
for all finite values of $a$. We then take the normalization $\epsilon$
such that $\epsilon F(a,0) = 1$ for any $a$, and assume that this is true
in the limit $a \rightarrow 0$ as well.}
such that $\epsilon \delta (0) = 1$. The factor ${\dot X}^2/\kappa |f|$
comes from the inverse of the metric \ref{II2.47}.
Apart
from the `covariant derivative terms', which I shall not discuss here,
(\ref{II2.60}) are just the {\it Maxwell equations} in flat spacetime.      

The gauge field  $A^{\mu} (x)$ is formally defined at all spacetime points
$x$, but because of the relation (\ref{II2.60a}) it has non-vanishing
values only at those points $x^{\mu}$ which are occupied by the membranes.
A continuous gauge field over the target space is an approximate concept in
this theory. Fundamentally, a gauge field is a functional of the system
configuration.

One can also insert (\ref{II2.59}) directly into the action (\ref{II2.48}).
We obtain
   $${\epsilon \oo {16 \pi}} \, 
   {\cal F}_{\mu (\phi) \nu (\phi')}
   {\cal F}^{\mu (\phi) \nu (\phi')}\hs{6cm}$$
   $$  = {{\epsilon \delta (0)} \oo {16 \pi}}
   \int {{\dot X}^2\oo {\kappa^2 |f|}} \, 
    {\cal F}_{\mu \nu} {\cal F}^{\mu \nu} (\phi) \, \dd \phi \hs{.8cm}$$ 
   $$\hs{1.06cm} = {{\epsilon \delta (0)} \oo {16 \pi}} \int 
    {\cal F}_{\mu \nu} {\cal F}^{\mu \nu} (\phi) \, \dd \phi \, \delta (x - 
   X (\phi) ) \, \dd^N x$$
\be   
    \hs{1.17cm} = {1\oo {16 \pi}} \int F_{\mu \nu} F^{\mu \nu} (x)
   \, {\dd}^N x \hs{3cm}
\lbl{II2.60aa}
\ee
where we have  used the definition (\ref{II2.60a}) for $A^{\mu}$ and introduced
\be
         F_{\mu \nu} = \p_{\mu} A_{\nu} - \p_{\nu} A_{\mu} .
\lbl{II2.60b}
\ee

Similarly
\be
    {\dot X}_{\mu (\phi)} {\dot X}^{\mu (\phi)} =
    \int {\dd} \phi \, \kappa \sqrt{{\dot X}^2} \sqrt{|f|}
\lbl{II2.60c}
\ee
and 
\be
   q {\cal A}_{\mu (\phi)} {\dot X}^{\mu (\phi)} = \int {\dd} \phi \,
   e^A \p_A X_{\mu} A^{\mu} .
\lbl{II2.60d}
\ee

The action (\ref{II2.39}) is thus shown to be a special case of the
${\cal M}$-space action (\ref{II2.48}). This justifies use 
of (\ref{II2.48}) as the
${\cal M}$-space action. Had we used the form (\ref{II2.45})
with a $\delta$-functional, then in the matter term the volume
$\int {\cal D} X$ would not occur, while it would still occur in the field
term.

\upperandlowertrue
\section[Dynamical metric field in ${\cal M}$-space]
{DYNAMICAL METRIC FIELD IN ${\cal M}$-SPACE}

So far I have considered the metric tensor of the membrane space ${\cal M}$
as a fixed background field. In previous section, for instance, I considered
a fixed metric (\ref{II2.47}). As shown in Section 4.2, a consequence of
such a metric is that a moving $p$-dimensional membrane sweeps a 
$(p+1)$-dimensional worldsheet, which is a solution to the usual minimal
surface action principle. Such a worldsheet  is a $(p+1)$-dimensional manifold
parametrized by $\phi^A = (\tau, \sigma^a)$. All those parameters are on an
equal footing, therefore we have used the compact notation when constructing the
action (\ref{II2.48}).

Now I am going to ascribe the dynamical role to the ${\cal M}$-space metric
as well. If the metric of ${\cal M}$-space is dynamical eq.\ (\ref{II2.47})
no longer holds.
From the point of view of the infinite-dimensional ${\cal M}$-space
we have the motion of a point ``particle" in the presence of the metric
field $\rho_{\mu (\phi) \nu (\phi')}$ which itself is dynamical. As a model
for such a dynamical system let us consider the action
\be
    I[\rho] = \int {\cal D} X \sqrt{|\rho |} \, \left 
    ( \rho_{\mu (\phi) \nu (\phi')} 
    {\dot X}^{\mu (\phi)} {\dot X}^{\nu (\phi')} +
    {\epsilon \oo {16 \pi}} {\cal R} \right )  .
\lbl{II2.61}
\ee
where $\rho$ is the determinant of the metric $\rho_{\mu (\phi) \nu (\phi')}$.
Here ${\cal R}$ is the Ricci scalar in ${\cal M}$-space, defined according to
\inxx{Ricci scalar,in ${\cal M}$-space}
\be
     {\cal R} = \rho^{\mu (\phi) \nu (\phi')} {\cal R}_{\mu (\phi) \nu (\phi')},
\lbl{II2.62}
\ee
where the Ricci tensor \inxx{Ricci tensor,in ${\cal M}$-space}
\bear
       {\cal R}_{\mu (\phi) \nu (\phi)} &=& \Gamma_{\mu (\phi) \rho (\phi''), 
       \nu (\phi')}^{\rho (\phi'')} - \Gamma_{\mu (\phi) \nu (\phi'),
       \rho (\phi'')}^{\rho (\phi'')} \hs{1cm} \lbl{II2.63} \\
       \nonumber \\
      & & + \, \Gamma_{\sigma (\phi''') \nu (\phi')}^{
       \rho (\phi'')} \Gamma_{\mu (\phi) \rho (\phi'')}^{\sigma (\phi''')}
       - \Gamma_{\mu (\phi) \nu (\phi')}^{\sigma (\phi''')} \Gamma_{\sigma
       (\phi''') \rho (\phi '')}^{\rho (\phi'')} \nonumber
\ear
is a contraction of the Riemann tensor in ${\cal M}$-space:

Variation of (\ref{II2.61}) with respect to $X^{\mu (\phi)}$ 
gives
     $$2 \left ( {{\dd {\dot X}^{\mu (\phi)}}\oo {\dd \tau}} + 
     \Gamma_{\alpha (\phi')  \beta (\phi'')}^{\mu(\phi)}
     {\dot X}^{\alpha (\phi')} {\dot X}^{\beta (\phi'')} \right ) \hs{1.7cm}$$
\be      
    \hs{.5cm} - \, {\epsilon \oo {16 \pi}} 
    \p^{\mu (\phi)} {\cal R} - {{\p^{\mu (\phi)}
      \sqrt{|\rho |}}\oo {\sqrt{|\rho|}}} \left ( {\dot X}^2 + {\epsilon\oo
      {16 \pi}} {\cal R} \right ) = 0
\lbl{II2.65}
\ee

Variation of (\ref{II2.61}) with respect to the metric 
gives the Einstein equations in ${\cal M}$-space:
\be
       {\epsilon \oo {16 \pi}} {\cal G}_{\mu (\phi) \nu (\phi')} + 
       {\dot X}_{\mu (\phi)}{\dot X}_{\nu (\phi')} - \mbox{$1\oo 2$}
        \rho_{\mu (\phi)
       \nu (\phi')} {\dot X}^2 = 0
\lbl{II2.66}
\ee
where 
\be
{\cal G}_{\mu (\phi) \nu (\xi')} \equiv {\cal R}_{\mu (\xi) 
\nu (\xi')} - \mbox{$1\oo 2$} \rho_{\mu (\xi) \nu (\xi')} {\cal R}
\lbl{II2.66a}
\ee 
is the Einstein tensor in ${\cal M}$-space. 
\inxx{Einstein tensor,in ${\cal M}$-space}

Contracting eq.\ (\ref{II2.66}) with $\rho^{\mu (\phi) \nu (\phi')}$ we have
\be
         {\dot X}^2 + {\epsilon \oo {16 \pi}} {\cal R} = 0 ,
\lbl{p1}
\ee
where ${\dot X}^2 \equiv {\dot X}^{\mu (\phi)} {\dot X}_{\mu (\phi)}$.
Because of (\ref{p1}) the equations of motion (\ref{II2.65}), (\ref{II2.66})
can be written as
\be
    {{\dd {\dot X}^{\mu (\phi)}}\oo {\dd \tau}} + \Gamma_{\alpha (\phi')
    \beta (\phi'')}^{\mu (\phi)}
     {\dot X}^{\alpha (\phi')} {\dot X}^{\beta (\phi'')}
    - {\epsilon \oo {32 \pi}} \p^{\mu (\phi)} {\cal R} = 0
\lbl{p2}
\ee
\be
    {\dot X}^{\mu (\phi)} {\dot X}^{\nu (\phi)} + {\epsilon \oo {16 \pi}}
    {\cal R}^{\mu (\phi) \nu (\phi')} = 0
\lbl{p3}
\ee
Using
\bear
   {\dot X}^{\nu (\phi')} {\rm D}_{\nu (\phi')} {\dot X}^{\mu (\phi)} &=&
   {{{\DD} {\dot X}^{\mu (\phi)}} \oo {\DD \tau}} \nonumber \\
    &\equiv& {{{\dd} {\dot X}^{\mu (\phi)}} \oo {\dd \tau}} + 
    \Gamma_{\alpha (\phi') \beta (\phi'')}^{\mu (\phi)}
    {\dot X}^{\alpha (\phi')} {\dot X}^{\beta (\phi'')},
\lbl{p4}\\
     {\dot X}_{\mu (\phi)} \p^{\mu (\phi)} {\cal R} &=& {{\dd {\cal R}}\oo
      {\dd \tau}},
\lbl{p5}
\ear
and multiplying (\ref{p2}) with ${\dot X}_{\mu (\phi)}$ we have
\be
    {1\oo 2} {{\dd {\dot X}^2}\oo {\dd \tau}} - {\epsilon\oo {32 \pi}}
    {{\dd {\cal R}}\oo {\dd \tau}} = 0.
\lbl{p6}
\ee
Equations (\ref{p1}) and (\ref{p6}) imply
\be
   {{\dd {\dot X}^2}\oo {\dd \tau}} = 0 \; , \qquad 
   {{\dd {\cal R}}\oo {\dd \tau}} = 0 .
\lbl{p7}
\ee

By the Bianchi identity
\be
          {\rm D}_{\nu (\phi')} {\cal G}^{\mu (\phi) \nu (\phi')} = 0
\lbl{II2.67}
\ee
eq.\ (\ref{II2.66}) implies
\be
     \left ( {\rm D}_{\nu (\phi')} {\dot X}^{\mu (\phi)} \right )
     {\dot X}^{\nu (\phi')} + {\dot X}^{\mu (\phi)} 
     \left ( {\rm D}_{\nu (\phi')} {\dot X}^{\nu (\phi')} \right )
     - \mbox{${1\oo 2}$} \p^{\mu (\phi)} {\dot X}^2 = 0
\lbl{II2.68}
\ee
If we multiply the latter equation by ${\dot X}_{\mu (\phi)}$ (and
sum over $\mu (\phi)$) we obtain
\be
     \DD_{\nu (\phi)} {\dot X}^{\nu (\phi)} = 0 .
\lbl{p8}
\ee
After inserting this back into eq.\ (\ref{II2.68}) we have
\be
      {{\DD {\dot X}^{\mu (\phi)}}\oo {\DD \tau}} - 
      \mbox{${1\oo 2}$} \p^{\mu (\phi)} 
      {\dot X}^2 = 0 ,
\lbl{p9}
\ee
or, after using (\ref{p1}),
\be
    {{\DD {\dot X}^{\mu (\phi)}}\oo {\DD \tau}} + {\epsilon\oo {32 \pi}}
    \p^{\mu (\phi)} {\cal R} = 0 .
\lbl{p10}
\ee
The latter equation together with eq.\ (\ref{p2}) implies
\be
       \p^{\mu (\phi)} {\cal R} = 0 ,
\lbl{p11}
\ee
or equivalently
\be
      \p^ {\mu (\phi)} {\dot X}^2 = 0 .
\lbl{p12}
\ee
Equation of motion (\ref{p2}) or (\ref{p9}) thus becomes simply
\be
   {{\DD {\dot X}^{\mu (\phi)}}\oo {\DD \tau}} \equiv                            
  {{{\dd} {\dot X}^{\mu (\phi)}} \oo {\dd \tau}} + \Gamma_{\alpha (\phi')
    \beta (\phi'')}^{\mu (\phi)}
     {\dot X}^{\alpha (\phi')} {\dot X}^{\beta (\phi'')} = 0 ,
\lbl{p13}
\ee
which is {\it the geodesic equation} in ${\cal M}$-space.
\inxx{geodesic equation,in ${\cal M}$-space}

Equation (\ref{p12}) is, in fact, the statement that the functional derivative of
the quadratic form
\be
     {\dot X}^2 \equiv {\dot X}^{\mu (\phi)} {\dot X}_{\mu (\phi)} =
     \rho_{\mu (\phi) \nu (\phi')} {\dot X}^{\mu (\phi)} {\dot X}^{\nu (\phi')}
\lbl{p14}
\ee
with respect to $X^{\mu (\phi)}$ is zero. The latter statement is
{\it the geodesic equation}.

Equation (\ref{p11}) states that the functional derivative of
the curvature scalar does not matter. Formally, in our variation of the action
(\ref{II2.61}) we have also considered the term $\p^{\mu (\phi)} {\cal R}$, and
from the consistency of the equations of motion we have found that
$\p^{\mu (\phi)} {\cal R} = 0$. Our equations of motion are therefore
simply (\ref{p13}) and (\ref{p3}).

The metric $\rho_{\mu (\phi) \nu (\phi')}$ is a functional
of the variables $X^{\mu (\phi)}$,
and (\ref{p3}), (\ref{p13}) is a system of functional 
differential equations which \inxx{functional differential equations}
determine the set of possible solutions for $X^{\mu (\phi)}$ 
and $\rho_{\mu (\phi) \nu (\phi')}$.
Note that our membrane model is {\it independent of background }: 
there is no pre-existing
space with a pre-exi\-sting metric, neither curved nor flat. 
The metric comes out as a solution
of the system of functional differential equations, 
and at the same time we also obtain
a solution for $X^{\mu (\phi)}$.

We can imagine a universe which consists of a single 
($n$-dimensional) membrane
whose configuration changes with $\tau$ according 
to the system of functional equations\
(\ref{p3}), (\ref{p13}). The metric $\rho_{\mu (\phi) \nu (\phi')} 
[X]$, at a given
point $\phi^A = (\tau,\xi^a)$ on the membrane, depends on 
the membrane configuration described
by the variables $X^{\mu (\phi)}$. In other words, {\it in the theory 
considered here the the metric $\rho_{\mu (\phi) \nu (\phi')} [X]$ 
is defined only
on the membrane, and it is meaningless to speak about 
a metric ``outside" the membrane}.
Also there is no such thing as a metric of a (finite-dimensional) 
space into which
the membrane is embedded. Actually, there is no embedding space. 
All what exists \inxx{membrane configuration}
is the membrane configuration $X^{\mu (\phi)}$ and the
corresponding metric $\rho_{\mu (\phi) \nu (\phi')}$ 
which determines the distance \inxx{distance,between membrane configurations}
between two possible configu\-rations $X^{\mu (\phi)}$ and
$X'^{\mu (\phi)}$ in $\cal M$-space. The latter
space ${\cal M}$ itself together with the metric 
$\rho_{\mu (\phi) \nu (\phi')}$ is a solution to our dynamical
system of equations  (\ref{II2.65}),(\ref{II2.66}).

Instead of a one membrane universe we can imagine a many membrane universe
as a toy model. \inxx{many membrane universe}
If $X^{\mu (\phi)}$, with the set of 
continuous indices $\phi \equiv \phi^A$,
represents a single membrane then $X^{\mu (\phi ,k)}$ with 
an extra discrete index
$k = 1,2,..., Z$ represents a system of Z membranes. 
If we replace $ (\phi)$ with
$(\phi , k)$, or, alternatively, if we interpret $(\phi)$ 
to include the index $k$, then {\it all equations} 
(\ref{II2.61})--(\ref{p13}) {\it are also
valid  for a system of membranes}. The metric 
$\rho_{\mu (\phi ,k) \nu (\phi' , k')}$
as a solution to the system of functional equations of
motion (\ref{II2.65}), (\ref{II2.66})
is defined only on those membranes. If there are many such membranes then the
points belonging to all those membranes approximately 
sample a finite-dimensional manifold $V_N$ into which 
the system of membranes is embedded. The metric
$\rho_{\mu (\phi ,k) \nu (\phi' , k')}$, although 
defined only on the set of points $(\phi ,
k)$ belonging to the system of membranes, 
approximately sample the metric of a
continuous finite-dimensional space $V_N$. 
See also Fig. 5.1 and the accompanying text.

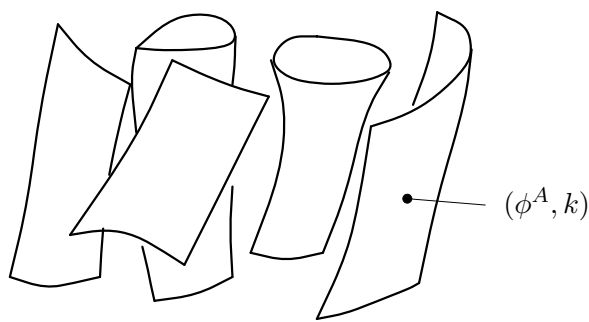
\begin{figure}[h]
\setlength{\unitlength}{.8mm}
\begin{picture}(120,80)(-18,0)

\put(80,35){\circle*{1.5}}
\put(80,35){\line(12,-1){13}}
\put(96,32.6){$(\phi^A,k)$}

\thicklines
\spline(14,22)(20,20)(25,21)(29,22)
\spline(29,22)(29.5,30)
\spline(30.5,39)(32,46)(34,54)
\spline(34,54)(27,58)(22,64)
\spline(22,64)(19,50)(17,35)(14,22)

\spline(24,29)(30,30)(35,29)(40,26)(43,24)
\spline(43,24)(46,30)(52,42)(57,52)
\spline(57,52)(50,54)(41,58)
\spline(41,58)(35,45)(24,29)

\spline(35,60)(40,59.5)(50,60)(52,62)(50,65)(45,66)(40,64.5)(35,60)
\spline(35,60)(34,55)(35,46)
\spline(36,27)(37,23)(38,18)
\spline(38,18)(45,19)(51,22)
\spline(51,22)(50.5,30)(51,37)
\spline(51,54)(51.5,58)(51.5,62)

\spline(65,62)(60,62)(57,58)(60,55)(65,54)(70,54)(75,55)(78,57)(75,61)(65,62)
\spline(57.4,58)(60,50)(59,39)(54,26)
\spline(54,26)(60,24.5)(66,26)
\spline(66,26)(70,36)(73,49)(77,56)

\spline(74,47)(80,49)(85,52)(90,57)(91,60)(90,64)(85,66)
\spline(85,66)(84,59)(81,50.6)
\spline(74,47)(72,35)(70,26)(65,15)
\spline(65,15)(70,16)(75,18)(80,20)(82,21)
\spline(82,21)(85,35)(90,52)(90.7,59.8)

\end{picture}

\caption{The system of membranes is represented as being 
embedded in a finite-dimensional
space $V_N$. The concept of a continuous 
embedding space is only an approximation
which, when there are many membranes, becomes good 
at large scales (i.e., at the
``macroscopic" level). The metric is defined 
only at the points $(\phi ,k)$ situated on
the membranes. At large scales (or equivalently, 
when the membranes are ``small"
and densely packed together) the set of all the 
points $(\phi ,k)$ is a good approximation
to a continuous metric space $V_N$.}

\end{figure}

The concepts introduced above are very important
and I shall now rephrase them.
Suppose we have a system of membranes. Each membrane is described by the
variables\footnote{
Other authors also considered a class of membrane theories in which the
embedding space has no prior existence, but is instead coded completely
in the degrees of freedom that reside on the membranes. They take for
granted that, as the background is not assumed to exist, there are no
embedding coordinates (see e.g., \ci{Smolin}). This seems to be
consistent with our usage of $X^{\mu (\phi)}$ which, at the fundamental
level, are not considered as the embedding coordinates, but as the
${\cal M}$-space coordinates. Points of ${\cal M}$-space are
described by coordinates $X^{\mu (\phi)}$, and the distance between
the points is determined by the metric $\rho_{\mu (\phi) \nu (\phi')}$,
which is dynamical.}
 $X^{\mu} (\phi^A) = X^{\mu} (\xi^a, \tau),
 \;  \mu = 1,2,...,N$; 
$a = 1,2,...,n ; \, \xi^a \in [0, \pi] , \; \tau \in [- \infty, \infty]$.
In order to distinguish different 
membranes we introduce an extra index $k$
which assumes discrete values. The system of membranes 
is then described by variables
$X^{\mu} (\phi^A,k)$,  $\; k = 1,2,...,Z$. 
Each configuration of the system is considered as
a point, with coordinates $X^{\mu (\phi ,k)} \equiv X^{\mu} 
(\phi^A , k)$, in an infinite-dimensional space ${\cal M}$ with metric 
$\rho_{\mu (\phi ,k) \nu (\phi' , k')}$. The
quantities $X^{\mu} (\phi^A , k)$ and $\rho_{\mu 
(\phi ,k) \nu (\phi' , k')}$ are not given
{\it a priori}, they must be a solution to a dynamical 
system of equations. As a preliminary
step towards a more realistic model I 
consider the action (\ref{II2.61}) which
leads to the system of equations (\ref{p3}), (\ref{p13}). 
In this model everything
that can be known about our toy universe is in the 
variables $X^{\mu (\phi ,k)}$ and
$\rho_{\mu (\phi ,k) \nu (\phi' , k')}$. The
metric is defined for all possible values of the 
parameters $(\phi^A, k)$ assigned to the points
on the membranes. At every chosen point $(\phi, k)$ the metric
$\rho_{\mu (\phi ,k) \nu (\phi' , k')}$ is a functional 
of the system's configuration,
represented by $X^{\mu (\phi, k)}$.

It is important to distinguish between the points, 
represented by the coordinates $X^{\mu (\phi,k)}$, 
of the infinite-dimensional ${\cal M}$-space,  and the points, represented by
$(\phi^A, k)$, of the finite-dimensional space 
spanned by the membranes. In the limit of
large distances the latter space becomes the 
continuous target space into which a
membrane is embeded. In this model the embedding 
space (or the ``target space") is \inxx{target space}
inseparable from the set of membranes. Actually, 
it is identified with the set of points
$(\phi^A ,k)$ situated on the membranes. Without membranes 
there is no target space.
With membranes there is also the target space, and 
if the membranes are ``small", and
if their density (defined roughly as the number of 
membranes per volume) is high,
then the target space becomes a good approximation to 
a continuous manifol $V_N$.

\upperandlowertrue
\subsection[\ \ Metric of $V_N$ from the metric of ${\cal M}$-space]
{METRIC OF $V_N$ FROM THE METRIC OF ${\cal M}$-SPACE}

${\cal M}$-space is defined here as the space of all possible 
configurations of a system of membranes. The
metric $\rho_{\mu (\phi ,k) \nu (\phi' , k')}$ 
determines {\it the distance} between the
points $X^{\mu (\phi,k)}$ and $X^{\mu (\phi,k)} + 
\dd X^{\mu (\phi,k)}$ belonging to
two different configurations (see Fig. 5.2) according to the relation
\be
      {\dd} {\ell}^2 = \rho_{\mu (\phi ,k) \nu (\phi' , k')} 
    \dd X^{\mu (\phi,k)} \dd X^{\nu (\phi',k')} ,
\lbl{II2.81}
\ee
where
\be
       \dd X^{\mu (\phi,k)} = X'^{\mu (\phi ,k)} - X^{\mu (\phi,k)} .
\lbl{II2.82}
\ee

Besides (\ref{II2.82}), which is an ${\cal M}$-space vector joining two
different membrane's configurations (i.e., two points 
of ${\cal M}$-space), we can define
another quantity which connects two different 
points, in the usual sense of the word,
within the same membrane configuration:
\be
    {\widetilde \Delta} X^{\mu} (\phi ,k) \equiv X^{\mu (\phi' , k')} - 
    X^{\mu (\phi ,k)} .
\lbl{II2.83}
\ee

By using the ${\cal M}$-space metric let us define the quantity
\be
        \Delta s^2 = \rho_{\mu (\phi ,k) \nu (\phi' , k')} 
        {\widetilde \Delta} X^{\mu}
    (\phi ,k) {\widetilde \Delta} X^{\nu} (\phi' , k') .
\lbl{II2.84}
\ee
In the above formula summation over the repeated indices $\mu$ and $\nu$ is
assumed, but no integration over $\phi$, $\phi'$ and no summation over 
$k$, $k'$.

\begin{figure}[h]
\setlength{\unitlength}{.8mm}
\begin{picture}(120,80)(-15,0)

\put(26.5,30){\circle*{1.5}}
\put(36,39.6){\circle*{1.5}}
\put(26.5,30){\line(1,1){9.5}}
\put(15,42){\vector(2,-1){16}}

\put(14,28){$(\phi,k)$}
\put(2,42){$\dd X^{\mu (\phi,k)}$}
\put(21,0){$X^{\mu (\phi,k)}$}
\put(41,0){$X'^{\mu (\phi,k)}$}

\put(72,32){\circle*{1.5}}
\put(79,39){\circle*{1.5}}
\put(72,32){\line(1,1){7}}
\put(75,61){\vector(0,-1){26}}

\put(57,30){$(\phi',k')$}
\put(77,61){$\dd X^{\mu (\phi',k')}$}
\put(69,0){$X^{\mu (\phi',k')}$}
\put(88,0){$X'^{\mu (\phi',k')}$}

\curvedashes[1mm]{0,1,1}
\curve(29,56, 34,46, 37,36, 41,20, 42,9)
\curve(64,64, 67,55, 72,48, 77,42, 82,35, 87,26, 89,11)

\thicklines

\spline(17,56)(20,46)(25,36)(29,20)(30,9)
\spline(52,64)(55,55)(60,48)(65,42)(70,35)(75,26)(77,11)

\end{picture}

\caption{Two different membrane configurations, represented by
the coordinates $X^{\mu (\phi,k)}$ and $X'^{\mu (\phi,k)}$, respectively.
The infinitesimal ${\cal M}$-space vector is $\dd X^{\mu (\phi,k)} =
X'^{\mu (\phi,k)} - X^{\mu (\phi,k)}$.
}

\end{figure}
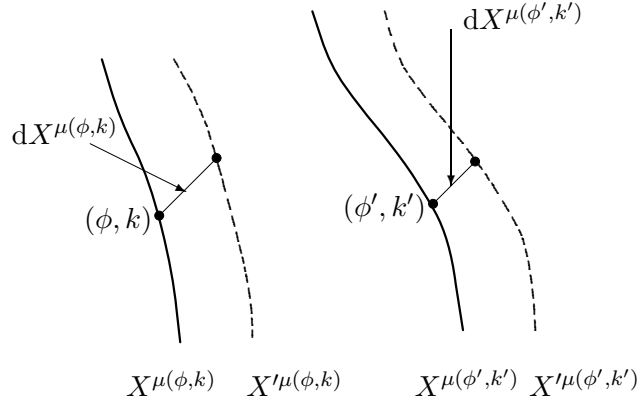

What is the geometric meaning of the quantity $\Delta s^2$ defined 
in (\ref{II2.84})?
To a point on a $k$-th membrane we can  assign parameters, or 
{\it intrinsic coordinates} \inxx{intrinsic coordinates},
$(\phi ,k)$. Alternatively, we can assign to the same points the
{\it extrinsic coordinates}\footnote{
Obviously $X^{\mu (\phi,k)} \equiv X^{\mu} (\phi,k)$ has a 
double role. On the one
hand it represents a {\it set} of all points with support 
on the system of membranes.
On the other hand it represents a {\it single point} 
in an infinite-dimensional $\cal M$-space.} \inxx{extrinsic coordinates}
$X^{\mu (\phi,k)} \equiv X^{\mu} (\phi,k)$, $\mu = 
1,2,...,N$, with $\phi , k$ fixed. It then seems
very natural to interpret $\Delta s^2$ as {\it the distance} 
between the points
$X^{\mu} (\phi',k')$ and $X^{\mu} (\phi,k)$ within a given 
membrane configutations.
The quantity $\rho_{\mu (\phi,k) \nu (\phi',k')}$ then has
the role of a prototype of the
{\it parallel propagator} \cite{Synge} in $V_N$. 
Note that we do not yet have a \inxx{parallel propagator,prototype of}
manifold $V_N$, but we already have a {\it skeleton} $S$ 
of it, formed by the set of
all the points with support on our membranes. 
The quantity ${\widetilde \Delta } X^{\mu} (\phi',k')$
is a prototype of a {\it vector} in $V_N$, and if we act 
on a vector with the parallel \inxx{vector,prototype of}
propagator we perform a {\it parallel transport} of the vector 
along the geodesic
joining the points $(\phi',k')$ and $(\phi,k)$. The latter 
terminology holds for a continuum \inxx{skeleton space}
space $V_N$, but we may retain it for our skeleton space $S$ as well. The
quantity $\rho_{\mu (\phi,k) \nu (\phi',k')} {\widetilde \Delta} X^{\nu}(\phi',k')$
is a vector, obtained from ${\widetilde \Delta} X^{\nu}(\phi',k')$ 
by parallel transport.

We define
\be
     \rho_{\mu (\phi,k) \nu (\phi',k')} {\widetilde \Delta} X^{\nu}(\phi',k') = 
     {\widetilde \Delta} X_{\mu} (\phi,k) .
\lbl{II2.85}
\ee
Then
\be
     \Delta s^2 = {\widetilde \Delta} X_{\mu} (\phi,k) {\widetilde \Delta} 
     X^{\mu} (\phi,k) .
\lbl{II2.86}
\ee
On the other hand, if in eq.\ (\ref{II2.84}) and (\ref{II2.85}) we take 
the coincidence limit $(\phi,k) \rightarrow (\phi',k')$, then
\be
    \Delta s^2 = \rho_{\mu \nu} (\phi,k) {\widetilde \Delta} x^{\mu} (\phi,k)
    {\widetilde \Delta} X^{\nu}(\phi,k) ,
\lbl{II2.87}
\ee
\be
    \rho_{\mu  \nu} (\phi,k) {\widetilde \Delta} X^{\nu}(\phi,k)  =  
  {\widetilde \Delta} x^{\mu} (\phi,k) ,
\lbl{II2.88}
\ee
with
\be
      \rho_{\mu \nu} (\phi,k) \equiv \rho_{\mu (\phi,k) \nu (\phi,k)} .
\lbl{II2.89}
\ee
We see that the quantity $\rho_{\mu (\phi,k) \nu (\phi,k)}$ has 
the property of the metric in the skeleton target space $S$.

Now we can envisage a situation in which the index $k$ is not discrete, but
consists of a set of continiuous indices such that instead of the discrete 
membranes we have a {\it fluid} of membranes.\inxx{fluid,of membranes} 
We can arrange a situation such that the fluid
fills an $N$-dimensional manifold $V_N$, and $\rho_{\mu \nu} (\phi,k)$ 
is the metric of $V_N$ defined at the points $(\phi,k)$ of the fluid. 
Here the fluid is a reference\inxx{reference system} 
system, or reference fluid  (being itself a part of the dynamical system under
consideration), with respect to which the points of the manifold $V_N$ 
are defined.\inxx{reference fluid}

The metric $\rho_{\mu \nu} (\phi,k)$ satisfies the functional 
differental equations \inxx{functional differential equations}
(\ref{II2.65}) which presumably reduce to the usual Einstein equations, 
at least \inxx{Einstein equations}
as an approximation. Later we shall show that this is indeed the case.
The description with a metric tensor will be surpassed in Sec.\ 6.2, 
when we shall discuss
a very promising description in terms of the Clifford algebra
equivalent of the
tetrad field which simplifies calculation significantly.
The specific theory based on the action (\ref{II2.61}) 
should be
considered as serving the purpose of introducing us to the
concept of mathematical objects, based on ${\cal M}$-space, representing
physical quantities at the fundamental level.

To sum up, we have taken {\it the membrane space} ${\cal M}$ seriously as an
arena for physics. The arena itself is also a part of the dynamical system, 
it is not \inxx{membrane space,as an arena for physics}
prescribed in advance. The membrane space ${\cal M}$ 
is the space of all kinematically
possible membrane configurations. Which particular 
configuration is actually realized is
selected by the functional equations of motion together with the initial 
and boundary conditions. A configuration may consist of many 
membranes. The points located on
all those membranes form a space which, at a sufficiently large scale, is a 
good approximation
to a continuous $N$-dimensional space. The latter space
is just {\it the embedding
space} for a particular membrane.

We have thus formulated a theory in which an embedding space {\it per se} 
does not exist, but is intimately connected to the existence of  membranes. 
Without membranes there is no embedding space. This approach
is {\it background independent}.
There is no pre-existing space and metric: they appear 
dynamically as solutions to the\inxx{background independent approach}
equations of motion.

The system, or condensate of membranes (which, in particular, may be so dense
that the corrseponding points form a continuum), represents a 
{\it reference system}
or {\it reference fluid} with respect to which 
the points of the target space are
defined. Such a system was postulated by DeWitt \ci{15}, and recently 
reconsidered by Rovelli \ci{16}.

The famous {\it Mach principle} states that the motion of 
matter at a given location \inxx{Mach principle}
is determined by the contribution of all the matter in the universe and this
provides an explanation for inertia (and inertial mass). 
Such a situ\-ation is implemented
in the model of a universe consisting of a system of membranes: 
the motion of a $k$-th
membrane, including its inertia, is determined by the presence of all the other
membranes.

In my opinion the ${\cal M}$-space approach is also somehow 
related to {\it loop quantum gravity} proposed by Rovelli \ci{55} and Smolin
\ci{56}. \inxx{loop quantum gravity}

It seems to me that several concepts which have been around in 
theoretical physics for some time, and especially those sketched above, 
all naturally emerge from ${\cal M}$-space physics.

\upperandlowerfalse
\chapter[Extended objects and Clifford algebra]
{Extended objects\\
and Clifford algebra}

We have seen that geometric calculus based on Clifford algebra is
\inxx{geometric calculus}
a very useful tool for a description of geometry and point particle
physics in {\it flat spacetime}. In Sec. 4.2 we have encountered
a generalization of the concepts of vectors and polyvectors in
an infinite-dimensional ${\cal M}$-space, which is in general a
curved space. What we need now is a more complete description,
firstly, of the usual finite-dimensional vectors and polyvectors in
{\it curved space}, and secondly, of the corresponding objects in
an infinite-dimensional space (which may be ${\cal M}$-space,
in particular).

After such mathematical preliminaries I shall discuss a membrane
description which employs ${\cal M}$-space basis vectors. 
A background independent\inxx{basis vectors,in ${\cal M}$-space}
 approach will be achieved by proclaiming
the basis vectors themselves as dynamical objects. The corresponding
action has its parallel in the one considered in the last section
of previous chapter. However, now the dynamical object is not
the ${\cal M}$-space metric, but its ``square root".

Until the techniques of directly solving the functional differential
equations are more fully developed we are forced to make the transition
from functional to the usual partial differential equations, i.e., from
the infinite-dimensional to the finite-dimensional differential
equations. Such a transition will be considered in Sec. 6.2, where
a description in terms of position-dependent target space vectors
will be provided. We shall observe that a membrane is a sort of a
\inxx{reference fluid}
fluid localized in the target space, and exploit this concept in relation
to the DeWitt-Rovelli reference fluid necessary for localization of
spacetime points. Finally we shall realize that we have just touched
the tip of an iceberg which is the full polyvector description of
the membrane together with the target space in which the membrane
is embedded.

\upperandlowerfalse
\section{Mathematical preliminaries}

I will now provide an intuitive description of vectors in curved spaces
\inxx{vectors,in curved spaces }
and their generalization to the infinite-dimensional (also curved) spaces.
The important concept of the vector and multivector, or polyvector,
derivative will also be explained. The usual functional derivative is
just a component description of the vector derivative 
in an infinite-dimensional space. My aim is to introduce 
the readers into those very
elegant mathematical concepts and give them a feeling about their
practical usefulness. To those who seek a more complete mathematical
rigour I advise consulting the literature \cite{13}. In the case in which
the concepts
discussed here are not found in the existing literature the interested reader
is invited to undertake the work and to develop the ideas initiated
here further in order to put them into a more rigorous mathematical envelope.
Such a development is beyond the scope of this book which aims to
point out how various pieces of physics and mathematics are starting
to merge before our eyes into a beautiful coherent picture.

\upperandlowerfalse
\subsection[\ \ Vectors in curved spaces]
{Vectors in Curved Spaces}

Basis vectors need not be equal at all point of a manifold $V_N$. They may
be position-dependent. Let $\gamma_\mu (x)$, $\mu = 1,2,...,N$, be $N$ linearly
independent coordinate vectors which depend on position $x^{\mu}$. The inner
products of  the basis vectors form the metric tensor which is 
also position-dependent:
\be
      \gamma_{\mu} \cdot \gamma_{\nu} = g_{\mu \nu} ,
\lbl{II3.1}
\ee
\be
       \gamma^{\mu} \cdot \gamma^{\nu} = g^{\mu \nu} .
\lbl{II3.1a}
\ee

What is the relation between vectors at two successive infinitesimally
separated points $x^{\mu}$ and $x^{\mu} + {\dd} x^{\mu}$? Clearly coordinate
vectors at $x^{\mu} +{\dd} x^{\mu}$ after the parallel transport to $x^\mu$
cannot be linearly independent from the vectors at $x^{\mu}$, since 
they both belong to the same manifold.
Therefore the derivative of a basis vector must be a linear
combination of basis vectors:
\be
    \p_{\alpha} \gamma^{\mu} = - \Gamma_{\alpha \beta}^{\mu} \gamma^{
   \beta} .
\lbl{II3.2}
\ee
Denoting by $\gamma^{\mu} (x + \dd x)$ a vector
transported from $x+\dd x$ to $x$, we have
$$\gamma^{\mu} (x + \dd x) = \gamma^{\mu} (x) + \p_{\alpha} \gamma^{\mu}
(x) \dd x^{\alpha} = ({\delta^{\mu}}_{\alpha} - \Gamma_{\alpha \beta}^{\mu}
  \dd x^{\beta} ) \gamma^{\alpha},$$
which is indeed a linear combination of
$\gamma^{\alpha} (x)$.

The coefficients $\Gamma_{\alpha \beta}^{\mu}$ form {\it the connection} of 
manifold $V_N$.
From (\ref{II3.2}) we have \inxx{connection}
\be
      \Gamma_{\alpha \beta}^{\mu} = - \gamma_{\beta} \cdot \p_{\alpha} 
    \gamma^{\mu} .
\lbl{II3.2a}
\ee
The relation $\gamma^{\mu} \cdot \gamma_{\beta} = {\delta^{\mu}}_{\beta}$
implies
\be
\p_{\alpha} \gamma^{\mu} \cdot \gamma_{\beta} + \gamma^{\mu} \cdot
\p_{\alpha} \gamma_{\beta} = 0,
\lbl{II3.2aa}
\ee
 from which it follows that
\be
      \Gamma_{\alpha \beta}^{\mu} = \gamma^{\mu} \cdot \p_{\alpha} 
      \gamma_{\beta} .
\lbl{II3.2b}
\ee
Multiplying  the latter expression by $\gamma_{\mu}$, summing over ${\mu}$ and
using
\be
   \gamma_{\mu} (\gamma^{\mu} \cdot a) = a ,
\lbl{II3.2bb}
\ee
which holds for any vector $a$, we obtain
\be
      \p_{\alpha} \gamma_{\beta} = \Gamma_{\alpha \beta}^{\mu} \gamma_{\mu} .
\lbl{II3.2c}
\ee
In general the connection is {\it not symmetric}. In particular, when it is {\it
symmetric}, we have
\be
     \gamma^{\mu} \cdot (\p_{\alpha} \gamma_{\beta} - \p_{\beta} 
     \gamma_{\alpha}) = 0
\lbl{II3.2d}
\ee
and also
   \be
          \p_{\alpha} \gamma_{\beta} - \p_{\beta} \gamma_{\alpha} = 0
\lbl{II3.2e}
\ee
in view of the fact that (\ref{II3.2d}) holds for any $\gamma^{\mu}$. For a
symmetric connection, after using (\ref{II3.1}),\ (\ref{II3.1a}),\ (\ref{II3.2e})
we find
\be
       \Gamma_{\alpha \beta}^{\mu} = 
       \mbox{${1\oo 2}$} g^{\mu \nu} (g_{\nu \alpha,
   \beta} + g_{\nu \beta, \alpha} - g_{\alpha \beta, \nu}).
\lbl{II3.2f}
\ee

Performing the second derivative we have
\be
    \p_{\beta} \p_{\alpha} \gamma^{\mu} = - \p_{\beta} 
    \Gamma_{\alpha \sigma}^{\mu}
   \gamma^{\sigma} - \Gamma_{\alpha \sigma}^{\mu} \p_{\beta} \gamma^{\sigma} =
 - \p_{\beta} \Gamma_{\alpha \sigma}^{\mu} \gamma^{\sigma} +
   \Gamma_{\alpha \sigma}^{\mu} \Gamma_{\beta \rho}^{\sigma} \gamma^{\rho}
\lbl{II3.3}
\ee
and
\be
   [\p_{\alpha}, \p_{\beta}] \gamma^{\mu} = {R^{\mu}}_{\nu \alpha \beta} 
  \gamma^{\nu} \; ,
\lbl{II3.4}
\ee
where
\be
       {R^{\mu}}_{\nu \alpha \beta} = \p_{\beta} 
       \Gamma_{\nu \alpha}^{\mu} - \p_{\alpha} \Gamma_{\nu \beta}^{\mu} + 
       \Gamma_{\beta \rho}^{\mu}
  \Gamma_{\alpha \nu}^{\rho} - \Gamma_{\alpha \rho}^{\mu} \Gamma_{\beta
  \nu}^{\rho}
\lbl{II3.5}
\ee
is {\it the curvature tensor}. In general the latter tensor does not 
vanish and we have a curved space. \inxx{curvature tensor}

\subsubsection{Some Illustrations}

\paragraph{Derivative of a vector} \inxx{derivative,of a vector}
Let $a$ be an arbitrary  position-dependent  vector, expanded according to
\be
       a = a^{\mu} \gamma_{\mu} .
\lbl{II3.8}
\ee
Taking the derivative with respect to coordinates $x^{\mu}$ we have
\be
       \p_{\nu} a = \p_{\nu} a^{\mu} \, \gamma_{\mu} + a^{\mu} \p_{\nu}
    \gamma_{\mu} .
\lbl{II3.9}
\ee
Using (\ref{II3.2c}) and renaming the indices we obtain
\be
    \p_{\nu} a = \left ( \p_{\nu} a^{\mu} + \Gamma_{\nu \rho}^{\mu} a^{\rho}
   \right ) \gamma_{\mu} ,
\lbl{II3.10}
\ee
or
\be
     \gamma^{\mu} \cdot \p_{\nu} a = \p_{\nu} a^{\mu} + \Gamma_{\nu \rho}^{\mu} 
    a^{\rho} \equiv {\DD}_{\nu} a^{\mu} ,
\lbl{II3.10a}
\ee
which is the well known \inxx{covariant derivative}
{\it covariant derivative}. The latter derivative is the
projection of $\p_{\nu} a$ onto one of the basis vectors.

\paragraph{Locally inertial frame}
At each point of a space $V_N$ we can define a set of $N$ linearly independent
vectors $\gamma_a$, $a = 1,2,...,N$, satisfying
\be
       \gamma_a \cdot \gamma_b = \eta_{a b} ,
\lbl{II3.11} 
\ee
where $\eta_{a b}$ is the Minkowski tensor. The set of vector fields 
$\gamma_a (x)$ will be called the {\it inertial or Lorentz 
(orthonormal) frame field}.
\inxx{frame field}.

A coordinate basis vector can be expanded in terms of local basis vector
\be
         \gamma^{\mu} = {e^{\mu}}_a \, \gamma^a ,
\lbl{II3.12}
\ee
where the expansion coefficients ${e^{\mu}}_a$ form the so called {\it fielbein
field} (in 4-dimensions ``fielbein" becomes ``vierbein" or ``tetrad"). 
\inxx{tetrad}
\be
          {e^{\mu}}_a = \gamma^{\mu} \cdot \gamma_a .
\lbl{II3.13}
\ee
Also
\be
    \gamma_a = {e^{\mu}}_a \gamma_{\mu} ,
\lbl{II3.14}
\ee
and analogous relations for the inverse vectors $\gamma_{\mu}$ and
$\gamma^a$ satisfying
\be
           \gamma^{\mu} \cdot \gamma_{\nu} = {\delta^{\mu}}_{\nu} ,
\lbl{II3.15}
\ee
\be
          \gamma^a \cdot \gamma_b = {\delta^a}_b ,
\lbl{II3.16}
\ee
From the latter relations we find
\be
      \gamma^{\mu} \cdot \gamma^{\nu} = ({e^{\mu}}_a \gamma^a) 
   \cdot ({e^{\nu}}_b \gamma^b) = {e^{\mu}}_a {e_{\nu}}^a = g^{\mu \nu} ,
\lbl{II3.16a}
\ee
\be
    \gamma^a \cdot \gamma^b = ({e_{\mu}}^a \gamma^{\mu}) \cdot
  ({e_{\nu}}^b \gamma^{\nu}) = {e_{\mu}}^a {e^{\mu b}} = \eta^{a b} .
\lbl{II3.16b}
\ee

A vector $a$ can be expanded either in terms of $\gamma_{\mu}$ or
$\gamma_a$:
\be
      a = a^{\mu} \gamma_{\mu} = a^{\mu} {e_{\mu}}^a \gamma_a = a^a \gamma_a
   \; , \qquad a^a = a^{\mu} {e_{\mu}}^a .
\lbl{II3.17}
\ee
Differentiation gives
\be
    \p_{\mu} \gamma^a = {\omega^a}_{b \mu} \gamma^b,
\lbl{II3.18} 
\ee
where ${\omega^a}_{b \mu}$ is the connection for the orthonormal frame 
field $\gamma^a$.
Inserting (\ref{II3.12}) into the relation (\ref{II3.2}) we obtain
\be
    \p_{\nu} \gamma^{\mu} = \p_{\nu} ({e^{\mu}}_a \gamma^a) = \p_{\nu}
  {e^{\mu}}_a \gamma^a + {e^{\mu}}_a \p_{\nu} \gamma^a = - \Gamma_{\nu \sigma}^{
  \mu} \gamma^{\sigma} ,
\lbl{II3.19}
\ee
which, in view of (\ref{II3.18}), becomes
\be
     \p_{\nu} e^{\mu a} + \Gamma_{\nu \sigma}^{\mu} e^{\sigma a} +
     {\omega^{ab}}_{\nu} {e^{\mu}}_b = 0 .
\lbl{II3.20}
\ee
Because of (\ref{II3.16a}),\, (\ref{II3.16b}) we have
\be
    \p_{\nu} {e_{\mu}}^a - \Gamma_{\nu \mu}^{\sigma} {e_{\sigma}}^a +
   {\omega^a}_{b \nu} {e_{\mu}}^b = 0 .
\lbl{II3.20a}
\ee
These are the well known relations for differentiation of the fielbein field.

\paragraph{Geodesic equation in $V_N$} Let $p$ be the momentum vector
satisfying the equation of motion \inxx{geodesic equation}
\be
     {\dd p\oo {\dd \tau}} = 0 .
\lbl{II3.5aa}
\ee
Expanding $p = p^{\mu} \gamma_{\mu}$, where $p^{\mu} = m {\dot X}^{\mu}$,
we have
\be
    {\dot p}^{\mu} + p^{\mu} {\dot \gamma}_{\mu} = {\dot p}^{\mu}
   \gamma_{\mu} + p^{\mu} \p_{\nu} \gamma_{\mu} \, {\dot X}^ {\nu} .
\lbl{II3.6}
\ee
Using (\ref{II3.2c}) we obtain, after suitably renaming the indices,
\be
     ({\dot p}^{\mu}
      +\Gamma_{\alpha \beta}^{\mu} p^{\alpha} {\dot X}^{
   \beta}) \gamma_{\mu} = 0 ,
\lbl{II3.7}
\ee
which is {\it the geodesic equation} in component notation.
 The equation of motion (\ref{II3.5aa}) says that the vector $p$ does not
change during the motion. This means that vectors $p(\tau)$ for all values
of the parameter $\tau$ remain parallel amongst themselves (and, of course,
retain the same magnitude square $p^2$). After using the expansion $p
= p^{\mu} \gamma_{\mu}$ we find that the change of the components
$p^{\mu}$ is compensated by the change of basis vectors $\gamma_{\mu}$.

\paragraph{Geometry in a submanifold $V_n$} In the previous example we
considered a geodesic equation in spacetime $V_N$. Suppose now that a
submanifold  --- a surface --- $V_n$, parametrized by $\xi^a$, \inxx{submanifold}
is embedded in $V_N$. Let\footnote{
Notice that the index $a$ has now a different meaning from that in the case of
a locally inertial frame considered before.} \inxx{tangent vectors}
$e^a$, $a = 1,2,...,n$, be a set of tangent vectors to $V_n$. They can be
expanded in terms of basis vectors of $V_N$:
\be
    e_a = \p_a X^{\mu} \, \gamma_{\mu},
\lbl{II3.21}
\ee
where
\be
     \p_a X^{\mu} = e_a \cdot \gamma^{\mu}
\lbl{II3.22}
\ee
are derivatives of the embedding functions of $V_n$. They satisfy
\be
    \p_a X^{\mu} \, \p_b X_{\mu} = (e_a \cdot \gamma^{\mu})
    (e_b \cdot \gamma_{\mu}) = e_a \cdot e_b = \gamma_{ab} ,
\lbl{II3.22a}
\ee
which is the expression for the induced metric of $V_n$. Differentiation
of $e_a$ gives
\be
   \p_b e_a = \p_b \p_a X^{\mu} \, \gamma_{\mu} + \p_a X^{\mu} \p_b  .
  \gamma_{\mu}
\lbl{II3.23}
\ee
Using
\be
    \p_b \gamma_{\mu} = \p_{\nu} \gamma_{\mu} \p_b X^{\nu}
\lbl{II3.24}
\ee
and the relation (\ref{II3.2c}) we obtain from (\ref{II3.23}), after performing
the inner product with $e^c$,
\be
    (e^c \cdot \p_b e_a) = \p_a \p_b X^{\mu}\, \p^c X_{\mu} +
  \Gamma_{\mu \nu}^{\sigma} \p_a X^{\mu} \p_b X^{\nu} \p^c X_{\sigma} .
  \lbl{II3.25}
\ee
On the other hand, the left hand side of eq.\ (\ref{II3.23}) involves the connection
of $V_n$:
\be    
    e^c \cdot \p_b e_a = \Gamma_{ba}^c ,
\lbl{II3.26}
\ee
and so we see that eq.\ (\ref{II3.25}) is a relation between the connection of
$V_n$ and $V_N$. Covariant derivative in the submanifold $V_n$ is
defined in terms of $\Gamma_{ba}^d$.

An arbitrary vector $P$ in $V_N$ can be expanded in terms of $\gamma_{\mu}$:
\be
      P = P_{\mu} \gamma^{\mu} .
\lbl{II3.27}
\ee
It can be projected onto a tangent vector $e_a$: \inxx{tangent vectors}
\be
    P \cdot e_a = P_{\mu} \gamma^{\mu} \cdot e_a  = P_{\mu} \p_a X^{\mu}
   \equiv P_a
\lbl{II3.28}
\ee
\hs{4mm}In particular, a vector of $V_N$ can be itself a tangent vector of a subspace
$V_n$. Let $p$ be such a tangent vector. It can be expanded either in terms
of $\gamma_{\mu}$ or $e_a$:
\be
     p = p^{\mu} \gamma^{\mu} = p_a e^a ,
\lbl{II3.29}
\ee
where
    $$p_a = p_{\mu} \gamma^{\mu} \cdot e_a = p_{\mu} \p_a X^{\mu} ,$$ 
\be   
    p_{\mu} = p_a e^a \cdot \gamma_{\mu} = p_a \p^a X_{\mu} .
\lbl{II3.31}
\ee
Such symmetric relations between $p_{\mu}$ and $p_a$ hold only for a vector
$p$ which is tangent to $V_n$.

Suppose now that $p$ is tangent to a geodesic of $V_n$. Its derivative with 
respect to an invariant parameter $\tau$ along the geodesic is
\bear
      {\dd \oo {\dd \tau}} (p^a e_a) &=& {\dd \oo {\dd \tau}} 
      \left ( p^a \p_a X^{\mu}
     \gamma_{\mu} \right ) \nonumber \\
      &=& ({\dot p}^a \p_a X^{\mu} + \p_a \p_b X^{\mu} p^a
   {\dot \xi}^b + p^a \p_a X^{\alpha} \Gamma_{\alpha \beta}^{\mu}
   {\dot X}^{\beta} ) \gamma_{\mu}
\lbl{II3.32}
\ear
where we have used eq.\ (\ref{II3.2c}) and 
    $${\dd \oo {\dd \tau}} \p_a X^{\mu} =
                           \p_a  \p_b X^{\mu}  {\dot \xi}^b .$$
The above derivative, in general, does not
vanish: a vector $p$ of $V_N$ that is tangent to a geodesic in a subspace $V_n$
changes with $\tau$.

Making the inner product of the left and the right side of eq.\ (\ref{II3.32}) 
with $e^c$ we obtain
\be
   {{\dd p}\oo {\dd \tau}} \cdot e^c = {\dot p}^c + \Gamma_{ab}^c p^a 
    {\dot \xi}^b = 0 .
\lbl{II3.33}
\ee
Here $\Gamma_{ab}^c$ is given by eq.\ (\ref{II3.25}). For a vector $p$
tangent to a geodesic of $V_n$ the right hand side of eq.\ (\ref{II3.33})
vanish.

On the other hand, starting from $p = p^{\mu} \gamma_{\mu}$ and using
(\ref{II3.2c}) the left hand side of eq.\ (\ref{II3.33}) gives
\be
     {{\dd p}\oo {\dd \tau}} \cdot e^c = \left ( {{\dd p^{\mu}}\oo {\dd \tau}} +
   \Gamma_{\alpha \beta}^{\mu} p^{\alpha} {\dot X}^{\beta} \right )
  \p^c X_{\mu} = 0 .
\lbl{II3.34}
\ee
Eqs.(\ref{II3.33}),\ (\ref{II3.34}) explicitly show that in general a geodesic
of $V_n$ is not a geodesic of $V_N$.

A warning is necessary. We have treated tangent vectors $e_a$ to a subspace
$V_n$ as vectors in the embedding space $V_N$. As such they do {\it not} form
a complete set of linearly independent vectors in $V_N$. An arbitrary vector
of $V_N$, of course, cannot be expanded in terms of $e_a$; only a tangent
vector to $V_n$ can be expanded so. Therefore, the object 
$\left ( {{\dd p}\oo {\dd \tau}} \cdot e^c \right ) e_c$ 
should be distinguished from the object 
$\left ( {{\dd p}\oo {\dd \tau}} \cdot 
\gamma^{\mu} \right ) \gamma_{\mu} = \dd p/\dd \tau$. 
The vanishing of the former object
does not imply the vanishing of the latter object.

\subsubsection{Derivative with Respect to a Vector}
\inxx{derivative,with respect to a vector}

So far we have considered derivatives of position-dependent vectors with respect
to (scalar) coordinates. We shall now consider the derivative with respect to a
vector. Let $F(a)$ be a polyvector-valued function of a vector valued 
argument $a$
which belongs to an $n$-dimensional vector space ${\cal A}_n$. For an arbitrary
vector $e$ in ${\cal A}_n$ the derivative of $F$ in the direction $e$ is given by
\be
    \left ( e \cdot {\p \oo {\p a}} \right ) F(a) = 
    \lim_{\tau \to 0}{{F(a + e \tau)
   - F(a)}\oo \tau} = {{\p F(a + e \tau)}\oo {\p \tau}} \Biggl\vert_{\tau=0}.
\lbl{II3.35}
\ee
For $e$ we may choose one of the basis vectors. Expanding $a = a^{\nu} e_{\nu}$,
we have
    $$ \left ( e_{\mu}  \cdot {\p \oo {\p a}} \right ) F(a) \equiv {{\p F}\oo 
     {\p a^{\mu}}}
   = \lim _{\tau \to 0} {{F(a^{\nu} e_{\nu} + e_{\mu} \tau) - F(a^{\nu} 
   e_{\nu})}\oo   \tau} \hs{2cm}$$
\be   
    \hs{2.5cm} = \lim_{\tau \to 0} {{F \left ((a^{\nu} + 
   {\delta^{\nu}}_{\mu} 
   \tau) e_{\nu} \right ) - F(a^{\nu} e_{\nu})}\oo \tau} .
\lbl{II3.36}
\ee
The above derivation holds for an arbitrary function $F(a)$. For instance, for
$F(a) = a = a^{\nu} e_{\nu}$ eq.\ (\ref{II3.36}) gives
\be
    {{\p F}\oo {\p a^{\mu}}} = {\p \oo {\p a^{\mu}}} (a^{\nu} e_{\nu}) = 
    e_{\mu} = {\delta_{\mu}}^{\nu} e_{\nu}
\lbl{II3.37}
\ee
For the components $F \cdot e^{\alpha} = a^{\alpha}$ it is
\be
     {{\p a^{\alpha}}\oo {\p a^{\mu}}} = {\delta_{\mu}}^{\alpha}
\lbl{II3.38}
\ee
The derivative in the direction $e_{\mu}$, as derived in (\ref{II3.36}), is the
partial derivative with respect to the component $a^{\mu}$ of the vector 
argument $a$. (See Box 6.1 for some other examples.)

In eq.\ (\ref{II3.36}) we have derived the operator
\be
        e_{\mu} \cdot {\p \oo {\p a}} \equiv {\p \oo {\p a^{\mu}}} .
\lbl{II3.39}
\ee
\nnnn For a running index $\mu$ these are components (or projections) 
of the operator
\be
      {\p \oo {\p a}} = e^{\mu} \, \left ( e_{\mu} \cdot {\p \oo {\p a}}
      \right ) = e^{\mu} {\p \oo {\p a^{\mu}}} 
\lbl{II3.39a}
\ee
which is {\it the derivative with respect to a vector} $a$.

The above definitions (\ref{II3.35})--(\ref{II3.39}) hold for any vector $a$
of ${\cal A}_n$. Suppose now that all those vectors are defined at a point
$a$ of an $n$-dimensional manifold $V_n$. They are said to be {\it tangent}
to a point $x$ in $V_n$ \cite{13}. If we allow the point $x$ to vary
we have thus a vector field $a(x)$. \inxx{vector field}
In components it is
\be
      a(x) = a^{\mu} (x) e_{\mu} (x) ,
\lbl{II3.40}
\ee
where $a^{\mu} (x)$ are arbitrary functions of $x$. A point $x$ is parametrized
by a set of $n$ coordinates $x^{\mu}$, hence $a^{\mu} (x)$ are functions of
$x^{\mu}$. In principle $a^{\mu} (x)$ are arbitrary functions of $x^{\mu}$.
In particular, we may choose
\be
      a^{\mu} (x) = x^{\mu} .
\lbl{II3.40a}
\ee
\setlength{\unitlength}{1cm}

\nnnn \begin{picture}(12.5,9.6)

\put(0,0){\framebox(12.5,9.6){
\begin{minipage}[t]{11.5cm}

\centerline{\bf Box 6.1: Examples of differentiation by a vector }

\vs{2mm}

1) {\it Vector valued function}:
$$F = x = x^{\nu} e_{\nu}\; ; \quad {{\p F}\oo {\p x^{\mu}}} = e_{\mu} =
 {\delta_{\mu}}^{\nu} e_{\nu} \; ; \quad {{\p F}\oo {\p x}} = e^{\mu} {{\p F}\oo
 {\p x^{\mu}}} = e^{\mu} e_{\mu} = n .$$

2) {\it Scalar valued function}
  $$F = x^2 = x^{\nu} x_{\nu} \; ; \quad {{\p F}\oo {\p x^{\mu}}} = 2 x_{\mu} \; ;
  \quad {{\p F}\oo {\p x}} = e^{\mu} 2 x_{\mu} = 2 x .$$

3) {\it Bivector valued function}
     $$ F = b \wedge x = (b^{\alpha} e_{\alpha}) \wedge (x^{\beta} e_{\beta}) =
     b^{\alpha} x^{\beta} \, e_{\alpha} \wedge e_{\beta} ,$$
     $$ {{\p F}\oo {\p x^{\mu}}} = \lim_{\tau \to 0} 
     {{b^{\alpha} e_{\alpha} \wedge
   (x^{\beta} e_{\beta} + e_{\mu} \tau) - 
   b^{\alpha} e_{\alpha} \wedge x^{\beta}
  e_{\beta}}\oo \tau} = b^{\alpha} e_{\alpha} \wedge e_{\mu},$$
  $${{\p F} \oo {\p x}} = e^{\mu} {{\p F}\oo {\p x^{\mu}}} = b^{\alpha}
  e^{\mu} (e_{\alpha} \wedge e_{\mu})
  = b^{\alpha} e^{\mu} \cdot (e_{\alpha} \wedge e_{\mu}) + b^{\alpha}
    e^{\mu} \wedge e_{\alpha} \wedge e_{\mu},$$ 
    $$ \hs{2cm}= b^{\alpha} ( {\delta_{\alpha}}^{\mu}
  e_{\mu} - {\delta_{\mu}}^{\mu} e_{\alpha}) = b^{\alpha} e_{\alpha} (1 - n)
  = b (1 - n) .$$

\end{minipage} }}
\end{picture}

\vs{4mm}

\nnnn Then
\be
       a(x) = x^{\mu} e_{\mu} (x) .
\lbl{II3.40b}
\ee
Under a passive coordinate transformation the components $a^{\mu} (x)$ change
according to
\be
   a'^{\mu} (x') = {{\p x'^{\mu}}\oo {\p x^{\nu}}} a^{\nu} (x) .
\lbl{II3.41}
\ee
This has to be accompanied by the corresponding (active) change of basis
vectors,
\be
      e'^{\mu} (x') = {{\p x^{\nu}}\oo {\p x'^{\mu}}} e_{\nu} (x) ,
\lbl{II3.41a}
\ee
in order for a vector $a (x)$ to remain unchanged. In the case in which
the components fields $a^{\mu} (x)$ are just coordinates themselves,
the transformation (\ref{II3.41}) reads
\be
      a'^{\mu} (x') = {{\p x'^{\mu}}\oo {\p x^{\nu}}} x^{\nu} .
\lbl{II3.42}
\ee
In new coordinates $x'^{\mu}$ the components $a'^{\mu}$ of a vector $a(x) =
x^{\mu} e_{\mu}$ are, of course, not equal to the new coordinates $x'^{\mu}$.
The reader can check by performing some explicit transformations (e.g., from
the Cartesian to spherical coordinates) that an object as defined 
in (\ref{II3.40a}),\ (\ref{II3.40b})
is quite a legitimate geometrical object and has, indeed, the required
properties of \inxx{vector field,in curved space}
a vector field, even in a curved space. The set of points of a curved space can
then, at least locally\footnote{Globally this canot be true in general, since a
single coordinate system cannot cover all the space.}, \inxx{vector field}
be considered as a vector field, such that its components in a certain basis
are coordinates. In a given space, there are infinitely many fields with such a
property, one field for every possible choice of coordinates. As an illustration
I provide the examples of two such fields, denoted $X$ and $X'$:

      $$a(x) = x^{\mu} e_{\mu} (x) = a'^{\mu} (x') e'_{\mu} (x') = X ,$$
\be      
      b(x) = b^{\mu}(x) e_{\mu} (x) = x'^{\mu} e'_{\mu} (x') = X' .
\lbl{II3.42a}
\ee

Returning to the differential operator (\ref{II3.35}) we can consider 
$a$ as a vector \inxx{vector field}
field $a(x)$ and the definition (\ref{II3.35}) is still valid at every 
point $x$ of
$V_n$. In particular we can choose
\be
      a(x) = x = x^{\nu} e_{\nu} .
\lbl{II3.43}
\ee
Then (\ref{II3.36}) reads
\be
   \left ( e_{\mu} \cdot {\p \oo {\p x}} \right ) F(x) \equiv {{\p F} \oo
   {\p x^{\mu}}} = \lim_{\tau \to 0} {{F \left ( (x^{\nu} + {\delta_{\mu}}^{\nu}
   \tau ) e_{\nu} \right ) - F(x^{\nu} e_{\nu})}\oo \tau} .
\lbl{II3.44}
\ee
This is the partial derivative of a multivector valued function $F(x)$ of
position $x$. The derivative with respect to the polyvector $x$ is
\be
     {{\p}\oo {\p x}} = e^{\mu} \left ( e_{\mu}'\cdot {\p \oo {\p x}}
     \right ) = e^{\mu}
     {\p \oo {\p x^{\mu}}}
\lbl{II3.44a}
\ee

Although we have denoted the derivative as $\p/\p x$ or $\p/\p a$, this
notation should not be understood as implying that $\p/\p a$ can be defined as
the limit of a difference quotient. The partial derivative (\ref{II3.36}) can be
so defined, but not the derivative with respect to a vector.

\subsubsection{Derivative with Resepct to a Polyvector}
\inxx{derivative,with resepct to a polyvector}

The derivative with respect to a vector can be generalized to  polyvectors. 
Definition (\ref{II3.35}) is then replaced by
\be
   \left (E * {\p \oo {\p A}} \right ) F(A) = 
   \lim_{\tau \to 0} {{F(A + E \tau) - F(A)}\oo \tau}
= {{\p F (A + E \tau)}\oo {\p \tau}} .
\lbl{II3.45}
\ee

Here $F(A)$ is a polyvector-valued function of a polyvector $A$, and $E$ is an
arbitrary polyvector . The star `` * " denotes the scalar product
\inxx{scalar product}
\be
      A * B = \langle A B \rangle_0
\lbl{II3.45b}
\ee      
of two polyvectors $A$ and $B$, where $\langle A B \rangle_0$ is the scalar
part of the Clifford product $AB$. Let $e_J$ be a
complete set of basis vector of Clifford algebra satisfying\footnote{
Remember that the set $\lbrace e_J \rbrace = \lbrace 1,e_{\mu}, e_{\mu},
e_{\mu} e_{\nu}, ... \rbrace$.}
\be
      e_J * e_K = \delta_{JK} ,
\lbl{II3.45a}
\ee
so that any polyvector can be expanded as $A = A^J e_J$. For $E$ 
in eq.\ (\ref{II3.45})
we may choose one of the basis vectors. Then
\be
\left (e_K * {\p \oo {\p A}} \right ) F(A) \equiv {{\p F}\oo {\p A^K}} = 
\lim_{\tau \to 0}
{{F(A^J e_J + e_K \tau) - F(A^J e_J)}\oo \tau} .
\lbl{II3.46}
\ee
This is the partial derivative of $F$ with respect to the multivector components
$A_K$. The derivative with respect to a polyvector $A$ is the sum
\be
  {{\p F}\oo {\p A}} = e^J \left ( e_J * {\p \oo {\p A}} \right ) F = 
  e^J {{\p F} \oo {\p A^J}} .
\lbl{II3.47}
\ee
The polyvector $A$ can be a polyvector field \inxx{polyvector field}
$A(X)$ defined over the position \inxx{position polyvector field}
polyvector field $X$ which \inxx{position vector field}
is a generalizatin of the position vector field $x$
defined in (\ref{II3.42a}). In particular, the field $A(X)$ can be $A(X) = X$. 
Then (\ref{II3.46}),\ (\ref{II3.47}) read
\be
\left (e_K * {\p \oo {\p X}} \right ) F(X) \equiv {{\p F}\oo {\p X^K}} = 
\lim_{\tau \to 0}
{{F(X^J e_J + e_K \tau) - F(X^J e_J)}\oo \tau} ,
\lbl{II3.48}
\ee
\be
  {{\p F}\oo {\p X}} = e^J \left ( e_J * {\p \oo {\p X}} \right ) F = 
  e^J {{\p F} \oo {\p X^J}}
\lbl{II3.49}
\ee
which generalizes eqs.\ (\ref{II3.44}),{\ref{II3.44a}).

\subsection[\ \ Vectors in an infinite-dimensional space]
{Vectors in an Infinite-Dimensional Space}

In functional analysis functions are considered as vectors in  infinite
dimensional spaces. As in the case of finite-dimensional spaces one can
introduce a basis $h(x)$ in an infinite-dimensional space $V_{\infty}$
and expand a vector of $V_{\infty}$ in terms of the basis vectors.
The expansion coefficients form a function $f(x)$:
\be
      f = \int \dd x \, f(x) h(x) .
\lbl{II3.50}
\ee
The basis vectors $h(x)$ are elements of the Clifford algebra ${\cal C}_{\infty}$
of $V_{\infty}$. The Clifford or geometric product of two vectors  is
\inxx{geometric product,in infinite-dimensional space}
\inxx{Clifford product,in infinite-dimensional space}
\be
    h(x) h(x') = h(x) \cdot h(x') + h(x) \wedge h(x') .
\lbl{II3.51}
\ee

{\it The symmetric part}
\be
      h(x) \cdot h(x') =  \mbox{$1\oo2$}  \left ( h(x) h(x') + h(x') h(x) \right )
\lbl{II3.52}
\ee
is {\it the inner product} and {\it the antisymmetric part}
\inxx{inner product,in infinite-dimensional space}}
\be
    h(x) \wedge h(x') = \mbox{$1\oo2$} \left ( h(x) h(x') - h(x') h(x) \right )
\lbl{II3.53}
\ee    
is {\it the outer, or wedge, product} of two vectors.
\inxx{outer product,in infinite-dimensional space}
\inxx{wedge product,in infinite-dimensional space}

The inner product defines the metric $\rho (x, x')$ of  $V_{\infty}$:
\be
      h(x) \cdot h(x') = \rho (x, x')
\lbl{II3.53a}
\ee
The square or the norm of $f$ is
\be
    f^2 = f \cdot f = \int \dd x \, \dd x' \, \rho (x, x') f(x) f(x') .
\lbl{II3.54}
\ee

It is convenient to introduce notation with upper and lower indices and
assume the convention of the integration over the repeated indices. Thus
\be
       f = f^{(x)} h_{(x)}  , \hs{6mm}
\lbl{II3.55}
\ee
\be
     f^2 = f^{(x)} f^{(x')} h_{(x)} \cdot h_{(x')} = f^{(x)} f^{(x')} 
     \rho_{(x) (x')} = f^{(x)} f_{(x)} ,
\lbl{II3.56}
\ee
where $(x)$ is the continuous index.

It is worth stressing here that $h_{(x)} \equiv h(x)$ are abstract elements
satisfying the Clifford algebra relation (\ref{II3.53a}) for a chosen metric
$\rho (x,x') \equiv \rho_{(x) (x')}$. We do not need to worry here about 
providing an explicit representation of $h(x)$; the requirement that they
satisfy the relation (\ref{II3.53a}) is all that matters for 
our purpose\footnote{
Similarly, when introducing the imaginary number $i$, we do not provide an
explicit representation for $i$. We remain satisfied by knowing that $i$
satisfies the relation $i^2 = -1$.}.

The basis vectors $h_{(x)}$ are generators of Clifford algebra ${\cal C}_{\infty}$
of $V_{\infty}$. An arbitrary element $F \in {\cal C}_{\infty}$, called a
{\it polyvector}, can be expanded as \inxx{polyvector,in infinite-dimensional
space}
\be
    F = f_0 + f^{(x)} h_{(x)} + f^{(x) (x')} h_{(x)} \wedge h_{(x')} +
  f^{(x) (x') (x'')} h_{(x)} \wedge h_{(x')} \wedge h_{(x'')} + . . . ,
\lbl{II3.57}
\ee
i.e.,
    $$F = f_0 + \int \dd x \, f(x) h(x) + \int \dd x \,  \dd x' \, 
    f(x,x') h(x) h(x') \hs{1cm}$$
\be
    \hs{2.6cm} + \int \dd x \, \dd x' \, \dd x'' \, f(x,x',x'') h(x) 
    h(x') h(x'') + . . . ,
\lbl{II3.58}
\ee
where the wedge product can be replaced by the Clifford product, if
$f(x,x')$, $f(x,x',x'')$ are antisymmetric in arguments $x$, $x'$, . . .\ .

We see that once we have a space $V_{\infty}$ of functions $f(x)$
and basis vectors $h(x)$, we also automatically have a larger space
of antisymmetric functions $f(x,x')$, $f(x,x',x'')$. This has far reaching
consequences which will be discussed in Sec.7.2.

\subsubsection{Derivative with Respect to an Infinite-Dimensional Vector}

The definition (\ref{II3.35}), (\ref{II3.36}) of the derivative can be
straightforwardly  generalized to the case of polyvector-valued functions
$F(f)$ of an infinite-dimensional vector argument $f$.
The derivative in the direction of a vector $g$ is defined according to
\be
   \left ( g \cdot {\p \oo {\p f}} \right ) F(f) = \lim_{\tau \to 0} 
   {{F(f + g \tau) - F(f)}\oo {\tau}} .
\lbl{II3.59}
\ee
If $g = h_{(x')}$ then
  $$\left ( h_{(x')} \cdot {\p \oo {\p f}} \right ) F = {{\p F}\oo {\p f^{(x')}}}
  \equiv {{\delta F}\oo {\delta f(x')}} \hs{6.5cm}$$
   $$\hs{2.9cm} = \lim_{\tau \to 0} {{F (f^{(x)} h_{(x)}
  + h_{(x')} \tau) - F(f^{(x)} h_{(x)})}\oo \tau}$$
\be
       \hs{3.8cm} = \, \lim_{\tau \to 0} {{F \left [(f^{(x)} + 
        \tau {\delta^{(x)}}_{(x')}) h_{(x')}
  \right ] - F [f^{(x)} h_{(x)}]}\oo \tau} ,
\lbl{II3.60}
\ee
where ${\delta^{(x)}}_{(x')} \equiv \delta (x - x')$. This is a definition of the
{\it functional derivative}. \inxx{functional derivative}
A polyvector $F$ can have a definite grade, e.g.,
\begin{eqnarray}
    r = 0 &:& F(f) = F_0 [f^{(x)} h_{(x)}] = \Phi_0 [f(x)] ,\nonumber \\
    \nonumber \\
    r = 1 &:&  F(f) = F^{(x)} [f^{(x)} h_{(x)}] h_{(x)} = 
    \Phi^{(x)} [f(x)] h_{(x)} \lbl{II3.60a} ,\\
    \nonumber \\
    r = 2 &:& F(f) = F^{(x) (x')} [f^{(x)} h_{(x)}] h_{(x)} h_{(x')} = 
   \Phi^{(x)(x')} [f(x)] h_{(x)} h_{(x')} .\nonumber
\end{eqnarray}

For a scalar-valued $F(f)$ the derivative (\ref{II3.60}) becomes
\be
       {{\delta \Phi_0}\oo {\delta f(x')}} = \lim_{\tau \to 0} 
       {{\Phi_0 [f(x) + \delta (x - x')
    \tau] - \phi_0 [f(x)]}\oo \tau} ,
\lbl{II3.61}
\ee
which is the ordinary definition of the functional derivative. 
Analogously for an
arbitrary $r$-vector valued field $F(f)$.

The derivative with respect to a vector $f$ is then
\be
     h^{(x)} \left ( h_{(x)} \cdot {\p \oo {\p f}} \right ) F =
     {{\p F}\oo {\p f}} = h^{(x)} {{\p F}\oo {\p f^{(x)}}} .
\lbl{II3.61a} 
\ee     

\subsubsection{Inclusion of Discrete Dimensions}

Instead of a single function $f(x)$ of a single argument $x$ we can consider
a discrete set of functions $f^a (x^{\mu})$, $a = 1,2,..., N$, of a multiple
argument $x^{\mu}$, \ $\mu = 1,2,...,n$. These functions can be 
considered as components
of a vector $f$ expanded in  terms of the basis vectors $h_a (x)$ according to
\be
    f = \int \dd x \, f^a (x) h_a (x) \equiv f^{a (x)} h_{a (x)} .
\lbl{II3.62}
\ee
Basis vectors $h_{a (x)} \equiv h_a (x)$ and components 
$f^{a (x)} \equiv f^a (x)$
are now labeled by a set of continuous numbers 
$x^{\mu}, \, \mu = 1,2,...,n$, and by a
set of discrete numbers $a$, such as, e.g., $a = 1,2,...,N$. 
All equations (\ref{II3.50})--(\ref{II3.61}) considered before 
can be straightforwardly generalized  by replacing the index $(x)$ with $a (x)$.

\upperandlowertrue
\section[Dynamical vector field in ${\cal M}$-space]
{DYNAMICAL VECTOR FIELD IN ${\cal M}$-SPACE}
\inxx{vector field,in ${\cal M}$-space}

We shall now reconsider the action (\ref{II2.61}) which describes a membrane
coupled to its own metric field in ${\cal M}$-space. The first term is
the square of the velocity vector
\be
    \rho_{\mu (\phi) \nu (\phi)} {\dot X}^{\mu (\phi)} {\dot X}^{\nu (\phi')}
    \equiv {\dot X}^2 .
\lbl{II3.63}
\ee

The velocity vector can be expanded in terms of ${\cal M}$-space basis
vectors $h_{\mu (\phi)}$ (which are a particular example of generic
basis vectors $h_a (x)$ considered at the end of Sec. 6.1):
\be
       {\dot X} = {\dot X}^{\mu (\phi)} h_{\mu (\phi)} .
\lbl{II3.64}
\ee

Basis vectors $h_{\mu (\phi)}$ are not fixed, but they depend on the
membrane configuration. We must therefore include in the action not only
\inxx{membrane configuration} \inxx{curvature scalar,in ${\cal M}$-space}
a kinetic term for ${\dot X}^{\mu (\phi)}$, but also for $h_{\mu (\phi)}$.\
One possibility is just to rewrite the ${\cal M}$-space curvature scalar in 
terms of $h_{\mu (\phi)}$ by exploiting the relations between the metric
and the basis vectors, \inxx{basis vectors,in ${\cal M}$-space}
\be
     \rho_{\mu (\phi) \nu (\phi')} = h_{\mu (\phi)} \cdot h_{\nu (\xi')} ,
\lbl{II3.65}
\ee
and then perform variations of the action with respect to $h_{\mu (\phi)}$ 
instead of $\rho_{\mu (\phi) \nu (\phi')}$.

A more direct procedure is perhaps to exploit the formalism of Sec. 6.1.
There we had the basis vectors $\gamma_{\mu} (x)$ which were functions
of the coordinates $x^{\mu}$. Now we have the basis vectors $h_{\mu (\phi)}
[X]$ which are functionals of the membrane configuration $X$ (i.e., of
the ${\cal M}$-space coordinates $X^{\mu (\phi)}$).
The relation (\ref{II3.2c}) now generalizes to
\be
    \p_{\alpha (\phi')} h_{\beta (\phi'')} = \Gamma_{\alpha (\phi') \beta 
    (\phi'')}^{\mu (\phi)} h_{\mu (\phi)}
\lbl{II3.66}
\ee
where 
$$\p_{\alpha (\phi')} \equiv {\p\oo {\p X^{\alpha (\phi')}}} \equiv
{\delta\oo {\delta X^{\alpha} (\phi')}}$$
is the functional derivative. The
\inxx{functional derivative} \inxx{curvature tensor,in ${\cal M}$-space}
commutator of two derivatives gives the curvature tensor in ${\cal M}$-space
\be
     [ \p_{\alpha (\phi')} , \p_{\beta (\phi'')}] h^{\mu (\phi)} =
     {{\cal R}^{\mu (\phi)}}_{\nu ({\bar \phi}) \alpha (\phi') \beta (\phi'')}
     h^{\nu ({\bar \phi})} .
\lbl{II3.67}
\ee
The inner product of the left and the right hand side of the above equation 
with $h_{\nu ({\bar \phi}')}$ gives (after renaming the indices)
\be
   {{\cal R}^{\mu (\phi)}}_{\nu ({\bar \phi}) \alpha (\phi') \beta (\phi'')}
   = \left ( [ \p_{\alpha (\phi')} , \p_{\beta (\phi'')}] h^{\mu (\phi)}
   \right ) \cdot h_{\nu ({\bar \phi})} .
\lbl{II3.68}
\ee
The Ricci tensor is then \inxx{Ricci tensor,in ${\cal M}$-space}
\be
    {\cal R}_{\nu ({\bar\phi}) \beta (\phi'')} = {{\cal R}^{\mu (\phi)}}_{
    \nu ({\bar \phi}) \mu (\phi) \beta (\phi'')} =
    \left ( [\p_{\mu (\phi)}, \p_{\beta (\phi'')}] h^{\mu (\phi)} \right )
    \cdot h_{\nu ({\bar \phi})} ,
\lbl{II3.69}
\ee
and the curvature scalar is \inxx{curvature scalar,in ${\cal M}$-space}
\bear
    {\cal R}  &=& \rho^{\nu ({\bar \phi}) \beta (\phi'')} {\cal R}_{\nu
    ({\bar \phi}) \beta (\phi'')} \nonumber \\
    \nonumber \\
     &=& \left ([\p_{\mu (\phi)}, \p_{\nu (\phi')}]
    \right ) \cdot h^{\nu (\phi')} \nonumber \\
    \nonumber \\
    &=& \left ( \p_{\mu (\phi)} \p_{\nu (\phi')} h^{\mu (\phi)} \right ) 
    \cdot h^{\nu (\phi')} -
    \left ( \p_{\nu (\phi')} \p_{\mu (\phi)} h^{\mu (\phi)} \right )
     \cdot h^{\nu (\phi')} .
\lbl{II3.69a}
\ear

A possible action is then
\be
     I[X^{\mu (\phi)}, h_{\mu (\phi)}] = \int {\cal D} X \sqrt{|\rho|} \left (
     {\dot X}^2 + {\epsilon \oo {16 \pi}} {\cal R} \right ) ,
\lbl{II3.70}
\ee
where $\rho$ is the determinant of ${\cal M}$-space metric, and where
${\cal R}$ and $\rho$ are now expressed in terms of $h_{\mu (\phi)}$.
Variation of (\ref{II3.70}) with respect to $h_{\alpha (\phi)}$ gives
their equations of motion. In order to perform such a variation we first
notice that $\rho = {\rm det} \, \rho_{\mu (\phi) \nu (\phi')}$, in view
of the relation (\ref{II3.65}), is now a function of $h_{\alpha (\phi)}$.
Differentiation of $\sqrt{|\rho|}$ with respect to a vector $h_{\alpha (\phi)}$
follows the rules given in Sec. 6.1:
\bear
     {{\p \sqrt{|\rho|}}\oo {\p h^{\alpha (\phi)}}} &=& {{\p \sqrt{|\rho|}}\oo
     {\p \rho^{\mu (\phi') \nu (\phi'')}}} {{\p \rho^{\mu (\phi') \nu (\phi'')}}
     \oo {\p h^{\alpha (\phi)}}} \hs{4cm} \nonumber \\
     \nonumber \\
    &=&  - \mbox{$1\oo2$} \sqrt{|\rho|} \rho_{\mu (\phi')
     \nu (\phi'')} ( {\delta^{\mu (\phi')}}_{\alpha (\phi)} h^{\nu (\phi'')}
     + {\delta^{\nu (\phi'')}}_{\alpha (\phi)} h^{\mu (\phi')}) \nonumber \\
     \nonumber \\
      &=& - \sqrt{|\rho|} h_{\alpha (\phi)} .
\lbl{II3.71}
\ear
Since the vectors $h_{\alpha (\phi)}$ are functionals of the membrane's configuration
$X^{\mu (\phi)}$, instead of the derivative we take the functional derivative
\be
     {{\delta \sqrt{|\rho [X]|}}\oo {\delta h^{\alpha (\phi)} [X']}} =
     - \sqrt{|\rho [X]} h_{\alpha (\phi)} [X] \delta^{({\cal M})} (X - X') ,
\lbl{II3.72}
\ee
where  
\be \delta^{({\cal M})} (X - X') \equiv \prod_{\mu (\phi)} (X^{\mu (\phi)} -
X'^{\mu (\phi)})
\lbl{II3.72a}
\ee
is the $\delta$-functional in ${\cal M}$-space.

Functional deriative of ${\cal R}$ with respect to $h_{\alpha (\phi)}[X]$
gives
     $${{\delta {\cal R}[X]}\oo {\delta h^{\alpha (\phi)} [X']}} =
     \left ( [\p_{\mu (\phi')}, \p_{\nu (\phi'')}] {\delta^{\mu (\phi')}}_{
     \alpha (\phi)} \delta^{({\cal M})} (X - X') \right ) 
     h^{\nu (\phi'')} [X] \hs{2cm}$$
\be     
    \hs{2.5cm}  + \, [\p_{\mu (\phi')}, \p_{\nu (\phi'')}] h^{\mu (\phi')}[X]
     {\delta^{\nu (\phi'')}}_{\alpha (\phi)} \delta^{({\cal M})} (X - X') .
\lbl{II3.73}
\ee
For the velocity term we have
    $${\delta \oo {\delta h^{\alpha (\phi)} [X']}} \left ( {\dot X}_{\mu (\phi')}
    h^{\mu (\phi')} {\dot X}_{\nu (\phi'')} h^{\nu (\phi'')} \right ) \hs{5cm}$$
\be     
    \hs{3.8cm} = \, 2 {\dot X}_{\alpha (\phi)} {\dot X}_{\nu (\phi'')} 
    h^{\nu (\phi'')} \delta^{({\cal M})} (X - X')  .
\lbl{II3.74}
\ee
We can now insert eqs.\ (\ref{II3.71})--(\ref{II3.74}) into
\be
   {{\delta I}\oo {\delta h^{\alpha (\phi)} [X']}} = \int {\cal D} X \,
   {{\delta}\oo {\delta h^{\alpha (\phi)} [X']}} \left [ \sqrt{|\rho|}
   \left ( {\dot X}^2 + {\epsilon\oo {16 \pi}} {\cal R} \right ) \right ]
   = 0 .
\lbl{II3.76}
\ee
We obtain
    $${\dot X}_{\alpha (\phi)} {\dot X}_{\nu (\phi')} h^{\nu (\phi')} -
    \mbox{$1\oo2$} h_{\alpha (\phi)} {\dot X}^2 \hs{7cm}$$
\be    
    \hs{3.1cm} + \, {\epsilon \oo {16 \pi}}
    \left ( - \mbox{$1\oo2$} h_{\alpha (\phi)} {\cal R} + [\p_{\nu (\phi')},
    \p_{\alpha (\phi'')} ] h^{\nu (\phi')} \right ) = 0 .
\lbl{II3.77}
\ee
These are the equations of ``motion" for the variables $h_{\mu (\phi)}$.
The equations for $X^{\mu (\phi)}$ are the same equations (\ref{II2.65}).

After performing the inner product of eq.\ (\ref{II3.77}) with a basis
vector $h_{\beta (\phi')}$ we obtain the ${\cal M}$-space Einstein
\inxx{Einstein equations,in ${\cal M}$-space}
equations (\ref{II2.66}). This justifies use of $h_{\mu (\phi)}$ as
dynamical variables, since their equations of motion contain the equations
for the metric $\rho_{\mu (\phi) \nu (\phi')}$.

\upperandlowerfalse
\subsection[\ \ Description with the vector field in spacetime]
{Description with the Vector Field in Spacetime}

The set of ${\cal M}$-space basis vectors $h_{\mu (\phi)}$
\inxx{basis vectors,in ${\cal M}$-space}
is an arbitrary solution to the dynamical equations (\ref{II3.77}). In order
to find a connection with the usual theory which is formulated, not in
${\cal M}$-space, but in a finite-dimensional (spacetime) manifold $V_N$,
we now assume a particular Ansatz for $h^{\mu (\phi)}$:
\be
    h^{\mu (\phi)} = h^{(\phi)} \gamma^{\mu} (\phi) ,
\lbl{II3.78}
\ee
where
\be
       h^{(\phi)} \cdot h^{(\phi')} = {{\sqrt{{\dot X}^2}}\oo {\kappa
       \sqrt{|f|}}} \, \delta (\phi - \phi')
\lbl{II3.79}
\ee
and
\be
          \gamma^{\mu} (\phi) \cdot \gamma^{\nu} (\phi) = g^{\mu \nu} .
\lbl{II3.79a}
\ee
Altogether the above Ansatz means that
\be
     h^{\mu (\phi)} \cdot h^{\nu (\phi')} = {{ \sqrt{{\dot X}^2}}\oo
     {\sqrt{|f|}}} \, g^{\mu \nu} (\phi) \delta (\phi - \phi') .
\lbl{II3.79b}
\ee
Here $g^{\mu \nu} (\phi)$ is {\it the proto-metric}, \inxx{proto-metric}
and $\gamma^{\mu} (\phi)$ \inxx{proto-vectors}
are {\it the proto-vectors} of {\it spacetime}. The symbol $f$ now meens
\be
    f \equiv {\rm det} \, f_{ab} \; , \qquad f_{ab} \equiv \p_a X^{\mu}
    \p_b X^{\mu} \gamma_{\nu} \gamma_{\nu} = \p_a X_{\mu} \, \p_b X_{\nu} \,
    \gamma^{\mu} \gamma^{\nu} .
\lbl{II3.79c}
\ee
Here the ${\cal M}$-space basis vectors are factorized into the vectors
$h^{(\phi)}$, $\phi = (\phi^A, k)$, $\phi^A \in [0, 2 \pi] \; , \; k = 
1,2,...,Z$, which are independent for all values of $\mu$, and are
functions of parameter $\phi$. Loosely speaking, $h(\phi)$ bear the task
of being basis vectors of the infinite-dimensional part (index $(\phi)$
of ${\cal M}$-space, while $\gamma^{\mu} (\phi)$ are basis vectors of the
finite-dimensional part (index $\mu$) of ${\cal M}$-space. As in Sec. 6.1,
$\gamma^{\mu}$ here also are functions of the membrane's parameters $\phi =
(\phi^A, k)$.

The functional derivative of $h^{\mu (\phi)}$ with respect to the membrane
coordinates $X^{\mu (\phi)}$ is
\be
      \p_{\nu (\phi')} h^{\mu (\phi)} = {{\delta h (\phi)}\oo {\delta 
      X^{\nu}(\phi')}} \gamma^{\mu} (\phi) + h (\phi) {{\delta \gamma^{\mu}
      (\phi)}\oo {\delta X^{\nu} (\phi')}} .
\lbl{II3.80}
\ee
Let us assume a particular case where
\bear
        [\p_{\mu (\phi)},\p_{\nu (\phi')}] \gamma^{\mu} (\phi'') &=&
        \left [ {\p \oo {\p X^{\mu}}}, {\p \oo {\p X^{\nu}}} \right ]
        \gamma^{\mu} (\phi'') \delta (\phi - \phi'') \delta (\phi' - \phi'')
        \nonumber \\
        \nonumber \\
        &\neq& 0 
\lbl{II3.81}
\ear     
and
\be
    [\p_{\mu (\phi)},\p_{\nu (\phi')} ] h(\phi'') = 0 .
\lbl{II3.82}
\ee
Inserting the relations (\ref{II3.78})--(\ref{II3.82}) into the action 
(\ref{II3.70}), and omitting the integration over ${\cal D} X$ we obtain
 $$\epsilon {\cal R} = \epsilon \int
   {{\sqrt{{\dot X}^2}}\oo {\kappa \sqrt{|f(\phi)|}}}
   \delta^2 (0) {\dd} \phi \left ( \left [ {\p \oo {\p X^{\mu}}}, 
   {\p \oo {\p X^{\nu}}} \right ] \gamma^{\mu} (\phi) \right ) \cdot 
   \gamma^{\nu} (\phi)\hs{1.1cm}$$
   $$= \epsilon \int
    {{\sqrt{{\dot X}^2}}\oo {\kappa \sqrt{|f(\phi)|}}}
   \delta^2 (0) {\dd} \phi \hs{5.5cm}$$
   $$\hs{2.6cm} \times \, \left ( \left [ {\p \oo {\p X^{\mu}}}, 
   {\p \oo {\p X^{\nu}}} \right ] \gamma^{\mu} (\phi) \right ) \cdot 
   \gamma^{\nu} (\phi){{\delta \left ( x - X(\phi) \right )}\oo {\sqrt{|g|}}}
   \, \sqrt{|g|} \, {\dd}^N x$$
\be
     = {1\oo G} \int {\dd}^N x \, \sqrt{|g|} \, {\widetilde {\cal R}}, 
     \hs{5.1cm}
\lbl{II3.83}
\ee
where we have set
  $$\epsilon \int \dd \phi \, {\delta}^2 (0) \, {{\sqrt{{\dot X}^2}}\oo 
{\kappa \sqrt{|f(\phi)|}}} \left ( \left [ {\p \oo {\p X^{\mu}}}, 
   {\p \oo {\p X^{\nu}}} \right ] \gamma^{\mu} (\phi) \right ) \cdot 
   \gamma^{\nu} (\phi){{\delta \left ( x - X(\phi) \right )}\oo {\sqrt{|g|}}}$$
\be   
   = {1\oo G} {\widetilde {\cal R}} (x) , \hs{5cm}
\lbl{II3.84}
\ee
$G$ being the gravitational constant. The expression 
${\widetilde {\cal R}} (x)$
is defined formally at all points $x$, but because of $\delta (x - X(\phi))$
it is actually different from zero only on the set of membranes. If we have a
set of membranes filling spacetime, then ${\widetilde {\cal R}} (x)$ becomes a
continuous function of $x$:
\be
    {\widetilde {\cal R}} (x) = \left ( [ \p_{\mu}, \p_{\nu}] \gamma^{\mu} (x)
    \right ) \cdot \gamma^{\nu} (x) = R(x)
\lbl{II3.85}
\ee
which is actually a {\it Ricci scalar}. \inxx{Ricci scalar,in spacetime}

For the first term in the action (\ref{II3.70}) we obtain
  $$h_{\mu (\phi)} h_{\nu (\phi')} {\dot X}^{
  \mu (\phi)} {\dot X}^{\nu (\phi')} \hs{8cm}$$
  $$ = \int {{\kappa \sqrt{|f|}}\oo {\sqrt{{\dot X}^2}}} \, \dd \phi \, 
  \gamma_{\mu} (\phi) \gamma_{\nu} (\phi) {\dot X}^{\mu} (\phi) {\dot X}^{\nu} 
  (\phi)\hs{2.44cm}$$
  $$\hs{.7cm} = 
  \int {\dd}^n \phi \, {{\kappa \sqrt{|f|}}\oo {\sqrt{{\dot X}^2}}} \, 
  \gamma_{\mu} (x) \gamma_{\nu} (x) {\dot X}^{\mu} (\phi) {\dot X}^{\nu} 
  (\phi) \, \delta \left ( x - X(\phi) \right ) {\dd}^N x$$
\be
    = \int {\dd}^n \phi \, {{\kappa \sqrt{|f|}}\oo {\sqrt{{\dot X}^2}}} \, 
  g_{\mu \nu} (x) {\dot X}^{\mu} (\phi) {\dot X}^{\nu} 
  (\phi) \, \delta \left ( x - X(\phi) \right ) {\dd}^N x  .
\lbl{II3.86}
\ee
Altogether we have
    $$I[X^{\mu} (\phi), \gamma^{\mu} (x)] = \kappa \int {\dd}\phi \, \sqrt{|f|}
    \sqrt{\gamma_{\mu} (x) \gamma_{\nu} (x) {\dot X}^{\mu} (\phi) {\dot X}^{\nu}
    (\phi)} \, \delta^N \left ( x - X(\phi) \right ) {\dd}^N x$$
\be
      \hs{3.1cm} + \, {1\oo {16 \pi G}} \int {\dd}^N x \sqrt{|g|} \, 
      \left ( [\p_{\mu},
      \p_{\nu}] \gamma^{\mu} (x) \right ) \cdot \gamma^{\nu} (x) .
\lbl{II3.86a}
\ee
This is an action for the spacetime vector 
field $\gamma^{\mu} (x)$ in the\inxx{vector field,in spacetime}
presence\inxx{membrane configuration,spacetime filling} of a  membrane 
configuration filling spacetime. It was derived from
the action (\ref{II3.70}) in which we have omitted the integration over
${\cal D} X \sqrt{\rho}$. 

Since $\gamma_{\mu} \cdot \gamma_{\nu} = g_{\mu \nu}$, and since 
according to (\ref{II3.4})
\be
      \left ( [\p_{\mu},\p_{\nu} ] \gamma^{\mu} \right ) \cdot
      \gamma^{\nu} = R ,
\lbl{II3.87}
\ee
the action (\ref{II3.86a}) is equivalent to
   $$I[X^{\mu} (\phi), g_{\mu \nu}] = \kappa \int {\dd} \phi \, \sqrt{|f|}
   \sqrt{{\dot X}^2} \, \delta \left ( x - X(\phi) \right ) \, {\dd}^N x$$
\be   
    \hs{5mm} + \, {1\oo {16 \pi G}} \int {\dd}^N x \sqrt{|g|} \, R
\lbl{II3.87a}
\ee
which is an action for the gravitational field $g_{\mu \nu}$ in the presence
of membranes.

Although (\ref{II3.87a}) formally looks the same as the usual gravitational
action in the presence of matter, there is a significant difference. In the
conventional general relativity the matter part of the action may vanish
and we thus obtain the \inxx{Einstein equations}
Einstein equations in vacuum. On the contrary, in the
theory based on ${\cal M}$-space, the metric of ${\cal M}$-space is intimately
connected to the existence of a membrane configuration. Without membranes
there is no ${\cal M}$-space and no ${\cal M}$-space metric. When considering
the ${\cal M}$-space action (\ref{II2.61}) or (\ref{II3.70}) from the point
of view of an effective spacetime (defined in our case by the Ansatz
\inxx{effective spacetime}
(\ref{II3.78})--(\ref{II3.79b}), we obtain  the spacetime action (\ref{II3.87a})
in which the matter part cannot vanish. There is always present a set of
membranes filling spacetime. Actually, the points of spacetime are identified
with the points on the membranes.

The need to fill spacetime with a reference fluid 
(composed of a set of\inxx{reference fluid}
reference particles) has been realized recently by Rovelli \ci{16}, following
an earlier work by DeWitt \ci{15}. According to Rovelli and DeWitt, because of
the Einstein \inxx{hole argument}
``hole argument'' \ci{30}, spacetime points cannot be identified at all.
This is a consequence of the invariance of the Einstein equations under
active diffeomorphisms. One can identify spacetime points if there exists
a material reference fluid with respect to which spacetime points are
identified. \inxx{diffeomorphisms,active} \inxx{reference fluid}

\vs{2mm}

We shall now vary the action (\ref{II3.86a}) with respect to the vector
field $\gamma^{\alpha} (x)$:
\bear
&&{{\delta I}\oo {\delta \gamma^{\alpha} (x)}} = \int \dd \phi \,
    {{\sqrt{|f|}}\oo {\sqrt{{\dot X}^2}}} \, {\dot X}_{\alpha} {\dot X}_{
    \nu} \, \gamma^{\nu} \delta \left (x - X(\phi) \right ) \hs{1cm}
    \lbl{II3.88}\\
    \nonumber \\
    &&\hs{1.8cm} + \, {1\oo {16 \pi G}} \, 
    \int \dd x' \left [ {{\delta \sqrt{|g(x')|}}\oo {\delta \gamma^{\alpha}
    (x)}} \, R(x') + \sqrt{|g(x')|} \, {{\delta R(x')}\oo {\delta
    \gamma^{\alpha} (x)}} \right ] . \nonumber
\ear
For this purpose we use
\be
     {{\delta \gamma^{\mu} (x)}\oo {\delta \gamma^{\nu} (x')}} =
     {\delta^{\mu}}_{\nu} \delta (x - x')
\lbl{II3.88a}
\ee
and
\be
     {{\delta \sqrt{|g(x)|}}\oo {\delta \gamma^{\nu} (x')}} = - \, \sqrt{|g(x)|}
     \, \gamma_{\nu} (x) \delta (x - x') .
\lbl{II3.89}
\ee
In (\ref{II3.89}) we have taken into account that $g \equiv 
{\rm det} \, g_{\mu \nu}$ and $g_{\mu \nu} = \gamma_{\mu} \cdot \gamma_{\nu}$. 
Using (\ref{II3.88a}), we have for the gravitational part
    $${{\delta I_g}\oo {\delta \gamma^{\alpha} (x)}} = {1\oo {16 \pi G}} \int
    {\dd}^N x' \, \sqrt{|g(x')|} \Biggl[ - \gamma_{\alpha} (x') R(x')
    \delta (x - x') \hs{2.5cm}$$    
    $$ \hs{1.9cm} + \left ( [\p'_{\mu}, \p'_{\nu}] {\delta^{\mu}}_{\alpha}
    \delta (x - x') \right ) \gamma^{\nu} (x') + [\p'_{\mu}, \p'_{\nu}]
    \gamma^{\mu} (x') {\delta^{\nu}}_{\alpha} \delta (x - x') \Biggr]$$
\be
      = {1\oo {16 \pi}} \sqrt{|g|} \left ( - \gamma_{\alpha} R + 
     2 [\p_{\mu} , \p_{\alpha}] \gamma^{\mu} \right ) . \hs{2.5cm}
\lbl{II3.89a}
\ee
The equations of motion for $\gamma_{\alpha} (x)$ are thus
\newpage
    $$[\p_{\mu} , \p_{\alpha}] \gamma^{\mu} - \mbox{$1\oo2$} R \, \gamma_{\alpha}
    \hs{9cm}$$     
   $$= - 8 \pi G \int \dd \phi \sqrt{|f|} \sqrt{{\dot X}^2} \left ( 
    {{{\dot X}_{\alpha} {\dot X}_{\nu}}\oo {{\dot X}^2}} + \p^a X_{\alpha} \, 
    \p_a X_{\nu} \right )\hs{5mm}$$
\be    
    \hs{3cm} \times \, \gamma^{\nu} {{\delta \left ( x - X(\phi) \right )}
    \oo {\sqrt{|g|}}} \; ,
\lbl{II3.90}
\ee
where now ${\dot X}^2 \equiv {\dot X}^{\mu} {\dot X}_{\mu}$. 

After performing the inner product with $\gamma_{\beta}$ the latter equations
become the Einstein equations \inxx{Einstein equations}
\be
      R_{\alpha \beta} - \mbox{$1\oo2$} R \, g_{\alpha \beta} = - 
      8 \pi G T_{\alpha \beta} \; ,
\lbl{II3.91}
\ee
where
\be
    R_{\alpha \beta} = ([\p_{\mu}, \p_{\alpha}]) \gamma^{\mu}
    \cdot \gamma_{\beta}
\lbl{II3.92}
\ee
and
\be
     T_{\alpha \beta} = \kappa \int \dd \phi \, \sqrt{|f|} \sqrt{{\dot X}^2}
     \left ( {{{\dot X}_{\alpha} {\dot X}_{\beta}}\oo {{\dot X}^2}} +
     \p^a X_{\alpha} \p_a X_{\beta} \right ) {{\delta (x - X(\phi)}\oo 
     \sqrt{|g|}}        
\lbl{II3.93}
\ee
is the stress--energy tensor of the membrane configuration.
\inxx{stress--energy tensor,of the membrane configuration} It is the ADW
split version of the full stress--energy tensor 
\be
    T_{\mu \nu} = \kappa \int \dd \phi \, ({\rm det} \p_A X^{\alpha} 
    \p_B X_{\alpha} )^{1/2} \, 
    \p^A X_{\mu} \p_A X_{\nu} \, {{\delta \left ( x - X(\phi)\right ) } \oo
    {\sqrt{|g|}}}  .
\lbl{II3.94}
\ee

The variables $\gamma^{\mu} (x)$ appear much easier to handle than
the variables $g_{\mu \nu}$. The expressions for the curvature scalar
(\ref{II3.87}) and the Ricci tensor (\ref{II3.92}) are very simple, and
it is easy to vary the action (\ref{II3.86a}) with 
respect to $\gamma (x)$.\footnote{
At this point we suggest the interested reader study the Ashtekar
\inxx{Ashtekar variables}
variables \ci{57a}, and compare them with $\gamma (x)$.}

We should not forget that the matter stress--energy tensor on the right hand
side of the Einstein equations (\ref{II3.91}) is present everywhere in
spacetime and it thus represents a sort of\inxx{background matter}
background matter\footnote{
It is tempting to speculate that this background is actually {\it the
hidden mass} or {\it dark matter} postulated in astrophysics and cosmology
\ci{57}.} \inxx{hidden mass}\inxx{dark matter}
whose origin is in the original ${\cal M}$-space formulation of the theory.
The ordinary matter is then expected to be present in addition to the
background matter. This will be discussed in Chapter 8.

\upperandlowertrue
\section[Full covariance in the space ot parameters $\phi^A$]
{FULL COVARIANCE IN THE SPACE OF PARAMETERS $\phi^A$}

So far we have exploited the fact that, according to (\ref{II1.1}), 
a membrane
moves as a point particle in an infinite-dimensional ${\cal M}$-space.
For a special choice of ${\cal M}$-space metric (\ref{II2.4}) which involves
the membrane velocity we obtain equations of motion which are identical
to the equations of motion of the conventional constrained membrane,
known in the literature as the $p$-{\it brane}, \inxx{$p$-brane} or simply 
the {\it brane}\footnote{For this reason I reserve the name $p$-{\it brane}
or {\it brane} \inxx{brane}
for the extended objetcs described by the conventional theory,
while the name {\it membrane} stands for the extended objects of the
more general, ${\cal M}$-space based theory studied in this book (even if
the same name ``membrane" in the conventional theory denotes 2-branes,
but this should not cause confusion).}.
A moving brane sweeps a surface $V_n$ which incorporates not only the brane
parameters $\xi^a , \, a = 1,2,...,p$, but also an extra, time-like
parameter $\tau$. Altogether there are $n = p + 1$ parameters $\phi^A =
(\tau, \xi^a)$ which denote a point on the surface $V_n$. The latter surface
is known in the literature under names such as {\it world surface},
{\it world volume} (now the most common choice) and {\it world sheet}
(my favorite choice).\inxx{world surface}

Separating\inxx{world volume} the parameter $\tau$ from the rest of the parameters turns out
to be very useful\inxx{world sheet} in obtaining {\it the unconstrained membrane} out of the
Clifford algebra based polyvector formulation of the theory.

On the other hand, when studying interactions, the separate treatment of
$\tau$ was a nuisance, therefore in Sec. 5.1 we switched to a description in
terms of the variables $X^{\mu} (\tau, \xi^a) \equiv X^{\mu} (\phi^A)$
which were considered as ${\cal M}$-space coordinates $X^{\mu} (\phi^A)
\equiv X^{\mu (\phi)}$. So ${\cal M}$-space was enlarged from that
described by coordinates $X^{\mu (\xi)}$ to that described by
\inxx{coordinates,of ${\cal M}$-space}
$X^{\mu (\phi)}$. In the action and in the equations of motion there occurred
the  ${\cal M}$-space velocity vector ${\dot X}^{\mu (\phi)} \equiv
\p X^{\mu (\phi)}/\p \tau$. Hence manifest covariance with respect to
reparametrizations of $\phi^A$ was absent in our formulation. In my opinion
such an approach was good for introducing the theory and fixing the development
of the necessary concepts. This is now to be superceded. We have learnt enough
to be able to see a way how a fully reparametrization covariant theory should
possibly be formulated.

\upperandlowerfalse
\subsection[\ \ Description in spacetime]
{Description in spacetime}
\upperandlowertrue
As a first step I now provide a version of the action (\ref{II3.86a}) which
is invariant under arbitrary reparametrizations $\phi^A \rightarrow
\phi'^A = f^A (\tau)$. In the form as it stands (\ref{II3.86a}) (more
precisely, its ``matter" term) is invariant under reparametrizations of
$\xi^a$ and $\tau$ separately. A fully invariant action (without a kinetic
term for the gravitational field) is
\be
     I[X^{\mu}] = \kappa \int {\dd}^n \phi \, ({\rm det} \, 
     \p_A X^{\mu} \p_B X^{\nu} g_{\mu \nu} 
     )^{1/2} .
\lbl{II3.95}
\ee
An equivalent form (considered in Sec. 4.2) is   
\be
   I[X^{\mu},\gamma^{AB}] = {\kappa \oo 2} \int {\dd}^n \phi \, \sqrt{|\gamma|}
   (\gamma^{AB} \p_A X^{\mu} \p_B X^{\nu} g_{\mu \nu} + 2 - n) ,
\lbl{II3.96}
\ee
where $X^{\mu}$ and $\gamma^{AB}$ are to be varied independently. Remember
that the variation of (\ref{II3.96}) with respect to $\gamma^{AB}$
gives the relation
\be
    \gamma_{AB} = \p_A X^{\mu} \p_B X_{\mu} \; ,
\lbl{II3.96a}
\ee
which says that $\gamma_{AB}$ is the induced metric on $V_n$.

\subsubsection{DESCRIPTION IN TERMS OF VECTORS $e^A$}

Let us now take into account the basic relation
\be
     \gamma^{AB} = e^A \cdot e^B \; ,
\lbl{II3.97}
\ee
where $e^A$, \  $A = 1,2,...,n$, form a set of basis vectors 
in an $n$-dimensional
space $V_n$ whose points are described by coordinates $\phi^A$. After inserting
(\ref{II3.97}) into (\ref{II3.96}) we obtain the action
\be
    I[X^{\mu} , e^A] = {\kappa \oo 2} \int {\dd}^n \phi \, \sqrt{|\gamma|}
    (e^A \p_A X^{\mu} e^B \p_B X^{\nu} g_{\mu \nu} + 2 -n)
\lbl{II3.98}
\ee
which is a functional of $X^{\mu} (\phi^A)$ and $e^A$. The vectors $e^A$
now serve as the auxiliary variables. The symbol $\gamma \equiv {\rm det}
\gamma_{AB}$ now depends on $e_A$. We have
\bear        {{\p \sqrt{|\gamma|}}\oo {\p e^A}} = {{\p \sqrt{|\gamma|}}\oo {\p
  \gamma^{CD} }} \, {{\p \gamma^{CD}}\oo {\p e^A}} &=& 
  - \mbox{$1\oo 2$} \sqrt{|\gamma|}\,
   \gamma_{CD} \,  ({\delta^C}_A e^D + 
   {\delta^D}_A e^C ) \nonumber \\
    &=& - \sqrt{|\gamma|} e_A .
\lbl{II3.99}
\ear
Variation of (\ref{II3.98}) with respect to $e^C$ gives
\be
    - {1\oo 2} e_C ( e^A \p_A X^{\mu} e^B \p_B X_{\mu} + 2 - n) + 
    \p_C X^{\mu} \p_D X_{\mu} e^D = 0 .
\lbl{II3.100}
\ee
Performing the inner product with $e^C$ and using $e^C \cdot e_C = n$ we
find
\be
     e^A \p_A X^{\mu} e^B \p_B X_{\mu}  = n .
\lbl{II3.101}
\ee
Eq.\ (\ref{II3.100}) then becomes
\be
    e_C =  \p_C X^{\mu} \p_D X_{\mu} e^D \; , 
\lbl{II3.102}
\ee
or
\be
      e_D ({\delta^D}_C - \p_C X^{\mu} \p_D X_{\mu}) = 0 .
\lbl{II3.103}
\ee
Forming the inner product of the left hand side and the right hand side
of eq.\ (\ref{II3.102}) with $e_A$ we obtain
\be
   e_C \cdot e_A = \p_C X^{\mu} \p_D X_{\mu} \, e^D  \cdot e_A .
\lbl{II3.104}
\ee
Since $e_C \cdot e_A = \gamma_{CA}$ and $e^D \cdot e_A = {\delta^D}_A$
the equation (\ref{II3.104}) becomes (after renaming the indices)
\be
     \gamma_{AB} = \p_A X^{\mu} \p_B X_{\mu} .
\lbl{II3.105}
\ee
This is the same relation as resulting when varying $I[X^{\mu}, \gamma^{AB}]$
(eq.\ (\ref{II3.96})) with respect to $\gamma^{AB}$. So we have verified
that $I[X^{\mu}, e^A]$ is equivalent to $I[X^{\mu}, \gamma^{AB}]$.

In the action (\ref{II3.98}) $e^A (\phi)$ are auxiliary fields serving as
Lagrange multipliers. If we add a kinetic term for $e^A$ then $e^A$
become dynamical variables:  
\bear  
      I[X^{\mu}, e^A] &=& {\kappa\oo 2} \int \dd^n \phi \, \sqrt{|\gamma|}
      (e^A \p_A X^{\mu} \, e^B \p_B X_{\mu} + 2 - n)\nonumber \\
      \nonumber \\
    & & + \, {1\oo {16 \pi G^{(n)}}}
      \int \dd^n \phi \, \sqrt{|\gamma|} \, R^{(n)} 
\lbl{II3.106}
\ear
Here, according to the theory of Sec. 6.2,
\be
    R^{(n)} = \left ( [\p_A, \p_B ] e^A \right ) \cdot e^B .
\lbl{II3.106a}
\ee

Variation of (\ref{II3.106}) with respect to $e^C$ leads to the Einstein
equations in $V_n$ which can be cast in the form
\be
    \gamma_{AB} = \p_A X^{\mu} \p_B X_{\mu} + {1\oo {16 \pi G^{(n)}}} 
    R^{(n)}_{AB} .
\lbl{II3.107}
\ee
In general, therefore, $\gamma_{AB}$, satisfying (\ref{II3.106}), 
is not the induced metric on $V_n$.

\vs{2mm}

\subsubsection{DESCRIPTION IN TERMS OF $\gamma^{\mu}$}
 
In eqs.\ (\ref{II3.98})--(\ref{II3.106}) we have written the metric of $V_n$
as $\gamma^{AB} = e^A \cdot e^B$ and considered  the action as a functional
of the vectors $e^A$, but we have kept the metric $g_{\mu \nu}$ of the
embedding space $V_N$ fixed. Now let us write
\be
    g_{\mu \nu} = \gamma_{\mu} \cdot \gamma_{\nu} \; ,
\lbl{II3.108}
\ee
where $\gamma_{\mu}$ , $\mu = 1,2,...,N$, form a complete set of basis vectors
in $V_N$, and write an action
\newpage
    $$I[X^{\mu},\gamma^{AB}, \gamma^{\mu}] \hs{10cm}$$
    $$ \hs{.8cm} =  {\kappa\oo 2} \int \dd^n \phi\, 
    \sqrt{|\gamma|} \, (\gamma^{AB} \p_A X^{\mu} \p_B X^{\nu} \gamma_{\mu}
    \gamma_{\nu} + 2 - n) \delta^N (x - X(\phi)) \dd^N x$$
\be
     + \, {1\oo {16 \pi G^{(N)}}} \int {\dd}^N x \sqrt{|g|} \, R^{(N)}. 
     \hs{3.5cm}
\lbl{II3.109}
\ee        
Variation of the latter action with respect to $\gamma^{\alpha}$ goes along
similar lines to eqs.\ (\ref{II3.88})--(\ref{II3.89a}). We obtain
\bear
    &&[\p_{\mu}, \p_{\alpha}] \gamma^{\mu} - {1\oo 2} R^{(N)} \gamma_{\alpha}
    \lbl{II3.110} \\
    \nonumber \\   
    &&\hs{1.77cm}= - \, 8 \pi G^{(N)} \int {\dd}^n \phi \, \sqrt{|\gamma|} \, 
   \p^A X_{\alpha}
    \p_A X_{\nu} \gamma^{\nu} \, {{\delta (x - X(\phi))}\oo {\sqrt{|g|}}} .
  \nonumber
\ear
These equations of motion for $\gamma^{\mu}$ are a fully reparametrization
covariant form of eq.\ (\ref{II3.90}). The inner product with $\gamma_{\beta}$
gives the Einstein equations (\ref{II3.91}) with \inxx{Einstein equations}
\be
     T_{\alpha \beta} = \kappa \int {\dd}^n \phi \, \sqrt{|\gamma|} \,
     \p^A X_{\alpha} \p_A X_{\beta} {{\delta (x - X(\phi))}\oo {\sqrt{|g|}}} 
     \; ,
\lbl{II3.111}
\ee
which is a fully reparametrization covariant form of the stress--energy tensor
(\ref{II3.93}). \inxx{stress--energy tensor}

\subsubsection{DESCRIPTION IN TERMS OF $e^A$ AND $\gamma^{\mu}$}

Combining the descriptions as given in eqs.\ (\ref{II3.106}) and (\ref{II3.109})
we obtain the following action:
   $$I[X^{\mu}, e^A, \gamma^{\mu}] = \int \dd \phi \sqrt{|\gamma|} \Biggl(
   {\kappa\oo 2} e^A e^B \p_A X^{\mu} \p_B X^{\nu} \gamma_{\mu} \gamma_{\nu}
   + {\kappa \oo 2} (2 - n) \hs{1cm}$$   
    $$\hs{4.5cm} + \, {1\oo {16 \pi G^{(n)}}} R^{(n)} \Biggr)
   {\dd}^N x \, \delta \left ( x - X(\phi) \right )$$
\be   
     \hs{3.1cm} + \, {1\oo {16 \pi G^{(N)}}} \int {\dd}^N x \sqrt{|g|} R^{(N)} .
\lbl{II3.112}
\ee
The first term in (\ref{II3.112}) can be written as
\be
     (e^A \cdot e^B + e^A \wedge e^B ) (\gamma_{\mu} \cdot \gamma_{\nu}
     + \gamma_{\mu} \wedge \gamma_{\nu})
     \p_A X^{\mu} \p_B X^{\nu} \; ,
\lbl{II3.113}
\ee
which gives the usual scalar part and a multivector part:
\be
(e^A \cdot e^B) \p_A X^{\mu} \p_B X^{\nu} (\gamma_{\mu} \cdot \gamma_{\nu}) +
 (e^A \wedge e^B) \p_A X^{\mu} \p_B X^{\nu} (\gamma_{\mu} \wedge \gamma_{\nu}) ,
\lbl{II3.114}
\ee
Now, considering {\it the string} (worldsheet dimension $n = 2$), 
we have \newline
$e^A\wedge e^B = I_{(2)} \epsilon^{AB}$, where $I_{(2)}$ is the pseudoscalar
unit, $I_{(2)} = e^1 e^2$, in 2 dimensions, and $\epsilon^{AB}$ is the
antisymmetric tensor density in 2 dimensions. So we have
\bear
     (e^A \wedge e^B)(\gamma_{\mu} \wedge \gamma_{\nu}) &=& \epsilon^{AB}
     I_{(2)} (\gamma_{\mu} \wedge \gamma_{\nu}) \lbl{II3.115}\\
     \nonumber \\
 \langle e^1 e^2 (\gamma_{\mu} \wedge \gamma_{\nu} \rangle_0 &=&
    e^1 \left [ (e^2 \cdot \gamma_{\mu}) \gamma_{\nu} - (e^2 \cdot
    \gamma_{\nu}) \gamma_{\mu} \right ] \\
    \nonumber \\
   &=& \, (e^2 \cdot \gamma_{\mu})(e^1 \cdot \gamma_{\nu}) - 
     (e^2 \cdot \gamma_{\nu})(e^1 \cdot \gamma_{\mu}) \nonumber \\
     \nonumber \\
      &\equiv& B_{\mu \nu}
\lbl{II3.116}
\ear
where $\langle \, \rangle_0$ means the scalar part. For the string
case the \inxx{antisymmetric field}
antisymmetric field $B_{\mu \nu}$ thus naturally occurs in the
action (\ref{II3.112}) whose Lagrangian contains the terms
\be
    \gamma^{AB} \p_A X^{\mu} \p_B X^{\nu} g_{\mu \nu} +
    \epsilon^{AB} \p_A X^{\mu} \p_B X^{\nu} B_{\mu \nu} \; ,
\lbl{II3.117}
\ee
well known in the usual theory of strings. However, in the theory of $p$-branes
(when $n = p+1$ is arbitrary), instead of a 2-form field there occurs an
$n$-form field and the action contains the term \inxx{$n$-form field}
\be
    \epsilon^{A_1 A_2 ... A_n} \p_{A_1} X^{\mu_1} ... \p_{A_n} X^{\mu_n}
    B_{\mu_1 .... \mu_n} \; .
\lbl{II3.117a}
\ee
If we generalize the action (\ref{II3.112}) to a corresponding {\it
polyvector action} \inxx{polyvector action,for $p$-branes}

\vs{1mm}
 
   $$I[X^{\mu},e^A,\gamma^{\mu}]\hs{9cm}$$
   $$ = {\kappa\oo 2} \int {\dd}^n \phi
   \sqrt{|\gamma|} \Biggl( e^{A_1} e^{A_2} \p_{A_1} X^{\mu_1} \p_{A_2} X^{\mu_2}
   \gamma_{\mu_1} \gamma_{\mu_2}\hs{2cm}$$
   $$\hs{3.1cm} +\,  e^{A_1} e^{A_2} e^{A_3} 
   \p_{A_1} X^{\mu_1} \p_{A_2} X^{\mu_2} \p_{A_3} x^{\mu_3} \gamma_{\mu_1}
   \gamma_{\mu_2} \gamma_{\mu_3}$$
   $$\hs{2.4cm}$$
   $$\hs{3cm} + ... + e^{A_1} e^{A_2} ... e^{A_n}
   \p_{A_1} X^{\mu_1} \p_{A_2} x^{\mu_2} ...\p_{A_n} X^{\mu_n} \gamma_{\mu_1}
   \gamma_{\mu_2} ... \gamma_{\mu_n}$$
    $$\hs{3.5cm} + \, (2 - n) + {1\oo {16 \pi G^{(n)}}} 
    {2\oo {\kappa}} R^{(n)} \Biggr)
     {\dd}^N x \, \delta \left ( x - X(\phi) \right )$$
\be     
     + \, {1\oo {16 G^{(N)}}} \int {\dd}^N x \sqrt{|g|} \, R^{(N)}\; , 
     \hs{3.1cm}
\lbl{II3.118}
\ee

\vs{1mm}

\nnnn then in addition to the other multivector terms we also obtain a term  
(\ref{II3.117a}),
since the highest rank multivector term gives
\newpage
    $$\langle (e^{A_1} \wedge ... \wedge e^{A_n}) (\gamma_{\mu_1} \wedge ...
    \wedge \gamma_{\mu_n}) \rangle_0 = \langle \epsilon^{A_1 ... A_n}
    I_{(n)} (\gamma_{\mu_1} \wedge ... \wedge \gamma_{\mu_n}) \rangle_0
    \hs{1cm}$$
\be    
   \hs{2.6cm}\equiv \, \epsilon^{A_1 ... A_n} B_{\mu_1 ... \mu_n} .
\lbl{II3.119}
\ee

Alternatively, for the $p$-brane Lagrangian we can take 
\be
    {\cal L} (X^{\mu}) = \kappa \sqrt{|\gamma|} \left [ \left \langle
    \sum_{r=1}^{2n} \prod_{i=1}^r e^{A_i} \p_{A_i} X^{\mu_i} \gamma_{\mu_i}
    \right \rangle_0 \right ]^{1/2}
\lbl{II3.118a}
\ee    
The latter expression is obtained if we square the brane part of the
Lagrangian (\ref{II3.118}), take the scalar part, and apply the square root.

Although the action (\ref{II3.118}) looks quite impressive, it is not the
most general polyvector action for a membrane. Some preliminary suggestions
of how to construct such an action will now be provided.

\upperandlowerfalse
\subsubsection{Polyvector Action}

\paragraph{Polyvector action in terms of $e^A$}

The terms $e^A \p_A X^{\mu} , \, \mu = 1,2,...,N$ which occur in the action
(\ref{II3.106}) can be generalized to {\it world sheet polyvectors}:
\be
     V^{\mu} = s^{\mu} + e^A \p_A X^{\mu} + e^{A_1} \wedge e^{A_2}
     v_{A_1 A_2}^{\mu} + ... + e^{A_1} \wedge ... \wedge e^{A_n}  .
     v_{A_1 .... A_n}^{\mu}
\lbl{II3.120}
\ee
The action can then be
\be
   I[X^{\mu}, e^A, g_{\mu \nu}] = {\kappa\oo 2} \int {\dd}^n \phi \sqrt{
   |\gamma|} V^{\mu} V^{\nu} g_{\mu \nu} + I[e^A] + I[g_{\mu \nu}] .
\lbl{II3.121}
\ee
The kinetic term for $e^A$ includes, besides $R^{(n)}$, the term with
$(2-n)$ and possible other terms required for consistency.

\paragraph{Polyvector action in terms of $\gamma^{\mu}$}

Similarly we can generalize the terms $\p_A X^{\mu} \gamma_{\mu}$, $A = 1,2,
...,n$, in eq.\ (\ref{II3.109}), so as to become spacetime polyvectors:
\inxx{spacetime polyvectors}
\be
    C_A = c_A + \p_A X^{\mu} \gamma_{\mu} + c_A^{\mu_1 \mu_2} \gamma_{\mu_1}
    \wedge \gamma_{\mu_2} + ... + c_A^{\mu_1 ... \mu_N} \gamma_{\mu_1}
    \wedge \gamma_{\mu_1} \wedge ... \wedge \gamma_{\mu_N}
\lbl{II3.122}
\ee
A corresponding action is then\
\be
   I[X^{\mu}, \gamma^{AB}, \gamma^{\mu}] = \mbox{$1\oo 2$} \int {\dd}^n \phi 
   \sqrt{|\gamma|} \gamma^{AB} C_A C_B + I[\gamma^{AB}]+ I[\gamma^{\mu}] .
\lbl{II3.123}
\ee
where
\be
       I[\gamma^{\mu}] = {1\oo {16 \pi G^{(N)}}} \int {\dd}^N x \sqrt{
       |g|} \left ( [\p_{\mu},\p_{\nu}] \gamma^{\mu} (x) \right ) \cdot 
       \gamma^{\nu} (x) .
\lbl{II3.123a}
\ee

\paragraph{Polyvector action in terms of $e^A$ and $\gamma^{\mu}$}

From the expression $V^{\mu}$ in eq.\ (\ref{II3.120}) we can form another
expression 
        $$V^{\mu} \gamma_{\mu} = s^{\mu} \gamma_{\mu} + 
        e^A \p_A X^{\mu} \gamma_{\mu}+ 
        e^{A_1} \wedge e^{A_2} 
     v_{A_1 A_2}^{\mu} \gamma_{\mu} + ... \hs{2cm} $$
\be     
    \hs{7cm} + \, e^{A_1} \wedge ... \wedge e^{A_n} 
     v_{A_1 .... A_n}^{\mu}\gamma_{\mu} .
\lbl{II3.123b}     
\ee
The latter expression can be generalized accoridng to
\be
    V = s + V^{\mu} \gamma_{\mu} + V^{\mu_1 \mu_2} \gamma_{\mu_1} \wedge
    \gamma_{\mu_2} + ... + V^{\mu_1 ... \mu_N} \gamma_{\mu_1} \wedge ...
    \wedge \gamma_{\mu_N} \; ,
\lbl{II3.124}
\ee
where we assume
\be
     V^{\mu_1 \mu_2 ... \mu_k} \equiv V^{\mu_1} V^{\mu_2} ...V^{\mu_k}
\lbl{II3.125}
\ee
and where $V^{\mu_1}$, $V^{\mu_2}$, etc., are given by (\ref{II3.120}).
For instance,
    $$V^{\mu_1 \mu_2} = V^{\mu_1} V^{\mu_2}\hs{9cm}$$
    $$\hs{1cm} = \,
    (s^{\mu_1} + e^A \p_A X^{\mu_1} + e^{A_1} \wedge e^{A_2}
     v_{A_1 A_2}^{\mu_1} + ... + e^{A_1} \wedge ... \wedge e^{A_n} 
     v_{A_1 .... A_n}^{\mu_1})$$
     $$\times \, (s^{\mu_2} + e^{A'} \p_{A'} X^{\mu_2} + e^{{A'}_1} \wedge e^{{A'}_2}
     v_{{A'}_1 {A'}_2}^{\mu_2} + ... \hs{2cm}$$
\be     
     \hs{4.5cm} + \, e^{{A'}_1} \wedge ... \wedge e^{{A'}_n} 
     v_{{A'}_1 .... {A'}_n}^{\mu_2}) .     
\lbl{II3.126}
\ee

The action is then
\be
     I[X^{\mu},v_{A_1...A_r},e^A,\gamma^{\mu}] = \mbox{$1\oo 2$} \int {\dd}^n
     \phi \sqrt{|\gamma|} \, V^2 + I[e^A] + I[\gamma^{\mu}] .
\lbl{II3.127}
\ee

Another possibility is to use (\ref{II3.123}) and form the expression
    $$e^A C_A = e^A c_A + e^A \p_A X^{\mu} \gamma_{\mu} + e^A c_A^{\mu_1
    \mu_2} \gamma_{\mu_1} \wedge \gamma_{\mu_2} + ...$$
\be    
     \hs{.8cm} + \, e^A
    c_A^{\mu_1...\mu_N} \gamma_{\mu_1} \wedge ... \wedge \gamma_{\mu_N} \; ,
\lbl{II3.128}
\ee
which is a spacetime polyvector, but a worldsheet 1-vector. We can generalize
(\ref{II3.128}) so as to become a world sheet polyvector: 
\inxx{world sheet polyvector}
\be
    C = c + e^A C_A + e^{A_1} \wedge e^{A_2} C_{A_1 A_2} + ... +
    e^{A_1} \wedge ... \wedge e^{A_n} C_{A_1...A_n} \; ,
\lbl{II3.129}
\ee
where
\be
       C_{A_1 A-2 ... A_k} = C_{A_1} C_{A_2} ... C_{A_k}
\lbl{II3.130}
\ee
and where $C_{A_1}$, $C_{A_2}$,... are  given by the expression (\ref{II3.128}).

A corresponding action is
\be    
    I[X^{\mu}, c_A^{\mu_1...\mu_k},e^A, \gamma^{\mu}] = \mbox{$1\oo 2$} 
    \int {\dd}^n \phi \sqrt{|\gamma|} \, C^2 + I[c_A^{\mu_1...\mu_k}] +
    I[e^A] + I[\gamma^{\mu}]
\lbl{II3.132}
\ee

Whichever action, (\ref{II3.127}) or (\ref{II3.132}), we choose, we have
an action which is a {\it polyvector both in the target space and on the
world sheet}. According to the discussion of Chapter 2, Section 5, particular
polyvectors are {\it spinors}. Our action therefore contains
the target space and \inxx{target space}
and the worldsheet spinors.\inxx{the worldsheet spinors} 
In order to obtain dynamics of such spinors the
coefficients $c_A^{\mu_1...\mu_k}$ or $v_{A_1...A_k}^{\mu}$ and the
corresponding kinetic terms should be written explicitly. This will come
out naturally within an ${\cal M}$-space description of our membrane.

\upperandlowertrue
\subsection[\ \ Description in ${\cal M}$-space]
{DESCRIPTION IN  ${\cal M}$-SPACE}

So far in this section we have considered various possible formulations of
the membrane action, but without the powerful ${\cal M}$-space formalism.
In Sec. 6.2, throughout eqs. (\ref{II3.63})--(\ref{II3.77}), 
we have used a description
with ${\cal M}$-space basis vectors $h_{\mu (\phi)}$, but the equations
were not fully invariant or covariant with respect to arbitrary 
reparametrizations of the worldsheet parameters $\phi^A \equiv (\tau, \xi^a)$.
One parameter, namely $\tau$, was singled out, and the expressions contained
the velocity ${\dot X}^{\mu (\phi)}$. On the contrary, throughout 
eqs.\ (\ref{II3.96})--(\ref{II3.132}) of this section we have used a 
fully reparametrization covariant formalism.

We shall now unify both formalisms. First we will consider a straightforward
reformulation of the action (\ref{II3.106}) and then proceed towards a
``final" proposal for a polyvector action in ${\cal M}$-space.
\inxx{polyvector action,in ${\cal M}$-space}

By introducing the metric
\be
    \rho_{\mu (\phi) \nu (\phi')} = \kappa \sqrt{|\gamma|} \, 
    \delta (\phi - \phi') g_{\mu \nu}
\lbl{II3.133}
\ee
the first term in the action (\ref{II3.106}) can be written as the quadratic
form in $\cal M$-space
\be
     {\kappa\oo 2} \int {\dd}^n \phi \, \sqrt{|\gamma|} \, g_{\mu \nu}
     e^A \p_A X^{\mu} e^B \p_B X^{\nu} = {1\oo 2} \rho_{\mu (\phi) \nu (\phi')}
     v^{\mu (\phi)} v^{\nu (\phi')} \; ,
\lbl{II3.134}
\ee
where
\be 
    v^{\mu (\phi)} = e^A \p_A X^{\mu (\phi)} .
\lbl{II3.135}
\ee
Here $v^{\mu (\phi)}$ is the uncountable (infinite) set of vectors in the
$n$-dimensional worldsheet manifold $V_n$. 

The projection
\be
    e_0 \cdot v^{\mu (\phi)} = \p_0 X^{\mu (\phi)} \equiv {\dot X}^{\mu (\phi)}
\lbl{II3.136}
\ee
is the membrane velocity considered in Chapter 4.             

More generally,
\be
    e_B \cdot v^{\mu (\phi)} = \p_B X^{\mu (\phi)} .
\lbl{II3.137}
\ee

\subsubsection{POLYVECTOR ACTION IN TERMS OF $e^A$}

The vectors (\ref{II3.135}) can be promoted to {\it polyvectors} in $V_n$:
\be
     V^{\mu (\phi)} = s^{\mu (\phi)} + e^A \p_A X^{\mu (\phi)} +
     e^{A_1}\wedge  e^{A_2} v_{A_1 A_2}^{\mu (\phi)} + ... + e^{A_1} \wedge ...
     \wedge v_{A_1 ... A_n}^{\mu (\phi)} .
\lbl{II3.140}
\ee
This is the same expression (\ref{II3.120}), $V^{\mu (\phi)} \equiv V^{\mu} 
(\phi) = V^{\mu}$.

The first term in the action (\ref{II3.121}) can be written as
\be
        V^{\mu (\phi)} V_{\mu (\phi)} = \rho_{\mu (\phi) \nu (\phi')}
        V^{\mu (\phi)} V^{\mu (\phi')} .
\lbl{II3.141}
\ee
Instead of the action (\ref{II3.121}) we can consider its ${\cal M}$-space
generalization
\bear
   &&I[X^{\mu (\phi)}, e^A, \rho_{\mu (\phi) \nu (\phi')}] \hs{8cm}
\lbl{II3.142} \\
\nonumber \\     
   &&\hs{8mm} = \int {\cal D} X
   \sqrt{|\rho|} \left ( {1\oo 2} \rho_{\mu (\phi) \nu (\phi')}
        V^{\mu (\phi)} V^{\mu (\phi')} + {\cal L} [e^A] + {\cal L} [\rho_{
        \mu (\phi) \nu (\phi')}] \right ) , \nonumber
\ear
where
\be
      {\cal L} [e^A] = \int {\dd}^n \phi \sqrt{|\gamma|} \left ( {\kappa\oo 2}
      (2 -n) + {1\oo {16 \pi G^{(n)}}} R^{(n)} \right )
\lbl{II3.143}
\ee
and
\be
     {\cal L} [\rho_{\mu (\phi) \nu (\phi')}] = {\epsilon \oo {16 \pi}} {\cal R}
\lbl{II3.144}
\ee
are now ${\cal M}$-space Lagrangians.

\subsubsection{POLYVECTOR ACTION IN TERMS OF $e^A$ AND $h_{\mu (\phi)}$}

The ${\cal M}$-space metric can be written as the inner product of two
${\cal M}$-space basis vectors: \inxx{basis vectors,in ${\cal M}$-space}
\be
   \rho_{\mu (\phi) \nu (\phi')} = h_{\mu (\phi)} \cdot h_{\nu (\phi')} .
\lbl{II3.138}
\ee
Hence (\ref{II3.134}) becomes
\bear
    v^{\mu (\phi)} v_{\mu (\phi)} &=& v^{\mu (\phi)} v^{\nu (\phi')}
    h_{\mu (\phi)} h_{\nu (\phi')} \nonumber \\
    \nonumber \\
     &=& e^A \p_A X^{\mu (\phi)} \, e^B \p_B
    X^{\nu (\phi')} \, h_{\mu (\phi)} h_{\nu (\phi')} .
\lbl{II3.139}
\ear
It is understood that $e^A \p_A X^{\mu (\phi)}$ is considered as a single object
with index $\mu (\phi)$.

Using (\ref{II3.138}) the quadratic form (\ref{II3.141}) becomes
\be 
     V^{\mu (\phi)} V_{\mu (\phi)} = V^{\mu (\phi)} h_{\mu (\phi)}
     V^{\nu (\phi')} h_{\mu (\phi')} .
\lbl{II3.145}
\ee
The object  $V^{\mu (\phi)} h_{\mu (\phi)}$ is a {\it vector} from the 
${\cal M}$-space point of view and a {\it polyvector} from the point of view of
the worldsheet $V_n$. We can promote $V^{\mu (\phi)} h_{\mu (\phi)}$ to
an ${\cal M}$-space polyvector: \inxx{polyvectors,in ${\cal M}$-space}

\vs{1mm}

     $$V = s + V^{\mu (\phi)} h_{\mu (\phi)} + V^{\mu_1 (\phi_1) \mu_2 (\phi_2)}
     h_{\mu_1 (\phi_1)} \wedge h_{\mu_2 (\phi_2)} + ... $$
\be   
    + V^{\mu_1 (\phi_1) ... \mu_N (\phi_N)} h_{\mu_1 (\phi_1)} \wedge ... 
     \wedge h_{\mu_N (\phi_N)} \; , \hs{1cm}
\lbl{II3.146}
\ee

\vs{1mm}

\nnnn where $V^{\mu_1 (\phi_1) ... \mu_k (\phi_k)}$ are generalizations of
$V^{\mu (\phi)}$:
   $$V^{\mu_1 (\phi_1) ... \mu_k (\phi_k)}  =    
   s^{\mu_1 (\phi_1) ... \mu_k (\phi_k)} + e^A \p_A 
   X^{\mu_1 (\phi_1) ... \mu_k (\phi_k)} + ...$$
\be    
   \hs{1cm} + \, e^{A_1} \wedge ... \wedge e^{A_n}
   v_{A_1 ... A_n}^{\mu_1 (\phi_1) ... \mu_k (\phi_k)}  .
\lbl{II3.147a}
\ee

A corresponding action is
   $$I[X^{\mu (\phi)}, v_{A_1 ... A_p}^{\mu_1 (\phi_1) ... \mu_k (\phi_k)},
   e^A, h^{\mu (\phi)}] \hs{7cm}$$
\be   
   = \, \int {\cal D} X \sqrt{|\rho|} \left ( 
  {1\oo 2} V^2 + {\cal L} [v_{A_1 ... A_p}^{\mu_1 (\phi_1) ... \mu_k (\phi_k)}]
  + {\cal L} [e^A] + {\cal L} [h^{\mu (\phi)}] \right ) .
\lbl{II3.148}
\ee

Although in any single term in an expression which involves products of
$e^A$ and $h_{\mu (\phi)}$ (or $\gamma_{\mu}$) there appears an 
ordering ambiguity because $e^A$ and $h_{\mu (\phi)}$ 
(or $\gamma_{\mu}$) do not commute, there is no such ordering ambiguity
in any expression which invloves worldsheet and target space polyvectors.
A change of order results in a reshuffling of various multivector terms,
so that only the coefficients become different from those in the
original expression, but the form of the expression remains the same.

\vs{1mm}

A question now arises of what are the coefficients
$v_{A_1 ... A_p}^{\mu_1 (\phi_1) ... \mu_k (\phi_k)}$ in the expansion
(\ref{II3.147a}). There is no simple and straightforward answer to this
question if $\phi \equiv \phi^A$ is a set of scalar coordinate functions
(parameters) on the worldsheet $V_n$. In the following we shall find out
that the theory becomes dramatically clear if we generalize $\phi$ to
polyvector coordinate functions. \inxx{coordinate functions,polyvector}

\vs{2mm}

\subsection[\ \ Description in the enlarged ${\cal M}$-space]
{DESCRIPTION IN THE ENLARGED ${\cal M}$-SPACE}

So far we have considered a set of $N$ scalar-valued functions 
$X^{\mu} (\phi^A)$ of parameters $\phi^A$. This can be written as a set of
scalar-valued functions of a vector valued parameter $\phi = \phi^A e_A$,
where $\phi^A$ are coordinate functions in the coordinate basis $e_A$
(see also Sec. 6.1).

Let us now promote $\phi$ to a polyvector $\Phi$ in $V_n$:
\be
   \Phi = \phi_0 + \phi^A e_A + \phi^{A_1 A_2} e_{A_1} e_{A_2} + ... \phi^{A_1
   ... A_n} e_{A_1} ... e_{A_n} \equiv \phi^J e_J ,
\lbl{II3.145r}
\ee
where $e_J$ are basis vectors of the Clifford algebra generated by the vectors
$e_A$. The worldsheet polyvector $V^{\mu (\phi)} \equiv V^{\mu} (\phi)$, 
considered in (\ref{II3.140}), is an extension of the velocity vector
$e^A \p_A X^{\mu}$. The latter expression is nothing but the derivative
of $X^{\mu} (\phi)$ with respect to the coordinate vector:
\be
    v^{\mu} (\phi) \equiv {{\p X^{\mu}}\oo {\p \phi}} = 
    e^A {{\p X^{\mu}}\oo {\p \phi^A}} .
\lbl{II3.145ra}
\ee
If instead of $X^{\mu} (\phi)$ we have $X^{\mu} (\Phi)$, i.e., a function
of the polyvector (\ref{II3.145r}), then (\ref{II3.145ra}) generalizes to
    $$V^{\mu} (\Phi) \equiv {{\p X^{\mu}}\oo {\p \Phi}} = {{\p X^{\mu}}\oo
    {\p \phi_0}} + e^A {{\p X^{\mu}}\oo {\p \phi^A}} + e^{A_1}e^{A_2}
    {{\p X^{\mu}}\oo {\p \phi^{A_1 A_2}}} + ...$$
    $$+ \, e^{A_1} ... e^{A_n}
    {{\p X^{\mu}}\oo {\p \phi^{A_1 ... A_n}}}$$
\be
  \equiv\,  e^J \p_J X^{\mu} . \hs{2.4cm}   
\lbl{II3.146r}
\ee
This is just like (\ref{II3.140}), except that the argument is now
$\Phi$ instead of $\phi$.

\vs{2mm}

In eq.\,(\ref{II3.140}) $V^{\mu} (\Phi) , \, \mu = 1,2,...,N$, are 
{\it polyvectors} in the worldsheet $V_n$, but {\it scalars} in target space
$V_N$. By employing the basis vectors $\gamma_{\mu} , \, \mu = 1,2,...,N$,
we can form a {\it vector} $V^{\mu} \gamma_{\mu}$ in $V_N$ (and also
generalize it to a polyvector in $V_N$).

\vs{2mm}

What about ${\cal M}$-space? Obviously {\it ${\cal M}$-space is now enlarged}:
instead of $X^{\mu} (\phi) \equiv X^{\mu (\phi)}$ we have 
$X{\mu} (\Phi) \equiv X^{\mu (\Phi)}$. Our basic object is no longer a simple
worldsheet ($n$-dimensional membrane or surface) described by 
$X^{\mu (\phi)}$, but a more involved geometrical object. It is described by
$X^{\mu (\Phi)}$ which are coordinates of a point in an infinite-dimensional
space which I will denote ${\cal M}_{(\Phi)}$, whereas the ${\cal M}$-space
considered so far will from now on be denoted ${\cal M}_{(\phi)}$.
Whilst the basis vectors in ${\cal M}_{(\phi)}$-space are $h_{\mu (\phi)}$,
the basis vectors in ${\cal M}_{(\Phi)}$-space are $h_{\mu (\Phi)}$. The
objects $V^{\mu} (\Phi) \equiv V^{\mu (\Phi)}$ are {\it components} of a vector
$V^{\mu (\Phi)} h_{\mu (\Phi)}$ in ${\cal M}_{(\Phi)}$-space.

\vs{2mm}

The generalization of (\ref{II3.146r}) is now at hand:
\be
     V = v_0 + e^J \p_J X^{\mu (\Phi)} h_{\mu (\Phi)} + e^J \p_J X^{\mu_1
     (\Phi_1) \mu_2 (\Phi_2)} h_{\mu_1 (\Phi_1)} h_{\mu_2 (\Phi_2)} + ... \, .
\lbl{II3.147r}
\ee
It is instructive to write the latter object more explicitly
\newpage
    $$V = v_0 + \biggl(  {{\p X^{\mu (\Phi)}}\oo
    {\p \phi_0}} + e^A {{\p X^{\mu (\Phi)}}\oo {\p \phi^A}} + e^{A_1}e^{A_2}
    {{\p X^{\mu (\Phi)}}\oo {\p \phi^{A_1 A_2}}} + ...\hs{3cm}$$
    $$\hs{3.5cm} + \,  e^{A_1} ... e^{A_n}
    {{\p X^{\mu (\Phi)}}\oo {\p \phi^{A_1 ... A_n}}}  \biggr) h_{\mu (\Phi)}$$
    $$+ \, \biggl( {{\p X^{\mu_1 (\Phi_1) \mu_2 (\Phi_2)}}\oo
    {\p \phi_0}} + e^A {{\p X^{\mu_1 (\Phi_1) \mu_2 (\Phi_2)}}\oo {\p \phi^A}} 
    + e^{A_1}e^{A_2}
    {{\p X^{\mu_1 (\Phi_1) \mu_2 (\Phi_2)}}\oo 
    {\p \phi^{A_1 A_2}}} + ...$$
\be    
    \hs{1.5cm}+ e^{A_1} ... e^{A_n}
    {{\p X^{\mu_1 (\Phi_1) \mu_2 (\Phi_2)}}\oo {\p \phi^{A_1 ... A_n}}} \biggr)
    h_{\mu_1 (\Phi_1)} h_{\mu_2 (\Phi_2)} + ...,
\lbl{II3.148r}
\ee
and more compactly,
\be
    V = e^J \p_J X^{[Z]} h_{[Z]} \; ,
\lbl{II3.149}
\ee
where $h_{[Z]} = h_{\mu (\phi)}, \, h_{\mu_1 (\phi_1)} h_{\mu_2 (\Phi_2)}, ... $
form a basis of Clifford algebra in ${\cal M}_{(\Phi)}$-space.
\inxx{Clifford algebra,in ${\cal M}_{(\Phi)}$-space}

A repeated index $(\Phi)$ means the integration:
\be
    V^{\mu (\Phi)} h_{\mu (\Phi)} = \int V^{\mu} (\Phi) h_{\mu} (\Phi)  \dd
    \Phi
\lbl{II3.150}
\ee
where
\bear
    \dd \Phi &\equiv& \prod_J \dd \phi^J \lbl{II3.151}\\
     &=& \dd \phi_0 \, \left ( \prod_{A}
    \dd \phi_A \right ) \left ( \prod_{A_1<A_2} \dd \phi^{A_1 A_2} \right ) 
    ... \left ( \prod_{A_1<A_2<...<A_n} \dd \phi^{A_1...A_n} \right ).
    \nonumber
\ear
We do not need to worry about the measure, since the measure is assumed to be
included in the ${\cal M}_{(\Phi)}$-space metric which we define as the
inner product
\be
    \rho_{\mu (\Phi) \nu (\Phi)} = h_{\mu (\Phi)} \cdot h_{\nu (\Phi')} .
\lbl{II3.151a}
\ee
All expressions of the form $\rho_{\mu (\Phi) \nu (\Phi)} A^{\mu (\Phi)}
B^{\nu (\Phi')}$ are then by definition invariant under reparametrizations of
$\Phi$. (See also the discussion in Sec. 4.1.)

The object of our study is a
\inxx{world sheet,generalized} generalized world sheet. It is described by
an ${\cal M}_{(\Phi)}$-space polyvector
\be
   X = X^{[Z]} h_{[Z]} = x_0 + X^{\mu (\Phi)} h_{\mu (\Phi)} + X^{\mu_1
   (\Phi_1) \mu_2 (\Phi_2)} h_{\mu_1 (\Phi_1)} h_{\mu_2 (\Phi_2)} + ...,
\lbl{II3.155}
\ee
where $X^{\mu (\Phi)}$, $X^{\mu_1 (\Phi_1) \mu_2 (\Phi_2)}$, etc., are
1-vector, 2-vector, etc., coordinate functions. In a finite-dimensional
\inxx{coordinate functions,generalized}
sector $V_N$ of ${\cal M}_{(\Phi)}$, i.e., at a fixed value of $\Phi =
\Phi_1 = \Phi_2 = ... = \Phi_p$, the quantities $X^{\mu_1 (\Phi) \mu_2
(\Phi)...\mu_p (\Phi)} \equiv X^{\mu_1 \mu_2 ... \mu_p}$, $p = 1,2,...,N$,
\inxx{holographic projections}
are holographic projections of points, 1-loops, 2-loops,..., $N$-loop onto
the embedding space planes spanned by $\gamma_{\mu}$, $\gamma{\mu_1} \wedge
\gamma_{\mu_2}$, ... , $\gamma_{\mu_1} \wedge ... \wedge \gamma_{\mu_N}$.
Infinite loops, manifesting themselves as straight $p$-lines, are also
included. 

On the other hand, the polyvector parameters
$\Phi = \phi^J e_J$, where $\phi^J$ are scalar parameters (coordinates) 
$\phi^A, \, \phi^{A_1 A_2}, ... , \, \phi^{
A_1 ... A_n}$, represent holographic projections of points,
1-loops, ..., $p$-loops, ... $n$-loops onto the planes, spanned
by $e_A,\, e_{A_1}\wedge e_{A_2}, ... , e_{A_1} \wedge ... \wedge e_{A_n}$
in the {\it world sheet} $V_n$.

If we let the polyvector parameter $\Phi$ vary, then $X^{\mu_1 (\Phi) \mu_2
(\Phi) ... \mu_p (\Phi)}$ become infinite sets.

For instance, if we let only $\phi^A$ vary, and keep $\phi^0, \, \phi^{A_1
A_2}, ... \phi^{A_1 ... A_n}$ fixed, then:
\begin{list}{}{}
 \item $X^{\mu (\Phi)}$ becomes a set of 0-loops (i.e., points), constituting
   an ordinary $n$-dimensional world sheet;

  \item $X^{\mu_1 (\Phi) \mu_2 (\Phi)}$ becomes a set of 1-loops, constituting
   a bivector type worldsheet;

  \item \hs{2cm} $\vdots$

   \item $X^{\mu_1 (\Phi) \mu_2 (\Phi)... \mu_N (\Phi)}$ becomes a set of 
   $(N-1)$-loops, constituting an $N$-vector type worldsheet.
\end{list}

To an $(r-1)$-loop is associated an $r$-vector type worldsheet, since
the differential of an $(r-1)$-loop, i.e., 
$\dd X^{\mu_1 (\Phi) \mu_2 (\Phi)... \mu_{r-1} (\Phi)}$ is an $r$-vector.
The multivector type of an ordinary worldsheet is 1.
\inxx{world sheet,$r$-vector type}

If we let all compnents $\phi^J = \phi^0, \, \phi^A, \, \phi^{A_1 A_2},
..., \phi^{A_1 ... A_n}$ of $\Phi$ vary, then 
$X^{\mu_1 (\Phi) \mu_2 (\Phi)... \mu_r (\Phi)}$, $1 \le r \le N$, become
even more involved infinite sets, whose precise properties will not be
studied in this book.

In general, $\Phi_1, \Phi_2, ..., \Phi_n$ can be different, and
$X^{\mu_1 (\Phi_1) \mu_2 (\Phi_2)... \mu_r (\Phi_r)}$ represent even more
complicated sets of $p$-loops in ${\cal M}_{(\Phi)}$-space. In 
(\ref{II3.155}) all those objects of different multivector type are
mixed together and $X$ is a Clifford algebra valued position in the
infinite-dimensional {\it Clifford manifold} which is just
the Clifford algebra of ${\cal M}_{(\Phi)}$-space. Here we have an
\inxx{pandimensional continuum}
infinite-dimensional generalization of the concept of ``{\it pandimensional
continuum}" introduced by Pezzaglia \ci{14} and used by Castro 
\ci{58} under the name  ``{\it Clifford manifold}". \inxx{Clifford manifold}

Let an action encoding the dynamics of such a generalized membrane be
\bear
   {\cal I}[X] &=& {\cal I} [X^{[Z]}, e^A, h^{\mu (\Phi)}] \nonumber \\
   \nonumber \\
    &=& \int {\cal D} X \, \sqrt{|\rho|} \left ( I[X^{[Z]}] + I[e^A] + 
   I[h^{\mu (\Phi)}] \right ) ,
\lbl{II3.152}
\ear
where
\be
      {\cal D} X \equiv \prod_{[Z]} \dd X^{[Z]}
\lbl{II3.153}
\ee
and
\be 
     \rho \equiv {\rm det} \rho_{[Z][Z']} \; \; , \; \; \;
     \rho_{[Z][Z']} = h_{[Z]} \ast h_{[Z']} = \langle h_{[Z]} h_{[Z']}
     \rangle_0 \; ,
\lbl{II3.154}              
\ee
\begin{eqnarray}
    I[X^{[Z]}] &=& \mbox{$1\oo 2$} V^2 \lbl{II3.152a} ,\\
    I[e^A] &=& \int \dd \Phi \left [ {1\oo 2} \sqrt{|\gamma|} + {1\oo {16 \pi
   G^{(n)}}} \left ( [\p_A, \p_B] e^A \right ) \cdot e^B \right ] ,
\lbl{II3.152b} \\
   I[h^{\mu (\Phi)}] &=& {1\oo {16 \pi G^{(N)}}} \left ( [\p_{\mu (\Phi)},
   \p_{\nu (\Phi')}] h^{\mu (\Phi)} \right ) \cdot h^{\nu (\Phi')} .
\lbl{II3.152c}
\end{eqnarray}

If we consider a fixed background ${\cal M}_{(\phi)}$-space then the last
term, as well as the integration over $X^{[Z]}$, can be omitted and we have
\be
    I[X] = I[X^{[Z]}] + I[e^A] .
\lbl{II3.152d}
\ee

In eq.\ (\ref{II3.152}) we have a generalization of the action (\ref{II3.148}). 
The
dynamical system described by (\ref{II3.152}) has no fixed background space with
a fixed metric. Actually, instead of a metric, we have the basis vectors
$h^{\mu (\Phi)}$ as dynamical variables (besides ${\cal M}_{(\Phi)}$-space
coordinates $X^{[Z]}$ and $V_n$ basis vectors $e^A$). The inner product
$h_{\mu (\Phi)} \cdot h_{\nu (\Phi')} = \rho_{\mu (\Phi) \nu (\Phi')}$ gives
the metric, but the latter is not prescribed. It comes out after a solution
to the equations of motion for $h_{\mu (\Phi)}$ is found. The target space is
an infinite-dimensional ${\cal M}_{(\Phi)}$-space. A finite-dimensional
spacetime $V_N$, whose points can be identified, is associated with a
particular worldsheet configuration $X^{\mu (\Phi)}$ (a system of many
worldsheets).

The first term in (\ref{II3.152}), $\, I[X^{[Z]}]$, 
is the square of $V$, which is
a worldsheet and a target space polyvector. We have seen in Sec. 2.5 that any
polyvector can be written as a superposition of 
basis spinors which are\inxx{basis spinors}
elements of left and right ideals of the Clifford algebra. World sheet and
target space spinors are automatically included in our action (\ref{II3.152}).
There must certainly be a connection with descriptions which employ
Grassmann odd variables. A step in this direction is indicated in
Sec. 7.2, where it is shown that a polyvector can be written as a
superposition of Grassmann numbers.

\inxx{Clifford algebra valued line}
A theory of the generalized point-particle tracing a Clifford algebra-valued
line $X(\Sigma)$ ``living" in the Clifford manifold ($C$-space) has been 
convincingly outlined by \inxx{Castro}
Castro \cite{58}. The polyvector 
$$X = \Omega_{p+1} + \Lambda^p x_{\mu} \gamma^{\mu} + 
\Lambda^{p-1} \sigma_{\mu \nu} \gamma^{\mu}
\gamma^{\nu} + ... $$
encodes at one single stroke the point history
represented by the ordinary $x_{\mu}$ coordinates and the nested family of
1-loop, 2-loop, ..., $p$-loop histories. Castro \cite{59} also investigated
some important consequences for string theory and the conjectured $M$-theory
\cite{59a}. Castro's work built on important contributions 
of Aurilia et al. \ci{60}, \inxx{Castro}
who formulated $p$-branes in terms of $(p+1)$-tangents to a $p$-brane's
worldvolumes (which I call worldsheets), 
$$X_{\nu_1 \nu_2 ... \nu_{p+1}}
(\sigma) = \epsilon^{\sigma^1 \sigma^2 ... \sigma^{p+1}} \p_{\sigma^1}
X_{\nu_1} ... \p_{\sigma^{p+1}} X_{\nu_{p+1}}.$$
The action they considered
contains the square of the latter expression. The same expression also enters 
the term of the highest degree in the actions (\ref{II3.118}) or 
(\ref{II3.118a}). It also occurs in (\ref{II3.148}) if we identify the 
coefficients $v_{A_1 ... A_n}^{\mu_1 ... \mu_n}$ with $\p_{A_1} X^{\mu_1}...
\p_{A_n} X^{\mu_n}$, and analogously for the coefficients $\p X^{\mu_1 (\Phi)
... \mu_2 (\Phi)}/ \p \phi^{A_1 ... A_n}$ in (\ref{II3.148r}).

Here we have considered a natural\inxx{loop quantum gravity} 
generalization of the approaches investigated
by Aurilia et al., and Castro.\inxx{spin foams} 
Our object of study is a mess which encompasses\inxx{spinors}  
the set of points, 1-lines (1-loops), 2-lines (2-loops), ..., $N$-lines 
($N$-loops).\inxx{spin networks} 
Spinors are automatically present because of the poly\-vector
character of the basic quantities $X$ and $V$. There is very probably a
deep relationship between this approach and the ones known as loop quantum
gravity \ci{56}, spin networks \ci{56a}, and spin foams \ci{56b}, pioneered 
by Smolin, Rovelli and Baez, 
to name just few. I beleive that in this book we have
an outline of a possible general, unifying, background independent geometrical
principle behind many of the approaches incorporating strings, $p$-branes,
and quantum gravity.

\upperandlowerfalse
\chapter{Quantization}

Quantization of extended objects has turned out to be extremely
difficult, and so far it has not yet been competely solved.
In this Chapter I provide some ideas of how quantization might
be carried out technically. First I demonstrate how quantization of
the unconstrained membrane is more or less straightforward.
Then I consider the subject from a much broader perspective,
involving the Clifford algebra approach. Section 1
is largely based on a published paper \cite{43}
(and also \ci{42}) with \inxx{membrane,unconstrained}
some additional material, whilst Section 2 is completely novel.

\section{The quantum theory of unconstrained membranes}

When studying the classical membrane we arrived in Sec. 4.3
at the {\it unconstrained action} (\ref{II1.74}) or (\ref{II1.78b}).
The latter action
can be written in terms of the ${\cal M}$-space tensor notation
\be
    I[X^{\mu (\xi)}, p_{\mu (\xi)}] = \int \dd \tau \left [p_{\mu (\xi)}
    {\dot X}^{\mu (\xi)} - \mbox{$1\oo 2$} (p^{\mu (\xi)} p_{\mu (\xi)} - K) -
    \Lambda^a \p_a X^{\mu (\xi)} p_{\mu (\xi)} \right ] ,
\lbl{II4.1}
\ee
where the metric is
\be
     \rho_{\mu (\xi) \nu (\xi')} = {1\oo \Lambda} \eta_{\mu \nu} \delta
     (\xi - \xi')
\lbl{II4.2}
\ee
and
\be
      K \equiv \int {\dd}^n \xi \, \Lambda \kappa^2 |f| = K[X] = 
      \alpha^{(\xi)} \alpha_{(\xi)} ,
\lbl{II4.3}
\ee
where $\alpha^{(\xi)} \equiv \Lambda \kappa \sqrt{|f|}$
In principle the metric need not be restricted to the above form. In this
section we shall often use, when providing specific examples, the metric
\be
     \rho_{\mu (\xi) \nu (\xi')} = {{\kappa \sqrt{|f|}}\oo \Lambda} \,
     \eta_{\mu \nu} \delta (\xi - \xi') ,
\lbl{II4.5}
\ee
which is just the metric (\ref{II1.20}) with fixed 
$\lambda = \Lambda$. eq.\,(\ref{II4.1})
holds for any fixed background metric of ${\cal M}$-space.

The Hamiltonian belonging to (\ref{II4.1}) is 
\inxx{Hamiltonian,for membrane} \ci{41}-\ci{43}
\be
     H = \mbox{$1\oo 2$} (p^{\mu (\xi)} p_{\mu (\xi)} - K[X]) + 
     \Lambda^a \p_a X^{\mu (\xi)} p_{\mu (\xi)} .
\lbl{II4.4}
\ee

In the following we are going to quantize the system described by the
action (\ref{II4.1}) and the Hamiltonian (\ref{II4.4}).

\subsection[\ \ The commutation relations and the Heisenberg equations of motion]
{The commutation relations and the Heisenberg equations of motion}

Since there are no constraints on the dynamical variables $X^{\mu} (\xi)$
and the corresponding canonical momenta $p_{\mu} (\xi)$, quantization of
the theory is straightforward. The classical variables become operators
and the Poisson brackets are replaced by commutators, taken at equal $\tau$:
\be
    [X^{\mu} (\xi), \, p_{\nu} (\xi')] = i\, {\delta^{\mu}}_{\nu}\, 
    \delta (\xi - \xi') ,
\label{II4.47}
\ee
\be
     [X^{\mu} (\xi), \, X^{\nu} (\xi')] = 0 \; \; , \quad \quad 
     [p_{\mu} (\xi), \, p_{\nu} (\xi')] = 0 .
\label{II4.48}
\ee
The Heisenberg equations of motion for an operator $A$ read
\inxx{Heisenberg equations}
\be
      \p_a A = - \, i \, [A, \, H_a] ,
\label{II4.50}
\ee
\be
     {\dot A} = -\, i \, [A, \, H] .
\label{II4.51}
\ee
 In
particular, we have
\be
    {\dot p}_{\mu}(\xi) = - \, i \, [p_{\mu}(\xi), \, H] ,
\label{II4.52a}
\ee
\be
      {\dot X}^{\mu}(\xi) = - \, i \, [X^{\mu}(\xi), \, H] .
\label{II4.52b}
\ee

We may use the representation in which the operators $X^{\mu} (\xi)$ are
diagonal, and
\be
         p_{\mu}(\xi) = - \, i \, \left ({{\delta} \over 
         {\delta X^{\mu} (\xi)}} + {{\delta F} \over {\delta X^{\mu} (\xi)}}
         \right ) ,
\label{II4.52c}
\ee
where $F$ is a suitable functional of $X^{\mu}$ (see eq.\,(\ref{II4.65})).
The extra term in $p_{\mu}(\xi)$ is introduced in order to take into
account the metric (\ref{II4.5}) of the membrane space ${\cal M}$.
In such a representation it is straightforward to calculate the
useful commutators
\be
    [p_{\mu} (\xi), \, \sqrt{|f|}] = - \, i \, \, {{\delta \sqrt{|f(\xi')|}}
    \over {\delta X^{\mu} (\xi)}} = i \, \p_a  \left ( \sqrt{|f|} \,
    \p^a X_{\mu} \, \delta (\xi - \xi') \right ) ,
\label{II4.54a}
\ee
\be
      [p_{\mu} (\xi), \p_a X^{\nu} (\xi') ] = - \, i \, {{
      \delta \p_a X^{\nu} (\xi')} \over {\delta X^{\mu} (\xi)}} = -\, i \,
      {\delta_{\mu}}^{\nu} \, \p_a \, \delta (\xi - \xi') .
\label{II4.54b}
\ee
For the Hamiltonian (\ref{II4.4})  the explicit form of eqs.
(\ref{II4.52a}) and (\ref{II4.52b}) can be obtained straightforwardly
by using the commutation relations (\ref{II4.47}), (\ref{II4.48}):
\bear
   {\dot p}_{\mu} (\xi) &=& - \, \p_a \left [ {{\Lambda} \over {2 \kappa}} \,
   \p^a X_{\mu} \, \sqrt{|f|} ( {{p^2} \over {|f|}} + \kappa^2) \right ] ,
\label{II4.54}\\
\nonumber \\
     {\dot X}^{\mu} (\xi) &=& - \, i \, \int {\dd} \xi' \, \Biggl[ X^{\mu}
     (\xi), \, {{\Lambda}\over {2 \kappa}} \sqrt{|f|} \left (
     {{p^{\alpha} (\xi') p_{\alpha} (\xi')} \over {|f|}} - \kappa^2 
     \right ) \nonumber \\
     & & \hs{4.5cm} - 
     \Lambda^a \p_a X^{\alpha} (\xi') p_{\alpha} (\xi') \Biggr]\nonumber \\
    &=& {\Lambda \over {\sqrt{|f|} \kappa}} \,
     p^{\mu} (\xi') - \Lambda^a \p_a X^{\mu} .
\label{II4.55}
\ear
We recognise that the operator equations (\ref{II4.54}),\,(\ref{II4.55})
have the same form as the classical equations of motion belonging to the
action (\ref{II4.1}).

\subsection[\ \ The Schr\" odinger representation]
{The Schr\" odinger representation}
\inxx{Schr\" odinger representation}

The above relations (\ref{II4.47})--(\ref{II4.52b}),\,
(\ref{II4.54}),\,(\ref{II4.55}) are valid
regardless of representation. A possible representation is one in
which the basic states $|X^{\mu} (\xi) \rangle$ have definite
values $X^{\mu} (\xi)$ of the membrane's position operators\footnote{
When necessary we use symbols with a hat in order to distinguish operators
from their eigenvalues.} \inxx{position operators,for membrane}
${\widehat X}^{\mu} (\xi)$. An arbitrary state $|a \rangle$ can be expressed
as
\be
    |a \rangle = \int |X^{\mu} (\xi) \rangle {\cal D} X^{\mu} (\xi) \langle
    X^{\mu} (\xi) | a \rangle
\label{II4.56}
\ee
where the measure ${\cal D} X^{\mu} (\xi)$ is given in eq.\,(\ref{A16})
with $\alpha = \kappa/\Lambda$.

We shall now write the equation of motion for the wave functional
$\psi \equiv \langle X^{\mu} (\xi)|a \rangle$. We
adopt the requirement that, in the classical limit, the wave functional
equation should reproduce the Hamilton--Jacobi equation (see ref.\ \cite{43})
\inxx{Hamilton--Jacobi equation,for membrane}
\be
    -{{\p S}\oo {\p \tau}} = H .
\lbl{II4.56a}
\ee    
The supplementary equation
\be
    {{\p S}\oo {\p X^{\mu (\xi)}}} = p_{\mu (\xi)}
\lbl{II4.56b}
\ee    
has to arise from
the corresponding quantum equation. For this goal we admit that $\psi$
evolves with the evolution parameter $\tau$ and is a
functional of $X^{\mu} (\xi)$ :
\be
        \psi = \psi [\tau, X^{\mu} (\xi)] .
\label{II4.63}
\ee
It is normalized according to
\be
     \int {\cal D} X \, \psi^* \psi = 1 ,
\label{II4.F1}
\ee
which is a straightforward extension of the corresponding relation
\be
   \int {\dd}^4 x \psi^* \psi = 1
\lbl{II4.F1a}
\ee   
for the unconstrained point particle
in Minkowski spacetime \cite{1}--\ci{11a}, \ci{41}--\ci{43}. It is important
to stress again \cite{41}--\cite{43} that, since (\ref{II4.F1}) is
satisfied at any $\tau$, the evolution operator $U$ which brings
$\psi (\tau) \rightarrow \psi (\tau') = U \psi (\tau)$ is
{\it unitary}. \inxx{evolution operator}

The following equations are assumed to be satisfied ($\rho =
{\rm Det} \, \rho_{\mu(\xi) \nu(\xi')}$):
\bear
    - \, i \hbar \,{1\over {|\rho|^{1/4}}} 
    \p_{\mu (\xi)} (|\rho|^{1/4} \psi) &=& {\hat p}_{\mu (\xi)} \psi ,
\label{II4.64a} \\
\nonumber \\
     i \hbar \, {1\over {|\rho|^{1/4}}}
     {{\p (|\rho|^{1/4} \psi)} \over {\p \tau}} &=& H \psi ,
\label{II4.64c}
\ear
where\footnote{
Using the commutator (\ref{II4.54b}) we find that
$$\Lambda^a [p_{\mu (\xi)}, \p_a X^{\mu (\xi)}] = \hs{1mm}
\int {\dd} \xi \, {\dd} \xi' \, \Lambda^a \, [p_{\mu} (\xi),
\p_a X^{\nu} (\xi')] \, {\delta_{\nu}}^{\mu} \, \delta(\xi - \xi') = 0,$$
therefore the order of operators in the second term of eq.\,(\ref{II4.64c1})
does not matter.}
\be
          H = \mbox{$1\oo 2$} ({\hat p}^{\mu (\xi)} {\hat p}_{\mu (\xi)}
          - K ) + \Lambda^a \p_a X^{\mu (\xi)} {\hat p}_{\mu (\xi)} .
\label{II4.64c1}
\ee         
When the metric $\rho_{\mu(\xi) \nu(\xi')}$ in ${\cal M}$ explicitly
depends on $\tau$ (which is the case when ${\dot \Lambda} \ne 0$) such a
modified $\tau$-derivative in eq.\,(\ref{II4.64c}) is required \cite{23}
in order to assure conservation of probability, as expressed by the
$\tau$-invariance of the integral (\ref{II4.F1}).

The momentum operator given by \inxx{momentum operator,for membrane}
\be
       {\hat p}_{\mu (\xi)} = -i \hbar \left (\p_{\mu (\xi)} +
       \mbox{$1\oo 2$} \Gamma_{\mu(\xi) \nu(\xi')}^{\nu(\xi')} \right ) ,
\label{II4.65}
\ee
where 
$${1\over 2} \Gamma_{\mu(\xi) \nu(\xi')}^{\nu(\xi')} =
|\rho|^{-1/4} \, \p_{\mu(\xi)} |\rho|^{1/4} ,$$ satisfies the commutation
relations (\ref{II4.47}),\,(\ref{II4.48}) and is Hermitian with respect to the
scalar product $\int {\cal D} X \, \psi^* 
{\hat p}_{\mu(\xi)}\psi$ in ${\cal M}$.
   
The expression (\ref{II4.64c1}) for the Hamilton operator
\inxx{Hamilton operator,for membrane}
$H$ is obtained from the corresponding classical
expression (\ref{II4.4}) in which the quantities
$X^{\mu (\xi)}$, $p_{\mu (\xi)}$ are replaced by the operators
${\widehat X}^{\mu (\xi)}$, ${\hat p}_{\mu (\xi)}$.
There is an ordering ambiguity in the definition of
\inxx{ordering ambiguity} 
${\hat p}^{\mu (\xi)} {\hat p}_{\mu (\xi)}$. Following the convention
in a finite-dimensional curved space \cite{23}, we 
use the identity 
   $$|\rho|^{1/4} {\hat p}_{\mu(\xi)}|\rho|^{-1/4} =
- \, i \hbar \, \p_{\mu(\xi)}$$
and define
\bear
    {\hat p}^{\mu (\xi)} {\hat p}_{\mu (\xi)} \psi &=&
    |\rho|^{-1/2} |\rho|^{1/4} {\hat p}_{\mu(\xi)}|\rho|^{-1/4}
    |\rho|^{1/2} \rho^{\mu(\xi) \nu(\xi')} |\rho|^{1/4} {\hat p}_{\nu(\xi')}
    |\rho|^{-1/4} \psi \nonumber \\
    \nonumber \\
     &=& - |\rho|^{-1/2} \p_{\mu(\xi)}\left ( |\rho|^{1/2}
    \rho^{\mu(\xi) \nu(\xi')} \p_{\nu(\xi')} \psi \right ) \nonumber \\ 
    \nonumber \\
     &=& - {\rm D}_{\mu(\xi)} {\rm D}^{\mu(\xi)} \psi .
\label{II4.65a}
\ear
  
Let us derive the classical limit of equations 
(\ref{II4.64a}),\,(\ref{II4.64c}).
For this purpose we write \inxx{classical limit}
\be
    \psi = A[\tau, X^{\mu} (\xi)] \; \mbox{\rm exp} \left [
    {i \over {\hbar}} S[\tau, X^{\mu}(\xi) \right ]
\label{II4.66}
\ee
with real $A$ and $S$.

Assuming (\ref{II4.66}) and taking the limit $\hbar \rightarrow 0$ eq.\,(\ref
{II4.64a}) becomes
\be
     {\hat p}_{\mu (\xi)} \psi = \p_{\mu (\xi)} S \, \psi  .   
\label{II4.67}
\ee
If we assume that in eq.\,(\ref{II4.66}) $A$ is a slowly varying and $S$
a quickly varying functional of $X^{\mu} (\xi)$ we find that 
$\p_{\mu (\xi)} S$ is the expectation value of the momentum operator
${\hat p}_{\mu (\xi)}$.

Let us insert (\ref{II4.66}) into eq.\,(\ref{II4.64c}). Taking the limit
$\hbar \rightarrow 0$, and writing separately the real and imaginary
part of the equation, we obtain
\be
      - \, {{\p S} \over {\p \tau}} = \mbox{$1\oo 2$} ( \p_{\mu (\xi)} S \,
      \p^{\mu (\xi)} S - K) + \Lambda^a \p_a X^{\mu (\xi)} \, \p_{\mu (\xi)} S ,
\label{II4.70}
\ee
\be
      {1\over {|\rho|^{1/2}}} 
      {{\p \ } \over {\p \tau}} (|\rho|^{1/2} A^2)
      + {\mbox{\rm D}}_{\mu (\xi)} [A^2
      (\p^{\mu (\xi)} S + \Lambda^a \, \p_a X^{\mu (\xi)}) ] = 0 .
\label{II4.71}
\ee
eq.\,(\ref{II4.70}) is just the functional Hamilton--Jacobi (\ref{II4.56a}) 
equation of the classical theory. Eq.\ (\ref{II4.71}) is the continuity
equation, where $\psi^* \psi = A^2 $ is {\it the probability density}
and \inxx{probability density}
\be
     A^2 \, (\p^{\mu (\xi)} S - \Lambda^a \, \p_a X^{\mu (\xi)}) 
      = j^{\mu (\xi)}
\label{II4.71a}
\ee
is {\it the probability current}.\inxx{probability current} Whilst the 
covariant components
$\p_{\mu (\xi)} S$ form a momentum vector $p_{\mu}$,
the contravariant components form a vector $\p X^{\mu}$ (see also Sec. 4.2):
\bear
   \p^{\mu (\xi)} S &=& \rho^{\mu (\xi) \nu (\xi')} \, \p_{\nu (\xi')} S
   \nonumber \\
   \nonumber \\
    &=& \int {\dd} \xi' {{\Lambda} \over {\kappa \sqrt{|f|}}} \, \eta^{\mu \nu} \,
   \delta (\xi - \xi') {{\delta S} \over {\delta X^{\nu} (\xi')}}\nonumber \\
   \nonumber \\
     &=& {{\Lambda} \over {\kappa \sqrt{|f|}}} \eta^{\mu \nu} 
   {{\delta S} \over {\delta X^{\nu} (\xi)}} = \p X^{\mu} \; ,
\label{II4.72}
\ear
where we have taken 
$$\delta S/\delta X^{\nu} (\xi) =
p_{\nu} (\xi) = {{\kappa \sqrt{|f|}} \over {\Lambda}} \, 
\p X_{\nu} (\xi),$$
and raised the index by
$\eta^{\mu \nu}$, so that 
\be
\p X^{\mu} (\xi) = \eta^{\mu \nu} \p X_{\nu} (\xi) .
\lbl{II4.72a}
\ee
So we have 
$$\p^{\mu (\xi)} S + \Lambda^a \,
\p_a X^{\mu (\xi)} = {\dot X}^{\mu (\xi)} ,$$ 
and the current (\ref{II4.71a})
is proportional to the velocity, as it should be.

Since eq.\,(\ref{II4.64c}) gives the correct classical limit it is consistent
\inxx{classical limit}
and can be taken as the equation of motion for the wave functional 
$\psi$. We shall call (\ref{II4.64c}) {\it the (functional) Schr\" odinger
equation}. In general, it admits the following {\it continuity
equation}: \inxx{Schr\" odinger equation} \inxx{continuity equation}
\be
     {1\over {|\rho|^{1/2}}} {{\p \ } \over {\p \tau}} 
     (|\rho|^{1/2} \psi^* \psi) 
    + {\mbox{\rm D}}_{\mu (\xi)} j^{\mu (\xi)}
    = 0 ,
\label{II4.73}
\ee
where
\begin{eqnarray}
     j^{\mu (\xi)} & = & \mbox{$1\oo 2$} \psi^* ({\hat p}^{\mu (\xi)} + 
     \Lambda^a \p_a X^{\mu (\xi)}) \psi + \mbox{\rm h.c.} \nonumber \\
     & = & - \, {i \over 2} (\psi^* \, \p^{\mu (\xi)} \psi -
     \psi \, \p^{\mu (\xi)} \psi^*) + \Lambda^a \p_a X^{\mu (\xi)}
     \psi^* \psi .
\label{II4.74}
\end{eqnarray}

For exercise we prove below that the probability current (\ref{II4.74})
satisfies (\ref{II4.73}). First we observe that
\bear
      {{\delta \, {\p'}_a X^{\mu} (\xi')} \over {\delta X^{\nu} (\xi)}}
      &=& {\p '}_a \left ( {{\delta X^{\mu} (\xi')} \over {\delta
      X^{\nu} (\xi')}} \right ) \nonumber \\
      \nonumber \\
       &=& {\delta ^{\mu}}_{\nu} \, {\p '}_a
      \delta (\xi' - \xi) = - \, {\delta^{\mu}}_{\nu}
      \p_a \delta (\xi' - \xi) .
\label{II4.75}
\ear
Then we calculate
   $$\p_{\nu (\xi)} (\Lambda^a \p_a X^{\mu (\xi')} \psi^* \psi) =
   {{\delta } \over {\delta X^{\nu} (\xi)}} (\Lambda^a \p_a X^{\mu}
   (\xi') \psi^* \psi)\hs{4cm}$$
\be
     = - \, \Lambda^a {\delta^{\mu}}_{\nu} \, \p_a \delta (\xi' - \xi)
     \psi^* \psi + \Lambda^a \p_a X^{\mu} (\xi') \left ( 
     \psi \, {{\delta \psi^*} \over {\delta X^{\nu} (\xi)}} +
     \psi^* \, {{\delta \psi} \over {\delta X^{\nu} (\xi)}} \right ) .
\label{II4.76}
\ee
Multiplying (\ref{II4.76}) by ${\delta^{\mu}}_{\nu} \delta (\xi' - \xi)
{\dd} \xi' {\dd} \xi$, summing over $\mu$, $\nu$ and integrating over 
$\xi'$, $\xi$, we obtain
\begin{eqnarray}
    \p_{\mu (\xi)} (\Lambda^a \p_a X^{\mu (\xi)} \psi^* \psi ) &=& 
     - N \int {\dd} \xi' \, {\dd} \xi \, \Lambda^a \, \delta (\xi' -
    \xi) \, \p_a \delta (\xi' - \xi) \psi^* \psi \nonumber \\
    \nonumber \\
     & &  + \int {\dd} \xi \,
    \Lambda^a \, \p_a X^{\mu} (\xi) \left ( 
    \psi \, {{\delta \psi^*} \over {\delta X^{\mu} (\xi)}} +
     \psi^* \, {{\delta \psi} \over {\delta X^{\mu} (\xi)}} \right )
     \nonumber \\
     \nonumber \\ 
     &=& \Lambda^a \, \p_a X^{\mu (\xi)} (\psi \, \p_{\mu (\xi)} \psi^* +
    \psi^* \, \p_{\mu (\xi)} \psi) .
\label{II4.77}
\end{eqnarray}
In eq.\,(\ref{II4.77}) we have taken ${\delta^{\mu}}_{\nu} {\delta^{\nu}}_
{\mu} = N$ and $\int {\dd} \xi \, \Lambda^a \delta (\xi' -
\xi) \, \p_a \delta (\xi' - \xi) = 0$. Next we take into account
$${\mbox{\rm D}}_{\mu (\xi)} (\Lambda^a \p_a X^{\mu (\xi)} \psi^* \psi )
= ({\mbox{\rm D}}_{\mu (\xi)} \, \p_a X^{\mu (\xi)}) \Lambda^a
\psi^* \psi + \p_a X^{\mu (\xi)}\, \p_{\mu (\xi)} (\Lambda^a \psi^* \psi)$$
and
\be
   {\mbox{\rm D}}_{\mu (\xi)}  \p_a X^{\mu (\xi)} = 
   \p_{\mu (\xi)} \p_a X^{\mu (\xi)} +
   \Gamma_{\mu (\xi) \nu (\xi')}^{\mu (\xi)} \, {\p '}_a X^{\nu (\xi')} .
\label{II4.78}
\ee
From (\ref{II4.74}),\,(\ref{II4.77}) and (\ref{II4.78}) we have
\bear
   {\DD}_{\mu (\xi)} j^{\mu (\xi)} &=& - \, {i \over 2} (
   \psi^* {\DD}_{\mu (\xi)} {\DD}^{\mu (\xi)} \psi -
   \psi {\DD}_{\mu (\xi)} {\DD}^{\mu (\xi)} \psi^*) \label{II4.84} \\
   \nonumber \\   
   & & \; + \, \Lambda^a \p_a
   X^{\mu (\xi)} (\psi^* \p_{\mu (\xi)} \psi + \psi \p_{\mu (\xi)} \psi^* 
   + \Gamma_{\mu(\xi) \nu(\xi')}^{\nu(\xi')}\, \psi^* \psi) . \nonumber
\ear
Using the Schr\" odinger equation \inxx{Schr\" odinger equation}
\be
    i |\rho|^{-1/4} \p (|\rho|^{1/4}\psi)/\p \tau = H \psi
\lbl{II4.84a}
\ee    
and the complex conjugate equation 
\be
    -i |\rho|^{-1/4}\p  (|\rho|^{1/4}\psi^*)/\p \tau = H^* \psi^* ,
\lbl{II4.84b}
\ee    
where
\inxx{continuity equation} \inxx{probability density} \inxx{probability current}
$H$ is given in (\ref{II4.64c1}), we obtain that the continuity equation
(\ref{II4.73}) is indeed satisfied by the probability density $\psi^* \psi$
and the current (\ref{II4.74}). For the Ansatz (\ref{II4.66}) 
the current (\ref{II4.74})
becomes equal to the expression (\ref{II4.71a}), as it should.

We notice that the term with 
$\Gamma_{\mu(\xi) \nu(\xi')}^{\nu(\xi')}$ in eq.\,(\ref{II4.84})
is canceled by the same type of the term in $H$. The latter term in $H$
comes from the definition (\ref{II4.65}) of the momentum operator in a (curved)
membrane space ${\cal M}$, whilst the analogous term in eq.\,(\ref{II4.84})
results from the covariant differentiation. The definition (\ref{II4.65}) of
${\hat p}_{\mu(\xi)}$, which is an extension of the definition introduced
by DeWitt \cite{23} for a finite-dimensional curved space,
is thus shown to be consistent also with the conservation
of the current (\ref{II4.74}).

\subsection[\ \ The stationary Schr\" odinger equation for a membrane]
{The stationary Schr\" odinger equation for a membrane}
\inxx{Schr\" odinger equation,for membrane,stationary}

The evolution of a generic membrane's state $\psi[\tau, X^{\mu}(\xi)]$
is given by the $\tau$-dependent functional Schr\" odinger equation
(\ref{II4.64c}) and the Hamiltonian (\ref{II4.64c1}). We are now going to
consider solutions which have the form
\be
    \psi[\tau, X^{\mu}(\xi)] = e^{-iE \tau} \phi[X^{\mu}(\xi)] ,
\label{II4.85}
\ee
where $E$ is a constant. We shall call it {\it energy}, since it has a role
analogous to energy in non-relativistic quantum mechanics.
Considering the case ${\dot \Lambda} = 0$, ${\dot \Lambda}^a = 0$ and
inserting
the Ansatz (\ref{II4.85}) into eq.\,(\ref{II4.64c}) we obtain
\be
     \left ( -\, \mbox{$1\oo 2$} {\DD}^{\mu (\xi)} {\DD}_{\mu (\xi)} +
     i\, \Lambda^a \p_a X^{\mu (\xi)} (\p_{\mu (\xi)} +
     \mbox{$1\oo 2$} \Gamma_{\mu(\xi) \nu(\xi')}^{\nu(\xi')})
     - {1\over 2} K \right ) \phi = E \, \phi .
\label{II4.86}
\ee
So far the membrane's dimension and signature have not been specified. Let us
now consider {\it Case 2} of Sec.\,4.3.     
All the dimensions of our membrane have the same
signature, and the index $a$ of a membrane's coordinates assumes
the values $a = 1,2,..., n = p$. Assuming a real $\phi$ eq.\,(\ref{II4.86})
becomes
\bear
   \left ( -\, \mbox{$1\oo 2$} {\DD}^{\mu (\xi)} {\DD}_{\mu (\xi)} -
     \mbox{$1\oo 2$} K - E \right ) \phi &=& 0 ,
\label{II4.88} \\
\nonumber \\
    \Lambda^a \p_a X^{\mu (\xi)} (\p_{\mu (\xi)} + 
     \mbox{$1\oo 2$} \Gamma_{\mu(\xi) \nu(\xi')}^{\nu(\xi')}) \phi &=& 0 .
\label{II4.89}
\ear
These are equations for a {\it stationary state}. They remind us of
\inxx{stationary state} 
the well known $p$-brane equations \cite{63}.

In order to obtain from (\ref{II4.88}),\,(\ref{II4.89}) the conventional
$p$-brane equations we have to assume that eqs.\,(\ref{II4.88}),\,(\ref{II4.89})
hold for any $\Lambda$ and $\Lambda^a$, which is indeed the case.
Then instead of eqs.\,(\ref{II4.88}),\,(\ref{II4.89}) 
in which we have the integration
over $\xi$, we obtain the equations without the integration over $\xi$:
\bear
    \left ( - \, {{\Lambda} \over {2 \kappa |f|}} \eta^{\mu \nu} {{{\DD}^2}
    \over {{\DD} X^{\mu} (\xi) {\DD} X^{\nu} (\xi)}} - {{\Lambda} \over {2}} -
    {\cal E} \right ) \phi &=& 0 ,
\label{II4.90}\\
\nonumber \\ 
     \p_a  X^{\mu} (\xi)\, \left ( {{\delta \ } \over 
     {\delta X^{\mu} (\xi)}} + 
     \mbox{$1\oo 2$} \Gamma_{\mu(\xi) \nu(\xi')}^{\nu(\xi')} 
     \right ) \phi &=& 0 .
\label{II4.91}
\ear
The last equations are obtained from (\ref{II4.88}),\,(\ref{II4.89}) after
writing the energy as the integral of the energy density ${\cal E}$ over
the membrane, $E = \int {\dd}^n \xi \, \sqrt{|f|} \, {\cal E}$, taking into
account that $K = \int {\dd}^n \xi \, \sqrt{|f|} \, \kappa \Lambda$, and
omitting the integration over $\xi$. 

Equations (\ref{II4.90}),\,(\ref{II4.91}), with ${\cal E} = 0$,
are indeed the quantum analogs of the classical $p$-brane constraints
used in the literature \cite{63,64} and
their solution $\phi$ represents states of a conventional, constrained,
$p$-brane with a tension $\kappa$. When ${\cal E} \neq 0$ the
preceding statement still holds, provided that ${\cal E} (\xi) $ is
proportional to $\Lambda (\xi)$, so that the quantity 
$\kappa (1 - 2 {\cal E}/\Lambda)$
is a constant, identified with the effective tension. Only the particular
stationary states (as indicated above)
correspond to the conventional, Dirac--Nambu--Goto $p$-brane states, but in
general they correspond to a sort of wiggly membranes \cite{41,65}.
\inxx{wiggly membrane}

\subsection[\ \ Dimensional reduction of the Schr\" odinger equation]
{Dimensional reduction of the Schr\" odinger equation}

Let us now consider the {\it Case 1}.
Our membrane has signature $(+ - - - ... )$ and is actually an
$n$-dimensional worldsheet. The index $a$ of the worldsheet coordinates
$\xi^a$ assumes the values $a = 0,1,2,...,p$, where $p = n - 1$.

Amongst all possible wave functional
satisfying eqs.\,(\ref{II4.64c}) there are also the special ones
for which it holds (for an example see see eqs.\,(\ref{II4.A50a}),
(\ref{II4.117})--(\ref{II4.119}))
\be
     {{\delta \psi} \over {\delta X^{\mu} (\xi^0, \xi^i)}} = 
     \delta (\xi^0 - \xi_{\Sigma}^0 ) (\p_0 X^{\mu} \p_0 X_{\mu})^{1/2}
     {{\delta \psi} \over {\delta X^{\mu} (\xi_{\Sigma}^0, \xi^i)}} ,
\label{II4.98}
\ee
    $$i = 1,2,...,p=n-1 ,$$
where $\xi_{\Sigma}^0$ is a fixed value of the time-like coordinate
$\xi^0$. In the compact tensorial notation in the membrane space
${\cal M}$ eq.\,(\ref{II4.98}) reads
\be
     \p_{\mu (\xi^0,\xi^i)} \phi = \delta (\xi^0 - \xi_{\Sigma}^0) \, 
     (\p_0 X^{\mu} \p_0 X_{\mu})^{1/2} 
     \p_{\mu (\xi_{\Sigma}^0, \xi^i)} \psi .
\label{II4.99}
\ee
Using (\ref{II4.99}) we find that the dimension of the Laplace operator in
\bear
    {\DD}^{\mu (\xi)}{\DD}_{\mu (\xi)} \psi &=& \int {\dd}^n \xi \,
    {{\Lambda} \over {\kappa \sqrt{|f|}}} \, \eta^{\mu \nu} {{{\DD}^2 \psi}
    \over
    {{\DD} X^{\mu} (\xi) {\DD} X^{\nu} (\xi)}} \nonumber \\
    \nonumber \\
     &=& \int {\dd} \xi^0 \, {\dd}^p \xi
    {{\Lambda} \over {\kappa \sqrt{|f|}}} (\p_0 X^{\mu} \p_0 X_{\mu})^{1/2}
    \, \eta^{\mu \nu}
    \delta (\xi^0 - \xi_{\Sigma}^0) \times \nonumber \\
    \nonumber \\
    & & \hs{3cm} \times \, {{{\DD}^2 \psi} \over
    {{\DD} X^{\mu} (\xi^0, \xi^i) {\DD} X^{\nu} (\xi_{\Sigma}^0, \xi^i)}}
    \nonumber \\
     &=& \int {\dd}^p \xi
    {{\Lambda} \over {\kappa \sqrt{|{\bar f}|}}} \eta^{\mu \nu}
    {{{\DD}^2 \psi} \over
    {{\DD} X^{\mu} (\xi_{\Sigma}^0, \xi^i) X^{\nu} (\xi_{\Sigma}^0, \xi^i)}}
    \nonumber \\
    \nonumber \\
     &=& \int {\dd}^p \xi
    {{\Lambda} \over {\kappa \sqrt{|{\bar f}|}}} \eta^{\mu \nu} {{{\DD}^2 \psi} \over
    {{\DD} X^{\mu} (\xi^i) {\DD} X^{\nu} (\xi^i)}} .
\label{II4.100}
\ear
Here ${\bar f} \equiv \mbox{\rm det} {\bar f}_{ij}$ is
the determinant of the
induced metric ${\bar f}_{ij} \equiv \p_i X^{\mu} \p_j X_{\mu}$ on $V_p$,
and it is related to the determinant $f \equiv \mbox{\rm det} f_{ab}$
of the induced metric $f_{ab} = \p_a X^{\mu} \p_b X_{\mu}$ on $V_{p+1}$
according to $f = {\bar f} \, \p_0 X^{\mu} \p_0 X_{\mu}$ 
(see refs.\,\cite{49,50}).

The differential
operator in the last expression of eq.\,(\ref{II4.100}) (where we have
identified
$X^{\mu (\xi_{\Sigma}^0, \xi^i)} \equiv X^{\mu (\xi^i)}$) acts in the space of
$p$-dimensional membranes, although the original ope\-rator we started from
acted in the space of $(p+1)$-dimensional membranes ($n=p+1$). This comes
from the fact that our special functional, satisfying (\ref{II4.98}), has
vanishing functional derivative $\delta \psi/ \delta X^{\mu} (\xi^0, \xi^i)$
for all values of $\xi^0$, except for $\xi^0 = \xi_{\Sigma}^0$. The xpression
(\ref{II4.98}) has its finite-dimensional analog in the relation
$\p \phi /\p x^A = {\delta_A}^{\mu} \, \p \phi / \p x^\mu$,
$A = 0,1,2,3,...,3 + m$, $\mu = 0,1,2,3$, which says that the field $\phi (
x^A)$ is constant along the extra dimensions. For such a field the
($4+m$)-dimensional Laplace expression $\eta^{AB} {{\p^2 \phi} \over
{\p x^A \p x^B}}$ reduces to the 4-dimensional expression
$\eta^{\mu \nu} {{\p^2 \phi} \over {\p x^\mu \p x^\nu}}$.

The above procedure can be performed not only for $\xi^0$ but also for any of
the coordinates $\xi^a$; it applies both to {\it Case 1} and {\it Case 2}.

Using (\ref{II4.99}),\,(\ref{II4.100}) we thus find that, for such a special
wave functional $\psi$, the equation (\ref{II4.64c}),
which describes a state of a $(p+1)$-dimensional membrane,
reduces to the equation for a $p$-dimensional membrane. This is an
important finding. Namely, at the beginning we may assume a certain dimension
of a membrane and then consider lower-dimensional membranes as
particular solutions of the higher-dimensional equation. This means
that the point particle theory (0-brane), the string theory (1-brane), and
in general a $p$-brane theory for arbitrary $p$, are all contained in
the theory of a $(p+1)$-brane.

\subsection[\ \ A particular solution to the covariant Schr\" odinger equation]
{A particular solution to the covariant Schr\" odinger equation}

Let us now consider the covariant functional Schr\" odinger equation
(\ref{II4.64c}) with the Hamiltonian operator (\ref{II4.64c1}). The quantities
$\Lambda^a$ are arbitrary in principle. For simplicity we now take 
$\Lambda^a = 0$. Additionally we also take a $\tau$-independent
$\Lambda$, so that ${\dot \rho} = 0$. Then eq.\,(\ref{II4.64c}) becomes
simply ($\hbar = 1$)
\be
   i \, {{\p \psi} \over {\p \tau}} = -\, \mbox{$1\oo 2$} \,
   ({\DD}^{\mu (\xi)} \DD_{\mu (\xi)} + K ) \psi .
\label{II4.101}
\ee
The operator on the right hand side is the infinite-dimensional
analog of the covariant Klein--Gordon operator. Using the definition of the
covariant derivative (\ref{A17}) and the corresponding affinity  we have
\cite{42,43}
\bear    
    {\DD}_{\mu (\xi)} {\DD}^{\mu (\xi)} \psi
    &=&  \rho^{\mu (\xi) \nu (\xi')}
   {\DD}_{\mu (\xi)}{\DD}_{\nu (\xi')} \psi \nonumber \\
   \nonumber \\
   &=& \rho^{\mu (\xi) \nu (\xi')} \left (\p_{\mu(\xi)} 
  \p_{\nu(\xi')} \psi -
     \Gamma_{\mu(\xi) \nu(\xi')}^{\alpha(\xi'')} \, \p_{\alpha(\xi'')}
     \psi \right ) .
\label{II4.102} 
\ear
The affinity is explicitly
\be
      \Gamma_{\mu(\xi) \nu(\xi')}^{\alpha(\xi'')} = \mbox{$1\oo 2$}
      \rho^{\alpha(\xi'') \beta( \xi''')}\,
      \left ( \rho_{\beta(\xi''') \mu(\xi),\nu(\xi')} +
      \rho_{\beta(\xi''') \nu(\xi'), \mu(\xi)} - \rho_{\mu(\xi) \nu(\xi')
      , \beta(\xi''')} \right ) ,
\label{II4.102a}
\ee
where the metric is given by (\ref{II4.5}).
Using
\bear
     \rho_{\beta(\xi'') \mu(\xi),\nu(\xi')} &=& \eta_{\mu \nu}\,
     \alpha (\xi) \delta(\xi - \xi'') {{\delta \sqrt{|f(\xi)|}} \over
     {\delta X^{\nu} (\xi')}} \hs{7mm} \label{II4.102b}\\
     \nonumber \\
     &=& \eta_{\mu \nu}\,
     \alpha (\xi) \delta(\xi - \xi'') \sqrt{|f(\xi)|} \p^a X_{\nu} (\xi)
     \p_a \delta (\xi - \xi')
\nonumber
\ear     
the equation (\ref{II4.102}) becomes
\bear
      &&{\DD}_{\mu (\xi)} {\DD}^{\mu (\xi)} \psi = \rho^{\mu (\xi) \nu (\xi')}
   {{\p^2 \psi} \over {\p X^{\mu (\xi)} X^{\nu (\xi')}}} \hs{6cm} 
   \label{II4.107} \\
   \nonumber \\
    &&\hs{2.5cm} - {{\delta (0)} \over {\kappa}} \int {\dd}^n \xi \, 
    {{\delta \psi} \over {\delta X^{\mu} (\xi)}}   
    \Biggl[ {N\over 2}
    {{\Lambda} \over {\sqrt{|f|}}} \, {1\over {\sqrt{|f|}}} \,
    {\p}_{a} (\sqrt{|f(\xi)|} \, \, {\p}^{a} X^{\mu}) \nonumber \\ 
    \nonumber \\       
    &&  \hs{4cm}+ \, ({N\over 2} + 1)
    \Lambda {\p}^{a} X^{\mu}\, \p_a ({1\over {\sqrt{|f(\xi)|}}}) 
    +{{{\p}^{a} X^{\mu} \p_a \Lambda} \over {\sqrt{|f(\xi)|}}} \Biggr] ,
    \nonumber
\ear
where $N = \eta^{\mu \nu} \eta_{\mu \nu}$ is the dimension of spacetime.
In deriving eq.\,(\ref{II4.107}) we encountered the expression $\delta^2
(\xi - \xi')$ which we replaced by the corresponding approximate expression
$F(a, \xi-\xi') \delta(\xi - \xi')$, where$F(a, \xi-\xi')$ is any finite
function, e.g., $(1/\sqrt{\pi} a) {\rm exp} [-(\xi-\xi')2/a^2]$, which in the
limit $a \rightarrow 0$ becomes $\delta(\xi-\xi')$. The latter limit was
taken after performing all the integrations, and $\delta(0)$ should be
considered as an abbreviation for $\lim_{a\to 0} F(a,0)$.

As already explained in footnote 5 of Chapter 5, the infinity 
$\delta(0)$ in an expression such as (\ref{II4.107}) can be
regularized by taking into account the plausible assumption that a
generic physical object is actually a fractal (i.e., an object with 
\inxx{fractal}
detailed structure on all scales). The objects $X^{\mu} (\xi)$ which we are
using in our calculations are well behaved functions with definite
derivatives and should be considered as approximations of the actual
physical objects. This means that a description with a given $X^{\mu} (\xi)$
below a certain scale $a$ has no physical meaning. In order to make
physical sense of the expression (\ref{II4.107}), $\delta (0)$ should
therefore be replaced by $F(a,0)$. Choice of the scale $a$ is arbitrary
and determines the precision of our description. (Cf., the length of a coast
depends on the scale of the units with which it is measured.)

Expression (\ref{II4.102}) is analogous to the corresponding finite
dimensional expression. In analogy with a finite-dimensional case,
a metric tensor ${\rho '}_{\mu(\xi)\nu(\xi')}$ obtained from
${\rho }_{\mu(\xi)\nu(\xi')}$ by a coordinate 
transformation (\ref{A5})\inxx{coordinate transformations}
belongs to the same space ${\cal M}$  and is equivalent to
${\rho }_{\mu(\xi)\nu(\xi')}$. Instead of a finite number of
coordinate conditions which
express a choice of coordinates, we now have  infinite coordinate
conditions. The second term in eq. (\ref{II4.107}) becomes zero if we take
\bear
  &&{N\over 2} {{\Lambda} \over {\sqrt{|f|}}} \, {1\over {\sqrt{|f|}}} \,
    {\p}_{a} \left (\sqrt{|f(\xi)|} \, \, {\p}^{a} X_{\mu} \right ) \hs{6cm}
    \label{II4.DD1} \\
        &&\hs{3cm} + \left (
    ({N\over 2} + 1) \Lambda \p_a ({1\over {\sqrt{|f(\xi)|}}}) +
    {{\p_a \Lambda} \over {\sqrt{|f(\xi)|}}} \right ) {\p}^{a} X_{\mu} = 0 ,
    \nonumber 
\ear
and these are just possible coordinate 
conditions in the\inxx{coordinate conditions,in membrane space}
membrane space ${\cal M}$. eq.\,(\ref{II4.DD1}), together with boundary
conditions, determines a family of functions $X^{\mu} (\xi)$   
for which the functional $\psi$ is defined; in the operator theory such
a family is called the domain of an operator. Choice of a family
of functions $X^{\mu} (\xi)$ is, in fact, a choice of  coordinates (a gauge)
in ${\cal M}$.

If we contract eq.\,(\ref{II4.DD1}) by $\p_b X_{\mu}$ and take into account
the identity $\p_c X^{\mu}\, {\DD}_a {\DD}_b X_{\mu} = 0$ we find
\be
    ({N\over 2} + 1) \Lambda \p_a ({1\over {\sqrt{|f(\xi)|}}}) +
    {{\p_a \Lambda} \over {\sqrt{|f(\xi)|}}} = 0 .
\label{II4.DD2}
\ee
From (\ref{II4.DD2}) and (\ref{II4.DD1}) we have
\be
   {1 \over { \sqrt{|f(\xi)|}}} \,
 {\p}_{a} \left ( \sqrt{|f(\xi)|} \, \, {\p}^{a} X_{\mu} \right) = 0 .
\label{II4.DD3}
\ee
Interestingly, the gauge condition (\ref{II4.DD1}) in ${\cal M}$
\inxx{gauge condition}
automatically implies the gauge condition (\ref{II4.DD2}) in $V_n$. The
latter condition is much simplified if we take $\Lambda \ne 0$
satisfying $\p_a \Lambda = 0$; then for $|f| \ne 0$ eq.\,(\ref{II4.DD2}) becomes
\be
                        \p_a \sqrt{|f|} = 0
\label{II4.DD4}
\ee
which is just the gauge condition considered by Schild \cite{51}
and Eguchi \cite{66}. \inxx{Schild action}
    
In the presence of the condition (\ref{II4.DD1}) (which is just a gauge,
or coordinate, condition in the function space ${\cal M}$) the
functional Schr\" odinger equation (\ref{II4.101}) can be written in the
form \inxx{Schr\" odinger equation}
\be
          i \, {{\p \psi} \over {\p \tau}} = - {1 \over 2} \int {\dd}^n \xi
     \left ( {{\Lambda} \over {\sqrt{|f|} \kappa}} \, {{{\delta}^2}
    \over {\delta X^{\mu} (\xi) \delta X_{\mu} (\xi)}} + \sqrt{|f|} \,
    \Lambda \kappa \right ) \psi .
\label{II4.109}
\ee
A particular solution to eq.\,(\ref{II4.109}) can be obtained by considering the
following eigenfunctions of the momentum operator
\be
     {\hat p}_{\mu (\xi)} \psi_p [X^{\mu (\xi)}] = 
     p_{\mu (\xi)} \psi_p [X^{\mu (\xi)}] ,
\label{II4.110}
\ee
with
\be    
    \psi_p [X^{\mu (\xi)}] = {\cal N} \mbox{\rm exp} \left [i \int_{X_0}^X
    p_{\mu (\xi)} \, {\dd} {X'}^{\mu (\xi)} \right ] .
\label{II4.111a}
\ee
This last expression is invariant under reparametrizations of $\xi^a$
(eq.\,(\ref{A2})) and of $X^{\mu} (\xi^a)$ (eq.\,(\ref{A5})). The momentum
field $p_{\mu (\xi)}$, in general,  functionally depends on $X^{\mu (\xi)}$
and satisfies (see \cite{43})
\be
     \p_{\mu (\xi)} p_{\nu (\xi')} -  \p_{\nu (\xi')} p_{\mu (\xi)}  = 0 \; ,
     \qquad {\DD}_{\mu (\xi)} p^{\mu (\xi)} = 0 .
\lbl{II4.111aaa}
\ee     
In particular $p_{\mu (\xi)}$
may be just a constant field, such that $\p_{\nu (\xi')} p_{\mu (\xi)} = 0$.
Then (\ref{II4.111a}) becomes
\bear
   \psi_p [X^{\mu (\xi)}] &=& {\cal N} \mbox{\rm exp} \left [i \int 
   {\dd}^n \xi \,
    p_{\mu} (\xi) (X^{\mu} (\xi) - X_0^{\mu} (\xi)) \right ]\nonumber \\
    \nonumber \\
     &=& {\cal N} \mbox{\rm exp} \left [i p_{\mu (\xi)} (X^{\mu (\xi)} -
    X_0^{\mu (\xi)}) \right ]  .
\label{II4.111}
\ear
The latter expression holds only in a particular parametrization\footnote{
{\it Solutions} to the equations of motion are always written in a
particular parametrization (or gauge). For instance, a plane wave
solution ${\rm exp} \, [i p_{\mu} x^{\mu}]$ holds in Cartesian coordinates,
but not in spherical coordinates.}
 (eq.\,(\ref
{A5})) of ${\cal M}$ space, but it is still invariant with respect to
reparametrizations of $\xi^a$. 

Let the $\tau$-dependent wave functional be \inxx{wave functional}   
\bear   
    & &\psi_p [\tau, X^{\mu} (\xi)] \hs{8cm} \nonumber \\
    \nonumber \\ 
    & & \hs{7mm} = {\cal N} \mbox{\rm exp} \left [i \int {\dd}^n \xi \,
    p_{\mu} (\xi) \left ( X^{\mu} (\xi) - X_0^{\mu} (\xi) \right ) \right ]
    \times \nonumber \\
    \nonumber \\    
    & &\hs{11mm} \times \, \mbox{\rm exp} \left [- {{i \tau} \over 2} 
    \int \Lambda {\dd}^n \xi
    \left ( {{p^{\mu} (\xi) p_{\mu} (\xi)} \over {\sqrt{|f|} \kappa}} -
    \sqrt{|f|} \kappa \right ) \right ] \nonumber \\ 
    \nonumber \\ 
    & &\hs{7mm} \equiv {\cal N} \mbox{\rm exp} \left [ i 
     \int {\dd}^n \xi \, {\cal S} \right ] .
\label{II4.112}
\ear
where $f \equiv {\rm det} \, \p_a X^{\mu} (\xi) \p_b X_{\mu} (\xi)$ should
be considered as a functional of $X^{\mu} (\xi)$ and independent of $\tau$.
From (\ref{II4.112}) we find
\bear
   - \, i \, {{\delta \psi_p [\tau,X^{\mu}]} \over {\delta X^{\alpha}}}
   &=& - \, i \, \left ( {{\p {\cal S}} \over {\p X^{\alpha}}} - \p_a
   {{\p {\cal S}} \over {\p \p_a X^{\alpha}}} \right ) 
   \psi_p [\tau, X^{\mu}] \nonumber \\ 
   \nonumber \\
   &=& p_{\alpha} \psi_p
    - \, \tau \p_a \left ( \sqrt{|f|} \, \p^a X_{\mu} \right ) 
    {{\Lambda} \over {2 \kappa}}
   \, \left ( {{p^2} \over {|f|}} + \kappa^2 \right ) \psi_p \nonumber \\
   \nonumber \\
    & & - \, \tau \, 
   \sqrt{|f|}
   \, \p^a X_{\mu} \p_a \left [ {{\Lambda} \over {2 \kappa}}
   \, \left ( {{p^2} \over {|f|}} + \kappa^2 \right ) \right ] \psi_p .
\label{II4.113}
\ear
Let us now take the gauge condition (\ref{II4.DD3}). Additionally let us
assume
\be
    \p_a \left [ {{\Lambda} \over {2 \kappa}}
   \, \left ( {{p^2} \over {|f|}} + \kappa^2 \right ) \right ] \equiv
   \kappa \, \p_a \mu = 0 .
\label{II4.114}
\ee
By inspecting the classical equations of motion with
$\Lambda^a = 0$  we see \cite{43} that eq.\,(\ref{II4.114}) is satisfied
when the momentum of a classical membrane does not change with $\tau$, i.e.,
\be
    { {{\dd} p_{\mu}} \over {{\dd} \tau} } = 0.
\label{II4.115}
\ee
Then the membrane satisfies
the minimal surface equation
\be
           {\DD}_a {\DD}^a X_{\mu} = 0 ,
\lbl{II4.115aa}
\ee           
which is just our gauge condition\footnote
  {The reverse is not necessarily true: the imposition of the gauge
  condition (\ref{II4.DD3}) does not imply 
  (\ref{II4.114}),\,(\ref{II4.115}). The latter
  are additional assumptions which fix a possible congruence of
  trajectories (i..e., of $X^{\mu (\xi)} (\tau)$) over which the wave
  functional is defined.} 
\inxx{gauge condition,in membrane space}
(\ref{II4.DD3}) in the membrane space ${\cal M}$. 
When ${\dot \Lambda} = 0$
 the energy $E \equiv \int \Lambda {\dd}^n \xi
    \left ( {{p^{\mu} p_{\mu}} \over {\sqrt{|f|} \kappa}} -
    \sqrt{|f|} \kappa \right )$ is a constant of motion. Energy conservation
in the presence of eq.\,(\ref{II4.115}) implies
\be
    {{{\dd} \sqrt{|f|}}\over {{\dd} \tau}} = 0 .
\label{II4.115a}
\ee
        
Our stationary state (\ref{II4.112}) is thus defined over a congruence of
classical trajectories satisfying \inxx{stationary state}
(\ref{II4.115aa}) and (\ref{II4.115}), which imply also
(\ref{II4.114}) and (\ref{II4.115a}). eq.\,(\ref{II4.113}) then becomes simply
\be
     - \, i \, {{\delta \psi_p} \over {\delta X^{\alpha}}} =
       p_{\alpha} \psi_p  .
\label{II4.116}
\ee
Using (\ref{II4.116}) it is straightforward to verify that (\ref{II4.112})
is a particular solution to the Schr\" odinger eqation (\ref{II4.101}). In
ref.\ \cite{42} we found the same relation (\ref{II4.116}), but by using a
different, more involved procedure.

\subsection[\ \ The wave packet]
{The wave packet}

From the particular solutions (\ref{II4.112}) we can compose a wave packet
\inxx{wave packet,for membrane}
\be
      \psi [\tau, X^{\mu} (\xi)] = \int {\cal D} p \, c[p] \,
      \psi_p [\tau, X^{\mu} (\xi)]
\label{II4.116a}
\ee
which is also a solution to the Schr\" odinger equation (\ref{II4.101}).
However, since  all $\psi_p [\tau, X^{\mu} (\xi)]$ entering (\ref{II4.116a})
are assumed to belong
to a restricted class of particular solutions with $p_{\mu} (\xi)$
which does not functionally depend on $X^{\mu} (\xi)$, a wave packet 
of the form (\ref{II4.116a}) cannot represent every possible state of the 
membrane.
This is just a particular kind of wave packet; a general wave packet can be
formed from a complete set of particular solutions which are not restricted to
momenta $p_{\mu (\xi)}$ satisfying $\p_{\nu (\xi')} p_{\mu (\xi)} = 0$,
but allow for $p_{\mu (\xi)}$ which do depend on $X^{\mu (\xi)}$.
Treatment of such a general case is beyond the scope of the present book.
Here we shall try to demonstrate some illustrative properties of the wave
packet (\ref{II4.116a}). 

In the definition of the invariant measure in \inxx{measure in momentum space}
momentum space we use the metric (\ref{A13}) with $\alpha = \kappa/\Lambda$:
\be
    {\cal D} p \equiv \prod_{\xi ,\mu} \mbox{$ {\left ( 
   {{\Lambda} \over {\sqrt{|f|} \, \kappa }} \right ) }^{1/2} $}
   {\dd } p_{\mu} (\xi) .
\label{II4.A48}
\ee
Let us take\footnote{
This can be written compactly as 
$$c[p] ={\cal B} \,\mbox{\rm exp}
[ -{1 \over 2} \rho_{\mu (\xi) \nu (\xi'')}
(p^{\mu (\xi)} - p_0^{\mu (\xi)}) (p^{\nu (\xi')} - p_0^{\nu (\xi')})
{\sigma_{(\xi')}}^{(\xi'')}],$$
where ${\sigma_{(\xi')}}^{(\xi'')} =
\sigma (\xi') \delta(\xi', \xi'')$. Since the covariant derivative of
the metric is zero, we have that ${\DD}_{\mu (\xi)} c[p] = 0$.
Similarly, the measure ${\cal D} p = (\mbox{\rm Det} \,
\rho_{\mu (\xi) \nu (\xi')})^{1/2} \prod_{\mu , \xi} {\dd} p_{\mu} (\xi)$,
and the covariant derivative of the determinant is zero. Therefore
$${\DD}_{\mu (\xi)} \int {\cal D} p \, c[p] \, \psi_p [\tau, X^{\mu} (\xi)] =
\int {\cal D} p \, c[p] \, {\DD}_{\mu (\xi)}\psi_p [\tau, X^{\mu} (\xi)].$$
This confirms that the superposition (\ref{II4.116a}) is a solution if
$\psi_p$ is a solution of (\ref{II4.64c}).}

\be
    c[p] = {\cal B} \, \mbox{\rm exp} \left [ -{1 \over 2} \int d^n \xi \,
    {{\Lambda} \over {\sqrt{|f|} \kappa}} (p^{\mu} - p_0^{\mu})^2 \,
     \sigma (\xi) \right ] ,
\label{II4.A49}
\ee
where
\be 
    {\cal B} = \lim_{\Delta \xi \to 0} \prod_{\xi,\mu}
{\left ( {{\Delta \xi \, \sigma (\xi)} \over {\pi}} \right ) }^{1/4}
\lbl{II4.A49a}
\ee
is the
normalization constant, such that $\int {\cal D} p \, c^{*} [p] c[p] = 1$.
For the
normalization constant ${\cal N}$ occurring in (\ref{II4.112}) we take
\be
{\cal N} = \lim_{\Delta \xi \to 0} \, \prod_{\xi,\mu} {\left ( 
{{\Delta \xi} \over {2 \pi}} \right )}^{1/2}.
\lbl{II4.A49b}
\ee
From (\ref{II4.116a})--(\ref{II4.A49}) and (\ref{II4.112}) we have
$$
   \psi[\tau, X (\xi)] = \hs{10cm} $$
  $$\lim_{\Delta \xi \to 0} \prod_{\xi,\mu} \int
   {\left ( {{\Delta \xi } \over {2 \pi}} \right )}^{1/2}
   {\left ( {{\Delta \xi \, 
   \sigma (\xi)} \over {\pi}} \right )}^{1/4}
   { \left ({{\Lambda} \over {\sqrt{|f|} \kappa}}\right )}^{1/2}
    \, \dd p_{\mu} (\xi) \hs{4cm}
$$    

$$
    \times \, \mbox{\rm exp} \left [- {{\Delta \xi} \over 2}
   {{\Lambda} \over {\sqrt{|f|} \kappa}}
   \left ( (p^{\mu} - p_0^{\mu})^2 \sigma (\xi)
   - 2 i {{\sqrt{|f|} \kappa} \over {\Lambda}} p_{\mu}
   (X^{\mu} - X_0^{\mu}) + i \tau p_{\mu} p^{\mu} \right ) \right ]$$

\be
   \times \, \mbox{\rm exp} \left [ {{i \tau} \over 2} \int 
  {\dd}^n \xi \, \sqrt{|f|} \Lambda \kappa \right ] . \hs{7.6cm}
\label{II4.A54}
\ee
We assume no summation over $\mu$ in the exponent of the above
expression and no integration (actually summation) over $\xi$,
because these operations are now already included in the product which
acts on the whole expression.
Because of the factor $
{ \left ( {{\Delta \xi \Lambda} \over {\sqrt{|f|} \kappa}} \right ) }^{1/2}$
occurring in the measure and the same factor in the exponent, the integration
over $p$ in eq.\,(\ref{II4.A54}) can be performed straightforwardly. The result is
   $$\psi[\tau , X] = 
   \left [ \lim_{\Delta \xi \to 0} {\prod}_{\xi,\mu}
   {\left ( {{\Delta \xi \, \sigma } \over {\pi}} \right )}^{1/4}
    {\left ( {1 \over {\sigma + i \tau }} \right ) }^{1/2} \right ] \hs{4cm}$$    
  $$\hs{1.6cm} \times \, \mbox{\rm exp} \left [ \int d^n \xi
   {{\Lambda} \over {\sqrt{|f|} \kappa}}
 \left (
 {{\left ( i {{\sqrt{|f|} \kappa} \over {\Lambda}} 
  (X^{\mu} - X_0^{\mu}) + p_0^{\mu} \sigma \right )^2} \over
  {2(\sigma + i \tau)}} - {{p_0^2 \sigma} \over 2} \right ) \right ] \times $$
\be  
    \times \, \mbox{\rm exp} \left[
   {{i \tau} \over 2} \int {\dd}^n \xi \, \sqrt{|f|} \Lambda \kappa \right ] .
  \hs{4cm}  
\label{II4.A50}
\ee
Eq.\ (\ref{II4.A50}) is a generalization of the familiar Gaussian wave packet.
\inxx{wave packet,Gaussian}
At $\tau = 0$ eq.\,(\ref{II4.A50}) becomes
$$
      \psi[0, X] =  \mbox{$ \left [ \lim_{\Delta \xi \to 0} {\prod}_{\xi,\mu}
   {\left ( {{\Delta \xi} \over {\pi \sigma}} \right )}^{1/4}
   \right ] $}
   \mbox{\rm exp} \left [ - \int \dd^n \xi {{\sqrt{|f|} \kappa} \over {\Lambda}}
   \, {{\left ( X^{\mu} (\xi) - X_0^{\mu} (\xi) \right )^2} \over 
   {2 \sigma (\xi)}} \right ]$$
\be
      \times \mbox{\rm exp} \left [i \int p_{0 \mu} (X^{\mu} - X_0^{\mu}) \,
      {\dd}^n \xi \right ] . \hs{3.2cm}
\label{II4.A50a}
\ee

The probability density is given by \inxx{probability density}
   $$|\psi [\tau , X ] |^2  =    
    \left [ \lim_{\Delta \xi \to 0} \prod_{\xi,\mu}
   {\left ( {{\Delta \xi \, \sigma } \over {\pi}} \right )}^{1/2}
    {\left ( {1 \over {{\sigma}^2 + {\tau }^2}}\right ) }^{1/2} \right ]
    \hs{3.2cm}$$
\be    
   \hs{2.4cm} \times \, \mbox{\rm exp} \left [ {- \int d^n \xi
     {{\sqrt{|f|} \kappa} \over {\Lambda}}
   {{( X^{\mu} - X_0^{\mu}  -  {{\Lambda} \over {\sqrt{|f|} \kappa}}
    p_0^{\mu} \tau)^2} \over {({\sigma}^2  + {\tau }^2)/ \sigma}}} \right ] ,
\label{II4.A51}
\ee
and the normalization constant, although containing the infinitesimal
$\Delta \xi$, gives precisely $\int |\psi|^2 {\cal D} X = 1$.

From (\ref{II4.A51})
we find that the motion of the centroid membrane of our particular wave
packet is determined by the equation
\be
   X_{\mbox{\rm c}}^{\mu} (\tau, \xi) = X_0^{\mu} (\xi) + 
   {{\Lambda} \over {\sqrt{|f|} \kappa}} \, p_0^{\mu} (\xi) \tau .
\label{II4.A52}
\ee
From the classical equation of motion \ci{43} derived from (\ref{II4.1}) 
(see also (\ref{II4.115}),
(\ref{II4.115a})) we indeed obtain a solution of the form (\ref{II4.A52}).
At this point it is interesting to observe that the classical null
strings considered, \inxx{null strings}
within different theoretical frameworks, by Schild \cite{51} and
Roshchupkin et al. \cite{67} also move according to 
equation (\ref{II4.A52}).

\paragraph{A special choice of wave packet}
The function $\sigma (\xi)$ in eqs.\,(\ref{II4.A49})--(\ref{II4.A51})
is arbitrary; the choice of $\sigma (\xi)$
determines how the wave packet is prepared. In particular, we may
consider {\it Case 1} of Sec.\ 4.3 and take
$\sigma (\xi)$ such that the wave packet of a $(p+1)$-dimensional
membrane ${\cal V}_{p+1}$ 
is peaked around a space-like $p$-dimensional membrane ${\cal V}_p$.
This means that the wave functional localizes ${\cal V}_{p+1}$ much more
sharply around ${\cal V}_p$ than in other regions of spacetime.
Effectively, such a wave packet describes the $\tau$-evolution of
${\cal V}_p$ (although formally it describes the $\tau$-evolution of
${\cal V}_{p+1}$). This can be clearly seen by taking the following
limiting form of the wave packet (\ref{II4.A50a}), such that

\vs{1mm}

\be
    {1 \over {\sigma (\xi)}} = {{\delta (\xi^0 - \xi_{\Sigma}^0)} \over
    {\sigma (\xi^i) (\p_0 X^{\mu} \p_0 X_{\mu})^{1/2} }} \; \; ,
    \quad \quad i = 1,2,...,p \; ,
\label{II4.117}
\ee

\vs{1mm}

\nnnn and choosing

\vs{1mm}
\be
     p_{0 \mu} (\xi^a) = {\bar p}_{0 \mu} (\xi^i) \delta (\xi^0 -
     \xi_{\Sigma}^0) .
\label{II4.118}
\ee

\vs{1mm}

\nnnn Then the integration over the $\delta$-function gives in the exponent
of eq.\,(\ref{II4.A50a}) the expression
\be
     \int \dd^p \xi {{\sqrt{|{\bar f}|} \kappa} \over {\Lambda}}
      \, (X^{\mu} (\xi^i) - X_0^{\mu} (\xi^i)
     )^2 {1 \over {2 \sigma (\xi^i)}} + i \int {\dd}^p \xi \, {\bar p}_{0 \mu}
     (X^{\mu} (\xi^i) (\xi^i) - X_0^{\mu} (\xi^i) ) ,
\label{II4.119}
\ee
so that eq.\,(\ref{II4.A50a}) becomes a wave functional 
of a $p$-dimensional membrane
$X^{\mu} (\xi^i)$. Here again ${\bar f}$ is the determinant of the induced
metric on $V_p$, while $f$ is the determinant of the induced metric on
$V_{p+1}$. One can verify that such a wave functional (\ref{II4.A50a}),
(\ref{II4.119}) satisfies the relation (\ref{II4.98}).

Considerations analogous to those described above hold for 
{\it Case 2} as well,
so that a wave functional of a $(p-1)$-brane can be considered as a
limiting case of a $p$-brane's wave functional.

\vs{2mm}

\subsection[\ \ The expectation values]
{The expectation values}\inxx{expectation value}

The expectation value of an operator ${\widehat A}$ is defined by
\be
   \langle {\widehat A} \rangle = \int {\psi}^* [\tau, X(\xi)] \, {\widehat A}\,
   \psi [\tau, X(\xi)] \, {\cal D} X \; ,
\label{II4.A53}
\ee
where the invariant measure (see (\ref{A15}) in the membrane space
${\cal M}$  with the metric (\ref{II4.5}) is
\be
   {\cal D} X = \prod_{\xi,\mu} 
   \mbox{$ {\left ({{\sqrt{|f(\xi)|} \, \kappa} \over 
   {\Lambda}} \right )}^{1/2} $} \, \dd X^{\mu} (\xi) .
\label{II4.A54a}
\ee
Using (\ref{II4.116a}) we can express eq.\,(\ref{II4.A53}) in the momentum space
\be
    \langle {\widehat A} \rangle = \int {\cal D} X \, {\cal D} p \,
    {\cal D} p' \, c^* [p'] c[p] {\psi}_{p'}^* {\widehat A} \, {\psi}_p \; ,
\label{II4.A55}
\ee
where ${\psi}_p$ is given in eq.\,(\ref{II4.112}). For the momentum
operator\footnote{
In order to distinguish operators from their eigenvalues we use here the
hatted notation.}
we have ${\hat p}_{\mu} \psi = p_{\mu} \, \psi$, and eq.\,(\ref{II4.A55}) becomes
\newpage
\be
   \langle {\hat p}_{\mu} (\xi) \rangle  
   = \hs{10cm} \label{II4.A56}
\ee  
    $$\lim_{\Delta \xi' \to 0} \prod_{\xi',\nu} \int
   \left ({{\Delta \xi'}
   \over {2 \pi}} \right ) {\left ( {{\sqrt{|f|} \kappa} \over {\Lambda}}
   \right ) }^{1/2}
    \, \dd X^{\nu} (\xi') \hs{4cm}$$
    $$ $$
    $$ \times \, {\rm exp} \biggl[ i \Delta \xi
    (p_{\nu} - p_{0 \nu})
   (X^{\nu} - X_0^{\nu}) \biggr] \hs{4.9cm} \,$$
   $$ $$   
   $$ \hspace{1cm} \times \, {\rm exp} \left [ -i \Delta \xi
   \left ( {{\sqrt{|f|} \kappa} \over {\Lambda}} \right )^{-1}  
   ({p'}_0^{\nu} {p'}_{0 \nu}- {p}_0^{\nu} {p}_{0 \nu} ) \tau \right ]
   \, p_{\mu} (\xi) c^* [p'] c[p] {\cal D} p \, {\cal D} p' . $$
The following relations are satisfied:
 $$\prod_{\xi,\mu} \int
   \mbox{$
   \left ( {{\Delta \xi}
   \over {2 \pi}} \right ) {\left ( {{\sqrt{|f|} \kappa} \over {\Lambda}}
   \right ) }^{1/2} $}
   \, \dd X^{\mu} (\xi) \, {\rm exp} \left [ i \Delta \xi
   ({p}_{\mu}^{'} - {p}_{0 \mu})
   (X^{\mu} - X_0^{\mu} ) \right ]$$
\be   
   \hs{5cm} = \prod_{\xi \mu}       
   {{\delta \left ({p '}_{\mu} (\xi) - {p}_{\mu} (\xi) \right )} \over
   {{\left ( {{\sqrt{|f|} \kappa} \over {\Lambda}} \right ) }^{-1/2}}} \; ,
\label{II4.A57a}
\ee
   $$\prod_{\xi,\mu} \int
   \mbox{$
   \left ( {{\Delta \xi}
   \over {2 \pi}} \right ) {\left ( {{\sqrt{|f|} \kappa} \over {\Lambda}}
   \right ) }^{-1/2} $}
   \, \dd {p}_{\mu} (\xi) \, {\rm exp} \left [ i \Delta \xi
   \, p_{\mu}
   ({X '}^{\mu} - X^{\mu} ) \right ]$$
\be   
   \hs{5cm} = \prod_{\xi, \mu}       
   {{\delta \left ({X '}^{\mu} (\xi) - X^{\mu} (\xi) \right ) } \over
   {{\left ( {{\sqrt{|f|} \kappa} \over {\Lambda}} \right ) }^{1/2}}} .
\label{II4.A57b}
\ee

Using (\ref{II4.A57b}) in eq.\,(\ref{II4.A56}) we have simply
\be
  \langle {\hat p}_{\mu} (\xi) \rangle = \int \, {\cal D} p \, p_{\mu} (\xi)
   c^* [p] c[p] .
\label{II4.A58}
\ee
Let us take for $c[p]$ the expression (\ref{II4.A49}). From (\ref{II4.A58}), using
the measure (\ref{II4.A48}), we obtain after the straightforward integration that
the expectation value of the momentum is equal to the centroid momentum
$p_{0 \mu}$ of the Gaussian wave packet (\ref{II4.A49}):
\inxx{wave packet,Gaussian}
\be
  \langle {\hat p}_{\mu} (\xi) \rangle = p_{\mu} (\xi) .
\label{II4.A60}
\ee
\par Another example is the expectation value of the membrane position
$X^{\mu} (\xi)$
with respect to the evolving wave packet (\ref{II4.A50}):
\be
    \langle {\widehat X}^{\mu} (\xi) \rangle = \int {\cal D} X \, {\psi}^*
    [\tau, X (\xi)] X^{\mu} (\xi) \psi [\tau, X (\xi)] .
\label{II4.A61}
\ee
Inserting the explicit expression (\ref{II4.A50}), into (\ref{II4.A61}) we find
\be
   \langle {\widehat X}^{\mu} \rangle = X_0^{\mu} (\xi) + 
   {{\Lambda} \over {\sqrt{|f|} \kappa}} \, p_0^{\mu} (\xi) \tau
   \equiv X_{\mbox{\rm c}}^{\mu} (\tau , \xi) ,
\label{II4.A62}
\ee
which is the correct expectation value, in agreement with that
obtained from the probability density (\ref{II4.A51}).

Although the normalization constants in eqs.\,(\ref{II4.A49})-(\ref{II4.A51}) 
contain the
infinitesimal $\Delta \xi$, it turns out that, when calculating the expectation
values, $\Delta \xi$ disappears from the expressions. Therefore our wave
functionals (\ref{II4.A49}) and (\ref{II4.A50}) with such normalization 
constants are well defined operationally.

\subsection[\ \ Conclusion]
{Conclusion}

We have started to elaborate a theory of relativistic $p$-branes which is more
general than the theory of conventional, constrained, $p$-branes. In
the proposed generalized theory, $p$-branes are unconstrained, but amongst
the solutions to the classical and quantum equations of motion there
are also the usual, constrained, $p$-branes. A strong motivation for such
a generalized approach is the elimination of the well known difficulties
due to the presence of constraints. Since the $p$-brane theories are
still at the stage of development and have not yet been fully confronted
with observations, it makes sense to consider an enlarged set of
classical and quantum $p$-brane states, such as, e.g., proposed in the
present and some previous works \cite{41}--\ci{43}.
What we gain is a theory without
constraints, still fully relativistic, which is straightforward both at
the classical and the quantum level, and is not in conflict with the
conventional $p$-brane theory.

Our approach might shed more light on some very interesting
developments concerning the duality \cite{68}, such as one of
strings and 5-branes,
and the interesting interlink between $p$-branes of various dimensions
$p$. In this chapter we have demonstrated how a higher-dimensional
$p$-brane equation naturally contains lower-dimensional $p$-branes as
solutions.

The highly non-trivial concept of unconstrained membranes enables us
\inxx{membranes,unconstrained}
to develop the elegant formulation of ``point particle" dynamics in the
infinite-dimensional space ${\cal M}$. It is fascinating that the action,
canonical and Hamilton formalism, and, after quantization, the
Schr\" odinger equation all look like nearly trivial extensions of the
correspondings objects in the elegant Fock--Stueckelberg--Schwinger--DeWitt
proper time formalism for a point particle in curved space.
Just this ``triviality", or better, simplicity, is
a distinguished feature of our approach and we have reasons to
expect that also the $p$-brane gauge field theory ---not yet a
completely solved problem--- can be straightforwardly formulated along
the lines indicated here.

\section{Clifford algebra and quantization}

\subsection[\ \ Phase space]
{Phase space}

Let us first consider the case of a 1-dimensional coordinate variable $q$
and its conjugate momentum $p$. The two quantities can be considered as 
coordinates of a point in the 2-dimensional {\it phase space}. Let $e_q$
and $e_p$ be the basis vectors satisfying the Clifford algebra relations
\inxx{basis vectors,in phase space}
\be         
         e_q \cdot e_p \equiv \mbox{$1\oo 2$} (e_q e_p + e_p e_q) = 0 ,
\lbl{II4clif.81a}
\ee         
\be
      e_q^2 = 1 \; , \qquad e_p^2 = 1 .
\lbl{II4clif.81b}
\ee
An arbitrary vector in phase space is then
\be
     Q = q e_q + p e_p .
\lbl{II4clif.81}
\ee

The product of two vectors $e_p$ and $e_q$ is the unit bivector in phase
space and it behaves as the imaginary unit \inxx{bivector,in phase space}
\inxx{imaginary unit,as a Clifford number}
\be
         i = e_p e_q \; , \qquad i^2 = - 1 .
\lbl{II4clif.82}
\ee
The last relation immediately follows from 
(\ref{II4clif.81a}),\,(\ref{II4clif.81b}):
$i^2 = e_p e_q e_p e_q = - e_p^2 e_q^2 = - 1$.

Multiplying (\ref{II4clif.81}) respectively from the right and from the left
by $e_q$ we thus introduce the quantities $Z$ and $Z^*$:
\be
     Q e_q = q + p e_p e_q = q + p i = Z ,
\lbl{II4clif.83}
\ee
\be
e_q Q = q + p e_q e_p = q - p i = Z^* .
\lbl{II4clif.84}
\ee
For the square we have
\be
     Q e_q e_q Q = Z Z^* = q^2 + p^2 + i (p q - q p) ,
\lbl{II4clif.85}
\ee
\be
    e_q Q Q e_q = Z^* Z =  q^2 + p^2 - i (p q - q p) .
\lbl{II4clif.86}
\ee
Upon quantization $q$, $p$ do not commute, but satisfy
\be
        [q,p] = i ,
\lbl{II4clif.87}
\ee
therefore (\ref{II4clif.85}),\,(\ref{II4clif.86}) become
\be
       Z Z^* = q^2 + p^2 + 1 , 
\lbl{II4clif.88}
\ee
\be
       Z^* Z = q^2 + p^2 - 1 ,
\lbl{II4clif.89}
\ee
\be
         [Z, Z^*] = 2 .
\lbl{II4clif.90}
\ee

Even before quantization the natural variables for describing physics
are {\it the complex quantity} $Z$ and its conjugate $Z^*$. {\it The
imaginary unit is the bivector of the phase space, which is 2-dimensional}.

Writing $q = \rho \, {\rm cos} \, \phi$ and $p = \rho \, {\rm sin} \, \phi$
we find
\be
    Z = \rho ({\rm cos} \, \phi + i \, {\rm sin} \, \phi) = \rho \, {\rm e}^{
    i \phi} \; ,
\lbl{II4clif.91}
\ee
\be
  Z^* = \rho ({\rm cos} \, \phi - i \, {\rm sin} \, \phi) = \rho \, {\rm e}^{
    - i \phi} \; ,
\lbl{II4clif.92}
\ee
where $\rho$ and $\phi$ are real numbers. Hence taking into account that
physics takes place in the phase space and that the latter can be described by
complex numbers, we automatically introduce complex numbers into both
the classical and quantum physics. And what is nice here is that {\it
the complex numbers are nothing but the Clifford numbers of the 2-dimensional
phase space}.

What if the configuration space has more than one dimension, say $n$? Then
with each spatial coordinate is associated a 2-dimensional phase space.
The dimension of the total phase space is then $2 n$. A phase space vector
then reads
\be
    Q = q^{\mu} e_{q \mu} + p^{\mu} e_{p \mu} .
\lbl{II4clif.93}
\ee
The basis vectors have now two indices $q$, $p$ (denoting the direction in the
2-dimensional phase space) and $\mu = 1,2,...,n$ (denoting the direction in
the $n$-dimensional configuration space). 

The basis vectors can be written as the product of the configuration space
basis vectors $e_{\mu}$ and the 2-dimensional phase space basis vectors
$e_q$, $e_p$: \inxx{basis vectors,in phase space}
\be
    e_{q \mu} = e_q e_{\mu} \; , \qquad e_{p \mu} = e_p e_{\mu} \; .
\lbl{II4clif.94}
\ee
A vector $Q$ is then 
\be
      Q = (q^{\mu} e_q + p^{\mu} e_p) e_{\mu} = Q^{\mu} e_{\mu} \; ,
\lbl{II4clif.95}
\ee
where
\be
       Q^{\mu} = q^{\mu} e_q + p^{\mu} e_p .
\lbl{II4clif.96}
\ee
Eqs.\ (\ref{II4clif.83}),\,(\ref{II4clif.84}) generalize to
\be
     Q^{\mu} e_q = q^{\mu} + p^{\mu} e_p e_q = q^{\mu} + p^{\mu} i = Z^{\mu} \; ,
\lbl{II4clif.97}
\ee
\be
    e_q Q^{\mu} = q^{\mu} + p^{\mu} e_q e_p =  q^{\mu} - p^{\mu} i = 
    Z^{* \mu}  .                   
\lbl{II4clif.98}
\ee

Hence, even if configuration space has many dimensions, the imaginary unit
$i$ in the variables $X^{\mu}$ comes from the bivector $e_q e_p$ of the
2-dimensional phase space which is associated with every direction $\mu$
of the configuration space.

When passing to quantum mechanics it is then natural that in general the
wave function is complex-valued. {\it The imaginary unit is related
to the phase space which is the direct product of the configuration space
and the 2-dimensional phase space.}

At this point let us mention that Hestenes was one of the first to point
out clearly that imaginary and complex numbers need not be postulated 
separately, but they are automatically contained in the geometric calculus
based on Clifford algebra. When discussing quantum mechanics Hestenes
ascribes the occurrence of the imaginary unit $i$ in the Schr\" odinger
and especially in the Dirac equation to a chosen configuration space 
Clifford number which happens to have the square $-1$ and which commutes with
all other Clifford numbers within the algebra. This brings an ambiguity as to
which of several candidates should serve as the imaginary unit $i$. In this
respect Hestenes had changed his point of view, since initially he proposed that
one must have a 5-dimensional space time whose pseudoscalar unit
$I = \gamma_0 \gamma_1 \gamma_2 \gamma_3 \gamma_4$ commutes with all the 
Clifford numbers of ${\cal C}_5$ and its square is $I^2 = - 1$.
Later he switched to 4-dimensional space time and chose the bivector
$\gamma_1 \gamma_2$ to serve the role of $i$. I regard this as unsatisfactory,
since $\gamma_2 \gamma_3$ or $\gamma_1 \gamma_3$ could be given such a role
as well.
In my opinion it is more natural to ascribe the role of $i$ to the bivector of
the 2-dimensional phase space sitting at every coordinate of the configuration
space. A more detailed discussion about the relation between the
geometric calculus in a generic 2-dimensional space 
(not necessarily interpreted as phase 
space)  and complex number is to be found in Hestenes' books \cite{13}.

\subsection[\ \ Wave function as a polyvector]
{Wave function as a polyvector}

We have already seen in Sec. 2.5 that a wave function can in general
be considered as a polyvector, i.e., as a Clifford number or Clifford
aggregate generated by a countable set of basis vectors $e_{\mu}$. Such a
wave function contains spinors, vectors, tensors, etc., all at once.
In particular, it may contain only spinors, or only vectors, etc.\ .

Let us now further generalize this important procedure.
In Sec. 6.1 we have discussed vectors in an infinite-dimensional
space $V_{\infty}$ from the point of view of geometric calculus based
on the Clifford algebra generated by the uncountable set of basis vectors
\inxx{basis vectors,uncountable set of} \inxx{wave function,complexe-valued}
$h(x)$ of $V_{\infty}$. We now apply that procedure to the case of the
wave function which, in general, is complex-valued.

For an arbitrary complex function we have
\be
      f(x) = {1\oo \sqrt{2}} (f_1 (x) + i f_2 (x)) \;  , \qquad
      f^*(x) = {1\oo \sqrt{2}} (f_1 (x) - i f_2 (x)) \; ,
\lbl{II4clif.99}
\ee
where $f_1(x)$, $f_2 (x)$ are real functions. From (\ref{II4clif.99}) we find
\be
    f_1 (x) = {1\oo \sqrt{2}} (f(x) + f^* (x)) \; , \qquad
    f_2 (x) = {1\oo {i \sqrt{2}}} (f(x) - f^* (x)) .
\lbl{II4clif.100}
\ee
Hence, instead of a complex function we can consider a set of two
independent real functions $f_1 (x)$ and $f_2 (x)$.

Introducing the basis vectors $h_1 (x)$ and $h_2 (x)$ satisfying
the Clifford algebra relations
\be
     h_i (x) \cdot h_j (x') \equiv \mbox{$1\oo 2$} (h_i (x) h_j (x') + h_j (x')
     h_i (x)) = \delta_{ij} \delta (x - x') \;  , \quad i, j = 1, 2,
\lbl{II4clif.101}
\ee
we can expand an arbitrary vector $F$ according to
\be
    F = \int \dd x (f_1 (x) h_1 (x) + f_2 (x) h_2 (x)) = f^{i (x)} h_{i (x)}\; ,
\lbl{II4clif.102}
\ee
where $h_{i (x)} \equiv h_i (x)$, $f^{i (x)} \equiv f_i (x)$.
Then
\be
      F \cdot h_1 (x) = f_1 (x) \; , \qquad F \cdot h_2 (x) = f_2 (x)
\lbl{II4clif.103}
\ee
are components of $F$.

Introducing the imaginary unit $i$ which commutes\footnote
{Now, the easiest way to proceed is in forgetting how we have obtained
the imaginary unit, namely as a bivector in 2-dimensional phase space, and
define all the quantities $i$, $h_1 (x)$, $h_2 (x)$, etc., in such a way
that $i$ commutes with everything. If we nevertheless persisted in
maintaining the geometric approach to $i$, we should then take
$h_1 (x) = e(x)$, $\; h_2 (x) = e(x) e_p e_q$, satisfying
\bear
h_1 (x) \cdot h_1 (x') &=& e (x) \cdot e(x') = \delta (x - x'), \nonumber \\
h_2 (x) \cdot h_2 (x) &=& - \delta (x - x'), \nonumber \\ 
h_1 (x) \cdot h_2 (x') &=& \delta (x - x') 1 \cdot (e_p e_q) = 0, \nonumber
\ear
where according to Hestenes
the inner product of a scalar with a multivector is zero. Introducing
$h = (h_1 + h_2)/\sqrt{2}$  and $ h^* = (h_1 - h_2)/\sqrt{2}$ one
finds $h(x) \cdot h^* (x') = \delta (x - x')$, $h(x) \cdot h(x') = 0$,
$h^* (x) \cdot h^* (x') = 0$.}
with $h_i (x)$ we can form a new set of basis vectors
\be
   h(x) = {{h_1 (x) + i h_2 (x)}\oo \sqrt{2}} \; , \qquad
   h^*(x) = {{h_1 (x) - i h_2 (x)}\oo \sqrt{2}} \; ,
\lbl{II4clif.104}
\ee
the inverse relations being
\be
   h_1 (x) = {{h(x) + h^* (x)}\oo \sqrt{2}} \;  , \qquad
   h_2 (x) = {{h(x) - h^* (x)}\oo {i \sqrt{2}}} .
\lbl{II4clif.105}
\ee
Using (\ref{II4clif.100}),\,(\ref{II4clif.105}) and (\ref{II4clif.102})
we can re-express $F$ as
\be
    F = \int \dd x (f(x) h^* (x) + f^* (x) h(x)) = f^{(x)} h_{(x)} +
    f^{*(x)} h_{(x)}^* \; ,
\lbl{II4clif.106}
\ee
where
\be
    f^{(x)} \equiv f^* (x) \; , \; \quad f^{*(x)} \equiv f(x) \;  , \qquad
    h_{(x)} \equiv h(x) \; , \quad h_{(x)}^* \equiv h^* (x) .
\lbl{II4clif.106a}
\ee

From (\ref{II4clif.101}) and (\ref{II4clif.105}) we have
\be
    h (x) \cdot h^* (x) \equiv \mbox{$1\oo 2$} (h(x) h^* (x') + h^* (x') h(x)) =
    \delta (x - x') \; ,
\lbl{II4clif.107}
\ee
\be
     h (x) \cdot h(x') = 0 \;  , \qquad h^* (x) \cdot h^* (x') = 0 \; ,
\lbl{II4clif.108}
\ee
which are {\it the anticommutation relations for a fermionic field}.
\inxx{anticommutation relations,for a fermionic field}

A vector $F$ can be straightforwardly generalized to a {\it polyvector}:
\bear
   F &=& f^{i (x)} h_{i (x)} + f^{i (x) j (x')} h_{i (x)} h_{j (x')} 
   + f^{i (x) j (x') k (x'')} h_{i (x)} h_{j (x')} h_{k (x'')} + . . .
   \nonumber \\
   \nonumber \\
   &=& f^{(x)} h_{(x)} + f^{(x)(x')} h_{(x)} h_{(x')} + f^{(x)(x')(x'')}
   h_{(x)} h_{(x')} h_{(x'')} + ... \nonumber \\
   \nonumber \\
    & & + \, f^{* (x)} h_{(x)}^* + f^{*(x)(x')} h_{(x)}^* h_{(x')}^* 
     +  f^{* (x)(x')(x'')} h_{(x)}^* h_{(x')}^* h_{(x'')}^* + ... \nonumber \\
\lbl{II4clif.109}
\ear
where $f^{(x)(x')(x'')...}$ are scalar coefficients, {\it antisymmetric}
in $(x)(x')(x'')... $

We have exactly the same expression (\ref{II4clif.109}) in the usual quantum 
field theory (QFT), where $f^{(x)}$, $f^{(x)(x')}$,..., are 1-particle, 
2-particle,..., wave functions (wave packet profiles). Therefore a natural
interpretation of the polyvector $F$ is that it represents a superposition
of multi-particle states.\inxx{wave packet profile}
\inxx{multi-particle states,superposition of}

{\it In the usual formulation} of QFT one
introduces a vacuum state $|0 \rangle$, and interprets $h (x)$, $h^* (x)$ as
the operators which create or annihilate  a particle or an antiparticle at $x$,
so that (roughly speaking) e.g. $h^* (x) |0 \rangle$ is a state with
a particle at position $x$. \inxx{creation operators,for fermions}

{\it In the geometric calculus formulation} (based on the Clifford algebra
of an infinite-dimensional space) the Clifford numbers $h^* (x)$, $h (x)$ 
already represent vectors. At the same time $h^* (x)$, $h (x)$ also behave
as operators, satisfying (\ref{II4clif.107}),\,(\ref{II4clif.108}). When we say that
a state vector is expanded in terms of $h^* (x)$, $h (x)$ we mean that
it is a superposition of states in which a particle has a definite position $x$.
The latter states are just $h^* (x)$, $h (x)$. Hence the Clifford numbers
(operators) $h^* (x)$, $h (x)$ need not act on a vacuum state in order 
to give the one-particle states. They are already the one-particle states.
Similarly the products $h(x) h(x')$, $h (x) h(x') h(x'')$, $h^* (x) h^* (x')$,
etc., already represent the multi-particle states.

When performing quantization of a classical system we arrived at {\it the
wave function}. The latter can be considered as an uncountable (infinite)
set of scalar components of a vector in an infinite-dimensional space,
spanned by the basis vectors $h_1 (x)$, $h_2 (x)$. Once we have basis
vectors we automatically have not only arbitrary vectors, but also
arbitrary polyvectors which are {\it Clifford numbers} generated by
$h_1 (x)$, $h_2 (x)$ (or equivalently by $h (x)$, $h^* (x)$. Hence
{\it the procedure in which we replace infinite-dimensional vectors
with polyvectors is equivalent to the second quantization}.

If one wants to consider {\it bosons} instead of {\it fermions} one needs
to introduce a new type of fields $\xi_1 (x)$, $\xi_2 (x)$, satisfying
{\it the commutation relations}
\be
    \mbox{$1\oo 2$} [\xi_i (x), \xi_j (x')] \equiv 
    \mbox{$1\oo 2$} [\xi_i (x) \xi_j (x')
    - \xi_j (x') \xi_i (x) ] = \epsilon_{ij} \Delta (x - x') \mbox{$1\oo 2$} ,
\lbl{II4clif.110}
\ee
with $\epsilon_{ij} = - \epsilon _{ji}$, $\Delta (x - x') = - \Delta (x' - x)$,
which stay instead of the anticommutation relations (\ref{II4clif.101}). Hence the
numbers $\xi (x)$ are not Clifford numbers. By (\ref{II4clif.110})
the $\xi_i (x)$
generate a new type of algebra, which could be called an {\it anti-Clifford
algebra}. 

Instead of $\xi_i (x)$ we can introduce the basis vectors
\be
    \xi (x) = {{\xi_1 (x) + i \xi_2}\oo \sqrt{2}} \; , \; \qquad 
    \xi^* (x) = {{\xi_1 (x) - i \xi_2}\oo \sqrt{2}}
\lbl{II4clif.111}
\ee
which satisfy the commutation relations
\bear
         &[\xi (x) , \xi^* (x')] = - i \Delta (x - x') , & \\
\lbl{II4clif.112}
\nonumber\\
    &[\xi (x), \xi (x')] = 0 \; , \qquad  \; [\xi^* (x), \xi^* (x')] = 0 .&
\lbl{II4clif.113}
\ear

A polyvector representing a superposition of bosonic multi-particle states 
is then expanded as follows:
\bear
    B &=& \phi^{i (x)} \xi_{i (x)} + \phi^{i (x) j (x')} \xi_{i (x)} \xi_{j (x')}
     + ... 
    \lbl{II4clif.114} \\
    \nonumber \\
      &=&\phi^{(x)} \xi_{(x)} + \phi^{(x)(x')} \xi_{(x)} \xi_{(x')} +
      \phi^{(x)(x')(x'')} \xi_{(x)}\xi_{(x')} \xi_{(x'')} + ... \nonumber\\
\nonumber \\      
      & &+ \, \phi^{*(x)} \xi_{(x)}^* + \phi^{*(x)(x')} \xi_{(x)}^* 
      \xi_{(x')}^* +\phi^{*(x)(x')(x'')} \xi_{(x)}^* \xi_{(x')}^* \xi_{(x'')}^*
       + ... \, , \nonumber
\ear
where $\phi^{i (x) j (x')...}$ and $\phi^{(x)(x')...}$, $\phi^{*(x)(x')(x'')...}$
are scalar coefficients, {\it symmetric} in $i (x) j (x')...$ and $(x)(x')...$,
respectively. They can be interpreted as representing 1-particle, 2-particle,...,
wave packet profiles. Because of (\ref{II4clif.112}) 
$\, \xi (x)$ and $\xi^* (x)$ can
\inxx{wave packet profile} \inxx{creation operators,for bosons}
be interpreted as creation operators for {\it bosons}. Again, 
{\it a priori} we do not
need to introduce a vacuum state. However, whenever convenient we may, 
of course, define a vacuum state and act on it by the operators
$\xi (x)$, $\xi^* (x)$.

\subsection[\ \ Equations of motion for basis vectors]
{Equations of motion for basis vectors}

In the previous subsection we have seen how the geometric calculus na\-turally
leads to the second quantization which incorporates superpositions of
multi-particle states. We shall now investigate what are the equations
of motion that the basis vectors satisfy.
\inxx{equations of motion,for basis vectors}

For illustration let us consider the action for a {\it real scalar
field} $\phi (x)$: \inxx{scalar field,real}
\be
     I[\phi] = \mbox{$1\oo 2$} \int {\dd}^4 x \, 
     (\p_{\mu} \phi \p^{\mu} \phi - m^2) .
\lbl{II4clif.115}
\ee
Introducing the metric
\be
    \rho(x,x') = h (x) \cdot h (x') \equiv 
    \mbox{$1\oo 2$} (h(x) h(x') + h(x') h(x))
\lbl{II4clif.116}
\ee
we have
\be
    I[\phi] = \mbox{$1\oo 2$} \int \dd x \, \dd x' \, \left ( \p_{\mu} \phi(x) 
    \p'^{\mu} \phi (x') - m^2 \phi (x) \phi (x') \right ) h (x) h(x') .
\lbl{II4clif.116a}
\ee
If, in particular,
\be
     \rho(x,x') = h(x) \cdot h(x') = \delta (x - x')
\lbl{II4clif.117}
\ee
then the action (\ref{II4clif.116a}) is equivalent to (\ref{II4clif.115}).

In general, $\rho (x,x')$ need not be equal to $\delta (x - x')$, and 
(\ref{II4clif.116a}) is then a generalization of the usual action 
(\ref{II4clif.115})
for the scalar field. An action which is invariant under field
redefinitions (`coordinate' transformations in the space of fields) has
been considered by Vilkovisky \cite{69}.
Integrating (\ref{II4clif.116a}) {\it per partes} over $x$ and $x'$ and omitting the
surface terms we obtain
\be
    I[\phi] = \mbox{$1\oo 2$} \int \dd x \, \dd x' \, \phi (x) \phi (x')
    \left ( \p_{\mu} h(x) \p'^{\mu} h(x') - m^2 h(x) h(x') \right ) .
\lbl{II4clif.118}
\ee
Derivatives no longer act on $\phi (x)$, but on $h(x)$. If we fix $\phi(x)$
then instead of an action for $\phi(x)$ we obtain an action for $h(x)$.

For instance, if we take
\be
    \phi (x) = \delta (x - y)
\lbl{II4clif.119}
\ee
and integrate over $y$ we obtain
\be
    I[h] = {1\oo 2} \int \dd y \, \left ({{\p h(y)}\oo {\p y^{\mu}}} 
    {{\p h(y)}\oo {\p y_{\mu}}} - m^2 h^2 (y) \right ) .
\lbl{II4clif.120}
\ee

The same equation (\ref{II4clif.120}), of course, follows directly from 
(\ref{II4clif.116a}) in which we fix $\phi(x)$ according to (\ref{II4clif.119}).

On the other hand, if instead of $\phi(x)$ we fix $h(x)$ according to
(\ref{II4clif.117}), then we obtain the action (\ref{II4clif.115}) which governs
the motion of $\phi (x)$.

Hence the same basic expression (\ref{II4clif.116a}) can be considered either as
an action for $\phi(x)$ or an action for $h(x)$, depending on which
field we consider as fixed and which one as a variable. If we consider
the basis vector field $h(x)$ as a variable and $\phi (x)$ as fixed according
to (\ref{II4clif.119}), then we obtain the action (\ref{II4clif.120}) for $h(x)$.
The latter field is actually an operator. The procedure from now on coincides
with the one of quantum field theory.

Renaming $y^{\mu}$ as $x^{\mu}$ (\ref{II4clif.120}) becomes an action for
a bosonic field: \inxx{bosonic field}
\be
     I[h] = \mbox{$1\oo 2$} \int \dd x \, (\p_{\mu} h \p^{\mu} h - m^2 h^2) .
\lbl{II4clif.121} 
\ee
The canonically conjugate variables are
 $$h(t,{\bf x}) \qquad \mbox{\rm and} \qquad \pi (t, {\bf x}) =
\p {\cal L}/\p {\dot h} = {\dot h} (t,{\bf x}).$$
They satisfy the
{\it commutation relations}
\be
    [h(t,{\bf x}), \pi (t, {\bf x'}) = i \delta^3 ({\bf x} - {\bf x'})
    \; , \; \qquad [h(t,{\bf x}), h(t,{\bf x'})] = 0 .
\lbl{II4clif.122}
\ee

At different times $t' \neq t$ we have
\be
    [h(x), h(x')] = i \Delta (x - x') \; ,
\lbl{II4clif.122a} 
\ee    
where $\Delta (x - x')$ is the well known covariant function, antisymmetric
under the exchange of $x$ and $x'$.

The geometric product of two vectors can be decomposed as
\be
      h(x) h(x') = \mbox{$1\oo 2$} \left ( h(x) h(x') + h(x') h(x) \right ) +
       \mbox{$1\oo 2$} \left ( h(x) h(x') - h(x') h(x) \right )  .
\lbl{II4clif.123}
\ee
In view of (\ref{II4clif.122a}) we have that {\it the role of the inner product
is now given to the antisymmetric part}, whilst {\it the role of the outer
product is given to the symmetric part}. This is characteristic for
{\it bosonic vectors}; they generate what we shall call the 
{\it anti-Clifford algebra}. In other words, when the 
basis vector field $h(x)$ happens to satisfy the {\it commutation relation}
\be
     [h(x), h(x')] = f(x,x') ,
\lbl{II4clif.123a}
\ee
where $f(x,x')$ is a scalar two point function (such as $i \Delta (x-x')$), it
behaves as a {\it bosonic field}. On the contrary, when $h(x)$ happens to
satisfy the {\it anticommutation relation}
\be
       \lbrace h(x), h(x') \rbrace = g(x,x') ,
\lbl{II4clif.124}
\ee
where $g(x,x')$ is also a scalar two point function, then it behaves as
a {\it fermionic field}\footnote{
  In the previous section the bosonic basis vectors were given a separate name
  $\xi(x)$. Here we retain the same name $h(x)$ both for bosonic and
  fermionic basis vectors.}

The latter case occurs when instead of (\ref{II4clif.115}) we take the action for
the {\it Dirac field}: \inxx{Dirac field}
\be
   I[\psi, {\bar \psi}] = \int {\dd}^4 x \, {\bar \psi} (x) 
   (i \gamma^{\mu} \p_{\mu} - m) \psi(x) .
\lbl{II4clif.125}
\ee
Here we are using the usual spinor representation in which the spinor
field $\psi(x) \equiv \psi_{\alpha} (x)$ bears the spinor index $\alpha$.
A generic vector is then
\bear
   \Psi &=& \int \dd x \left ( {\bar \psi}_{\alpha} (x) h_{\alpha} (x) +
   \psi_{\alpha} (x) {\bar h}_{\alpha} (x) \right ) \nonumber \\
   &\equiv&\int \dd x \left ( {\bar \psi} (x) h (x) +
   \psi (x) {\bar h} (x) \right )  .
\lbl{II4clif.125a}
\ear
Eq.\ (\ref{II4clif.125}) is then equal to the scalar part of the action
\bear
     I[\psi, {\bar \psi}] &=& \int \dd x \, \dd x' \, {\bar \psi} (x')
     {\bar h} (x) h (x') (i \gamma^{\mu} \p_{\mu} - m) \psi (x) \nonumber \\
     &=& \int \dd x \, \dd x' \, {\bar \psi} (x') \left [ {\bar h} (x)
     (i \gamma^{\mu} - m) h(x') \right ] \psi (x) ,
\lbl{II4clif.126}
\ear
where $h (x)$, ${\bar h} (x)$ are assumed to satisfy
\be
    {\bar h} (x) \cdot h (x') \equiv 
    \mbox{$1\oo 2$} \left ( {\bar h} (x) h (x')
    +h (x') {\bar h} (x) \right ) = \delta (x - x')
\lbl{II4clif.126a}
\ee
The latter relation follows from the Clifford algebra relations amongst the
basis fields $h_i (x)$, $i = 1,2$,
\be
    h_i (x) \cdot h_j (x') = \delta_{ij} (x - x')
\lbl{II4clif.126b}
\ee
related to ${\bar h} (x)$, $h (x)$ accroding to
\be
      h_1 (x) = {{h (x) + {\bar h} (x)}\oo \sqrt{2}} \;  , \qquad \;
      h_2 (x) = {{h (x) - {\bar h} (x)}\oo {i\sqrt{2}}} .
\lbl{II4clif.127}
\ee

Now we relax the condition (\ref{II4clif.126a}) and ({\ref{II4clif.126}) becomes a
generalization of the action (\ref{II4clif.125}).

Moreover, if in (\ref{II4clif.126}) we fix the field $\psi$ according to
\be 
     \psi (x) = \delta (x - y) ,
\lbl{II4clif.128}
\ee
integrate over $y$, and rename $y$ back into $x$, we find
\be
     I[h,{\bar h}] = \int \dd x \, {\bar h} (x) (i \gamma^{\mu} \p_{\mu} - m) 
     h (x) .
\lbl{II4clif.129}
\ee
This is an action for the basis vector field $h (x)$, ${\bar h} (x)$,
which are {\it operators}.

The canonically conjugate variables are now
 $$h(t,{\bf x})   \qquad {\rm and} \qquad \pi(t,{\bf x})
= \p {\cal L}/\p {\dot h} = i {\bar h} \gamma^0 = i h^{\dagger}.$$ 
They satisfy the anticommutation relations
\bear
     &\lbrace h(t, {\bf x}), h^{\dagger} (t,{\bf x'} \rbrace = \delta^3
     ({\bf x} - {\bf x'}) ,&
\lbl{II4clif.130}\\
\nonumber \\
    &\lbrace h(t,{\bf x}),h(t,{\bf x'}) \rbrace = \lbrace  
    h^{\dagger} (t,{\bf x}),h^{\dagger} (t,{\bf x'}) \rbrace  = 0 .&
\lbl{II4clif.130a}
\ear

At different times $t' \neq t$ we have
\bear
    &\lbrace h(x), {\bar h} (x') \rbrace = (i \gamma^{\mu} + m) i \Delta
    (x - x') ,&
\lbl{II4clif.131}\\
\nonumber \\
       &\lbrace h (x), h(x') \rbrace = \
       \lbrace {\bar h} (x), {\bar h} (x') \rbrace = 0 .&
\ear
The basis vector fields $h_i (x)$, $i = 1,2$, defined in (\ref{II4clif.127})
then satisfy
\be
        \lbrace h_i (x), h_j (x') \rbrace = \delta_{ij}   
         (i \gamma^{\mu} + m) i \Delta (x - x') ,
\lbl{II4clif.132}
\ee
which can be written as {\it the inner product}
\be
    h_i (x) \cdot h_j (x') = \mbox{$1\oo 2$} \delta_{ij}   
         (i \gamma^{\mu} + m) i \Delta (x - x') = \rho (x,x')
\lbl{II4clif.133}
\ee
with the metric $\rho (x,x')$. We see that our procedure leads us to a
metric which is different from the metric assumed in (\ref{II4clif.126b}).

Once we have basis vectors we can form an arbitrary {\it vector} according to
\be
    \Psi = \int \dd x \, \left ( \psi (x) {\bar h} (x) + 
    {\bar \psi} (x) h (x)
    \right ) = \psi^{(x)} h_{(x)} + {\bar \psi}^{(x)} {\bar h}_{(x)} .
\lbl{II4clif.134}
\ee
Since the $h (x)$ generates a Clifford algebra we can form not only a vector
but also an arbitrary multivector and a superposition of multivectors,
i.e., a {\it polyvector} (or {\it Clifford aggregate}):
\bear
     \Psi &=&\int \dd x \, \left ( \psi (x) {\bar h} (x) + 
    {\bar \psi} (x) h (x) \right ) \nonumber \\
    & &+ \int \dd x\, \dd x' \, \left ( \psi(x,x') {\bar h} (x) {\bar h} (x')
    + {\bar \psi} (x,x') h (x) h (x') \right ) + ...  \nonumber
\ear    
    $$= \psi^{(x)} h_{(x)} + \psi^{(x)(x')} h_{(x)} h_{(x')} + ... \hs{2cm}$$
\be   
    + \, {\bar \psi}^{(x)} {\bar h}_{(x)} + {\bar \psi}^{(x)(x')} 
   {\bar h}_{(x)} {\bar h}_{(x')} + ... \, , \hs{9mm}
\lbl{II4clif.135}
\ee
where $\psi (x,x',...) \equiv {\bar \psi}^{(x)(x')...}$, 
${\bar \psi} (x,x',...) \equiv \psi^{(x)(x')...}$ are {\it antisymmetric}
functions, interpreted as wave packet profiles for a system of free
{\it fermions}.

Similarly we can form an arbitrary polyvector
\bear     
     \Phi &=& \int \dd x \, \phi (x) h (x) + \int \dd x \, \dd x' \, \phi(x,x')
     h (x) h(x') + ... \nonumber \\
      &\equiv& \phi^{(x)} h_{(x)} + \phi^{(x)(x')} h_{(x)} h_{(x')} + ...
\lbl{II4clif.136}
\ear    
generated by {\it 
the basis vectors} which happen to satisfy the {\it commutation relations}
(\ref{II4clif.122}). In such a case the uncountable set of basis vectors
behaves as a {\it bosonic field}. The corresponding multi-particle wave packet
profiles $\phi(x,x',...)$ are {\it symmetric} functions of $x$, $x'$,... .
If one considers a complex field, then the equations 
(\ref{II4clif.121})--(\ref{II4clif.122}) and (\ref{II4clif.136}) 
are generalized in an obvious way.

As already mentioned, within the conceptual scheme of Clifford algebra and
hence also of anti-Clifford algebra we do not need, if we wish so, to introduce
a vacuum state\footnote{
Later, when discussing the states of the quantized $p$-brane, we nevertheless
introduce a vacuum state and the set of orthonormal basis states spanning
the Fock space.}, since the operators $h (x)$, ${\bar h} (x)$ already
represent states. From the actions (\ref{II4clif.121}),\,(\ref{II4clif.125}) 
we can
derive the corresponding Hamiltonian, and other relevant operators (e.g.,
the generators of spacetime translations, Lorentz transformations, etc.).
In order to calculate their expectation values in a chosen multi-particle
state one may simply sandwich those operators between the state and
its Hermitian conjugate (or Dirac conjugate) and take the scalar part
of the expression. For example, the expectation value of the
\inxx{Hamiltonian,expectation value of} 
Hamiltonian $H$ in a bosonic 2-particle state is
\be
   \langle H \rangle = \langle \int \dd x \, \dd x' \, \phi (x,x')
   h(x) h(x') H \int \dd x'' \, \dd x''' \, \phi (x'',x''') h(x'') h(x''')
   \rangle_0 ,
\lbl{II4clif.137}
\ee
where, in the case of the real scalar field,
\be
H = \mbox{$1\oo 2$} \int {\dd}^3 {\bf x} ({\dot h}^2 (x) - \p^i h \p_i h +
m^2 h^2) .
\lbl{II4clif.137aa}
\ee

Instead of performing the operation $\langle \; \; \rangle_0$ (which means
taking the scalar part), in the conventional approach to quantum field theory one
performs the operation $\langle 0|....|0 \rangle$ (i.e., taking the vacuum
expectation value). However, instead of writing, for instance,
\be
 \langle 0|a({\bf k}) a^* ({\bf k'})|0 \rangle = 
 \langle 0|[a({\bf k}), a^* ({\bf k'})]|0 \rangle = 
\delta^3 ({\bf k} - {\bf k'}) ,
\lbl{II4clif.137a}
\ee
 we can write
\bear
   \langle a({\bf k}) a^* ({\bf k'}) \rangle_0  & = & \mbox{$1\oo 2$} \langle 
   a({\bf k}) a^* ({\bf k'}) + a^* ({\bf k'}) a({\bf k}) \rangle_0 \nonumber \\
   \nonumber \\
    & & + \, \mbox{$1\oo 2$} \langle 
   a({\bf k}) a^* ({\bf k'}) - a^* ({\bf k'}) a({\bf k}) \rangle_0 \nonumber \\
   \nonumber \\  
   & = & \mbox{$1\oo 2$} \delta^3 ({\bf k} - {\bf k'}) ,
\lbl{II4clif.137b}
\ear
where we have taken into account that for a bosonic operator the symmetric
part is not a scalar. Both expressions (\ref{II4clif.137a}) and 
(\ref{II4clif.137b})
give the same result, up to the factor $1\oo 2$ which can be absorbed into the
normalization of the states.

We leave to the interested reader to explore in full detail (either as an
exercise or as a research project), for various operators and kinds of field, 
how much the results of the above procedure (\ref{II4clif.137}) deviate, 
if at all, 
from those of the conventional approach. Special attention should be
paid to what happens with the vacuum energy (the cosmological constant problem)
and what remains of the anomalies. According to a very perceptive explanation 
provided by Jackiw \ci{70}, anomalies are the true physical 
effects related to the
choice of vacuum (see also Chapter 3). So they should be present,
at least under certain circumstances, in the procedure like (\ref{II4clif.137})
which does not explicitly require a vacuum. I think that, e.g. for the
Dirac field our procedure, in the language of QFT means dealing with
\inxx{bare vacuum}
bare vacuum. In other words, the momentum space Fourier transforms of
the vectors $h(x)$, ${\bar h} (x)$ represent states which in QFT are
created out of the bare vacuum. For consistency reasons, in QFT the bare
vacuum is replaced by the Dirac vacuum, and the creation and annihilation
operators are redefined accordingly. Something analogous should also be done 
in our procedure.

\subsubsection{Quantization of the Stueckelberg field}
\inxx{Stueckelberg field,quantization of}

In Part I we have paid much attention to the unconstrained theory which
involves a {\it Lorentz invariant evolution parameter} $\tau$. We have also
seen that such an unconstrained Lorentz invariant theory is embedded in a 
polyvector generalization of the theory. Upon quantization we obtain
\inxx{Schr\" odinger equation}
the Schr\" odinger equation for the wave function $\psi (\tau, x^{\mu})$:
\be
      i {{\p \psi}\oo {\p \tau}} = {1\oo {2 \Lambda}} (- \p_{\mu} \p^{\mu}
      - \kappa^2) \psi .
\lbl{II4clif.138}
\ee
The latter equation follows from the action
\be
    I[\psi, \psi^*] = \int \dd \tau \, {\dd}^4 x \, \left ( i \psi^* 
    {{\p \psi}\oo {\p \tau}} - {\Lambda\oo 2} (\p_{\mu} \psi^* \p^{\mu} \psi
    - \kappa^2 \psi^* \psi) \right ) .
\lbl{II4clif.139}
\ee
This is equal to the scalar part of
   $$I[\psi,\psi^*] = \int \dd \tau \, \dd \tau' \, \dd x \, \dd x' \, 
   \Biggl[i \psi^* (\tau',x') {{\p \psi (\tau,x)}\oo {\p \tau}}
   - {\Lambda\oo 2} (\p'_{\mu} \psi^*(\tau',x') \p^{\mu} \psi (\tau,x)$$
   $$\hs{3cm} - \kappa^2 \psi^* (\tau',x') \psi (\tau,x) \Biggr] h^* 
   (\tau,x) h (\tau',x')$$
   $$= \, \int \dd \tau \, \dd \tau' \, \dd x \, \dd x' \,
   \psi^*(\tau',x') \psi (\tau,x) \hs{3.6cm}$$
   $$\times \Biggl[-i {{\p h^*(\tau,x)}
   \oo {\p \tau}} h(\tau',x') \hs{3.6cm}$$
\be
    \hs{2.8cm}- \, {\Lambda\oo 2} \left ( \p_{\mu} h^* (\tau,x) \p'^{\mu} h (\tau',x') -
    \kappa^2 h^* (\tau,x) h(\tau',x') \right ) \Biggr]   
\lbl{II4clif.140}
\ee
where $h(\tau',x')$, $h^* (\tau,x)$ are assumed to satisfy 
\bear   
   h(\tau',x') \cdot h^* (\tau,x) &\equiv& \mbox{$1\oo 2$} \left ( h(\tau',x') h^* (\tau,
   x) + h^* (\tau,x) h(\tau',x') \right ) , \nonumber \\ 
   &=& \delta (\tau - \tau')
   \delta^4 (x - x') \nonumber
\ear   
\be
    h(\tau,x) \cdot h(\tau',x') = 0  \; , \; \qquad h^* (\tau,x) \cdot h^* (
    \tau',x') = 0 .
\lbl{II4clif.141}
\ee
The relations above follow, as we have seen in previous subsection
(eqs.\,(\ref{II4clif.101})--(\ref{II4clif.108})) from the Clifford
algebra relations amongst the basis fields $h_i (x)$, $i = 1,2$,
\be
    h_i (\tau,x) \cdot h_j (\tau',x') = \delta_{ij} \delta (\tau - \tau') 
    \delta (x - x')
\lbl{II4clif.142}
\ee
related to $h(\tau,x)$, $h^*(\tau,x)$ according to
\be
     h_1 = {{h + h^*}\oo \sqrt{2}} \;  , \; \qquad h_2 = {{h - h^*}\oo {i
     \sqrt{2}}}.
\lbl{II4clif.143}
\ee

Let us now relax the condition (\ref{II4clif.141}) so that (\ref{II4clif.140})
becomes a generalization of the original action (\ref{II4clif.139}).

Moreover, if in (\ref{II4clif.140}) we fix the field $\psi$ according to
\be
   \psi (\tau,x) = \delta (\tau - \tau') \delta^4 (x - x') ,
\lbl{II4clif.144}
\ee
integrate over $\tau'$, $x'$, and rename $\tau'$, $x'$ back into $\tau$, $x$,
we find
\be
    I[h,h^*] = \int \dd \tau \, {\dd}^4 x \, \left [ i h^* {{\p h}\oo {\p \tau}} -
    {\Lambda \oo 2} ( \p_{\mu} h^* \p^{\mu} h - \kappa^2 h^* h) \right ] ,
\lbl{II4clif.145}
\ee
which is an action for basis vector fields $h(\tau,x)$, $h^* (\tau,x)$.
\inxx{basis vector fields,action}
The latter fields are {\it operators}.

The usual canonical procedure then gives that the field $h(x)$ and its
conjugate momentum $\pi = \p {\cal L}/\p {\dot h} = i h^*$, where
${\dot h} \equiv \p h/\p \tau$, satisfy {\it the commutation relations}
\bear
     &[h (\tau,x), \pi (\tau',x')]|_{\tau' = \tau} = i \delta (x - x') ,&
\lbl{II4clif.146}
\nonumber \\
         &[h(\tau,x),h(\tau',x')]_{\tau' = \tau} = 
          [h^* (\tau,x),h^* (\tau',x')]_{\tau' = \tau} = 0 .&
\lbl{II4clif.147}
\ear
From here on the procedure goes along the same lines as discussed in
Chapter 1, Section 4.

\subsubsection{Quantization of the parametrized Dirac field}
\inxx{Dirac field,parametrized,quantization of}

In analogy with the Stueckelberg field we can introduce an invariant evolution
parameter for the Dirac field $\psi (\tau,x^{\mu})$. Instead of the 
usual Dirac equation we have
\be
     i {{\p \psi}\oo {\p \tau}} = - i \gamma^{\mu} \p_{\mu} \psi .
\lbl{II4clif.147a}
\ee
The corresponding action is
\be
    I[\psi, {\bar \psi}] = \int \dd \tau \, {\dd}^4 x \, \left ( i {\bar \psi}
    {{\p \psi}\oo {\p \tau}}+ i {\bar \psi} \gamma^{\mu} \p_{\mu} 
    \psi \right ) .
\lbl{II4clif.148}
\ee

Introducing a basis $h(\tau, x)$ in function space so that a generic vector
can be expanded according to
\bear
    \Psi &=& \int \dd \tau \,  \dd x \,  \left ( 
    {\bar \psi}_{\alpha} (\tau,x) h_{\alpha} (\tau,x) + \psi_{\alpha} (\tau,x)
    {\bar h}_{\alpha} (\tau,x) \right ) \nonumber \\
    &\equiv& \int \dd \tau \, \dd x \,  \left ( 
    {\bar \psi} (\tau,x) h (\tau,x) + \psi (\tau,x)
    {\bar h} (\tau,x) \right ) .
\lbl{II4clif.149}
\ear
We can write (\ref{II4clif.148}) as the scalar part of
\be
  I[\psi, {\bar \psi}] =  \int \dd \tau \, \dd \tau' \, \dd x \, \dd x' \, 
   i {\bar \psi} (\tau',x') {\bar h} (\tau,x) h(\tau',x')
  \left ( {{\p \psi (\tau,x)}\oo {\p \tau }} + i \gamma^{\mu} \p_{\mu}
  \psi (\tau,x) \right ) ,
\lbl{II4clif.150}
\ee
where we assume 
\bear     {\bar h} (\tau,x) \cdot h (\tau',x') &\equiv& 
       \mbox{$1\oo 2$} \left (
       {\bar h} (\tau,x) h (\tau',x')  + h (\tau',x') {\bar h} (\tau,x) 
       \right ) \nonumber \\
        &=& \delta (\tau - \tau') \delta (x - x')
\lbl{II4clif.151}
\ear
For simplicity, in the relations above we have suppressed the spinor indices.

Performing partial integrations in (\ref{II4clif.150}) we can switch the
derivatives from $\psi$ to $h$, as in (\ref{II4clif.126}):
\bear
    & &I = \int \dd \tau \, \dd \tau' \, \dd x \, \dd x' \, 
{\bar \psi} (\tau,x) \hs{3cm} \lbl{II4clif.152} \\
& &\hs{18mm} \times \, \Biggl[ - i {{{\bar h} 
(\tau,x)}\oo {\p \tau}} h(\tau',x') - i \gamma^{\mu}
\p_{\mu} {\bar h} (\tau,x) h (\tau',x') \Biggr] \psi (\tau',x')  . \nonumber    
\ear

We now relax the condition (\ref{II4clif.151}). Then 
eq. (\ref{II4clif.152}) is no longer equivalent to the action (\ref{II4clif.148}).
Actually we shall no more consider (\ref{II4clif.152}) as an action for $\psi$.
Instead we shall fix\footnote{Taking also the spinor indices into account,
instead of (\ref{II4clif.154}) we have
$$\psi_{\alpha} (\tau,x) = \delta_{\alpha,
\alpha'} \delta (\tau - \tau'') \delta^4 (x - x'').$$}
$\psi$ according to
\be 
       \psi (\tau,x) = \delta (\tau - \tau') \delta^4 (x - x') .
\lbl{II4clif.154}
\ee
Integrating (\ref{II4clif.152}) over $\tau''$, $x''$ and renaming  
$\tau''$, $x''$ 
back into  $\tau$, $x$, we obtain an action for basis vector fields
$h (\tau,x)$, ${\bar h} (\tau,x)$:
\be
    I[h,{\bar h}] = \int \dd \tau \, {\dd}^4 x \left [ i {\bar h} {{\p h}\oo
    {\p \tau}} + i {\bar  h} \gamma^{\mu} \p_{\mu} h\right ] .
\lbl{II4clif.155}
\ee
Derivatives now act on $h$, since we have performed  additional partial 
integrations and have omitted the surface terms.

Again we have arrived at an action for field operators $h$, ${\bar h}$. The
equations of motion (the field equations) are
\be
    i {{\p h}\oo {\p \tau}} = - i \gamma^{\mu} \p_{\mu} h .
\lbl{II4clif.156}
\ee
The canonically conjugate variables are $h$ and $\pi = \p {\cal L}/\p {\dot h}
= i {\bar h}$, and they satisfy {\it the anticommutation relations}
\be
      \lbrace h(\tau,x), \pi (\tau',x') \rbrace \vert_{\tau'=\tau} =
      i \delta (x - x') ,
\lbl{II4clif.157}
\ee
or
\be
      \lbrace h(\tau,x), {\bar h} (\tau',x') \rbrace \vert_{\tau'=\tau} =
      \delta (x - x') , 
\lbl{II4clif.159}
\ee
and
\be
     \lbrace h(\tau,x), h (\tau',x') \rbrace \vert_{\tau'=\tau} =                
           \lbrace {\bar h} (\tau,x), {\bar h} (\tau',x') 
           \rbrace \vert_{\tau'=\tau}
       = 0 .
\lbl{II4clif.159a}
\ee

The anticommutation relations above being satisfied, the Heisenberg equation
\inxx{Heisenberg equation,for Dirac field}
\be
    {{\p h}\oo {\p \tau}} = i  [h, H] \; , \; \qquad H = \int 
    \dd x \, i {\bar h} \gamma^{\mu} \p_{\mu} h \; ,
\lbl{II4clif.160}
\ee
is equivalent to the field equation (\ref{II4clif.156}).    

\upperandlowertrue  
\subsection[\ \ Quantization of the $p$-brane: a geometric approach]
{QUANTIZATION OF THE $p$-BRANE: \\
A GEOMETRIC APPROACH}
\inxx{quantization of the $p$-brane,a geometric approach}
\upperandlowerfalse

We have seen that a field can be considered
as an uncountable set of components of an infinite-dimensional vector. Instead
of considering the action which governs the dynamics of components, 
we have considered the action which governs the dynamics of the basis vectors.
The latter behave as operators satisfying the Clifford algebra. The
quantization consisted of the crucial step in which we abolished the 
requirement that the basis vectors satisfy the Clifford algebra relations for a
``flat" metric in function space (which is proportional to the $\delta$-function).
We admitted an arbitrary metric in principle. The action itself suggested
which are the (commutation or anti commutation) relations the basis vectors
(operators) should satisfy. Thus we arrived at the conventional procedure of the
field quantization.

Our geometric approach brings a new insight about the nature of field
quantization. In the conventional 
approach classical fields are replaced by\inxx{field quantization,nature of}
operators which satisfy the canonical commutation or anti-commutation
relations. In the proposed geometric approach we observe that {\it the
field operators are, in fact, the basis vectors $h(x)$.} By its very definition
a basis vector $h(x)$, for a given $x$, ``creates" a particle at the position
$x$. Namely, an arbitrary vector $\Phi$ is written as a superposition
of basis vectors
\be
    \Phi = \int \dd x' \, \phi (x') h(x')
\lbl{II4clif.161}
\ee
and $\phi(x)$ is ``the wave packet" profile. If in particular $\phi(x') =
\delta (x' - x)$, i.e., if the ``particle" is located at $x$, then
\be
     \Phi = h(x) .
\lbl{II4clif.162}
\ee

We shall now explore further the possibilities brought by such a geometric
approach to quantization. Our main interest is to find out how it could be
applied to the quantization of strings and $p$-branes in general. In Sec. 4.2,
we have found out that a conventional $p$-brane can be described
by the following action
\be
    I[X^{\alpha (\xi)} (\tau)] = \int \dd \tau' \, \rho_{\alpha(\xi') 
    \beta(\xi'')} {\dot X}^{\alpha(\xi')} {\dot X}^{\beta (\xi'')} =
    \rho_{\alpha (\phi')\beta(\phi'')} {\dot X}^{\alpha(\phi')}
    {\dot X}^{\beta (\phi'')} ,
\lbl{II4clif.163}
\ee
where
\be
    \rho_{\alpha (\phi')\beta(\phi'')} = {{\kappa \sqrt{|f|}}\oo {\sqrt{{\dot 
    X}^2}}} \delta (\tau' - \tau'') \delta (\xi' - \xi'') g_{\alpha \beta} .
\lbl{II4clif.164}
\ee
Here ${\dot X}^{\alpha (\phi)} \equiv {\dot X}^{\alpha (\tau,\xi)} \equiv
{\dot X}^{\alpha (\xi)} (\tau)$, where $\xi \equiv \xi^a$ are the
$p$-brane coordinates, and $\phi \equiv \phi^A = (\tau,\xi^a)$ are coordinates
of the world surface which I call worldsheet.

If the ${\cal M}$-space metric $\rho_{\alpha (\phi')\beta(\phi'')}$ is
different from (\ref{II4clif.164}), then we have a deviation from the usual
Dirac--Nambu--Goto $p$-brane theory. Therefore in the classical theory
$\rho_{\alpha (\phi')\beta(\phi'')}$ was made dynamical by adding a
suitable kinetic term to the action.

Introducing the basis vectors $h_{\alpha (\phi)}$ satisfying
\be
   h_{\alpha (\phi')} \cdot h_{\beta (\phi'')} = \rho_{\alpha (\phi') \beta
   (\phi'')}
\lbl{II4clif.165}
\ee
we have
\be
    I[X^{\alpha (\phi)}] = h_{\alpha (\phi')} h_{\beta (\phi'')} 
    {\dot X}^{\alpha (\phi')} {\dot X}^{\beta (\phi'')} .
\lbl{II4clif.166}
\ee
Here $h_{\alpha (\phi')}$ are fixed while $X^{\alpha (\phi')}$ are variables.
If we now admit that $h_{\alpha (\phi')}$ also change with $\tau$, we can
perform the partial integrations over $\tau'$ and $\tau''$ so that
eq.\,(\ref{II4clif.166}) becomes
\be
      I = {\dot h}_{\alpha (\phi')} {\dot h}_{\beta (\phi'')} X^{\alpha (\phi')}
      X^{\beta (\phi'')} .
\lbl{II4clif.167}
\ee

We now assume that $X^{\alpha (\phi')}$ is an arbitrary configuration, not
necessarily the one that solves the variational principle (\ref{II4clif.163}).
In particular, let us take
\be
     X^{\alpha (\phi')} = {\delta^{\alpha (\phi')}}_{\mu (\phi)} \; \; ,
     \qquad X^{\beta (\phi'')} = {\delta^{\beta (\phi'')}}_{\mu (\phi)} ,
\lbl{II4clif.168}
\ee
which means that our $p$-brane is actually a {\it point} at the values of the
parameters $\phi \equiv (\tau, \xi^a)$ and the value of the index $\mu$.
So we have
\be
    I_0 = {\dot h}_{\mu (\phi)} {\dot h}_{\mu (\phi)} \quad {\rm no \; sum \;
    and \; no \; integration} .
\lbl{II4clif.169}
\ee
By taking (\ref{II4clif.168}) we have in a sense ``quantized" the classical
action. The above expression is a ``quantum" of (\ref{II4clif.163}).

Integrating (\ref{II4clif.169}) over $\phi$ and summing over $\mu$ we obtain
\be
    I[h_{\mu (\phi)}] = \int \dd \phi \, \sum_{\mu} {\dot h}_{\mu} (\phi)
    {\dot h}_{\mu} (\phi) .
\lbl{II4clif.169a}
\ee
The latter expression can be written as\footnote{
Again summation and integration convention is assumed.}
\bear
     I[h_{\mu (\phi)}] &=& \int \dd \phi \, \dd \phi' \, \delta (\phi - \phi')
     \eta^{\mu \nu} {\dot h} (\phi) {\dot h}_{\nu} (\phi') \nonumber \\
     &\equiv& \eta^{\mu (\phi) \nu (\phi')} {\dot h}_{\mu (\phi)}
     {\dot h}_{\nu (\phi')} , 
\lbl{II4clif.169b}
\ear
where
\be
     \eta^{\mu (\phi) \nu (\phi')} = \eta^{\mu \nu} \delta (\phi - \phi')
\lbl{II4clif.169c}
\ee
is the flat ${\cal M}$-space metric. In general, of course, ${\cal M}$-space
is not flat, and we have to use arbitrary metric. Hence (\ref{II4clif.169b})
generalizes to
\be
      I[h_{\mu (\phi)}] =  \rho^{\mu (\phi) \nu (\phi')} {\dot h}_{\mu (\phi)}
     {\dot h}_{\nu (\phi')} ,
\lbl{II4clif.169d}
\ee
where
\be
        \rho^{\mu (\phi) \nu (\phi')} = h^{\mu (\phi)} \cdot h^{\nu (\phi')} =
        \mbox{$1\oo 2$} (h^{\mu (\phi)} h^{\nu (\phi')} +  h^{\nu (\phi')}
         h^{\mu (\phi)}) .
\lbl{II4clif.169e}
\ee
Using the expression (\ref{II4clif.169e}), the action becomes
\be
        I[h_{\mu (\phi)}] =  h^{\mu (\phi)} h^{\nu (\phi')} 
        {\dot h}_{\mu (\phi)} {\dot h}_{\nu (\phi')} . 
\lbl{II4clif.169f}
\ee                

\subsubsection{Metric in the space of operators}

In the definition of the action (\ref{II4clif.169d}), or (\ref{II4clif.169f}),
we used the relation (\ref{II4clif.169e}) in which the ${\cal M}$-space
metric is expressed in terms of the basis vectors $h_{\mu (\phi)}$.
In order to allow for a more general case, we shall introduce the {\it 
metric} $Z^{\mu (\phi) \nu (\phi')}$ in the ``space" of operators.
In particular it can be
\be
        \mbox{$1\oo 2$} Z^{\mu (\phi) \nu (\phi')} = 
        h^{\mu (\phi)} h^{\nu (\phi')} ,
\lbl{II4clif.169g}
\ee
or
\be
        \mbox{$1\oo 2$} Z^{\mu (\phi) \nu (\phi')} = 
        \mbox{$1\oo 2$} (h^{\mu (\phi)} h^{\nu (\phi')} + h^{\nu (\phi')}
        h^{\mu (\phi)}) ,
\lbl{II4clif.169h}
\ee
but in general, $Z^{\mu (\phi) \nu (\phi')}$ is expressed arbitrarily in
terms of $h^{\mu (\phi)}$. Then instead of (\ref{II4clif.169f}), we have
\be
    I[h] = \mbox{$1\oo 2$} Z^{\mu (\phi) \nu (\phi')}
    {\dot h}_{\mu (\phi)} {\dot h}_{\nu (\phi')}
    = \mbox{$1\oo 2$} \int \dd \tau \, Z^{\mu (\xi) \nu (\xi')}
    {\dot h}_{\mu (\xi)} {\dot h}_{\nu (\xi')} .
\lbl{II4clif.170}
\ee
The factor $1\oo 2$ is just for convenience; it does not influence the equations
of motion.

Assuming $Z^{\mu (\phi) \nu (\phi')} = Z^{\mu (\xi) \nu (\xi')} \delta
(\tau - \tau')$ we have
\be
  I[h] = \mbox{$1\oo 2$} \int \dd \tau \, Z^{\mu (\xi) \nu (\xi')}
    {\dot h}_{\mu (\xi)} {\dot h}_{\nu (\xi')} .
\lbl{II4clif.170a}
\ee

Now we could continue by assuming the validity of the scalar product relations
(\ref{II4clif.165}) and explore the equations of motion 
derived from (\ref{II4clif.170})
for a chosen  $Z^{\mu (\phi) \nu (\phi')}$. This is perhaps a possible
approach to geometric quantization, but we shall not pursue it here.

Rather we shall forget about (\ref{II4clif.165}) and start directly from the
action (\ref{II4clif.170a}), considered as an action for the operator field
$h_{\mu (\xi)}$ where the commutation relations should now be determined.
The canonically conjugate variables are
\be
   h_{\mu (\xi)} \; ,  \qquad \pi^{\mu (\xi)} =
\p L/\p {\dot h}_{\mu (\xi)} = Z^{\mu (\xi) \nu (\xi')} 
{\dot h}_{\nu (\xi')} .
\lbl{II4clif.170b}
\ee
They are assumed to satisfy the equal $\tau$
commutation relations
\bear
     &[h_{\mu (\xi)}, h_{\nu (\xi')}] = 0 \; , \; \qquad [\pi^{\mu (\xi)},
     \pi^{\nu (\xi')}] = 0 \; ,&
\nonumber \\
      &[h_{\mu (\xi)}, \pi^{\nu (\xi')}] = 
      i {\delta_{\mu (\xi)}}^{\nu (\xi')} .&
\lbl{II4clif.171b}
\ear
By imposing (\ref{II4clif.171b}) we have abolished the
Clifford algebra relation (\ref{II4clif.165}) in which the inner product
(defined as the symmetrized Clifford product) is
equal to a {\it scalar valued} metric.

The Heisenberg equations of motion are \inxx{Heisenberg equations}
\bear
    &{\dot h}_{\mu (\xi)} = - i [h_{\mu (\xi)}, H] ,&
\lbl{II4clif.172}\\
   &{\dot \pi}^{\mu (\xi)} = -i [\pi^{\mu (\xi)}, H] , &
\lbl{II4clif.173} 
\ear    
where the Hamiltonian is \inxx{Hamiltonian}
\be
     H = \mbox{$1\oo 2$} Z_{\mu (\xi) \nu (\xi')} 
     \pi^{\mu (\xi)} \pi^{\nu (\xi')} .
\lbl{II4clif.172a}
\ee                   

In particular, we may take a trivial metric which does not contain 
$h_{\mu (\xi)}$, e.g., 
\be
    Z^{\mu (\xi) \nu (\xi')} = \eta^{\mu \nu} \delta (\xi - \xi') .
\lbl{II4clif.174}
\ee
Then the equation of motion resulting from (\ref{II4clif.173}) or directly from the
action (\ref{II4clif.170a}) is
\be
        {\dot \pi}_{\mu (\xi)} = 0 .
\lbl{II4clif.175}
\ee
Such a dynamical system cannot describe the usual $p$-brane, since the 
equations of motion is too simple. It serves here for the purpose of
demonstrating the procedure. In fact, in the quantization procedures for the 
Klein--Gordon, Dirac, Stueckelberg field, etc., we have in fact used a
fixed prescribed metric which was proportional to the $\delta$-function.

In general, the metric $Z^{\mu (\xi) \nu (\xi')}$ is an expression containing
$h_{\mu (\xi)}$. The variation of the action (\ref{II4clif.170}) with respect
to $h_{\mu (\xi)}$ gives
\be
    {\dd \oo {\dd \tau}} (Z^{\mu (\phi) \nu (\phi')} {\dot h}_{\nu (\phi)})
    - {1\oo 2} {{\delta Z^{\alpha (\phi') \beta (\phi'')}}\oo {\delta
    h_{\mu (\phi)}}} {\dot h}_{\alpha (\phi')}{\dot h}_{\beta (\phi'')} = 0 .
\lbl{II4clif.176}
\ee
Using (\ref{II4clif.171b}) one finds 
that the Heisenberg equation
(\ref{II4clif.173}) is equivalent to (\ref{II4clif.176}).

\subsubsection{The states of the quantized brane}
\inxx{quantized $p$-brane}

According to the traditional  approach to QFT one would now introduce a vacuum
{\it state vector} $|0 \rangle$ and define
\be
    h_{\alpha (\xi)} |0 \rangle \; \; , \quad h_{\alpha (\xi)} h_{\beta (\xi)}
    |0 \rangle \; , ...
\lbl{II4clif.177}
\ee
as {\it vectors} in Fock space. Within our geometric approach we can
\inxx{Fock space}
do something quite analogous. First we realize that because of the
commutation relations (\ref{II4clif.171b}) $h_{\mu (\xi)}$
are in fact not
elements of the Clifford algebra. Therefore they are not vectors in the
usual sense. In order to
obtain vectors we introduce an object $v_0$ which, by definition, is a
Clifford number satisfying\footnote{
The procedure here is an alternative to the one considered when discussing
quantization of the Klein--Gordon and other fields.}
\be
       v_0 v_0 = 1
\lbl{II4clif.177a}
\ee
and has the property that the products
\be
    h_{\mu (\xi)} v_0 \; \;, \quad h_{\mu (\xi)} h_{\nu (\xi)} v_0\; ...
\lbl{II4clif.178}
\ee
are also Clifford numbers. Thus $h_{\mu (\xi)} v_0$ behaves as a vector.
The inner product between such vectors is defined as usually in Clifford 
algebra:
\bear
    (h_{\mu (\xi)} v_0) \cdot (h_{\nu (\xi')} v_0) &\equiv& 
    \mbox{$1\oo 2$} \left [
    (h_{\mu (\xi)}) v_0)(h_{\nu (\xi')} v_0) + (h_{\nu (\xi')} v_0)
    (h_{\mu (\xi)}) v_0) \right ] \nonumber \\
     &=& {\rho_0}_{\mu (\xi) \nu (\xi')} \; ,
\lbl{II4clif.179}\
\ear
where ${\rho_0}_{\mu (\xi) \nu (\xi')}$ is a scalar-valued metric.
The choice of ${\rho_0}_{\mu (\xi) \nu (\xi')}$ is determined by the choice of 
$v_0$. We see that the vector $v_0$ corresponds to the vacuum state vector
\inxx{vacuum state vector}
of QFT. The vectors (\ref{II4clif.178}) correspond to the other basis vectors of 
Fock space. \inxx{Fock space}
And we see here that choice of the vacuum vector $v_0$ determines the 
metric in Fock space. Usually basis vector of Fock space are orthonormal,
hence we take
\be
     {\rho_0}_{\mu (\xi) \nu (\xi')} = \eta_{\mu \nu} \delta (\xi - \xi') .
\lbl{II4clif.179a}
\ee 

In the conventional field-theoretic notation the relation (\ref{II4clif.179})
reads
\be
      \langle 0| \mbox{$1\oo 2$} 
      (h_{\mu (\xi)} h_{\nu (\xi')} +  h_{\nu (\xi')} 
       h_{\mu (\xi)}) |0 \rangle =  {\rho_0}_{\mu (\xi) \nu (\xi')} .
\lbl{II4clif.179aa}
\ee
This is the vacuum expectation value of the operator
\inxx{vacuum expectation value,of the metric operator}
\be
       {\hat \rho}_{\mu (\xi) \nu (\xi')} =  \mbox{$1\oo 2$} (h_{\mu (\xi)} 
       h_{\nu (\xi')} +  h_{\nu (\xi')} h_{\mu (\xi)}) ,
\lbl{II4clif.179b}
\ee
which has the role of the ${\cal M}$-{\it space metric operator}.

In a generic state $|\Psi \rangle$ of Fock space the expectation value
of the operator ${\hat \rho}_{\mu (\xi) \nu (\xi')}$ is
\be
    \langle \Psi |{\hat \rho}_{\mu (\xi) \nu (\xi')}|\Psi \rangle
    = \rho_{\mu (\xi) \nu (\xi')} .
\lbl{II4clif.180}
\ee
{\it Hence, in a given state, for the expectation value of the metric operator
we obtain a certain {\it scalar valued}  ${\cal M}$-space metric
$\rho_{\mu (\xi) \nu (\xi')}$.}

In the geometric notation (\ref{II4clif.180}) reads
\be
        \langle V h_{\mu (\xi)} h_{\nu (\xi')} V \rangle_0 =
         \rho_{\mu (\xi) \nu (\xi')} .
\lbl{II4clif.181}
\ee
This means that we choose a Clifford number (Clifford aggregate) $V$ formed
from (\ref{II4clif.178})
\be
    V = (\phi^{\mu (\xi)} h_{\mu (\xi)} + \phi^{\mu (\xi) \nu (\xi')}
      h_{\mu (\xi)} h_{\nu (\xi')} + ... \, ) v_0 ,
\lbl{II4clif.181a}
\ee
where $\phi^{\mu (\xi)}$, $\phi^{\mu (\xi) \nu (\xi')}$, ... , are
the wave packet profiles, then we write the expression 
$V h_{\mu (\xi)} h_{\nu (\xi')} V$ and take its {\it scalar part}.

Conceptually our procedure appears to be very clear. We have an action
(\ref{II4clif.170a}) for operators $h_{\mu (\xi)}$ satysfying the commutation
relations (\ref{II4clif.171b}) and the Heisenberg equation
of motion (\ref{II4clif.173}). With the aid of those operators we form a
Fock space of states, and then we calculate the expectation value of
the metric operator ${\hat \rho}_{\mu (\xi) \nu (\xi')}$ in a chosen
state. {\it We interpret this expectation value as the classical metric of
${\cal M}$-space.} This is justified because there is a correspondence
between the operators $h_{\mu (\xi)}$ and the ${\cal M}$-space basis
vectors (also denoted $h_{\mu (\xi)}$, but obeying the Clifford algebra
relations (\ref{II4clif.169})). The operators create, when acting on $|0 \rangle$
or $v_0$, the many brane states and it is natural to interpret the
expectation value of ${\hat \rho}_{\mu (\xi) \nu (\xi')}$ 
as the classical ${\cal M}$-space metric for such a many brane configuration.

\subsubsection{Which choice for the operator metric $Z^{\mu (\xi) \nu (\xi')}$?}

A question now arise of how to choose $Z^{\mu (\xi) \nu (\xi')}$. In principle
any combination of operators $h_{\mu (\xi)}$ is good, provided that
$Z^{\mu (\xi) \nu (\xi')}$ has its inverse defined according to
\be
     Z^{\mu (\xi) \alpha (\xi'')} Z_{\alpha (\xi'') \nu (\xi')} =
     {\delta^{\mu (\xi)}}_{\nu (\xi')} \equiv {\delta^{\mu}}_{\nu}
     \delta (\xi - \xi') .
\lbl{II4clif.182}
\ee
Different choices of $Z^{\mu (\xi) \nu (\xi')}$  mean different membrane
theories, and hence different expectation values $\rho_{\mu (\xi) \nu (\xi')}
= \langle {\hat \rho}_{\mu (\xi) \nu (\xi')} \rangle$ of the ${\cal M}$-space
metric operator ${\hat \rho}_{\mu (\xi) \nu (\xi')}$. We have already
observed that different choices of $\rho_{\mu (\xi) \nu (\xi')}$ correspond 
to different classical membrane theories. In order to get rid of a fixed 
background we have given $\rho_{\mu (\xi) \nu (\xi')}$ the status of a
dynamical variable and included a kinetic term for $\rho_{\mu (\xi) \nu (\xi')}$
in the action (or equivalently for $h_{\mu (\xi)}$, which is the ``square root"
of $\rho_{\mu (\xi) \nu (\xi')}$, since classically $\rho_{\mu (\xi) \nu (\xi')}
= h_{\mu (\xi)} \cdot h_{\nu (\xi')}$). In performing the quantization we have
seen that to the classical vectors $h_{\mu (\xi)}$ there correspond quantum
operators\footnote{
Now we use hats to make a clear distinction between the classical vectors
$h_{\mu (\xi)}$, satisfying the Clifford algebra 
relations (\ref{II4clif.165}), and
the quantum operators ${\hat h}_{\mu (\xi)}$ satisfying
(\ref{II4clif.171b}).}
${\hat h}_{\mu (\xi)}$ which obey the equations of motion determined by
the action (\ref{II4clif.170a}). Hence in the quantized theory we do not need a
separate kinetic term for ${\hat h}_{\mu (\xi)}$. But now we have something
new, namely, $Z^{\mu (\xi) \nu (\xi')}$ which is a background metric in
the space of operators ${\hat h}_{\mu (\xi)}$. In order to obtain a
background independent theory we need a kinetic term for 
$Z^{\mu (\xi) \nu (\xi')}$. The search for  such a kinetic term will remain
a subject of future investigations. It has its parallel in the attempts
to find a background independent string or $p$-brane theory.
However, it may turn out that we do not need a kinetic term for
$Z^{\mu (\xi) \nu (\xi')}$ and that it is actually given by the
expression (\ref{II4clif.169g}) or (\ref{II4clif.169h}), so that 
(\ref{II4clif.169f}) is already the ``final" action for the quantum
$p$-brane.

In the field theory discussed previously we also have a fixed metric
$Z^{\mu (\xi) \nu (\xi')}$, namely, $Z^{(\xi) (\xi')} = 
\delta (\xi - \xi')$ for the Klein--Gordon and similarly for the Dirac field.
Why such a choice and not some other choice? This clearly points to the 
plausible possibility that the usual QFT is not complete. That QFT is not yet
a finished story is clear from the occurrence of infinities and the need
for ``renormalization"\footnote{An alternative approach to the quantization
of field theories, also based on Clifford algebra, has been pursued by
Kanatchikov \ci{Kanatchikov}.}.

\part{Brane World}

\chapter[Spacetime as a membrane in a \ \\ higher-dimensional space]
{Spacetime as a membrane \\ in a higher-dimensional space}

When studying dynamics of a system of membranes, as seen from the
${\cal M}$-space point of view, we have arrived in Chapter 5 at a
fascinating conclusion that all that exists in such a world model is
\inxx{membrane configuration}
a membrane configuration. {\it The membrane configuration itself is
a `spacetime'.} Without membranes there is no spacetime. According to
our basic assumption, at the fundamental level we have an ${\cal M}$-space ---
the space of all possible membrane configurations --- and nothing else.
If the membrane configuration consists of the membranes of various
dimensions $n$, lower and higher than the dimension of our observed
word ($n = 4$), then we are left with a model in which our 4-dimensional
spacetime is one of those (4-dimensional) membranes (which I call
{\it worldsheets}).

What is the space our worldsheet is embedded in? It is just the space
formed by the other $n$-dimensional $(n = 0,1,2,...\, )$ extended objects
(say membranes) entering the membrane configuration. If all those other
membranes are sufficiently densely packed together, then as an
approximation a concept of a continuous embedding space can be used.
Our spacetime can then be considered as a 4-dimensional worldsheet
embedded into a higher-dimensional space.

\section{The brane in a curved embedding space}

We are now going to explore a brane moving in 
a curved background\inxx{brane} \inxx{brane,in curved background}
embedding space $V_N$. Such a brane sweeps an $n$-dimensional surface
which I call {\it worldsheet}\footnote{
Usually, when $n>2$ such a surface is called a {\it world volume}. Here I
prefer to retain the name {\it worldsheet}, by which we can vividly
imagine a surface in an embedding space.}
The dynamical principle governing motion of the brane requires that its
worldsheet is a minimal surface. Hence the action is
\be
    I[\eta^a] = \int \sqrt{|{\tilde f}|} \, {\dd}^n x ,
\lbl{III1.1}
\ee
where
\be
    {\tilde f} = {\rm det} {\tilde f}_{\mu \nu}  \; , \qquad \; 
    {\tilde f}_{\mu \nu} = \p_{\mu} \eta^a \p_{\nu} \eta^b \gamma_{ab} .
\lbl{III1.2}
\ee
Here $x^{\mu}$, $\mu = 0,1,2,...,n-1$, are coordinates on the worldsheet
$V_n$, whilst $\eta^a (x)$ are the embedding functions. The metric of the
embedding space (from now on also called {\it bulk}) is $\gamma_{ab}$, and
the induced metric on the worldsheet is ${\tilde f}_{\mu \nu}$. \inxx{bulk}
\inxx{induced metric}

In this part of the book I shall use the notation which is adapted to the
idea that our world is a {\it brane}. Position coordinates in our world
are commonly denoted as $x^{\mu}$, $\mu = 0,1,2,...,n-1$, and usually it is
assumed that $n = 4$ (for good reasons, of course, unless one considers
Kaluza--Klein theories). The notation in (\ref{III1.1})\,(\ref{III1.2}) is
"the reverse video" of the notation used so far. The correspondence
between the two notations is the following
\begin{description}
\item worldsheet coordinates \hs{2.8cm} $\xi^a, \; \xi^A , \phi^A$
      \hs{2cm} $x^{\mu}$
\item embedding space coordinates \hs{2.3cm} $x^{\mu}$ \hs{2.8cm} $\eta^a$
\item embedding functions \hs{2.5cm} $X^{\mu} (\xi^a) \, , 
  X^{\mu}(\phi^A)$ \hs{1.5cm} $\eta^a (x^{\mu})$
\item worldsheet metric \hs{3.5cm} $\gamma_{ab} \, , \gamma_{AB}$ 
        \hs{2.2cm} $g_{\mu \nu}$
\item embedding space metric \hs{3cm} $g_{\mu \nu}$ \hs{2.5cm} $\gamma_{ab}$
\end{description}
Such a reverse notation reflects the change of role given to spacetime.
So far `spacetime' has been associated with the embedding space, whilst 
the brane
has been an object in spacetime. Now spacetime is associated with a brane, so
spacetime itself is an object in the embedding space\footnote{Such a
distinction is only manifest in the picture in which we already have an
effective embedding space. In a more fundamental picture the embedding space
is inseparable from the membrane configuration, and in general is not a
manifold at all.}.

For the extended object described by the minimal 
surface action (\ref{III1.1})\inxx{brane}
I use the common name {\it brane}. For a more general extended object
described by a Clifford algebra generalization of the action (\ref{III1.1})
I use the name {\it membrane} (and occasionally also {\it worldsheet}, 
when
I wish to stress that the object of investigation is a direct 
generalization of the object $V_n$ described by (\ref{III1.1}) which is now
understood as a special kind of (generalized) worldsheet).

Suppose now that the metric of $V_N$ is conformally flat (with $\eta_{ab}$
being the Minkowski metric tensor in $N$-dimensions):
\be
    \gamma_{ab} = \phi \, \eta_{ab} .
\lbl{III1.4}
\ee
Then from (\ref{III1.2}) we have
\be
    {\tilde f}_{\mu \nu} = \phi \, \p_{\mu} \eta^a \p_{\nu} \eta^b \eta_{ab}
    \equiv \phi f_{\mu \nu} \; ,
\lbl{III1.5}
\ee
\be
     {\tilde f} \equiv {\rm det} {\tilde f}_{\mu \nu} = \phi^n \, {\rm det}
     f_{\mu \nu} \equiv \phi^n f \; 
\lbl{III1.6}
\ee
\be
     \sqrt{|{\tilde f}|} = \omega |f| \; , \quad \omega \equiv \phi^{n/2} .
\lbl{III1.7}
\ee
Hence the action (\ref{III1.1}) reads
\be
      I[\eta^a] = \int \omega (\eta) \, \sqrt{|f|} {\dd}^n x
\lbl{III1.8} ,
\ee
which looks like an action for a brane in a flat embedding space, except for
a function $\omega (\eta)$ which depends on the position\footnote{
We use here the same symbol $\eta^a$ either for position coordinates in
$V_N$ or for the embedding functions $\eta^a (x)$.}
$\eta^a$ in the embedding space $V_N$.

Function $\omega (\eta)$ is related to the fixed background metric which is
arbitrary in principle. Let us now assume \ci{71} that $\omega (\eta)$ consists
of a constant part $\omega_0$ and a singular part with support on another
brane's worldsheet ${\widehat V}_m$:\
\be
    \omega (\eta) = \omega_0 + \kappa \int {\dd}^m {\hat x} \, \sqrt{|{\hat f}|}
    \, {{\delta^N (\eta - {\hat \eta})}\oo {\sqrt{|\gamma|}}} .
\lbl{III1.8a}
\ee
Here ${\hat \eta}^a ({\hat x})$ are the embedding functions of the
$m$-dimensional worldsheet ${\widehat V}_m$, ${\hat f}$ is the determinant of
the induced metric on ${\widehat V}_m$, and $\sqrt{|\gamma|}$ allows for
taking curved coordinates in otherwise flat $V_N$.

The action for the brane which sweeps a worldsheet $V_n$ is then given by
(\ref{III1.8}) in which we replace $\omega (\eta)$ with the specific
expression (\ref{III1.8a}):
\be
    I[\eta] = \int \omega_0 \, {\dd}^n x \, \sqrt{|f|} + \kappa \int {\dd}^n x
    \, {\dd}^m {\hat x} \, \sqrt{|f|} \, \sqrt{|{\hat f}|} \, {{\delta^N (\eta - 
    {\hat \eta})}\oo {\sqrt{|\gamma|}}}  .            
\lbl{III1.9}
\ee

If we take the second brane as dynamical too, then the kinetic term for 
${\hat \eta}^a$
should be added to (\ref{III1.9}). Hence the total action for both branes is
\be
    I[\eta, {\hat \eta}] = \int \omega_0 \, {\dd}^n x \, \sqrt{|f|} + \int
    \omega_0 \, {\dd}^m {\hat x} \sqrt{|{\hat f}|} + \kappa \int {\dd}^n x \, 
    {\dd}^m {\hat x} \, \sqrt{|f|} \, \sqrt{|{\hat f}|} \, {{\delta^N (\eta - 
    {\hat \eta})}\oo {\sqrt{|\gamma|}}} . 
\lbl{III1.10}
\ee
The first two terms are the actions for free branes, whilst the last term
represents the interaction between the two branes. The interaction occurs
when the branes intersect. If we take $m = N - n + 1$ then the
\inxx{intersection,of branes}
{\it intersection} of $V_n$ and ${\widehat V}_m$ can be a (one-dimensional) line,
i.e., a {\it worldline} $V_1$. In general, when $m = N - n + (p+1)$, the
intersection can be a $(p+1)$-dimensional worldsheet representing the motion of
a $p$-brane.

\vs{1mm}

\figIIIA

In eq.\,(\ref{III1.10}) we assume contact interaction between the branes
\inxx{interaction,between the branes}
(i.e., the interaction at the intersection). This could be understood by
imagining that gravity decreases so quickly in the transverse direction from
the brane that it can be approximated by a $\delta$-function. More about this
will be said in Section 4.

\vs{1mm}

The equations of motion derived from the (\ref{III1.10}) by varying
respectively $\eta^a$ and ${\hat \eta}$ are:
\bear
     \p_{\mu} \left [ \sqrt{|f|} \p^{\mu} \eta_a \left (\omega_0 + 
      \kappa \int {\dd}^m {\hat x} \sqrt{|{\hat f}|}
    \, {{\delta^N (\eta - {\hat \eta})}\oo {\sqrt{|\gamma|}}}  \right ) \right ]
    = 0 & &
\lbl{III1.11}\\
\nonumber \\
     {\hat \p}_{\hat \mu} \left [ \sqrt{|{\hat f}|} {\hat \p}^{\hat \mu}
     {\hat \eta}_a \left (\omega_0 + 
      \kappa \int {\dd}^n x \sqrt{|f|}
    \, {{\delta^N (\eta - {\hat \eta})}\oo {\sqrt{|\gamma|}}}  \right ) \right ]
    = 0 & &
\lbl{III1.12}
\ear
where $\p_{\mu} \equiv \p/\p x^{\mu}$ and ${\hat \p}_{\hat \mu} \equiv \p/ \p
{\hat x}^{\hat \mu}$. When deriving eq.\,(\ref{III1.11}) we have taken into account
that
\newpage
\bear
    {{\p}\oo {\p \eta^a}} \int \kappa \, {\dd}^m {\hat x} \, \sqrt{|{\hat f}|}
      \, \delta^N (\eta - {\hat \eta}) &=& - \int \kappa \, {\dd}^m {\hat x} \,  
      \sqrt{|{\hat f}|} {{\p}\oo {\p {\hat \eta}^a}} \, \delta^N (\eta -
      {\hat \eta})\hs{1.5cm}\lbl{III1.13}\\
      &=& \, \kappa \int {\dd}^m {\hat x} \, {{\p \sqrt{|{\hat f}|}}
        \oo {\p {\hat \eta}^a}} \, \delta^N (\eta - {\hat \eta}) = 0 , \nonumber
\ear
since
 \be
     {{\p \sqrt{|{\hat f}|}} \oo {\p {\hat \eta}^a}} = 
     {{\p \sqrt{|{\hat f}|}} \oo {\p {\hat f}}} \, {{\p {\hat f}_{\mu \nu}}\oo
     {\p {\hat \eta}^a}} = 0 ,
\lbl{III1.14}
\ee
because
\be
       {\hat f}_{\mu \nu} = {\hat \p}_{\hat \mu} {\hat \eta}^a \, 
        {\hat \p}_{\hat \nu} {\hat \eta}_a \; , \quad {{\p {\hat f}_{\mu \nu}}\oo
     {\p {\hat \eta}^a}} = 0 .
\lbl{III1.15}
\ee
Analogous holds for eq.\,(\ref{III1.12}).

Assuming that the intersection $V_{p+1} = V_n \bigcap {\widehat V}_m$ does exist,
and, in particular, that it is a worldline (i.e., $p = 0$), then we can write
\be     
  \int {\dd}^m {\hat x} \sqrt{|{\hat f}|}
    \, {{\delta^N (\eta - {\hat \eta})}\oo {\sqrt{|\gamma|}}} =
    \int \dd \tau {{\delta^n (x - X(\tau))}\oo {\sqrt{|f|}}} \, ({\dot X}^{\mu}
    {\dot X}_{\mu})^{1/2} .
\lbl{III1.16}
\ee
The result above was obtained by writing\newline
 $${\dd}^m {\hat x} = {\dd}^{m-1} {\hat x} \, \dd \tau, \qquad
  \sqrt{|{\hat f}|} =\sqrt{|{\hat f}^{(m-1)}|}
({\dot X}^{\mu} {\dot X}_{\mu})^{1/2}$$ 
and taking the coordinates $\eta^a$
such that $\eta^a = (x^{\mu}, \eta^n, \eta^{n+1},...,\eta^{N-1})$, where
$x^{\mu}$ are (curved) coordinates on $V_n$. The determinant of the metric
of the embedding space $V_N$ in such a curvilinear coordinates is then
$\gamma = {\rm det} \, \p_{\mu} \eta^a \p_{\nu} \eta_a = f$.

In general, for arbitrary intersection we have
\be
    \int {\dd}^m {\hat x} \sqrt{|{\hat f}|}
    \, {{\delta^N (\eta - {\hat \eta})}\oo {\sqrt{|\gamma|}}} =
    \int {\dd}^{p+1} \xi \, {{\delta^n (x - X{\xi})}\oo {\sqrt{|f|}}} 
    \, ({\rm det} \, 
    \p_A X^{\mu} \p_B X_{\mu})^{1/2} ,
\lbl{III1.17}
\ee
where $X^{\mu} (\xi^A)$, $\mu = 0,1,2,..., n-1$, $A = 1,2,...,p$, are the
embedding functions of the $p$-brane's worlsdsheet $V_{p+1}$ in $V_n$.

Using (\ref{III1.16}) the equations of motion become
\be
      \p_{\mu} \left [ \sqrt{|f|} (\omega_0 f^{\mu \nu} + T^{\mu \nu}) \p_{\nu}
      \eta_a \right ] = 0 ,
\lbl{III1.18}
\ee
\be
      {\hat \p}_{\hat \mu} \left [ \sqrt{|{\hat f}|} (\omega_0 
      {\hat f}^{{\hat \mu} {\hat \nu}} + 
      {\widehat T}^{{\hat \mu} {\hat \nu}}) {\hat \p}_{\hat \nu}
      {\hat \eta}_a \right ] = 0 ,
\lbl{III1.19}
\ee
where
\be
      T^{\mu \nu} = \int {\kappa \oo {\sqrt{|f|}}} \, \delta^n (x - X(\tau))
      \, {{{\dot X}^{\mu} {\dot X}^{\nu}}\oo {({\dot X}^{\alpha} {\dot X}_{
      \alpha})^{1/2}}} \, \dd \tau
\lbl{III1.20}
\ee
and
\be
      \widehat T^{{\hat \mu} {\hat \nu}} = \int {\kappa \oo {\sqrt{|{\hat f}|}}}
       \, \delta^n ({\hat x} - {\hat X}(\tau))       
     \, {{{\dot {\hat X}}^{\hat \mu} {\dot {\hat X}}^{\hat \nu}}
      \oo {({\dot {\hat X}}^{\hat \alpha} {\dot {\hat X}}_{\hat 
      \alpha})^{1/2}}} \, \dd \tau
\lbl{III1.20a}
\ee
are the stress--energy tensors of the point particle on $V_n$ and ${\widehat V}_m$,
respectively. \inxx{stress--energy tensor,of the point particle}

If dimensions $m$ and $n$ are such that the intersection $V_{p+1}$ is
a worldsheet with a dimension $p\geq 1$, then using (\ref{III1.17}) we obtain
the equations of motion of the same form (\ref{III1.18}),(\ref{III1.19}),
but with the stress--energy tensor
\inxx{stress--energy tensor,of an extended object}
\be
      T^{\mu \nu} = \int {\kappa \oo {\sqrt{|f|}}} \, \delta^n (x - X(\xi)) \, 
      \p_A X^{\mu} \p^A X^{\nu} \, ({\rm det} \, 
      \p_C X^{\alpha} \p_D X_{\alpha})^{1/2} \, {\dd}^{p+1} \xi ,
\lbl{III1.21}
\ee
\be
      {\widehat T}^{{\hat \mu} {\hat \nu}} = \int {\kappa \oo {\sqrt{|{\hat f}|}}}
       \, \delta^n ({\hat x} - {\hat X}(\xi)) \, 
      \p_A {\hat X}^{\hat \mu} \p^A {\hat X}^{\hat \nu} \, ({\rm det} \, 
      \p_C {\hat X}^{\hat \alpha} \p_D {\hat X}_{\hat \alpha})^{1/2}
      \, {\dd}^{p+1} \xi .
\lbl{III1.22}
\ee

This can also be seen directly from the action (\ref{III1.9}) in which we
substitute eq.\,(\ref{III1.16})
\be
    I[\eta^a,X^{\mu}]
     = \omega_0 \int {\dd}^n x \, \sqrt{|f|} + \kappa \int {\dd}^n x \, \dd
    \tau \, \delta^n (x - X(\tau)) (f_{\mu \nu} {\dot X}^{\mu} {\dot X}^{\nu}
    )^{1/2}
\lbl{III1.23}
\ee
or if we substitute (\ref{III1.17})
\bear
    I[\eta^a,X^{\mu}]
     &=& \omega_0 \int {\dd}^n x \, \sqrt{|f|} \lbl{III1.24} \\
     & & + \, \kappa \int {\dd}^n x \, 
    {\dd}^{p+1} \xi \, \delta^n (x - X(\xi)) \, ({\rm \det}\, \p_A X^{\mu} \p_B
    X^{\nu} f_{\mu \nu})^{1/2} . \nonumber 
\ear
Remembering that
\be
      f_{\mu \nu} = \p_{\mu} \eta^a \p_{\nu} \eta^b \eta_{ab}
\lbl{III1.25}
\ee
we can vary (\ref{III1.23}) or (\ref{III1.24}) with respect to $\eta^a (x)$
and we obtain (\ref{III1.18}).

Eq.(\ref{III1.18}) can be written as
\be
   \omega_0 {\DD}_{\mu} \DD^{\mu} \eta_a + {\DD}_{\mu}(T^{\mu \nu} \p_{\nu}
   \eta_a) = 0 .
\lbl{III1.26}
\ee
where ${\DD}_{\mu}$ denotes covariant derivative in $V_n$.
If we multiply the latter equation by $\p^{\alpha} \eta^a$, sum over $a$, and
take into account the identity
\be
     \p^{\alpha} \eta^a {\DD}_{\mu} {\DD}_{\nu} \eta_a = 0 ,
\lbl{III1.27}
\ee
which follows from ${\DD}_{\mu} (\p_{\rho} \eta^a \p_{\sigma} \eta_a) =
{\DD}_{\mu} f_{\rho \sigma} = 0$, we obtain
\be
       {\DD}_{\mu} T^{\mu \nu} = 0 .
\lbl{III1.28}
\ee

Equation (\ref{III1.28}) implies that $X^{\mu} (\tau)$ is a geodesic equation in
a space with metric $f_{\mu \nu}$, i.e., $X^{\mu} (\tau)$ is a geodesic on
\inxx{geodesic equation}
$V_n$. This can be easily shown by using the relation
\be
      {\DD} T^{\mu \nu} = {1\oo {\sqrt{|f|}}} \, \p_{\mu} (\sqrt{|f|} \,
      T^{\mu \nu}) + \Gamma_{\rho \mu}^{\nu} T^{\rho \mu} = 0 .
\lbl{III1.29}
\ee
Taking (\ref{III1.20}) we have
      $$\int \dd \tau {\p \oo {\p x^{\mu}}} \delta^n (x - X(\tau))
      {{{\dot X}^{\mu} {\dot X}^{\nu}}\oo {({\dot X}^{\alpha} {\dot X}_{
      \alpha})^{1/2}}} {\dd}^n x \hs{4cm}$$
\be
       \hs{3cm} + \, \Gamma_{\rho \mu}^{\nu} \int \dd \tau
       \delta^n (x - X(\tau)) 
       {{{\dot X}^{\mu} {\dot X}^{\nu}}\oo {({\dot X}^{\alpha} {\dot X}_{
      \alpha})^{1/2}}} \dd^n x = 0 .
\lbl{III1.30}
\ee
The first term in the latter equation gives
\bear
   & & - \int \dd \tau \, {\p \oo {\p X^{\mu} (\tau)}} \, \delta^n (x - X(\tau)) 
   \, {{{\dot X}^{\mu} {\dot X}^{\nu}}\oo {({\dot X}^{\alpha} {\dot X}_{
      \alpha})^{1/2}}}  \, {\dd}^n x \hs{2cm} \nonumber \\
\nonumber \\      
    & & \hs{1.5cm}= - \int \dd \tau \, {\dd \oo {\dd \tau}}
      \delta^n (x - X(\tau)) \, {{{\dot X}^{\nu}}\oo {({\dot X}^{\alpha} 
      {\dot X}_{
      \alpha})^{1/2}}} {\dd}^n x \nonumber \\
\nonumber \\
     & & \hs{1.5cm} = \int \dd \tau \, {\dd \oo {\dd \tau}}
      \left ( {{{\dot X}^{\nu}}\oo {({\dot X}^{\alpha} {\dot X}_{
      \alpha})^{1/2}}} \right ) .
\lbl{III1.31}
\ear

Differentiating eq.\,(\ref{III1.30}) with respect to $\tau$ we indeed obtain
the geodesic equation.

In a similar way we find for $T^{\mu \nu}$, as given in eq.\,(\ref{III1.21}),
 that (\ref{III1.29}) implies
\be
    {1\oo {\sqrt{|{\rm det} \p_C X^{\alpha} \p_D X_{\alpha}}| }} \p_A (
    \sqrt{|{\rm det} \p_C X^{\alpha} \p_D X_{\alpha}}| \,  \p^A X^{\nu}) +
    \Gamma_{\rho \mu}^{\nu} \p_A X^{\rho} \p^A X^{\mu} = 0,
\lbl{III1.32}
\ee
which is the equation of motion for a $p$-brane in a background metric
\inxx{equation of motion,for a $p$-brane}
$f_{\mu \nu} = \p_{\mu} \eta^a \p_{\nu} \eta_a$. Do not forget that the latter
$p$-brane is the intersection between two branes:
\be
       V_{p+1} = V_n \cap {\widehat V}_m \; .
\lbl{III1.33}
\ee

It is instructive to integrate (\ref{III1.26}) over ${\dd}^n x$. We find
\bear
    &&\omega_0 \oint \sqrt{|f|} \, \dd \Sigma_{\mu} \, \p^{\mu} \eta_a 
    \lbl{III1.33a} \\
    && \hs{.5cm} = \, - \kappa \int {\dd}^{p+1} \xi (|{\rm det}\,  
     \p_C X^{\alpha} \p_D X_{\alpha}|)^{1/2} \, \p_A X^{\mu} \p^A X^{\nu} \, 
     {\DD}_{\mu} {\DD}_{\nu} \eta_a\biggl\vert_{x=X(\xi)} \nonumber
\ear
where  $\dd \Sigma_{\mu}$ is an element of an $(n-1)$-dimensional hypersurface
$\Sigma$ on $V_n$. Assuming that the integral over the time-like part of
$\Sigma$ vanishes (either because $\p^{\mu} \eta_a \rightarrow 0$ at the
infinity or because $V_n$ is closed) we have    
\bear   
    &&\omega_0 \int \sqrt{|f|} \, \dd \Sigma_{\mu} \, \p^{\mu} \eta_a 
    \Biggl\vert_{\tau_2} - \, \omega_0 \int \sqrt{|f|} \, \dd 
    \Sigma_{\mu} \, \p^{\mu} \eta_a \Biggl\vert_{\tau_1} \hs{4.5cm}\lbl{III1.33b} \\
    \nonumber \\
   &&\hs{0.5cm} = - \kappa \int \dd \tau \, {\dd}^p \xi (|{\rm det} \, 
    \p_C X^{\alpha} \p_D X_{\alpha}|)^{1/2} \, \p_A X^{\mu} \p^A X^{\nu} \,
    {\DD}_{\mu} {\DD}_{\nu} \eta_a\biggl\vert_{x=X(\xi)} \nonumber
\ear
or
\be
      {{\dd P_a}\oo {\dd \tau}} = - \kappa  \int {\dd}^p \xi (|{\rm det} \, 
    \p_C X^{\alpha} \p_D X_{\alpha}|)^{1/2} \, \p_A X^{\mu} \p^A X^{\nu} \,
    {\DD}_{\mu} {\DD}_{\nu} \eta_a\biggl\vert_{x=X(\xi)}
\lbl{III1.33c}
\ee
where
\be
     P_a \equiv  \omega_0 \int \sqrt{|f|} \, \dd \Sigma_{\mu} \, 
     \p^{\mu} \eta_a \; .
\lbl{III1.33d}
\ee
When $p = 0$, i.e., when the intersection is  a worldline, eq.\,(\ref{III1.33c})
reads
\be
    {{\dd P_a}\oo {\dd \tau}} = - \kappa  \, 
    {{{\dot X}^{\mu} {\dot X}^{\nu}}\oo {({\dot X}^{\alpha} {\dot X}_{
      \alpha})^{1/2}}} \, 
    {\DD}_{\mu} {\DD}_{\nu} \eta_a\biggl\vert_{x=X(\xi)} \; .
\lbl{III1.33e}
\ee
         
\section{A system of many intersecting branes}
\inxx{system,of many intersecting branes}

Suppose we have a sytem of branes of various dimensionalities.
They may intersect
and their intersections are the branes of lower dimensionality.
The action governing the dynamics of such a system is a generalization of 
(\ref{III1.10}) and consists of the free part plus the interactive part
($i, j = 1,2,...\, $):
\be
    I[\eta_i] = \sum_i \int \omega_0 \sqrt{|f_i|} \, \dd x_i + {1\oo 2} 
    \sum_{i \neq j} \int \omega_{i j} \, {{\delta^N (\eta_i - \eta_j)}\oo
    {\sqrt{|\gamma|}}} \sqrt{|f_i|} \sqrt{|f_j|}\, \dd x_i \dd x_j .
\lbl{III1.34}
\ee
The equations of motion for the $i$-th brane are
\be
    \p_{\mu} \left [\sqrt{|f_i|} \p^{\mu} \eta_i^a \left ( \omega_0 +  
    \sum_{i \neq j} \int \omega_{i j} \, {{\delta^N (\eta_i - \eta_j)}\oo
    {\sqrt{|\gamma|}}} \,  \sqrt{|f_j|} \dd x_j \right ) \right ] = 0 .
\lbl{III1.35}
\ee
Neglecting the kinetic term for all other branes the action leading to 
(\ref{III1.35}) is (for a fixed $i$)
\be
     I[\eta_i] = \int \omega_0 \sqrt{|f_i|} \, \dd x_i + 
     \sum_{i \neq j} \int \omega_{i j} \, {{\delta^N (\eta_i - \eta_j)}\oo
    {\sqrt{|\gamma|}}} \sqrt{|f_i|} \sqrt{|f_j|}\, \dd x_i \dd x_j
\lbl{III1.36}
\ee
or
\be
      I[\eta_i] = \int \omega_i (\eta) \sqrt{|f_i|} \, \dd x_i \; ,
\lbl{III1.37}
\ee
with
\be
     \omega_i (\eta) = \omega_0 + \sum_j \kappa_j {{\delta^N (\eta - \eta_j)}
     \oo {\sqrt{|\gamma|}}} \, \sqrt{|f_j|} \dd x_j \; ,
\lbl{III1.37a}
\ee
where $\kappa_j \equiv \omega_{i j}$.

\vs{1mm}

Returning now to eqs.\,(\ref{III1.4})--(\ref{III1.7}) we see that 
$\omega_i (\eta)$ is related to the conformally flat background metric as
experienced by the $i$-th brane. The action (\ref{III1.37}) is thus the
action for a brane in a background metric $\gamma_{a b}$, which is
conformally flat:
\be
     I[\eta_i] = \int \sqrt{|{\tilde f}|} \, \dd x_i .
\lbl{III1.38}
\ee
Hence the interactive term in (\ref{III1.34}) can be interpreted as a 
\inxx{interactive term,for a system of branes}
contribution to the background metric in which the $i$-th brane moves.
Without the interactive term the metric is simply a flat metric (multiplied
by $\omega$); with the interactive term the background metric is
singular on all the branes within our system. \inxx{background metric}

\vs{1mm}

The total action (\ref{III1.34}), which contains the kinetic terms for all
the other branes, renders the metric of the embedding space $V_N$ dynamical.
The way in which other branes move depends on the dynamics of the whole system.
It may happen that for a system of many branes, densely packed together,
\inxx{effective metric} \inxx{system,of many branes}
the effective (average) metric could no longer be conformally flat. We have
already seen in Sec 6.2 that the effective metric for a system of generalized
branes (which I call {\it membranes}) indeed satisfies the Einstein equations.

\vs{1mm}

Returning now to the action (\ref{III1.36}) as experienced by one of the
branes whose worldsheet $V_n$ is represented by $\eta_i^a (x_i) \equiv 
\eta^a (x)$ we find, after integrating out $x_j$, $j \neq i$, that
\newpage
\bear
    I[\eta^a,X^{\mu}] &=& \omega_0 \int \dd^n x \, \sqrt{|f|} \hs{7cm}
    \lbl{III1.39}  \\   
    && + \, \sum_j \kappa_j \int
    \dd^n x \, {\dd}^{p_j + 1} \xi \, ({\rm det} \p_A X_j^{\mu} \p_B X_j^{\nu}
    f_{\mu \nu} )^{1/2} \delta^n (x - X_j (\xi)) . \nonumber
\ear
For various $p_j$ the latter expression is an action for a system of point
particles $(p_j = 0)$, strings $(p_j = 1)$, and higher-dimensional branes
$(p_j = 2,3,...\,)$ moving in the background metric $f_{\mu \nu}$, which is
the induced metric on our brane $V_n$ represented by $\eta^a (x)$. Variation
of (\ref{III1.39}) with respect to $X_k^{\mu}$ gives the equations of motion 
(\ref{III1.32}) for a $p$-brane with $p = p_k$. Variation of (\ref{III1.39})
with respect to $\eta^a (x)$ gives the equations of motion (\ref{III1.18}
for the $(n-1)$-brane. If we vary (\ref{III1.39}) with respect to $\eta^a (x)$
then we obtain the equation of motion ({III1.18}) for an $(n-1)$-brane.
The action (\ref{III1.39}) thus describes the dynamics
of the $(n-1)$-brane (world sheet $V_n$) and the dynamics of the $p$-branes
living on $V_n$.

We see that the interactive term in (\ref{III1.34}) manifests itself in
various ways, depending on how we look at it. It is a manifestation of
the metric of the embedding space being curved (in particular, 
the metric is singular on the system of branes). From the point of view of
a chosen brane $V_n$ the interactive term becomes the action for a system of
$p$-branes (including point particles) moving on $V_n$. If we now adopt
the {\it brane world view}, where $V_n$ is our spacetime, we see that {\it
matter on $V_n$ comes from other branes' worldsheets which happen to intersect 
our worldsheet $V_n$.} Those other branes are responsible for the non trivial
metric of the embedding space, also called the {\it bulk}.

\subsection[\ \ The brane interacting with itself]
{The brane interacting with itself}

In (\ref{III1.36}) or (\ref{III1.39}) we have a description of a brane
interacting with other branes. What about self-interaction? In the second term
\inxx{self-interaction}
of the action (\ref{III1.34})\, (\ref{III1.36}) we have excluded self-interaction.
In principle we should not exclude self-interaction, since there is no reason why
a brane could not interact with itself.

Let us return to the action (\ref{III1.9}) and let us calculate 
$\omega (\eta)$,\
this time assuming that ${\widehat V}_m$ coincides with our brane $V_n$. Hence
the intersection is the brane $V_n$ itself, and according to (\ref{III1.17})
we have
\bear    
    \omega (\eta) &=& \omega_o + \kappa \int {\dd}^n {\hat x} \,
    \sqrt{|{\hat f}|}
    \, {{\delta^N (\eta - {\hat \eta} ({\hat x})}\oo {\sqrt{|\gamma|}}}
    \nonumber \\
     &=&
    \omega_0 + \kappa \int {\dd}^n \xi \, {{\delta^n (x - X(\xi))}\oo 
    {\sqrt{|f|}}} \, \sqrt{|{\hat f}|} \nonumber \\
   &=& \, \omega_0 + \kappa \int {\dd}^n x \, \delta^n (x - X(x)) =
      \omega_0 + \kappa .
\lbl{III1.39b}
\ear  
Here the coordinates $\xi^A$, $A = 0,1,2,..., n-1$, cover the manifold $V_n$,
and ${\hat f}_{AB}$ is the metric of $V_n$ in coordinates $\xi^A$. The other
coordinates are $x^{\mu}$, $\mu = 0,1,2,...,n-1$. In the last step in
(\ref{III1.39b}) we have used the property that the measure is invariant,
${\dd}^n \xi \sqrt{|{\hat f}|} = {\dd}^n x \sqrt{|f|}$.

The result (\ref{III1.39b}) demonstrates that we do not need to separate a
constant term $\omega_0$ from the function $\omega (\eta)$. For a brane moving
\inxx{background of many branes}
in a background of many branes we can replace (\ref{III1.37a}) with
\be
      \omega (\eta) = \sum_j \kappa_j {{\delta^N (\eta - \eta_j)}
     \oo {\sqrt{|\gamma|}}} \, \sqrt{|f_j|} \dd x_j ,
\lbl{III1.39c}
\ee
where $j$ runs over {\it all} the branes within the system. Any brane feels
the same background, and its action for a fixed $i$ is
\be
   I[\eta_i] = \int \omega (\eta_i) \sqrt{|f|} \dd x = 
   \sum_j \int \kappa_j {{\delta^N (\eta_i - \eta_j)}\oo {\sqrt{|\gamma|}}}
   \, \sqrt{|f_i|} \sqrt{|f_j|} \dd x_i \, \dd x_j .
\lbl{III1.39d}
\ee
However the background is self-consistent: it is a solution to the 
variational principle given by the action
\be
    I[\eta_i] = \sum_{i \ge j} \omega_{ij} \, \delta^N (\eta_i - \eta_j) 
    \, \sqrt{|f_i|} \sqrt{|f_j|} \dd x_i \, \dd x_j ,
\lbl{III1.39e}
\ee
where now also $i$ runs over {\it all} the branes within the system;
the case $i = j$ is also allowed.

In (\ref{III1.39e}) the self-interaction or self coupling occurs whenever
$i = j$. The self coupling term of the action is
\begin{eqnarray}
      I_{\rm self} [\eta_i] &=& \sum_i \kappa_i \int \delta^N (\eta_i (x_i) -
      \eta_i (x'_i)) \sqrt{|f_i (x_i)|} \sqrt{|f_i (x'_i)|} \dd x_i \dd x'_i
      \nonumber \\
      &=& \sum_i \kappa_i \int \delta^N (\eta - \eta_i (x_i)) 
      \delta^N (\eta - \eta_i (x'_i)) \nonumber \\
      & &\hs{1.4cm} \times \sqrt{|f_i (x_i)|} \sqrt{|f_i (x'_i)|} 
      \dd x_i \dd x'_i {\dd}^N \eta \nonumber \\
      \nonumber \\
      &=& \sum_i \kappa_i \int \delta^N (\eta - \eta_i (x_i)) \delta^{n_i}
      (x'_i - x''_i) \sqrt{|f_i (x_i)|} \dd x_i \dd x'_i {\dd}^N \eta
      \nonumber \\
      &=& \sum_i \kappa_i \, \sqrt{|f_i (x_i)|} {\dd}^{n_i} x_i \; ,
\lbl{III1.39f}
\ear
where we have used the same procedure which led us to eq.\,(\ref{III1.17}) or
(\ref{III1.39b}). We see that the interactive action (\ref{III1.39e})
automatically contains the minimal surface terms as well, so they do
not need to be postulated separately.

\subsection[\  \ A system of many branes creates the bulk and its metric]
{A system of many branes creates the bulk and its metric}
\inxx{system,of many branes}

We have seen several times in this book (Chapters 5,\,6) that a system of
membranes (a membrane configuration) can be identified with the embedding space
in which a single membrane moves. Here we have a concrete realization of that 
idea. We have a system of branes which intersect. The only interaction between
the branes is owed to intersection (`contact' interaction). The interaction at
\inxx{contact interaction} \inxx{test brane}
the intersection influences the motion of a (test) brane: it feels a potential
because of the presence of other branes. If there are many branes and a test
brane moves in the midst of them, then on average it feels a metric field
which is approximately continuous. Our test brane moves in the effective metric
of the embedding space. \inxx{effective metric}

A single brane or several branes give the singular conformal metric. Many 
branes give, on average, an arbitrary metric.

There is a close inter-relationship between the presence of branes and
the bulk metric. In the model we discuss here the bulk metric is singular on
the branes, and zero elsewhere. Without the branes there is no metric and no
bulk. Actually the bulk consists of the branes which determine
its metric.

\vs{4mm}

\figIIIB

\section{The origin of matter in the brane world}

Our principal idea is that\inxx{embedding space} we have a 
system of branes \inxx{bulk}(a brane configuration).
With all the branes in the system we associate\inxx{brane configuration}
the embedding space (bulk). One of the branes (more precisely, its worldsheet)
represents our spacetime. Interactions between the branes (occurring at the
intersections) represent {\it matter} in spacetime.\

\subsection[\ \ Matter from the intersection of our brane with other branes]
{Matter from the intersection of our brane with other branes}

We have seen that matter in $V_n$ naturally occurs as a result of the
intersection of our worldsheet $V_n$ with other worldsheets. We obtain
exactly the stress--energy tensor for a dust of point particles, or $p$-branes
\inxx{stress--energy tensor,for dust of point particles}
\inxx{stress--energy tensor,for dust of $p$-branes}
in general. Namely, varying the action (\ref{III1.39}) with respect to
$\eta^a (x)$ we obtain
\be
    \omega_0 {\DD}_{\mu} {\DD}^{\mu} \eta_a + {\DD}_{\mu} (T^{\mu \nu} \p_{\nu}
    \eta_a) = 0 ,
\lbl{III1.40}
\ee
with
\be
    T^{\mu \nu} = \sum_j \kappa_j \int {\dd}^{p_j +1} \xi \, ({\rm det} \, \p_A
    X_j^{\mu} \p_B X_j^{\nu} \, f_{\mu \nu})^{1/2}
     \, {{\delta^n (x - X_j(\xi))}\oo {\sqrt{|f|}}}
\lbl{III1.41}
\ee
being the stress--energy tensor for a system of $p$-branes (which are the
intersections of $V_n$ with the other worldsheets). The above expression
for $T^{\mu \nu}$ holds if the extended objects have any dimensions $p_j$.
In particular, when {\it all} objects have $p_j = 0$ (point particles)
eq.\,(\ref{III1.41}) becomes
\be
     T^{\mu \nu} = \sum_j \kappa_j \int \dd \tau \, {{{\dot X}^{\mu} {\dot X}^{
     \nu}}\oo {\sqrt{{\dot X}^2}}} \, {{\delta (x - X(\tau))}\oo {\sqrt{|f|}}} .
\lbl{III1.42}
\ee
From the equations of motion (\ref{III1.41}) we obtain 
(see eqs.\,(\ref{III1.26})--(\ref{III1.28}))
\be
    {\DD}_{\mu} T^{\mu \nu} = 0 ,
\lbl{III1.42a}
\ee
which implies (see (\ref{III1.29})--(\ref{III1.32})) that any of the objects
follows a geodesic in $V_n$. \inxx{geodesic} \inxx{self-intersecting brane}
                 
\subsection[\ \ Matter from the intersection of our brane with itself]
{Matter from the intersection of our brane with itself}

Our model of intersecting branes allows for the possibility that a brane
intersects with itself, as is schematically illustrated in Fig. 8.3.
The analysis used so far is also valid for situations like that
in Fig. 8.3, if we divide the worldsheet $V_n$ into two pieces which are glued
together at a submanifold $C$ (see Fig. 8.4).

\figIIIC

\figIIID

There is a variety of ways a worldsheet can self-intersect. Some of them are
sketched in Fig. 8.5.

\vs{5mm}

\figIIIE

In this respect some interesting new possibilities occur, waiting to be
explored in detail. For instance, it is difficult to imagine how the
three particles entangled in the topology of the situation (a) in Fig. 8.5
could be separated to become asymptotically free. Hence this might be
a possible classical model for hadrons composed of quarks; the extra dimensions
of $V_n$ would bring, via the Kaluza--Klein mechanism, the chromodynamic force
into the action.\inxx{matter configurations,on the brane}

To sum up, it is obvious that a self-intersecting 
brane can provide a variety\inxx{self-intersecting brane}
of matter configurations on the brane. This is a fascinating and
intuitively clear mechanism for the origin of matter in a brane world.
\newpage

\section{Comparison with the Randall--Sundrum model}
\inxx{Randall--Sundrum model}

In our brane world model, which starts from the ${\cal M}$-space Einstein
equations, we have assumed that gravity is localized on the brane. This was
formally represented by the $\delta$-function. In a more conventional
approach the starting point is Einstein's equations in the ordinary
space, not in ${\cal M}$-space. Let us therefore explore a little what
such an approach has to say about gravity around a brane embedded in a ``bulk".
\inxx{gravity around a brane}

Randall and Sundrum \ci{72} have considered a model in which a 3-brane with tension
$\kappa$ is coupled to gravity, the cosmological constant $\Lambda$ being
different from zero. After solving the Einstein equations they found that the
metric tensor decreases exponentially with the distance from the brane.
Hence gravity is localized on the brane.

More precisely, the starting point is the action
\be
     I = \kappa \int {\dd}^n x \, \sqrt{|f|} \, \delta^N (\eta - \eta (x))
     \, {\dd}^N \eta + {1\oo {16 \pi G^{(N)}}} \int {\dd}^N \eta \sqrt{|f|} (
     2 \Lambda + R),
\lbl{III1.43}
\ee
which gives the Einstein equations
\be
    G_{ab} \equiv R_{ab} - \mbox{$1\oo 2$} R \, \gamma_{ab}
     = - \Lambda \gamma_{ab} - 8 \pi G^{(N)} T_{ab} \; ,
\lbl{III1.44}
\ee
\be
    T_{ab} = \int \kappa \, {\dd}^n x \, \sqrt{|f|} \, f^{\mu \nu} \, \p_{\mu}
    \eta_a \p_{\nu} \eta_b \, \delta^N (\eta - \eta (x)) .
\lbl{III1.45}
\ee

Let us consider a 3-brane ($n=4$) embedded in a 5-dimensional bulk ($N=5$).
In a particular gauge the worldsheet embedding functions are $\eta^{\mu} =
x^{\mu}$, $\eta^5 = \eta^5 (x^{\mu})$. For a flat worldsheet $\eta^5 (x^{\mu})
= y_0$, where $y_0$ is independent of $x^{\mu}$, it is convenient to take
$y_0 = 0$. For such a brane located at $\eta^5 \equiv y = 0$ the appropriate
Ansatz for the bulk metric respecting the symmetry of the brane configuration
is
\be
    \dd s^2 = a^2 (y) \eta_{\mu \nu} \, \dd x^{\mu} \, \dd x^{\nu} - \dd y^2 .
\lbl{III1.46}
\ee
The Einstein equations read
\bear
     {G^0}_0 &=& {G^1}_1 = {G^2}_2 = {G^3}_3 \nonumber \\
      &=& {{3 a''}\oo a} + {{3 a'^2}\oo 
     {a^2}} = - \Lambda - 8 \pi G^{(N)} {T^0}_0 \; , \lbl{III1.47}\\
     {G^5}_5 &=& {{6 a'^2}\oo {a^2}} = - \Lambda \; ,\lbl{III1.48}
\ear
where
\be
      T_{\alpha \beta} = \int \kappa \, {\dd}^4 x \, \sqrt{|f|} 
      f_{\alpha \beta} \, \delta^4 (\eta^{\mu} - x^{\mu}) \delta (y)
      = \kappa \sqrt{|f|} f_{\alpha \beta} \delta (y) ,
\lbl{III1.49}
\ee
whilst $T_{\alpha 5} = 0$, $T_{55} = 0$. The induced metric is
$$f_{\alpha \beta} = \p_{\alpha} \eta^a \p_{\beta} \eta_a = \eta_{\alpha
\beta} \, a^2 (y).$$
Hence
\be
     {T^0}_0 = {T^1}_1  = {T^2}_2 = {T^3}_3 = \kappa\, a^4 \delta (y) .
\lbl{III1.49a}
\ee
From eq.\,(\ref{III1.48}), which can be easily integrated, we obtain
\be
        a = a_0 \, {\rm e}^{- |y| \sqrt{-\Lambda/6}} .
\lbl{III1.50}
\ee
Such a solution makes sense if $\Lambda < 0$ and it respects the symmetry
$a(y) = a(-y)$, so that the bulk metric is the same on both sides of the
brane.

Introducing $\alpha' = a'/a$ (where $a' \equiv \dd a/\dd y$) eq. (\ref{III1.47})
can be written as
\be
    3 \alpha'' = - 8 \pi G^{(N)} \kappa \, a^4 \delta (y) .
\lbl{III1.51}
\ee
Integrating both sides of the latter equation over $y$ we find
\be
    3 (\alpha' (0^+) - \alpha' (0^-)) = - 8 \pi G^{(N)} \, \kappa a^4 (0) .
\lbl{III1.52}
\ee
Using (\ref{III1.50}) we have
     $${{a'(0^+)}\oo a} = \alpha' (0^+) = - \sqrt{ {{- \Lambda} \oo 6}} \; , $$
\be
     {{a'(0^-)}\oo a} = \alpha' (0^-) =  \sqrt{ {{- \Lambda} \oo 6}} \; ,
\lbl{III1.53}
\ee
    $$a(0) = a_0 = 1.$$
Hence (\ref{III1.52}) gives
\be
     6 \sqrt{ {{- \Lambda} \oo 6}  } = 8 \pi G^{(N)} \kappa 
\lbl{III1.54}
\ee
which is a relation between the cosmological constant $\Lambda$ and the brane
tension $\kappa$.

From (\ref{III1.46}) and (\ref{III1.50}) it is clear that the metric tensor
is localized on the brane's worldsheet and falls quickly when the transverse
coordinates $y$ goes off the brane.

\paragraph{An alternative Ansatz}

We shall now consider an alternative Ansatz in which the metric is
conformally flat:
\be
     \dd s^2 = b^2 (z) (\eta_{\mu \nu} \dd x^{\mu} \dd x^{\nu} - \dd z^2) .
\lbl{III1.55}
\ee
The Einstein equations read
\be
     {G^0}_0 = {{3 b''}\oo {b^3}} = - \Lambda - 8 \pi G^{(N)} \kappa b^4 (z)
\lbl{III1.56}
\ee
\be
     {G^5}_5 = {{6 b'^2}\oo {b^4}} = - \Lambda
\lbl{III1.57}
\ee
The solution of (\ref{III1.57}) is
\be
     b = - {1 \oo {C +  \sqrt{ {{- \Lambda} \oo 6}} |z|}} .
\lbl{III1.58}
\ee
From (\ref{III1.56})\,(\ref{III1.57}) we have
\be
     3 \left ( {b'' \oo b^2} - {b'^2 \oo b^3} \right ) = 
     - 8 \pi G^{(N)} \kappa b^5 \delta (z) .
\lbl{III1.59}
\ee
Introducing $\beta' = b'/b^2$ the latter equation becomes
\be
      3 \beta'' =  - 8 \pi G^{(N)} \kappa b^5 \delta (z) . 
\lbl{III1.60}
\ee
After integrating over $z$ we have
\be
      3(\beta' (0^+) - \beta' (0^-)) = - 8 \pi G^{(N)} \kappa b^5 (0) ,
\lbl{III1.61}
\ee
where
\be
        \beta' (0^+) = C^{-4} \sqrt{ {{- \Lambda} \oo 6}}  \; , \quad
        \beta' (0^-) = - C^{-4} \sqrt{ {{- \Lambda} \oo 6}}  \; , \quad
         b(0) = - C^{-1} .
\lbl{III1.62}
\ee
Hence

\hs{1mm}

\be
        6 \sqrt{ {{- \Lambda} \oo 6}}  = 8 \pi G^{(N)} \kappa C^{-1} .
\lbl{III1.63}
\ee

\hs{1mm}

\nnnn If we take $C = 1$ then the last relation coincides with (\ref{III1.54}).
The metric in the Ansatz (\ref{III1.55}) is of course obtained from that
in (\ref{III1.46}) by a coordinate transformation.
\newpage

\subsection[\ \ The metric around a brane in a higher-dimensional bulk]
{The metric around a brane in a \\ higher-dimensional bulk}

It would be very interesting to explore what happens to the gravitational
field around a brane embedded in more than five dimensions. One could set
an appropriate Ansatz for the metric, rewrite the Einstein equations
and attempt to solve them. My aim is to find out whether in a space of
sufficiently high dimension the metric --- which is a solution to the 
Einstein equations --- can be approximated with the metric (\ref{III1.4}),
the conformal factor being localized on the brane. \inxx{conformal factor}

Let us therefore take the Ansatz
\be
    \gamma_{ab} = \Omega^2 {\bar \gamma}_{ab} \; .
\lbl{III1.64}
\ee
We then find
    $${R_a}^b = \Omega^{-2} {{\bar R}_a}^b + (N-2) \Omega^{-3} {\Omega_{;a}}^{;b}
    - 2 (N-2) \Omega^{-4} \Omega_{,a} \Omega^{,b} \hs{2.3cm}$$
\be
    \hs{2.5cm} + \, \Omega^{-3} {\delta_a}^b
    {\Omega_{;c}}^{;c} + (N-3) \Omega^{-4} {\delta_a}^b \Omega_{,c} \Omega^{,c}
    \; , \hs{1.3cm}
\lbl{III1.65}
\ee
\be
     R = \Omega^{-2} {\bar R} + 2 (N-1) \Omega^{-3} {\Omega_{;c}}^{;c} +
     (N-1)(N-4) \Omega^{-4} \Omega_{,c} \Omega^{,c} .
\lbl{III1.66}
\ee

Splitting the coordinates  according to
\be
     \eta^a = (x^{\mu}, y^{\bar \mu}) ,
\lbl{III1.67}
\ee
where $y^{\bar \mu}$ are the transverse coordinates and assuming that $\Omega$
depends on  $y^{\bar \mu}$ only, the Einstein equations become
\bear
      {G_{\mu}}^{\nu} &=& \Omega^{-2} {{\bar G}_{\mu}}^{\nu} + \Omega^{-3}
      {\delta_{\mu}}^{\nu} \left [ (2-N) {\Omega{;{\bar \mu}}}^{;{\bar \mu}}
      +(N-3) \Omega^{-1} \Omega{,{\bar \mu}} \Omega^{,{\bar \mu}} \right ]
      \nonumber \\
      \nonumber \\
      &=& - 8 \pi G^{(N)} {T_{\mu}}^{\nu} - \Lambda {\delta_{\mu}}^{\nu} \; ,
\lbl{III1.68}\\
\nonumber \\
{G_{\bar \mu}}^{\bar \nu} &=& \Omega^{-2} {{\bar G}_{\bar \mu}}^{\bar \nu}
  + \Omega^{-3} \Bigl[ (N-2) {\Omega_{; {\bar \mu}}}^{; {\bar \nu}} -
  2 (N-2) \Omega^{-1} \Omega_{, {\bar \mu}} \Omega^{, {\bar \nu}}\nonumber \\
  \nonumber \\
   & &+ \, {\delta_{\bar \mu}}^{\bar \nu} \left ( (2-N) {\Omega_{; {\bar \alpha}}}^{;
  {\bar \alpha}} + \Omega^{-1} \Omega_{,{\bar \alpha}} \Omega^{,{\bar
  \alpha}}\left ( (N-3) + (N-1)(N-4) \right ) \right ) \Bigr] \nonumber \\
  \nonumber \\
        &=&   - 8 \pi G^{(N)} {T_{\bar \mu}}^{\bar \nu} - 
        \Lambda {\delta_{\bar \mu}}^{\bar \nu} \, .
\lbl{III1.69}
\ear
Let ${T_a}^b$ be the stress--energy tensor of the brane itself. Then
${T_{\bar \mu}}^{\bar \nu} = 0$ (see eq.\,(\ref{III1.49}). Using (\ref{III1.69})
we can express $\Omega^{-1} \Omega_{,{\bar \alpha}} \Omega^{,{\bar \alpha}}$
in terms of ${\Omega_{; {\bar \alpha}}}^{; {\bar \alpha}}$ and insert it into
(\ref{III1.68}). Taking\footnote{
If dimension $N = 5$ then $\Lambda$ must be different from zero, otherwise
eq.\,(\ref{III1.69}) gives $\Omega_{,5} \Omega^{,5} = 0$, which is inconsistent
with eq.\,(\ref{III1.68}).}
$N>5$, $\Lambda = 0$ and assuming that close to the brane the term $\Omega^{-2}
{{\bar G}_a}^b$ can be neglected we obtain the Laplace equation
for $\Omega$
\be
    {\Omega_{;{\bar \mu}}}^{;{\bar \mu}} = 16 \pi G^{(N)} T \, \Omega^3 A
\lbl{III1.70}
\ee
where
\bear
   A &=& (N-2)(N-1)\lbl{III1.71} \\
   \nonumber \\
   & & \times \left [ 2 - (N-4) {{(N-2) + {\bar n} (2-N)}\oo
   {-2 (N-2) + {\bar n} (N-3) - {{\bar n}\oo 2} (N-1)(N-4)}} \right ] ,
\nonumber 
\ear
${\bar n}$ being the dimension of the transverse space, ${\bar n} = {\delta_{\bar
\mu}}^{\bar \mu}$, and $T \equiv {T_a}^a = {T_{\mu}}^{\mu}$.

The above procedure has to be taken with reserve. Neglect of the term       
$\Omega^{-2} {{\bar G}_a}^b$ in general is not expected to be consistent
with the Bianchi identities. Therefore equation (\ref{III1.71}) is merely
an approximation to the exact equation. Nevertheless it gives an
idea about the behavior of the function $\Omega( y^{\bar \mu})$.

The solution of eq.\,(\ref{III1.70}) has the form
\be
    \Omega = - {k \oo {r^{{\bar n} -2}}} ,
\lbl{III1.72}
\ee
where $r$ is the radial coordinate in the transverse space.
For a large transverse dimension ${\bar n}$ the function $\Omega$ falls
very quickly with $r$. The gravitational field around the brane is very strong
close to the brane, and negligible anywhere else. The interaction is
\inxx{contact interaction}
practically a {\it contact interaction} and can be approximated by the
$\delta$-function. Taking a cutoff $r_{\rm c}$ determined by the thickness
of the brane we can normalize $\Omega$ according to
\be
        \int_{r_{\rm c}}^\infty \Omega (r) \, \dd r = 
        {k\oo {({\bar n}-1) r_{\rm c}^{{\bar n}-1}}} = 1
\lbl{III1.73}
\ee
and take $\Omega (r) \approx \delta (r)$.

The analysis above is approximate and requires a more rigorous study. But
intuitively it is clear that in higher dimensions gravitational interaction
falls very quickly. For a {\it point particle} the gravitational potential
has the asymptotic behavior
 $\gamma_{00} - 1 \propto r^{-(N-3)}$, and for a
sufficiently high spacetime dimension $N$ the interaction is practically
contact (like the Van der Waals force). Particles then either do not feel
each other, or they form bound states upon contact. Network-like
configurations are expected to be formed, as shown in Fig.8.6. 
Such configurations
mimic very well the intersecting branes considered in Secs. 8.1--8.3.

\figIIIF

In this section we have started from the conventional theory of gravitation
and found strong arguments that in a space of very high dimension the
gravitational force is a contact force. Various {\it network-like 
configurations} are then possible and they are stable. Effectively there is
no gravity outside such a network configuration. Such a picture matches
very well the one we postulated in the previous three sections of this
chapter, and also the picture we considered when studying the ${\cal M}$-space
formulation of the membrane theory.

\chapter[The Einstein--Hilbert action on the brane as \\ the effective action]
{The Einstein--Hilbert action \\ 
on the brane \\ 
as the effective action}

After so many years of intensive research the quantization of gravity is
still an 
unfinished project. Amongst many approaches followed, there is the
one which seems to be especially promising.\inxx{quantization of gravity}
This is the so called induced\inxx{induced gravity}
gravity proposed by Sakharov \cite{73}. His idea was to treat the metric not
as a fundamental field but as one induced from more basic fields. The idea
has been pursued by numerous authors \cite{74}; especially illuminating are
works by Akama, Terazawa and Naka \cite{75}. Their basic action contains $N$
scalar fields and it is formally just a slight generalization of the
well-known Dirac--Nambu--Goto action for an $n$-dimensional world sheet swept
by an $(n-1)$-dimensional membrane.

Here we pursue such an approach and give a concrete physical 
interpretation of the $N$ scalar fields which we denote ${\eta}^a (x)$. We
assume that the spacetime is a surface $V_4$, called the spacetime sheet,
embedded in a higher-dimensional space $V_N$, and ${\eta}^a (x)$ are
the embedding functions. An embedding model has been
first proposed by Regge and Teitelboim \cite{76} and investigated by
\inxx{spacetime sheet}others \cite{77}. In that model the action 
contains the Ricci scalar
expressed in terms of the embedding functions. In our present 
model \cite{71,71a,71b}, on the contrary, we
start from an action which is essentially the minimal surface action\footnote{
Recently Bandos \ci{71c} has considered a string-like description of gravity
by considering bosonic $p$-branes coupled to an antisymmetric tensor field.}
weighted
with a function $\omega (\eta)$ in $V_N$. For a suitably chosen $\omega$,
such that it is singular ($\delta$-function like) on certain surfaces
${\widehat V}_m$, also embedded in $V_N$, we obtain on $V_4$ a set of world
lines. In the previous chapter it was shown that these worldlines are 
geodesics of $V_4$, provided
that $V_4$ described by ${\eta}^a (x)$ is a solution to our variational
procedure.  I will show that after performing functional integrations over
${\eta}^a (x)$ we obtain two contributions to the path integral. One
\inxx{path integral}\inxx{effective action}contribution comes from all 
possible ${\eta}^a (x)$ not intersecting 
${\widehat V}_m$, and the other from those ${\eta}^a (x)$ which
do intersect the surfaces ${\widehat V}_m$. In the effective action so obtained
the first contribution gives the Einstein--Hilbert term $R$ plus higher-order
terms like $R^2$. The second contribution can be cast into the form of
a path integral over all possible worldlines $X^{\mu} (\tau)$. Thus we
obtain an action which contains matter sources and a kinetic term for
the metric field (plus higher orders in $R$). So in the proposed approach
both the metric field and the matter field  are induced from more basic
fields ${\eta}^a (x)$.

\section{The classical model}

Let us briefly summarize the discussion of Chapter 8.
We assume that the arena where physics takes place is an $N$-dimensional
space $V_N$ with $N \geq 10$. Next we assume that an $n$-dimensional
surface $V_n$ living in $V_N$ represents a possible spacetime. The parametric
equation of such a `spacetime sheet' $V_n$ is given by the embedding
functions ${\eta}^a (x^{\mu})$, $a = 0,1,2,...,N$, where $x^{\mu}$,
$\mu = 0,1,2, \ldots ,n-1$, are coordinates (parameters) on $V_n$. We assume
that the action is just that for a minimal surface $V_n$
\inxx{minimal surface action}
\be
   I[{\eta}^a (x)] = \int (\mbox{det} \, {\p}_{\mu} {\eta}^a \, {\p}_{\nu}
   {\eta}^b
  {\gamma}_{ab})^{1/2} {\dd}^n x ,
\label{III2.1}
\ee
where ${\gamma}_{ab}$ is the metric tensor of $V_N$. The dimension of the
\inxx{spacetime sheet}
spacetime sheet $V_n$ is taken here to be arbitrary, in order to allow
for the Kaluza--Klein approach. In particular, we may take $n = 4$. We admit
that the embedding space is curved in general. In particular, let us consider
 the case of a conformally flat $V_N$, such that ${\gamma}_{ab} =
{\omega}^{2/n} {\eta}_{ab}$, where ${\eta}_{ab}$ is the $N$-dimensional
Minkowski tensor. Then Eq.\,(\ref{III2.1}) becomes
\be
   I[{\eta}^a (x)] = \int  \omega (\eta )
   (\mbox{det} \, {\p}_{\mu} {\eta}^a \, {\p}_{\nu} {\eta}^b
  {\eta}_{ab})^{1/2} {\dd}^n x .
\label{III2.2}
\ee
From now on we shall forget about the origin of $\omega (\eta )$ and
consider it as a function of position in a {\it flat} embedding space.
The indices $a, b, c$ will be raised and lowered by ${\eta}^{ab}$ and
${\eta}_{ab}$, respectively.

In principle $\omega (\eta )$ is arbitrary. But it is very instructive to
choose the following function:
\be
    \omega (\eta ) = {\omega}_0 + \sum_i \int m_i {{{\delta}^N (\eta -
   {\hat \eta}_i)}
    \over {\sqrt{|\gamma|}}} \, \, {\dd}^m {\hat x} \sqrt{|{\hat f}|} \; ,
\label{III2.3}
\ee
where ${\eta}^a = {\hat \eta}_i^a ({\hat x})$ is the parametric equation of
an $m$-dimensional surface ${\widehat V}_m^{(i)}$, called
the {\it matter sheet}, also \inxx{matter sheet}
embedded in $V_N$, ${\hat f}$ is the determinant of the induced metric
on ${\widehat V}_m^{(i)}$, and $\sqrt{|\gamma|}$ allows for taking curved
coordinates in otherwise flat $V_N$. If we take $m = N - n + 1$ then the
intersection of $V_n$ and ${\widehat V}_m^{(i)}$ can be a (one-dimensional)
line, i.e., a worldline $C_i$ on $V_n$. In general, when $m = N - n + (p+1)$
the intersection can be a $(p+1)$-dimensional world sheet representing
the motion of a $p$-dimensional membrane (also called $p$-brane). In this
chapter we confine our consideration to the case $p = 0$, that is, to
the motion of a point particle.

Inserting (\ref{III2.3}) into (\ref{III2.2}) and writing $f_{\mu \nu} \equiv
{\p}_{\mu} {\eta}^a {\p}_{\nu} {\eta}_a$,
$f \equiv \mbox{det} \, f_{\mu \nu}$
we obtain
\be
  I[\eta] = {\omega}_0 \, \int {\dd}^n x \, \sqrt{|f|}
  \ +  \int {\dd}^n x \sum_i m_i \,
  \delta^n (x - X_i) (f_{\mu \nu} {\dot X}_i^{\mu} {\dot X}_i^{\nu})^{1/2}
  \, {\dd}\tau.                                        \label{III2.4}
\ee
As already explained in Sec. 8.1,
the result above was obtained by writing 
$${\dd}^m {\hat x} = {\dd}^{m-1}
{\hat x} \, {\dd} \tau \; , \qquad
{\hat f} = {\hat f}^{(m-1)} ({\dot X}_i^{\mu}
{\dot X}_{i \mu})^{1/2}$$ 
and taking the coordinates ${\eta}^a$
such that ${\eta}^a = (x^{\mu}, {\eta}^n, ... , {\eta}^{N-1})$,
where $x^{\mu}$ are (curved) coordinates on $V_n$. The determinant of the
metric of the
embedding space $V_N$ in such curvilinear coordinates is then
$\gamma = \mbox{det} \, {\p}_{\mu} {\eta}^a \, {\p}_{\nu} {\eta}_a$ $= f$.

If we vary the action (\ref{III2.4}) with respect to ${\eta}^a (x)$ we obtain
\be
   {\p}_{\mu} \left [ \sqrt{|f|} \, ({\omega}_0 f^{\mu \nu} + T^{\mu \nu})
   {\p}_{\nu} {\eta}_a \right ] = 0 ,                        \label{III2.5}
\ee
where
\bearr
   T^{\mu \nu} = {1 \over {\sqrt{|f|}}} \, \sum_i \int {\dd}^n x \, m_i \,
   {\delta}^n (x {-} X_i ) \, {{{\dot X}_i^{\mu} {\dot X}_i^{\nu}}
   \over {({\dot X}_i^{\alpha} {\dot X}_{i \alpha} )^{1/2}}} \, {\dd} \tau
   \nnn \label{III2.6}
\ear
is the stress--energy tensor of dust. Eq.\,(\ref{III2.5}) can  be rewritten in
\inxx{stress--energy tensor,for dust}
terms of the covariant derivative $D_{\mu}$ on $V_n$:
\be
D_{\mu} \left[({\omega}_0 f^{\mu \nu} + T^{\mu \nu}) {\p}_{\nu} {\eta}_a
        \right ] = 0 .                                           \label{III2.7}
\ee
The latter equation gives
\be
   {\p}_{\nu} {\eta}_a \, D_{\mu} T^{\mu \nu} + ({\omega}_0 f^{\mu \nu} +
  T^{\mu \nu}) D_{\mu} D_{\nu} {\eta}_a = 0  ,                \label{III2.8}
\ee
where we have taken into account that a covariant derivative of metric
is zero,
i.e., $D_{\alpha} f_{\mu \nu} = 0$ and $D_{\alpha} f^{\mu \nu} = 0$, which
implies also ${\p}_{\alpha} \eta^c \, D_{\mu} D_{\nu} {\eta}_c = 0$, since
$f_{\mu \nu} \equiv$ ${\p}_{\mu} {\eta}^a {\p}_{\nu} {\eta}_a $. Contracting
Eq.(\ref{III2.8}) by ${\p}^{\alpha} {\eta}^a $ we have
\be
    D_{\mu} T^{\mu \nu} = 0   .                               \label{III2.9}
\ee
The latter are the well known equations of motion for the sources. In the
case of dust (\ref{III2.9}) implies that dust particles move along geodesics of
\inxx{geodesic,as the intersection of two branes}
the spacetime $V_n$. We have thus obtained the very interesting result that
the worldlines $C_i$ which are obtained as intersections $V_n \cap
{\widehat V}_m^{(i)}$ are geodesics of the spacetime sheet $V_n$. The same
result is also obtained directly by varying the action (\ref{III2.4}) with
respect to the variables $X_i^{\mu} (\tau)$.

A solution to the equations of motion (\ref{III2.5}) (or (\ref{III2.7})) gives both:
the spacetime sheet ${\eta}^a (x)$ and the worldlines $X_i^{\mu}
(\tau)$.  Once ${\eta}^a (x)$ is determined the induced metric $g_{\mu
\nu} = {\p}_{\mu} {\eta}^a {\p}_{\nu} {\eta}_a$ is determined as well. But
such a metric in general does not satisfy the Einstein equations. In the
next section we shall see that quantum effects induce the necessary
Einstein--Hilbert term $(-g)^{1/2} R$.

\section{The quantum model}

For the purpose of quantization we shall use the classical action \cite{71},
that is, a generalization of the well known Howe--Tucker action
\inxx{Howe--Tucker action}
\cite{21}, which is equivalent to (\ref{III2.2}) (see also Sec. 4.2):
\be
    I[{\eta}^a , g^{\mu \nu}] = \mbox{$1\oo 2$} \int {\dd}^n x
 \sqrt{|g|}\omega
  (\eta ) (g^{\mu \nu} {\p}_{\mu} {\eta}^a {\p}_{\nu} {\eta}_a + 2 - n) .
\label{III2.10}
\ee
It is a functional of the embedding functions ${\eta}^a (x)$ and the
Lagrange multipliers $g^{\mu \nu}$. Varying (\ref{III2.10}) with respect to
$g^{\mu \nu}$ gives the constraints
\be
  - {{\omega} \over 4} \, \sqrt{|g|} \, g_{\alpha \beta}
  (g^{\mu \nu} {\p}_{\mu} {\eta}^a {\p}_{\nu} {\eta}_a + 2 - n) 
   +  {{\omega}  \over 2} \, \sqrt{|g|} \,
  {\p}_{\alpha} {\eta}^a {\p}_{\beta} {\eta}_a = 0 .        \label{III2.11}
\ee
Contracting (\ref{III2.11}) with $g^{\alpha \beta}$, we find
$g^{\mu \nu} {\p}_{\mu} {\eta}^a {\p}_{\nu} {\eta}_a = n$, and after inserting
the latter relation back into (\ref{III2.11}) we find
\be
   g_{\alpha \beta} = {\p}_{\alpha} {\eta}^a {\p}_{\beta} {\eta}_a ,
   \label{III2.12}
\ee
which is the expression for an induced metric on the surface $V_n$. In the
following paragraphs we shall speci\-fy $n = 4$; however, whenever necessary,
we shall switch to the generic case of arbitrary $n$.

In the classical theory we may say that a 4-dimensi\-onal spacetime sheet is
swept by a 3-dimensional space-like hypersurface $\Sigma$ which moves
forward in time. The latter surface is specified by initial conditions, and
the equations of motion then determine $\Sigma$ at every value of a time-like
coordinate $x^0 = t$. Knowledge of a particular hypersurface $\Sigma$
implies knowledge of the corresponding intrinsic 3-geometry specified by
the 3-metric $g_{ij} = {\p}_i {\eta}^a {\p}_j {\eta}_a$ induced on $\Sigma$
$(i, j = 1,2,3)$. However, knowledge of the data ${\eta}^a (t, x^i)$ on
an entire infinite $\Sigma$ is just a mathematical idealization which cannot
be realized in a practical situation by an observer because of the finite
speed of light.

In quantum theory a state of a surface $\Sigma$ is not specified by
\inxx{wave functional}
the coordinates ${\eta}^a (t, x^i)$, but by a wave functional $\psi [t,
{\eta}^a (x^i)]$. The latter represents the probability amplitude that at
time $t$ an observer would obtain, as a result of measurement, a particular
surface $\Sigma$.

\vs{1mm}

The probability amplitude for the 
transition from a state with definite ${\Sigma}_1$\inxx{probability amplitude}
 at time $t_1$ to a state\inxx{Feynman} ${\Sigma}_2$ at 
 time $t_2$ is given by the Feynman
path integral\inxx{path integral}
\be
   K(2,1) = \langle {\Sigma}_2 , t_2 | {\Sigma}_1 , t_1 \rangle =
   \int \e^{i \, I[\eta , g]} {\cal D} \eta {\cal D} g .       \label{III2.13}
\ee
Now, if in Eq.(\ref{III2.13}) we perform integration only over the embedding
functions ${\eta}^a (x^{\mu})$, then we obtain the so called {\it effective
action} $I_{\rm eff}$ \inxx{effective action}
\be
    \e^{i I_{\rm eff} [g]} \equiv \int \e^{i \, I[\eta , g]}  {\cal D} \eta ,
\label{III2.14}
\ee
which is a functional of solely the metric $g^{\mu \nu}$.
From eq.\ (\ref{III2.14}) we obtain by functional differentiation
\be
  {{\delta I_{\rm eff} [g]} \over {\delta g^{\mu \nu}}} = 
  {{\int  {{\delta I_ [\eta,
   g]} \over {\delta g^{\mu \nu}}} \, \e^{i \, I[\eta , g]}  {\cal D} \eta }
   \over {\int \e^{i \, I[\eta , g]}  {\cal D} \eta }}
\equiv \left \langle {{\delta I [\eta, g]} \over {\delta g^{\mu \nu}}} 
\right \rangle= 0 .
\label{III2.15}
\ee
On the left hand side of eq.\,(\ref{III2.15}) we have taken into account the
constraints $\delta I [\eta, g] / \delta g^{\mu \nu} = 0$
(explicitly given in eq.\,(\ref{III2.11})).

\vs{1mm}

The expression $\delta I_{\rm eff}[\eta, g]/\delta g^{\mu\nu} = 0$
gives the classical equations for the metric $g_{\mu \nu}$, derived from
the effective action.

\vs{1mm}

Let us now consider a specific case in which we take for $\omega (\eta) $ the
expression (\ref{III2.3}). Then our action (\ref{III2.10}) splits into two terms
\be
I[\eta,g] = I_0 [\eta, g] + I_m [\eta , g]                 \label{III2.16}
\ee
with
\bearr
 I_0 [\eta , g] = {{{\omega}_0} \over 2} \int {\dd}^n \, x \sqrt{|g|} \,
  (g^{\mu \nu} {\p}_{\mu} {\eta}^a {\p}_{\nu} {\eta}_a {+}2{-}n), \label{III2.17}
                                                                \\  \lal 
   I_m [\eta , g] = {1 \over 2} \, \int {\dd}^n x \, \sqrt{|g|} \,
  \sum_i \, m_i \,
  {{{\delta}^N (\eta - {\hat \eta}_i )} \over {\sqrt{|\gamma|}}} \,
  {\dd}^m {\hat x}                                                  \nnn
\qquad \hs{2cm} \times \sqrt{|{\hat f}|}\,(g^{\mu \nu} {\p}_{\mu} 
{\eta}^a{\p}_{\nu}  {\eta}_a + 2 - n) .
 \label{III2.18}
\ear
The last expression can be integrated over \ $m{-}1$ \ coordinates
${\hat x}^{\mu}$, while ${\hat x}^0$ is chosen so to coincide with
the parameter $\tau$ of a worldline $C_i$.
We also split the metric as $g^{\mu \nu} =
{{n^{\mu} n^{\nu}}/{n^2}} + {\bar g}^{\mu \nu}$, where $n^{\mu}$ is
a time-like vector and ${\bar g}^{\mu \nu}$ is the projection tensor, giving
${\bar g}^{\mu \nu} {\p}_{\mu} {\eta}^a {\p}_{\nu} {\eta}_a = n - 1$.
So we obtain
\newpage
\bearr
   I_m [\eta , g] =  {1 \over 2} \int {\dd}^n x \, \sqrt{|g|} \, \sum_i \,
    {{{\delta}^n (x - X_i (\tau ))} \over   { \sqrt{|g|}}}           \nnn
\qquad  \hs{2cm}  \times \left ( {{g_{\mu \nu} {\dot X}_i^{\mu}
   {\dot X}_i^{\nu}} \over {{\mu}_i}}
    + {\mu}_i \right ) {\dd} \tau = I_m [X_i , g] .   \label{III2.19}
\ear
Here ${\mu}_i \equiv 1/ \sqrt{n^2} |_{C_i}$ are the Lagrange multipliers
gi\-ving, after variation, the worldline constraints ${\mu}_i^2 = {\dot
X}_i^2$.  Eq.\,(\ref{III2.19}) is the well known Howe-Tucker action
\cite{21} for point particles. \inxx{Howe--Tucker action}

Now let us substitute our specific action 
(\ref{III2.16})--(\ref{III2.19}) into the expression
(\ref{III2.14}) for the effective action. The functional integration now runs over
two distinct classes of spacetime sheets $V_n$ [represented by
${\eta}^a (x)$]:

\begin{description}
\item{(a)} those $V_n$ which either {\it do not intersect} the matter sheets
${\widehat V}_m^{(i)}$ [represented by ${\hat \eta}_i^a ({\hat x})$] and do not
self-intersect, or, if
they do, the intersections are just single points, and

\item{(b)} those $V_n$ which {\it do intersect} ${\widehat V}_m^{(i)}$ and/or
self-intersect, the intersections being worldlines $C_i$.
\end{description}

The sheets $V_n$ which correspond to the case (b) have two
distinct classes of points (events):

\medskip
(b1) {\it the points outside the intersection}, i.e., outside the worldlines
        $C_i$, and

\medskip
(b2) {\it the points on the intersection}, i.e., the events belonging to
        $C_i$.

\medskip
The measure ${\cal D} \eta^a(x)$ can be factorized into the contribution
which corresponds to the case (a) or (b1) ($x \notin C_i$), and the
contribution which corresponds to the case (b2) ($x \in C_i$):
\bear
  {\cal D} \eta &=& \prod_{a,x} (|g(x)|)^{1/4} {\dd} {\eta}^a (x) \nonumber \\
   &=& \prod_{a,x \notin C_i} (|g(x)|)^{1/4} {\dd} {\eta}^a (x)
   \prod_{a,x \in C_i} (|g(x)|)^{1/4} {\dd} {\eta}^a (x) \nonumber \\
   \nonumber \\
      &\equiv&  {\cal D}_0 \eta \, {\cal D}_m  .      \label{III2.20}
\ear
The additional factor $(|g(x)|)^{1/4}$ comes from the requirement that the
measure be invariant under repara\-metrizations of $x^{\mu}$ (see Ref.
\cite{79} and Sec. 4.1
for details).  From the very definition of $\prod_{a,x \in C_i}
(|g(x)|)^{1/4} {\dd} {\eta}^a (x)$ as the measure of the set of points on
the worldlines $C_i$ (each $C_i$ being represented by an equation $x = 
X_i^{\mu} (\tau)$) we conclude that
\be
  {\cal D}_m {\eta}^a (x) = {\cal D} X_i^{\mu} (\tau)  . \label{III2.21}
\ee
The effective action then satisfies [owing to (\ref{III2.16})--(\ref{III2.21})]
\inxx{effective action}
\bear
 \nq \e^{i I_{\rm eff} [g]}\eql \int\nhq \e^{iI_0 [\eta, g]}\,{\cal D}_0
   \eta \,\e^{i I_m [ X_i , g ]} \, {\cal D} X_i \equiv \e^{i
   W_0}\, \e^{i W_{\rm m}},   \nnn
                                   \label{III2.22}           \\
   I_{\rm eff} \eql  W_0 + W_{\rm m}.    \label{III2.23}
\ear
The measure ${\cal D}_0 \eta$ includes all those sheets $V_n$ that
do not intersect the matter sheet and do not self-intersect [case (a)], 
and also all those sheets which do intersect and/or self-intersect [case (b1)],
apart from the points on the intersections.

The first factor in the product (\ref{III2.22}) contains the action (\ref{III2.17}).
The latter has the same form as the action for $N$ scalar fields in a curved
background spacetime with the metric $g_{\mu \nu}$. The corresponding
\inxx{scalar fields,in curved spacetime}
effective action has been studied and derived in Refs. \cite{80}. Using
the same procedure and substituting our specific constants
${\omega}_0/2$ and $(n - 1)$ occurring in eq.\,(\ref{III2.17}), we find the
following expression for the effective action: \inxx{effective action}
\bear
   &&I_{\rm eff} =
   \lim_{\mu^2 \to 0} \int {\dd}^n x \sqrt{|g|} \, \Biggl( 
   N\,{\omega}_0^{-1} (4 \pi)^{-n/2} \sum_{j=0}^{\infty}
   (\mu^2)^{n/2 -j} \, a_j (x) \, \Gamma (j {-} {n \over 2}) 
   \hs{1cm}\nonumber \\
  & & \hs{3.9cm} + {\omega_0 \oo 2} (2 - n) \Biggr)
\label{III2.24}
\ear
with
\bear
  a_0 (x) \eql 1     \label{III2.25} \\
  a_1 (x) \eql R/6   \label{III2.26} \\
  a_2 (x) \eql {1 \over 12} R^2 + {1 \over 180} (R_{\alpha \beta \gamma
  \delta} R^{\alpha \beta \gamma  \delta} - R_{\alpha \beta} R^{\alpha
  \beta}) - {1 \over 30} D_{\mu} D^{\mu} R \label{III2.27}
\ear
where $R,\ R_{\alpha \beta}$ and $R_{\alpha \beta \gamma \delta} $
\inxx{Ricci scalar} \inxx{Ricci tensor} \inxx{Riemann tensor}
are the Ricci scalar, the Ric\-ci tensor and the Riemann tensor, respectively.
The function $\Gamma (y) = \int_0^{\infty} \e^{-t} \, t^{y - 1} \, {\dd} t$;
it is divergent at negative integers $y$ and finite at $y = {3 \over 2},
{1 \over 2}, - {1 \over 2}, - {3 \over 2}, - {5 \over 2}, ...$. The effective
action (\ref{III2.24}) is thus divergent in even-dimensional spaces $V_n$.
For instance, when $n = 4$, the argument 
in eq.\,(\ref{III2.24}) is $j - 2$, which
for $j = 0, 1, 2, ...$ is indeed a negative integer. Therefore,
in order to obtain a finite effective action one needs to
introduce a suitable cut-off parameter $\Lambda$ and replace
     $$(\mu^2)^{n/2-j} \Gamma(j-{n\oo 2}) = \int_0^\infty t^{j-1-n/2}
     e^{-\mu^2 t} \dd t$$
with
     $$\int_{1/\Lambda^2}^\infty t^{j-1-n/2} e^{-\mu^2 t} \dd t$$
Then we find     
        $$I_{\rm eff} = \int {\dd}^n x \sqrt{|g|} \, \Bigl(
        {\lambda}_0 + {\lambda}_1 R + {\lambda}_2 R^2  
    + \, {\lambda}_3   (R_{\alpha \beta \gamma\delta}
           R^{\alpha \beta \gamma  \delta}
                 - R_{\alpha \beta} R^{\alpha  \beta})$$
\be                 
                 + \, {\lambda}_4 D_{\mu} D^{\mu} R 
                 + O(\Lambda^{n-6}) \Bigr) \; , \hs{5mm}           
\label{III2.28}
\ee
where\footnote{In the case $n=4$ it is convenient to use Akama's
regularization \ci{75}. Similarly for the problematic coefficients
in higher dimensions.}, for $n > 4$,
\bearr
{\lambda}_0 = {{N \omega_0}\oo {4 (4 \pi)^{n/2}}} 
  {{2 \Lambda^n}\oo n} + {\omega_0 \oo 2} (2 - n)  , \nnn
   {\lambda}_1 = {{N \omega_0}\oo {4 (4 \pi)^{n/2}}} 
   {{2 \Lambda^{n-2}}\oo {6  (n - 2)}} , \nnn
   {\lambda}_2 =  {{N \omega_0}\oo {4 (4 \pi)^{n/2}}} 
   {{2 \Lambda^{n-4}}\oo {12 (n - 4)}}   ,\nnn
   {\lambda}_3 = {{N \omega_0}\oo {4 (4 \pi)^{n/2}}}
   {{2 \Lambda^{n-4}}\oo {180 (n - 4)}}, \nnn
   \lambda_4 = -\, {{N \omega_0}\oo {4 (4 \pi)^{n/2}}}
    {{2 \Lambda^{n-4}}\oo {30 (n - 4)}} . \label{III2.29}
\ear
Here ${\lambda}_0 $ is the cosmological constant, whilst ${\lambda}_1$
\inxx{cosmological constant} \inxx{gravitational constant,induced}
is related to the gravitational constant $G$ in $n$-dimensions according
to
\be
    {\lambda}_1 \equiv (16 \pi G)^{-1} .
\lbl{III2.30}
\ee     
This last relation shows how the induced 
gravitational constant is calculated in terms of ${\omega}_0$, which is
a free parameter of our embedding model, and the cutoff parameter $\Lambda$.
According to Akama \ci{75}, et al., and Sugamoto \ci{Sugamoto}, we
consider the cutoff $\Lambda$ to be a physical quantity, the inverse
thickness of a membrane, because the original action (\ref{III2.17})
describes an idealized theory of extended objects with vanishing
thickness, but the real extended objects have non-vanishing thickness
playing a role of the ultraviolet cutoff. In the case of thin extended objects,
we can ignore the $O(\Lambda^{n-4})$ terms in eq.\,(\ref{III2.28}).

In the above calculation of the effective action we have treated all
functions ${\eta}^a (x)$ entering the path integral (\ref{III2.14}) as
those representing \inxx{effective action}
distinct spacetime sheets $V_n$. However, owing to the
reparametrization invariance, there exist equivalence classes of functions
representing the same $V_n$. This complication must be taken into account
when calculating the entire amplitude (\ref{III2.13}).
The conventional approach is to introduce ghost fields which cancel
the non-physical degrees of freedom. An alternative approach, first explored in
ref.\, \cite{41}--\ci{43} and much discussed in this book, is to assume that all 
possible embedding functions ${\eta}^a
(x)$ can nevertheless be interpreted as describing physically distinct
spacetime sheets ${\cal V}_n$. This is possible if the extra degrees
of freedom in ${\eta}^a (x)$ describe tangent deformations of ${\cal V}_n$.
Such
a deformable surface ${\cal V}_n$ is then a different concept from a
non-deformable surface $V_n$. The path integral can be 
performed in a straightforward
way in the case of ${\cal V}_n$, as was done in arriving at the
result (\ref{III2.24}). However, even from the standard point of view
\ci{Floreanini}, gauge fixing is not required for the calculation of the
effective action, since in the $\eta^a$ integration, $g_{\mu \nu}$ is
treated as a fixed background.

Let us now return to eq.\,(\ref{III2.22}). In the second factor of 
eq.\,(\ref{III2.22})
the functional integration runs over all possible worldlines $X^{\mu}
(\tau)$.  Though they are obtained as intersections of {\it various} $V_n$
with ${\widehat V}_m^{(i)}$, we may consider all those worldlines to be lying in
{\it the same} effective spacetime $V_n^{(\rm eff)}$ with the intrinsic
\inxx{effective spacetime}
metric $g_{\mu \nu}$. In other words, in the effective theory we identify
all those various $V_n$'s, having the same induced (intrinsic) metric
$g_{\mu \nu}$, as one and the same spacetime. If one considers the
embedding space $V_N$ of sufficiently high dimension $N$, then there is
enough freedom to obtain as an intersection any possible worldline in the
effective spacetime $V_n^{\rm (eff)}$.

This is even more transparent when the spacetime sheet self-intersects.
Then we can have a situation in which various sheets coincide in all
the points, apart from the points in the vicinity of the intersections

When the condition for the classical approximation is satisfied, i.e., when
$I_m \gg \hbar = 1$, then only those trajectories $X_i^{\mu}$ which
are close to the classically allowed ones effectively contribute:
\be
   \e^{i W_{\rm m}} = \e^{i I_m [X_i, g]} \; , \quad \; \quad W_{\rm m} = I_m
\label{III2.31}
\ee

The effective action is then a sum of the gravitational kinetic term
\inxx{effective action}
$W_0$ given in eq.\,(\ref{III2.28}) and the source term $I_m$ 
given in (\ref{III2.19}).
Variation of $I_{\rm eff}$ with respect to $g^{\mu \nu}$ then gives the
gravitational field equation in the presence of point particle sources
\inxx{stress--energy tensor}with the stress--energy 
tensor $T^{\mu \nu}$ as given in Eq.\,(\ref{III2.6}):
\be
    R^{\mu\nu} - {1 \over 2} g^{\mu \nu} + {\lambda}_0 g^{\mu \nu} +
   (\mbox{\rm higher order terms})
    = - 8 \pi G \, T^{\mu \nu}  \, .                \label{III2.32}
\ee

However, in general the classical approximation is not satisfied, and in the
eva\-luation of the matter part $W_{\rm m}$ of the effective action one must take
into account the contributions of all possible paths $X_i^{\mu} (\tau)$.
So we have (confining ourselves to the case of only one particle, omitting the
subscript $i$, and taking $\mu = 1$)
\bear
\e^{i W_{\rm m}} &=&\int_{x_a}^{x_b} {\cal D} X \, \mbox{exp} \left ( {i \over 2}
 \int_{{\tau}_a}^{{\tau}_b} {\dd} \tau \, m \, (g_{\mu \nu} {\dot X}^{\mu}
{\dot X}^{\nu} + 1) \right ) \nonumber \\
\nonumber \\
 &=&  {\cal K} (x_b , {\tau}_b ; x_a , {\tau}_a) \equiv {\cal K} (b,a) 
\label{III2.33}
\ear
which is a propagator or a Green's function satisfying (for
${\tau}_b \ge {\tau}_a$) the equation \inxx{Green function}
\be
 \left ( i\, {{\p} \over {\p {\tau}_b}} - H \right )
  {\cal K} (x_b , {\tau}_b ; x_a , {\tau}_a) 
   =  - {1 \over {\sqrt{|g|}}} \,
    {\delta}^n (x_b - x_a ) \, \delta ({\tau}_b - {\tau}_a ) \; ,
\label{III2.34}
\ee
where $H = (|g|)^{-1/2} {\p}_{\mu} ((|g|)^{1/2} {\p}^{\mu} )$. From (\ref{III2.34})
it follows that
\be
    {\cal K} (b,a) = - \, {\left [ i {{\p} \over {\p \tau}} -
     H \right ]}_
{x_b , {\tau}_b ; x_a , {\tau}_a}^{-1}
\label{III2.35}             \ee
where the inverse Green's function is treated as a matrix in the $(x, \tau)$
space. \inxx{Green function,inverse}

Using the following relation \cite{22} for Gaussian integration
\be
    \int y_m y_n \prod_{i = 1}^N {\dd} y_i \, \e^{- \sum_{ij} y_i A_{ij} y_j}
  \propto {{(A^{-1})_{mn}} \over {(\mbox{\rm det} \, |A_{ij}|)^{1/2}}},
\label{III2.36}
\ee
we can rewrite the Green's function in terms of the second quantized field:
\inxx{second quantized field}
  $${\cal K}(a,b) =\int {\psi }^* (x_b,\tau_b) \, \psi (x_b,\tau_b)  \,
  {\cal D} {\psi}^* \, {\cal D} \psi \hs{3cm}$$
\be  
  \hs{3.6cm} \times \exp \left[-i \int {\dd} \tau \,
  {\dd}^n x \, \sqrt{|g|} \, {\psi}^* (i\p_\tau- H)\psi \right] .\label{III2.37}
\ee

If the conditions for the ``classical" approximation are satisfied, so that
the phase in (\ref{III2.37}) is much greater than $\hbar = 1$, then only those
paths $\psi (\tau, x)\, , \quad {\psi}^* (\tau , x )$ which are close to the
extremal path, along which the phase is zero, effectively contribute to
${\cal K} (a , b)$. Then the propagator is simply \inxx{propagator}
\be
  {\cal K} (b , a) \propto \exp \biggl[ -i \int {\dd} \tau \,
  {\dd}^n x  \sqrt{|g|} \, {\psi}^* (i\p_\tau{-} H) \psi \biggr].\label{III2.38}
\ee
The effective one-particle ``matter" action $W_{\rm m}$ is then
\be
   W_{\rm m} = -  \int {\dd} \tau \,
  {\dd}^n x \, \sqrt{|g|} \, {\psi}^* (\tau , x) (i {\p}_\tau {-} H) \psi
  (\tau , x).                                                 \label{III2.39}
\ee

\noi
If we assume that the $\tau$-dependence of the field $\psi (\tau , x) $ is
specified\footnote{
By doing so we in fact project out the so called physical states from
the set of all possible states. Such a procedure, which employs
a ``fictitious" evolution parameter $\tau$ is often used (see Chapter 1).
When gauge fixing the action (\ref{III2.33}) one pretends that such
an action actually represents an ``evolution" in parameter $\tau$. Only 
later, when all the calculations (e.g., path integral) are performed, one
integrates over $\tau$ and thus projects out the physical quantities.}
by $\psi (\tau , x) = \e^{i m \tau} \, \phi (x)$, then
eq.\,(\ref{III2.39}) simplifies
to the usual well known expression for a scalar field:
\bearr
\nq W_{\rm m} = \int {\dd}^n x \sqrt{|g|} {\phi}^* (x) \biggl[{1\over {\sqrt{|g|}}}
  {\p}_{\mu} (\sqrt{|g|} {\p}^{\mu} ) + m^2 \biggr) \phi (x)       \nnn
  \hs{3mm}  = - \mbox{$1\oo 2$} \int {\dd}^n x \, \sqrt{|g|} \, (g^{\mu \nu} \,
  {\p}_{\mu} {\phi}^* \, {\p}_{\nu} \phi - m^2 ) \; , 
\label{III2.40}
\ear
where the surface term has been omitted.

Thus, starting from our basic fields ${\eta}^a (x)$, which are the embedding
functions for a spacetime sheet $V_n$, we have arrived at the effective
action $I_{\rm eff}$ which contains the kinetic term $W_0$ for the metric field
\inxx{source term} \inxx{bosonic matter field}
$g^{ \mu \nu}$ (see Eq.\,(\ref{III2.28})) and the source term $W_{\rm m}$ (see
eq.\,(\ref{III2.32}) and (\ref{III2.18}), or (\ref{III2.40})). 
Both the metric field $g_{\mu
\nu}$ and the bosonic matter field $\phi $ are induced from the basic fields
${\eta}^a (x)$.

\section{Conclusion}                                                  

We have investigated a model which seems to be very promising in attempts
to find a consistent relation between quantum theory and gravity. Our
model exploits the approach of induced gravity and the concept of
spacetime embedding in a higher-dimensional space and has the following
property of interest: what appear as worldlines in, e.g., a 4-dimensional
spacetime,
are just the intersections of a spacetime sheet $V_4$ with ``matter"
sheets  ${\widehat V}_{\rm m}^{(i)}$, or self-intersections of $V_4$. 
Various choices of spacetime sheets then give
various configurations of worldlines. Instead of $V_4$ it is convenient to
consider a spacetime sheet $V_n$ of arbitrary dimension $n$. When passing
to the quantized
theory a spacetime sheet is no longer definite. All possible alternative
spacetime sheets are taken into account in the expression for a wave
functional or a Feynman path integral. The intersection points of $V_n$
with itself or 
with a matter sheet ${\widehat V}_m^{(i)}$ are treated specially, and it is found
that their contribution to the path integral is identical to the
contribution of a point particle path. We have paid special attention to the
effective action which results from functionally integrating out all
possible embeddings with the same induced metric tensor. We have
found that the effective action, besides the Einstein--Hilbert term and the
corresponding higher-order terms, also contains a source term. The
expression for the latter is equal to that of a classical (when such an
approximation can be used) or quantum point particle source described by a
scalar (bosonic) field.

In other words, we have found that the ($n$-dimensi\-onal) Einstein
equations (including the $R^2$ and higher derivative terms) with classical
or quantum point-particle sources are effective equations resulting from
performing a quantum average over all possible embeddings of the
spacetime. Gravity ---as described by Einstein's general relativity--- is
thus considered not as a fundamental phenomenon, but as something
induced quantum-mechanically from more fundamental phenomena.

In our embedding model of gravity with bosonic sources, new and
interesting possibilities are open. For instance, instead of a 4-dimensional
spacetime sheet we can consider a sheet which possesses additional
dimensions, parametrized either with usual or Grassmann
coordinates. In such a way we expect to include, on the one hand, via the
Kaluza--Klein mechanism, other interactions as well besides the gravitational
one, and, on the other hand, fermionic sources. The latter are expected
also to result from the polyvector generalization of the action, as discussed
in Sec. 2.5.

It is well known that the quantum field theory based on the
action (\ref{III2.40}) implies the infinite vacuum energy density
\inxx{vacuum energy density} and consequently the infinite (or, more
precisely, the Planck scale cutoff) cosmological constant.
\inxx{cosmological constant}This is only a part of the total cosmological
constant predicted by the complete theory, as formulated here. The other
part comes from eqs.\,(\ref{III2.28}),\ (\ref{III2.29}). The parameters
$N$, $\omega_0$, $\Lambda$ and $n$ could in principle be adjusted so to
give a small or vanishing total cosmological constant. If carried out
successfully this would be then an alternative (or perhaps a complement)
to the solution of the cosmological constant problem as suggested in
Chapter 3.

Finally, let us observe that within the ``brane world" model of Randall and 
Sundrum \cite{72}, in which matter fields are localized on a 3-brane, 
whilst gravity
propagates in the bulk, it was recently proposed \cite{80a} to treat the 
Einstein--Hilbert term on the brane as being induced in the quantum theory
of the brane. It was found that the localized matter on a brane can induce
via loop correction a 4D kinetic term for gravitons. This also happens in
our quantum model if we consider the effective action obtained after
integrating out the second quantized field in (\ref{III2.37}). (The procedure
expounded in refs. \ci{12} is then directly applicable.) 
In addition,
in our model we obtain, as discussed above, a kinetic term for gravity
on the brane after functionally integrating out the embedding functions.

\chapter[On the resolution of time problem \\ in quantum gravity]
{On the resolution of \\ time problem \\ in quantum gravity}

Since the pioneering works of Sakharov \cite{73} and Adler \cite{74}
there has been increasing
interest in various models of induced gravity \cite{75}.
A particularly interesting \inxx{induced gravity}
and promising model seems to be the one in which spacetime is a 4-dimensional
manifold
(a `spacetime sheet') $V_4$ embedded in an $N$-dimensional space $V_N$
\cite{78,71},\,\ci{41}--\ci{43}.  The
dynamical variables are the embedding functions ${\eta}^a(x)$ which determine
positions (coordinates) of points on $V_4$ with respect to $V_N$.  The action
is
a straightforward generalization 
of the Dirac--Nambu--Goto action. The latter can be
written in an equivalent form in which there appears the induced metric $g_{\mu
\nu}(x)$ and ${\eta}^a(x)$ as variables which have to be varied independently.
Quantization of such an action enables one to express {\it an effective action} as
\inxx{effective action}
a functional of $g_{\mu \nu}(x)$.  The effective action is obtained in the
Feynman path integral in which we functionally integrate over the embedding
functions ${\eta}^a(x)$ of $V_4$, so that what remains is a functional
dependence on $g_{\mu \nu}(x)$.  Such an effective action contains the Ricci
curvature scalar $R$ and its higher orders.
This theory was discussed in more detail in the previous chapter.

In this chapter we are going to generalize the above approach.  The main
problem with any reparametrization invariant theory is the presence of
constraints relating the dynamical variables.  Therefore there exist
equivalence classes of functions ${\eta}^a (x)$ ---related by reparametrizations
of the coordinates $x^{\mu}$--- such that each member of an equivalence class
represents the same spacetime sheet $V_4$.  This must be taken into account in
the quantized theory, e.g., when performing, for instance, a functional
integration over ${\eta}^a (x)$.  Though elegant solutions to such problems
were
found in string theories \cite{81}, the technical difficulties accumulate
in the case of a $p$-dimensional membrane ($p$-brane)
with $p$ greater than 2 \cite{64}.

In Chapters  4--6 we discussed the possibility of removing constraints from a
membrane ($p$-brane) theory.  Such a generalized theory possesses additional
degrees of freedom and contains the usual $p$-branes of the Dirac--Nambu--Goto
type \ci{82}
as a special case.  It is an extension, from a point particle to a
$p$-dimensional membrane, of a theory which treats a relativistic particle
without constraint, so that all coordinates $x^{\mu}$ and the conjugate
momenta
$p_{\mu}$ are independent dynamical variables which evolve along the invariant
evolution parameter $\tau$ \cite{1}--\ci{11a}.
A membrane is then considered as a continuum of
such point particles and has no constraints.  It was shown \cite{41}
that the extra
degrees of freedom are related to variable stress and fluid velocity on the
membrane, which is therefore, in general, a ``{\it wiggly membrane"}. We shall
now apply the concept of a relativistic membrane without constraints to the
embedding \inxx{wiggly membrane} \inxx{induced gravity}
model of induced gravity in which the whole spacetime is considered as a
membrane in a flat embedding space.

\vs{1mm}

In Sec.\ 10.1 we apply the theory of unconstrained membranes to the concept of 
an $(n-1)$-dimensional membrane ${\cal V}_{n-1}$  moving in an
$N$-dimensional embedding space $V_N$ and thus sweeping a spacetime 
sheet $V_n$. \inxx{spacetime sheet}

\vs{1mm}

In Sec.\ 10.2 we consider the theory in which
the whole space-time is an $n$-dimensional\footnote{Usually $n=4$, 
but if we wish
to consider a Kaluza--Klein like theory then $n > 4$}
unconstrained membrane ${\cal V}_n$.  The theory allows for {\it motion} of
${\cal V}_n$ in the embedding space $V_N$.  When considering the quantized
theory it \inxx{unconstrained membrane}
turns out that a particular wave packet functional exists such that:

\begin{description}
 
\item{(i)} it approximately represents evolution of
a simultaneity surface ${\cal V}_{n-1}$ (also denoted
${\cal V}_{\Sigma}$), and

\item{(ii)} all possible space-time membranes ${\cal V}_n$ composing
the wave packet
are localized near an average space-time membrane ${\cal V}_n^{(c)} $
which corresponds to a classical space-time unconstrained membrane.

\end{description}

This approach gives both:  the evolution of a state (to which classically there
corresponds the progression of time slice) and a fixed spacetime as the
expectation value.  The notorious problem of time, as it occurs in a
reparametrization invariant theory (for instance in general relativity), does
not exist in our approach. \inxx{problem of time}
\newpage

\section{Space as a moving 3-dimensional membrane in $V_N$}

The ideas which we have developed in Part II may be used to describe elementary
particles as extended objects ---unconstrained $p$-dimensional membranes 
${\cal V}_p$--- living in spacetime.
In this Part we are following yet another application of $p$-branes:
to represent spacetime itself!  Spacetime is considered as a surface
---also called a
{\it spacetime sheet}--- $V_n$ embedded in a higher-dimensional space $V_N$.
For
details about some proposed models of such a kind see refs. \ci{76}--\ci{78},
\ci{71,71a,42}. 
In this section we
consider a particular model in which an $(n-1)$-dimensional membrane, called a
{\it simultaneity surface} ${\cal V}_{\Sigma}$, moves in the embedding
space according to
the unconstrained theory of Part II and sweeps an
$n$-dimensional spacetime sheet $V_n$. In particular, the moving membrane
is 3-dimensional and it sweeps a 4-dimensional surface.

Since we are now talking about spacetime which is conventionally
para\-metrized by
coordinates $x^{\mu}$, the notation of Part II is not appropriate. For this
particular application of the membrane theory we use different notation.
Coordinates denoting position of a spacetime sheet $V_n$ (alias worldsheet) in
the embedding space $V_N$ are
\be {\eta}^a \; , \quad \quad a = 0,1,2,...,N-1  , \label{III3.13a} \ee
whilst parameters denoting positions of points on $V_n$ are
\be x^{\mu} \; , \quad \quad\mu = 0,1,2,...,n-1 .  \label{III3.14} \ee
The parametric
equation of a spacetime sheet is\footnote{
To simplify notation we use the same
symbol ${\eta}^a$ to denote coordinates of an arbitrary point in $V_N$ and also
to to denote the embedding variables (which are functions of $x^{\mu}$).}
\be {\eta}^a = {\eta}^a (x)  \label{III3.15} \ee
Parameters on a simultaneity surface ${\cal V}_{\Sigma}$ are
\inxx{simultaneity surface}
\be {\sigma}^i \; , \quad \quad i =1,2,...,n-1 ,  \label{III3.16} \ee
and its parametric equation is ${\eta}^a = {\eta}^a (\sigma)$. A moving
${\cal V}_{\Sigma}$ is described by the variables ${\eta}^a (\tau,\sigma)$.

The formal theory goes along the similar lines as in Sec. 7.1.  The action is
given by
\be
  I[{\eta}^a(\tau,\sigma)] = \mbox{$1\oo 2$} \int \omega \, {\mbox{\rm d}} \tau \,
  {\mbox{\rm d}}^{n-1}
	\sigma \sqrt{\bar f} \left ( {{ {\dot \eta}^a \,
	{\dot
	\eta}_a} \over {\Lambda}} + \Lambda \right ) \; ,
\label{III3.17}
\ee
where 
\be
{\bar f} \equiv \mbox{\rm det} {\bar f_{ij}} \; , \hs{1cm} {\bar f}_{ij}
\equiv {{\partial {\eta}^a} \over {\partial {\sigma}^i}} 
{{\partial {\eta}_a} \over {\partial
{\sigma}^j}} \; ,
\lbl{III3.17a}
\ee
is the determinant of the induced metric on
${\cal V}_{\Sigma}$, and
$\Lambda = \Lambda(\tau,\sigma)$ a fixed function.  The tension $\kappa$ is now
replaced by the symbol $\omega$.  The latter may be a constant.
However, in the proposed
embedding model of spacetime 
we admit $\omega$ to be a function of position in
$V_N$:  
   \be \omega = \omega (\eta) .  \label{III3.18} \ee In the case in which
$\omega$ is a constant we have a spacetime without ``matter" sources.  When
$\omega$ is a function of ${\eta}^a$, we have in general
a spacetime with sources (see Chapter 8)

A solution to the equations of motion derived from (\ref{III3.17}) represents a
motion of a simultaneity surface ${\cal V}_{\Sigma}$.  This is analogous
to the motion of an unconstrained membrane \inxx{unconstrained membrane}
discussed in Part II.  Here again we see a big advantage
of such an {\it
unconstrained} theory:  it predicts actual {\it motion} of
${\cal V}_{\Sigma}$ and {\it
evolution} of a corresponding quantum state with $\tau$ being the evolution
parameter or historical time \cite{5a}.  The latter is a distinct concept 
from the coordinate time $t \equiv x^0$.  The existence (and progression) of a
time slice\inxx{historical time}\inxx{time slice}
is automatically incorporated in our unconstrained theory.  It not need be
separately postulated, as it is in the usual, constrained
relativistic theory\footnote{
More or less explicit assumption of the existence of a time slice
(associated with the perception of ``now") is manifest in conventional
relativistic theories from the very fact that the talk is about
'`point particles" or ``strings" which are objects in three dimensions.}.
Later, when discussing the quantized theory, we shall show how to take
into account that setting up data on an entire (infinite) spacelike
hypersurface  is an idealistic situation, since it would require an
infinite time span which, in practice, is not available to an observer.

The theory based on the action (\ref{III3.17}) is satisfactory in several respects.
However, it still cannot be considered as a complete theory, because it is not
manifestly invariant with respect to general coordinate transformations of
spacetime coordinates (which include Lorentz transformations).  In the next
section we shall ``improve" the theory and explore some of its consequences.
We shall see that the theory of the motion of a time slice
${\cal V}_{\Sigma}$, based on the
action (\ref{III3.17}), comes out as a particular case (solution) of the
generalized
theory which is fully relativistic, i.e., invariant with respect to
reparametrizations of $x^{\mu}$.  Yet it incorporates the concept of state
evolution.

\vs{1cm}

\section{Spacetime as a moving 4-dimensional membrane in $V_N$ }

\subsection[\ \ General consideration]
{General consideration}

The experimental basis\footnote{
Crucial is the fact that simultaneity of events is relative to
an observer. Different observers
(in relative motion) determine as simultaneous different sets of 
events.  Hence all those events must exist in a 4-dimensional spacetime 
in which time is just one of the coordinates.}
on which rests the special relativity and its
generalization to curved spacetime clearly indicates that spacetime is a
continuum in which events are existing.  On the contrary, our subjective
experience clearly tells us that not the whole 4-dimensional\footnote{
When
convenient, in order to specify the discussion, let us specify the dimension of
spacetime and take it to be 4.}
spacetime, but only a 3-dimensional section of it is
accessible to our immediate experience.  How to reconcile those seemingly
contradictory observations?

It turns out that this is naturally achieved by joining the formal theory of
membrane motion (7.1) with the concept of spacetime embedded 
in a higher-dimensional space $V_N$ (Chapter 8).  Let us assume that the spacetime is
an unconstrained 4-dimensional membrane ${\cal V}_4$ which evolves
(or moves) in the embedding space $V_N$. Positions of points on ${\cal V}_4$
at a given instant of the evolution time $\tau$ are described by 
embedding variables ${\eta}^a(\tau,x^{\mu})$.  The latter now depend not 
only on the spacetime sheet
parameters (coordinates) $x^{\mu}$, but also on $\tau$.  Let us, for the moment,
just accept such a possibility that ${\cal V}_4$ evolves, and we shall
later see how the
quantized theory brings a physical sense to such an evolution.

The action, which is analogous to that of eq.\,(\ref{III3.17}), is
\be
    I[{\eta}^a(\tau,x)] = {1 \over 2} \int\omega {\mbox{\rm d}} \tau
    {\mbox{\rm d}}^4 x \sqrt{|f|} \left ( {{ {\dot \eta}^a
    {\dot \eta}_a} \over {\Lambda}} + \Lambda \right ) \; , 
    \label{III3.19}
\ee
\be    
    f \equiv \mbox{\rm det} f_{\mu \nu} \; \; , \quad f_{\mu \nu} \equiv
    {\partial}_{\mu} {\eta}^a {\partial}_{\nu} {\eta}_a \; , \label{III3.19aa}
\ee
where $\Lambda = \Lambda(\tau,x)$ is a fixed function of $\tau$ and
$x^{\mu}$
(like a ``background field") and $\omega = \omega (\eta)$.

The action (\ref{III3.19}) is invariant with respect to 
arbitrary transformations of the
spacetime coordinates $x^{\mu}$.  But it is not invariant under
reparametrizations of the evolution parameter $\tau$.  Again we use analogous
reasoning as in Chapter 4. Namely, the freedom of choice of parametrization 
on a given initial ${\cal V}_4$ is trivial and it does not impose
any constraints on the
dynamical variables ${\eta}^a$ which depend also on $\tau$.  In other words, we
consider spacetime $V_4$ as a physical continuum, the points of which can be
identified and their $\tau$-evolution in the embedding space $V_N$ followed.
For a chosen parametrization $x^{\mu}$ of the points on ${\cal V}_4$ different
functions ${\eta}^a (x)$, ${\eta '}^a (x)$
(at arbitrary $\tau$) represent different {\it
physically deformed} spacetime continua ${\cal V}_4$, ${\cal V}_4^{'}$.
Different functions
${\eta}^a (x)$, ${\eta '}^a (x)$,
even if denoting positions on the same mathematical surface
$V_4$ will be interpreted as describing physically distinct spacetime
continua, ${\cal V}_4$, ${\cal V}_4^{'}$,
locally deformed in different ways.  An evolving physical spacetime continuum
${\cal V}_4$ is not a concept identical to a mathematical surface $V_4$.
\footnote{A strict notation would then require a new symbol,
for instance ${\tilde \eta}^a (x)$ for the variables of the physical
continuum ${\cal V}_4$,
to be distinguished from the embedding functions ${\eta}^a (x)$ of a
mathematical surface $V_4$.  We shall not use this distinction in notation,
since the meaning will be clear from the context.}

If, on the other hand, we focus attention to the mathematical manifold
$V_4$ belonging to ${\cal V}_4$, then different functions
${\eta}^a (x)$, ${\eta '}^a (x)$ which represent the same manifold $V_4$
are interpreted as belonging to an equivalence class due to
reparametrizations of $x^{\mu}$. There are thus two interpretations 
(see also Sec. 4.1) of
the transformations relating such functions ${\eta}^a (x)$, ${\eta '}^a (x)$:
(i) the passive transformations, due to reparametrizations of $x^{\mu}$,
and (ii) the active transformations, due to deformations of the physical
continuum ${\cal V}_4$ into ${\cal V}_4^{'}$. In the first case, the induced
metric on $V_4$, which is isometrically embedded in $V_N$, is given by
$g_{\mu \nu} = {\p}_{\mu} {\eta}^a {\p}_{\nu} {\eta}_a$ . The same expression
also holds for the metric on the physical continuum ${\cal V}_4$ . If the
latter is deformed into ${\cal V}_4^{'}$ , then the metric tensor of
${\cal V}_4^{'}$ is $g_{\mu \nu}^{'} = {\p}_{\mu} {\eta}^{' a} {\p}_{\nu}
{\eta}_a^{'}$ , which is still the expression for the induced metric on
a mathematical manifold isometrically embedded in $V_N$. 

Let us now start developing some basic formalism.  The canonically conjugate
variables belonging to the action (\ref{III3.19}) are
\inxx{canonically conjugate variables}
\be
  {\eta}^a (x)\; , \qquad p_a (x) = {{\partial L} \over {\partial {\dot \eta}^a}}
   =\omega \sqrt{|f|} \; {{{\dot \eta}_a} \over {\Lambda}}\, .
\label{III3.20}
\ee

The Hamiltonian is
\be
    H = {1 \over 2} \int \, {\dd}^4 x \sqrt{|f|} \; {{\Lambda} \over
    {\omega}} \left ( {{p^a p_a} \over {|f|}} - {\omega}^2 \right ) \, .
\label{III3.21}
\ee

The theory can be straightforwardly quantized by considering ${\eta}^a (x) ,
\; p_a (x)$ as operators satisfying the equal $\tau$ commutation relations
\be 
   [{\eta}^a (x) , \;p_b (x')] = {{\delta}^a}_b \delta (x - x') .
\label{III3.22}
\ee
In the representation in which ${\eta}^a (x)$ are diagonal the momentum
operator is given by the functional derivative \inxx{momentum operator}
\be p_a = -\, i \, {{\delta} \over {{\delta} {\eta}^a (x)}} \, .
\label{III3.23}
\ee
A quantum state is represented by a wave functional $\psi[\tau,
{\eta}^a(x)]$ which depends on the evolution parameter $\tau$ and the
\inxx{wave functional} \inxx{evolution parameter} \inxx{Schr\" odinger equation}
coordinates ${\eta}^a(x)$ of a physical spacetime sheet ${\cal V}_4$,
and satisfies the functional Schr\" odinger equation\footnote
{This is the extension, from point particle to membrane, of
the equation proposed by Stueckelberg \cite{2}. Such an equation has been
discussed several times throughout this book.}
\be
   i \, {{\partial \psi} \over {\partial \tau}} = H \psi .
\label{III3.24}
\ee
The Hamiltonian operator is given by eq.\,(\ref{III3.21}) in which $p_a$
\inxx{Hamilton operator}
are now operators (\ref{III3.23}). According to the analysis given in Sec.\,7.1
a possible solution to eq.\,(\ref{III3.24}) is a linear superposition of
states with definite momentum $p_a(x)$ which are taken as constant
functionals of ${\eta}^a (x)$, so that $\delta p_a / \delta {\eta}^a = 0$:
   $$\psi[\tau, \eta (x)] = \int {\cal D} p \, c(p) \, 
   {\rm exp} \left [ i \int {\dd}^4 x \, p_a (x) ({\eta}^a (x) - 
   \eta_0^a (x)) \right ] \hs{1.8cm}$$
\be   
   \hs{3.2cm} \times \, {\rm exp} \left [-{i \over 2} \, 
   \int {\dd}^4 x \sqrt{|f|} \, 
   {{\Lambda} \over {\omega}} \left (
   {{p^a p_a} \over {|f|}} - {\omega}^2 \right ) \tau \right ] \, ,
\label{III3.25}
\ee
where $p_a (x)$ are now eigenvalues of the momentum operator,
and ${\cal D} p$ is the invariant measure in momentum space.

\subsection[\ \ A physically interesting solution]
{A physically interesting solution}

Let us now pay attention to eq.\,(\ref{III3.25}). It defines a wave functional
packet spread over a continuum of functions ${\eta}^a (x)$. As discussed
in more detail in Section\,7.1, the expectation value of ${\eta}^a (x)$ is
\be
       \langle {\eta}^a (x) \rangle = {\eta}_{\mbox{\rm c}}^a (\tau,x)
\label{III3.27}
\ee
where ${\eta}_{\mbox{\rm c}}^a (\tau,x)$ represents motion of the
{\it centroid} spacetime sheet ${\cal V}_4^{(c)}$ which is the ``centre"
of the wave  functional packet. This is illustrated in Fig. 10.1.
\inxx{spacetime sheet,centroid} \inxx{wave  functional,packet}

In general, the theory admits an arbitrary motion
${\eta}_{\mbox{\rm c}}^a (\tau,x)$ which is a solution of the
classical equations of motion derived from the action (\ref{III3.19}).
But, in particular, a wave packet (\ref{III3.25}) which is a solution of the
Schr\" odinger equation (\ref{III3.24}) can be such that its centroid
spacetime sheet is either

\begin{description}
  \item {\  \rm (i)} \  at rest in the embedding space $V_N$, i.e.
  ${\dot \eta}_{\mbox{\rm c}}^a = 0$;
\end{description}
or, more generally:

\begin{description}
  \item {\rm (ii)} \  it moves ``within itself" so that its shape
  does not change
  with $\tau$. More precisely, at every $\tau$ and $x^{\mu}$ there exists
  a displacement $\Delta X^{\mu}$ along a curve $X^{\mu} (\tau)$ such that
  ${\eta}_{\mbox{\rm c}}^a (\tau +
  \Delta \tau,x^{\mu}) = {\eta}_{\mbox{\rm c}}^a (\tau,x^{\mu} +
  \Delta X^{\mu})$ , which implies
  ${\dot \eta}_{\mbox{\rm c}}^a = {\partial}_{\mu} {\eta}_{\mbox{\rm c}}^a \,
  {\dot X}^{\mu}$ . Therefore ${\dot \eta}_{\mbox{\rm c}}^a$ is always
  tangent to a fixed mathematical surface $V_4$ which does
  not depend on $\tau$.
\end{description}

Existence of a membrane ${\eta }^a (\tau, x^{\mu})$, satisfying the requirement
(ii) and the classical equations of motion, can be demonstrated
as follows. The relation ${\dot \eta}^a = {\dot X}^{\mu} {\p}_{\mu} {\eta}^a$
can be
written (after expressing ${\dot X}^{\mu} = {\dot \eta}_a {\p}^{\mu} {\eta}^a$)
also in the form ${\dot \eta}^c {b_c}^a = 0$, where ${b_c}^a \equiv 
{{\delta}_c}^a - {\p}^{\mu} {\eta}_c \, {\p}_{\mu} {\eta}^a$.
At an initial time $\tau = 0$, the membrane
${\eta }^a (0, x^{\mu})$ and its velocity ${\dot \eta }^a (0, x^{\mu})$
can be specified arbitrarily. Let us choose the initial data such that
\be
   {\dot \eta}^c (0) {b_c}^a (0) = 0 ,
\label{III3.B1} \ee
\be
  {\ddot \eta}^c (0) {b_c}^a (0) + {\dot \eta}^c (0) {{\dot b}_c}^{\,a} (0) = 0,
\label{III3.B2}
\ee
where ${\ddot \eta}^a (0)$ can be expressed in terms of ${\eta}^a (0)$ and
${\dot \eta}^a (0)$ by using the classical equations of motion \ci{42,43}.
At an infinitesimally later
instant $\tau = \Delta \tau$ we have ${\eta}^a (\Delta \tau) =
{\eta}^a (0) + {\dot \eta}^a (0) \Delta \tau$, ${\dot \eta}^a (\Delta \tau) =
{\dot \eta}^a (0) + {\ddot \eta}^a (0) \Delta \tau$, and
\be
    {\dot \eta}^c (\Delta \tau) {b_c}^a (\Delta \tau) =
    {\dot \eta}^c (0) {b_c}^a (0) + \left [ {\ddot \eta}^c (0) {b_c}^a (0) +
    {\dot \eta}^c (0) {{\dot b}_c}^{\,a} (0) \right ] \, \Delta \tau .
\label{III3.B3}
\ee
Using (\ref{III3.B1}), (\ref{III3.B2}) we find that eq.\ (\ref{III3.B3}) becomes
\be
    {\dot \eta}^c (\Delta \tau) {b_c}^a (\Delta \tau) = 0 .
\label{III3.B4}
\ee
From the classical equations of motion \ci{42,43} it follows that also
\be
{\ddot \eta}^c (\Delta \tau) {b_c}^a (\Delta \tau) + 
{\dot \eta}^c (\Delta \tau) {{\dot b}_c}^{\,a} (\Delta \tau) = 0.
\label{III3.B4a}
\ee
By continuing
such a procedure step by step we find that the relation
${\dot \eta}^c {b_c}^a = 0$ holds at every $\tau$. This proves that a membrane
satisfying the initial conditions (\ref{III3.B1}),(\ref{III3.B2}) and the equations
of motion indeed moves ``within itself".

\figIIIG

Now let us consider a special form of 
the wave packet as illustrated in\inxx{wave packet}
Fig.\,10.2. Within the effective boundary $B$ a given function ${\eta}^a (x)$
is admissible with high probability, outside $B$ with low probability.
Such is, for instance, a Gaussian wave packet which, at the initial $\tau
= 0$, is given by \inxx{wave packet,Gaussian}
  $$\psi [0, {\eta}^a (x)] = A \, {\rm exp} \left [- \, \int \, d^4 x \sqrt{|f|} \,
  {{\omega} \over {\Lambda}} \left ({\eta}^a (x) -
  {\eta}_0^a (x) \right )^2 {1 \over {2 \sigma (x)}} \right ] \hs{1cm}$$
\be  
   \times \, {\rm exp} \left [ i \int d^4 x \, p_{0a} (x) 
  \left ({\eta}^a (x) - {\eta}_0^a (x) \right ) \right ] \hs{1mm} \, ,
\label{III3.28}
\ee
where $p_{0c} (x)$ is momentum and ${\eta}_0^a (x)$ position
of the ``centre" of the wave packet at $\tau = 0$. The function
$\sigma (x)$ vary with $x^{\mu}$ so that the wave packet corresponds to 
Fig. 10.2. \inxx{wave packet,centre of}

\figIIIH

Of special interest in Fig. 10.2  is the region $P$ around a spacelike
hypersurface $\Sigma$ on ${\cal V}_4^{(c)}$. In that region the wave functional
\inxx{wave functional}
is much more sharply localized than in other regions (that is, at other
values of $x^{\mu}$). This means that in the neighborhood of $\Sigma$
a spacetime sheet ${\cal V}_4$ is relatively well defined. On the
contrary, in the regions that we call {\it past} or {\it future},
space-time is not so well defined, because the wave packet is spread over a
relatively large range of functions ${\eta}^a (x)$ (each representing a
possible spacetime sheet ${\cal V}_4$).

The above situation holds at a certain, let us say initial, value of the
evolution parameter $\tau$. Our wave packet satisfies the Schr\" odinger
\inxx{evolution parameter}\inxx{localization,within a space like region}equation
and is therefore subjected to evolution. The region of sharp
localization depends on $\tau$, and so it moves as $\tau$ increases.
In particular, it can move within the mathematical spacetime
surface $V_4$ which corresponds to such a ``centroid" (physical) spacetime
sheet ${\cal V}_4^{(c)} = \langle {\cal V}_4 \rangle$
which ``moves within itself" (case (ii) above). Such a solution of the
Schr\" odinger equation provides, on the one hand,
the existence of a fixed spacetime
$V_4$, defined within the resolution of the wave packet (see Fig.\,10.2), and,
on the other hand, the existence of a moving region $P$ in which the
wave packet is more sharply localized.
The region $P$ represents the ``present" of an observer.
We assume
that an observer measures, in principle, the embedding positions ${\eta}^a (x)$
of the entire spacetime sheet. Every ${\eta}^a (x)$ is
possible, in principle. However, in the practical situations available
to us a possible measurement procedure is expected to be such that
only the embedding positions
${\eta}^a (x_{\Sigma}^{\mu})$ of a simultaneity hypersurface $\Sigma$ are
measured with high precision\footnote
{Another possibility is to measure the induced metric 
$g_{\mu \nu}$ on ${\cal V}_4$, and measure ${\eta}^a (x)$ merely with 
a precision at a cosmological scale or not measure it at all.},
whereas the embedding positions ${\eta}^a (x)$ of all other
regions of spacetime sheet are measured with low precision. As a
consequence of such a measurement a wave packet like one of Fig.\,10.2
and eq.\,(\ref{III3.28}) is formed and it is then subjected to the unitary
$\tau$-evolution given by the covariant functional Schr\" odinger
equation (\ref{III3.24}).

Using our theory, which is fully covariant with respect to
reparametrizations of spacetime coordinates $x^{\mu}$, we have thus
arrived in a natural way at the existence of a time slice $\Sigma$
\inxx{time slice}
which corresponds to the ``present" experience and which progresses
forward in spacetime. The theory of Sec.10.1 is just a particular
case of this more general theory. This can be seen by taking in
the wave packet (\ref{III3.28}) the limit
\be
    {1 \over {\sigma(x)}} = {{\delta (x^0 - x_{\Sigma}^0)} \over
    {\sigma (x^i) ({\p}_0 {\eta}^a {\p}_0 {\eta}_a)^{1/2}}} \;  ,
    \quad \quad i = 1,2,3\; ,
\label{III3.limit1}
\ee
and choosing
\be
   p_{0a} (x^{\mu}) = p_{0a}^{(3)} (x^i)
   \delta (x^0 - x_{\Sigma}^0) .
\label{III3.limit2}
\ee   
Then the integration over the
$\delta$-function gives in the exponent the expression
$$\int {\dd}^3 x \sqrt{|{\bar f}|} {{\omega} \over {\Lambda}}\,
   {\left ({\eta}^a (x^i) - 
   {\eta}_0^a (x^i) \right )}^2 /2 \sigma(x^i) + 
   i \int {\dd}^3 x \, p_{0a} (x^i) \left ({\eta}^a (x^i) -
  {\eta}_0^a (x^i) \right )$$
so that eq.\,(\ref{III3.28}) becomes a wave functional of a 3-dimensional membrane
${\eta}^a (x^i)$

So far we have taken the region of sharp localization $P$ of a
wave functional packet situated around a spacelike surface $\Sigma$, and
so we have obtained a time slice. But there is a difficulty with the
concept of ``time slice" related to the fact that an observer in
\inxx{time slice}
practice never has access to the experimental data on an
entire spacelike hypersurface. Since the signals travel with the
final velocity of light there is a delay in receiving information.
Therefore, the greater is a portion of a space-like hypersurface,
the longer is the delay. This imposes limits on the extent of
a space-like region within which the wave functional packet (\ref{III3.28})
can be sharply localized. The situation in Fig. 10.2 is just an
idealization. A more realistic wave packet, illustrated in Fig. 10.3, is
sharply localized around an finite region of spacetime. 
It can still be represented by the expression (\ref{III3.28}) with a
suitable width function $\sigma (x)$.

A possible interpretation is that such a wave packet of Fig. 10.3 represents
a state which is relative (in the Everett sense\footnote
{The Everett interpretation \cite{Everett} of quantum mechanics
which introduces\inxx{Everett interpretation}
the concept\inxx{quantum gravity} of a wave function relative to a registering device is not so
unpopular among cosmologists. Namely, in quantum gravity the whole
universe is considered to be described (of course, up to a sensible
approximation) by a single wave function, and there is no outside
observer who could measure it.})
to a state of a registering device with memory of past records. The
region of sharp localization of such a relative state is centered within
the registering device.

\figIIII
 
In summary, our model predicts: 

\medskip
\ (i) existence of a spacetime continuum $V_4$ 
without evolution in $\tau$ (such is a spacetime of the conventional
special and general relativity);

(ii) a region $P$ of spacetime which changes its position on $V_4$ while
the evolution time $\tau$ increases.

\medskip
Feature (i) comes from the expectation value of our wave packet, and
feature (ii) is due to its peculiar shape (Fig.\ 10.2 or Fig.\ 10.3)
which evolves in $\tau$.

\subsection[\ \ Inclusion of sources]
{Inclusion of sources}

In the previous chapter we have included point particle sources in the
embedding model of gravity (which was based on the usual constrained
$p$-brane theory). This was achieved by including in the action for
a spacetime sheet a function $\omega (\eta)$ which consists of a
constant part and a $\delta$-function part. In an analogous way we
can introduce sources into our unconstrained embedding model which
has explicit $\tau$-evolution.

For $\omega$ we can choose the following function of the embedding
space coordinates ${\eta}^a$:
\be
     \omega (\eta ) = {\omega}_0 + \sum_i \int m_i \, {\delta}^N
     (\eta - {\hat \eta}_i ) \sqrt{|{\hat f}|} \, {\dd}^m {\hat x} \, ,
\label{III3.29}
\ee
where ${\eta}^a =  {\hat \eta}_i^a (\hat x)$ is the
parametric equation of an $m$-dimensional surface
${\widehat V}_m^{(i)}$, called {\it matter sheet}, also embedded in
$V_N$; ${\hat x}^{\hat \mu}$ are parameters (coordinates) on
${\widehat V}_m^{(i)}$ and ${\hat f}$ is the determinant of the induced
metric tensor on ${\widehat V}_m^{(i)}$\,. If we take $m = N - 4 +1$,
then the intersection of $V_4$ and ${\widehat V}_m^{(i)}$ can be a
(one-dimensional) line, i.e., a worldline $C_i$ on $V_4$. If $V_4$
\inxx{world line,as an intersection} \inxx{spacetime sheet,moving}
moves in $V_N$, then the intersection $C_i$ also moves. A moving
spacetime sheet was denoted by ${\cal V}_4$ and described by
$\tau$-dependent coordinate functions ${\eta}^a (\tau,x^{\mu})$.
Let a moving worldline be denoted ${\cal C}_i$. It can be described
either by the coordinate functions ${\eta}^a (\tau, u)$ in the
embedding space $V_N$ or by the coordinate functions
$X^{\mu} (\tau, u)$ in the moving spacetime sheet ${\cal V}_4$.
Besides the evolution parameter $\tau$ we have also a
1-dimensional worldline parameter $u$ which has an analogous
role as the spacetime sheet parameters $x^{\mu}$ in
${\eta}^a (\tau ,x^{\mu})$. At a fixed $\tau$, $\, X^{\mu} (\tau, u)$
gives a 1-dimensional worldline $X^{\mu} (u)$. If $\tau$
increases monotonically, then the worldlines continuously change or
{\it move}. In the expression (\ref{III3.29}) $m-1$ coordinates
${\hat x}^{\hat \mu}$ can be integrated out and we obtain
\be
     \omega = {\omega}_0 + \sum_i \int m_i {{{\delta}^4 (x - X_i)}
     \over {\sqrt{|f|}}} \left ( {{{\dd} X_i^{\mu}} \over {{\dd}{u}}}
     {{{\dd} X_i^{\nu}} \over {{\dd}{u}}} \, f_{\mu \nu} \right )
     ^{1/2} {\dd} u \, ,
\label{III3.30}
\ee
where $x^{\mu} = X_i^{\mu} (\tau,u)$ is the parametric equation
of a ($\tau$-dependent) worldline ${\cal C}_i$ , $u$ an
arbitrary parameter on ${\cal C}_i \, ,  \quad f_{\mu \nu} \equiv
{\partial }_{\mu} {\eta}^a {\partial }_{\nu} {\eta}_a $ the induced metric
on ${\cal V}_4$ and $f \equiv \mbox{\rm det} f_{\mu \nu}$. By
inserting eq.\ (\ref{III3.30}) into the membrane's action (\ref{III3.19}) we obtain
the following action
\bear
   I[X^{\mu} (\tau, u)] &=&  I_0 + I_{\rm m}  \label{III3.31} \\
   \nonumber \\
   &=& {{{\omega}_0} \over 2}  \int {\mbox{\rm d}} \tau \, 
  {\mbox{\rm d}}^4 x \, \sqrt{|f|} \left ({{ {\dot \eta}^a
  {\dot \eta}_a} \over {\Lambda}} + \Lambda \right ) \nonumber \\
  \nonumber \\
   & & + {1 \over 2} \int {\mbox{\rm d}}
  \tau \, {\mbox{\rm d}}^4 x \sqrt{|f|} \sum_i
  \left (
  {{{\dot X}_i^{\mu} {\dot X}_i^{\nu} f_{\mu \nu}} \over {\Lambda}} +
  \Lambda \right ) \nonumber \\
  \nonumber \\
  & &\hs{11mm} \times m_i \, {{{\delta}^4 (x - X_i)} \over {\sqrt{|f|}}}
  {\left ( {{{\dd} X_i^{\mu}} \over {{\dd}{u}}}
  {{{\dd} X_i^{\nu}} \over {{\dd} {u}}} \, f_{\mu \nu} \right ) }^{1/2}
  {\dd} \lambda . \nonumber
\ear
In a special case where the membrane ${\cal V}_4$ is static with
respect to the evolution in $\tau$, i.e., all $\tau$ derivatives are
zero, then we obtain, taking $\Lambda = 2$,
the usual Dirac--Nambu--Goto 4-dimensional membrane
coupled to point particle sources
\be
   I[X^{\mu} (u)] = {\omega}_0 \int {\dd}^4 x \, \Lambda \, \sqrt{|f|} +
   \int {\dd} u \sum_i m_i 
   \left ( {{{\dd} X_i^{\mu}} \over {{\dd}{u}}}
   {{{\dd} X_i^{\nu}} \over {{\dd} {u}}} \, f_{\mu \nu} \right )
 ^{1/2} .
 \label{III3.32}
 \ee
However, the action (\ref{III3.31}) is more general than (\ref{III3.32}) and
 it allows for solutions which evolve in $\tau$. The first part
 $I_0$ describes a 4-dimensional membrane which evolves in $\tau$,
 whilst the second part $I_{\rm m}$ describes a system of (1-dimensional)
 worldlines which evolve in $\tau$. After performing the integration
 over $x^{\mu}$, the `matter' term $I_{\rm m}$ becomes ---in the case of
 one particle--- analogous to the membrane's term $I_0$:
 \be
     I_{\rm m} = {1 \over 2} \int {\dd} \tau \, {\dd} u \, m
     \left ( {{{\dd} X^{\mu}} \over {{\dd}{u}}}
     {{{\dd} X^{\nu}} \over {{\dd} {u}}} \, f_{\mu \nu} \right )
     ^{1/2}
     \left (
    {{{\dot X}^{\mu} {\dot X}^{\nu} f_{\mu \nu}} \over {\Lambda}} +
     \Lambda \right ) .
\label{III3.33}
\ee
Instead of 4 parameters (coordinates) $x^{\mu}$ we have in (\ref{III3.33})
a single parameter $u$, instead of the variables ${\eta}^a (\tau,x^{\mu})$
we have $X^{\mu} (\tau,u)$, and instead of the determinant of the
4-dimensional induced metric $f_{\mu \nu} \equiv {\partial}_{\mu}
{\eta}^a {\partial}_{\nu} {\eta}_a$ we have
$({\dd} X^{\mu}/{\dd} u) ({\dd} X_{\mu}/{\dd} u)$.
All we have said about the theory of an unconstrained $n$-dimensional
membrane evolving in $\tau$ can be straightforwardly applied to a
worldline (which is a special membrane with $n=1$).

After inserting the matter function $\omega (\eta )$ 
of eq.\ (\ref{III3.29}) into
the Hamilton\-ian (\ref{III3.21}) we obtain \inxx{Hamiltonian}
\bear
   &&H = {{1 \over 2} \int {\dd}^4 x \sqrt{|f|} {{\Lambda} \over {{\omega}_0}}
    \left ( {{p^{(0) a} (x) p_a^{(0)} (x)} \over {|f|}} -
    {\omega}_0^2 \right ) \Biggl\vert }_{x \not\in {\cal C}_i}\hs{1.5cm}
    \nonumber \\    
    && \hs{7mm} + \, {1 \over 2} \sum_i \dd u
     \left ( {{{\dd} X_i^{\mu}} \over {{\dd} u}}
     {{{\dd} X_{i \mu}} \over {{\dd} u}} \right )
     ^{1/2} {{\Lambda} \over {m_i}} \left ( P_{\mu}^{(i)} P^{(i) \mu}
     - m_i^2 \right ) \, ,
\label{III3.34}
\ear
where $p_a^{(0)} = {\omega}_0 {\Lambda}^{-1} \sqrt{|f|} \,
{\dot \eta}_a$ is
the membrane's momentum everywhere except on the intersections
$V_4 \cap {\widehat V}_m^{(i)}$, and $P_{\mu}^{(i)} = m_i {\dot X}_{i \mu}/
{\Lambda}$ is the membrane momentum on the intersections
$V_4 \cap {\widehat V}_m^{(i)}$. In other words, $P_{\mu}^{(i)}$ is the
momentum of a worldline ${\cal C}_i$ . The contribution of the
wordlines is thus explicitly separated out in the Hamiltonian (\ref{III3.34}).

In the quantized theory a membrane's state is represented by a wave
functional which satisfies the Schr\" odinger equation (\ref{III3.24}).
\inxx{wave packet} \inxx{wave functional}A wave packet 
(e.g., one of eq.\ (\ref{III3.28})) contains, in the case of
$\omega (\eta)$ given by eq.\ (\ref{III3.29}), a separate contribution of
the membrane's portion outside and on the intersection
$V_4 \cap {\widehat V}_m^{(i)}$:
\bear
& &\hs{3mm}\psi [0,\eta(x)] = {\psi}_0 [0,\eta(x)] {\psi}_m [0,X(u)] 
\label{III3.35a} \\
\nonumber \\
 & &{\psi}_0 [0, {\eta}^a (x)] = A_0 \, {\rm exp} \left [- \, \int \, d^4 x 
 \sqrt{|f|} \,
  {{\omega} \over {\Lambda}} \left ({\eta}^a (x) -
  {\eta}_0^a (x) \right )^2 {1 \over {2 \sigma (x)}} \right ]
  \hs{1.2cm} \nonumber \\
  \nonumber \\
  &&  \hs{2.4cm} \times \, {\rm exp} \left [ i \int d^4 x \, p_{0a} (x) 
  \left ({\eta}^a (x) -
  {\eta}_0^a (x) \right ) \right ] {\Biggl\vert }_{x \not\in
  {\cal C}_i}
\label{III3.35b} \\
\nonumber \\  
&&{\psi}_m [0,X(u)] = A_m \, {\rm exp} \Biggl[- \sum_i \int d u {{m_i} 
  \over {\Lambda}} 
  {\left ( {{d X_i^{\mu}} \over {d u}}
  {{d X_i^{\nu}} \over {d u}} \, f_{\mu \nu} \right ) }^{1/2} 
   \hs{1cm} \nonumber \\
  \nonumber \\  
  &&\hs{5.5cm} \times \, {\left ( X_i^{\mu} (u) - X_{0i}^{\mu} (u) \right ) }^2
  {1 \over {2 {\sigma}_i (u)}}\Biggr] \nonumber \\
  \nonumber \\
  &&  \hspace{2.5cm} \times \, {\rm exp} \Biggl[{i \int d u \left ( {{d X_i^{\mu}} 
  \over {d u}}
  {{d X_i^{\nu}} \over {d u}} \, f_{\mu \nu} \right ) }^{1/2} \nonumber \\
  \nonumber \\
  && \hs{4.4cm} \times \, P_{0 \mu}^{(i)} (u)
  \left ( X_i^{\mu} (u) - X_{0i}^{\mu} (u) \right ) \Biggr] . 
  \label{III3.35c}
\ear
In the second factor of eq.\ (\ref{III3.35a}) the wave packets of worldlines are
expressed explicitly (\ref{III3.35c}).
For a particular $\sigma (x)$, such that a wave
packet has the form as sketched in Fig. 10.2 or Fig. 10.3, there exists a region
$P$ of parameters $x^{\mu}$ at which the membrane ${\cal V}_4$ is much more
sharply localized than outside $P$. The same is true for the intersections
(which are wordlines): any such worldline ${\cal C}_i$  is much more
sharply localized in a certain interval of the worldline parameter
$u$. With the passage of the evolution time $\tau$ the region of
\inxx{sharp localization} \inxx{evolution time}
sharp localization on a worldline moves in space-time $V_4$.
In the limit (\ref{III3.limit1}),\ (\ref{III3.limit2})
 expression (\ref{III3.35c})
becomes a wave packet of a point particle
(event) localized in space-time. The latter particle
is just an unconstrained point particle,
a particular case (for $p=0$) of a generic unconstrained $p$-dimensional
membrane described in Sec. 7.1.

The unconstrained theory of point particles has a long history. It was
\inxx{unconstrained theory}
considered by Fock, Stueckelberg, Schwinger, Feynman, Horwitz, Fanchi,
Enatsu, and many others \cite{1}--\ci{11a}.
Quantization of the theory had appeared under various
names, for instance the Schwinger proper time method or the parametrized
relativistic quantum theory. The name unconstrained theory is
used in Ref. \cite{11,11a}
both for the classical and the quantized theory. In the last few years
similar ideas in relation to canonical quantization of
gravity and Wheeler DeWitt equation were proposed by Greensite and
Carlini \cite{29}. In a very lucid paper Gaioli and Garcia-Alvarez
\ci{29a} have
convincingly explained why the invariant evolution parameter is necessary
in quantum gravity (see also ref. \ci{29b}).

\subsubsection{Conclusion}

We have formulated a reparametrization invariant and Lorentz invariant theory
of $p$-dimensional membranes without constraints on the dynamical variables.
This is possible if we assume a generalized form of the Dirac--Nambu--Goto
action, such that the dependence of the dynamical variables on an extra
parameter, the evolution time $\tau$, is admitted. In Sec. 4.2 we have
seen that the unconstrained theory naturally arises within the larger
framework of a constrained theory formulated by means of geometric
calculus based on Clifford algebra which incorporates polyvectors (Clifford
aggregates) as representing physical quantities.

Such a membrane theory manifests its full power in the embedding model
of gravity, in which space-time is treated as a 4-dimensional unconstrained
membrane, evolving in an $N$-dimensional embedding space. The embedding
model was previously discussed within the conventional theory of
constrained membranes \cite{71}. Release of the constraints and introduction of
the $\tau$ evolution brings new insight into the quantization of the model.
Particularly interesting is a state represented by a functional of
4-dimensional membranes ${\cal V}_4$, localized around an average space-time
membrane ${\cal V}_4^{(c)}$, and even more sharply localized around a
finite segment of a
space-like surface $\Sigma$ on ${\cal V}_4^{(c)}$. Such a state incorporates
the existence of a classical space--time continuum and the evolution in it. The
notorious problem of time \cite{time} is thus resolved in our approach 
to quantum \inxx{problem of time}
gravity. The space-time coordinate $x^0 = t$ is not time\footnote{
In this sentence `time' stands for the parameter of evolution. 
Such is the meaning of the word `time' adopted by the authors who discuss
the problem of \inxx{parameter of evolution} \inxx{evolution time}
time in general relativity. What they say is essentially just that
there is a big problem because the coordinate $x^0$  does not 
appear at all in the Wheeler DeWitt equation and hence cannot have the role
of an evolution parameter (or `time,' in short).  In our work, following Horwitz
\cite{5a},
we make explicit distinction between the coordinate $x^0$ and the
parameter of evolution $\tau$ (which indeed takes place in our classical and
quantum equations of motions). These two distinct concepts are usually
mixed and given the same name `time'. In order to distinguish them, we use
the names `{\it coordinate time}' and `{\it evolution time}'.}
at all! Time must
be separately introduced, and this has been achieved in our theory in which the
action depends on the evolution time $\tau$. \inxx{evolution time}
The importance of the evolution time has been considered, in the case
of a point particle,
by many authors \cite{1}--\ci{11a}. 

Our embedding model incorporates sources in a natural way.
Worldlines occur in our embedding model as intersections of space--time
membranes ${\cal V}_4$ with $(N-4+1)$-dimensional ``matter" sheets or as
the intersections of $V_4$ with itself. In the
quantized theory the state of a worldline can be represented by a wave
functional ${\psi}_m [\tau,X^{\mu}(u)]$ ,
which may be localized around an average worldline (in the
quantum mechanical sense of the expectation value). Moreover, at a certain
value $u = u_P$ of the worldline parameter the wave functional
may be much more sharply localized than at other values of $u$,
thus approximately imitating the wave function of a point particle
(or event) localized in space-time.  And since ${\psi}_m [\tau,X^{\mu}(u)]$
evolves with $\tau$, the point $u_P$ also changes with $\tau$.

The embedding model, based on the theory of unconstrained
membranes satisfying the action (\ref{III3.19}), appears to be a promising
candidate for the theoretical formulation of quantum gravity including
bosonic sources. Incorporation of fermions is expected to be
achieved by taking into account the Grassmann coordinates or by employing
the polyvector generalization of the underlying formalism.

\part{Beyond the Horizon}

\chapter[The landscape of theoretical physics: \ \\ a global view]
{The Landscape of theoretical \\
physics: a global view}

In the last Part, entitled ``Beyond the Horizon", I am going to discuss
conceptual issues and the foundations of theoretical physics. I shall try
to outline a broader\footnote{
The outline of the view will in many respects be indeed ``broader" and
will go beyond the horizon.}
view of the theoretical physics landscape as I see it, and, as seems to me, is
becoming a view of an increasing number of researchers. The introductory
chapter of Part IV, bearing the same title as the whole book, is an
overview aimed at being understandable to the widest possible circle
of readers. Therefore use of technical terminology and jargon will be
avoided. Instead, the concepts and ideas will be explained by analogies and
illustrative examples. The cost, of course, is a reduced scientific
rigor and precision of expression. The interested reader who seeks
a more precise scientific explanation will find it (but without much
maths and formulas) in the next chapters, where many concepts will be
discussed at a more elaborate level.

Throughout history people have been always inventing various cosmological
models, and they all have always turned out to be wrong, or at least
incomplete. Can we now be certain that a similar fate does not await
the current widely accepted model, according to which the universe was
born in a ``big bang"? In 1929 an american astronomer Edwin Hubble
discovered that light coming from galaxies is shifted towards the
red part of the spectrum, and the shift increases with galactic distance.
If we ascribe the red shift to galactic velocity, then Hubble's discovery
means that the universe is expanding, since the more distant a galaxy is from
us the greater is its velocity; and this is just a property of expansion.
Immediately after that discovery Einstein recognized that his equation
for gravity admitted precisely such a solution which represented the expansion
of a universe uniformly filled with matter. In fact, he had already come to just
such a result in 1917, but had rejected it because he had
considered it a nonsense, since an expanding universe was in disagreement
with the static model of the universe widely accepted at that time.
In 1917 he had preferred to modify his equation by adding an extra term containing
the so called ``cosmological constant". He had thus missed the opportunity
of predicting Hubble's discovery, and later he proclaimed his episode
with the cosmological constant as the biggest blunder in his life.

General relativity is one of the most successful physical theories.
It is distinguished by an extraordinary conceptual elegance, simplicity
of the basic postulates, and an accomplished mathematical apparatus, whilst
numerous predictions of the theory have been tested in a variety of
important and well known experiments. No experiment of whatever kind
has been performed so far that might cast doubt on the validity of
general relativity. The essence of the theory is based on the
assumption (already well tested in special relativity) that space and time
form a four-dimensional continuum named {\it spacetime}. In distinction
with special relativity, which treats spacetime as a {\it flat} continuum,
in general relativity spacetime can be curved, and curvature is
responsible for gravitational phenomena. How spacetime is curved is
prescribed by Einstein's equation. Strictly speaking, Einstein's equations
determine only in which many different possible ways spacetime can be
curved; how it is  actually curved we have to find out at ``the very place".
But how do we find this? By observing particles in their motion. If we
are interested in spacetime curvature around the Sun, then such particles
are just planets, and if we are interested in the curvature of the Universe
as the whole, then such particles are galaxies or clusters of galaxies.
In flat spacetime, in the absence of external forces, all particles move
uniformly along straight lines, whilst in a curved spacetime particles
move non-uniformly and in general along curved lines. By measuring
the relative acceleration and velocity of one particle with respect to another,
nearby, particle we can then calculate the curvature of spacetime in a given
point (occupied by the particle). Repeating such a procedure we can
determine the curvature in all sample points in a given region of spacetime.
The fact that a planet does not move along a straight line, but along
an elliptic trajectory, is a consequence of the curvature of spacetime 
around the Sun. The gravitational ``force" acting on a planet is a consequence
of the curvature. This can be illustrated by an example of a curved
membrane onto which we throw a tiny ball. The ball moves along a curved
trajectory, hence a force is acting on the ball. And the latter force results
from the membrane's curvature.

Another very successful theory is quantum mechanics. Without quantum mechanics
we would not be able to explain scattering of electrons by crystals,
nor the ordered stable crystal structure itself, nor the properties
of electromagnetic waves and their interactions with matter. The widely
known inventions of today, such as the laser, semiconductors, and transistors,
have developed as a result of understanding the implications of
quantum mechanics. Without going into too much detail, the
essence of quantum mechanics, or at least one of its essential points,
can be summarized in the following simplified explanation. There exists
a fundamental uncertainty about what the universe will be like at a
future moment. This uncertainty is the bigger, as more time passes after
a given moment. For instance, it is impossible to predict precisely at
which location an electron will be found, after leaving it to move undisturbed
for some time. When we finally measure its position it will be,
in principle, anywhere in space; however, the probability
of finding the electron will be greater at some places than at others.
To everyone of those possible results of measurements there corresponds
a slightly different universe. In classical, Newtonian, physics the
uncertainty about the future evolution of the universe is a consequence of
the uncertainty about the present state of the universe. If the present state 
could be known precisely, then also the future evolution of 
the universe could be precisely calculated. The degree of precision about
the prediction of the future is restricted by the degree of precision with
which the initial conditions are determined. (I am intentionally speaking
about the whole universe, since I wish to point out that the size
of the observed system and its complexity here does not, in principle,
play any role.) In quantum mechanics, on the contrary, such uncertainty is
of quite a different kind from that in classical mechanics. No matter how
precisely the present state of an observed system is known, the uncertainty
about what position of the particles we shall measure in the future remains.
A generic state of a system can be considered as a superposition of a certain 
set of basis
states. It can be described by the {\it wave function} which enables calculation
of {\it the probability} to observe a definite
quantum state upon measurement. The latter state is just one amongst the states belonging
to the set of basis states, and the latter {\it set} itself is determined
by the measurement situation. Such a probability or statistical 
interpretation of quantum mechanics was unacceptable for Einstein, who
said that ``God does not play dice". And yet everything points 
to him having been
wrong. So far no experiment, no matter how sophisticated, 
has disproved
the probability interpretation, whilst many experiments have 
eliminated various rival 
interpretations which assume the existence of some ``hidden variables"
supposedly responsible for the unpredictable behavior of quantum systems.

\vs{5mm}

\centerline{* * *}

\vs{4mm}

We thus have two very successful theories, {\it general relativity} on the
one hand, and {\it quantum mechanics} on the other, which so far have
not been falsified by any experiment. What is then more natural than to
unify those two theories into a single theory? And yet such a unification
has not yet been successfully achieved. The difficulties are conceptual
as well as mathematical and technical. As it appears now, final success
will not be possible without a change of paradigm. Some of the basic principles
the two theories rest on will have to be changed or suitably generalized.
Certain significant moves in this direction have already been made. In
the following I will briefly, and in a simplified way, discuss some of those,
in my opinion, very important approaches. Then I will indicate how those
seemingly unconnected directions of research lead towards a possible solution
of the problem of quantum gravity, and hence towards an even more profound
understanding of the universe and the role of an intelligent observer in it.

Before continuing, let me point out that some epochs in history are
more ready for changes, other less. The solution of a certain basic
scientific problem or a significantly improved insight into the nature
of Nature is nearly always a big shock for those who have been used 
to thinking in the old terms,
and therefore do their best to resist the changes, while regretfully they do
not always use the methods of scientific argument and logic only.
Copernicus did not publish his discoveries until coming close to his death,
and he had reason for having done so. The idea that the whole Earth, together
with the oceans, mountains, cities, rivers, is moving around the Sun, was
too much indeed! Just as were Wegener's theory about the relative motions of 
the continents, Darwin's theory about the origin and evolution of the species,
and many other revolutionary theories. I think that we could already have 
learned something from the history of science and be now slightly more
prudent while judging new ideas and proposals. At least the ``arguments" that
a certain idea is much too fantastic or in disagreement with common sense
should perhaps not be used so readily. The history of science has taught us
so many times that many successful ideas were just such, namely 
at first sight crazy, therefore in the future  we should avoid such a ``criterion"
of judging the novelties and rather rely less on emotional, and more on
scientific criteria. The essence of the latter is a cold, strictly rational
investigation of the consequences of the proposed hypotheses and verification
of the consequences by experiments. However, it is necessary to have in mind
that a final elaboration of a successful theory takes time. Many researchers
may participate in the development and every contribution is merely a piece
of the whole. Today it is often stressed that a good theory has to able
to incorporate all the known phenomena and predict new ones, not yet
discovered. This is, of course, true, but it holds for a finished
theory, and not for the single contributions of scientists who enabled
the development of the theory.

In 1957 the American physicist Hugh Everett \ci{Everett} 
successfully defended his PhD thesis
and published a paper in which he proposed that all the possibilities,
implicit in the wave function, actually exist. In other words, all the possible
universes incorporated in the wave function actually exist, together with
all the possible observers which are part of those universes. 
In addition to that,
Everett developed the concept of {\it relative state}.\inxx{relative state}
Namely, if a given
physical system consists of two mutually interacting subsystems, then each
of them can be described by a wave function which is relative to
the possible states of the other subsystem. As one subsystem we can take,
for example, an intelligent observer, and as the other subsystem the rest
of the universe. The wave function of the remaining universe is relative
to the possible states of the observer. The quantum mechanical
correlation, also known under the name ``entanglement", is established
amongst the possible quantum states of the observer and the possible quantum
states of the remaining universe. As an example let us consider an observer
who measures the radioactive gamma decay of a low activity source with
short life time. A Geiger counter which detects the particles (in our example
these are photons, namely gamma rays) coming from the source will then
make only single sounds, e.g., one per hour. Imagine now that we have
isolated a single atom containing the nucleus of our radioactive source.
At a given moment the wave function is a superposition of two quantum
states: the state with photon emission and the state without the photon
emission. The essence of Everett's thesis (for many still unacceptable today)
lies in assuming that the states of the Geiger counter, namely
the state with the sound and the state without the sound, also enter
the superposition. Moreover, even the states of the observer, i.e., the
state in which the observer has heard the sound and the state in which
the observer has not heard the sound, enter the superposition. In this example 
the quantum correlation manifests itself in the following. To the state
in which the observer became aware\footnote{In this example we are using 
a {\it male} observer and the source of gamma rays. In some other example 
we could use a {\it female} observer and laser beams instead. In fact,
throughout the book I am using female or male
observers interchangely for doing experiments for my illustrations.
So I avoid using rather
cumbersome (especially if frequently repeated) ``he or she", but use 
``he" or ``she" instead. When necessary, ``he" may stand for a generic
observer. Similarly for ``she".}
that he has heard the sound there corresponds the state in which the detector
has detected a photon, and to the latter state, in turn, there correspond
the state in which the excited nucleus has emitted the photon. And similarly,
to the state in which the observer has not heard the sound, there corresponds
the state in which the detector has not detected and the source
has not emitted a photon. Each of those two chains of events belongs to
a different universe: in one universe the decay has happened and the observer
has perceived it, whilst in the other universe at the given moment there was
no decay and the observer has not perceived the decay. The total wave function
of the universe is a superposition of those two chains of events. In
any of the chains, from the point of view of the observer, there is no
superposition.

The Everett interpretation of quantum mechanics was strongly supported
by John Archibald Wheeler \ci{83}. Somewhat later he was joined by many others,
among them also Bryce DeWitt who gave the name 
``many worlds interpretation",\inxx{many worlds interpretation}
that is, the interpretation with many worlds or universes. Today the majority
of physicists is still opposed to the Everett interpretation, but it is
becoming increasingly popular amongst cosmologists.

Later on, Wheeler distanced himself from the Everett interpretation and
developed his own theory, in which he put the quantum principle as the
basis on which rests the creation and the functioning 
of the universe \ci{85}.
The {\it observer} is promoted to the {\it participator}, who not only
perceives, but is actively involved in, the development of the universe.
He illustrated his idea as follows. We all know the game ``twenty
questions". Person $A$ thinks of an object or a concept---and person B
poses questions to which the answer is {\it yes} or {\it no}. Wheeler
slightly changed the rules of the game, so that $A$ may decide what
the object is {\it after} $B$ asks the first question. 
After the second question $A$ may change the idea and choose another object,
but such that it is in agreement with his first answer. This continues
from question to question. The object is never completely determined,
but is only determined within the set of possible objects which are in
agreement with the questions posed (and the answers obtained) so far. 
However, with every new question the set of possible objects is narrowed,
and at the end it may happen that only one object remains. The player who
asked  questions,  with the very choice of her questions, has herself
determined
the set of possible answers and thus the set of possible objects.
In some way reality is also determined by the question we ask it.
The observer observes the universe by performing various measurements
or experiments. With the very choice of experiment she determines what
the set of possible results of measurement is, and hence what the set of
possible universes at a given moment is. The observer is thus involved in
the very creation of the universe she belongs to. In my opinion Wheeler's
approach is not in disagreement with Everett's, but completes it, just as it
also completes the commonly accepted interpretation of quantum mechanics.

Nowadays a strong and influential supporter of the Everett interpretation
is an Oxford professor David Deutsch. In his book {\it The Fabric of Reality}
\ci{86}
he developed the concept of 
\inxx{multiverse}{\it multiverse}, which includes all possible
universes that are admitted by a wave function. In a 1991 {\it Physical
Review} article \ci{87} he proved that the paradoxes of so called {\it time
machines} can be resolved by means of the Everett interpretation of quantum
mechanics. Many theoretical physicists study in detail some special
kinds of solutions to the Einstein equation, amongst them the best known
are {\it wormwholes}\ci{88}. These are special, topologically non-trivial, 
configurations of spacetime which under certain conditions allow for
{\it causal loops}. Therefore 
\inxx{causal loops}such solutions are called {\it time machines}.
A particle which enters a time machine will go back in time and meet
itself in the past. Such a situation is normally considered paradoxical
and the problem is how to avoid it. On the one hand, if we believe
the Einstein equations such time machines are indeed possible. On the other
hand, they are in conflict with the principle of causality, according to
which it is impossible to influence the past. Some researchers, therefore,
have developed a hypothesis of a self-consistent arrangement of events
which prevents a particle from meeting itself in the past; the time
machine may exist and a particle may enter it and travel back into the past,
but there is no means by which it can arrive at a point in spacetime at which
it had already been.
Others, with Stephen Hawking as the leader, on the contrary, 
are proving that quantum mechanics forbids the formation of time machines,
since the quantum fluctuations in the region of the supposed formation of
a time machine are so strong that they prevent the formation of the
time machine. However, Deutsch has shown that, exactly because of quantum
mechanics and the Everett interpretation, causal loops are not paradoxical
at all! Namely, a particle never travels a well defined trajectory, but its
quantum mechanical motion is spread around an average trajectory. According
to the Everett interpretation this means that there exist many copies of the
particle, and hence many universes which distinguish between themselves 
by the slightly
different positions the particle occupies in each and every 
of those universes. If a
particle travels in a time machine and meets its copy in the past, the result
of such a collision will be quantum mechanically undetermined within the
range of spreading of the wave function. To every possible pair of
directions to which the two particles can recoil  after the collision there 
corresponds a different universe. We have a causal paradox only if we assume
the existence of a single universe. Then the collision of a particle
with its own copy in the past necessarily changes {\it the initial history},
which is the essence of the causal paradox. But if we assume that a set of
universes exists, then there also exists a set of histories, and hence
a journey of a particle into the past does not imply any paradox at all.
A similar resolution \ci{89} of the causal paradox has also been proposed for
{\it tachyons}. Tachyons
\inxx{tachyon}are so far unobserved particles moving with
a  speed faster than light . The equations of relativity in principle admit
not only the existence of 
\inxx{bradyon}{\it bradyons} (moving slower than light) and
{\it photons} (moving with the speed of light), 
\inxx{tachyon}but also of tachyons.
But tachyons appear problematic in several respects\footnote{Some more
discussion about tachyons is provided in Sec. 13.1.},
mainly because they allow for the formation of causal loops.\inxx{causal loops} 
This is one of the main
arguments employed against the possibility that 
\inxx{tachyon}tachyons could be found in
nature. However, the latter argument no longer holds after assuming
the validity of the Everett interpretation of quantum mechanics, since then
causal loops are not paradoxical, and in fact are not ``loops" at all.

We have arrived at the following conclusion. If we take seriously the equations
of general relativity, then we have also to take seriously their solutions.
Amongst the solutions there are also such configurations of spacetime
which allow for the formation of causal loops. We have mentioned wormholes.
Besides, there also exists the well known G\" odel solution for spacetime
around a rotating mass. If such solutions are in fact realized in nature,
then we have to deal with time machines and such experimental situations,
which enables us to {\it test} the Everett interpretation of quantum mechanics.
In this respect the Everett interpretation distinguishes itself from the
other interpretations, including the conventional {\it Copenhagen interpretation}.
In the other experimental situations known so far the Everett interpretation
gives the same predictions about the behavior of physical systems as the
rival interpretations (including the Copenhagen interpretation).

\vs{5mm}

\centerline{* * *}

\vs{4mm}

Life, as we know it, requires the fulfilment of certain strict conditions.
It can develop only within a restricted temperature interval, and this can be
realized only on a planet which is at just the right distance from a star with
just the right activity and sufficiently long life time. If the fundamental
constants determining the strength of the gravitational, electromagnetic,
weak and strong forces were slightly different those conditions would
not have been met, the universe would be different to the extent that
a life of our kind would not be possible in it. In physics so far no
a reliable principle or law has been discovered according to which the
values of the fundamental constant could be determined. Just the contrary,
all values of those constants are possible in principle. The fact that they
are ``chosen" just as they are, has been attempted to be explained by 
the so called\inxx{anthropic principle}
{\it anthropic principle} \ci{90}. According to that principle there exists a 
fundamental relationship between the values of the fundamental constants and
our existence; our existence in the universe conditions the values of those 
constants. Namely, the world must be such that we the observers can exist in it
and observe it. However, by this we have not explained much, since the
question remains, why is the universe just such that it enables our life.
Here we can again help ourselves by employing the Everett 
interpretation which says
that everything which physically can happen actually does happen---in
some universe. The physical reality consists of a collection of universes.
There exist all sorts of universes, with various values of the fundamental
constants. In a vast majority of the universes life is not possible, but in 
few of them life is, nevertheless, possible, and in some universes life actually
develops. In one such universe we live. We could say as well that ``in
one of {\it those} universes we live". The small probability of the occurrence
of life is not a problem at all. It is sufficient that the emergence of life
is possible, and in some universes life would have actually developed.
Hence in the Everett interpretation the 
\inxx{anthropic principle}anthropic principle is automatically
contained.

\vs{5mm}

\centerline{* * *}

\vs{4mm}

It is typical for general relativity that it deals merely with the intrinsic
properties of spacetime, such as its metric and the intrinsic curvature. It
disregards how spacetime looks ``from the outside". The practitioners of 
general relativity are
not interested in an eventual existence of an embedding space in which our
spacetime is immersed. At the same time, paradoxically, whenever they wish to
illustrate various solutions of the Einstein equations they actually draw
spacetime as being embedded in a higher-dimensional space. Actually
they draw spacetime as a 2-dimensional surface in 3-dimensional space.
If they had known how to do it they would have drawn it as a 4-dimensional 
surface in a higher-dimensional space, but since this is not possible\footnote{
By using suitable projection techniques this might be in fact possible, but
such drawings would not be understandable to an untrained person.}
they help themselves by suppressing two dimensions of spacetime.

How can we talk at all about a fourth, fifth, or even higher dimension, if we
are unable to perceive them. For a description of a point in a three dimensional
space we need three numbers, i.e., coordinates. In order to describe its motion,
that is the trajectories, we need three equations. There is an isomorphism
between the algebraic equations and geometric objects, for instance curves
in space. This we can generalize, and instead of three equations take
four or more equations; we then talk about four- or higher-dimensional spaces.

Instead of considering the embedding of spacetime in a higher-dimensional
space merely as a usefull tool for the illustration of Einstein's equations, some
physicists take the embedding space seriously as an ``arena" in which
lives the 4-dimensional surface representing spacetime. Distribution of
matter on this surface is determined by the distribution of matter in
the embedding space\footnote{
In Chapter 8 we have developed a model in which our spacetime surface is a
worldsheet of a brane. Assuming that there are many other similar branes
of various dimensionality which can intersect our world brane we obtain,
as a result of the intersection, the matter on our world brane in the form
of point particles, strings, 2-branes and 3-branes (i.e., space filling
branes). All those other branes together with our world brane form 
the matter in the embedding space. Moreover, we have shown that the embedding
space is actually identified with all those branes. Without the branes there
is no embedding space.}.
The motion of the latter surface (actually the motion of a 3-brane which
sweeps a 4-dimensional surface, called a {\it worldsheet})  can be 
considered as being a classical motion, which
means that the surface and its position in the embedding space are
well determined at every moment. However, such a classical description
does not correspond to the reality. The motion of the 3-brane has to obey
the laws of quantum mechanics, hence a generic state of the brane is
represented by a wave function. The latter function in general does not 
represent a certain well determined brane's worldsheet, but is ``spread" 
over various
worldsheets. More precisely, a  wave function is, in general, a
superposition of the particular wave functions, every one of them representing
some well defined worldsheet. Such a view automatically implies that our
spacetime worldsheet is not the only possible one, but that there exist
other possible worldsheets which represent other possible universes, with
different configurations of geometry and matter, and thus with different
possible observers. But they all stay in a quantum mechanical superposition!
How can we then reconcile this with the fact that at the macroscopic level we
observe a well determined spacetime, with a well determined matter 
configuration? We again employ the Everett interpretation. According to Everett
all those spacetime worldsheets together with the corresponding observers,
which enter the superposition, are not merely {\it possible}, but they
actually exist in the multiverse. Relative to every one of those observers
the wave function represents a state with a well determined universe, of
course up to the accuracy with which the observer monitors the rest of
his universe. This is the ``objective" point of view. From a ``subjective"
point of view the situation looks as follows. If I ``measure" the position
of a single atom in my surroundings, then the positions of all the other
atoms, say in a crystal, will be irrevocably determined forever and I
shall never observe a superposition of that crystal. Moreover, since
the crystal is in the interaction with its surroundings and indirectly also
with the entire universe, I shall never be able to observe a superposition
of the universe, at least not at the macroscopic level\footnote{
In fact, I measure the position of an atom in a crystal 
by the very act of looking at
it. So my universe actually is no longer in such a macroscopic superposition 
after the moment I looked at it (or even touched it) for the first time.
{\it Relative to me} the universe certainly was in a superposition (and
consequently I was not aware of anything) before my embryo started to evolve,
and will be again in a superposition after my death. The latter metaphor
attempts to illustrate that a conscious observer and the corresponding
definite macroscopic universe are in a tight relationship.}.
Of course, a superposition of the universe at the microscopic level remains,
but is reduced every time we perform a corresponding measurement.

According to the conventional 
\inxx{copenhagen interpretation}Copenhagen interpretation of quantum
mechanics it is uncertain which of the {\it possible} universes will be
realized after a measurement of a variable. According to the Everett
interpretation, however, all those universes actually exist. This is an
`objective' point of view. By introducing the concept of {\it relative
wave function} Everett explains that from a ``subjective" point of view
it is uncertain in which of those universes the observer will happen
to ``find himself". In this respect the Everett interpretation coincides
with the Copenhagen interpretation. The questions ``what universe?" and
``which universe?" are intertwined in the Everett interpretation, depending
on whether we look at it from an ``objective" or a ``subjective" point of
view.

\vs{5mm}

\centerline{* * *}

\vs{4mm}

The quantum theory of the spacetime worldsheet in an embedding space, 
outlined in
rough contours in this chapter, is in my opinion one of the most
promising candidates for the quantum description of gravity. In its future
development it will be necessary to include the other interactions,
such as the electromagnetic, weak and strong interactions.
This could be achieved
by following the Kaluza--Klein idea and extend the dimensionality of the
spacetime sheet from four to more dimensions. Also fermions could be
included by performing a supersymmetric generalization of the theory, that is
by extending the description to the anticommuting Grassmann coordinates,
or perhaps by taking a polyvector generalization of the theory.

\chapter[Nobody really understands quantum \ \\ mechanics]
{Nobody really understands \\ quantum mechanics}

\prologue{Quantum mechanics is a theory about the relative information that
subsystems have about each other, and this is a complete description
about the world}
{Carlo Rovelli}

The motto from a famous sentence by Feynman \ci{91} will guide us through this
chapter. There are many interpretations of quantum mechanics (QM) described in
some excellent books and articles. No consensus about which one is ``valid",
if any, has been established so far. My feeling is that each interpretation
has its own merits and elucidates certain aspects of QM. Let me briefly
discuss the essential points (as I see them) of the 
three main interpretations\footnote{
Among modern variants of the interpretations let me mention the {\it relational
quantum mechanics} of Rovelli \ci{91a}, and the {\it many mind interpretation}
of Butterfield \ci{91b}}.

\paragraph{Conventional 
(Copenhagen) interpretation}\inxx{copenhagen interpretation}
The wave function $\psi$ evolves according to a certain evolution law
(the Schr\" odinger equation). $\psi$ carries the information about {\it possible}
outcomes of a measurement process. Whenever a measurement is performed
the wave function collapses 
\inxx{collaps of the wave function}into one of its eigenstates. The absolute
square of the scalar product of $\psi$ with its eigenfunctions are the
probabilities (or probability densities) of the occurrence of these
particular eigenvalues in the measurement process \cite{92,93}.

\paragraph{Collapse or the reduction of the wave function occurs in an
observer's mind} 
In order to explain how the collapse, which is extraneous to the Schr\" odinger 
evolution of $\psi$, happens at all, one needs something more. If one
postulates that the collapse occurs in a (say, macroscopic) measuring
apparatus the problem is not solved at all, since also the interaction of
our original system (described by $\psi$) with the measuring apparatus is
governed by the Schr\" odinger evolution for the combined system--apparatus
wave function. Therefore also the measuring apparatus is in a state which
is a superposition of different eigenstates corresponding to different
results of measurement\footnote{
For a more detailed description of such a superposition and its duration
see the section on decoherence.}.
This is true even if the result of measurement is registered by a magnetic
tape, or punched tape, etc.\,. A conscious observer has to look at the result
of measurement; only at that moment is it decided which of various
possibilities actually occurs \cite{94}. Meanwhile, the tape has been in a state
which is a superposition of states corresponding to the eigenvalues in 
question.

\paragraph{Everett, Wheeler, Graham 
many worlds interpretation}\inxx{many worlds interpretation}
Various quantum possibilities actually occur, but in different
branches of the world \cite{Everett,83,84}. Every time a measurement is performed
the observed world splits into several (often many) worlds corresponding
to different eigenvalues of the measured quantities. All those worlds
coexist in a higher universe, the {\it multiverse}.\inxx{multiverse} 
In the multiverse there exists\inxx{multiverse}
a (sufficiently complicated) subsystem (e.g., an automaton) with
memory sequences. To a particular bran\-ching path there corresponds a
particular memory sequence in the automaton, and vice versa, to
a particular memory sequence there belongs a particular branching path.
No collapse of the wave function is needed. All one needs is to decide
which of the possible memory sequences is the one to follow. (My interpretation
is that there is no collapse in the multiverse, whilst a particular memory 
sequence or stream of consciousness experiences the collapse at each
branching point.) A particular memory sequence in the automaton
actually defines a possible life history of an observer (e.g., a human
being). Various well known paradoxes like that of Einstein--Podolsky--Rosen,
which are concerned with correlated, non-interacting systems, or that
of Schr\" odinger's cat, etc., are\inxx{Schr\" odinger's cat}
easily investigated and clarified
in this scheme \cite{Everett}.

Even if apparently non-related the previous three interpretations in fact
illuminate QM each from its own point of view. In order to introduce the
reader to my way of looking at the situation I am now going to describe
some of my earlier ideas. Although not being the final word I have to
say about QM, these rough ideas might provide a conceptual background which will
facilitate understanding the more advanced discussion (which will also
take into account the modern {\it decoherence} approach) provided later
in this chapter. A common denominator to the three views of QM discussed above
we find in the assumption that a 3-dimensional simultaneity hypersurface
$\Sigma$ moves in a higher-dimensional space of {\it real} events\footnote{
We shall be more specific about what the ``higher-dimensional space" is later.
It can be either the usual higher-dimensional configuration space, or,
if we adopt the brane world model then there also exists an
infinite-dimensional membrane space ${\cal M}$. 
The points of ${\cal M}$-space correspond to the ``coordinate" basis vectors
of a Hilbert space which span an arbitrary brane state.}.
Those events which are intersected by a certain $\Sigma$-motion are observed
by a corresponding observer. Hence we no longer have a conflict between realism
and idealism. There exists a certain physical reality, i.e., the world of
events in a higher-dimensional space. In this higher universe there exist
many 4-dimensional worlds corresponding to different quantum possibilities
(see also Wheeler \cite{95}). A particular observer, or, better, his mind
chooses by an act of free will one particular $\Sigma$-surface, in the
next moment another $\Sigma$-surface, etc.\,. A sequence of $\Sigma$-surfaces
describes a 4-dimensional world\footnote{This is elaborated in Sec. 10.1.}.
A consequence of the act of free choice which happens in a particular mind 
is the wave function reduction (or collapse). 
\inxx{collaps of the wave function}Before the observation the
mind has
certain information about various {\it possible} outcomes of measurement;
this information is incorporated in a certain wave function. Once the
measurement is performed (a measurement procedure terminates in one's mind),
one of the {\it possible} outcomes has become the {\it actual} outcome;
the term actual is relative to a particular stream of consciousness (or
memory sequence in Everett's sense). Other possible outcomes are actual
relative to the other possible streams of consciousness.\inxx{consciousness}

So, which of the possible quantum outcomes will happen 
is--as I assume--indeed decided by mind (as Wigner had 
already advocated). But this fact does
not require from us to accept an idealistic or even solipsistic interpretation
of the world, namely that the external worlds is merely an illusion of
a mind. The duty of mind is merely {\it a choice of a path} in a
higher-dimensional space, i.e., a choice of a sequence of $\Sigma$-hypersurfaces
(the three dimensional ``nows"). But various possible sequences exist
independently of a mind; they are real and embedded in a timeless
higher-dimensional world.

However,  a strict realism alone, independent\inxx{consciousness}
 of mind or consciousness
is also no more acceptable. There does not exist a {\it motion} of a real external
object. The external ``physical" world is a static, higher-dimensional structure of
events. One gets a dynamical (external) 4-dimensional world by postulating
the existence of a new entity, a {\it mind}, with the property of {\it
mo\-ving} the simultaneity surface $\Sigma$ into any permissible direction in the
higher space. This act of $\Sigma$-{\it motion} must be separately postulated; 
a consequence of this motion is the subjective experience that the
(3-dimensional) external world is continuously changing. {\it The change}
of the (3-dimensional) external world is in fact an illusion; what really
changes with time is an observer's mind, while the external world ---which
is more than (3+1)-dimensional)--- is real and static (or timeless).

Let us stress: only {\it the change} of an external (3-dimensional) world is
an illusion, not {\it the existence} of an external world as such. Here
one must be careful to distinguish between the concept of time as a coordinate
(which enters the equations of special and general relativity) and the
concept of time as a subjective experience of change or becoming.
Unfortunately we often use the same word `time' when speaking about the
two different concepts\footnote{One of the goals of the present book is
to formalize such a distinction; see the previous three parts of the book.}.

One might object that we are introducing a kind of metaphysical or non physical
object ---mind or consciousness--- into the theory,\inxx{consciousness}
 and that a physical
theory should be based on {\it observable} quantities only. I reply:
how can one dismiss mind and consciousness\inxx{consciousness}
 as something non-observable
or irrelevant to nature, when, on the contrary, our own consciousness
is the most obvious and directly observable of all things in nature;
it is through our consciousness that we have contacts with the external
world (see also Wigner \cite{94}).

\section{The `I' intuitively understands quantum mechanics}

If we think in a really relaxed way and unbiased with preconcepts, we realize
the obvious, that the wave function is consciousness. In the following
I will elaborate this a little. But before continuing let me say something
about the role of extensive verbal explanations and discussions, especially
in our attempts to clarify the meaning of quantum mechanics. My point
is that we actually need as much such discussion as possible,
in order to develop our inner, intuitive, perception of what 
quantum mechanics is about. In the case of Newtonian (classical) mechanics
we already have such an intuitive perception. We have been developing our
perception since we are born. Every child intuitively understands how
objects move and what the consequences are  of his actions, for instance what
happens if he throws a ball. Imagine our embarrassment, 
if, since our birth, we had 
no direct contact with  the physical environment, but we had 
nevertheless been indirectly taught about the existence of such an 
environment. The precise situation is not important for the argument,
just imagine that we are born in a space ship on a journey to a nearby
galaxy, and remain  fixed in our beds with eyes closed all the time
and learning only by listening. Even if not seeing and touching the objects
around us, we would eventually nevertheless learn indirectly about the
functioning of the physical world, and perhaps even master Newtonian
mechanics. We might have become very good at solving all sorts of
mechanical problem, and thus be real experts in using rigorous techniques.
We might even be able to perform experiments by telling the computer 
to ``throw" a stone and then to tell us about what has happened.
And yet such an expertise would not help us much in {\it understanding}
what is behind all the theory and ``experiments" we master so well.  
Of course, what is needed is a direct contact with the environment 
we model so well. In the absence of such a direct contact, however, it
will be indispensable for us to discuss as much as possible the
functioning of the physical environment and the meaning of the theory we
master so well. Only then would we have developed to a certain extent an
intuition, although indirect, about the physical environment.

An analogous situation, of course, should be true for quantum mechanics.
The role of extensive verbalization when we try to understand quantum mechanics
can now be more appreciated. We have to read, discuss,
and think about quantum mechanics as much as we are interested. 
When many people are doing so
the process will eventually crystallize into a very clear and obvious
picture. At the moment we see only some parts of the picture. I am now
going to say something about how I see my part of the picture.

Everything we know about\inxx{consciousness}
 the world we know through consciousness. We are
describing the world by a wave function. Certain simple phenomena can
be described by a simple wave function which we can treat mathematically. In
general, however, phenomena are so involved that a mathematical treatment
is not possible, and yet conceptually we can still talk about
the wave function. The latter is our information about the world. Information
does not exist {\it per se}, information is relative to consciousness \ci{96}. 
Consciousness has information about something. This could be pushed to
its extreme and it be asserted that information is consciousness, especially when
information refers to itself (self-referential information). On the other
hand, a wave function is information (which is at least a certain very important
aspect of wave function). Hence we may conclude that a wave function has a
very close relation with consciousness. In the strongest version we cannot
help but conclude that a wave function should in fact be identified with 
consciousness. Namely, {\it if, on the one hand, the wave function 
is everything I can know
about the world, and, on the other, the content of my consciousness
is everything I can know about the world, then consciousness is a
wave function}. In certain particular cases the content of my consciousness
can be very clear: after having prepared an experiment I {\it know} that an
electron is localized in a given box. This situation can be described
precisely by means of a mathematical object, namely, the wave function. If I
open the box then I know that the electron is no longer localized within 
the box,
but can be anywhere around the box. Precisely how the probability of finding
it in some place evolves with time I can calculate by means of quantum
mechanics. Instead of an electron in a box we can consider electrons
around an atomic nucleus. We can consider not one, but many atoms. Very
soon we can no longer do maths and quantum mechanical calculation, but the
fact remains that our knowledge about the world is encoded in the
wave function. We do not know any longer a precise mathematical expression
for the wave function, but we still have a perception of the wave
function. The very fact that we see definite macroscopic objects around
us is a signal of its existence: so we know that the atoms of the
objects are localized at the locations of the object. Concerning
single atoms, we know that electrons are localized in a well defined way
around the nuclei, etc.\,. Everything I know about the external world is
encoded in the wave function. However, consciousness is more than that.
It also knows about its internal states, about the memories of past events,
about its thoughts, etc.\,. It is, indeed, a very involved self-referential
information system. I cannot touch upon such aspects of consciousness here, 
but the
interesting reader will profit from reading some good works \cite{97,98}.

The wave function of an isolated system evolves freely according to the
Schr\" odinger evolution. After the system interacts with its surroundings,
the system and its surroundings then become entangled and they are in
a quantum mechanical superposition. However,
there is, in principle, a causal connection with my brain. For a distant system 
it takes some time until the information about the interaction reaches me. 
The collapse of the wave function happens at the moment when the information 
arrives in my brain. Contrary to what we often read, 
\inxx{collaps of the wave function}the collapse of the
wave function does not spread with infinite speed from the place of
interaction to the observer. There is no collapse until the signal reaches
my brain. Information about the interaction need not be explicit, as 
it usually is when we perform a controlled experiment, e.g., with laser beams. 
Information can
be implicit, hidden in the many degrees of freedom of my environment,
and yet the collapse happens, since my brain is coupled to the environment.
But why do I experience the collapse of the wave function? Why does the wave
function not remain in a superposition? The collapse occurs 
because the information about the content of my
consciousness about the measured system cannot be in superposition.
Information about an external degree of freedom can be in superposition.
{\it Information about the degrees of freedom which are the carriers of the very
same information cannot remain in a superposition.} This would be a logical
paradox, or the G\" odel knot \ci{97,98}:\inxx{G\" odel knot}
 it is resolved by the collapse of the
wave function. My consciousness ``jumps" into one of the possible universes,
each one containing a different state of the measured system and my different
knowledge about the measurement result. However, from the viewpoint of
an external observer no collapse has happened until the information has
arrived in his brain. Relative to him the measured system and my brain 
have both remained in a superposition.

In order to illustrate the situation it is now a good point to provide 
a specific example.

\paragraph{A single electron plane wave hits the screen}

Suppose an electron described by a wide wave packet hits a screen. Before
hitting the screen the electron's position was undetermined within the
wave packet's localization. What happens after the collision with the screen?
If we perform strictly quantum mechanical calculations by taking into account
the interaction of the electron with the material in the screen we find that
the location of the traces the interaction has left within the screen is also
undetermined. This means that the screen is in a superposition of the states
having a ``spot" at different places of the screen. Suppose now that an
observer ${\cal O}$ looks at the screen. Photons reflected from the screen 
bear the information about the position of the spot. They are, according to
quantum mechanical calculations, in a superposition. The same is true for an
observer who looks at the screen. His eyes' retinas are in a superposition
of the states corresponding to different positions of the spot, and the signal
in the nerves from the retina is in a superposition as well.
Finally, the signal reaches the visual
center in the observer's brain, which is also in the superposition. Before
the observer has looked at the screen the latter has been in a superposition
state. After having looked, the screen state is still in a superposition, 
but at the same time there is also a superposition of the brain states
representing different states of consciousness of the observer ${\cal O}$.

Read carefully again: different brain (quantum mechanical) states
{\it represent} different {\it consciousness} states. And what is the
content of those consciousness state? Precisely the information about the
location of the spot on the screen. But the latter information is, in fact,
{\it the wave function} of the screen, 
\inxx{collaps of the wave function}more precisely the collapsed wave
function. So we have a direct piece of evidence about the relation between the
wave function about an external state and a conscious state. The external
state is {\it relative} to the brain state, and the latter state in turn
represents  a state of consciousness. At this point it is economical to
identify the relative ``external" state with the corresponding consciousness
state.

Relative to the observer ${\cal O}$'s consciousness states there is no
superposition of the screen states. ``Subjectively", a collapse of the
wave function has occurred relative to the observer's consciousness
state, but ``objectively" there is no collapse.

The term {\it objective} implies that there should exist an ``objective"
wave function of the universe which never collapses. We now ask ``is such
a concept of an objective, universal, wave function indeed necessary?"
Or, put it differently, what is ``the universal wave function"? Everett himself
introduced the concept of {\it the relative wave function}, i.e., the wave
function which is relative to another wave function. In my opinion the
relative wave function suffices, and there is no such a thing as an
objective or universal wave function. This will become more clear after
continuing with our discussion.

Now let us investigate how I experience the situation described above.
Before I measure the position of the electron, it was in a superposition state.
Before I had any contact with the screen, the observer ${\cal O}$, or their
environment, they were altogether in a superposition state. After looking
at the screen, or after communicating with the observer ${\cal O}$, there
was no longer superposition\inxx{consciousness}
 relative to my consciousness. However, relative
to another
observer ${\cal O'}$ the combined state of the screen $S$, ${\cal O}$, and
my brain can remain in superposition until ${\cal O'}$ himself gets in
contact with me, ${\cal O}$, $S$, or the environment of $S$, ${\cal O}$, and me.
A little more thought in such a direction should convince everybody that
a wave function is always {\it relative} to something, or, better, to
somebody. There can be no ``objective" wave function. 

If I contemplate the electron wave packet hitting the screen I know that
the wave packet implies the\inxx{multiverse} 
existence of the multiverse, but I also know, after
looking at the screen, that I have found myself in one of those many
universes. I also know that according to some other observer my brain state
can be a superposition. But I do not know how my brain state could 
{\it objectively} be a superposition. Who, then is this objective observer?
Just think hard enough about this and you will start to realize
that there can be no objective wave function, and if so, then a wave function,
being always relative to someone's consciousness, can in fact be identified
with someone's consciousness. The phrase ``wave function is relative
to someone's consciousness" could be replaced by ``wave function is
(someone's) consciousness". All the problems with quantum mechanics,
also the difficulties concerning the Everett interpretation, then disappear
at once. 

I shall, of course, elaborate this a little bit more in due course.
At the moment let me say again that the difficulties concerning 
the understanding of QM can be avoided if we consider a wave function as a
measure of the information an observer has about the world. A wave function,
in a sense, {\it is} consciousness. We do not yet control all the variables
which are relevant to consciousness. But we already understand some 
of those variables, and we are able to define them strictly by employing
mathematics: for instance, those variables of the consciousness which are
responsible for the perception of physical experiments by which we measure
quantum observables, such as a particle's position, spin, etc.\,.

\section{Decoherence}\inxx{decoherence}

Since the seminal work by Zurek \ci{Zurek} and Zeh \ci{Zeh}
it has becomes very clear why
a macroscopic system cannot be in a superposition state. A system $S$ which
we study is normally coupled to its environment $E$. As a consequence
$S$ no longer behaves as a quantum system. More precisely, the
partial wave function of $S$ relative to $E$ is no longer a superposition
of $S$'s eigenstates. The combined system $SE$, however, still behaves
as a quantum system, and is in a superposition state. Zurek and Zeh have
demonstrated this by employing the description with {\it density matrices}.

\paragraph{The density matrix}\inxx{density matrix}
A quantum state is a vector $|\psi \rangle$ in {\it Hilbert space}. The
projection of a generic state onto the position eigenstates $|x \rangle$ is
{\it the wave function}
\be
    \psi (x) \equiv \langle x |\psi \rangle .
\lbl{IV.1}
\ee

Instead of $|\psi \rangle $ we can take the product
\be
      |\psi \rangle \langle \psi | = {\hat \rho} \, ,
\lbl{IV.2}
\ee
which is called {\it the density operator}. The description of a quantum
system by means of $|\psi \rangle$ is equivalent to description by
means of ${\hat \rho}$.

Taking the case of a single particle we can form the sandwich
\be
     \langle x|{\hat \rho}|x' \rangle \equiv \rho (x,x') =
     \langle x|\psi \rangle \langle \psi |x' \rangle = \psi(x) \psi^* (x') .
\lbl{IV.3}
\ee
This is {\it the density matrix} in the coordinate representation. Its diagonal
elements
\be
     \langle x|{\hat \rho}|x \rangle = \rho (x,x) \equiv \rho(x)
     = |\psi (x)|^2
\lbl{IV.4}
\ee
form {\it the probability density} of finding the particle at the position
$x$. However, the off-diagonal elements are also different from zero, and
they are responsible for interference phenomena. If somehow the 
off-diagonal terms vanish, then the interference also vanishes.

Consider, now, a state $|\psi \rangle$ describing a spin $1\oo 2$ particle
coupled to a detector:
\be
    |\psi \rangle = \sum_i \alpha_i |i \rangle \langle d_i| \, ,
\lbl{IV.5}
\ee
where
\be
    |i \rangle = |\mbox{$1\oo 2$} \rangle , \; |-\mbox{$1\oo 2$} \rangle              
\lbl{IV.6}
\ee
are spin states, and
\be
    |d_i \rangle = |d_{1/2} \rangle , \; |d_{-1/2} \rangle
\lbl{IV.7}
\ee
are the detector states.

The density operator is
\be
     |\psi \rangle \langle \psi | = \sum_{ij} \alpha_i \alpha_j^* |i \rangle
     |d_i \rangle \langle j|\langle d_j| .
\lbl{IV.8}
\ee
It can be represented in some set of basis states $|m \rangle$ which are
rotated relative to $|i \rangle$:
\be
    |m \rangle = \sum_k |k \rangle \langle k|m \rangle \; , \quad |d_m \rangle =
    \sum_{d_k} |d_k \rangle  \langle d_k |d_m \rangle
\lbl{IV.9}
\ee
We then obtain the density matrix
\be
    \langle d_m , m|\psi \rangle \langle \psi |n, d_n \rangle = \sum_{ij}
    \alpha_i \alpha_J^* \langle d_m , m|i, d_i \rangle \langle j , d_j |
    n, d_n \rangle .
\lbl{IV.10}
\ee
which has non-zero off diagonal elements. Therefore the combined system
{\it particle--detector} behaves quantum mechanically.

Let us now introduce yet another system, namely, the environment. After
interacting with the environment  the evolution brings the system to the
state
\be
     | \psi \rangle = \sum_i \alpha_i |i \rangle |d_i \rangle |E_i \rangle \, ,
\lbl{IV.11}
\ee
where
\be
    |E_i \rangle = |E_{1/2} \rangle \, , \; |E_{-1/2} \rangle
\lbl{IV.12}
\ee
are the environment states after the interaction with the 
{\it particle--detector} system.

The density operator is
\be
     |\psi \rangle \langle \psi | = \sum_{ij} \alpha_i \alpha_j^* |i \rangle
     |d_i \rangle |E_i \rangle \langle j|\langle d_j|\langle E_j |
\lbl{IV.13}
\ee
The combined system {\it particle--detector--environment} is also in a
{\it superposition state}.\inxx{density matrix}
 The density matrix has-non zero off-diagonal 
elements.

Whilst the degrees of freedom of the particle and the detector are under
the control of an observer, those of the environment are not. The observer 
cannot distinguish $|E_{1/2} \rangle$ from $|E_{-1/2} \rangle$, therefore
he cannot know the total density matrix. We can define {\it the reduced
density operator} which takes into account the observer's ignorance of
$|E_i \rangle$. This is achieved by summing over the environmental degrees
of freedom:
\be
     \sum_k \langle E_k|\psi \rangle \langle \psi|E_k \rangle = \sum_i
     |\alpha_i|^2 |i \rangle |d_i \rangle  \langle i| \langle d_i| .
\lbl{IV.14}
\ee
We see that the reduced density operator, when represented as a matrix in
the states $|i \rangle $, has only the diagonal terms different from zero.
This property is preserved under rotations of the states $|i \rangle$.

We can paraphrase this as follows. With respect to the environment the
density matrix is diagonal. Not only with respect to the environment, but
with respect to any system, the density matrix is diagonal. This has
already been studied by Everett \ci{Everett}, who introduced 
the concept of {\it
relative state}.\inxx{relative state} {\it The reduced density matrix} 
indeed describes\inxx{reduced density matrix} 
the relative state. In the above specific case the state of the system
{\it 
particle--detector}  is relative to the environment. 
Since the observer is also a
part of the environment the state of the system {\it particle--detector}
is relative to
the observer. The observer {\it cannot} see a superposition (\ref{IV.5}),
since very soon the system evolves into the state ({\ref{IV.11}), where
$|E_i \rangle$ includes the observer as well. After the interaction with
environment the system {\it particle--detector} loses the interference
properties and behaves as a classical system. However, the total system
{\it particle--detector--environment} remains in a superposition, but nobody
who is coupled to the environment can observe such a superposition
after the interaction reaches him. This happens very soon on the Earth, but it
may take some time for an observer in space.

The famous {\it Schr\" odinger's cat} 
experiment \ci{99}\inxx{Schr\" odinger's cat} 
can now be easily clarified.
In order to demonstrate that the probability interpretation of quantum
mechanics leads to paradoxes Schr\" odinger envisaged a box in which a
macroscopic object ---a cat--- is linked to\inxx{Schr\" odinger's cat} 
a quantum system, such as
a low activity radioactive source. At every moment the source is in a
superposition of the state in which a photon has been emitted and the
state in which no photon has been emitted. The photons are detected by a 
Geiger counter connected to a device which triggers the release of a poisonous
gas. Schr\" odinger considered the situation as paradoxical, as the cat
should remain in a superposition state, until somebody looks into the box.
According to our preceding discussion, however, the cat could have
remained in a superposition only if completely isolated from the
environment. This is normally not the case, therefore the cat remains in a
superposition for a very short time, thereafter the combined system
{\it cat--environment} is in a superposition state. 
The environment includes
me as well. But I cannot be\inxx{consciousness}
 in a superposition, therefore my consciousness
jumps into one of the two branches of the superposition (i.e., the cat alive
and the cat dead). This happens even {\it before} I look into the box.
Even before I look into the box it is already decided into which of the two
branches my consciousness resides. This is so because I am coupled to
the environment, to which also the cat is coupled. Hence, I am already
experiencing one of the branches. My consciousness, or, better subconsciousness,
has already decided to choose one of the branches, even before I became aware 
of the cat's state by obtaining the relevant information (e.g., by looking
into the box). What counts here is that the necessary information is
available in principle: it is implicit in the environmental degrees
of freedom. The latter are different if the cat is alive or dead.

\section{On the problem of basis in the Everett interpretation}

One often encounters an objection against the Everett interpretation
of quantum mechanics that is known under a name such as ``the problem of
basis". In a discussion group on internet (Sci.Phys., 5 Nov.,1994) 
I have found a very
lucid discussion by Ron Maimon (Harvard University, Cambridge, MA)
which I quote below.

\begin{quotation}

It's been about half a year since I read Bell's analysis, and I don't have it
handy. I will write down what I remember as being the main point of
his analysis and demonstrate why it is incorrect.

Bell claims that Everett is introducing a new and arbitrary assumption into
quantum mechanics in 
\inxx{collaps of the wave function}order to establish collapse, namely the ``pointer basis".
His claim is that it is highly arbitrary in what way you split up the
universe into a macroscopic superposition and the way to do it is in no way
determined by quantum mechanics. For example, if I have an electron in a spin
eigenstate, say $|+ \rangle$ then I measure it with a device which has a
pointer, the pointer should (if it is a good device) be put into an
eigenstate of its position operator.

This means that if we have a pointer which swings left when the electron has
spin up, it should be put into the state ``pointer on the left" if the
electron was in the state $|+ \rangle$. If it similarly swings right when
the electron is in the state $|- \rangle$ then if the electron is in the
state $|-\rangle$ the pointer should end up in the state ``pointer on the
right".

Now, says Bell, if we have the state $(1/\sqrt{2})(|+\rangle + |-\rangle)$
then the pointer should end up in the state $(1/\sqrt{2})(|{\rm right}\rangle
+|{\rm left}\rangle)$. According to Bell, Everett says that this is to be
interpreted as two universes, distinct and non interacting, one in which
the pointer is in the state ``right" and one in which the pointer is in the state
"left".

But aha! says Bell, this is where that snaky devil Everett gets in an
extra hypothesis! We don't have to consider the state 
$1/\sqrt{2}(|{\rm right}\rangle +|{\rm left}\rangle)$ as 
a superposition---I mean it is a state in its own right. Why not say that there has been no split 
at all, or that the split is into two universes, one in which the pointer
is in the state
\be
         a_1 |{\rm right} \rangle + a_2 |{\rm left} \rangle
\lbl{IV.a}
\ee
and one where it is in the state
\be
        b_1 |{\rm right} \rangle + b_2 |{\rm left} \rangle 
\lbl{IV.b}
\ee
So long as $a_1 + b_1 = a_2 + b_2 = 1/\sqrt{2}$ this is allowed. Then if we
split the universe along these lines we again get those eerie macroscopic
superpositions.

In other words, Everett's unnatural assumption is that the splitting of the
universes occurs along the eigenstates of the pointer position operator.
Different eigenstates of the pointer correspond to different universes,
and this is arbitrary, unnatural, and just plain ugly.

Hence Everett is just as bad as anyone else.

Well this is WRONG.

The reason is that (as many people have mentioned) there is no split of
the universe in the Everett interpretation. The state
\be
      {1\oo \sqrt{2}} (|{\rm right}\rangle + |{\rm left}\rangle)
\lbl{IV.c}
\ee
is no more of a pair of universe than the state $1/\sqrt{2}(|+\rangle
+ |- \rangle)$ of spin for the electron.

Then how comes we never see eerie superpositions of position eigenstates?

Why is it that the ``pointer basis" just happens to coincide with him or
her self. This is the ``state of mind" basis. The different states of this
basis are different brain configurations that correspond to different
states of mind, or configurations of thoughts.

Any human being, when thrown into a superposition of state of mind will
split into several people, each of which has a different thought. Where 
before there was only one path of mind, after there are several paths. These
paths all have the same memories up until the time of the experiment, and these
 all believe different events have occurred. This is the basis along which the
universe {\it subjectively seems to split}.

There is a problem with this however---what guarantees that eigenstates of my
state of mind are the same as eigenstates of the pointer position. If
this wasn't the case, then a definite state of mind would correspond to an
eerie neither here nor there configuration of the pointer.

The answer is, NOTHING. It is perfectly possible to construct a computer with
sensors that respond to certain configurations by changing the internal
state, and these configurations are not necessarily eigenstates of position of
a needle. They might be closer to eigenstates of momentum of the needle.
Such a computer wouldn't see weird neither-here-nor-there needles, it
would just ``sense" momenta, and won't be able to say to a very high accuracy
where the needle is.

So why are the eigenstates of our thoughts the same as the position
eigenstates of the needle?

They aren't!

They are only very approximately position eigenstates of the needle.

This can be seen by the fact that when we look at a needle it doesn't start
to jump around erratically, it sort of moves on a smooth trajectory.
This means that when we look at a 
\inxx{collaps of the wave function}needle, we don't ``collapse" it into
a position eigenstate, we only ``collapse it into an approximate position
eigenstate. In Everett's language, we are becoming correlated with a state
that is neither an eigenstate of the pointer's position, nor its momentum,
but approximately an eigenstate of both, constrained by the uncertainty
principle. This means that we don't have such absurdly accurate eyes that 
can see the location of a pointer with superhigh accuracy.

If we were determining the {\it exact} position of the needle, we would
have gamma ray sensor for eyes and these gamma rays would have enough energy
to visibly jolt the needle whenever we looked at it.

In order to determine exactly what state we are correlated with, or if you
like, the world (subjectively) 
\inxx{collaps of the wave function}collapses to, you have to understand the
mechanism of our vision.

A light photon bouncing off a needle in a superposition
\be
      {1\oo \sqrt{2}} (|{\rm right}\rangle + |{\rm left}\rangle)
\lbl{IV.d}
\ee
will bounce into a superposition of the states $|1\rangle$ or $|2\rangle$
corresponding to the direction it will get from either state. The same photon 
may then
interact with our eyes. The way it does this is to impinge upon a certain
place in our retina, and this place is highly sensitive to the direction of the
photon's propagation. The response of the pigments in our eyes is both highly 
localized in position (within the radius of a cell) and in momentum (the
width of the aperture of our pupil determines the maximal resolution of
our eyes). So it is not surprising that our pigment excitation states become
correlated with approximate position and approximate momentum eigenstates
of the needle. Hence we see what we see.

If we had good enough mathematical understanding of our eye we could say
in the Everett interpretation {\it exactly} what state we seem to collapse the
needle into. Even lacking such information it is easy to see that we will put
it in a state resembling such states where Newton's laws are seen to hold,
and macroscopic reality emerges.

\end{quotation}

A similar reasoning holds for other information channels that connect the outside
world with our brane (e.g., ears, touch, smell, taste). The problem of
choice of basis in the Everett interpretation is thus 
nicely clarified by the above quotation from Ron Maimon.

\vs{6mm}

\section{Brane world and brain world}

Let us now consider the model in which our world is a 3-brane moving in
a higher-dimensional space. How does it move? According to the laws
of quantum mechanics. A brane is described by a wave packet and the latter
is a solution of the Schr\" odinger equation. This was more precisely
discussed in Part III. Now I will outline the main ideas and concepts.
An example of a wave packet is sketched in Fig. 12.1}.

If the brane self-intersects we obtain {\it matter} on the brane
(see Sec. 8.3). When the brane moves it sweeps a surface
of one dimension more. A 3-brane sweeps a 4-dimensional surface,
called a {\it world sheet} or a {\it spacetime sheet}.

We have seen in Sec. 10.2 that instead of considering a 3-brane we
can consider a 4-brane. The latter brane is assumed to be a possible 
spacetime sheet (and thus has three space-like and one time-like intrinsic
dimensions). Moreover, it is assumed that the 4-brane is subjected to
dynamics along an invariant evolution parameter $\tau$. It is one of the
main messages of this book to point out that such a dynamics naturally
arises within the description of geometry and 
physics based on Clifford algebra. 
Then a scalar and a pseudoscalar parameter appear naturally, and
evolution proceeds with respect to such a parameter.

\figIVA

A 4-brane state is represented by a wave packet localized around an average
4-surface (Fig. 12.2)

It can be even more sharply localized within a region $P$, as shown in
Fig. 10.2 or Fig. 12.3. (For convenience we repeat Fig. 10.3.)

All these were mathematical possibilities. We have a Hilbert space of
4-brane kinematic states. We also have the Schr\" odinger equation
which a dynamically possible state has to satisfy. As a dynamically possible
state we obtain a wave packet. A wave packet can be localized in a
number of possible ways, and one is that of Fig. 10.3, i.e., localization within
a region $P$. How do we interpret such a localization of a wave packet? 
What does
it mean physically that a wave packet is localized within a 4-dimensional
region (i.e., it is localized in 3-space and at ``time" $t \equiv x^0$)?
This means that the 4-brane configuration is better known within $P$
than elsewhere. Since the 4-brane represents spacetime and matter
(remember that the 4-brane's self-intersections yield matter on the 4-brane),
such a localized wave packet tells us that spacetime and matter 
configuration are
better known within $P$ than elsewhere. Now recall {\it when}, according to
quantum mechanics, a matter configuration 
(for instance a particle's
position) is better known than otherwise. It is better known after
a suitable measurement. But we have also seen that a measurement procedure
terminates in one's brain, where it is decided 
---relative to the brain state---
about the outcome of the measurement. Hence the 4-brane wave packet
is localized within $P$, because an observer has measured the 4-brane's
configuration. Therefore the wave packet (the wave function) is {\it relative}
to that observer.

\figIVB

The 4-brane configuration after the measurement is not well known at
every position on the 4-brane, but only at the positions within
$P$, i.e., within a certain 3-space region and within a certain (narrow)
interval of the coordinate $x^0$. Such a 4-brane configuration (encompassing
a matter configuration as well) can be very involved. It can be involved to
the extent that it forms the structure of an observer's brain contemplating
the ``external" world by means of sense organs (eyes, ears, etc.).

We have arrived at a very important observation. {\it A wave packet
localized within $P$ can represent the brain structure of an observer ${\cal
O}$ and his sense organs, and also the surrounding world}\,! Both the
observer and the surrounding world are represented by a single (very
complicated) wave packet. Such a wave packet represents the observer's
knowledge about his brain's state and the corresponding surrounding 
world---all together.\inxx{consciousness}
 It represents the observer's consciousness! This is the
most obvious conclusion; without explicitly adopting it, the whole picture
about the meaning of QM remains foggy.

One can now ask, ``does not the 4-brane wave packet represent the
brain structure of another observer ${\cal O'}$ too?" Of course it does, but
not as completely as the structure of ${\cal O}$. By ``brain" structure I mean
here also the content of the brain's thought processes. 
The thought processes of ${\cal O'}$ are not known very well to ${\cal O}$. 
In contrast,
his own thought processes are very well known to ${\cal O}$, at the first
person level of perception. Therefore, the 4-brane wave packet is well 
localized within ${\cal O}$'s head and around it.

Such a wave packet is relative to ${\cal O}$. There exists, of course,
another possible wave packet which is relative to the observer ${\cal O'}$,
and is localized around ${\cal O'}$'s head.

\figIIII

Different initial conditions for a wave function mean different initial
conditions for consciousness. A wave function can be localized in another
person's head: my body can be in a superposition state with respect to
that person\inxx{decoherence}
 (at least for a certain time allowed by decoherence). If I say
(following the Everett interpretation) that there are many Matejs writing
this page, I have in mind a wave function relative to another observer.
Relative to me the wave function is such that I am writing these words
right now. In fact, I am identical with the latter wave function. Therefore
at the basic level of perception I intuitively understand quantum
mechanics. An `I' intuitively understands quantum mechanics. After clarifying
this I think that I have acquired a deeper understanding of quantum
mechanics. The same, I hope, holds for the careful reader. 
I hope, indeed, that
after reading these pages the reader will understand quantum mechanics,
not only at the lowest, intuitive, level, but also at
a higher cognitive level of perception. An ultimate understanding, however,
of what is really behind quantum mechanics and consciousness will
probably never be reached by us, and according to
 the G\" odel incompletness\inxx{G\" odel incompleteness theorem}
theorem \cite{97,98} is even not possible.

\boksIVA

\section{Final discussion on quantum mechanics, and conclusion}

{\it In classical mechanics} different initial conditions give
different {\it possible} trajectories of a dynamical system. Differential
equations of motion tell us only what is a possible set of solutions,
and say nothing about which one is actually realized. Selection of
a particular trajectory (by specifying initial conditions) is an {\it ad hoc}
procedure.

The property of classical mechanics admitting many possible trajectories is
further developed by Hamilton--Jacobi theory. The latter theory naturally
suggests its generalization---quantum mechanics. In quantum mechanics
different possible trajectories, or better,
a particle's positions, are described by means of a wave function satisfying
the Schr\" odinger equation of motion.

{\it In quantum mechanics} different initial conditions give different
{\it possible} wave functions. In order to make discussion more
concrete it turns out to be convenient to employ a brane world model in which
spacetime together with matter in it is described by a self-intersecting
4-dimensional sheet, a worldsheet $V_4$. According to QM such a sheet is not
definite, but is described by a wave function\footnote{
For simplicity we call it a `wave function', but in fact it is
a wave functional---a functional of the worldsheet embedding
functions $\eta^a (x^{\mu})$.}.
It is spread around an average spacetime sheet, and is more sharply localized
around a 3-dimensional hypersurface $\Sigma$ on $V_4$. Not all the points
on $\Sigma$ are equally well localized. Some points are more sharply localized
within a region $P$ (Fig. 10.3), which can be a region around an observer on
$V_4$. Such a wave function then evolves in an invariant evolution
parameter $\tau$, so that the region of sharp localization $P$
{\it moves} on $V_4$.

Different {\it possible} wave functions are localized around different 
observers. QM is a mechanics of consciousness. Differently localized
wave functions give different possible consciousnesses and corresponding
universes (worlds).

My brain and body can be a part of somebody's else consciousness. The wave
function relative to an observer ${\cal O'}$ can encompass my body and my 
brain states. Relative to ${\cal O'}$ my brain states can be in a superposition
(at least until decoherence becomes effective). Relative to ${\cal O'}$
there are many Matejs, all in a superposition state. Relative to me, 
there is always one Matej only. All the others are already out of my reach
because the wave function has collapsed.\inxx{collaps of the wave function}

According to Everett a wave function never does collapse. Collapse is
subjective for an observer. My point is that subjectivity is the essence
of wave function. A wave function is always relative to some observer, and
hence is subjective. So there is indeed collapse, call it subjective, if you
wish. Relative to me a wave function is collapsing all the time:
whenever the information (direct or indirect---through the environmental
degrees of freedom) about the outcome of measurement reaches me.

There is no collapse\footnote{
There is no collapse until decoherence becomes effective. If I am very
far from an observer ${\cal O'}$, e.g., on Mars, then ${\cal O'}$ and the
states of his measurement apparatus are in a superposition relative to me for
a rather long time.}
if I contemplate other observers performing their experiments.

Let us now consider, assuming the brane world description, a wave packet
of the form given in Fig. 12.2.
There is no region of sharp localization for such a wave packet. It contains a
superposition of all the observers and worlds within an effective boundary $B$.
Is this then the {\it universal wave function}? If so, why is it not spread
a little bit more, or shaped slightly differently? The answer can only make
sense if we assume that such a wave function is relative to a super-observer
${\cal O}_S$ who resides in the embedding space $V_N$. The universe of the
observer ${\cal O}_S$ is $V_N$, and the wave packet of Fig. 12.2 
is a part of the
wave function, relative to ${\cal O}_S$, describing ${\cal O}_S$'s 
consciousness and the corresponding universe.\inxx{consciousness}

To be frank, we have to admit that the wave packet itself, as illustrated in
Fig. 10.3, is relative to a super-observer ${\cal O}_S$. 
In order to be specific in
describing our universe and a conscious observer ${\cal O}$ we have mentally
placed ourselves in the position of an observer ${\cal O}_S$ outside
our universe, and envisaged how ${\cal O}_S$ would have described the
evolution of the consciousness states of ${\cal O}$ and the universe
belonging to ${\cal O}$. The wave packet, relative to ${\cal O}_S$,
representing ${\cal O}$ and his world could be so detailed that the
super-observer ${\cal O}_S$ would have identified himself with the observer 
${\cal O}$ and his world, similarly as we identify ourselves 
with a hero of a novel or a movie.

At a given value of the evolution parameter $\tau$ the wave packet
represents in detail the state of the observer ${\cal O}$'s brain and
the belonging world. With evolution the wave packet spreads. At a later value
of $\tau$ the wave packet might spread to the extent that it no longer
represents a well defined state of ${\cal O}$'s brain.
Hence, after a while, such a wave packet could no longer represent
${\cal  O}$'s consciousness state, but a superposition of ${\cal O}$'s
consciousness states. This makes sense relative to some other observer
${\cal O'}$, but not relative to ${\cal O}$. From the viewpoint of ${\cal O}$
the wave packet which describes ${\cal O}$'s brain state cannot be in
a superposition. Otherwise ${\cal O}$ would not be conscious. Therefore
when the evolving wave packet\inxx{collaps of the wave function} 
spreads too much, it collapses relative to
${\cal O}$ into one of the well defined brain states representing well
defined states of ${\cal O}$'s consciousness . Relative to another observer
${\cal O'}$, however, no collapse need happen until decoherence becomes
effective.

If the spreading wave packet would not collapse from time to time, the observer
could not be conscious. The quantum states that represent ${\cal O}$'s
consciousness are given in terms of certain basis states. The same wave
packet can be also expanded in terms of some other set of basis states,
but those states need not represent (or support) consciousness states.
This explains why collapse happens with respect to a certain basis, and
not with respect to some other basis.

We have the following model. An observer's consciousness and the
world to which he belongs are defined as being represented
by an evolving wave packet. At
every moment $\tau$ the wave packet says which universes ($\equiv$
consciousness state + world belonged to) are at disposal. A fundamental
postulate is that from the first person viewpoint the observer (his
consciousness) necessarily finds himself in one of the available universes
implicit in the spreading wave packet.
During the observer's life his body and brain retain a well preserved
structure, which poses strict constraints on the set of possible universes:
a universe has to encompass one of the available consciousness states of
${\cal O}$ and the ``external" worlds coupled to those brain states.
This continues until ${\cal O}$'s death. At the moment of ${\cal O}$'s
death ${\cal O}$'s brain no longer supports consciousness states.
${\cal O}$'s body and brain no longer impose constraints on possible
universes. The set of available universes increases dramatically: every
possible world and observer are in principle available! If we retain
the fundamental postulate, and I see no logical reason why not to retain it,
then the consciousness has to find itself in one of the many available
universes. Consciousness jumps into one of the available universes and
continues to evolve. When I am dead I find myself born again! In fact,
every time my wave packets spreads too much, I am dead; such a spread
wave packet cannot represent my consciousness. But I am immediately
``reborn'', since I find myself in one of the ``branches'' of the wave
packet, representing my definite consciousness state and a definite
``external" world.

A skeptical reader might think that I have gone too far with my
discussion. To answer this I wish to recall how improbable otherwise is 
the fact that I exist. (From the viewpoint of the reader `I' refers
to himself, of course.) Had things gone slightly differently, for
instance if my parents had not met each other, I would not have been born,
and my consciousness would not not have existed. Thinking along such lines,
the fact that I exist is an incredible accident!. Everything before my
birth had to happen just in the way it did, in order to enable 
the emergence of my existence. Not only my parents,
but also my grandparents had to meet each other, and so on back in time until the first
organisms evolved on the Earth! And the fact that my parents had 
become acquainted
was not sufficient, since any slightly different course of their life
together would have led to the birth not of me, but of my brother or sister
(who do not exist in this world). Any sufficiently deep reasoning in such
a direction leads to an unavoidable
\inxx{multiverse} conclusion that (i) the multiverse
in the Everett--Wheeler--DeWitt--Deutsch sense indeed exist, and (ii) consciousness
is associated (or identified) with the wave function which is relative
to a sufficiently complicated information processing system (e.g., an
observer's brain), and  {\it evolves} according to (a) the Schr\" odinger
evolution and\inxx{collaps of the wave function} (b)
experiences {\it collapse} at every measurement situation.
In an extreme situation (death) available quantum states (worlds) can 
include those far away from the states (the worlds) I have experienced
so far. My wave function (consciousness) then collapses into one
of those states (worlds), and I start experiencing the evolution of my
wave functions representing my life in such a ``new world''.

All this could, of course, be put on a more rigorous footing, by providing
precise definitions of the terms used. However, I think that before
attempting to start a discussion on more solid ground a certain amount
of heuristic discussion, expounding ideas and concepts, is necessary.

A reader might still be puzzled at this point, since, according to
the conventional viewpoint, in Everett's 
\inxx{many worlds interpretation}many worlds interpretation of
quantum mechanics there is no collapse of the wave function. To
understand why I am talking both about the many worlds interpretation
(the multiverse) and collapse one has to recall that according to
Everett and his followers collapse is a subjective event. Precisely that! 
Collapse of the wave function is a subjective event for an observer,
but such also is  the wave function itself. The wave function is always relative
and thus subjective. Even the Everett ``universal" wave function has to
be relative to some (super-) observer.

In order to strengthen the argument that (my) consciousness is not
necessarily restricted to being localized just in my brain, imagine
the following example which might indeed be realized in a not so
remote future. Suppose that my brain is connected to another person's
brain in such a way that I can directly experience her perceptions. So I can
experience what she sees, hears, touches, etc.\,. Suppose that the information
channel is so perfect that I can also experience her thoughts and even her
memories. After experiencing her life in such a way for a long enough time
my personality would become split between my brain and her brain. The wave
function representing my consciousness\inxx{consciousness}
 would be localized not only in 
my brain but also in her brain. After long time my consciousness would
become completely identified with her life experience; at that moment
my body could die, but my consciousness would have continued to experience
the life of her body.

The above example is a variant of the following thought experiment which is
often discussed. Namely, one could gradually install into my brain small
electronic or bioelectronic devices which would resume the functioning of
my brain components. If the process of installation is slow enough my
biological brain can thus be replaced by an electronic brain, and I would
not have noticed much difference concerning my consciousness and my
experience of `I'.

Such examples (and many others which can be easily envisaged by the reader)
of the transfer of consciousness from one physical system to another clearly
illustrate the idea that (my) consciousness, although currently associated
(localized) in my brain, could in fact be localized in some other brain too.
Accepting this, there is no longer a psychological barrier to accepting
the idea that the wave function (of the universe) is actually closely related,
or even identified, with the consciousness of an observer who is part
of that universe. After becoming habituated with such, at first sight
perhaps strange, wild, or even crazy ideas, one necessarily starts to 
realize that quantum mechanics is not so mysterious after all. It is
a mechanics of consciousness.

With quantum mechanics the evolution of science has again united two
pieces, matter and mind, which have been put apart by the famous
Cartesian cut. By separating mind from matter\footnote{There is an
amusing play of words\cite{100}: 
\begin{verse}
{\it What is matter? --- Never mind!\\
What is mind?  --- No matter!}
\end{verse}} ---so that the natural sciences
have disregarded the question of mind and consciousness--- Decartes
set the ground for the unprecedented development of physics and other
natural sciences. The development has finally led in the 20th century
to the discovery of quantum mechanics, which cannot be fully understood
without bringing mind and consciousness into the game.

\chapter{Final discussion}

We are now at the concluding chapter of this book. I have discussed many
different topics related to fundamental theoretical physics. My
emphasis has been on the exposition of ideas and concepts rather than
on their further development and applications. However, even the formulation 
of the basic principles has often required quite involved formalism and
mathematics. I expect that that has been stimulating 
to mathematically oriented readers,
whilst others could have skipped the difficult passages, since the ideas 
and concepts can be grasped to certain extent also by reading many
non-technical descriptions, especially in Part IV. In the following
few sections I will discuss some remaining open questions, without
trying to provide precisely formulated answers.

\section{What is wrong with tachyons?}\inxx{tachyon}

In spite of many stimulating works \cite{101} tachyons nowadays 
have predominantly
the status of impossible particles. Such an impossibility has its roots in the 
current mainstream theoretical constructions, especially quantum field
theory and special relativity. Not only that tachyons are generally considered
as violating causality (a sort of the ``grandfather paradox") 
\ci{Tolman}, there are
also some other well known problems. It is often taken for granted that
the existence of tachyons, which may have negative energies, destabilizes
the vacuum and thus renders the theory unreasonable. Moreover, when one tries
to solve a relativistic equation for tachyons, e.g., the Klein--Gordon equation,
one finds that (i) localized tachyon disturbances are subluminal, and
(ii) superluminal disturbances\inxx{superluminal disturbances}
 are non-local, and supposedly cannot be used for
information transmission.

There is now a number of works \cite{102}, experimental and theoretical, which
point out that the superluminal disturbances
 (case (ii)) are X-shaped
and hence, in a sense, localized after all. Therefore they could indeed be used
to send information faster than light.

Moreover, by using the Stueckelberg generalization of the
Klein--Gordon equation, discussed at length
in this book, the question of tachyon localization and its speed acquires
a new perspective. The tachyon's speed is no longer defined with respect to the
coordinate time $x^0 \equiv t$, but with respect to the invariant evolution
parameter $\tau$. Also the problem of negative energy and vacuum destabilization
has to be reformulated and reexamined within the new theory. According to 
my experience
with such a Stueckelberg-like quantum field theory, the vacuum is not unstable
in the case of tachyons, on the contrary, tachyons are necessary for the
consistency of the theory.

Finally, concerning causality, by adopting the Everett interpretation of
quantum mechanics, transmission of information into the past or future,
is not paradoxical at all! A signal that reaches an observer in the past,
merely ``splits" the universe into two branches: in one branch no tachyon
signal is detected and the course of history is ``normal", whilst in the
other branch the tachyonic signal is observed, and hence the course of
history is altered. This was discussed at length by Deutsch \ci{87} 
in an example
where not tachyon signals, 
\inxx{tachyon}but ordinary matter was supposed sent\inxx{wormholes}
into the past by means of {\it time machines},\inxx{time machines}
 such as wormholes\footnote{
Although Deutsch nicely explains why traveling into the past is not
paradoxical from the multiverse
\inxx{multiverse} point of view, he nevertheless maintains
the view that tachyons are impossible.}.

My conclusion is that tachyons are indeed theoretically possible; actually 
they are predicted by the Stueckelberg theory \ci{103}.
Their discovery and usage
for information transmission will dramatically extend our perception of the 
world by providing us with a window into the multiverse.

\section[Is the electron indeed an event moving in spacetime?]
{Is the electron indeed an event moving in spacetime?}

First let me make it clear that by ``point-like object moving in spacetime"
I mean precisely the particle described by the Stueckelberg action
(Chapter 1). In that description a particle is considered as being an
``event" in spacetime, i.e., an object localized not only in 3-space, but
also in the time coordinate $x^0 \equiv t$. We have developed a classical
and quantum theory of the {\it relativistic} dynamics of such objects. That
theory was just a first step, in fact an idealization---a studying example.
Later we generalized the Stueckelberg theory to extended objects (strings
and membranes of any dimension) and assumed that in reality physical
objects are extended. We assumed that a generic membrane need not
be space-like. It can be time-like as well. For instance, a string may extend
into a time-like direction, or\inxx{string,time-like}
 into a space-like direction.\inxx{string,space-like}

\setlength{\unitlength}{.8mm}
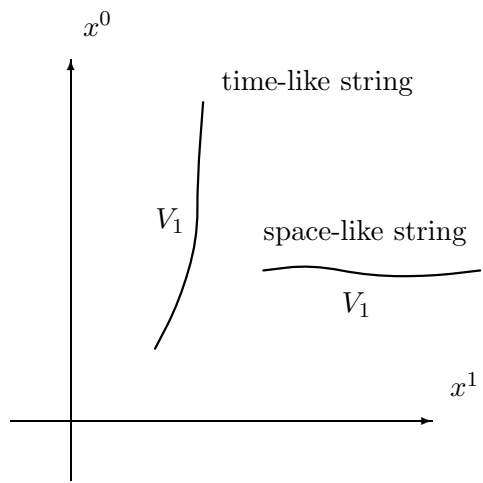
\begin{figure}[h]
\begin{picture}(120,90)(-25,0)

\put(10,0){\vector(0,1){70}}
\put(0,10){\vector(1,0){70}}
\thicklines
\spline(32,63)(31,50)(31,40)(28,30)(24,22)
\spline(42,35)(50,36)(60,34)(70,34)(78,35)

\put(12,74){$x^0$}
\put(73,14){$x^1$}
\put(35,65){time-like string}
\put(42,40){space-like string}
\put(24,42){$V_1$}
\put(55,28){$V_1$}

\end{picture}

\caption{Illustration of a time-like and a space-like string. 
They both move in $\tau$, and the picture is taken at a fixed 
value of $\tau$ (which is an invariant evolution parameter).}
\end{figure}

If we calculate ---at fixed $\tau$--- the electromagnetic interaction between
two time-like strings we find that it is just that of the familiar
Maxwell theory. A string is an extended object, is described by embedding
functions $X^{\mu} (u)$, and can be considered as a collection of
point-like objects (the Stueckelberg particles), each one being specified
by a value of the parameter $u$. The electromagnetic interaction
between the time-like strings is of the Maxwell type, and can be obtained by
summing the contributions of all the Stueckelberg particles constituting
the string. A pre-Maxwell field is a function of the parameter $u$
(telling us which Stueckelberg particle); the integration over $u$ gives
the corresponding Maxwell field\footnote{In the literature \cite{5a}
Maxwell fields are obtained by considering not strings, but
point-like Stueckelberg particles, and by integrating the corresponding
pre-Maxwell fields over the evolution parameter $\tau$. Such a procedure
is called concatenation, and is in fact a sort of averaging over $\tau$.}.
The latter field does not depend on the string parameter $u$, but it
still depends on the evolution parameter $\tau$.

Between the time-like strings we thus obtain the Maxwell interaction,
more precisely, a generalized Maxwell interaction since it depends on $\tau$.

The first and the second quantized theory of the Stueckelberg particle
(Chapter 1) is in fact an example for study. In order to apply it to
the electron the theory has to be generalized to strings, more precisely, to
time-like strings. Namely, the electromagnetic properties of electrons
clearly indicate that electrons cannot be point-like Stueckelberg particles,
but rather time-like strings, obeying the generalized Stueckelberg theory
(the unconstrained theory of strings, a particular case of the unconstrained
theory of membranes of any dimension discussed in Chapter 4). The wave function
$\psi(\tau,x^{\mu})$ is generalized to the string wave functional
$\psi [\tau, X^{\mu} (u)]$. A state of an electron can be represented
by $\psi$ which is a functional of the time-like string $X^{\mu} (u)$,
and evolves in $\tau$.

{\it The conventional field theory} of the electron is expected to be a
specific case when $\psi[\tau,X^{\mu}(u)]$ is {\it
stationary} in $\tau$. More generally, for a membrane of any dimension
we have obtained {\it the theory of conventional $p$-branes} as a particular
case, when the membrane wave functional is {\it stationary} in $\tau$
(see Sec. 7.1).

\section[Is our world indeed a single huge 4-dimensional membrane?]
{Is our world indeed a single huge 4-dimensional membrane?}

At first sight it may seem strange that the whole universe could be
a 4-dimensional membrane. The quantum principle resolves the problem:
at every scale it is undetermined what the membrane is. Besides that,
in quantum field theory there is not only one membrane, but a system
of (many) membranes. The intersections or self-intersections amongst those
membranes behave as bosonic sources (matter). In the case of supermembranes
the intersections presumably behave as fermionic sources (matter). A single
membrane can be very small, and also closed. Many such small membranes 
of various dimensionalities can
constitute a large 4-dimensional membrane of the size of the universe .

\setlength{\unitlength}{1mm}
\begin{figure}[h]
\begin{picture}(140,40)(-7,14)

\thicklines

\put(21,25){\circle{8}}\put(28,22){\circle{12}}\put(35,22){\circle{5}}
\put(42,36){\circle{5}}
\put(48,29){\circle{3}}\put(55,30){\circle{9}}\put(60,29){\circle{10}}
\put(73,31){\circle{6}}\put(88,33){\circle{8}}\put(91,28){\circle{6}}

\spline(35,32)(38,31)(41,32)(42,29)(39,27)(35,26)(34,29)(35,32)
\spline(62,28)(66,29)(69,26)(72,29)
\spline(75,28)(77,29)(79,25)(83,28)
\spline(80,28)(82,30)(85,29.5)(87,31)

\end{picture}

\caption{A system of many small membranes of various dimensionalities
constitutes a large membrane at the macroscopic scale.}
\end{figure}
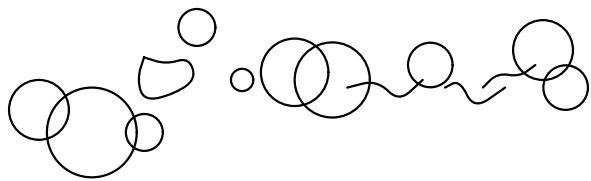

The universe is thus a system of intersecting and self-intersecting
membranes of various dimensions. Compactification of the embedding space is
not necessary. The wave functional can be localized around some
average (centroid) 4-dimensional membrane. The latter membrane is
our classical spacetime. The Einstein gravity described by the action
$I_{\rm eff} = \int \sqrt{|g|} {\dd}^4 x\, R$ is an approximation valid at
the scale of the solar system. 
At larger scales the effect of embedding becomes
significant and the effective action deviates from the Einstein--Hilbert action.

In short, the picture that our universe is a big 4-dimensional membrane is 
merely an idealization: the universe on average (the ``expectation value").
Actually the universe is a system of many membranes, described by a many
membranes wave functionals $f[\eta_1,\eta_2,...]$ forming a generic
state of the universe
\bear
    &&| A \rangle = \Biggl[ \int f[\eta] \Phi^{\dagger} [\eta]
    {\cal D} \eta + \int f[\eta_1,\eta_2] \Phi^{\dagger} [\eta_1] \Phi^{\dagger}
    [\eta_2] {\cal D} \, \eta_1 \, {\cal D} \eta_2 \hs{2cm} \lbl{IV.15} \\
    \nonumber \\
    && \hs{1.3cm} + \,\int f[\eta_1,\eta_2,\eta_3] 
    \Phi^{\dagger} [\eta_1] \Phi^{\dagger}
    [\eta_2]  \Phi^{\dagger} [\eta_3] {\cal D} \, \eta_1 \, {\cal D} \eta_2
    {\cal D} \eta_3
    + ... \Biggr] \, |0 \rangle , \nonumber
\ear
where $\phi^{\dagger} [\eta]$ is the creation operator for a membrane
$\eta \equiv \eta^a (x^{\mu})$.

\section{How many dimensions are there?}

In an important paper entitled {\it Space--time dimension from a
variational principle} \cite{104} D. Hochberg and J.T. Wheeler
generalize  the concept of coordinates and dimension. Instead of
coordinates $x^{\mu}$ with a discrete set of indices $\mu = 1,2,...,n$ they
take a set of coordinates
\be
        x^{\mu} \rightarrow x(z) ,
\lbl{IV.16}
\ee
where $z$ is a continuous real index. They also introduce a weight function
$\rho (z)$ and define the dimension $p$ as
\be
    p = \int \rho (z) \, \dd z .
\lbl{IV.17}
\ee
In general $\rho (z)$ is an arbitrary function of $z$. In particular, it
may favor a certain subset of $z$-values. For instance, when
\be
    \rho (z) = \sum_{\mu = 1}^n \delta (z - z_{\mu})
\lbl{IV.18}
\ee
the weight favors a set of discrete values $z_{\mu}$. We then have
\be
    p = \int \sum_{\mu=1}^n \delta (z - z_{\mu}) \, \dd z = n .
\lbl{IV.19}
\ee
A space of a discrete set of dimensions $\mu = 1,2,...,n$ is thus a particular 
case of a more general space which has a continuous set of dimensions $z$.

A problem then arises of what is $\rho (z)$. It should not be considered as a
fixed function, given once for all, but has to arise dynamically as a 
consequence of a chosen solution to certain dynamical equations.
Formulation of a theory in which dimensions are not {\it a priori} discrete and
their number fixed remains one of the open problems for the future.
A first step has been made in \cite{104}, but a generalization to membranes
is still lacking.

\section{Will it ever be possible to find solutions to the classical and
quantum brane equations of motion and make predictions?}

My answer is YES. Only it will be indispensable to employ the full
power of computers. Consider the equation of a minimal surface
\be
      {\DD}_{\mu} {\DD}^{\mu} \eta^a = 0.
\lbl{IV.20}
\ee
Choosing a gauge $\eta^{\mu} = x^{\mu}$ the above equation becomes
\be
      {\DD}_{\mu} {\DD}^{\mu} \eta^{\bar a} = 0 ,
\lbl{IV.21}
\ee
where $\eta^{\bar a} (x)$, ${\bar a} = n+1, n+2,...,N$, are the transverse
embedding functions.

What are solutions to the equation (\ref{IV.21})? We shall write a general
solution as
\be
     \eta^{\bar a} (x) = {\rm Surf}^{\bar a} (x^{\mu}, \alpha) ,
\lbl{IV.22}
\ee
where $\alpha$ is a set of parameters. But what is ${\rm Surf}^{\bar a}$?
It is {\it defined} as a general solution to the set of second order
non-linear differential equations (\ref{IV.21}). Yes, but what are its
numerical values, what are the graphs for various choices of parameters
$\alpha$\,? Well, ask a computer! A computer program is required which, at
a press of button will give you the required answer on the display. Fine,
but this is not an exact solution.

Consider again what you mean by an exact solution. Suppose we have studied a
dynamical system and found that its equation of motion is
\be
       {\ddot x} + \omega^2 x = 0 .
\lbl{IV.23}
\ee
What is the general solution to the second order differential equation 
(\ref{IV.23})? That is easy, it is
\be
    x(t) = A \, {\rm sin} \, \omega t + B \, {\rm cos} \, \omega t \equiv 
    {\rm Osc} (t,A,B,\omega) ,
\lbl{IV.24}
\ee
where $A$, $B$ are arbitrary parameters. Yes, but what are the numerical
values of ${\rm Osc}(t,A,B,\omega)$, what do the graphs for various
choices of $A$, $B$, $\omega$ look like? 
This is something we all know from school. We used
to look up the tables, but nowadays we obtain the answer easily by
computer or calculator. There is a number of algorithms which enable us 
to compute
${\rm sin}\, \omega t$ and ${\rm cos}\, \omega t$ to arbitrary precision. 
Similarly for solutions to other second order differential equations,
like ${\ddot x} - \omega^2 x = 0$, etc.\,. If so, why do we not develop
algorithms for computing ${\rm Surf}^{\bar a}$? Once the algorithm is installed
in the computer we could treat ${\rm Surf}^{\bar a}$ in an analogous
way as ${\rm Osc} (t,A,B,\omega)$ (i.e., as ${\rm sin}\, \omega t$ and
${\rm cos} \,\omega t$). Similarly we could develop algorithms for computing
numerical values and plotting the graphs corresponding to the
solutions of other differential
equations of motion describing various classical and quantum dynamical
systems of branes and their generalizations. Once we can simply visualize
a solution as a graph, easily obtained by means of a suitable computer
program, it is not so difficult to derive predictions of various competing
theories and to propose strategies for their experimental verification.

\section{Have we found a unifying principle?}

We have discussed numerous ideas, theories, techniques and approaches
related to fundamental theoretical physics. Can we claim that we
have found a unifying principle, according to which we could formulate
a final theory, incorporating all the fundamental interactions, including
gravity? No, we cannot. If anything, we have found a sort of meta-principle,
according to which, in principle, there is no limit to the process of
understanding the Nature. A final observer can never completely understand
and describe reality. Whenever we may temporarily have an impression of
understanding the relationship between gravity and quantum theory, some
new experiments and theoretical ideas will soon surpass it. In this
book I have aimed, amongst other things, to point out how rich and vast is
the arsenal of conceptual, theoretical, and technical possibilities. And
there seems to be no limit to the numerous ways in which our theoretical
constructions could be generalized and extended in a nontrivial and
elegant way. No doubt there is also no limit to surprises that cleverly
and ingeniously designed experiments can and will bring us. Theory and
experiment will feed each other and bring us towards greater and greater
progress. And this is a reason why we insist on our road of scientific
investigation, in spite of being aware that we shall never come to
the end, and never grasp the whole landscape. 

And yet, at the current stage of our progress a nice picture is starting
to emerge before our eyes. We have found that there is much more to
geometry than it is usually believed. By fully employing the powerful
language of Clifford algebra we have arrived at a very rich geometrical
structure which offers an elegant formulation of current 
fundamental theories in
a unified way, and also provides a natural generalization. I sincerely
hope that I have succeeded in showing the careful reader, if nothing more,
at least a glimpse of that magnificent picture without which, in my opinion,
there can be no full insight into contemporary fundamental theoretical
physics.

\appendix{The dilatationally invariant system of units}

\prologue{That an electron here has the same mass as an electron there
is
also a triviality or a miracle. It is a triviality in quantum 
electrodynamics because it is assumed rather than derived.
However, it is
a miracle on any view that regards the universe as being from
time to time
``reprocessed''.}
{Charles W. Misner, Kip S. Thorne and John Archibald
Wheeler\footnote{
See ref. \ci{106}} }

\normalsize
We shall show how all the equations of physics can be cast
in the system of units in which $\hbar = c = G = 4 \pi \epsilon_0
= 1$.
In spite of its usefulness for all sorts of calculations such
a sytem of units is completely unknown.

Many authors of modern theoretical works use the system of units
in which
either $\hbar = c = 1$ or $c = G = 1$, etc.\,. This significantly
simplifies
equations and calculations, since various inessential $\hbar^3$,
$c^2$, etc.,
are no longer present in formal expressions. But I have never
seen the use of the next
step, namely the units in which ``all" fundamental constant are 1,
that is $\hbar = c = G = 4 \pi \epsilon_0 = 1$. Let us call such
a
system {\it the dilatationally invariant system of units},
briefly,
the system D. It is introduced with the aid of the fine structure
constant $\alpha$, the Planck mass $M_{\rm P}$, the Planck time
$T_{\rm P}$ and the
Planck length $L_{\rm P}$ by setting $\hbar = c = G = 4 \pi
\epsilon_0 = 1$
in the usual MKSA expression for these quantities
(Table~\ref{T1}). That\inxx{system of units,dilatationally invariant}
is, in the system D all quantities are expressed relative to the
Planck\inxx{Planck units}
units, which are dimensionless; the unit is 1. For practical
reasons sometimes
we will formally add the symbol D: so there holds 1 = 1D.
With the aid of the formulas in Table {\ref{T1} we obtain the
relation between
the units MKSA and the units D (Table \ref{T2}).

\begin{table}[h]
\caption{Physical constants in two systems of units}
\begin{tabular*}{\textwidth}{@{\extracolsep{\fill}}lclc}
\hline
{\it Description}& {\it Symbol}& \hs{6mm} MKSA& D \cr
\hline
{\it Planck's constant/2$\pi$} & $\hbar$ & $1.0545887 \times
10^{-34}$ Js & 1 \cr
{\it Speed of light} & $c$ & $2.99792458 \times 10^8 {\rm m
s}^{-1}$ & 1 \cr
{\it Gravitational constant} & $G$ & $6.6720 \times 10^{-11} {\rm
kg}^{-1}
{\rm m}^3 {\rm s}^{-2}$ & 1 \cr
{\it Dielectric constant of vacuum} & $\epsilon_0$ & 
$8.8541876 \times 10^{-12} {\rm kg}^{-1} {\rm A}^2 {\rm s}^4 {\rm
m}^{-3}$ &
${1\oo {4 \pi}}$ \cr
{\it Induction constant of vacuum} & $\mu_0$ & $1.2566371 \times
10^{-6} {\rm kg}
\, {\rm m \, s}^{-2} {\rm A}^{-2}$ & $4 \pi$ \cr
{\it Electron's charge} & $e$ & $1.6021892 \times {\rm A} \, {\rm
s}$ &
$\alpha^{1/2}$ \cr
{\it Electron's mass} & $m_{\rm e}$ & $9.109534 \times 10^{-31}$
kg &
$\kappa_0 \alpha^{1/2}$ \cr
{\it Boltzman constant} &$k_{\rm B}$ & $1.380622 \times 10^{-23}$
J/$^o$K & 1 \cr 
\hline
\end{tabular*}
\begin{tabular*}{\textwidth}{@{\extracolsep{\fill}}llll}
$\hs{2cm}$ & {\it Fine structure constant} & $\alpha = 
1/137.03604$ & 
$\qquad \hs{2cm}$ \cr
$\hs{2cm}$ & {\it `Fundamental scale'} & $\kappa_0 = 0.489800
\times 10^{-21}$ &
$\qquad \hs{2cm}$ \\
\hline
\end{tabular*}
\begin{tabular*}{\textwidth}{@{\extracolsep{\fill}}llllll}
$\hs{5mm}$ & $e = \alpha^{1/2} (4 \pi \epsilon_0 \hbar c)^{1/2}$
&
$M_{\rm P} = (\hbar c/G)^{1/2}$ & $T_{\rm P} = (\hbar
G/c^5)^{1/2}$ &
$L_{\rm P} = (\hbar G/c^3)^{1/2}$ & \hs{5mm} \\
\hline
\end{tabular*}
\label{T1}
\end{table}

Let us now investigate more closely the system $D$. The Planck
length is just
the Compton wavelength of a particle with the Planck mass $M_{\rm
P}$
\be
    L_{\rm P} = {\hbar \oo {M_{\rm P} c}} \; ({\rm system \;
MKSA}), \qquad 
    L_{\rm P} = {1 \oo M_{\rm P}} = M_{\rm P} = 1 \; ({\rm system
\; D})
\lbl{Ap1}
\ee
In addition to $M_{\rm P}$ and $L_{\rm P}$ we can introduce
\be
     M = \alpha^{1/2} M_{\rm P} \; , \qquad L = \alpha^{1/2}
L_{\rm P} .
\lbl{Ap2}
\ee
In the system D it is
\be
     e = M = L = \alpha^{1/2} .
\lbl{Ap3}
\ee
The length $L$ is the classical radius that a particle with mass
$M$ and charge $e$ would have:
\be
     L = {e^2 \oo {4 \pi \epsilon_0 M c^2}} \; ({\rm system \;
MKSA}),
     \qquad L = {e^2 \oo M} \; ({\rm system \; D}) .
\lbl{Ap4}
\ee
The classical radius of electron (a particle with the mass
$m_{\rm e}$ and
the charge $e$) is
\be
     r_c = {e^2 \oo {4 \pi \epsilon_0 m_{\rm e} c^2}} \; ({\rm
system \; MKSA}),
     \qquad r_c = {e^2 \oo m_{\rm e}} \; ({\rm system \; D}).
\lbl{Ap5}
\ee
From (\ref{Ap4}) and (\ref{Ap5}) we have in consequence of
(\ref{Ap1})
and Table \ref{T1}
\be
    {r_{\rm c} \oo L} = {M\oo m_{\rm e}} = {e\oo m_{\rm e}} (4
\pi \epsilon_0 G)^{-1/2}
    \equiv \kappa_0^{-1}
\lbl{Ap6}
\ee
Therefore
\be
     m_{\rm e} = \kappa_0 M = \kappa_0 \alpha^{1/2} = \kappa_0 e
\;
     ({\rm system \; D}).
\lbl{Ap7}
\ee
The ratio $L/r_{\rm c}$ represents the scale of the electron's
classical 
radius relative to the length $L$. As a consequence of
(\ref{Ap4})--(\ref{Ap7}) we have $r_{\rm c} = \kappa_0^{-1} e$.

\begin{table}[h]
\caption{Translation between units D and units MKSA}

\begin{tabular*}{\textwidth}{@{\extracolsep{\fill}}l}
\\
\hs{2.5cm} 1D $= (\hbar c/G)^{1/2} = 2.1768269 \times 10^{-8}$ kg
\\
\hs{2.5cm} 1D $= (\hbar G/c^5)^{1/2} = 5.3903605 \times 10^{-44}$
s \\
\hs{2.5cm} 1D $= (\hbar G/c^3)^{1/2} = 1.6159894 \times 10^{-35}$
m \\
\hs{2.5cm} 1D $= (4 \pi \epsilon_0 \hbar c)^{1/2} = \alpha^{-1/2}
e
= 1.8755619 \times 10^{-18}$ As \\
\hs{2.5cm} 1D $= c^3 (4 \pi \epsilon_0/G)^{1/2} = 3.4794723 \times
10^{25}$ A \\
\hs{2.5cm} 1D $= c^2 (4 \pi \epsilon_0 G)^{-1/2} = 1.0431195
\times
10^{27}$ V \\
\hs{2.5cm} 1D $= c^2 (\hbar c/G)^{1/2} = 1.9564344 \times 10^{9}$
J \\
\hs{2.5cm} 1D $= 1.41702 \times 10^{32} {\ }^0$K \\
\end{tabular*}
\label{T2}
\end{table}

At this point let us observe that the fundamental constants
$\hbar$, $c$,
$G$ and $\epsilon$ as well as the quantities $L_{\rm P}$, $M_{\rm
P}$, $T_{\rm P}$, $L$, $M$,
$e$, are by definition invariant under dilatations. The effect of
a
dilatation on various physical quantities, such as the spacetime
coordinates
$x^{\mu}$, mass $m$, 4-momentum $p_{\mu}$, 4-force $f^{\mu}$, and
4-acceleration $a^{\mu}$ is \ci{107}--\ci{109}:
\bear
       x^{\mu} &\rightarrow& x'^{\mu} = \rho x^{\mu} ,\nonumber
\\
       p_{\mu} &\rightarrow& p'_{\mu} = \rho^{-1} p_{\mu},
\nonumber \\
       m  &\rightarrow& m' = \rho^{-1} m ,\nonumber \\
       f^{\mu} &\rightarrow& f'^{\mu} = \rho^{-2} f^{\mu},
\nonumber \\
       a^{\mu} &\rightarrow&  a'^{\mu} = \rho^{-1} a^{\mu} .
\lbl{Ap8}
\ear

Instead of the {\it inhomogeneous coordinates} $x^{\mu}$ one can
introduce 
\ci{107}--\ci{109}
the {\it homogeneous coordinates} ${\tilde x}^{\mu} = \kappa
x^{\mu}$ which
are invariant under dilatations provided that the quantity
$\kappa$
transforms as
\be
     \kappa \rightarrow \kappa' = \rho^{-1} \kappa .
\lbl{Ap9}
\ee
For instance, if initially $x^0 = 1$ sec, then after applying a
dilatation,
say by the factor $\rho = 3$, we have $x'^0 = 3$ sec, $\kappa =
{1\oo 3}$,
${\tilde x}'^0 = {\tilde x}^0 = 1$ sec. The quantity $\kappa$ is
the {\it scale} of the quantity $x^{\mu}$ relative to the
corresponding
invariant quantity ${\tilde x}^{\mu}$.

If we write a given equation we can check its consistency by
comparing
the dimension of its left hand and right hand side. In the MKSA
system
the dimensional control is in checking the powers of meters,
kilograms,
seconds and amp\` eres on both sides of the equation. In the
system D one has
to verify that both sides transform under dilatations as the same
power
of $\rho$. For instance, eq.\,(\ref{Ap7}) is consistent, since
$[m_{\rm e}] = \rho^{-1}$, $[\kappa_0] = \rho^{-1}$ and $[e] =
1$, where
$[A]$ denotes the dimension of a generic quantity $A$.

\begin{table}[h]
\caption{Some basic equations in the two systems of units}
\begin{tabular*}{\textwidth}{@{\extracolsep{\fill}}lccc}
\hline
{\it Description} & {\it Symbol} & MKSA & D \\
\hline
\begin{minipage} [t] {5cm}
     \vspace*{5mm}
     Electric force between two electrons 
     \end{minipage} &
\begin{minipage} [t] {.3cm}     
     \vspace*{5mm}
     $F_e$ \end{minipage} & 
\begin{minipage} [t] {1.5cm}
      $${e^2 \oo {4 \pi \epsilon_0 r^2}}$$ \end{minipage} & 
\begin{minipage} [t] {1.5cm}
      $${e^2 \oo r^2}$$ \end{minipage} \\   
\begin{minipage} [t] {5cm}
     \vspace*{5mm}
     Electric force between two electrons at the distance $r_c$
     \end{minipage} &
\begin{minipage} [t] {.3cm}     
     \vspace*{5mm}
     $F_e$ \end{minipage} & 
\begin{minipage} [t] {1.5cm}
      $${e^2 \oo {4 \pi \epsilon_0 r_c^2}}$$ \end{minipage} &
\begin{minipage} [t] {1.5cm}
      $$\kappa_0^2$$ \end{minipage} \\        
\begin{minipage} [t] {5cm}
     \vspace*{5mm}
     Gravitational force between two electrons at the distance $r_c$
     \end{minipage} &
\begin{minipage} [t] {4mm}     
     \vspace*{5mm}
     $F_G$ \end{minipage} &
\begin{minipage} [t] {1.5cm}
      $${{\kappa^{-2} G m_e^2}\oo r_c^2}$$ \end{minipage} &
\begin{minipage} [t] {1.5cm}
      $$\kappa^{-2} \kappa_0^4$$ \end{minipage} \\ 
\begin{minipage} [t] {5cm}
     \vspace*{5mm}
     Ratio between electric and gravitational force
     \end{minipage} &
\begin{minipage} [t] {4mm}     
     $${F_e\oo F_G}$$ \end{minipage} &
\begin{minipage} [t] {1.5cm}
      $${e^2 \oo {4 \pi \epsilon_0 G m_e^2}}$$ \end{minipage} &
\begin{minipage} [t] {1.5cm}
      $$\kappa^{2} \kappa_0^{-2}$$ \end{minipage} \\
\begin{minipage} [t] {5cm}
     \vspace*{5mm}
     Bohr radius
     \end{minipage} &
\begin{minipage} [t] {4mm}     
     \vspace*{5mm}
     $a_0$ \end{minipage} &
\begin{minipage} [t] {1.5cm}
      $${{4 \pi \epsilon_0 \hbar^2}\oo {m_e e^2}}$$ \end{minipage} &
\begin{minipage} [t] {1.5cm}
      $$\kappa_0^{-1} e^{-3}$$ \end{minipage} \\                
\begin{minipage} [t] {5cm}
     \vspace*{5mm}
     Potential energy of electron at the distance $a_0$ from the centre
     \end{minipage} &
\begin{minipage} [t] {4mm}     
     \vspace*{5mm}
     $E_c$ \end{minipage} &
\begin{minipage} [t] {1.5cm}
      $${e^2 \oo {4 \pi \epsilon_0 a_0}}$$ \end{minipage} &
\begin{minipage} [t] {1.5cm}
      $$\kappa_0 e^5$$ \end{minipage} \\
\begin{minipage} [t] {5cm}
     \vspace*{5mm}
     Rydberg constant
     \end{minipage} &
\begin{minipage} [t] {4mm}     
     \vspace*{5mm}
     $Ry$ \end{minipage} &
\begin{minipage} [t] {1.5cm}
      $${{m_e e^4}\oo {2 (4 \pi \epsilon_0 \hbar )^2}}$$ \end{minipage} & 
\begin{minipage} [t] {1.5cm}
      $${1\oo 2} \kappa_0 e^5$$ \end{minipage} \\                                      
\hline
\end{tabular*}
\label{T3}
\end{table}

In Table \ref{T3} some well known equations are written in both
systems of units.
They are all covariant under dilatations. Taking $G$ invariant, 
the equation $F = Gm^2/r^2 $ is
not dilatationally covariant, as one can directly check from
(\ref{Ap8}).
The same is true for the Einstein equations $G_{\mu \nu} = - 8
\pi G T^{\mu
\nu}$ with $T^{\mu \nu} = (\rho + p) u^{\mu} u^{\nu} - p\, 
g^{\mu \nu}$,
from which the Newtonian gravitation equation is derivable.
Usually this
non-covariance is interpreted as the fact that the gravitational
coupling
constant $G$ is not dimensionless.  One can avoid this difficulty
by using
the homogeneous coordinates ${\tilde x}^{\mu}$ and express the
Einstein
tensor $G^{\mu \nu}$, the rest mass density $\rho$ and all other
relevant
quantities in terms of these homogeneous coordinates
\ci{109,110}. 
Then the Einstein
equations become ${\widetilde G}^{\mu \nu} = - 8 \pi G 
{\widetilde T}^{\mu \nu}$ with ${\widetilde T}^{\mu \nu} =
({\tilde \rho} + {\tilde p}) 
{\tilde u}^{\mu} {\tilde u}^{\nu} - {\tilde p}\,  {\tilde g}^{\mu
\nu}$,
where the quantities with tildes are invariant
under dilatations.
If the homogeneous Einstein equations are written back in terms
of
the inhomogeneous quantities, we have $G^{\mu \nu} = - 8 \pi G
\kappa^{-2}
T^{\mu \nu}$, which is covariant with respect to dilatations. If
we choose
$\kappa = 1$ then the equations for this particular choice
correspond
to the usual Einstein equations, and we may use either the MKSA
system or
the D system. Further discussion of this interesting and
important
subject would go beyond the scope of this book. More about the
dilatationally and conformally covariant theories the reader will
find
in refs.\,\ci{107}--\ci{111}.

Using the relations of Table \ref{T2} all equations in the 
D system can be
transformed back into the MKSA system. Suppose we have an
equation
in the D system:
\be
    a_0 = {1\oo {m_{\rm e} e^2}} = \kappa_0^{-1} e^{-3} =
\kappa_0^{-1}
    \alpha^{-3/2} = \kappa_0^{-1} \alpha^{-3/2} \, {\rm D} . 
\lbl{Ap10}
\ee
We wish to know what form the latter equation assumes in the
MKSA system.
Using the expression for the fine structure constant
$\alpha = e^2 (4 \pi \epsilon_0 \hbar c)^{-1}$ and
eq.\,(\ref{Ap6}) we have
\be
    a_0 = {e\oo {m_{\rm e} (4 \pi \epsilon_0 G)^{1/2}}} \left
({e^2\oo {4 \pi 
    \epsilon_0 \hbar c}} \right )^{-3/2} {\rm D} .
\lbl{Ap11}
\ee
If we put 1D = 1 then
the right hand side of the latter equation is a dimensionless
quantity.
If we wish to obtain a quantity of the dimension of length we
have
to insert 1D $= (\hbar G/c^3)^{1/2}$, which represents
translation from
meters to the D units. So we obtain
\be
     a_0 = {{4 \pi \epsilon_0 \hbar^2} \oo {m_{\rm e} e^2}} ,
\lbl{Ap11a}
\ee
which is the expression for Bohr's radius.

Instead of rewriting equations from the D system in the MKSA
system, we
can retain equations in the D system and perform all the
algebraic
and numerical calculations in the D units. If we wish to know the
numerical
results in terms of the MKSA units, we can use the numbers of
Table~\ref{T2}.
For example,
\be
    a_0 = \kappa_0^{-1} e^{-3} = 2.04136 \times 10^{21} \times
    137.03604^{3/2} {\rm D} .
\lbl{Ap12}
\ee
How much is this in meters? From Table \ref{T2} we read 1 D $=
1.615989 \time
10^{-35}$~m. Inserting this into (\ref{Ap12}) we have $a_0 =
0.529177
\times 10^{-10}$m which is indeed the value of Bohr's radius.

Equations in the D system are very simple in comparison with those in
the
MKSA system. Algebraic calculations are much easier, since there
are 
no inessential factors like $\hbar^2$, $c^3$, etc., which obsure
legibility and
clarity of equations. The transformation into the familiar MKSA
units
is quick with the aid of Table \ref{T2} (and modern pocket
calculators, unknown in
the older times from which we inherit the major part of present
day
physics). However, I do not propose to replace the international
MKSA system
with the D system. I only wish to recall that most modern
theoretical works
do not use the MKSA system and that it is often very tedious to
obtain the results in meters, seconds, kilograms and amp\` eres.
What I wish
to point out here is that even when the authors are using the
units 
in which, for example, $\hbar = c =1$, or similar, we can easily
transform
their equations  into the units in which $\hbar=c=G=4 \pi
\epsilon_0
=1$ and use Table \ref{T2} to obtain the numerical results in the
MKSA system.

To sum up, besides the Planck length, Planck time and Planck
mass, which
are composed of the fundamental constants $\hbar$, $c$ and $G$,
we have also introduced (see Table \ref{T2}) the corresponding 
electromagnetic quantity,
namely the charge $E_{\rm P} = (4 \pi \epsilon_0 \hbar c)^{1/2}$
(or, equivalently, the current and the potential difference),
by bringing into play the  fundamental constant
$\epsilon_0$.
We have then extended {\it the Planck system } of units \ci{106}
in
which $c=\hbar=G=1$ to the system of units in which $c=\hbar=G=4
\pi
\epsilon_0=1$, in order to
incorporate all known sorts of physical quantities.

Finally, let me quote the beautiful paper by Levy-Leblond
\ci{112} 
in which it is
clearly stated that our progress in understanding the unity of
nature follows
the direction of eliminating from theories various (inessential)
numerical constants with the improper name of ``fundamental"
constants. In
fact, those constants are merely the constants which result from
our
unnatural choice of units, the choice due to our incomplete
understanding
of the unified theory behind.

\kluwerprintindex

\end{document}